\documentclass[11pt,a4paper,twoside,openany]{book}

\input{glyphtounicode}
\usepackage[T1]{fontenc}
\usepackage[utf8]{inputenc}
\usepackage{textcomp}
\usepackage{mathpazo}                 
\usepackage[scaled=0.90]{helvet}      

\usepackage{amssymb}
\usepackage{pmboxdraw}                
\usepackage{microtype}
\DeclareUnicodeCharacter{2192}{\ensuremath{\rightarrow}}
\DeclareUnicodeCharacter{2193}{\ensuremath{\downarrow}}
\DeclareUnicodeCharacter{25B6}{\ensuremath{\blacktriangleright}}
\DeclareUnicodeCharacter{25C0}{\ensuremath{\blacktriangleleft}}
\DeclareUnicodeCharacter{25BC}{\ensuremath{\blacktriangledown}}
\DeclareUnicodeCharacter{2713}{\ensuremath{\checkmark}}
\DeclareUnicodeCharacter{2717}{\ensuremath{\times}}
\DeclareUnicodeCharacter{23F8}{\textbf{||}}
\DeclareUnicodeCharacter{00B7}{\textperiodcentered}
\DeclareUnicodeCharacter{00A7}{\S}
\DeclareUnicodeCharacter{2020}{\textdagger}   

\usepackage[a4paper,inner=32mm,outer=28mm,top=25mm,bottom=30mm,
            headsep=7mm,footskip=12mm]{geometry}

\usepackage{xcolor}
\definecolor{deepindigo}{HTML}{0F1B33}   
\definecolor{indigo}{HTML}{1E2E56}       
\definecolor{brass}{HTML}{C9A227}        
\definecolor{brassdark}{HTML}{8C5A16}    
\definecolor{cream}{HTML}{F4F1E8}        
\definecolor{bodyink}{HTML}{16202B}      
\definecolor{paleindigo}{HTML}{DFE5F0}   
\definecolor{palebrass}{HTML}{F6E8CC}    
\definecolor{palegrey}{HTML}{ECEFF5}     
\definecolor{quoterule}{HTML}{6E7A8B}    
\definecolor{inkgrey}{HTML}{5A6472}      
\definecolor{codebg}{HTML}{EAEDF3}       

\usepackage{titlesec}
\usepackage{fancyhdr}
\usepackage{enumitem}
\usepackage{longtable}
\usepackage{array}
\usepackage{booktabs}
\usepackage{ragged2e}
\usepackage{needspace}
\usepackage{graphicx}
\usepackage{tikz}
\usepackage[most]{tcolorbox}
\usepackage{eso-pic}
\usepackage{placeins}
\usepackage{multicol}

\newenvironment{deflist}{%
  \begin{list}{}{%
    \setlength{\leftmargin}{6mm}\setlength{\itemindent}{-6mm}%
    \setlength{\listparindent}{0pt}\setlength{\labelwidth}{0pt}%
    \setlength{\labelsep}{0pt}\setlength{\topsep}{2mm}%
    \setlength{\itemsep}{0.9mm}\setlength{\parsep}{0pt}}}%
  {\end{list}}

\usepackage[hidelinks,bookmarksnumbered,pdfusetitle]{hyperref}
\expandafter\def\expandafter\UrlBreaks\expandafter{\UrlBreaks\do\-}
\hypersetup{
  pdfsubject={A verified practitioner's reference for agentic work. Version 1.1. Verification baseline: Claude Code 2.1.322, 12 Nov. 2026.},
  pdfkeywords={agentic artificial intelligence, AI-assisted software engineering,
    Claude Code, large language model agents, Model Context Protocol, prompt injection,
    zero trust, software supply-chain security, human oversight, AI governance,
    verification and validation}}
\usepackage{bookmark}

\newcommand{\handbookfoot}{%
  \fancyfoot[L]{\fontsize{7.5}{9}\selectfont\sffamily\color{inkgrey}\handbookfootertext}%
  \fancyfoot[R]{\fontsize{8.5}{10}\selectfont\sffamily\bfseries\color{deepindigo}\thepage}}
\handbookfoot

\fancypagestyle{plain}{\fancyhf{}\handbookfoot}

\newcommand{\rulestack}[1][brass]{%
  \noindent
  \begin{tikzpicture}[baseline=0pt]
    \fill[#1]        (0,0)     rectangle (74mm,1.5mm);
    \fill[#1!78!white] (0,-3.6mm) rectangle (52mm,-2.1mm);
    \fill[#1!56!white] (0,-7.2mm) rectangle (34mm,-5.7mm);
    \fill[#1!36!white] (0,-10.8mm) rectangle (20mm,-9.3mm);
  \end{tikzpicture}}

\newcommand{\partbackground}{%
  \AddToShipoutPictureBG*{%
    \AtPageLowerLeft{\tikz[remember picture,overlay]{%
      \fill[left color=deepindigo,right color=indigo]
        (0,0) rectangle (\paperwidth,\paperheight);}}}}

\newcommand{\parttitle}[2]{%
  \FloatBarrier\clearpage
  \thispagestyle{empty}%
  \partbackground
  \begingroup
  \sffamily\color{cream}
  \vspace*{0.62\textheight}
  \noindent\rulestack\par
  \vspace{7mm}
  {\fontsize{10}{12}\selectfont\bfseries\color{brass}\MakeUppercase{#1}\par}
  \vspace{5mm}
  {\fontsize{27}{31}\selectfont\bfseries\raggedright #2\par}
  \endgroup
  \clearpage
  \markboth{#1\ \ \textperiodcentered\ \ #2}{#1\ \ \textperiodcentered\ \ #2}}

\titleformat{\chapter}[display]
  {\sffamily\color{deepindigo}}
  {\vspace*{0.30\textheight}%
   \parbox{\textwidth}{\rulestack\\[6mm]%
     \fontsize{9.6}{12}\selectfont\bfseries\color{brassdark}%
     \MakeUppercase{\chaptertitlelabel}}}
  {4mm}
  {\fontsize{25}{29}\selectfont\bfseries\raggedright}
  [\vspace*{\fill}\clearpage]
\titlespacing*{\chapter}{0pt}{0pt}{0pt}
\newcommand{\chaptertitlelabel}{}

\newcommand{\frontchapter}[1]{%
  \FloatBarrier\clearpage
  \thispagestyle{plain}%
  \noindent\rulestack\par
  \vspace{7mm}
  {\sffamily\fontsize{21}{25}\selectfont\bfseries\color{deepindigo}\raggedright #1\par}
  \vspace{7mm}
  \addcontentsline{toc}{chapter}{#1}%
  \markboth{#1}{#1}}

\makeatletter
\newcommand{\handbooktoc}{%
  \FloatBarrier\clearpage
  \thispagestyle{plain}%
  \noindent\rulestack\par
  \vspace{7mm}
  {\sffamily\fontsize{21}{25}\selectfont\bfseries\color{deepindigo}\raggedright
    \contentsname\par}
  \vspace{7mm}
  \markboth{\contentsname}{\contentsname}%
  \@starttoc{toc}}
\makeatother

\newcommand{\appendixchapter}[2]{%
  \FloatBarrier\clearpage
  \thispagestyle{plain}%
  \noindent\rulestack\par
  \vspace{6mm}
  {\sffamily\fontsize{9.6}{12}\selectfont\bfseries\color{brassdark}%
    \MakeUppercase{Appendix #1}\par}
  \vspace{3.4mm}
  {\sffamily\fontsize{21}{25}\selectfont\bfseries\color{deepindigo}\raggedright #2\par}
  \vspace{7mm}
  \addcontentsline{toc}{chapter}{\protect\numberline{#1}#2}%
  \markboth{Appendix #1\ \ \textperiodcentered\ \ #2}{Appendix #1\ \ \textperiodcentered\ \ #2}}

\titleformat{\section}{\sffamily\large\bfseries\color{indigo}}{\thesection}{0.6em}{}
\titlespacing*{\section}{0pt}{5.2mm plus 1mm minus .6mm}{1.9mm}
\titleformat{\subsection}{\sffamily\normalsize\bfseries\color{indigo}}{}{0em}{}
\titlespacing*{\subsection}{0pt}{4.2mm plus .8mm minus .5mm}{1.4mm}

\usepackage{titletoc}
\titlecontents{part}[0pt]
  {\addvspace{6.5mm}\sffamily\fontsize{10.5}{13}\selectfont\bfseries\color{indigo}}
  {}{}{\hfill}
  [\vspace{0.6mm}{\color{indigo}\titlerule[0.6pt]}\vspace{1.2mm}]
\titlecontents{chapter}[3.4em]
  {\addvspace{2.2mm}\sffamily\fontsize{10}{12.5}\selectfont\bfseries\color{deepindigo}}
  {\contentslabel{3.4em}}{\hspace*{-3.4em}}
  {\hfill\contentspage}
\titlecontents{section}[6.6em]
  {\addvspace{0.15mm}\fontsize{9}{11.2}\selectfont\color{inkgrey}}
  {\contentslabel{3.2em}}{}
  {\,{\color{inkgrey!55}\titlerule*[0.55pc]{.}}\,\contentspage}

\setlist[itemize]{leftmargin=5.5mm,itemsep=0.5mm,topsep=1.4mm,parsep=0pt}
\setlist[enumerate]{leftmargin=6.5mm,itemsep=0.5mm,topsep=1.4mm,parsep=0pt}

\newtcolorbox{codeblock}[2]{breakable, enhanced, sharp corners, boxrule=0pt,
  colback=codebg, colframe=codebg,
  borderline west={0.7mm}{0pt}{inkgrey},
  left=3.4mm, right=1.5mm, top=2.2mm, bottom=2.2mm,
  before skip=2.8mm, after skip=2.8mm, parbox=false,
  fontupper=\fontsize{#1}{#2}\selectfont\ttfamily}

\newtcolorbox{codefig}[2]{breakable=false, enhanced, sharp corners, boxrule=0pt,
  colback=codebg, colframe=codebg,
  borderline west={0.7mm}{0pt}{inkgrey},
  left=3.4mm, right=1.5mm, top=2.2mm, bottom=2.2mm,
  before skip=0mm, after skip=0mm, parbox=false,
  fontupper=\fontsize{#1}{#2}\selectfont\ttfamily}
\newcommand{\cl}[1]{\noindent\mbox{#1}\par}

\newtcolorbox{calloutbox}[3]{breakable, enhanced, sharp corners, boxrule=0pt,
  colback=#1, colframe=#1, borderline west={#3}{0pt}{#2},
  left=3.4mm, right=3mm, top=2.4mm, bottom=2.4mm,
  before skip=3mm, after skip=3mm, fontupper=\small}

\newcommand{\tabcaption}[1]{\par\addvspace{2.6mm}\needspace{4\baselineskip}%
  {\sffamily\footnotesize\bfseries\color{deepindigo}#1\par}\addvspace{1.2mm}}
\newcommand{\boxcaption}[1]{%
  {\sffamily\footnotesize\bfseries\color{deepindigo}#1\par}\addvspace{1.4mm}}
\newcommand{\figcaption}[1]{\par\addvspace{2.6mm}%
  {\sffamily\footnotesize\color{deepindigo}\RaggedRight #1\par}}

\newcommand{\breakrule}{\par\addvspace{3mm}\noindent
  \hspace*{\fill}{\color{brass}\rule{18mm}{0.5mm}}\hspace*{\fill}\par\addvspace{3mm}}

\newsavebox{\tblbox}
\newsavebox{\kpbox}
\newcolumntype{L}[1]{>{\RaggedRight\arraybackslash\hyphenpenalty=2500\relax}p{#1}}
\newcolumntype{H}[1]{>{\RaggedRight\arraybackslash\hyphenpenalty=150\relax}p{#1}}

\makeatletter
\def\cleardoublepage{\clearpage\if@twoside\ifodd\c@page\else
  \null\thispagestyle{empty}\newpage\if@twocolumn\null\newpage\fi\fi\fi}
\renewcommand\mainmatter{\clearpage\@mainmattertrue\pagenumbering{arabic}}
\makeatother

\AtBeginDocument{\color{bodyink}}

\newenvironment{refslist}{\begin{list}{}{\leftmargin=8mm\labelwidth=6mm\labelsep=2mm\itemindent=-8mm\listparindent=0pt\itemsep=1.1mm\parsep=0pt\topsep=2mm}}{\end{list}}
\newcommand{\handbookfootertext}{Claude Code Complete User Handbook\ \ \textperiodcentered\ \ David Soldani}
\newenvironment{handbookimprint}{\par\begingroup\setlength{\parindent}{0pt}\sffamily}{\endgroup\par}
\title{Claude Code Complete User Handbook}
\author{David Soldani}
\date{12 November 2026}
\begin{document}
\frontmatter
\begingroup\let\cleardoublepage\clearpage
\begin{titlepage}
\thispagestyle{empty}
\begin{tikzpicture}[remember picture,overlay]
  \node[anchor=east,inner sep=0pt] at (current page.east)
    {\includegraphics[width=\dimexpr\paperwidth-20mm\relax,keepaspectratio]{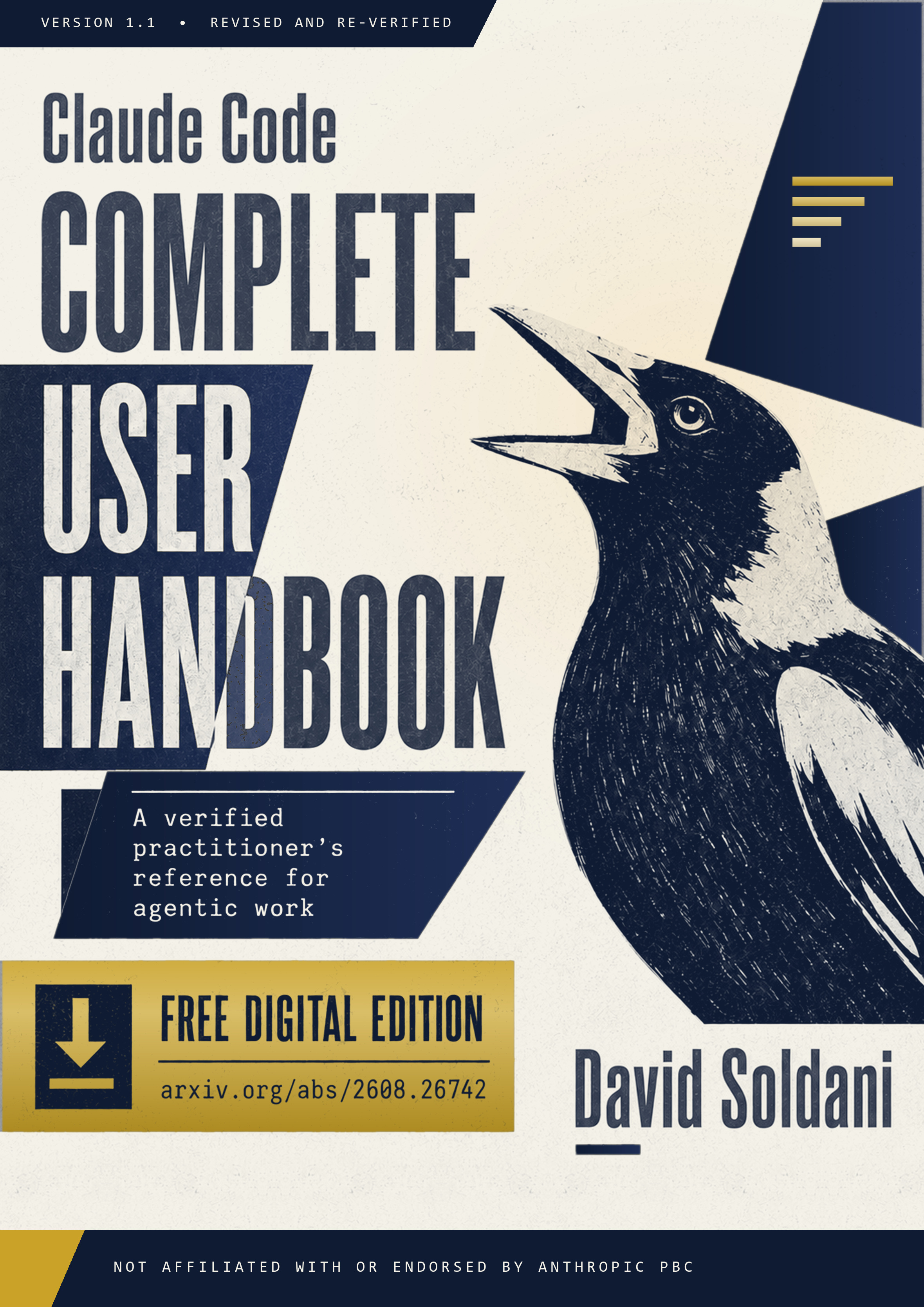}};
  \node[anchor=south west,inner sep=0pt,text opacity=0,text width=135mm,align=left]
    at ([xshift=47mm,yshift=18mm]current page.south west)
    {\sffamily\fontsize{6}{7}\selectfont Claude Code Complete User Handbook · David Soldani · Version 1.1 · Revised and re-verified · Free digital edition · Not affiliated with or endorsed by Anthropic PBC};
\end{tikzpicture}
\end{titlepage}
\endgroup
\clearpage
\pagestyle{fancy}
\handbooktoc
\clearpage
\begin{handbookimprint}
{\fontsize{15}{19}\selectfont\bfseries\color{deepindigo}\raggedright Claude Code Complete User Handbook\par}
\vspace{1.5mm}
{\fontsize{10.5}{14}\selectfont\color{inkgrey}\raggedright A verified practitioner's reference for agentic work\par}
\vspace{4mm}
{\fontsize{11}{14}\color{bodyink}\selectfont\bfseries David Soldani\par}
\vspace{1.5mm}
{\fontsize{9}{12}\selectfont\color{inkgrey}Version 1.1\ \ \textperiodcentered\ \ Verified 12 November 2026\ \ \textperiodcentered\ \ Claude Code 2.1.322\par}
\vspace{3mm}{\color{brassdark}\rule{\textwidth}{0.4pt}}\par\vspace{6mm}
\end{handbookimprint}

\textbf{Verification baseline:} Claude Code 2.1.322, 12 November 2026. Every statement of product behaviour in this book was adjudicated against the raw text of Anthropic's official documentation fetched on that date. The published documentation index [1] was enumerated: 203 pages, 41 added and 29 removed since Version 1.0. Its pages were fetched and hashed, one excepted, including all 92 this book cites. One cited page, the weekly digest, has left the index and was still being served. All thirty-four chapters, Part 0 and the appendices were re-adjudicated in full against those pages, and the effort vocabulary, the \texttt{-\allowbreak{}-\allowbreak{}permission-\allowbreak{}mode} validator and the \texttt{claude -\allowbreak{}-\allowbreak{}help} surface were re-established against a running installation. Of the 107 unique URLs in the text, 105 resolved; of the other two, one answered with an authentication challenge and one refused an automated request. The principal corrections are listed in Appendix J, and the statements Version 1.0 printed wrongly in “Changes since Version 1.0”. The adjudication covers what this book says about Claude Code. The other sources it cites — standards, regulations and papers, and the crosswalk built on them — were adjudicated for Version 1.0 on 23 August 2026. On 14 November 2026 each cited standard, regulation and paper was re-checked against its issuer for identity, edition and currency; what each requires, and the crosswalk's mapping to it, was not re-adjudicated.

\textbf{Version 1.0} was verified against Claude Code 2.1.241 on 23 August 2026, and ten of its most heavily cited pages were re-adjudicated on 26 August 2026 against 2.1.246.

Interfaces, limits, plan entitlements and preview features change on a weekly cadence; Appendix K states the re-verification procedure and should be run before this book is relied upon for a consequential decision.

\breakrule

\FloatBarrier
\section*{Abstract}
\markboth{Abstract}{Abstract}

Claude Code is an agentic work environment: a language model operating in a loop with filesystem access, shell execution, browser control, scheduled and cloud execution, external tool connections through the Model Context Protocol, and multi-agent orchestration [2], [3]. Its capability envelope now exceeds what a single practitioner can supervise by attention alone, and its failure modes are systemic rather than local: an unreviewed hook, an over-scoped connector, a stale completion condition, an autonomous routine inheriting every credential on an account.

This book is a task-oriented reference for operating that system safely and productively, written for practitioners who are accountable for the result. It advances four propositions. First, that capability without a defined and observable completion condition is not productivity. Second, that instruction, permission enforcement, sandboxing and operating-system isolation are four distinct layers of a control stack, the first advisory and the other three enforced, and that conflating them is the most common cause of loss of control. Third, that third-party skills, plugins, marketplaces, channels and MCP servers are software supply-chain dependencies and must be governed as such. Fourth, that the correct unit of trust in agentic work is observed evidence, not an agent's closing statement.

Thirty-four chapters take the reader from installation to a fully verified capstone, with a governance part addressing managed policy, data residency and retention, observability and accessibility. Every product claim carries a citation to a primary source, and the evidence ledger of Appendix J records the adjudication wherever a claim in circulation was found wrong, what this edition changed, and what remains unverified; controls are mapped to seventeen external frameworks in the crosswalk of Appendix H; and an organisational adoption maturity model is proposed in Appendix L. Claims that could not be confirmed from primary sources are labelled UNVERIFIED rather than softened.

\textbf{Keywords:} agentic artificial intelligence, AI-assisted software engineering, Claude Code, large language model agents, Model Context Protocol, prompt injection, zero trust, software supply-chain security, human oversight, AI governance, verification and validation.

\FloatBarrier
\section*{Contributions of this work}
\markboth{Contributions of this work}{Contributions of this work}

The product documentation this book cites is authoritative and extensive. This book exists because authoritative documentation answers \emph{what a feature does} and leaves four questions unanswered for anyone accountable for the outcome: what may be trusted, what must be enforced, what evidence would demonstrate correctness, and what an organisation must be able to show. Its original contributions are:

\begin{enumerate}
\item \textbf{The four-element brief} — Outcome, Context, Boundaries, Evidence — proposed in Chapter 1 as the invariant structure of an agentic instruction, and threaded through the plan contract (Chapter 6), the delegation brief (Chapter 16), the completion condition (Chapter 21) and the release gate (Chapter 34).
\item \textbf{The agentic control stack} (Figure 1) — a four-layer model separating advisory guidance from enforced control, which resolves the most persistent category error in current practice: the belief that an instruction in a project file constrains an agent.
\item \textbf{Evidence-Gated Delivery} (Chapter 34) — an eight-stage method with named gates for producing an agentic deliverable that can be released on evidence rather than assertion.
\item \textbf{A verification methodology for fast-decaying technical subjects} (“Scope and method”) — raw-source adjudication, a machine-checked citation and cross-reference apparatus, an explicit UNVERIFIED class, and a published re-verification procedure (Appendix K).
\item \textbf{A threat model and control map} (Appendix F) for agentic development environments, mapped to OWASP LLM Top 10 and MITRE ATLAS [4], [5].
\item \textbf{A standards and regulatory crosswalk} (Appendix H) mapping the book's controls to NIST AI RMF 1.0 and its generative profile, NIST SP 800-207, SP 800-218 and CSF 2.0, ISO/IEC 42001, 27001, 23894 and 25010, IEEE Std 1012-2024 and 7000-2021, the EU AI Act, and the ASD Essential Eight [6]–[18].
\item \textbf{An adoption maturity model} (Appendix L) — five levels of organisational capability, each with its enforced controls, required evidence and accountable owner.
\item \textbf{A consolidated correction of superseded guidance} (Part 0) — twenty-one widely repeated claims about Claude Code that are wrong, incomplete or inverted at the verification date, each with the primary source that settles it.
\end{enumerate}

\FloatBarrier
\section*{Executive summary}
\markboth{Executive summary}{Executive summary}

\emph{For readers who will decide whether and how an organisation adopts this class of tool, rather than operate it.}

\textbf{What the system is.} Claude Code is not a code-completion assistant. It is an agent that reads and writes files, executes shell commands, drives a browser using your authenticated sessions, connects to external services under your identity, publishes web pages, schedules its own future work, runs unattended in cloud infrastructure, and delegates to as many as a thousand copies of itself in a single run [2], [19]–[21].

\textbf{Where the risk concentrates.} Not in the model's competence, which is high and improving, but in four structural places. \emph{Delegation}, where a subagent created casually inherits the conversation it was meant to audit independently (Chapter 16). \emph{Untrusted input}, where a web page, a tool result, a pull-request comment or a peer message carries instructions to the agent (Chapter 20;Chapter 13) — the failure mode documented in the peer-reviewed literature as indirect prompt injection [22], [23]. \emph{Autonomy}, where a scheduled routine runs with every connected credential and, where auto mode is available, a classifier rather than a person approving its actions (Chapter 24). And \emph{supply chain}, where a plugin may place executables on the shell path, start background processes and replace the system prompt of the main session on enable (Chapter 18).

\textbf{What must be enforced rather than requested.} An instruction in a project file is a preference; a permission rule is a control; a sandbox is an operating-system boundary; a managed policy, drawn above the four layers of Figure 1, is the only one of these a user cannot unset. Organisations that write their standards into \texttt{CLAUDE.\allowbreak{}md} have documented an intention. Organisations that write them into managed settings have deployed a control (Chapter 31).

\textbf{What can be evidenced.} Permission decisions, mode changes, connector connections, plugin loads, authentication events and model refusals are all exportable through OpenTelemetry, and every event raised while processing a prompt carries its identifier, which reconstructs the full activity of a single instruction (Chapter 33). An organisation that enables this can answer an auditor. One that does not cannot.

\textbf{What the reader should decide.} Six questions, in this order: which permission mode is the default for each cohort and who may change it; whether sandboxing is enforced or optional; which connectors, plugins and marketplaces are permitted; whether autonomous execution — routines, goals, unattended loops — is allowed at all and under whose approval; what leaves the boundary and under which retention terms (Chapter 32); and who is accountable when an agent acts. Appendix L sequences these into five maturity levels; Appendix E is the operational checklist that follows from them.

\textbf{The single-sentence version.} Adopt the capability at the pace at which you can evidence its output, and not faster.

\FloatBarrier
\section*{How to cite this work}
\markboth{How to cite this work}{How to cite this work}

\begin{calloutbox}{palegrey}{quoterule}{1.2mm}
D. Soldani, \emph{Claude Code Complete User Handbook: A Verified Practitioner's Reference for Agentic Work}, Version 1.1. Sydney, Australia, Nov. 2026. Verification baseline: Claude Code 2.1.322, 12 Nov. 2026.
\end{calloutbox}

\begin{codeblock}{9.0}{10.6}
\cl{@book\{soldani2026claudecode,}
\cl{~~author~~~~=~\{Soldani,~David\},}
\cl{~~title~~~~~=~\{Claude~Code~Complete~User~Handbook:}
\cl{~~~~~~~~~~~~~~~A~Verified~Practitioner\textquotesingle{}s~Reference~for~Agentic~Work\},}
\cl{~~version~~~=~\{1.1\},}
\cl{~~year~~~~~~=~\{2026\},}
\cl{~~month~~~~~=~\{11\},}
\cl{~~address~~~=~\{Sydney,~Australia\},}
\cl{~~note~~~~~~=~\{Verification~baseline:~Claude~Code~2.1.322,~12~Nov.~2026\}}
\cl{\}}
\end{codeblock}

\FloatBarrier
\section*{Scope and method}
\markboth{Scope and method}{Scope and method}

\textbf{Scope.} The book covers Claude Code as delivered through the terminal CLI, the VS Code and JetBrains extensions, the desktop application, the web and mobile surfaces, and Remote Control, at the verification baseline stated above. It covers the extension mechanisms — MCP, skills, subagents, hooks, plugins, output styles — the autonomy mechanisms — goals, workflows, three distinct schedulers, agent teams — and the governance mechanisms — managed policy, data retention, telemetry and accessibility. It does not cover the Agent SDK as a development platform beyond the points at which it changes the behaviour of an interactive session, nor cloud-provider deployment topologies beyond their effect on feature availability.

\textbf{Method.} Four rules were applied, and are stated here so that a reader can judge the evidence rather than trust the author.

\emph{Raw-source adjudication.} Every atomic product claim was adjudicated against the raw text of the primary documentation page that governs it, never against a summary of that page. The rule earned its keep twice. An automated summarisation pass reported three real product capabilities as non-existent; and, during the reference audit, a second summarisation pass asserted that an international secure-AI guideline had been issued by a single agency, when the issuing bodies' own announcements record it as a joint publication of two national agencies with seventeen further international partners. Both were false negatives, and either would have introduced an error had it been accepted. Where a summary and the source disagree, the source is authoritative and the summary is discarded.

\emph{Machine-checked apparatus.} The manuscript is authored with symbolic citation keys and symbolic cross-references, resolved to IEEE numbering and section numbers by a build step. The build fails on an unresolved key, a dangling cross-reference, a reference cited but not listed, a reference listed but not cited, or a cited documentation page absent from the official index. Two hundred and thirty-three automated checks gate the released artefacts. This is a modest application of the principle behind IEEE Std 1012: verification is an activity with defined inputs, defined outputs and an independent result, not an assertion of diligence [15].

\emph{An explicit unverified class.} Claims that could be settled only against a running installation — and therefore could not be settled against documentation — are labelled UNVERIFIED and listed in Appendix J with the local check that closes them. They were not softened into plausible prose. The temptation to do so is precisely what produces confident, wrong technical writing.

\emph{Version stamping.} Behaviour that was introduced or changed at a specific release carries that release inline. Sixty-three release-specific behaviour notes appear in the pages this book cites, across a span of roughly 160 releases. A reference work on this subject has a half-life measured in weeks; the honest response is a stated procedure rather than a disclaimer, and that procedure is Appendix K.

\textbf{Independence.} This book is not published by, endorsed by, or affiliated with Anthropic PBC. "Claude", "Claude Code" and related marks are the property of their respective owners and are used descriptively. Quotations from official documentation are short, attributed, and used for technical criticism and review.

\textbf{Tools used in preparation.} This manuscript was drafted, critically reviewed, revised and typeset with the assistance of AI coding agents (Claude Code, Anthropic; Codex, OpenAI). Their use spans the writing of chapter prose from the author's outlines and source material; the critical review of that prose against the documentation it cites; the toolchain that resolves citations and cross-references and renders the four editions; the automated audit that gates every release; and the typographic revision of the typeset edition. Every claim of product behaviour is cited to the primary documentation that governs it, and the author has reviewed and accepted each one; the evidence ledger of Appendix J records the outcome. Generative AI tools are not authors of this work and are not listed as such. Responsibility for every statement, correction and error in it rests solely with the author, however the text was produced.

\FloatBarrier
\section*{Limitations}
\markboth{Limitations}{Limitations}

The method in the preceding section is stated so that a reader can judge the evidence. This section states what that evidence does not establish. It is placed in the front matter rather than in an appendix because a reader deciding how much weight to give the book should not have to reach page two hundred to find out.

\textbf{Documentation is not a running system.} The verification method adjudicates claims against primary documentation. Documentation describes intended behaviour; a running installation exhibits actual behaviour, and the two diverge — in defaults that differ by platform, in features gated by plan or organisation, and in the gap between a documented flag and its effect. Claims that could be settled only against a running installation are labelled UNVERIFIED and listed in Appendix J with the local check that closes each one. Six remain open at this release. Three narrow what the \texttt{/\allowbreak{}effort} selector offers on a surface or a plan that was not captured; the others are statements the page fetched for this edition no longer settles, or settles differently from the release notes of the same day. A separate class could not be settled from publisher sources rather than from a running installation: the ISO/IEC 27001:2022 Annex A control identifiers cited in the crosswalk. Two items once in that class are now settled from their issuers: the page range of one conference paper whose publisher refuses automated clients, confirmed from the publisher's metadata on 14 November 2026, and the expansion of the MITRE ATLAS acronym, taken from the maintainers' own data release on 15 November 2026. Both classes are recorded in Appendix J.

\textbf{Paywalled standards were verified structurally, not textually.} ISO and IEC standards are sold, not published. The ISO/IEC 42001 clause structure used in the crosswalk was corrected against the issuing body's own public catalogue record and clause listing, and the correction was material — seven crosswalk rows and one worked example carried wrong clause numbers before it. But the \textasciitilde{}25 Annex A control identifiers cited from ISO/IEC 27001:2022 were reachable only through secondary summaries, and a caveat to that effect is recorded in the crosswalk rather than implying they were checked. Any verification of ISO-family standards performed without purchased access has this limitation; stating it is more useful than concealing it.

\textbf{Link liveness is hygiene, not attestation.} A sweep from an unproxied network on 12 November 2026 of the 107 unique URLs in the text found 105 resolving, one answering with an authentication challenge and one refusing an automated request. That sweep is not part of the default check suite, which still treats reachability as UNCHECKED, and reachability is in any case not a claim about content: a URL can resolve and no longer say what was adjudicated. A reader who finds a dead link should treat the citation's bibliographic data as the authoritative record.

\textbf{The false-negative finding cuts both ways.} Two automated summarisation passes were caught asserting things that the raw sources contradict, and both were false negatives — a capability or a fact reported as absent when it was present. False negatives are detectable by this method, because the raw source is consulted and the discrepancy surfaces. False \emph{positives} — a summariser asserting a capability that does not exist, in a case where the raw source was also consulted and happened to be ambiguous — would not surface the same way. No such case was found, and the method provides no strong assurance that none occurred.

\textbf{One author, one reviewer.} Verification was performed by the author. There was no independent adjudicator, no second reader reproducing the checks from the sources, and no editorial review of the kind a publisher supplies. Drafting and review were AI-assisted, as recorded in “Scope and method”; that is not independent adjudication, because the same class of tool both produced and checked the prose, and the author remains the only reviewer. The apparatus compensates in one direction only: the automated checks catch mechanical failures — a dangling cross-reference, an uncited reference, a stale count — and cannot catch a claim that is well-formed and wrong. Readers with access to a running installation are invited to close the UNVERIFIED items and to report discrepancies; the re-verification procedure in Appendix K is written to make that reproducible.

\textbf{The maturity model and the exercises are proposals, not findings.} The five-level adoption model in Appendix L, the four-element brief, the eight-point plan contract and the fourteen exercises with their worked solutions in Appendix M are the author's own constructions, offered as sequencing and teaching aids. They are informed by practice but have not been validated empirically against a population of organisations or learners, and no claim of external validity is made for them. Where the book cites a standard, the standard is the authority; where the book proposes a framework, the proposal is the author's and should be treated as such.

\textbf{Half-life.} The verification baseline is a single release. Sixty-three release-specific behaviour notes appear in the cited pages across roughly 160 releases, which is the measured rate of change of the subject rather than an impression of it. Statements of product behaviour in this book should be assumed to decay from the baseline date, and Appendix K states the procedure and the cadence rather than offering a disclaimer in place of one.

\FloatBarrier
\section*{Disclaimer}
\markboth{Disclaimer}{Disclaimer}

This book describes the operation of a system that can modify files, execute commands, transact with external services and act autonomously under a user's own credentials. Procedures are provided for instruction. The author accepts no liability for data loss, disclosure of confidential information, financial loss, regulatory breach or service disruption arising from their use. Nothing here constitutes legal, financial, security-assurance or compliance advice; where a control is described in relation to a regulatory or standards framework, the description is informational and the reader remains responsible for their own obligations. Test every procedure in a disposable environment before applying it to systems that matter.

\FloatBarrier
\section*{Conventions used in this book}
\markboth{Conventions used in this book}{Conventions used in this book}

\subsection*{Evidence labels}

\begingroup
\def\tblrows{%
\textbf{VERIFIED} & Confirmed in Anthropic's official documentation, cited at the point of use, as of 12 November 2026. \\
\textbf{PRACTITIONER NOTE} & A technique from the author's practice or the wider community. Not a product guarantee. \\
\textbf{UNVERIFIED} & Plausible and in circulation, but not confirmable from primary sources at the verification date. Check locally before relying on it. \\
\textbf{CAUTION} & An action with material security, privacy, financial, regulatory or data-loss consequence. \\
\textbf{CHECKPOINT} & Evidence the reader should observe before continuing. \\
}%
\def\tblbody{\begin{minipage}{\textwidth}
{\footnotesize\begin{tabular}{L{24.6mm}L{118.4mm}}
\toprule
\textbf{Label} & \textbf{Meaning} \\
\midrule
\tblrows
\bottomrule\end{tabular}}\end{minipage}}%
\begingroup
\def\sloppy{\tolerance 9999\emergencystretch 3em\hfuzz 200pt\vfuzz 200pt}%
\hbadness=10000\vbadness=10000\hfuzz=200pt\vfuzz=200pt
\global\setbox\tblbox=\hbox{\tblbody}%
\endgroup
\par\addvspace{2.6mm}
\ifdim\dimexpr\ht\tblbox+\dp\tblbox\relax>0.55\textheight
  \typeout{HANDBOOK-TABLE broken \the\dimexpr\ht\tblbox+\dp\tblbox\relax}%
  
  {\footnotesize\begin{longtable}{L{24.6mm}L{118.4mm}}
  \toprule
\textbf{Label} & \textbf{Meaning} \\
\midrule\endfirsthead
  \multicolumn{2}{@{}l@{}}{%
  \sffamily\footnotesize\itshape\color{inkgrey}Continued}\\[1.2mm]
  \toprule
\textbf{Label} & \textbf{Meaning} \\
\midrule\endhead
  \bottomrule\endfoot
  \bottomrule\endlastfoot
  \tblrows
  \end{longtable}}%
\else
  \typeout{HANDBOOK-TABLE atomic \the\dimexpr\ht\tblbox+\dp\tblbox\relax}%
  \noindent\tblbody
\fi
\par\addvspace{2.6mm}
\endgroup

\subsection*{Citation, cross-reference and version conventions}

Citations are IEEE numeric: a bracketed number resolves to the reference list, numbered in order of first appearance. Cross-references name a chapter or a numbered section, for example Chapter 7 or Section 7.2. Figures and tables are numbered sequentially and listed after the contents.

Where a behaviour was introduced or changed at a specific release, the release is stated inline in the form \emph{(requires 2.1.219 or later)}. Where a behaviour is plan-dependent, surface-dependent or provider-dependent, that dependency is stated rather than assumed away.

Command syntax is shown as typed. A leading \texttt{/\allowbreak{}} marks an in-session command; a leading \texttt{claude} marks a shell invocation. The two are different interfaces and are indexed separately in Appendix A and Appendix B. Each chapter closes with \textbf{Key points} — the three to five statements a reader should retain.

\FloatBarrier
\section*{Release and re-verification}
\markboth{Release and re-verification}{Release and re-verification}

\begingroup
\def\tblrows{%
1.0 & 23 Aug. 2026 & Claude Code 2.1.241 & First release. 34 chapters, 14 appendices, 7 figures. \\
1.0, revised & 26 Aug. 2026 & Re-verified against 2.1.246 & Cited corpus re-fetched, hashed and reconciled against the published index; ten pages re-adjudicated; the effort and permission-mode vocabularies generated from one source. \\
1.1 & 12 Nov. 2026 & Claude Code 2.1.322 & A new baseline. The published index re-enumerated and its pages fetched and hashed; thirty-four chapters, Part 0 and the appendices re-adjudicated in full; six new sections; the command-line vocabularies re-observed; errata for Version 1.0; every cited standard, regulation and paper re-checked against its issuer on 14 Nov. 2026. 233 automated build checks, all passing. \\
}%
\def\tblbody{\begin{minipage}{\textwidth}
{\footnotesize\begin{tabular}{L{16.9mm}L{9.7mm}L{22.5mm}L{87.4mm}}
\toprule
\textbf{Version} & \textbf{Date} & \textbf{Baseline} & \textbf{Status} \\
\midrule
\tblrows
\bottomrule\end{tabular}}\end{minipage}}%
\begingroup
\def\sloppy{\tolerance 9999\emergencystretch 3em\hfuzz 200pt\vfuzz 200pt}%
\hbadness=10000\vbadness=10000\hfuzz=200pt\vfuzz=200pt
\global\setbox\tblbox=\hbox{\tblbody}%
\endgroup
\par\addvspace{2.6mm}
\ifdim\dimexpr\ht\tblbox+\dp\tblbox\relax>0.55\textheight
  \typeout{HANDBOOK-TABLE broken \the\dimexpr\ht\tblbox+\dp\tblbox\relax}%
  
  {\footnotesize\begin{longtable}{L{16.9mm}L{9.7mm}L{22.5mm}L{87.4mm}}
  \toprule
\textbf{Version} & \textbf{Date} & \textbf{Baseline} & \textbf{Status} \\
\midrule\endfirsthead
  \multicolumn{4}{@{}l@{}}{%
  \sffamily\footnotesize\itshape\color{inkgrey}Continued}\\[1.2mm]
  \toprule
\textbf{Version} & \textbf{Date} & \textbf{Baseline} & \textbf{Status} \\
\midrule\endhead
  \bottomrule\endfoot
  \bottomrule\endlastfoot
  \tblrows
  \end{longtable}}%
\else
  \typeout{HANDBOOK-TABLE atomic \the\dimexpr\ht\tblbox+\dp\tblbox\relax}%
  \noindent\tblbody
\fi
\par\addvspace{2.6mm}
\endgroup

Re-verify quarterly, or before relying on the book for a consequential decision, following Appendix K. Record what changed in the ledger of Appendix J.

\FloatBarrier
\section*{Changes since Version 1.0}
\markboth{Changes since Version 1.0}{Changes since Version 1.0}

Most of what changed between Version 1.0 and this edition is the product moving. Table 33 lists the principal corrections, and each chapter stamps the release that changed the behaviour. Six statements were different in kind: they were wrong when Version 1.0 was printed, and a reader who acted on them may still be carrying the consequence. They are listed here so that a reader of Version 1.0 does not have to find them.

\begin{enumerate}
\item \textbf{The \texttt{.\allowbreak{}gitignore} in the Chapter 8 solution ignored nothing.} Each pattern carried its comment on the same line. Git treats \texttt{\#\allowbreak{}} as a comment only at the start of a line, so every line was read as a pattern that matched no file, and \texttt{.\allowbreak{}env} was not ignored. If you copied it, run \texttt{git check-\allowbreak{}ignore -\allowbreak{}v .\allowbreak{}env} in that repository: no output means the file is not ignored. Replace the file with the form in Appendix M, and run \texttt{git log -\allowbreak{}-\allowbreak{}all -\allowbreak{}-\allowbreak{} .\allowbreak{}env} to see whether the file was ever committed. If it was, treat what it held as exposed and follow Section 14.3; removing it from history does not recall a copy already pushed [24].
\item \textbf{The Chapter 8 sequence inspected before it staged.} It ran \texttt{git diff -\allowbreak{}-\allowbreak{}staged} before \texttt{git add .\allowbreak{}}, so the inspection showed an empty diff and the commit that followed went unreviewed. Section 8.1 now stages first.
\item \textbf{The example enterprise policy in Chapter 31 did not do what it said.} It set \texttt{"strict\allowbreak{}Known\allowbreak{}Marketplace\allowbreak{}s":\allowbreak{} true}, which is not a value the key accepts, and \texttt{allow\allowbreak{}Managed\allowbreak{}Mcp\allowbreak{}Servers\allowbreak{}Only} with no \texttt{allowed\allowbreak{}Mcp\allowbreak{}Servers} beside it. The caution in Section 31.2 gives the consequence on each client and the corrected lines.
\item \textbf{The control stack disagreed with itself.} The abstract and Chapter 1 said two of its four layers are enforced; Figure 1 and the chapter's key points said three. The figure was right: the first layer is advisory and the other three are enforced.
\item \textbf{Chapter 2 gave \texttt{/\allowbreak{}doctor}'s \texttt{CLAUDE.\allowbreak{}md} trims to \texttt{claude doct\allowbreak{}or}.} The documentation already placed them in the in-session \texttt{/\allowbreak{}doctor} checkup when Version 1.0 was printed. \texttt{claude doct\allowbreak{}or} prints read-only diagnostics and applies nothing, so it never proposed a trim. To have a checked-in \texttt{CLAUDE.\allowbreak{}md} reviewed, run \texttt{/\allowbreak{}doctor} inside a session (Section 2.3) [25].
\item \textbf{The crosswalk cited a superseded standard.} Version 1.0 cited IEEE Std 1012-2016 for verification and validation. IEEE Std 1012-2024 had superseded it before Version 1.0 was printed: approved in November 2024 and published in August 2025. The reference check of 24 August 2026 confirmed the older edition's DOI and missed the supersession. The crosswalk in Appendix H now cites IEEE Std 1012-2024 [15]; if you cited the standard from Version 1.0, cite the current edition.
\end{enumerate}

Items 1, 4 and 6 are now tested on every build: each \texttt{.\allowbreak{}gitignore} block in the book is run through \texttt{git check-\allowbreak{}ignore}, every statement of the control stack must agree with the figure, and no superseded edition of IEEE Std 1012 may be cited.

\FloatBarrier
\section*{Acknowledgements}
\markboth{Acknowledgements}{Acknowledgements}

Anthropic's documentation team maintains an unusually candid public record of limitations, deprecations and version-specific behaviour [1]; a book of this kind would not be possible without it. Responsibility for every judgement, correction and error rests with the author.

\FloatBarrier
\frontchapter{List of figures and tables}

\FloatBarrier
\section*{Figures}
\markboth{Figures}{Figures}

\begingroup
\def\tblrows{%
1 & The agentic control stack. Layers 1 and 2 are the product's; layers 3 and 4 are the operating system's… & Ch. 1 \\
2 & How a tool call is decided. Deny is evaluated first and no rule or hook widens it; a blocking hook stops the call before the rules… & Ch. 7 \\
3 & What each context operation keeps of conversation and session state. Only the last row survives all of them. & Ch. 11 \\
4 & Where untrusted content enters an agentic session. Every arrow is a documented injection channel… & Part IV intro \\
5 & Five delegation topologies and what each one inherits. & Ch. 16 \\
6 & A cross-checked research workflow with a human release gate & Ch. 22 \\
7 & Choosing a scheduling mechanism. Start from what must be true when the work runs, not from how often it runs. & Ch. 24 \\
}%
\def\tblbody{\begin{minipage}{\textwidth}
{\footnotesize\begin{tabular}{L{9.7mm}L{110.8mm}L{19.4mm}}
\toprule
\textbf{Fig.} & \textbf{Caption} & \textbf{Location} \\
\midrule
\tblrows
\bottomrule\end{tabular}}\end{minipage}}%
\begingroup
\def\sloppy{\tolerance 9999\emergencystretch 3em\hfuzz 200pt\vfuzz 200pt}%
\hbadness=10000\vbadness=10000\hfuzz=200pt\vfuzz=200pt
\global\setbox\tblbox=\hbox{\tblbody}%
\endgroup
\par\addvspace{2.6mm}
\ifdim\dimexpr\ht\tblbox+\dp\tblbox\relax>0.55\textheight
  \typeout{HANDBOOK-TABLE broken \the\dimexpr\ht\tblbox+\dp\tblbox\relax}%
  
  {\footnotesize\begin{longtable}{L{9.7mm}L{110.8mm}L{19.4mm}}
  \toprule
\textbf{Fig.} & \textbf{Caption} & \textbf{Location} \\
\midrule\endfirsthead
  \multicolumn{3}{@{}l@{}}{%
  \sffamily\footnotesize\itshape\color{inkgrey}Continued}\\[1.2mm]
  \toprule
\textbf{Fig.} & \textbf{Caption} & \textbf{Location} \\
\midrule\endhead
  \bottomrule\endfoot
  \bottomrule\endlastfoot
  \tblrows
  \end{longtable}}%
\else
  \typeout{HANDBOOK-TABLE atomic \the\dimexpr\ht\tblbox+\dp\tblbox\relax}%
  \noindent\tblbody
\fi
\par\addvspace{2.6mm}
\endgroup

\FloatBarrier
\section*{Tables}
\markboth{Tables}{Tables}

\begingroup
\def\tblrows{%
1 & Superseded guidance and common misconceptions, with the verified position & front matter \\
2 & Claude Code surfaces and their trade-offs & Ch. 1 \\
3 & Effort levels accepted by surface, each column stamped with the release it was observed on & Ch. 4 \\
4 & Suggested starting posture by work type (practitioner guidance) & Ch. 4 \\
5 & Permission modes, the cycle position that reaches each, and appropriate use & Ch. 7 \\
6 & Instruction file locations, in load order from broadest to most specific & Ch. 9 \\
7 & Where a given piece of knowledge belongs & Ch. 10 \\
8 & Mechanisms that change the shape of a conversation & Ch. 11 \\
9 & Controls over transcript location, retention and writing & Ch. 12 \\
10 & MCP configuration scopes & Ch. 13 \\
11 & Three ways an organisation governs MCP servers & Ch. 13 \\
12 & What each secret control reaches, and what it does not & Ch. 14 \\
13 & Skill frontmatter fields with a security consequence & Ch. 15 \\
14 & Subagent definition precedence & Ch. 16 \\
15 & Choosing a delegation mechanism & Ch. 16 \\
16 & Plugin components, all at the plugin root & Ch. 18 \\
17 & Managed controls over mods & Ch. 18 \\
18 & Built-in output styles & Ch. 19 \\
19 & Dynamic workflow runtime constraints & Ch. 22 \\
20 & What \texttt{/\allowbreak{}loop} does with each combination of arguments & Ch. 23 \\
21 & Comparing the three scheduling mechanisms & Ch. 24 \\
22 & Where channels sit among the ways to reach a session & Ch. 27 \\
23 & Artifact page constraints & Ch. 28 \\
24 & Ways to run more than one thing at a time & Ch. 29 \\
25 & How a managed policy reaches a machine & Ch. 31 \\
26 & Which way an invalid managed policy fails & Ch. 31 \\
27 & Non-inference data flows and their opt-outs & Ch. 32 \\
28 & Retention by account type & Ch. 32 \\
29 & Exported metrics & Ch. 33 \\
30 & Book controls mapped to external frameworks & App. H \\
31 & Misconception adjudication, with verdict and primary source & App. J \\
32 & Corrections applied at the 26 August 2026 re-verification & App. J \\
33 & Principal corrections applied at the 12 November 2026 adjudication & App. J \\
34 & Items closed by direct observation, with the release each was observed on & App. J \\
35 & Adoption maturity levels for agentic development tooling & App. L \\
}%
\def\tblbody{\begin{minipage}{\textwidth}
{\footnotesize\begin{tabular}{L{12.1mm}L{108.4mm}L{19.4mm}}
\toprule
\textbf{Table} & \textbf{Caption} & \textbf{Location} \\
\midrule
\tblrows
\bottomrule\end{tabular}}\end{minipage}}%
\begingroup
\def\sloppy{\tolerance 9999\emergencystretch 3em\hfuzz 200pt\vfuzz 200pt}%
\hbadness=10000\vbadness=10000\hfuzz=200pt\vfuzz=200pt
\global\setbox\tblbox=\hbox{\tblbody}%
\endgroup
\par\addvspace{2.6mm}
\ifdim\dimexpr\ht\tblbox+\dp\tblbox\relax>0.55\textheight
  \typeout{HANDBOOK-TABLE broken \the\dimexpr\ht\tblbox+\dp\tblbox\relax}%
  
  {\footnotesize\begin{longtable}{L{12.1mm}L{108.4mm}L{19.4mm}}
  \toprule
\textbf{Table} & \textbf{Caption} & \textbf{Location} \\
\midrule\endfirsthead
  \multicolumn{3}{@{}l@{}}{%
  \sffamily\footnotesize\itshape\color{inkgrey}Continued}\\[1.2mm]
  \toprule
\textbf{Table} & \textbf{Caption} & \textbf{Location} \\
\midrule\endhead
  \bottomrule\endfoot
  \bottomrule\endlastfoot
  \tblrows
  \end{longtable}}%
\else
  \typeout{HANDBOOK-TABLE atomic \the\dimexpr\ht\tblbox+\dp\tblbox\relax}%
  \noindent\tblbody
\fi
\par\addvspace{2.6mm}
\endgroup

\breakrule

\FloatBarrier
\mainmatter
\parttitle{Part 0}{Superseded guidance and common misconceptions}
\addcontentsline{toc}{part}{Part 0 \textemdash\ Superseded guidance and common misconceptions}

Read this part before following any procedure in the book, and before following any procedure found elsewhere.

Claude Code changes weekly. Guidance written against a release three months old is not merely dated: in several documented cases it is inverted, and a reader who follows it does the opposite of what they intend. Twenty-one such claims are in wide circulation at the verification date.

This part is a register, not a treatment. Each row states the claim as it is commonly made, the position established from primary documentation, and the chapter that owns the correction and develops it in full. \textbf{The chapter is authoritative}; this table exists so that a reader who already holds one of these beliefs recognises the collision when they meet the correct statement. The four printed in bold are the ones whose consequences are operational rather than cosmetic.

Every entry appears again in the evidence ledger of Appendix J, with the verdict recording \emph{how} the claim failed and the primary source that settles it. Both tables are generated from one record, so the claim is written once.

\begingroup
\def\tblrows{%
\textbf{M01} & Run \texttt{/\allowbreak{}loop} with no arguments to see your scheduled tasks & A bare \texttt{/\allowbreak{}loop} runs the default loop prompt, which unless you have replaced it is autonomous maintenance. It is not the command that lists tasks: ask Claude in natural language or use \texttt{Cron\allowbreak{}List}. [26] & Chapter 23 \\
\textbf{M02} & Closing the session stops a loop & Exiting stops tasks firing; backgrounding does not. Backgrounding carries loop tasks into a background session with no terminal attached, and resuming restores unexpired recurring tasks and one-shots not yet due, though not a self-paced loop. Whether a fresh conversation still clears them is UNVERIFIED. [26] & Chapter 23 \\
\textbf{M03} & A subagent gives you an independent second opinion & Only if you make it one. Fork mode is on by default and a fork inherits the entire parent conversation; a parent in \texttt{accept\allowbreak{}Edits}, \texttt{bypass\allowbreak{}Permissions} or \texttt{auto} also overrides the subagent's own mode. [27] & Chapter 16 \\
\textbf{M04} & \texttt{.\allowbreak{}env} keeps credentials out of Claude's reach & False as a guarantee. Permission deny rules, sandbox credential controls, network restriction, output discipline and least-privilege credentials are the controls that hold. [25], [28], [29] & Chapter 14 \\
M05 & Scheduling jitter is unpredictable & It is specified: a recurring task runs late by up to half its interval, at most 30 minutes, and the same amount each time when its runs are evenly spaced. A task every five minutes, which \texttt{/\allowbreak{}loop 5m} creates, is not delayed; and only a one-shot avoids its shift by being scheduled off \texttt{:\allowbreak{}00} and \texttt{:\allowbreak{}30}. [26] & Chapter 23 \\
M06 & \texttt{CLAUDE\_\allowbreak{}CODE\_\allowbreak{}DISABLE\_\allowbreak{}AUTO\_\allowbreak{}MEMORY} is how you turn auto memory off & The \texttt{auto\allowbreak{}Memory\allowbreak{}Enabled} setting is the primary control, settable per project; the environment variable is the fallback. [25] & Chapter 10 \\
M07 & The 30-day cleanup period governs local Claude Code state & Session transcripts are swept, but auto-memory files are excluded from that sweep and persist until edited or deleted. [25], [30] & Chapter 10 \\
M08 & Session history is capped at one month & Thirty days is a configurable default, set by \texttt{cleanup\allowbreak{}Period\allowbreak{}Days}, and Claude Desktop and Cowork transcripts are kept at any age unless an age limit is set or, among other cases, managed settings set \texttt{cleanup\allowbreak{}Period\allowbreak{}Days} or the HIPAA configuration applies. [31], [32] & Chapter 12 \\
M09 & \texttt{bypass\allowbreak{}Permissions} is a reasonable productivity setting & Isolated containers or virtual machines only. It propagates to workflow agents and agent-team teammates, who cannot decline it. [19], [30], [33] & Chapter 7 \\
M10 & Every MCP tool definition always consumes context & Tool Search defers many definitions until they are needed. Measure occupancy with \texttt{/\allowbreak{}context} rather than assuming it. [34], [35] & Chapter 13 \\
M11 & Copying \texttt{.\allowbreak{}mcp.\allowbreak{}json} transfers an authenticated server & Shared project configuration must be credential-free. In an interactive session a project-scoped server also needs per-user approval and workspace trust; \texttt{claude -\allowbreak{}p}, the Agent SDK and cloud sessions outside a project's threads load it without asking, as does a session started in \texttt{bypass\allowbreak{}Permissions} with \texttt{skip\allowbreak{}Dangerous\allowbreak{}Mode\allowbreak{}Permission\allowbreak{}Prompt} set in user or managed settings. [34] & Chapter 13 \\
M12 & A popular third-party skill is effectively safe & Skills execute inline shell, grant tools for the invoking turn and can register hooks that persist for the session. Stars are not a security review. [36] & Chapter 15 \\
M13 & \texttt{/\allowbreak{}agents} opens an agent editor & Changed at release 2.1.198: it now directs you to ask Claude to manage agents, or to edit the agent files directly. [27], [37] & Chapter 16 \\
M14 & Workflows are beta and cannot be monitored in an IDE & Dynamic workflows run in the CLI, the desktop app, the IDE extensions, \texttt{claude -\allowbreak{}p} and the Agent SDK, with \texttt{/\allowbreak{}workflows} as the progress view. [19] & Chapter 22 \\
M15 & Workflow agents inherit the session's permission mode & They use your permission rules and take their mode by the rules for any subagent. Under a session in \texttt{bypass\allowbreak{}Permissions}, \texttt{accept\allowbreak{}Edits} or \texttt{auto} they run in that mode; under \texttt{plan} they stay in plan mode, and a \texttt{dont\allowbreak{}Ask} agent there is denied what plan mode would ask; under Manual or \texttt{dont\allowbreak{}Ask} an agent type's own \texttt{permission\allowbreak{}Mode} applies, unless it asks for \texttt{bypass\allowbreak{}Permissions}, or for \texttt{auto} where auto mode is unavailable, and an unset field inherits the session's mode. Version 1.0 said they always run in \texttt{accept\allowbreak{}Edits}, which the page no longer states. [19], [27] & Chapter 22 \\
M16 & Claude in Chrome is a beta feature & Generally available. It is unavailable under WSL and to sessions authenticated with an API key or a long-lived token, and its access to authenticated browser state raises the stakes on permission hygiene. [21], [38] & Chapter 20 \\
M17 & \texttt{/\allowbreak{}loop} is equivalent to operating-system cron & It is session-scoped, idle-dependent and jittered, capped at 50 tasks per session, expiring after seven days, with no catch-up for missed fires. [26] & Chapter 23 \\
M18 & Plugins are convenient feature packages & A plugin may ship executables placed on the Bash tool's \texttt{PATH}, background monitors, hooks, MCP and LSP servers, code that runs inside Claude Code, and a settings key that activates one of its agents as the main thread. [39], [40] & Chapter 18 \\
M19 & Routines are simply sessions on a schedule & Now called cloud scheduled tasks. A research preview that also triggers on API calls and GitHub events, includes all connected connectors by default, acts under your identity, and in a new cloud session starts in auto mode where it is available, so a classifier rather than a person approves actions. [20] & Chapter 24 \\
M20 & A green status on a routine run means the task succeeded & A run marked Succeeded means the session ended without Claude Code reporting a failure. Blocked requests, missing connector tools and task failures appear only in the transcript. [20] & Chapter 24 \\
M21 & Remote Control keeps everything local & Execution is local, but while connected the transcript is stored on Anthropic servers to synchronise across devices — which is why it is unavailable under Zero Data Retention or the HIPAA configuration. [41]–[43] & Chapter 25 \\
}%
\def\tblbody{\begin{minipage}{\textwidth}\boxcaption{Table 1 — Superseded guidance and common misconceptions, with the verified position}
{\scriptsize\begin{tabular}{L{5.4mm}L{55.5mm}L{60.9mm}L{14.8mm}}
\toprule
\textbf{} & \textbf{Claim as commonly made} & \textbf{The position at 12 November 2026} & \textbf{Treated in} \\
\midrule
\tblrows
\bottomrule\end{tabular}}\end{minipage}}%
\begingroup
\def\sloppy{\tolerance 9999\emergencystretch 3em\hfuzz 200pt\vfuzz 200pt}%
\hbadness=10000\vbadness=10000\hfuzz=200pt\vfuzz=200pt
\global\setbox\tblbox=\hbox{\tblbody}%
\endgroup
\par\addvspace{2.6mm}
\ifdim\dimexpr\ht\tblbox+\dp\tblbox\relax>0.55\textheight
  \typeout{HANDBOOK-TABLE broken \the\dimexpr\ht\tblbox+\dp\tblbox\relax}%
  \tabcaption{Table 1 — Superseded guidance and common misconceptions, with the verified position}
  {\scriptsize\begin{longtable}{L{5.4mm}L{55.5mm}L{60.9mm}L{14.8mm}}
  \toprule
\textbf{} & \textbf{Claim as commonly made} & \textbf{The position at 12 November 2026} & \textbf{Treated in} \\
\midrule\endfirsthead
  \multicolumn{4}{@{}l@{}}{%
  \sffamily\footnotesize\itshape\color{inkgrey}Table 1 — Superseded guidance and common misconceptions, with the verified position \textemdash\ continued}\\[1.2mm]
  \toprule
\textbf{} & \textbf{Claim as commonly made} & \textbf{The position at 12 November 2026} & \textbf{Treated in} \\
\midrule\endhead
  \bottomrule\endfoot
  \bottomrule\endlastfoot
  \tblrows
  \end{longtable}}%
\else
  \typeout{HANDBOOK-TABLE atomic \the\dimexpr\ht\tblbox+\dp\tblbox\relax}%
  \noindent\tblbody
\fi
\par\addvspace{2.6mm}
\endgroup

Several of these claims are traceable to a widely viewed introductory video course published on 10 July 2026 [44], which remains a reasonable orientation to the product's surface but was recorded against an earlier release. Where a reader has learned one of the positions above from that source or from material derived from it, the entry above supersedes it.

\begin{calloutbox}{palebrass}{brassdark}{2.0mm}
\textbf{CAUTION — non-negotiable human gates} Require explicit human confirmation before publication, payment, credential or account change, production deployment, destructive deletion, merge, or direct push to a protected branch. An agent reporting "done" is a claim, not evidence. This principle is operationalised as Evidence-Gated Delivery in Chapter 34 and as a checklist in Appendix E.
\end{calloutbox}

\breakrule

\FloatBarrier
\frontchapter{How to use this book}

The book has four reading paths.

The \textbf{essential path} is Chapters 1–12. It takes a reader from installation to a controlled working method: a project with bounded context, a plan reviewed before execution, permissions that hold, version control underneath, and durable project knowledge. Complete it before adding autonomous or third-party components. Nothing in Chapters 13 onward is safe without it.

The \textbf{extension path} is Chapters 13–19. It covers the mechanisms that change what Claude Code can reach: connectors, secrets, reusable procedures, delegated agents, event-driven automation, packaged plugins, and the system-prompt layer. Every chapter in this part is also a supply-chain chapter, and each carries a threat model.

The \textbf{autonomy path} is Chapters 20–30. Browser control, persistent goals, multi-agent workflows, three distinct scheduling mechanisms, remote and cloud execution, event push, shared artifacts, and team orchestration. The organising question in this part is not "can it do this" but "how will I know what it did".

The \textbf{governance path} is Chapters 31–33 and Appendices D, F and K. It is written for readers who must answer to somebody else: managed policy, data residency and retention, observability, audit, accessibility, and the re-verification procedure that keeps a document like this one honest.

Chapter 34 combines the whole system into a single verified deliverable. Appendix E is the release checklist that goes with it.

Use a disposable practice folder. Do not learn permissions, hooks, browser actions or scheduling in a production repository, and do not open a home directory as a project.

\FloatBarrier
\section*{The running project}
\markboth{The running project}{The running project}

Worked examples build a \textbf{Research Brief Workspace}. It is deliberately not a software project: the argument of this book is that agentic work is a general operating discipline, and a non-programming example makes the discipline visible without the distraction of a build system.

\begin{codeblock}{9.0}{10.6}
\cl{research-brief-workspace/}
\cl{\pmboxdrawuni{251C}\pmboxdrawuni{2500}\pmboxdrawuni{2500}~CLAUDE.md}
\cl{\pmboxdrawuni{251C}\pmboxdrawuni{2500}\pmboxdrawuni{2500}~README.md}
\cl{\pmboxdrawuni{251C}\pmboxdrawuni{2500}\pmboxdrawuni{2500}~sources/~~~~~~~~~~\#~immutable~source~material}
\cl{\pmboxdrawuni{251C}\pmboxdrawuni{2500}\pmboxdrawuni{2500}~notes/~~~~~~~~~~~~\#~working~extraction}
\cl{\pmboxdrawuni{251C}\pmboxdrawuni{2500}\pmboxdrawuni{2500}~drafts/~~~~~~~~~~~\#~unapproved~manuscripts}
\cl{\pmboxdrawuni{251C}\pmboxdrawuni{2500}\pmboxdrawuni{2500}~deliverables/~~~~~\#~human-approved~outputs~only}
\cl{\pmboxdrawuni{251C}\pmboxdrawuni{2500}\pmboxdrawuni{2500}~references/~~~~~~~\#~policies~the~agent~retrieves~on~demand}
\cl{\pmboxdrawuni{251C}\pmboxdrawuni{2500}\pmboxdrawuni{2500}~templates/~~~~~~~~\#~reusable~structures}
\cl{\pmboxdrawuni{2514}\pmboxdrawuni{2500}\pmboxdrawuni{2500}~.claude/}
\cl{~~~~\pmboxdrawuni{251C}\pmboxdrawuni{2500}\pmboxdrawuni{2500}~settings.json}
\cl{~~~~\pmboxdrawuni{251C}\pmboxdrawuni{2500}\pmboxdrawuni{2500}~rules/}
\cl{~~~~\pmboxdrawuni{251C}\pmboxdrawuni{2500}\pmboxdrawuni{2500}~skills/}
\cl{~~~~\pmboxdrawuni{2514}\pmboxdrawuni{2500}\pmboxdrawuni{2500}~agents/}
\end{codeblock}

By Chapter 34 the workspace will research a bounded question, keep an evidence ledger, draft a brief, review it with an agent that did not write it, render a deliverable, and stop at a human release gate, held by a deny rule on \texttt{deliverable\allowbreak{}s/\allowbreak{}} and by the procedure, not by either alone.

\breakrule

\FloatBarrier
\parttitle{Part I}{Foundations}
\addcontentsline{toc}{part}{Part I \textemdash\ Foundations}

\FloatBarrier
\renewcommand{\chaptertitlelabel}{Part I \textperiodcentered\ Chapter 1}
\setcounter{chapter}{0}
\chapter{What Claude Code is, and what it is not}

Claude Code is an agentic work environment. A chat interface receives messages and returns messages. Claude Code additionally inspects a working directory, edits files, executes tools, requests permissions, reaches configured services, publishes pages, schedules its own future turns, and delegates to other instances of itself [2], [3].

That capability does not make it an employee, and it does not make it an expert. A productive mental model is this:

\begin{calloutbox}{palegrey}{quoterule}{1.2mm}
Claude Code is a fast operator with variable judgement, broad tool access, no inherent knowledge of your unstated intent, and a strong disposition to report success.
\end{calloutbox}

The last clause is the operationally important one. Every control described in this book exists because the agent's own account of what it did is the least reliable artefact it produces.

\FloatBarrier
\setcounter{section}{0}
\section{Four things that must be explicit}

Work goes well when four elements are stated rather than inferred:

\begin{enumerate}
\item \textbf{Outcome} — the finished state you need, in terms someone else could check.
\item \textbf{Context} — files, constraints, prior decisions, and what is out of scope.
\item \textbf{Boundaries} — what may change without asking, and what requires approval.
\item \textbf{Evidence} — how both parties will know the outcome is correct.
\end{enumerate}

Chapters 5 and 6 turn these into a brief and a plan; Chapter 21 turns the fourth into a machine-checked completion condition; Appendix E turns all four into a release checklist.

\FloatBarrier
\setcounter{section}{1}
\section{The control stack}

Most loss of control in agentic work is a category error: treating a layer that \emph{shapes intent} as though it \emph{constrains action}. The four layers of Figure 1 are not interchangeable. The first is advisory; the other three are enforced, each by a different enforcer — the product, the operating system, and the boundary the whole session runs inside.

\begin{figure}[tbp]
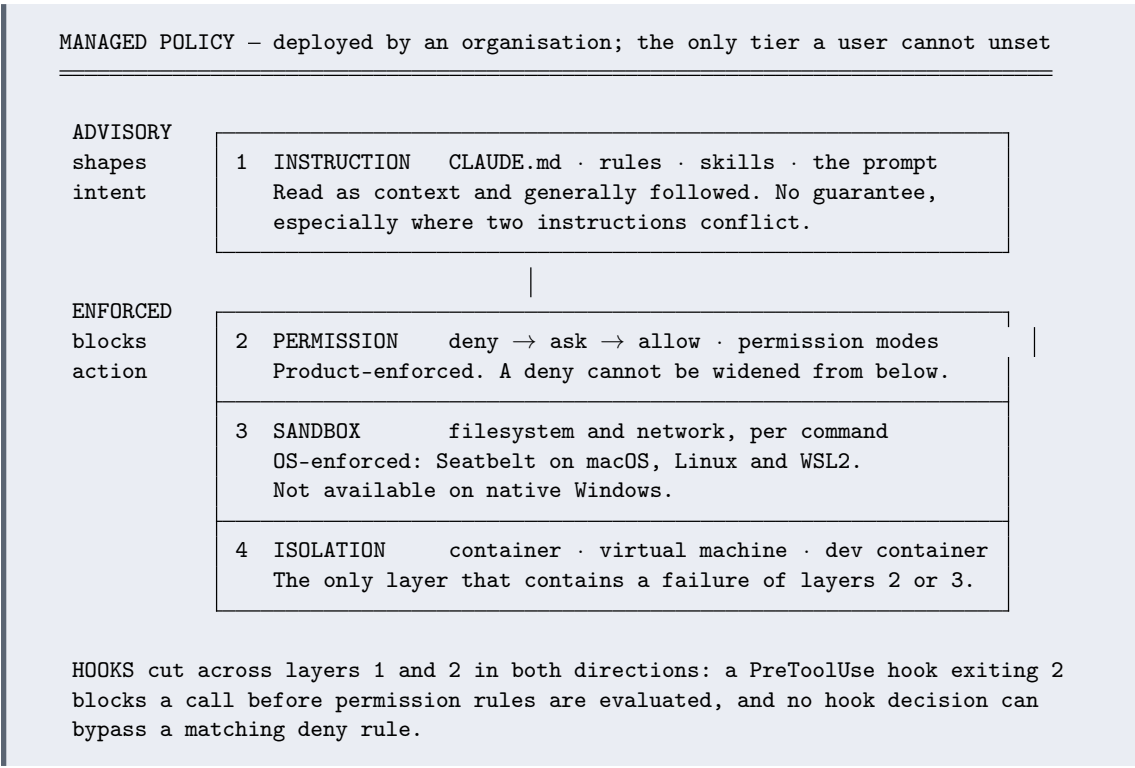

\begin{codefig}{9.0}{10.6}
\cl{~~MANAGED~POLICY~\pmboxdrawuni{2500}~deployed~by~an~organisation;~the~only~tier~a~user~cannot~unset}
\cl{~~\pmboxdrawuni{2550}\pmboxdrawuni{2550}\pmboxdrawuni{2550}\pmboxdrawuni{2550}\pmboxdrawuni{2550}\pmboxdrawuni{2550}\pmboxdrawuni{2550}\pmboxdrawuni{2550}\pmboxdrawuni{2550}\pmboxdrawuni{2550}\pmboxdrawuni{2550}\pmboxdrawuni{2550}\pmboxdrawuni{2550}\pmboxdrawuni{2550}\pmboxdrawuni{2550}\pmboxdrawuni{2550}\pmboxdrawuni{2550}\pmboxdrawuni{2550}\pmboxdrawuni{2550}\pmboxdrawuni{2550}\pmboxdrawuni{2550}\pmboxdrawuni{2550}\pmboxdrawuni{2550}\pmboxdrawuni{2550}\pmboxdrawuni{2550}\pmboxdrawuni{2550}\pmboxdrawuni{2550}\pmboxdrawuni{2550}\pmboxdrawuni{2550}\pmboxdrawuni{2550}\pmboxdrawuni{2550}\pmboxdrawuni{2550}\pmboxdrawuni{2550}\pmboxdrawuni{2550}\pmboxdrawuni{2550}\pmboxdrawuni{2550}\pmboxdrawuni{2550}\pmboxdrawuni{2550}\pmboxdrawuni{2550}\pmboxdrawuni{2550}\pmboxdrawuni{2550}\pmboxdrawuni{2550}\pmboxdrawuni{2550}\pmboxdrawuni{2550}\pmboxdrawuni{2550}\pmboxdrawuni{2550}\pmboxdrawuni{2550}\pmboxdrawuni{2550}\pmboxdrawuni{2550}\pmboxdrawuni{2550}\pmboxdrawuni{2550}\pmboxdrawuni{2550}\pmboxdrawuni{2550}\pmboxdrawuni{2550}\pmboxdrawuni{2550}\pmboxdrawuni{2550}\pmboxdrawuni{2550}\pmboxdrawuni{2550}\pmboxdrawuni{2550}\pmboxdrawuni{2550}\pmboxdrawuni{2550}\pmboxdrawuni{2550}\pmboxdrawuni{2550}\pmboxdrawuni{2550}\pmboxdrawuni{2550}\pmboxdrawuni{2550}\pmboxdrawuni{2550}\pmboxdrawuni{2550}\pmboxdrawuni{2550}\pmboxdrawuni{2550}\pmboxdrawuni{2550}\pmboxdrawuni{2550}\pmboxdrawuni{2550}\pmboxdrawuni{2550}\pmboxdrawuni{2550}\pmboxdrawuni{2550}\pmboxdrawuni{2550}\pmboxdrawuni{2550}\pmboxdrawuni{2550}}
\cl{}
\cl{~~~ADVISORY~~~\pmboxdrawuni{250C}\pmboxdrawuni{2500}\pmboxdrawuni{2500}\pmboxdrawuni{2500}\pmboxdrawuni{2500}\pmboxdrawuni{2500}\pmboxdrawuni{2500}\pmboxdrawuni{2500}\pmboxdrawuni{2500}\pmboxdrawuni{2500}\pmboxdrawuni{2500}\pmboxdrawuni{2500}\pmboxdrawuni{2500}\pmboxdrawuni{2500}\pmboxdrawuni{2500}\pmboxdrawuni{2500}\pmboxdrawuni{2500}\pmboxdrawuni{2500}\pmboxdrawuni{2500}\pmboxdrawuni{2500}\pmboxdrawuni{2500}\pmboxdrawuni{2500}\pmboxdrawuni{2500}\pmboxdrawuni{2500}\pmboxdrawuni{2500}\pmboxdrawuni{2500}\pmboxdrawuni{2500}\pmboxdrawuni{2500}\pmboxdrawuni{2500}\pmboxdrawuni{2500}\pmboxdrawuni{2500}\pmboxdrawuni{2500}\pmboxdrawuni{2500}\pmboxdrawuni{2500}\pmboxdrawuni{2500}\pmboxdrawuni{2500}\pmboxdrawuni{2500}\pmboxdrawuni{2500}\pmboxdrawuni{2500}\pmboxdrawuni{2500}\pmboxdrawuni{2500}\pmboxdrawuni{2500}\pmboxdrawuni{2500}\pmboxdrawuni{2500}\pmboxdrawuni{2500}\pmboxdrawuni{2500}\pmboxdrawuni{2500}\pmboxdrawuni{2500}\pmboxdrawuni{2500}\pmboxdrawuni{2500}\pmboxdrawuni{2500}\pmboxdrawuni{2500}\pmboxdrawuni{2500}\pmboxdrawuni{2500}\pmboxdrawuni{2500}\pmboxdrawuni{2500}\pmboxdrawuni{2500}\pmboxdrawuni{2500}\pmboxdrawuni{2500}\pmboxdrawuni{2500}\pmboxdrawuni{2500}\pmboxdrawuni{2500}\pmboxdrawuni{2500}\pmboxdrawuni{2510}}
\cl{~~~shapes~~~~~\pmboxdrawuni{2502}~1~~INSTRUCTION~~~CLAUDE.md~·~rules~·~skills~·~the~prompt~~~~~\pmboxdrawuni{2502}}
\cl{~~~intent~~~~~\pmboxdrawuni{2502}~~~~Read~as~context~and~generally~followed.~No~guarantee,~~~~~\pmboxdrawuni{2502}}
\cl{~~~~~~~~~~~~~~\pmboxdrawuni{2502}~~~~especially~where~two~instructions~conflict.~~~~~~~~~~~~~~~\pmboxdrawuni{2502}}
\cl{~~~~~~~~~~~~~~\pmboxdrawuni{2514}\pmboxdrawuni{2500}\pmboxdrawuni{2500}\pmboxdrawuni{2500}\pmboxdrawuni{2500}\pmboxdrawuni{2500}\pmboxdrawuni{2500}\pmboxdrawuni{2500}\pmboxdrawuni{2500}\pmboxdrawuni{2500}\pmboxdrawuni{2500}\pmboxdrawuni{2500}\pmboxdrawuni{2500}\pmboxdrawuni{2500}\pmboxdrawuni{2500}\pmboxdrawuni{2500}\pmboxdrawuni{2500}\pmboxdrawuni{2500}\pmboxdrawuni{2500}\pmboxdrawuni{2500}\pmboxdrawuni{2500}\pmboxdrawuni{2500}\pmboxdrawuni{2500}\pmboxdrawuni{2500}\pmboxdrawuni{2500}\pmboxdrawuni{2500}\pmboxdrawuni{2500}\pmboxdrawuni{2500}\pmboxdrawuni{2500}\pmboxdrawuni{2500}\pmboxdrawuni{2500}\pmboxdrawuni{2500}\pmboxdrawuni{2500}\pmboxdrawuni{2500}\pmboxdrawuni{2500}\pmboxdrawuni{2500}\pmboxdrawuni{2500}\pmboxdrawuni{2500}\pmboxdrawuni{2500}\pmboxdrawuni{2500}\pmboxdrawuni{2500}\pmboxdrawuni{2500}\pmboxdrawuni{2500}\pmboxdrawuni{2500}\pmboxdrawuni{2500}\pmboxdrawuni{2500}\pmboxdrawuni{2500}\pmboxdrawuni{2500}\pmboxdrawuni{2500}\pmboxdrawuni{2500}\pmboxdrawuni{2500}\pmboxdrawuni{2500}\pmboxdrawuni{2500}\pmboxdrawuni{2500}\pmboxdrawuni{2500}\pmboxdrawuni{2500}\pmboxdrawuni{2500}\pmboxdrawuni{2500}\pmboxdrawuni{2500}\pmboxdrawuni{2500}\pmboxdrawuni{2500}\pmboxdrawuni{2500}\pmboxdrawuni{2500}\pmboxdrawuni{2518}}
\cl{~~~~~~~~~~~~~~~~~~~~~~~~~~~~~~~~~~~~~~~\pmboxdrawuni{2502}}
\cl{~~~ENFORCED~~~\pmboxdrawuni{250C}\pmboxdrawuni{2500}\pmboxdrawuni{2500}\pmboxdrawuni{2500}\pmboxdrawuni{2500}\pmboxdrawuni{2500}\pmboxdrawuni{2500}\pmboxdrawuni{2500}\pmboxdrawuni{2500}\pmboxdrawuni{2500}\pmboxdrawuni{2500}\pmboxdrawuni{2500}\pmboxdrawuni{2500}\pmboxdrawuni{2500}\pmboxdrawuni{2500}\pmboxdrawuni{2500}\pmboxdrawuni{2500}\pmboxdrawuni{2500}\pmboxdrawuni{2500}\pmboxdrawuni{2500}\pmboxdrawuni{2500}\pmboxdrawuni{2500}\pmboxdrawuni{2500}\pmboxdrawuni{2500}\pmboxdrawuni{2500}\pmboxdrawuni{2500}\pmboxdrawuni{2500}\pmboxdrawuni{2500}\pmboxdrawuni{2500}\pmboxdrawuni{2500}\pmboxdrawuni{2500}\pmboxdrawuni{2500}\pmboxdrawuni{2500}\pmboxdrawuni{2500}\pmboxdrawuni{2500}\pmboxdrawuni{2500}\pmboxdrawuni{2500}\pmboxdrawuni{2500}\pmboxdrawuni{2500}\pmboxdrawuni{2500}\pmboxdrawuni{2500}\pmboxdrawuni{2500}\pmboxdrawuni{2500}\pmboxdrawuni{2500}\pmboxdrawuni{2500}\pmboxdrawuni{2500}\pmboxdrawuni{2500}\pmboxdrawuni{2500}\pmboxdrawuni{2500}\pmboxdrawuni{2500}\pmboxdrawuni{2500}\pmboxdrawuni{2500}\pmboxdrawuni{2500}\pmboxdrawuni{2500}\pmboxdrawuni{2500}\pmboxdrawuni{2500}\pmboxdrawuni{2500}\pmboxdrawuni{2500}\pmboxdrawuni{2500}\pmboxdrawuni{2500}\pmboxdrawuni{2500}\pmboxdrawuni{2500}\pmboxdrawuni{2500}\pmboxdrawuni{2510}}
\cl{~~~blocks~~~~~\pmboxdrawuni{2502}~2~~PERMISSION~~~~deny~→~ask~→~allow~·~permission~modes~~~~~~~\pmboxdrawuni{2502}}
\cl{~~~action~~~~~\pmboxdrawuni{2502}~~~~Product-enforced.~A~deny~outranks~every~ask~and~allow.~~~~\pmboxdrawuni{2502}}
\cl{~~~~~~~~~~~~~~\pmboxdrawuni{251C}\pmboxdrawuni{2500}\pmboxdrawuni{2500}\pmboxdrawuni{2500}\pmboxdrawuni{2500}\pmboxdrawuni{2500}\pmboxdrawuni{2500}\pmboxdrawuni{2500}\pmboxdrawuni{2500}\pmboxdrawuni{2500}\pmboxdrawuni{2500}\pmboxdrawuni{2500}\pmboxdrawuni{2500}\pmboxdrawuni{2500}\pmboxdrawuni{2500}\pmboxdrawuni{2500}\pmboxdrawuni{2500}\pmboxdrawuni{2500}\pmboxdrawuni{2500}\pmboxdrawuni{2500}\pmboxdrawuni{2500}\pmboxdrawuni{2500}\pmboxdrawuni{2500}\pmboxdrawuni{2500}\pmboxdrawuni{2500}\pmboxdrawuni{2500}\pmboxdrawuni{2500}\pmboxdrawuni{2500}\pmboxdrawuni{2500}\pmboxdrawuni{2500}\pmboxdrawuni{2500}\pmboxdrawuni{2500}\pmboxdrawuni{2500}\pmboxdrawuni{2500}\pmboxdrawuni{2500}\pmboxdrawuni{2500}\pmboxdrawuni{2500}\pmboxdrawuni{2500}\pmboxdrawuni{2500}\pmboxdrawuni{2500}\pmboxdrawuni{2500}\pmboxdrawuni{2500}\pmboxdrawuni{2500}\pmboxdrawuni{2500}\pmboxdrawuni{2500}\pmboxdrawuni{2500}\pmboxdrawuni{2500}\pmboxdrawuni{2500}\pmboxdrawuni{2500}\pmboxdrawuni{2500}\pmboxdrawuni{2500}\pmboxdrawuni{2500}\pmboxdrawuni{2500}\pmboxdrawuni{2500}\pmboxdrawuni{2500}\pmboxdrawuni{2500}\pmboxdrawuni{2500}\pmboxdrawuni{2500}\pmboxdrawuni{2500}\pmboxdrawuni{2500}\pmboxdrawuni{2500}\pmboxdrawuni{2500}\pmboxdrawuni{2500}\pmboxdrawuni{2524}}
\cl{~~~~~~~~~~~~~~\pmboxdrawuni{2502}~3~~SANDBOX~~~~~~~filesystem~and~network,~per~command~~~~~~~~~\pmboxdrawuni{2502}}
\cl{~~~~~~~~~~~~~~\pmboxdrawuni{2502}~~~~OS-enforced:~Seatbelt~on~macOS;~bubblewrap~on~Linux,~WSL2.\pmboxdrawuni{2502}}
\cl{~~~~~~~~~~~~~~\pmboxdrawuni{2502}~~~~Not~available~on~native~Windows.~~~~~~~~~~~~~~~~~~~~~~~~~~\pmboxdrawuni{2502}}
\cl{~~~~~~~~~~~~~~\pmboxdrawuni{251C}\pmboxdrawuni{2500}\pmboxdrawuni{2500}\pmboxdrawuni{2500}\pmboxdrawuni{2500}\pmboxdrawuni{2500}\pmboxdrawuni{2500}\pmboxdrawuni{2500}\pmboxdrawuni{2500}\pmboxdrawuni{2500}\pmboxdrawuni{2500}\pmboxdrawuni{2500}\pmboxdrawuni{2500}\pmboxdrawuni{2500}\pmboxdrawuni{2500}\pmboxdrawuni{2500}\pmboxdrawuni{2500}\pmboxdrawuni{2500}\pmboxdrawuni{2500}\pmboxdrawuni{2500}\pmboxdrawuni{2500}\pmboxdrawuni{2500}\pmboxdrawuni{2500}\pmboxdrawuni{2500}\pmboxdrawuni{2500}\pmboxdrawuni{2500}\pmboxdrawuni{2500}\pmboxdrawuni{2500}\pmboxdrawuni{2500}\pmboxdrawuni{2500}\pmboxdrawuni{2500}\pmboxdrawuni{2500}\pmboxdrawuni{2500}\pmboxdrawuni{2500}\pmboxdrawuni{2500}\pmboxdrawuni{2500}\pmboxdrawuni{2500}\pmboxdrawuni{2500}\pmboxdrawuni{2500}\pmboxdrawuni{2500}\pmboxdrawuni{2500}\pmboxdrawuni{2500}\pmboxdrawuni{2500}\pmboxdrawuni{2500}\pmboxdrawuni{2500}\pmboxdrawuni{2500}\pmboxdrawuni{2500}\pmboxdrawuni{2500}\pmboxdrawuni{2500}\pmboxdrawuni{2500}\pmboxdrawuni{2500}\pmboxdrawuni{2500}\pmboxdrawuni{2500}\pmboxdrawuni{2500}\pmboxdrawuni{2500}\pmboxdrawuni{2500}\pmboxdrawuni{2500}\pmboxdrawuni{2500}\pmboxdrawuni{2500}\pmboxdrawuni{2500}\pmboxdrawuni{2500}\pmboxdrawuni{2500}\pmboxdrawuni{2500}\pmboxdrawuni{2524}}
\cl{~~~~~~~~~~~~~~\pmboxdrawuni{2502}~4~~ISOLATION~~~~~container~·~virtual~machine~·~dev~container~\pmboxdrawuni{2502}}
\cl{~~~~~~~~~~~~~~\pmboxdrawuni{2502}~~~~The~only~layer~that~contains~a~failure~of~layers~2~or~3.~~\pmboxdrawuni{2502}}
\cl{~~~~~~~~~~~~~~\pmboxdrawuni{2514}\pmboxdrawuni{2500}\pmboxdrawuni{2500}\pmboxdrawuni{2500}\pmboxdrawuni{2500}\pmboxdrawuni{2500}\pmboxdrawuni{2500}\pmboxdrawuni{2500}\pmboxdrawuni{2500}\pmboxdrawuni{2500}\pmboxdrawuni{2500}\pmboxdrawuni{2500}\pmboxdrawuni{2500}\pmboxdrawuni{2500}\pmboxdrawuni{2500}\pmboxdrawuni{2500}\pmboxdrawuni{2500}\pmboxdrawuni{2500}\pmboxdrawuni{2500}\pmboxdrawuni{2500}\pmboxdrawuni{2500}\pmboxdrawuni{2500}\pmboxdrawuni{2500}\pmboxdrawuni{2500}\pmboxdrawuni{2500}\pmboxdrawuni{2500}\pmboxdrawuni{2500}\pmboxdrawuni{2500}\pmboxdrawuni{2500}\pmboxdrawuni{2500}\pmboxdrawuni{2500}\pmboxdrawuni{2500}\pmboxdrawuni{2500}\pmboxdrawuni{2500}\pmboxdrawuni{2500}\pmboxdrawuni{2500}\pmboxdrawuni{2500}\pmboxdrawuni{2500}\pmboxdrawuni{2500}\pmboxdrawuni{2500}\pmboxdrawuni{2500}\pmboxdrawuni{2500}\pmboxdrawuni{2500}\pmboxdrawuni{2500}\pmboxdrawuni{2500}\pmboxdrawuni{2500}\pmboxdrawuni{2500}\pmboxdrawuni{2500}\pmboxdrawuni{2500}\pmboxdrawuni{2500}\pmboxdrawuni{2500}\pmboxdrawuni{2500}\pmboxdrawuni{2500}\pmboxdrawuni{2500}\pmboxdrawuni{2500}\pmboxdrawuni{2500}\pmboxdrawuni{2500}\pmboxdrawuni{2500}\pmboxdrawuni{2500}\pmboxdrawuni{2500}\pmboxdrawuni{2500}\pmboxdrawuni{2500}\pmboxdrawuni{2500}\pmboxdrawuni{2518}}
\cl{}
\cl{~~~HOOKS~cut~across~layers~1~and~2~in~both~directions:~a~PreToolUse~hook~exiting~2}
\cl{~~~blocks~a~call~before~permission~rules~are~evaluated,~and~no~hook~decision~can}
\cl{~~~bypass~a~matching~deny~rule.~A~mod~is~the~exception~where~the~built-in~guard}
\cl{~~~is~not~loaded~(Chapter~18).}
\end{codefig}
\figcaption{Figure 1 — The agentic control stack. Layers 1 and 2 are the product's; layers 3 and 4 are the operating system's. Only layers 2 to 4 constrain what can happen.}
\end{figure}

An instruction that must hold belongs at layer 2 or below. An instruction that merely should hold belongs at layer 1, where it costs context and buys a tendency. Chapters 7, 17 and 31 develop the enforced layers; Chapters 9 and 10 develop the advisory one.

\FloatBarrier
\setcounter{section}{2}
\section{Where the work runs}

The same engine is reachable through several surfaces, and your repository's \texttt{CLAUDE.\allowbreak{}md}, settings and MCP servers apply across most of them [2], [45]. The exception is a cloud project's threads, which load each repository's \texttt{CLAUDE.\allowbreak{}md} but not its skills, \texttt{.\allowbreak{}mcp.\allowbreak{}json} servers or \texttt{.\allowbreak{}claude/\allowbreak{}settings.\allowbreak{}json} [46].

\begingroup
\def\tblrows{%
Terminal CLI & Earliest feature access, scripting, remote hosts, full command surface & Requires comfort with a terminal \\
VS Code extension & Inline diffs, \texttt{@}-mentions, plan review, conversation history beside files [47] & Some controls are terminal-only \\
JetBrains plugin & IntelliJ, PyCharm, WebStorm and related IDEs [48] & Requires the CLI installed separately \\
Desktop app & Parallel sessions with git isolation, visual diff review, scheduled local tasks, computer use [49] & A different interaction model from the terminal \\
Web (claude.ai/code) & Cloud sandboxes, work on repositories you do not have locally, long tasks you check later [50] & Runs off your machine; counts toward plan usage \\
Mobile & Starting, monitoring and steering tasks from a phone [51] & Small control surface; poor for reviewing diffs \\
Remote Control & Supervising a session that is executing on your own machine, started from the CLI, Desktop or VS Code [41], [45] & Transcript is relayed and stored server-side while connected \\
}%
\def\tblbody{\begin{minipage}{\textwidth}\boxcaption{Table 2 — Claude Code surfaces and their trade-offs}
{\footnotesize\begin{tabular}{L{28.7mm}L{69.3mm}L{41.8mm}}
\toprule
\textbf{Surface} & \textbf{Best for} & \textbf{Trade-off} \\
\midrule
\tblrows
\bottomrule\end{tabular}}\end{minipage}}%
\begingroup
\def\sloppy{\tolerance 9999\emergencystretch 3em\hfuzz 200pt\vfuzz 200pt}%
\hbadness=10000\vbadness=10000\hfuzz=200pt\vfuzz=200pt
\global\setbox\tblbox=\hbox{\tblbody}%
\endgroup
\par\addvspace{2.6mm}
\ifdim\dimexpr\ht\tblbox+\dp\tblbox\relax>0.55\textheight
  \typeout{HANDBOOK-TABLE broken \the\dimexpr\ht\tblbox+\dp\tblbox\relax}%
  \tabcaption{Table 2 — Claude Code surfaces and their trade-offs}
  {\footnotesize\begin{longtable}{L{28.7mm}L{69.3mm}L{41.8mm}}
  \toprule
\textbf{Surface} & \textbf{Best for} & \textbf{Trade-off} \\
\midrule\endfirsthead
  \multicolumn{3}{@{}l@{}}{%
  \sffamily\footnotesize\itshape\color{inkgrey}Table 2 — Claude Code surfaces and their trade-offs \textemdash\ continued}\\[1.2mm]
  \toprule
\textbf{Surface} & \textbf{Best for} & \textbf{Trade-off} \\
\midrule\endhead
  \bottomrule\endfoot
  \bottomrule\endlastfoot
  \tblrows
  \end{longtable}}%
\else
  \typeout{HANDBOOK-TABLE atomic \the\dimexpr\ht\tblbox+\dp\tblbox\relax}%
  \noindent\tblbody
\fi
\par\addvspace{2.6mm}
\endgroup

\textbf{PRACTITIONER NOTE:} the source course prefers VS Code because it puts an explorer, an editor, a terminal and several Claude tabs on one screen [44]. That is a reasonable default for a beginner. It is not a requirement, and several controls in Part V are easier to observe in the CLI.

\FloatBarrier
\setcounter{section}{3}
\section{Appropriate first tasks}

Choose work that is reversible and cheap to inspect:

\begin{itemize}
\item organise copies of documents;
\item produce a report outline from supplied notes;
\item compare two files and return a difference table;
\item research a bounded question and save a source ledger;
\item explain an unfamiliar repository without changing it.
\end{itemize}

Avoid, as a first task, anything that touches an account you cannot afford to have altered. Capability should expand only after verification habits do.

\subsection*{Exercise}

Write a four-line brief and keep it. You will run it in Chapter 5.

\begin{codeblock}{8.0}{9.4}
\cl{Outcome:~~~~A~one-page~brief~built~from~the~files~in~notes/.}
\cl{Context:~~~~Use~only~local~files.~Identify~evidence~gaps~rather~than~filling~them.}
\cl{Boundaries:~Do~not~edit~or~delete~anything~in~notes/~or~sources/.~No~network~access.}
\cl{Evidence:~~~Save~to~drafts/brief-v1.md~with~a~claim-to-source~table.}
\end{codeblock}

\emph{A worked solution is given in Appendix M.}

\begingroup
\def\kprows{%
\item Claude Code is an agent, not an assistant: it edits files, runs commands, reaches accounts and schedules its own work.
\item Four elements must be explicit in every instruction — Outcome, Context, Boundaries, Evidence. Omitting the fourth is what turns capability into rework.
\item The control stack has four layers and only the lower three are enforced. An instruction is a preference; a permission rule is a control.
\item Choose first tasks that are reversible and cheap to inspect. Capability should expand only after verification habits do.
}%
\def\kpbody{\begin{minipage}{\textwidth}\subsection*{Key points}\begin{itemize}\kprows\end{itemize}\end{minipage}}%
\begingroup
\def\sloppy{\tolerance 9999\emergencystretch 3em\hfuzz 200pt\vfuzz 200pt}%
\hbadness=10000\vbadness=10000\hfuzz=200pt\vfuzz=200pt
\global\setbox\kpbox=\hbox{\kpbody}%
\endgroup
\par\addvspace{4.2mm}
\ifdim\dimexpr\ht\kpbox+\dp\kpbox\relax>0.30\textheight
  \typeout{HANDBOOK-KEYPOINTS broken \the\dimexpr\ht\kpbox+\dp\kpbox\relax}%
  \subsection*{Key points}
  \begin{itemize}\kprows\end{itemize}
\else
  \typeout{HANDBOOK-KEYPOINTS atomic \the\dimexpr\ht\kpbox+\dp\kpbox\relax}%
  \noindent\kpbody
\fi
\par\addvspace{1.4mm}
\endgroup

\FloatBarrier
\renewcommand{\chaptertitlelabel}{Part I \textperiodcentered\ Chapter 2}
\setcounter{chapter}{1}
\chapter{Install, verify and authenticate}

\FloatBarrier
\setcounter{section}{0}
\section{Requirements}

As verified on 12 November 2026, Claude Code supports macOS 13 and later, Windows 10 1809 or Windows Server 2019 and later, Ubuntu 20.04+, Debian 10+ and Alpine 3.19+, on an x64 or ARM64 processor with at least 4 GB of RAM, network access from a supported country, and Bash, Zsh, PowerShell or CMD. A Pro, Max, Ultra, Team, Enterprise or Console account is required — the free claude.ai plan does not include Claude Code — or a configured cloud provider [2], [52].

Not every feature follows the platform matrix. Sandboxing does not run on native Windows [28]; Chrome integration does not run under WSL [21]; artifacts, cloud sessions, routines and channels are unavailable on third-party model providers [20], [50], [53], [54]. Appendix D lists the dependencies that most often surprise people.

\FloatBarrier
\setcounter{section}{1}
\section{Installation}

The native installer is the official default and updates itself in the background [2], [52].

\begin{codeblock}{9.0}{10.6}
\cl{\#~macOS,~Linux,~WSL}
\cl{curl~-fsSL~https://claude.ai/install.sh~|~bash}
\end{codeblock}

\begin{codeblock}{9.0}{10.6}
\cl{\#~Windows~PowerShell}
\cl{irm~https://claude.ai/install.ps1~|~iex}
\end{codeblock}

On Alpine and other musl-based distributions the command above fails until \texttt{bash} and \texttt{curl} are installed; the runtime also needs \texttt{libgcc}, \texttt{libstdc++} and \texttt{ripgrep} (\texttt{apk add bas\allowbreak{}h curl libg\allowbreak{}cc libstdc+\allowbreak{}+ ripgrep}), with \texttt{USE\_\allowbreak{}BUILTIN\_\allowbreak{}RIPGREP=\allowbreak{}0} set [52].

\begin{codeblock}{8.0}{9.4}
\cl{REM~Windows~CMD}
\cl{curl~-fsSL~https://claude.ai/install.cmd~-o~install.cmd~\&\&~install.cmd~\&\&~del~install.cmd}
\end{codeblock}

Package-manager installations — Homebrew, WinGet, and the Linux repositories — do \textbf{not} auto-update by default and must be upgraded explicitly [2], [52]:

\begin{codeblock}{8.0}{9.4}
\cl{brew~install~-{}-cask~claude-code~~~~~~~~\#~stable~channel;~claude-code@latest~tracks~latest}
\cl{winget~install~Anthropic.ClaudeCode~~~~\#~upgrade~with:~winget~upgrade~Anthropic.ClaudeCode}
\end{codeblock}

Signed \texttt{apt}, \texttt{dnf} and \texttt{apk} repositories serve Debian, Ubuntu, Fedora, RHEL and Alpine, each with a \texttt{stable} channel about a week behind and a \texttt{latest} channel [52].

\begin{calloutbox}{palebrass}{brassdark}{2.0mm}
\textbf{CAUTION:} a shell installer executes code on your machine with your privileges. Never substitute a command copied from a video description, an advertisement, a forum post or an unverified repository for the documented one. Verify the domain character by character.
\end{calloutbox}

\FloatBarrier
\setcounter{section}{2}
\section{Verify before you log in}

\begin{codeblock}{9.0}{10.6}
\cl{claude~-{}-version}
\cl{claude~doctor}
\end{codeblock}

\texttt{claude doct\allowbreak{}or} prints read-only installation and settings diagnostics without starting a session [52], [55]. Its in-session counterpart, \texttt{/\allowbreak{}doctor}, also proposes fixes and asks before applying them; from 2.1.206 that includes trims for a checked-in \texttt{CLAUDE.\allowbreak{}md}, and \texttt{/\allowbreak{}doctor prom\allowbreak{}pt-\allowbreak{}audit} \emph{(2.1.283 or later)} checks instruction files for outdated or conflicting content [25], [56], [57]. Version 1.0 gave the trims to \texttt{claude doct\allowbreak{}or}. Then start the product and complete browser authentication:

\begin{codeblock}{9.0}{10.6}
\cl{claude}
\end{codeblock}

Inside a session, \texttt{/\allowbreak{}login} reauthenticates or switches account and \texttt{/\allowbreak{}logout} signs out [37], [58]. Credentials are stored in the macOS Keychain where available and protected by file permissions on Windows and Linux [59].

\textbf{CAUTION:} several features require a claude.ai subscription login specifically and are hidden or refused when \texttt{ANTHROPIC\_\allowbreak{}API\_\allowbreak{}KEY}, \texttt{ANTHROPIC\_\allowbreak{}AUTH\_\allowbreak{}TOKEN} or an \texttt{api\allowbreak{}Key\allowbreak{}Helper} takes precedence — Chrome integration, artifacts, routines, cloud sessions and teleport among them [20], [21], [50], [53]. Some simply disappear — on a cloud provider \texttt{/\allowbreak{}schedule} answers \texttt{Unknown com\allowbreak{}mand}, while with a Console API key and feature-flag fetching on it says it needs a claude.ai subscription — and the cloud commands name the credential in their error [20], [60]. If a documented command appears to be missing, check the authentication path before you check the version.

\FloatBarrier
\setcounter{section}{3}
\section{Windows: native or WSL 2}

\begin{itemize}
\item \textbf{Native Windows} suits Windows-native projects and tooling. Sandboxing is unavailable [28]. Git for Windows is optional: it provides the Git Bash that the Bash and Monitor tools need, and without it Claude Code runs shell commands through PowerShell [52].
\item \textbf{WSL 2} suits Linux toolchains and is required for the sandboxed Bash tool; WSL 1 does not support sandboxing [28], [52]. Chrome integration is not supported under WSL [21].
\end{itemize}

\begin{calloutbox}{palebrass}{brassdark}{2.0mm}
\textbf{CAUTION — Windows WebDAV.} Anthropic recommends against enabling WebDAV or allowing access to paths such as \texttt{\textbackslash{}\textbackslash{}*} that may contain WebDAV subdirectories, because doing so can let Claude Code trigger network requests to remote hosts outside the permission system [59].
\end{calloutbox}

\subsection*{Checkpoint}

\begin{itemize}
\item \texttt{claude -\allowbreak{}-\allowbreak{}version} prints a version.
\item \texttt{claude doct\allowbreak{}or} reports no unresolved critical issue.
\item \texttt{claude} starts and shows the account you intended.
\item You know whether your installation auto-updates.
\item You know whether you are authenticated with a subscription login or an API key.
\end{itemize}

\begingroup
\def\kprows{%
\item The native installer auto-updates; package-manager installations do not by default (Homebrew and WinGet can opt in with \texttt{CLAUDE\_\allowbreak{}CODE\_\allowbreak{}PACKAGE\_\allowbreak{}MANAGER\_\allowbreak{}AUTO\_\allowbreak{}UPDATE=\allowbreak{}1}).
\item Verify with \texttt{claude -\allowbreak{}-\allowbreak{}version} and \texttt{claude doct\allowbreak{}or} before you log in, and know which authentication path you are on.
\item Several features — Chrome, artifacts, routines, cloud sessions, teleport — require a claude.ai subscription login and are hidden or refused behind an API key.
\item Sandboxing does not run on native Windows; Chrome integration does not run under WSL. Choose the Windows path accordingly.
}%
\def\kpbody{\begin{minipage}{\textwidth}\subsection*{Key points}\begin{itemize}\kprows\end{itemize}\end{minipage}}%
\begingroup
\def\sloppy{\tolerance 9999\emergencystretch 3em\hfuzz 200pt\vfuzz 200pt}%
\hbadness=10000\vbadness=10000\hfuzz=200pt\vfuzz=200pt
\global\setbox\kpbox=\hbox{\kpbody}%
\endgroup
\par\addvspace{4.2mm}
\ifdim\dimexpr\ht\kpbox+\dp\kpbox\relax>0.30\textheight
  \typeout{HANDBOOK-KEYPOINTS broken \the\dimexpr\ht\kpbox+\dp\kpbox\relax}%
  \subsection*{Key points}
  \begin{itemize}\kprows\end{itemize}
\else
  \typeout{HANDBOOK-KEYPOINTS atomic \the\dimexpr\ht\kpbox+\dp\kpbox\relax}%
  \noindent\kpbody
\fi
\par\addvspace{1.4mm}
\endgroup

\FloatBarrier
\renewcommand{\chaptertitlelabel}{Part I \textperiodcentered\ Chapter 3}
\setcounter{chapter}{2}
\chapter{Surfaces, projects and boundaries}

\FloatBarrier
\setcounter{section}{0}
\section{Set up the IDE}

Install Visual Studio Code, open \textbf{Extensions}, search for \textbf{Claude Code}, and confirm the publisher is \textbf{Anthropic} before installing [47]. From the Command Palette, \textbf{Claude Code: Open in New Tab} gives the conversation a full editor tab and \textbf{Open in Terminal} launches the CLI inside the IDE. In VS Code, Chrome integration needs the browser extension but no flag: a session connects when you type \texttt{@browser}, or as it starts once \textbf{Enabled by default} is on in \texttt{/\allowbreak{}chrome} \emph{(2.1.287 or later)} — a setting the CLI shares [21], [47].

\FloatBarrier
\setcounter{section}{1}
\section{Create the practice workspace}

Create \texttt{research-\allowbreak{}brief-\allowbreak{}workspace}, open it with \textbf{File → Open Folder}, and create the directories of “The running project”. Add two harmless text files to \texttt{notes/\allowbreak{}}.

The opened folder is the default working boundary. In Manual mode Claude Code asks before its file tools read or write outside that folder and its subfolders. \textbf{The boundary is a permission prompt, not a wall}: a Bash command you approve can still write anywhere your account can [59]. In auto mode, file reads outside the working directories run without a prompt, except reads from network paths. In an interactive session the first such read by a file tool asks for consent, and your answer decides whether it asks again; \texttt{-\allowbreak{}p} runs and background sessions never ask [33]. Answering that reads should be blocked sets \texttt{permissions.\allowbreak{}block\allowbreak{}Reads\allowbreak{}Outside\allowbreak{}Working\allowbreak{}Directories}, which makes the file tools \emph{refuse} those reads in every mode; its counterpart for writes refuses \texttt{Write}, \texttt{Edit} and \texttt{Notebook\allowbreak{}Edit} outside the working directories but does not limit shell commands, MCP tools or hooks. The sandbox limits sandboxed shell commands; hooks and local MCP servers run outside it, and only isolation around the Claude Code process itself contains them [28], [33], [61]. Version 1.0 said Manual mode could write only inside the folder. Extend the boundary deliberately with \texttt{-\allowbreak{}-\allowbreak{}add-\allowbreak{}dir} or \texttt{/\allowbreak{}add-\allowbreak{}dir}, not reflexively [29].

\begin{calloutbox}{palebrass}{brassdark}{2.0mm}
\textbf{CAUTION:} an opened folder is not a security sandbox. Claude Code can request additional directories, run commands, reach the network and use connected services subject to permissions. Do not open your home directory as a project: trust acceptance there is held for the session only and is never written to disk, so the prompt reappears at every launch, and the blast radius of a permission mistake is your entire personal filesystem [59].
\end{calloutbox}

\FloatBarrier
\setcounter{section}{2}
\section{Monorepos and large trees}

In a monorepo, nested \texttt{CLAUDE.\allowbreak{}md} files from other teams are picked up from every ancestor directory. Use \texttt{claude\allowbreak{}Md\allowbreak{}Excludes} to skip them, path-scoped rules to load instructions only for matching files, and per-package skills to keep guidance local [25], [62]. Sparse worktrees and code intelligence are covered in [62].

\subsection*{Exercise}

In Plan mode, ask:

\begin{codeblock}{7.0}{8.3}
\cl{Inspect~this~workspace~without~modifying~it.~Describe~its~structure,~identify~what~is~missing}
\cl{for~a~research-brief~workflow,~and~propose~the~smallest~safe~next~step.~Do~not~access~anything}
\cl{outside~this~folder.}
\end{codeblock}

\emph{A worked solution is given in Appendix M.}

\subsection*{Checkpoint}

The response names only files inside the practice folder and proposes — rather than performs — changes.

\begingroup
\def\kprows{%
\item The opened folder is the default working boundary, not a security sandbox.
\item In Manual mode the file tools ask before reading or writing outside it, but an approved Bash command can write anywhere.
\item Never open a home directory as a project: trust acceptance there is never written to disk and the blast radius is your whole filesystem.
\item In a monorepo, use \texttt{claude\allowbreak{}Md\allowbreak{}Excludes} and path-scoped rules before you use a bigger context window.
}%
\def\kpbody{\begin{minipage}{\textwidth}\subsection*{Key points}\begin{itemize}\kprows\end{itemize}\end{minipage}}%
\begingroup
\def\sloppy{\tolerance 9999\emergencystretch 3em\hfuzz 200pt\vfuzz 200pt}%
\hbadness=10000\vbadness=10000\hfuzz=200pt\vfuzz=200pt
\global\setbox\kpbox=\hbox{\kpbody}%
\endgroup
\par\addvspace{4.2mm}
\ifdim\dimexpr\ht\kpbox+\dp\kpbox\relax>0.30\textheight
  \typeout{HANDBOOK-KEYPOINTS broken \the\dimexpr\ht\kpbox+\dp\kpbox\relax}%
  \subsection*{Key points}
  \begin{itemize}\kprows\end{itemize}
\else
  \typeout{HANDBOOK-KEYPOINTS atomic \the\dimexpr\ht\kpbox+\dp\kpbox\relax}%
  \noindent\kpbody
\fi
\par\addvspace{1.4mm}
\endgroup

\FloatBarrier
\renewcommand{\chaptertitlelabel}{Part I \textperiodcentered\ Chapter 4}
\setcounter{chapter}{3}
\chapter{Modes, models, effort and usage}

\FloatBarrier
\setcounter{section}{0}
\section{The controls that matter first}

\begin{itemize}
\item \textbf{Permission mode} — switched with \texttt{Shift+Tab} in the CLI, the mode indicator in VS Code, or the mode selector in Desktop [33]. The cycle runs \texttt{default} → \texttt{accept\allowbreak{}Edits} → \texttt{plan} in every session, with two further modes slotting in after \texttt{plan} when they are enabled; \texttt{dont\allowbreak{}Ask} never appears in it at all. Table 5 gives the full picture. Covered in Chapter 7.
\item \textbf{Model} — \texttt{/\allowbreak{}model}, or the picker. Availability changes; do not memorise a model name from a recording [63].
\item \textbf{Effort} — \texttt{/\allowbreak{}effort}, with levels \texttt{low}, \texttt{medium}, \texttt{high}, \texttt{xhigh} and \texttt{max} on every surface, plus \texttt{auto}, which the command accepts and the launch flag refuses [37], [63]. Section 4.2 records what each surface accepted. \texttt{ultracode} is not an effort level but a setting the same command switches on; see Chapter 22.
\item \textbf{Fast mode} — \texttt{/\allowbreak{}fast}, up to 2.5 times faster output from a supported model at a higher cost per token; the model is the same [64].
\item \textbf{Advisor} — \texttt{/\allowbreak{}advisor} pairs the working model with a typically stronger model consulted at key moments. It is experimental, and needs the Anthropic API or Claude Platform on AWS [65].
\item \textbf{Context attachment} — \texttt{@} mentions and the attachment control.
\item \textbf{Interrupt} — \texttt{Esc} stops the current response or tool call and keeps the work done so far; \texttt{Ctrl+C} also interrupts, and with nothing running clears the input, then exits on a second press [66].
\item \textbf{Settings from the prompt} — \texttt{/\allowbreak{}config} opens settings or sets an individual key, including keys with no dedicated command [37].
\item \textbf{Command menu} — type \texttt{/\allowbreak{}} at the start of a message.
\end{itemize}

\FloatBarrier
\setcounter{section}{1}
\section{The effort vocabulary, by surface}

\texttt{/\allowbreak{}effort} and \texttt{claude -\allowbreak{}-\allowbreak{}effort} are two interfaces onto one control, and they do not accept the same words. The table records what each surface did when the value was offered to it, and each column carries the release it was observed on: the command, through \texttt{claude -\allowbreak{}p}, and the flag on 2.1.322 on 12 November 2026; the VS Code extension on an earlier release, not re-observed since. It is a configuration-stamped observation, not a product guarantee: the menu a given plan and model offer may differ, and the entry in Appendix J states the residual [37], [63].

\begingroup
\def\tblrows{%
\texttt{low} & yes & yes & yes \\
\texttt{medium} & yes & yes & yes \\
\texttt{high} & yes & yes & yes \\
\texttt{xhigh} & yes & yes & yes \\
\texttt{max} & yes & yes & yes \\
\texttt{auto} & yes & yes & no \\
}%
\def\tblbody{\begin{minipage}{\textwidth}\boxcaption{Table 3 — Effort levels accepted by surface, each column stamped with the release it was observed on}
{\footnotesize\begin{tabular}{L{26.6mm}L{34.6mm}L{38.1mm}L{37.3mm}}
\toprule
\textbf{Level} & \textbf{\texttt{/\allowbreak{}effort} (CLI) · 2.1.322} & \textbf{\texttt{/\allowbreak{}effort} (VS Code) · 2.1.246} & \textbf{\texttt{claude -\allowbreak{}-\allowbreak{}effort} · 2.1.322} \\
\midrule
\tblrows
\bottomrule\end{tabular}}\end{minipage}}%
\begingroup
\def\sloppy{\tolerance 9999\emergencystretch 3em\hfuzz 200pt\vfuzz 200pt}%
\hbadness=10000\vbadness=10000\hfuzz=200pt\vfuzz=200pt
\global\setbox\tblbox=\hbox{\tblbody}%
\endgroup
\par\addvspace{2.6mm}
\ifdim\dimexpr\ht\tblbox+\dp\tblbox\relax>0.55\textheight
  \typeout{HANDBOOK-TABLE broken \the\dimexpr\ht\tblbox+\dp\tblbox\relax}%
  \tabcaption{Table 3 — Effort levels accepted by surface, each column stamped with the release it was observed on}
  {\footnotesize\begin{longtable}{L{26.6mm}L{34.6mm}L{38.1mm}L{37.3mm}}
  \toprule
\textbf{Level} & \textbf{\texttt{/\allowbreak{}effort} (CLI) · 2.1.322} & \textbf{\texttt{/\allowbreak{}effort} (VS Code) · 2.1.246} & \textbf{\texttt{claude -\allowbreak{}-\allowbreak{}effort} · 2.1.322} \\
\midrule\endfirsthead
  \multicolumn{4}{@{}l@{}}{%
  \sffamily\footnotesize\itshape\color{inkgrey}Table 3 — Effort levels accepted by surface, each column stamped with the release it was observed on \textemdash\ continued}\\[1.2mm]
  \toprule
\textbf{Level} & \textbf{\texttt{/\allowbreak{}effort} (CLI) · 2.1.322} & \textbf{\texttt{/\allowbreak{}effort} (VS Code) · 2.1.246} & \textbf{\texttt{claude -\allowbreak{}-\allowbreak{}effort} · 2.1.322} \\
\midrule\endhead
  \bottomrule\endfoot
  \bottomrule\endlastfoot
  \tblrows
  \end{longtable}}%
\else
  \typeout{HANDBOOK-TABLE atomic \the\dimexpr\ht\tblbox+\dp\tblbox\relax}%
  \noindent\tblbody
\fi
\par\addvspace{2.6mm}
\endgroup

\textbf{yes} accepted · \textbf{no} refused · \textbf{—} not established on that surface · \textbf{†} absent from that surface's own help or usage text.

Invocation differs too. In an interactive terminal session a bare \texttt{/\allowbreak{}effort} shows the current level and a selector; in a \texttt{claude -\allowbreak{}p} run and in the VS Code extension it requires an explicit value and answers a bare call with its usage string. \texttt{/\allowbreak{}effort stat\allowbreak{}us} prints the level without changing it.

Two consequences matter in practice. \textbf{\texttt{claude -\allowbreak{}-\allowbreak{}effort auto} is refused}: the flag warns that \texttt{auto} is not a valid value and the session then runs at the default effort, so a script that sets it neither gets what it asked for nor fails. And \textbf{\texttt{ultracode} is accepted by the flag although \texttt{claude -\allowbreak{}-\allowbreak{}help} does not list it}, and is recognised by \texttt{/\allowbreak{}effort} although that command's usage string does not list it either — which is why this book treats a help string as evidence of what is supported and never as evidence of what is refused (Appendix J). \texttt{ultracode} is absent from the table because it is no longer an effort level: from 2.1.284 it is a setting, and switching it with \texttt{/\allowbreak{}effort} or the \texttt{ultracode} setting leaves the level where it is, while \texttt{claude -\allowbreak{}-\allowbreak{}effort ultr\allowbreak{}acode} also sets \texttt{xhigh} (Chapter 22) [63].

\FloatBarrier
\setcounter{section}{2}
\section{A decision rule for model and effort}

\begingroup
\def\tblrows{%
Locate a file; reformat known text & Lower-cost model, normal effort \\
Summarise supplied material & Normal effort \\
Design a project; resolve genuine ambiguity & Stronger model, high effort \\
Execute an already approved, explicit plan & Normal to high effort; verify completeness rather than creativity \\
Multi-source research; architecture trade-off & \texttt{xhigh}, or a dynamic workflow (Chapter 22) \\
}%
\def\tblbody{\begin{minipage}{\textwidth}\boxcaption{Table 4 — Suggested starting posture by work type (practitioner guidance)}
{\footnotesize\begin{tabular}{L{70.4mm}L{72.6mm}}
\toprule
\textbf{Work} & \textbf{Suggested starting posture} \\
\midrule
\tblrows
\bottomrule\end{tabular}}\end{minipage}}%
\begingroup
\def\sloppy{\tolerance 9999\emergencystretch 3em\hfuzz 200pt\vfuzz 200pt}%
\hbadness=10000\vbadness=10000\hfuzz=200pt\vfuzz=200pt
\global\setbox\tblbox=\hbox{\tblbody}%
\endgroup
\par\addvspace{2.6mm}
\ifdim\dimexpr\ht\tblbox+\dp\tblbox\relax>0.55\textheight
  \typeout{HANDBOOK-TABLE broken \the\dimexpr\ht\tblbox+\dp\tblbox\relax}%
  \tabcaption{Table 4 — Suggested starting posture by work type (practitioner guidance)}
  {\footnotesize\begin{longtable}{L{70.4mm}L{72.6mm}}
  \toprule
\textbf{Work} & \textbf{Suggested starting posture} \\
\midrule\endfirsthead
  \multicolumn{2}{@{}l@{}}{%
  \sffamily\footnotesize\itshape\color{inkgrey}Table 4 — Suggested starting posture by work type (practitioner guidance) \textemdash\ continued}\\[1.2mm]
  \toprule
\textbf{Work} & \textbf{Suggested starting posture} \\
\midrule\endhead
  \bottomrule\endfoot
  \bottomrule\endlastfoot
  \tblrows
  \end{longtable}}%
\else
  \typeout{HANDBOOK-TABLE atomic \the\dimexpr\ht\tblbox+\dp\tblbox\relax}%
  \noindent\tblbody
\fi
\par\addvspace{2.6mm}
\endgroup

\textbf{PRACTITIONER NOTE.} This table is judgement, not a product guarantee. Raising effort does not make an answer correct, and lowering it does not make an operation safe. Effort changes how much the model deliberates; it changes nothing about what it is permitted to do.

\FloatBarrier
\setcounter{section}{3}
\section{Measure usage and context separately}

\texttt{/\allowbreak{}usage} reports session cost, plan limits and activity, and can break usage down by skill, subagent, plugin and MCP server [37], [67]. \texttt{/\allowbreak{}cost} and \texttt{/\allowbreak{}stats} are both documented aliases of it, and \texttt{/\allowbreak{}stats} opens on the Stats tab [37]. \texttt{/\allowbreak{}context} answers a different question: what currently occupies the conversation window [35]. Allocation and occupancy are related to tokens but they are not the same quantity, and optimising one at the expense of the other is a common and expensive mistake.

Two further facts belong here because they explain most surprising bills. First, every request in a long session carries the whole conversation, re-read at the cached rate, so a one-line question late in the day still draws usage for everything before it [67]. Second, prompt caching means a model switch forces a fully uncached request, and a \texttt{/\allowbreak{}compact} rebuilds the conversation layer while the system prompt can stay cached; and an edit to a project-root or user \texttt{CLAUDE.\allowbreak{}md} that has already loaded does not apply until the next \texttt{/\allowbreak{}clear}, \texttt{/\allowbreak{}compact} or restart [68]. If a change to project instructions appears to have been ignored, that is usually why.

\FloatBarrier
\setcounter{section}{4}
\section{What to optimise first}

\begin{enumerate}
\item Stop passing irrelevant material.
\item Persist durable outputs to files instead of regenerating them.
\item Delegate only bounded, independent work (Chapter 16).
\item Remove unused plugins, connectors and agents — after measuring their contribution, not before.
\item Start a clean session when the task changes.
\end{enumerate}

Do not optimise by skipping verification. A cheaper wrong answer is still waste, and in this domain it is waste that arrives looking like success.

\begingroup
\def\kprows{%
\item Permission mode, model, effort and output style are four separate controls, with documented couplings: \texttt{opusplan} ties the model to plan mode, and a model change can change the effort levels on offer.
\item \texttt{/\allowbreak{}usage} measures your allocation; \texttt{/\allowbreak{}context} measures what occupies the window. They are different questions.
\item A long session costs more per question, because every request carries the whole conversation.
\item An edit to a loaded project-root or user \texttt{CLAUDE.\allowbreak{}md} does not take effect until \texttt{/\allowbreak{}clear}, \texttt{/\allowbreak{}compact} or a restart — it is read once at session start.
}%
\def\kpbody{\begin{minipage}{\textwidth}\subsection*{Key points}\begin{itemize}\kprows\end{itemize}\end{minipage}}%
\begingroup
\def\sloppy{\tolerance 9999\emergencystretch 3em\hfuzz 200pt\vfuzz 200pt}%
\hbadness=10000\vbadness=10000\hfuzz=200pt\vfuzz=200pt
\global\setbox\kpbox=\hbox{\kpbody}%
\endgroup
\par\addvspace{4.2mm}
\ifdim\dimexpr\ht\kpbox+\dp\kpbox\relax>0.30\textheight
  \typeout{HANDBOOK-KEYPOINTS broken \the\dimexpr\ht\kpbox+\dp\kpbox\relax}%
  \subsection*{Key points}
  \begin{itemize}\kprows\end{itemize}
\else
  \typeout{HANDBOOK-KEYPOINTS atomic \the\dimexpr\ht\kpbox+\dp\kpbox\relax}%
  \noindent\kpbody
\fi
\par\addvspace{1.4mm}
\endgroup

\breakrule

\FloatBarrier
\parttitle{Part II}{A safe working method}
\addcontentsline{toc}{part}{Part II \textemdash\ A safe working method}

\FloatBarrier
\renewcommand{\chaptertitlelabel}{Part II \textperiodcentered\ Chapter 5}
\setcounter{chapter}{4}
\chapter{Start and steer a session}

\FloatBarrier
\setcounter{section}{0}
\section{Begin with an inspectable brief}

A first message needs no special syntax. It needs enough structure that omissions are visible:

\begin{codeblock}{7.0}{8.3}
\cl{Objective}
\cl{Create~a~one-page~research~brief~from~the~files~in~notes/.}
\cl{}
\cl{Scope}
\cl{Use~only~the~supplied~notes.~Identify~evidence~gaps~instead~of~filling~them~from~memory~or}
\cl{the~web.}
\cl{}
\cl{Output}
\cl{Save~drafts/brief-v1.md~with~an~executive~summary,~three~findings,~and~a~claim-to-source~table.}
\cl{}
\cl{Boundaries}
\cl{Do~not~edit~source~files.~Do~not~access~other~folders~or~the~internet.}
\cl{}
\cl{Done~when}
\cl{Every~factual~statement~maps~to~a~named~source~note,~and~the~file~passes~a~completeness~review.}
\end{codeblock}

The headings serve the reader as much as the model. They make a missing boundary conspicuous before it becomes an incident.

\FloatBarrier
\setcounter{section}{1}
\section{Watch the beginning of unfamiliar work}

Claude Code renders progress and tool calls. Early in a task, read them and ask:

\begin{itemize}
\item Is it reading the files you intended?
\item Is it attempting an external action you did not request?
\item Is its interpretation of "done" the same as yours?
\item Did it delegate, and if so with what brief? (See Chapter 16.)
\end{itemize}

Once the direction is stable you need not watch every line. Unattended work, by contrast, needs tighter boundaries and a stronger completion test, not looser ones.

\FloatBarrier
\setcounter{section}{2}
\section{Steer without restarting}

Type a message and press Enter to queue it. If tools are running, Claude receives it when those calls finish and can adjust within the same turn; anything further becomes a subsequent turn [66]. \texttt{Ctrl+Enter} sends what you queued without waiting \emph{(2.1.275 or later)}; where the terminal does not report extended keys it arrives as plain \texttt{Enter}, so use \texttt{Ctrl+X Ctrl\allowbreak{}+S}, and in shell mode it only queues. Commands you queue wait for the turn to end, except those Claude Code runs at once, such as \texttt{/\allowbreak{}model}, \texttt{/\allowbreak{}effort} and \texttt{/\allowbreak{}fast} [66]. Desktop takes a typed correction the same way, and its stop button interrupts at once [49].

When the correction cannot wait, press \texttt{Esc} to stop the current response or tool call. Completed work remains; queued content is sent; \texttt{Up} from the first line of the input takes queued entries back for editing [66]. Say explicitly whether to resume:

\begin{codeblock}{8.0}{9.4}
\cl{Stop.~Do~not~use~the~internet;~I~omitted~that~boundary.~Keep~the~local~analysis~already}
\cl{completed,~discard~any~external~material,~and~resume~from~the~approved~notes~only.}
\end{codeblock}

\begin{calloutbox}{palebrass}{brassdark}{2.0mm}
\textbf{CAUTION:} interruption does not undo completed file changes and cannot recall an external action. Inspect the diff or the checkpoint state (Chapter 7).
\end{calloutbox}

\FloatBarrier
\setcounter{section}{3}
\section{Side questions, focus and recall}

Three small controls repay learning early. \texttt{/\allowbreak{}btw} asks a side question without adding it to the conversation history. \texttt{/\allowbreak{}focus} collapses the view to the last prompt and response, in fullscreen rendering only; VS Code has its own Focus view. \texttt{/\allowbreak{}recap} generates a one-line summary of the session, and \texttt{/\allowbreak{}export} writes a rendered plain-text transcript for the record [31], [37]. The last of these matters for audit: it is the supported way to retain evidence of what happened, and it is preferable to parsing the raw JSONL transcript, whose format is internal and changes between releases [31].

\FloatBarrier
\setcounter{section}{4}
\section{Voice input}

Voice dictation makes long briefs easier to compose [69]. Read the transcription before sending. Filenames, command flags, negations and numbers are where dictation fails, and a dropped "not" in a boundary statement is the most expensive single-word error in this book.

\subsection*{Exercise}

Run the brief from Chapter 1. Mid-task, queue one non-urgent steering message requiring the findings to be ordered by decision relevance. Interrupt only if a boundary is crossed.

\emph{A worked solution is given in Appendix M.}

\begingroup
\def\kprows{%
\item A brief with headings makes omissions visible before they become incidents.
\item Queue a correction and it applies after the active tool call; press \texttt{Esc} only when it cannot wait.
\item Interruption does not undo completed file changes and cannot recall an external action.
\item \texttt{/\allowbreak{}export} is the supported way to retain evidence of a session; the raw JSONL transcript format is internal and changes between releases.
}%
\def\kpbody{\begin{minipage}{\textwidth}\subsection*{Key points}\begin{itemize}\kprows\end{itemize}\end{minipage}}%
\begingroup
\def\sloppy{\tolerance 9999\emergencystretch 3em\hfuzz 200pt\vfuzz 200pt}%
\hbadness=10000\vbadness=10000\hfuzz=200pt\vfuzz=200pt
\global\setbox\kpbox=\hbox{\kpbody}%
\endgroup
\par\addvspace{4.2mm}
\ifdim\dimexpr\ht\kpbox+\dp\kpbox\relax>0.30\textheight
  \typeout{HANDBOOK-KEYPOINTS broken \the\dimexpr\ht\kpbox+\dp\kpbox\relax}%
  \subsection*{Key points}
  \begin{itemize}\kprows\end{itemize}
\else
  \typeout{HANDBOOK-KEYPOINTS atomic \the\dimexpr\ht\kpbox+\dp\kpbox\relax}%
  \noindent\kpbody
\fi
\par\addvspace{1.4mm}
\endgroup

\FloatBarrier
\renewcommand{\chaptertitlelabel}{Part II \textperiodcentered\ Chapter 6}
\setcounter{chapter}{5}
\chapter{Plan, comment, approve, execute}

\FloatBarrier
\setcounter{section}{0}
\section{When to plan}

Use Plan mode when the work has several stages, unclear requirements, expensive external actions, or broad file impact. Enter it with \texttt{/\allowbreak{}plan}, the mode selector, or \texttt{claude -\allowbreak{}-\allowbreak{}permission-\allowbreak{}mode plan} [33], [55]. In Plan mode Claude reads and explores but does not edit source, and edits stay blocked until you approve the plan — except in an interactive terminal session with bypass permissions available, where neither the classifier nor a prompt stands in front of a planning command [33]. Plan mode is not a browser sandbox: the documented prompts for browser calls during planning cover recording a GIF and opening a new tab, not clicks or typing (Section 20.3) [21].

A plan should reduce uncertainty before edits, not restate your request in longer form.

\FloatBarrier
\setcounter{section}{1}
\section{The minimum complete plan}

A plan is ready for approval when it identifies:

\begin{enumerate}
\item the intended outcome and the explicit exclusions;
\item inputs and the source-of-truth files;
\item every file or system that will change;
\item ordered implementation steps;
\item human decisions still outstanding;
\item tests or inspection evidence for each material result;
\item external side effects;
\item a rollback or recovery route.
\end{enumerate}

For a report, item 6 means source coverage, link checks, spelling and rendering. For software it means automated tests plus direct behavioural observation.

\FloatBarrier
\setcounter{section}{2}
\section{Review the plan like a contract}

Do not approve a plan because it is long or confident. Comment on specific weaknesses:

\begin{codeblock}{8.0}{9.4}
\cl{Step~3~is~too~broad.~Separate~source~extraction~from~interpretation,~and~require~one}
\cl{evidence-ledger~row~per~material~claim~before~any~drafting~begins.}
\end{codeblock}

\begin{codeblock}{9.0}{10.6}
\cl{Add~a~human~approval~gate~before~anything~is~published~or~sent~externally.}
\end{codeblock}

\begin{codeblock}{9.0}{10.6}
\cl{Define~what~"complete"~means~for~the~final~file,~and~how~you~will~verify~it.}
\end{codeblock}

In surfaces that render an editable plan, select text and attach comments; otherwise quote the step. Continue until the high-impact uncertainties are resolved.

Accepting a plan also gives the session a generated title based on the plan, unless you have already named it [31] — a small but useful audit artefact.

\FloatBarrier
\setcounter{section}{3}
\section{Approve the right scope}

Approving a plan is not blanket permission for any method, and \textbf{approval changes the mode}: it exits plan mode into the mode the option names. \textbf{Yes, and use auto mode} switches to auto mode; where auto mode is unavailable it reads \textbf{Yes, auto-accept edits} and switches to \texttt{accept\allowbreak{}Edits}; in a session started with bypass permissions enabled it switches to \texttt{bypass\allowbreak{}Permissions} instead. \textbf{Yes, manually approve edits} switches to Manual, where you review each edit. With \texttt{show\allowbreak{}Clear\allowbreak{}Context\allowbreak{}On\allowbreak{}Plan\allowbreak{}Accept} on, an extra first option approves and clears the planning context. \texttt{Ctrl+G} opens the plan in your editor before you approve [33]. Keep the permission mode you meant to be in. A good plan combined with auto mode can support long unattended execution, but auto mode is a classifier-assisted permission mode: it is not a sandbox and not a guarantee [33]. See Chapter 7.

\FloatBarrier
\setcounter{section}{4}
\section{Require a completion report, then check it}

\begin{codeblock}{8.0}{9.4}
\cl{Report~the~files~changed,~the~tests~or~checks~actually~executed~with~their~output,~unresolved}
\cl{limitations,~and~every~external~side~effect.~Do~not~describe~a~check~that~did~not~run.}
\end{codeblock}

Then inspect the evidence yourself: \texttt{/\allowbreak{}diff} opens the interactive diff viewer for uncommitted changes [37], and \texttt{git diff} is the version you can archive.

\subsection*{Exercise}

Plan a second brief that adds web research. Require a query strategy, source-quality criteria, a claim ledger, a source register, a contradiction pass, and explicit approval before any external publication. Do not execute until the plan satisfies all eight points of Section 6.2.

\emph{A worked solution is given in Appendix M.}

\begingroup
\def\kprows{%
\item Plan mode reduces uncertainty before edits; it is not a longer restatement of your request.
\item A plan is ready for approval when it names outcomes, inputs, changes, steps, open decisions, tests, side effects and a rollback route.
\item Approving a plan is not blanket permission for any method — keep the permission mode you intended.
\item Require a completion report that distinguishes checks that ran from checks that were described, then inspect the evidence yourself.
}%
\def\kpbody{\begin{minipage}{\textwidth}\subsection*{Key points}\begin{itemize}\kprows\end{itemize}\end{minipage}}%
\begingroup
\def\sloppy{\tolerance 9999\emergencystretch 3em\hfuzz 200pt\vfuzz 200pt}%
\hbadness=10000\vbadness=10000\hfuzz=200pt\vfuzz=200pt
\global\setbox\kpbox=\hbox{\kpbody}%
\endgroup
\par\addvspace{4.2mm}
\ifdim\dimexpr\ht\kpbox+\dp\kpbox\relax>0.30\textheight
  \typeout{HANDBOOK-KEYPOINTS broken \the\dimexpr\ht\kpbox+\dp\kpbox\relax}%
  \subsection*{Key points}
  \begin{itemize}\kprows\end{itemize}
\else
  \typeout{HANDBOOK-KEYPOINTS atomic \the\dimexpr\ht\kpbox+\dp\kpbox\relax}%
  \noindent\kpbody
\fi
\par\addvspace{1.4mm}
\endgroup

\FloatBarrier
\renewcommand{\chaptertitlelabel}{Part II \textperiodcentered\ Chapter 7}
\setcounter{chapter}{6}
\chapter{Permissions, sandboxing and checkpoints}

Permissions are product-enforced rules governing tool use. Instructions in a prompt or in \texttt{CLAUDE.\allowbreak{}md} influence what Claude attempts; they do not grant or revoke access [25], [29]. This distinction is the foundation of the whole book: an instruction is a preference, a permission rule is a control, and a hook is executable enforcement (Chapter 17).

\FloatBarrier
\setcounter{section}{0}
\section{Permission modes}

From Claude Code 2.1.283 the built-in starting mode of an interactive terminal or VS Code session is \textbf{auto mode}, on every plan; before that release it was auto only on Pro, Max, Ultra and Team plans [33]. That is a categorical statement in current documentation, not a tendency. Three cases start in Manual instead: \texttt{claude -\allowbreak{}p} and Agent SDK runs \emph{(requires 2.1.302 or later)}, where a call that needs approval is denied rather than asked; an organisation with the HIPAA configuration applied, in a session eligible for it \emph{(requires 2.1.285 or later)}; and any session where a settings file sets \texttt{disable\allowbreak{}Auto\allowbreak{}Mode} to \texttt{"disable"}.

Those are built-in defaults, consulted last. In a terminal a \texttt{-\allowbreak{}-\allowbreak{}permission-\allowbreak{}mode} flag comes first and a \texttt{permissions.\allowbreak{}default\allowbreak{}Mode} in a settings file second, and a repository cannot opt you into either risky mode: \texttt{auto} written into a project's \texttt{.\allowbreak{}claude/\allowbreak{}settings.\allowbreak{}json} or \texttt{.\allowbreak{}claude/\allowbreak{}settings.\allowbreak{}local.\allowbreak{}json} does not take effect and the built-in default is used instead — not the \texttt{default\allowbreak{}Mode} in your user settings — while \texttt{bypass\allowbreak{}Permissions} written there starts the session in Manual. The VS Code extension keeps its own order: \texttt{claude\allowbreak{}Code.\allowbreak{}initial\allowbreak{}Permission\allowbreak{}Mode}, then the mode last picked from the mode indicator if it was Manual, Edit automatically or Auto (picking Plan or Bypass permissions applies to that conversation only), then a \texttt{default\allowbreak{}Mode} from managed or user settings, then the built-in default; with \texttt{claude\allowbreak{}Code.\allowbreak{}claude\allowbreak{}Process\allowbreak{}Wrapper} set, the last two steps are skipped and the conversation starts in Manual. It never reads a project's settings files for the starting mode. Where auto mode is selected but is not available to the session, the session starts in Manual [33]. Inspect the selector rather than assuming.

\begingroup
\def\tblrows{%
Manual & \texttt{default} (alias \texttt{manual}, 2.1.200+) & 1 & Asks before most actions that edit files, run commands or reach the network & Learning, sensitive repositories, unfamiliar tasks \\
Accept Edits & \texttt{accept\allowbreak{}Edits} & 2 & Auto-approves file edits and a fixed set of filesystem Bash commands (\texttt{mkdir}, \texttt{touch}, \texttt{rm}, \texttt{rmdir}, \texttt{mv}, \texttt{cp}, \texttt{sed}) inside the working directory & Well-bounded local editing with review \\
Plan & \texttt{plan} & 3 & Reads and explores; source edits stay blocked until you approve the plan, except in an interactive terminal session with bypass permissions available. Shell commands are reviewed by the classifier where auto mode is available and \texttt{use\allowbreak{}Auto\allowbreak{}Mode\allowbreak{}During\allowbreak{}Plan} is on, and otherwise prompt & Analysis and design before execution \\
Bypass Permissions & \texttt{bypass\allowbreak{}Permissions} & 4 & Skips ordinary prompts & Isolated disposable containers or VMs only \\
Auto & \texttt{auto} & 5 & A classifier model reviews actions in place of you and blocks those it judges unsafe & Bounded work with sandboxing and verification \\
Don't Ask & \texttt{dont\allowbreak{}Ask} & — & Denies anything that would prompt; file reads inside the working directories and pre-approved tools still run, and an installed mod can approve a call that nothing but the mode would deny & Locked-down repeatable workflows \\
}%
\def\tblbody{\begin{minipage}{\textwidth}\boxcaption{Table 5 — Permission modes, the cycle position that reaches each, and appropriate use}
{\scriptsize\begin{tabular}{L{19.8mm}L{30.5mm}L{19.1mm}L{39.5mm}L{24.5mm}}
\toprule
\textbf{Mode} & \textbf{Config value} & \textbf{\texttt{Shift+Tab}} & \textbf{Behaviour} & \textbf{Sensible use} \\
\midrule
\tblrows
\bottomrule\end{tabular}}\end{minipage}}%
\begingroup
\def\sloppy{\tolerance 9999\emergencystretch 3em\hfuzz 200pt\vfuzz 200pt}%
\hbadness=10000\vbadness=10000\hfuzz=200pt\vfuzz=200pt
\global\setbox\tblbox=\hbox{\tblbody}%
\endgroup
\par\addvspace{2.6mm}
\ifdim\dimexpr\ht\tblbox+\dp\tblbox\relax>0.55\textheight
  \typeout{HANDBOOK-TABLE broken \the\dimexpr\ht\tblbox+\dp\tblbox\relax}%
  \tabcaption{Table 5 — Permission modes, the cycle position that reaches each, and appropriate use}
  {\scriptsize\begin{longtable}{L{19.8mm}L{30.5mm}L{19.1mm}L{39.5mm}L{24.5mm}}
  \toprule
\textbf{Mode} & \textbf{Config value} & \textbf{\texttt{Shift+Tab}} & \textbf{Behaviour} & \textbf{Sensible use} \\
\midrule\endfirsthead
  \multicolumn{5}{@{}l@{}}{%
  \sffamily\footnotesize\itshape\color{inkgrey}Table 5 — Permission modes, the cycle position that reaches each, and appropriate use \textemdash\ continued}\\[1.2mm]
  \toprule
\textbf{Mode} & \textbf{Config value} & \textbf{\texttt{Shift+Tab}} & \textbf{Behaviour} & \textbf{Sensible use} \\
\midrule\endhead
  \bottomrule\endfoot
  \bottomrule\endlastfoot
  \tblrows
  \end{longtable}}%
\else
  \typeout{HANDBOOK-TABLE atomic \the\dimexpr\ht\tblbox+\dp\tblbox\relax}%
  \noindent\tblbody
\fi
\par\addvspace{2.6mm}
\endgroup

\textbf{The keystroke does not reach every mode, and what it reaches depends on your session.} In the CLI the cycle runs \texttt{default} → \texttt{accept\allowbreak{}Edits} → \texttt{plan} and back, in every session. Two further modes slot in after \texttt{plan}, in this order: \textbf{\texttt{bypass\allowbreak{}Permissions}}, when the session was started with \texttt{-\allowbreak{}-\allowbreak{}permission-\allowbreak{}mode bypass\allowbreak{}Permissions}, \texttt{-\allowbreak{}-\allowbreak{}dangerously-\allowbreak{}skip-\allowbreak{}permissions}, \texttt{-\allowbreak{}-\allowbreak{}allow-\allowbreak{}dangerously-\allowbreak{}skip-\allowbreak{}permissions} or a \texttt{bypass\allowbreak{}Permissions} default in settings; then \textbf{\texttt{auto}}, when auto mode is available to the session. So the keystroke reaches three modes by default, and at most five in a session that has enabled both. \textbf{\texttt{dont\allowbreak{}Ask} never appears in the cycle at all}: it is set with \texttt{claude -\allowbreak{}-\allowbreak{}permission-\allowbreak{}mode dont\allowbreak{}Ask} [33], [55].

The numbering in the table is the cycle order, not a ranking. Two consequences follow for a reader who takes the table and reaches for the keystroke. If \texttt{bypass\allowbreak{}Permissions} and \texttt{auto} are not enabled, the keystroke moves between the first three rows only, and the fourth and fifth are reached by flag or by settings. And \texttt{dont\allowbreak{}Ask} is never reached by cycling in any configuration, which is the one case where the table and the keystroke disagree about what exists.

\texttt{claude -\allowbreak{}-\allowbreak{}permission-\allowbreak{}mode} names six values when it refuses one — \texttt{accept\allowbreak{}Edits}, \texttt{auto}, \texttt{bypass\allowbreak{}Permissions}, \texttt{dont\allowbreak{}Ask}, \texttt{manual} and \texttt{plan} — and additionally accepts \texttt{default}, which that enumeration omits. That is not two vocabularies but one: \texttt{manual} is a documented alias for \texttt{default}, from 2.1.200 onward, so the validator lists the alias and not the canonical value the settings files and the SDK use [33]. It is the plainest case of the rule in Appendix J — an enumeration is evidence of what is accepted, never of what is not.

Two mode facts have operational consequences. A terminal resume does \textbf{not} restore \texttt{bypass\allowbreak{}Permissions}: the session returns to the mode a new one would start in, and bypass has to be enabled again at launch. \texttt{plan}, by contrast, is restored, and \texttt{/\allowbreak{}resume} inside a session continues in the mode the current session is in, unless the conversation it switches to ended in plan mode [31]; Section 12.1 gives the rule route by route. And \texttt{bypass\allowbreak{}Permissions} propagates: agent-team teammates inherit a lead running without checks [30], and a child agent inherits its parent's mode, which a \texttt{permission\allowbreak{}Mode} of its own cannot override when the parent is in \texttt{bypass\allowbreak{}Permissions}, \texttt{accept\allowbreak{}Edits} or \texttt{auto} [27]. Workflow subagents follow the same rules. The subagents a workflow spawns use your permission rules and take their mode by the rules for any subagent [19], [27]. Under a session in \texttt{bypass\allowbreak{}Permissions}, \texttt{accept\allowbreak{}Edits} or \texttt{auto} they run in that mode; under \texttt{plan} they stay in plan mode, and a \texttt{dont\allowbreak{}Ask} agent there is denied what plan mode would ask; under Manual or \texttt{dont\allowbreak{}Ask} an agent type's own \texttt{permission\allowbreak{}Mode} applies, unless it asks for \texttt{bypass\allowbreak{}Permissions}, or for \texttt{auto} where auto mode is unavailable, and an unset field inherits the session's mode. Version 1.0 said they always run in \texttt{accept\allowbreak{}Edits}, which the page no longer states. The session's mode is therefore the first thing to check before any run that spawns agents; Chapter 22 treats the workflow case in full.

\FloatBarrier
\setcounter{section}{1}
\section{Rules, and the order they are evaluated in}

Run \texttt{/\allowbreak{}permissions} to inspect and edit Allow, Ask and Deny rules. \textbf{Rules are evaluated in the order deny, then ask, then allow; the first match in that order decides, and specificity does not change the order} [29]. A broad deny such as \texttt{Bash(aws *)} therefore blocks every matching call, including one that also matches a narrower allow like \texttt{Bash(aws s3\allowbreak{} ls)}. A deny rule cannot carry allowlist exceptions.

\begin{calloutbox}{palebrass}{brassdark}{2.0mm}
\textbf{CAUTION:} deny-first precedence holds against hooks, and not unconditionally against a \emph{mod}. A mod you have installed that handles the \texttt{tool.\allowbreak{}check} event answers after the \texttt{Pre\allowbreak{}Tool\allowbreak{}Use} hooks have run. Where the session has managed settings, a \texttt{managed-\allowbreak{}mcp.\allowbreak{}json} file, or a Team or Enterprise login as its credential, a deny rule still refuses, an ask rule still prompts and a hook's block still stands. On a setup with none of those — an individual account on an unmanaged machine, or a session running on an API key — the mod can approve a call that a deny rule refuses or that a \texttt{Pre\allowbreak{}Tool\allowbreak{}Use} hook blocked [29]. Treat an installed mod as part of the permission system, not as an extension beneath it (Section 18.3).
\end{calloutbox}

Two subtleties are worth committing to memory. A \textbf{bare tool name} in \texttt{deny} removes the tool from Claude's context entirely, so Claude never sees it; a \textbf{scoped rule} such as \texttt{Bash(rm *)} leaves the tool available and blocks matching calls [29]. And a deny or ask rule matches past any leading environment assignment, while an allow rule matches past only an assignment of certain known-safe variables, such as \texttt{NODE\_\allowbreak{}ENV=\allowbreak{}test}: \texttt{Bash(rm *)} in \texttt{deny} still matches \texttt{FOO=\allowbreak{}bar rm -\allowbreak{}rf tmp/\allowbreak{}} [29].

\begin{codeblock}{9.0}{10.6}
\cl{\{}
\cl{~~"permissions":~\{}
\cl{~~~~"deny":~[}
\cl{~~~~~~"Read(./.env)",}
\cl{~~~~~~"Read(./.env.*)",}
\cl{~~~~~~"Read(./secrets/**)",}
\cl{~~~~~~"Read(\textasciitilde{}/.ssh/**)",}
\cl{~~~~~~"Bash(git~push~*)",}
\cl{~~~~~~"Bash(curl~*)",}
\cl{~~~~~~"Bash(wget~*)"}
\cl{~~~~],}
\cl{~~~~"ask":~[}
\cl{~~~~~~"WebFetch",}
\cl{~~~~~~"WebSearch"}
\cl{~~~~],}
\cl{~~~~"allow":~[}
\cl{~~~~~~"WebFetch(domain:code.claude.com)"}
\cl{~~~~]}
\cl{~~\}}
\cl{\}}
\end{codeblock}

Note what that example does \emph{not} do: because deny precedes allow, the \texttt{Web\allowbreak{}Fetch} allow rule does not defeat any deny rule, and the \texttt{Bash(curl *\allowbreak{})} deny is the reason to route web access through the \texttt{Web\allowbreak{}Fetch} tool with a domain specifier rather than through the shell [29], [59]. Wildcards in a domain specifier match only between dots unless they lead — \texttt{Web\allowbreak{}Fetch(domai\allowbreak{}n:\allowbreak{}*.\allowbreak{}example.\allowbreak{}com)} matches subdomains but not \texttt{example.\allowbreak{}com}, and \texttt{Web\allowbreak{}Fetch(domai\allowbreak{}n:\allowbreak{}example.\allowbreak{}*)} matches \texttt{example.\allowbreak{}org} but not \texttt{example.\allowbreak{}evil.\allowbreak{}com} [29].

Symlinks are resolved against both allow and deny: a link inside an allowed directory pointing at \texttt{\textasciitilde{}/\allowbreak{}.\allowbreak{}ssh/\allowbreak{}id\_\allowbreak{}rsa} is blocked because the target fails the allow rule and matches the deny rule [29].

Share team rules in \texttt{.\allowbreak{}claude/\allowbreak{}settings.\allowbreak{}json}; keep personal overrides and saved approvals in \texttt{.\allowbreak{}claude/\allowbreak{}settings.\allowbreak{}local.\allowbreak{}json}, which is not committed; user-wide settings live in \texttt{\textasciitilde{}/\allowbreak{}.\allowbreak{}claude/\allowbreak{}settings.\allowbreak{}json} [70]. Managed settings sit above all of them and cannot be overridden by any other level, including command-line arguments [29], [71]. See Chapter 31. Figure 2 traces a single tool call through the whole chain, in the order the decision is actually taken.

\begin{figure}[tbp]
\begin{codefig}{9.0}{10.6}
\cl{~~~Claude~proposes~a~tool~call}
\cl{~~~~~~~~~~~~~\pmboxdrawuni{2502}}
\cl{~~~~~~~~~~~~~▼}
\cl{~~~\pmboxdrawuni{250C}\pmboxdrawuni{2500}\pmboxdrawuni{2500}\pmboxdrawuni{2500}\pmboxdrawuni{2500}\pmboxdrawuni{2500}\pmboxdrawuni{2500}\pmboxdrawuni{2500}\pmboxdrawuni{2500}\pmboxdrawuni{2500}\pmboxdrawuni{2500}\pmboxdrawuni{2500}\pmboxdrawuni{2500}\pmboxdrawuni{2500}\pmboxdrawuni{2500}\pmboxdrawuni{2500}\pmboxdrawuni{2500}\pmboxdrawuni{2500}\pmboxdrawuni{2500}\pmboxdrawuni{2500}\pmboxdrawuni{2500}\pmboxdrawuni{2500}\pmboxdrawuni{2500}\pmboxdrawuni{2500}\pmboxdrawuni{2510}~exit~2,~deny~~\pmboxdrawuni{250C}\pmboxdrawuni{2500}\pmboxdrawuni{2500}\pmboxdrawuni{2500}\pmboxdrawuni{2500}\pmboxdrawuni{2500}\pmboxdrawuni{2500}\pmboxdrawuni{2500}\pmboxdrawuni{2500}\pmboxdrawuni{2500}\pmboxdrawuni{2500}\pmboxdrawuni{2500}\pmboxdrawuni{2500}\pmboxdrawuni{2500}\pmboxdrawuni{2500}\pmboxdrawuni{2500}\pmboxdrawuni{2500}\pmboxdrawuni{2500}\pmboxdrawuni{2500}\pmboxdrawuni{2500}\pmboxdrawuni{2500}\pmboxdrawuni{2500}\pmboxdrawuni{2500}\pmboxdrawuni{2500}\pmboxdrawuni{2500}\pmboxdrawuni{2500}\pmboxdrawuni{2500}\pmboxdrawuni{2500}\pmboxdrawuni{2500}\pmboxdrawuni{2500}\pmboxdrawuni{2500}\pmboxdrawuni{2510}}
\cl{~~~\pmboxdrawuni{2502}~PreToolUse~hook~~~~~~~\pmboxdrawuni{2502}\pmboxdrawuni{2500}\pmboxdrawuni{2500}\pmboxdrawuni{2500}\pmboxdrawuni{2500}\pmboxdrawuni{2500}\pmboxdrawuni{2500}\pmboxdrawuni{2500}\pmboxdrawuni{2500}\pmboxdrawuni{2500}\pmboxdrawuni{2500}\pmboxdrawuni{2500}\pmboxdrawuni{2500}\pmboxdrawuni{2500}\pmboxdrawuni{2500}▶\pmboxdrawuni{2502}~BLOCKED~before~rules~are~~~~~\pmboxdrawuni{2502}}
\cl{~~~\pmboxdrawuni{2502}~(if~one~matches)~~~~~~\pmboxdrawuni{2502}~~~~~~~~~~~~~~~\pmboxdrawuni{2502}~evaluated~—~overrides~allow~~\pmboxdrawuni{2502}}
\cl{~~~\pmboxdrawuni{2514}\pmboxdrawuni{2500}\pmboxdrawuni{2500}\pmboxdrawuni{2500}\pmboxdrawuni{2500}\pmboxdrawuni{2500}\pmboxdrawuni{2500}\pmboxdrawuni{2500}\pmboxdrawuni{2500}\pmboxdrawuni{2500}\pmboxdrawuni{2500}\pmboxdrawuni{252C}\pmboxdrawuni{2500}\pmboxdrawuni{2500}\pmboxdrawuni{2500}\pmboxdrawuni{2500}\pmboxdrawuni{2500}\pmboxdrawuni{2500}\pmboxdrawuni{2500}\pmboxdrawuni{2500}\pmboxdrawuni{2500}\pmboxdrawuni{2500}\pmboxdrawuni{2500}\pmboxdrawuni{2500}\pmboxdrawuni{2518}~~~~~~~~~~~~~~~\pmboxdrawuni{2514}\pmboxdrawuni{2500}\pmboxdrawuni{2500}\pmboxdrawuni{2500}\pmboxdrawuni{2500}\pmboxdrawuni{2500}\pmboxdrawuni{2500}\pmboxdrawuni{2500}\pmboxdrawuni{2500}\pmboxdrawuni{2500}\pmboxdrawuni{2500}\pmboxdrawuni{2500}\pmboxdrawuni{2500}\pmboxdrawuni{2500}\pmboxdrawuni{2500}\pmboxdrawuni{2500}\pmboxdrawuni{2500}\pmboxdrawuni{2500}\pmboxdrawuni{2500}\pmboxdrawuni{2500}\pmboxdrawuni{2500}\pmboxdrawuni{2500}\pmboxdrawuni{2500}\pmboxdrawuni{2500}\pmboxdrawuni{2500}\pmboxdrawuni{2500}\pmboxdrawuni{2500}\pmboxdrawuni{2500}\pmboxdrawuni{2500}\pmboxdrawuni{2500}\pmboxdrawuni{2500}\pmboxdrawuni{2518}}
\cl{~~~~~~~~~~~~~~\pmboxdrawuni{2502}~exit~0:~no~decision,~"allow"~or~"ask"}
\cl{~~~~~~~~~~~~~~▼}
\cl{~~~\pmboxdrawuni{250C}\pmboxdrawuni{2500}\pmboxdrawuni{2500}\pmboxdrawuni{2500}\pmboxdrawuni{2500}\pmboxdrawuni{2500}\pmboxdrawuni{2500}\pmboxdrawuni{2500}\pmboxdrawuni{2500}\pmboxdrawuni{2500}\pmboxdrawuni{2500}\pmboxdrawuni{2500}\pmboxdrawuni{2500}\pmboxdrawuni{2500}\pmboxdrawuni{2500}\pmboxdrawuni{2500}\pmboxdrawuni{2500}\pmboxdrawuni{2500}\pmboxdrawuni{2500}\pmboxdrawuni{2500}\pmboxdrawuni{2500}\pmboxdrawuni{2500}\pmboxdrawuni{2500}\pmboxdrawuni{2500}\pmboxdrawuni{2510}~~~match~~~~~~~\pmboxdrawuni{250C}\pmboxdrawuni{2500}\pmboxdrawuni{2500}\pmboxdrawuni{2500}\pmboxdrawuni{2500}\pmboxdrawuni{2500}\pmboxdrawuni{2500}\pmboxdrawuni{2500}\pmboxdrawuni{2500}\pmboxdrawuni{2500}\pmboxdrawuni{2500}\pmboxdrawuni{2500}\pmboxdrawuni{2500}\pmboxdrawuni{2500}\pmboxdrawuni{2500}\pmboxdrawuni{2500}\pmboxdrawuni{2500}\pmboxdrawuni{2500}\pmboxdrawuni{2500}\pmboxdrawuni{2500}\pmboxdrawuni{2500}\pmboxdrawuni{2500}\pmboxdrawuni{2500}\pmboxdrawuni{2500}\pmboxdrawuni{2500}\pmboxdrawuni{2500}\pmboxdrawuni{2500}\pmboxdrawuni{2500}\pmboxdrawuni{2500}\pmboxdrawuni{2500}\pmboxdrawuni{2500}\pmboxdrawuni{2510}}
\cl{~~~\pmboxdrawuni{2502}~DENY~rules~~~~~~~~~~~~\pmboxdrawuni{2502}\pmboxdrawuni{2500}\pmboxdrawuni{2500}\pmboxdrawuni{2500}\pmboxdrawuni{2500}\pmboxdrawuni{2500}\pmboxdrawuni{2500}\pmboxdrawuni{2500}\pmboxdrawuni{2500}\pmboxdrawuni{2500}\pmboxdrawuni{2500}\pmboxdrawuni{2500}\pmboxdrawuni{2500}\pmboxdrawuni{2500}\pmboxdrawuni{2500}▶\pmboxdrawuni{2502}~BLOCKED.~A~hook~returning~~~~\pmboxdrawuni{2502}}
\cl{~~~\pmboxdrawuni{2502}~(managed~first)~~~~~~~\pmboxdrawuni{2502}~~~~~~~~~~~~~~~\pmboxdrawuni{2502}~"allow"~does~not~override~it~\pmboxdrawuni{2502}}
\cl{~~~\pmboxdrawuni{2514}\pmboxdrawuni{2500}\pmboxdrawuni{2500}\pmboxdrawuni{2500}\pmboxdrawuni{2500}\pmboxdrawuni{2500}\pmboxdrawuni{2500}\pmboxdrawuni{2500}\pmboxdrawuni{2500}\pmboxdrawuni{2500}\pmboxdrawuni{2500}\pmboxdrawuni{252C}\pmboxdrawuni{2500}\pmboxdrawuni{2500}\pmboxdrawuni{2500}\pmboxdrawuni{2500}\pmboxdrawuni{2500}\pmboxdrawuni{2500}\pmboxdrawuni{2500}\pmboxdrawuni{2500}\pmboxdrawuni{2500}\pmboxdrawuni{2500}\pmboxdrawuni{2500}\pmboxdrawuni{2500}\pmboxdrawuni{2518}~~~~~~~~~~~~~~~\pmboxdrawuni{2514}\pmboxdrawuni{2500}\pmboxdrawuni{2500}\pmboxdrawuni{2500}\pmboxdrawuni{2500}\pmboxdrawuni{2500}\pmboxdrawuni{2500}\pmboxdrawuni{2500}\pmboxdrawuni{2500}\pmboxdrawuni{2500}\pmboxdrawuni{2500}\pmboxdrawuni{2500}\pmboxdrawuni{2500}\pmboxdrawuni{2500}\pmboxdrawuni{2500}\pmboxdrawuni{2500}\pmboxdrawuni{2500}\pmboxdrawuni{2500}\pmboxdrawuni{2500}\pmboxdrawuni{2500}\pmboxdrawuni{2500}\pmboxdrawuni{2500}\pmboxdrawuni{2500}\pmboxdrawuni{2500}\pmboxdrawuni{2500}\pmboxdrawuni{2500}\pmboxdrawuni{2500}\pmboxdrawuni{2500}\pmboxdrawuni{2500}\pmboxdrawuni{2500}\pmboxdrawuni{2500}\pmboxdrawuni{2518}}
\cl{~~~~~~~~~~~~~~\pmboxdrawuni{2502}~no~match}
\cl{~~~~~~~~~~~~~~▼}
\cl{~~~\pmboxdrawuni{250C}\pmboxdrawuni{2500}\pmboxdrawuni{2500}\pmboxdrawuni{2500}\pmboxdrawuni{2500}\pmboxdrawuni{2500}\pmboxdrawuni{2500}\pmboxdrawuni{2500}\pmboxdrawuni{2500}\pmboxdrawuni{2500}\pmboxdrawuni{2500}\pmboxdrawuni{2500}\pmboxdrawuni{2500}\pmboxdrawuni{2500}\pmboxdrawuni{2500}\pmboxdrawuni{2500}\pmboxdrawuni{2500}\pmboxdrawuni{2500}\pmboxdrawuni{2500}\pmboxdrawuni{2500}\pmboxdrawuni{2500}\pmboxdrawuni{2500}\pmboxdrawuni{2500}\pmboxdrawuni{2500}\pmboxdrawuni{2510}~~~match~~~~~~~\pmboxdrawuni{250C}\pmboxdrawuni{2500}\pmboxdrawuni{2500}\pmboxdrawuni{2500}\pmboxdrawuni{2500}\pmboxdrawuni{2500}\pmboxdrawuni{2500}\pmboxdrawuni{2500}\pmboxdrawuni{2500}\pmboxdrawuni{2500}\pmboxdrawuni{2500}\pmboxdrawuni{2500}\pmboxdrawuni{2500}\pmboxdrawuni{2500}\pmboxdrawuni{2500}\pmboxdrawuni{2500}\pmboxdrawuni{2500}\pmboxdrawuni{2500}\pmboxdrawuni{2500}\pmboxdrawuni{2500}\pmboxdrawuni{2500}\pmboxdrawuni{2500}\pmboxdrawuni{2500}\pmboxdrawuni{2500}\pmboxdrawuni{2500}\pmboxdrawuni{2500}\pmboxdrawuni{2500}\pmboxdrawuni{2500}\pmboxdrawuni{2500}\pmboxdrawuni{2500}\pmboxdrawuni{2500}\pmboxdrawuni{2510}}
\cl{~~~\pmboxdrawuni{2502}~ASK~rules~~~~~~~~~~~~~\pmboxdrawuni{2502}\pmboxdrawuni{2500}\pmboxdrawuni{2500}\pmboxdrawuni{2500}\pmboxdrawuni{2500}\pmboxdrawuni{2500}\pmboxdrawuni{2500}\pmboxdrawuni{2500}\pmboxdrawuni{2500}\pmboxdrawuni{2500}\pmboxdrawuni{2500}\pmboxdrawuni{2500}\pmboxdrawuni{2500}\pmboxdrawuni{2500}\pmboxdrawuni{2500}▶\pmboxdrawuni{2502}~PROMPT,~even~when~a~narrower~\pmboxdrawuni{2502}}
\cl{~~~\pmboxdrawuni{2514}\pmboxdrawuni{2500}\pmboxdrawuni{2500}\pmboxdrawuni{2500}\pmboxdrawuni{2500}\pmboxdrawuni{2500}\pmboxdrawuni{2500}\pmboxdrawuni{2500}\pmboxdrawuni{2500}\pmboxdrawuni{2500}\pmboxdrawuni{2500}\pmboxdrawuni{252C}\pmboxdrawuni{2500}\pmboxdrawuni{2500}\pmboxdrawuni{2500}\pmboxdrawuni{2500}\pmboxdrawuni{2500}\pmboxdrawuni{2500}\pmboxdrawuni{2500}\pmboxdrawuni{2500}\pmboxdrawuni{2500}\pmboxdrawuni{2500}\pmboxdrawuni{2500}\pmboxdrawuni{2500}\pmboxdrawuni{2518}~~~~~~~~~~~~~~~\pmboxdrawuni{2502}~allow~rule~also~matches~~~~~~\pmboxdrawuni{2502}}
\cl{~~~~~~~~~~~~~~\pmboxdrawuni{2502}~no~match~~~~~~~~~~~~~~~~~~~\pmboxdrawuni{2514}\pmboxdrawuni{2500}\pmboxdrawuni{2500}\pmboxdrawuni{2500}\pmboxdrawuni{2500}\pmboxdrawuni{2500}\pmboxdrawuni{2500}\pmboxdrawuni{2500}\pmboxdrawuni{2500}\pmboxdrawuni{2500}\pmboxdrawuni{2500}\pmboxdrawuni{2500}\pmboxdrawuni{2500}\pmboxdrawuni{2500}\pmboxdrawuni{2500}\pmboxdrawuni{2500}\pmboxdrawuni{2500}\pmboxdrawuni{2500}\pmboxdrawuni{2500}\pmboxdrawuni{2500}\pmboxdrawuni{2500}\pmboxdrawuni{2500}\pmboxdrawuni{2500}\pmboxdrawuni{2500}\pmboxdrawuni{2500}\pmboxdrawuni{2500}\pmboxdrawuni{2500}\pmboxdrawuni{2500}\pmboxdrawuni{2500}\pmboxdrawuni{2500}\pmboxdrawuni{2500}\pmboxdrawuni{2518}}
\cl{~~~~~~~~~~~~~~▼}
\cl{~~~\pmboxdrawuni{250C}\pmboxdrawuni{2500}\pmboxdrawuni{2500}\pmboxdrawuni{2500}\pmboxdrawuni{2500}\pmboxdrawuni{2500}\pmboxdrawuni{2500}\pmboxdrawuni{2500}\pmboxdrawuni{2500}\pmboxdrawuni{2500}\pmboxdrawuni{2500}\pmboxdrawuni{2500}\pmboxdrawuni{2500}\pmboxdrawuni{2500}\pmboxdrawuni{2500}\pmboxdrawuni{2500}\pmboxdrawuni{2500}\pmboxdrawuni{2500}\pmboxdrawuni{2500}\pmboxdrawuni{2500}\pmboxdrawuni{2500}\pmboxdrawuni{2500}\pmboxdrawuni{2500}\pmboxdrawuni{2500}\pmboxdrawuni{2510}~~~match~~~~~~~\pmboxdrawuni{250C}\pmboxdrawuni{2500}\pmboxdrawuni{2500}\pmboxdrawuni{2500}\pmboxdrawuni{2500}\pmboxdrawuni{2500}\pmboxdrawuni{2500}\pmboxdrawuni{2500}\pmboxdrawuni{2500}\pmboxdrawuni{2500}\pmboxdrawuni{2500}\pmboxdrawuni{2500}\pmboxdrawuni{2500}\pmboxdrawuni{2500}\pmboxdrawuni{2500}\pmboxdrawuni{2500}\pmboxdrawuni{2500}\pmboxdrawuni{2500}\pmboxdrawuni{2500}\pmboxdrawuni{2500}\pmboxdrawuni{2500}\pmboxdrawuni{2500}\pmboxdrawuni{2500}\pmboxdrawuni{2500}\pmboxdrawuni{2500}\pmboxdrawuni{2500}\pmboxdrawuni{2500}\pmboxdrawuni{2500}\pmboxdrawuni{2500}\pmboxdrawuni{2500}\pmboxdrawuni{2500}\pmboxdrawuni{2510}}
\cl{~~~\pmboxdrawuni{2502}~ALLOW~rules~~~~~~~~~~~\pmboxdrawuni{2502}\pmboxdrawuni{2500}\pmboxdrawuni{2500}\pmboxdrawuni{2500}\pmboxdrawuni{2500}\pmboxdrawuni{2500}\pmboxdrawuni{2500}\pmboxdrawuni{2500}\pmboxdrawuni{2500}\pmboxdrawuni{2500}\pmboxdrawuni{2500}\pmboxdrawuni{2500}\pmboxdrawuni{2500}\pmboxdrawuni{2500}\pmboxdrawuni{2500}▶\pmboxdrawuni{2502}~RUNS~without~a~prompt,~save~~\pmboxdrawuni{2502}}
\cl{~~~\pmboxdrawuni{2514}\pmboxdrawuni{2500}\pmboxdrawuni{2500}\pmboxdrawuni{2500}\pmboxdrawuni{2500}\pmboxdrawuni{2500}\pmboxdrawuni{2500}\pmboxdrawuni{2500}\pmboxdrawuni{2500}\pmboxdrawuni{2500}\pmboxdrawuni{2500}\pmboxdrawuni{252C}\pmboxdrawuni{2500}\pmboxdrawuni{2500}\pmboxdrawuni{2500}\pmboxdrawuni{2500}\pmboxdrawuni{2500}\pmboxdrawuni{2500}\pmboxdrawuni{2500}\pmboxdrawuni{2500}\pmboxdrawuni{2500}\pmboxdrawuni{2500}\pmboxdrawuni{2500}\pmboxdrawuni{2500}\pmboxdrawuni{2518}~~~~~~~~~~~~~~~\pmboxdrawuni{2502}~protected-path~writes~and~a~~\pmboxdrawuni{2502}}
\cl{~~~~~~~~~~~~~~\pmboxdrawuni{2502}~no~match~~~~~~~~~~~~~~~~~~~\pmboxdrawuni{2502}~hook\textquotesingle{}s~"ask",~which~prompts~~\pmboxdrawuni{2502}}
\cl{~~~~~~~~~~~~~~\pmboxdrawuni{2502}~~~~~~~~~~~~~~~~~~~~~~~~~~~~\pmboxdrawuni{2514}\pmboxdrawuni{2500}\pmboxdrawuni{2500}\pmboxdrawuni{2500}\pmboxdrawuni{2500}\pmboxdrawuni{2500}\pmboxdrawuni{2500}\pmboxdrawuni{2500}\pmboxdrawuni{2500}\pmboxdrawuni{2500}\pmboxdrawuni{2500}\pmboxdrawuni{2500}\pmboxdrawuni{2500}\pmboxdrawuni{2500}\pmboxdrawuni{2500}\pmboxdrawuni{2500}\pmboxdrawuni{2500}\pmboxdrawuni{2500}\pmboxdrawuni{2500}\pmboxdrawuni{2500}\pmboxdrawuni{2500}\pmboxdrawuni{2500}\pmboxdrawuni{2500}\pmboxdrawuni{2500}\pmboxdrawuni{2500}\pmboxdrawuni{2500}\pmboxdrawuni{2500}\pmboxdrawuni{2500}\pmboxdrawuni{2500}\pmboxdrawuni{2500}\pmboxdrawuni{2500}\pmboxdrawuni{2518}}
\cl{~~~~~~~~~~~~~~\pmboxdrawuni{2502}~(a~hook\textquotesingle{}s~"ask"~prompts;~its~"allow"~skips~the~prompt,}
\cl{~~~~~~~~~~~~~~\pmboxdrawuni{2502}~~save~the~documented~exceptions)}
\cl{~~~~~~~~~~~~~~▼}
\cl{~~~\pmboxdrawuni{250C}\pmboxdrawuni{2500}\pmboxdrawuni{2500}\pmboxdrawuni{2500}\pmboxdrawuni{2500}\pmboxdrawuni{2500}\pmboxdrawuni{2500}\pmboxdrawuni{2500}\pmboxdrawuni{2500}\pmboxdrawuni{2500}\pmboxdrawuni{2500}\pmboxdrawuni{2500}\pmboxdrawuni{2500}\pmboxdrawuni{2500}\pmboxdrawuni{2500}\pmboxdrawuni{2500}\pmboxdrawuni{2500}\pmboxdrawuni{2500}\pmboxdrawuni{2500}\pmboxdrawuni{2500}\pmboxdrawuni{2500}\pmboxdrawuni{2500}\pmboxdrawuni{2500}\pmboxdrawuni{2500}\pmboxdrawuni{2500}\pmboxdrawuni{2500}\pmboxdrawuni{2500}\pmboxdrawuni{2500}\pmboxdrawuni{2500}\pmboxdrawuni{2500}\pmboxdrawuni{2500}\pmboxdrawuni{2500}\pmboxdrawuni{2500}\pmboxdrawuni{2500}\pmboxdrawuni{2500}\pmboxdrawuni{2500}\pmboxdrawuni{2500}\pmboxdrawuni{2500}\pmboxdrawuni{2500}\pmboxdrawuni{2500}\pmboxdrawuni{2500}\pmboxdrawuni{2500}\pmboxdrawuni{2500}\pmboxdrawuni{2500}\pmboxdrawuni{2500}\pmboxdrawuni{2500}\pmboxdrawuni{2500}\pmboxdrawuni{2500}\pmboxdrawuni{2500}\pmboxdrawuni{2500}\pmboxdrawuni{2500}\pmboxdrawuni{2500}\pmboxdrawuni{2500}\pmboxdrawuni{2500}\pmboxdrawuni{2500}\pmboxdrawuni{2500}\pmboxdrawuni{2500}\pmboxdrawuni{2500}\pmboxdrawuni{2500}\pmboxdrawuni{2500}\pmboxdrawuni{2500}\pmboxdrawuni{2500}\pmboxdrawuni{2500}\pmboxdrawuni{2500}\pmboxdrawuni{2500}\pmboxdrawuni{2500}\pmboxdrawuni{2500}\pmboxdrawuni{2500}\pmboxdrawuni{2500}\pmboxdrawuni{2510}}
\cl{~~~\pmboxdrawuni{2502}~PERMISSION~MODE~decides:~Manual~asks~·~auto~sends~it~to~the~~~~~~~~\pmboxdrawuni{2502}}
\cl{~~~\pmboxdrawuni{2502}~classifier~·~acceptEdits~auto-approves~edits~·~dontAsk~denies~·~~~~\pmboxdrawuni{2502}}
\cl{~~~\pmboxdrawuni{2502}~bypassPermissions~skips~the~prompt~entirely~~~~~~~~~~~~~~~~~~~~~~~~\pmboxdrawuni{2502}}
\cl{~~~\pmboxdrawuni{2514}\pmboxdrawuni{2500}\pmboxdrawuni{2500}\pmboxdrawuni{2500}\pmboxdrawuni{2500}\pmboxdrawuni{2500}\pmboxdrawuni{2500}\pmboxdrawuni{2500}\pmboxdrawuni{2500}\pmboxdrawuni{2500}\pmboxdrawuni{2500}\pmboxdrawuni{2500}\pmboxdrawuni{2500}\pmboxdrawuni{2500}\pmboxdrawuni{2500}\pmboxdrawuni{2500}\pmboxdrawuni{2500}\pmboxdrawuni{2500}\pmboxdrawuni{2500}\pmboxdrawuni{2500}\pmboxdrawuni{2500}\pmboxdrawuni{2500}\pmboxdrawuni{2500}\pmboxdrawuni{2500}\pmboxdrawuni{2500}\pmboxdrawuni{2500}\pmboxdrawuni{2500}\pmboxdrawuni{2500}\pmboxdrawuni{2500}\pmboxdrawuni{2500}\pmboxdrawuni{2500}\pmboxdrawuni{2500}\pmboxdrawuni{2500}\pmboxdrawuni{2500}\pmboxdrawuni{2500}\pmboxdrawuni{2500}\pmboxdrawuni{2500}\pmboxdrawuni{2500}\pmboxdrawuni{2500}\pmboxdrawuni{2500}\pmboxdrawuni{2500}\pmboxdrawuni{2500}\pmboxdrawuni{2500}\pmboxdrawuni{2500}\pmboxdrawuni{2500}\pmboxdrawuni{2500}\pmboxdrawuni{2500}\pmboxdrawuni{2500}\pmboxdrawuni{2500}\pmboxdrawuni{2500}\pmboxdrawuni{2500}\pmboxdrawuni{2500}\pmboxdrawuni{2500}\pmboxdrawuni{2500}\pmboxdrawuni{2500}\pmboxdrawuni{2500}\pmboxdrawuni{2500}\pmboxdrawuni{2500}\pmboxdrawuni{2500}\pmboxdrawuni{2500}\pmboxdrawuni{2500}\pmboxdrawuni{2500}\pmboxdrawuni{2500}\pmboxdrawuni{2500}\pmboxdrawuni{2500}\pmboxdrawuni{2500}\pmboxdrawuni{2500}\pmboxdrawuni{2500}\pmboxdrawuni{2500}\pmboxdrawuni{2518}}
\cl{~~~~~~~~~~~~~~\pmboxdrawuni{2502}~approved}
\cl{~~~~~~~~~~~~~~▼}
\cl{~~~\pmboxdrawuni{250C}\pmboxdrawuni{2500}\pmboxdrawuni{2500}\pmboxdrawuni{2500}\pmboxdrawuni{2500}\pmboxdrawuni{2500}\pmboxdrawuni{2500}\pmboxdrawuni{2500}\pmboxdrawuni{2500}\pmboxdrawuni{2500}\pmboxdrawuni{2500}\pmboxdrawuni{2500}\pmboxdrawuni{2500}\pmboxdrawuni{2500}\pmboxdrawuni{2500}\pmboxdrawuni{2500}\pmboxdrawuni{2500}\pmboxdrawuni{2500}\pmboxdrawuni{2500}\pmboxdrawuni{2500}\pmboxdrawuni{2500}\pmboxdrawuni{2500}\pmboxdrawuni{2500}\pmboxdrawuni{2500}\pmboxdrawuni{2500}\pmboxdrawuni{2500}\pmboxdrawuni{2500}\pmboxdrawuni{2500}\pmboxdrawuni{2500}\pmboxdrawuni{2500}\pmboxdrawuni{2500}\pmboxdrawuni{2500}\pmboxdrawuni{2500}\pmboxdrawuni{2500}\pmboxdrawuni{2500}\pmboxdrawuni{2500}\pmboxdrawuni{2500}\pmboxdrawuni{2500}\pmboxdrawuni{2500}\pmboxdrawuni{2500}\pmboxdrawuni{2500}\pmboxdrawuni{2500}\pmboxdrawuni{2500}\pmboxdrawuni{2500}\pmboxdrawuni{2500}\pmboxdrawuni{2500}\pmboxdrawuni{2500}\pmboxdrawuni{2500}\pmboxdrawuni{2500}\pmboxdrawuni{2500}\pmboxdrawuni{2500}\pmboxdrawuni{2500}\pmboxdrawuni{2500}\pmboxdrawuni{2500}\pmboxdrawuni{2500}\pmboxdrawuni{2500}\pmboxdrawuni{2500}\pmboxdrawuni{2500}\pmboxdrawuni{2500}\pmboxdrawuni{2500}\pmboxdrawuni{2500}\pmboxdrawuni{2500}\pmboxdrawuni{2500}\pmboxdrawuni{2500}\pmboxdrawuni{2500}\pmboxdrawuni{2500}\pmboxdrawuni{2500}\pmboxdrawuni{2500}\pmboxdrawuni{2500}\pmboxdrawuni{2510}}
\cl{~~~\pmboxdrawuni{2502}~SANDBOX~(if~enabled)~enforces~filesystem~and~network~at~the~OS~~~~~\pmboxdrawuni{2502}}
\cl{~~~\pmboxdrawuni{2514}\pmboxdrawuni{2500}\pmboxdrawuni{2500}\pmboxdrawuni{2500}\pmboxdrawuni{2500}\pmboxdrawuni{2500}\pmboxdrawuni{2500}\pmboxdrawuni{2500}\pmboxdrawuni{2500}\pmboxdrawuni{2500}\pmboxdrawuni{2500}\pmboxdrawuni{2500}\pmboxdrawuni{2500}\pmboxdrawuni{2500}\pmboxdrawuni{2500}\pmboxdrawuni{2500}\pmboxdrawuni{2500}\pmboxdrawuni{2500}\pmboxdrawuni{2500}\pmboxdrawuni{2500}\pmboxdrawuni{2500}\pmboxdrawuni{2500}\pmboxdrawuni{2500}\pmboxdrawuni{2500}\pmboxdrawuni{2500}\pmboxdrawuni{2500}\pmboxdrawuni{2500}\pmboxdrawuni{2500}\pmboxdrawuni{2500}\pmboxdrawuni{2500}\pmboxdrawuni{2500}\pmboxdrawuni{2500}\pmboxdrawuni{2500}\pmboxdrawuni{2500}\pmboxdrawuni{2500}\pmboxdrawuni{2500}\pmboxdrawuni{2500}\pmboxdrawuni{2500}\pmboxdrawuni{2500}\pmboxdrawuni{2500}\pmboxdrawuni{2500}\pmboxdrawuni{2500}\pmboxdrawuni{2500}\pmboxdrawuni{2500}\pmboxdrawuni{2500}\pmboxdrawuni{2500}\pmboxdrawuni{2500}\pmboxdrawuni{2500}\pmboxdrawuni{2500}\pmboxdrawuni{2500}\pmboxdrawuni{2500}\pmboxdrawuni{2500}\pmboxdrawuni{2500}\pmboxdrawuni{2500}\pmboxdrawuni{2500}\pmboxdrawuni{2500}\pmboxdrawuni{2500}\pmboxdrawuni{2500}\pmboxdrawuni{2500}\pmboxdrawuni{2500}\pmboxdrawuni{2500}\pmboxdrawuni{2500}\pmboxdrawuni{2500}\pmboxdrawuni{2500}\pmboxdrawuni{2500}\pmboxdrawuni{2500}\pmboxdrawuni{2500}\pmboxdrawuni{2500}\pmboxdrawuni{2500}\pmboxdrawuni{2518}}
\cl{}
\cl{~~~MOD:~one~that~handles~tool.check~answers~after~the~hooks.~Where~the}
\cl{~~~built-in~guard~is~not~loaded~it~can~approve~a~call~a~deny~rule~refused}
\cl{~~~or~a~hook~blocked~(Chapter~18).}
\end{codefig}
\figcaption{Figure 2 — How a tool call is decided. Deny is evaluated first and no rule or hook widens it; a blocking hook stops the call before the rules. A mod is the exception where the built-in guard is not loaded.}
\end{figure}

\FloatBarrier
\setcounter{section}{2}
\section{Understand a prompt before approving}

On supported Bash or PowerShell permission prompts, \texttt{Ctrl+E} asks for an explanation of the command and its risk without executing it. Deny is also an answer, and denying with a comment proposing a safer alternative is often the fastest route forward.

Before approving anything you did not write, ask:

\begin{itemize}
\item What exact files, account or host does it touch?
\item Is the target resolved, or does it rely on a broad wildcard?
\item Is the operation reversible?
\item Can it place a credential into output?
\item Can the same information be obtained read-only first?
\end{itemize}

Claude Code applies its own protections here — command-injection detection, fail-closed matching for unmatched commands in Manual mode, natural-language descriptions of complex commands, and, for most fetches, a separate model call over a fetched page so that Claude receives that call's answer rather than the raw page [59], [72]. That reduces exposure; the answer is still untrusted. None of them removes the need for the five questions.

\FloatBarrier
\setcounter{section}{3}
\section{Sandboxing}

Run \texttt{/\allowbreak{}sandbox} to open the sandbox panel [28]. The sandbox constrains filesystem and network access \textbf{below} the model layer, enforced by the operating system for the shell commands it wraps and their child processes. Not everything is wrapped: command hooks, local MCP servers, helper commands and mods run with your full access, as do the built-in file and web tools (which follow permission rules instead), commands matching \texttt{excluded\allowbreak{}Commands}, a command Claude retries unsandboxed (approved by your permission mode, or with no prompt by a matching allow rule), and most commands you type at the \texttt{!} prompt [28]. It runs on macOS (via Seatbelt), Linux and WSL2; native Windows is not supported [28].

The panel writes to \texttt{.\allowbreak{}claude/\allowbreak{}settings.\allowbreak{}local.\allowbreak{}json}; set \texttt{sandbox.\allowbreak{}enabled} in user settings to enable it everywhere, or deploy it through managed settings for an organisation [28]. The controls worth knowing:

\begin{itemize}
\item \texttt{sandbox.\allowbreak{}filesystem.\allowbreak{}deny\allowbreak{}Write}, \texttt{deny\allowbreak{}Read} and \texttt{allow\allowbreak{}Read}. Where read rules overlap, the more specific path wins, and \textbf{a deny holds inside a wider allow} — so \texttt{allow\allowbreak{}Read:\allowbreak{} ["\textasciitilde{}/\allowbreak{}"]} with \texttt{deny\allowbreak{}Read:\allowbreak{} ["\textasciitilde{}/\allowbreak{}**/\allowbreak{}.\allowbreak{}env"]} leaves every \texttt{.\allowbreak{}env} blocked while the rest of the home directory is readable [28].
\item \texttt{sandbox.\allowbreak{}credentials.\allowbreak{}files} and \texttt{sandbox.\allowbreak{}credentials.\allowbreak{}env\allowbreak{}Vars}, each entry \texttt{deny} or \texttt{mask}. Denied files are unreadable inside the sandbox; denied environment variables are unset before each sandboxed command runs. Masking shows the command a placeholder and has the sandbox proxy substitute the real value on requests to the hosts you allow; it is a grant of use, not a redaction, and its conditions are in Section 14.2 [28].
\item \texttt{sandbox.\allowbreak{}auto\allowbreak{}Allow\allowbreak{}Bash\allowbreak{}If\allowbreak{}Sandboxed} defaults to \texttt{true}, so sandboxed commands run without prompts, with exceptions. Content-scoped ask rules such as \texttt{Bash(git pu\allowbreak{}sh *)} still prompt, and so does a bare \texttt{Bash} ask rule in plan mode. \texttt{rm} of a critical path, Monitor commands, and forms such as \texttt{FOO=\allowbreak{}bar npm tes\allowbreak{}t}, \texttt{make CC=\allowbreak{}clang} and \texttt{eval "ls"} go through the regular permission flow, where a matching allow rule or the permission mode decides; plan mode, an auto-mode command carrying its own allowed domains, and server-side classifier review add further exceptions. \texttt{false} sends every command through the regular permission flow [28], [61].
\item Deny entries from every settings scope are merged, and a deny only ever narrows access: any scope can add one and no scope can remove one another scope added [28].
\end{itemize}

Sandboxing complements permissions. Neither is sufficient alone, and for a stronger boundary — untrusted code, unfamiliar dependencies, anything you would not run on your laptop — compare the options in [73] and consider a dev container [74].

\FloatBarrier
\setcounter{section}{4}
\section{Restricted mode for harnesses on shared machines}

\texttt{claude -\allowbreak{}-\allowbreak{}restricted} \emph{(requires 2.1.248 or later)} is for an evaluation harness that drives \texttt{claude} on a shared machine, where Claude Code must not run commands or read that machine's user and project settings [55]. It removes the built-in tools that run commands or code, and WebFetch, unless you name them individually in \texttt{-\allowbreak{}-\allowbreak{}tools} — the \texttt{default} preset does not bring them back. It confines the built-in file tools to the working directories, loads only managed settings and \texttt{-\allowbreak{}-\allowbreak{}settings}, and refuses \texttt{bypass\allowbreak{}Permissions} [33], [55]. It also refuses to create cloud sessions, on the client and before contacting the server, because a cloud session would run outside the restricted process and would not enforce it [60].

Two further effects are easy to miss. In auto mode the classifier cannot approve writes to protected paths in a restricted session, and the Markdown files in the auto-memory and subagent-memory directories lose the exemption from the protected \texttt{.\allowbreak{}claude} directory that they have in an ordinary session [33]. \texttt{CLAUDE\_\allowbreak{}CODE\_\allowbreak{}RESTRICTED=\allowbreak{}1} starts a session the same way, but Claude Code ignores it in a settings file's \texttt{env} block; \texttt{claude agen\allowbreak{}ts -\allowbreak{}-\allowbreak{}restricted} applies it to every session dispatched from agent view [75], [76]. Restricted mode is a launch condition, not a policy: nothing in it stops someone who can start \texttt{claude} without the flag. Enforce what must hold for everyone in managed settings (Chapter 31).

\FloatBarrier
\setcounter{section}{5}
\section{Checkpoints and rewind}

Claude Code creates a checkpoint for each prompt that starts a turn, tracking its file tools' edits, and \texttt{/\allowbreak{}rewind} restores conversation, code or both [77]. It is useful. It is not a backup, and the gaps are systematic:

\begin{itemize}
\item external actions are not reversible by a local rewind;
\item a file that only Bash commands or external programs changed is not tracked (once Claude has edited a file with its file tools, later shell changes to it are);
\item most edits made by subagents are not in the parent's checkpoint history;
\item writes through symlinks or hard links may fall outside coverage;
\item untracked or ignored files need separate care;
\item browser actions, messages, published artifacts and pushed commits cannot be un-sent.
\end{itemize}

Use Git for durable history before you enable anything in Part V. Chapter 8 is not optional preparation.

\subsection*{Checkpoint}

\begin{itemize}
\item Sensitive paths carry deny rules, and you have tested that they actually block.
\item External writes require ask, or a human gate.
\item You know whether sandboxing is active and what it covers.
\item You can display the current diff and the checkpoint state.
\item \texttt{bypass\allowbreak{}Permissions} is not in use outside an isolated environment.
\end{itemize}

\begingroup
\def\kprows{%
\item Rules are evaluated deny, then ask, then allow. Specificity does not change that order, and a deny cannot carry exceptions.
\item A bare tool name in \texttt{deny} removes the tool from Claude's context; a scoped rule leaves it available and blocks matching calls.
\item Sandboxing constrains filesystem and network below the model layer, and a deny holds inside a wider allow.
\item Checkpoints are not a backup: they do not cover files only the shell changed, most subagent edits, or anything already sent externally.
}%
\def\kpbody{\begin{minipage}{\textwidth}\subsection*{Key points}\begin{itemize}\kprows\end{itemize}\end{minipage}}%
\begingroup
\def\sloppy{\tolerance 9999\emergencystretch 3em\hfuzz 200pt\vfuzz 200pt}%
\hbadness=10000\vbadness=10000\hfuzz=200pt\vfuzz=200pt
\global\setbox\kpbox=\hbox{\kpbody}%
\endgroup
\par\addvspace{4.2mm}
\ifdim\dimexpr\ht\kpbox+\dp\kpbox\relax>0.30\textheight
  \typeout{HANDBOOK-KEYPOINTS broken \the\dimexpr\ht\kpbox+\dp\kpbox\relax}%
  \subsection*{Key points}
  \begin{itemize}\kprows\end{itemize}
\else
  \typeout{HANDBOOK-KEYPOINTS atomic \the\dimexpr\ht\kpbox+\dp\kpbox\relax}%
  \noindent\kpbody
\fi
\par\addvspace{1.4mm}
\endgroup

\FloatBarrier
\renewcommand{\chaptertitlelabel}{Part II \textperiodcentered\ Chapter 8}
\setcounter{chapter}{7}
\chapter{Protect work with Git and GitHub}

Git records local history. GitHub hosts a repository remotely and adds collaboration, review and access control. They are related but not interchangeable: a commit can exist entirely on your machine, and that is often exactly what you want while an agent is working.

\FloatBarrier
\setcounter{section}{0}
\section{A safe beginner sequence}

\begin{codeblock}{9.0}{10.6}
\cl{git~init}
\cl{git~status}
\end{codeblock}

Create \texttt{.\allowbreak{}gitignore} \textbf{before} the first commit:

\begin{codeblock}{9.0}{10.6}
\cl{.env}
\cl{.env.*}
\cl{!.env.example}
\cl{.claude/settings.local.json}
\cl{CLAUDE.local.md}
\cl{secrets/}
\cl{*.key}
\cl{*.pem}
\end{codeblock}

That pattern is a starting point, not a secret detector. Stage first, then inspect what you staged: \texttt{git diff -\allowbreak{}-\allowbreak{}staged} shows nothing until something is in the index, and it is the staged snapshot — not the working tree — that a commit records.

\begin{codeblock}{9.0}{10.6}
\cl{git~status~~~~~~~~~~~~~~~~~\#~what~is~untracked~or~modified}
\cl{git~add~.~~~~~~~~~~~~~~~~~~\#~stage}
\cl{git~status~-{}-short~~~~~~~~~\#~the~staged~file~list}
\cl{git~diff~-{}-staged~~~~~~~~~~\#~the~staged~content:~exactly~what~will~be~committed}
\cl{git~commit~-m~"Create~research~brief~workspace"}
\end{codeblock}

\FloatBarrier
\setcounter{section}{1}
\section{Add a remote deliberately}

\begin{codeblock}{9.0}{10.6}
\cl{gh~auth~login}
\cl{gh~repo~create~research-brief-workspace~-{}-private~-{}-source=.~-{}-remote=origin}
\cl{git~remote~-v}
\end{codeblock}

Confirm owner, repository name, \textbf{visibility} and remote URL, then push only after a second secret and diff review:

\begin{codeblock}{9.0}{10.6}
\cl{git~push~-u~origin~main}
\end{codeblock}

If the branch is not \texttt{main}, use its real name. For collaborative work prefer a feature branch and a pull request over direct pushes to a default branch. Claude can propose all of these commands; account creation, authentication, visibility and release approval belong to you.

\FloatBarrier
\setcounter{section}{2}
\section{An agent-assisted Git workflow}

\begin{enumerate}
\item Claude summarises the intended change.
\item You approve the local edit plan.
\item Claude edits and runs verification.
\item You inspect \texttt{git status} and \texttt{git diff}, and open any untracked file: bare \texttt{git diff} omits what is already staged and never shows the content of a new file.
\item The change is staged deliberately, and you inspect \texttt{git diff -\allowbreak{}-\allowbreak{}staged} — the exact snapshot that will be committed.
\item Claude proposes a commit message grounded in that staged diff.
\item You approve the commit.
\item Push or pull request is a \textbf{separate}, visible step that a human owns and the agent is not authorised to perform.
\end{enumerate}

\begin{calloutbox}{palebrass}{brassdark}{2.0mm}
\textbf{CAUTION:} an automatic "commit and push at session end" hook can publish secrets, broken work, generated noise, or changes from the wrong branch. If you must automate, use a dedicated branch, a protected default branch, scoped credentials, pre-push checks and failure reporting (Chapter 17).
\end{calloutbox}

\FloatBarrier
\setcounter{section}{3}
\section{Automated review, and its limits}

\texttt{/\allowbreak{}security-\allowbreak{}review} runs a security pass over the diff between the current branch and origin's default branch, so it needs an \texttt{origin} remote; \texttt{/\allowbreak{}code-\allowbreak{}review} (alias \texttt{/\allowbreak{}review}), a bundled skill, reviews the current diff or a pull request, branch or path for correctness bugs at an effort level you choose, and can apply or post its findings [37], [78]. The \texttt{security-\allowbreak{}guidance} plugin has Claude review its own changes for vulnerabilities in the same session, and the Claude Security plugin scans a codebase and turns findings into patches you review [79], [80]. For a deeper pass, \texttt{/\allowbreak{}code-\allowbreak{}review ultr\allowbreak{}a} runs a multi-agent review in the cloud — a research preview that needs a claude.ai sign-in [81].

None of these replaces a human reviewer or a specialist assessment: the security page leaves the review of proposed code and commands with you [59], and none should be a release gate on its own.

\FloatBarrier
\setcounter{section}{4}
\section{Managed Code Review on pull requests}

Separate from the commands above, \textbf{Code Review} is a service an Owner turns on for a Team or Enterprise organisation: Claude reviews pull requests in GitHub repositories, including GitHub Enterprise Server, and posts inline comments ranked by severity, a summary, and a check run named \textbf{Claude Code Review} [78]. Several agents examine the diff and the surrounding code in parallel, a verification step filters the candidate findings against the code, and what remains is deduplicated and ranked. Each repository is set to review when a pull request opens, after every push, or only on request. Automatic reviews skip drafts and pull requests from forks, and after some pushes the earlier result is kept instead of a new review starting; an \texttt{@claude rev\allowbreak{}iew} comment starts one only when it comes from a person with write access to the repository. An Owner sets the review level — Quick, Standard or Deep — and Code Review, not the Owner, chooses the models [78].

Its boundaries are the useful part. A review runs on Anthropic infrastructure, not in a self-hosted environment; inside it, access to the repository is read-only, it pushes no commits, and it works by reading the code without building or running it. When it follows \texttt{REVIEW.\allowbreak{}md} it reads the copy on the base or default branch, never from the pull request's own changes, so a pull request cannot rewrite the instructions it is reviewed against [78]. Whether it follows one at all depends on branch protection: where Code Review cannot confirm the protection settings it reads none, and it honours instructions that lower a finding's severity, exclude paths or cap nits only from the root file on a protected default branch [82]. Protect the default branch. It never approves or blocks a merge, and its check run does not conclude as a failure. It is unavailable to organisations with Zero Data Retention, the HIPAA configuration or customer-managed encryption keys [78]. The page's own caution is the right one to carry into a release process: a review with no findings does not mean the code is free of errors or secure.

\FloatBarrier
\setcounter{section}{5}
\section{Recovery is broader than the repository}

Git does not cover untracked files, external databases and services, secret-manager values, browser actions, published artifacts, or local configuration excluded by \texttt{.\allowbreak{}gitignore}. Plan recovery for the workflow, not only for the source tree.

\subsection*{Exercise}

Initialise Git in the practice workspace. Ask Claude to draft a conservative \texttt{.\allowbreak{}gitignore} and to explain every pattern. Show the staged diff. Commit only after confirming that no sensitive or unrelated file is included.

\emph{A worked solution is given in Appendix M.}

\begingroup
\def\kprows{%
\item Create \texttt{.\allowbreak{}gitignore} before the first commit, and inspect \texttt{git diff -\allowbreak{}-\allowbreak{}staged} before every one.
\item Push is a separate, human-gated action — never a step inside an agent's plan.
\item An automatic commit-and-push hook can publish secrets, broken work or the wrong branch.
\item Git does not cover untracked files, external services, secret managers, browser actions or published artifacts. Plan recovery for the workflow.
}%
\def\kpbody{\begin{minipage}{\textwidth}\subsection*{Key points}\begin{itemize}\kprows\end{itemize}\end{minipage}}%
\begingroup
\def\sloppy{\tolerance 9999\emergencystretch 3em\hfuzz 200pt\vfuzz 200pt}%
\hbadness=10000\vbadness=10000\hfuzz=200pt\vfuzz=200pt
\global\setbox\kpbox=\hbox{\kpbody}%
\endgroup
\par\addvspace{4.2mm}
\ifdim\dimexpr\ht\kpbox+\dp\kpbox\relax>0.30\textheight
  \typeout{HANDBOOK-KEYPOINTS broken \the\dimexpr\ht\kpbox+\dp\kpbox\relax}%
  \subsection*{Key points}
  \begin{itemize}\kprows\end{itemize}
\else
  \typeout{HANDBOOK-KEYPOINTS atomic \the\dimexpr\ht\kpbox+\dp\kpbox\relax}%
  \noindent\kpbody
\fi
\par\addvspace{1.4mm}
\endgroup

\breakrule

\FloatBarrier
\parttitle{Part III}{Project knowledge and context}
\addcontentsline{toc}{part}{Part III \textemdash\ Project knowledge and context}

\FloatBarrier
\renewcommand{\chaptertitlelabel}{Part III \textperiodcentered\ Chapter 9}
\setcounter{chapter}{8}
\chapter{Build an effective \texttt{CLAUDE.\allowbreak{}md}}

\texttt{CLAUDE.\allowbreak{}md} is durable project guidance loaded at the start of every session. It is delivered as a user message after the system prompt, not as part of the system prompt, so Claude reads it and tries to follow it but there is no guarantee of compliance — especially for vague or conflicting instructions [25]. An instruction that \emph{must} hold belongs in a permission rule (Chapter 7) or a hook (Chapter 17).

\FloatBarrier
\setcounter{section}{0}
\section{Where instructions live}

\begingroup
\def\tblrows{%
Managed policy & macOS \texttt{/\allowbreak{}Library/\allowbreak{}Application\allowbreak{} Support/\allowbreak{}Claude\allowbreak{}Code/\allowbreak{}CLAUDE.\allowbreak{}md}; Linux and WSL \texttt{/\allowbreak{}etc/\allowbreak{}claude-\allowbreak{}code/\allowbreak{}CLAUDE.\allowbreak{}md}; Windows \texttt{C:\allowbreak{}\textbackslash{}Program Fil\allowbreak{}es\textbackslash{}Claude\allowbreak{}Code\textbackslash{}CLAUDE.\allowbreak{}md}; or the \texttt{claude\allowbreak{}Md} key in managed settings & Organisation-wide standards, security policy, compliance reminders. Cannot be excluded with \texttt{claude\allowbreak{}Md\allowbreak{}Excludes}, but is not loaded in a session started with instruction loading turned off: guidance, not enforcement \\
User & \texttt{\textasciitilde{}/\allowbreak{}.\allowbreak{}claude/\allowbreak{}CLAUDE.\allowbreak{}md} & Personal defaults that genuinely apply everywhere \\
Project & \texttt{.\allowbreak{}/\allowbreak{}CLAUDE.\allowbreak{}md} or \texttt{.\allowbreak{}/\allowbreak{}.\allowbreak{}claude/\allowbreak{}CLAUDE.\allowbreak{}md} & Purpose, structure, commands, rules, verification. Shared through version control \\
Local project & \texttt{.\allowbreak{}/\allowbreak{}CLAUDE.\allowbreak{}local.\allowbreak{}md} & Private project preferences. Add to \texttt{.\allowbreak{}gitignore} \\
Subdirectory & \texttt{path/\allowbreak{}to/\allowbreak{}CLAUDE.\allowbreak{}md} & Instructions relevant only within that subtree; loaded on demand when Claude reads, writes or edits a file there \\
}%
\def\tblbody{\begin{minipage}{\textwidth}\boxcaption{Table 6 — Instruction file locations, in load order from broadest to most specific}
{\footnotesize\begin{tabular}{L{24.6mm}L{59.4mm}L{55.8mm}}
\toprule
\textbf{Scope} & \textbf{Location} & \textbf{Appropriate content} \\
\midrule
\tblrows
\bottomrule\end{tabular}}\end{minipage}}%
\begingroup
\def\sloppy{\tolerance 9999\emergencystretch 3em\hfuzz 200pt\vfuzz 200pt}%
\hbadness=10000\vbadness=10000\hfuzz=200pt\vfuzz=200pt
\global\setbox\tblbox=\hbox{\tblbody}%
\endgroup
\par\addvspace{2.6mm}
\ifdim\dimexpr\ht\tblbox+\dp\tblbox\relax>0.55\textheight
  \typeout{HANDBOOK-TABLE broken \the\dimexpr\ht\tblbox+\dp\tblbox\relax}%
  \tabcaption{Table 6 — Instruction file locations, in load order from broadest to most specific}
  {\footnotesize\begin{longtable}{L{24.6mm}L{59.4mm}L{55.8mm}}
  \toprule
\textbf{Scope} & \textbf{Location} & \textbf{Appropriate content} \\
\midrule\endfirsthead
  \multicolumn{3}{@{}l@{}}{%
  \sffamily\footnotesize\itshape\color{inkgrey}Table 6 — Instruction file locations, in load order from broadest to most specific \textemdash\ continued}\\[1.2mm]
  \toprule
\textbf{Scope} & \textbf{Location} & \textbf{Appropriate content} \\
\midrule\endhead
  \bottomrule\endfoot
  \bottomrule\endlastfoot
  \tblrows
  \end{longtable}}%
\else
  \typeout{HANDBOOK-TABLE atomic \the\dimexpr\ht\tblbox+\dp\tblbox\relax}%
  \noindent\tblbody
\fi
\par\addvspace{2.6mm}
\endgroup

Files are \textbf{concatenated}, not overridden. Across the tree, content is ordered from the filesystem root down to the working directory, so instructions closest to where you launched Claude are read last; within a directory, \texttt{CLAUDE.\allowbreak{}local.\allowbreak{}md} is appended after \texttt{CLAUDE.\allowbreak{}md} [25].

\textbf{\texttt{AGENTS.\allowbreak{}md} is read too, but only in the absence of a \texttt{CLAUDE.\allowbreak{}md}} \emph{(requires 2.1.277 or later)}. By default Claude Code reads a repository's \texttt{AGENTS.\allowbreak{}md} when there is no \texttt{CLAUDE.\allowbreak{}md}, \texttt{.\allowbreak{}claude/\allowbreak{}CLAUDE.\allowbreak{}md} or \texttt{CLAUDE.\allowbreak{}local.\allowbreak{}md} in the working directory or above it, and reads only the \texttt{CLAUDE.\allowbreak{}md} files when there is one [25]. Three consequences follow. Adding a private \texttt{CLAUDE.\allowbreak{}local.\allowbreak{}md} to a repository that relies on \texttt{AGENTS.\allowbreak{}md} silently stops \texttt{AGENTS.\allowbreak{}md} being read. \texttt{AGENTS.\allowbreak{}local.\allowbreak{}md}, \texttt{AGENTS.\allowbreak{}override.\allowbreak{}md} and anything under \texttt{.\allowbreak{}agents/\allowbreak{}} are never read. And an \texttt{AGENTS.\allowbreak{}md} read this way does not fire \texttt{Instruction\allowbreak{}s\allowbreak{}Loaded} hooks, so a log built on that hook does not show it. The \textbf{Project instructions} setting in \texttt{/\allowbreak{}config} changes the default: \texttt{claude-\allowbreak{}md-\allowbreak{}and-\allowbreak{}agents-\allowbreak{}md} reads both, \texttt{claude-\allowbreak{}md} reads only \texttt{CLAUDE.\allowbreak{}md}, and \texttt{managed-\allowbreak{}only} leaves out project, local and user instruction files at launch. Where both files exist and one source of truth is wanted, \texttt{@AGENTS.\allowbreak{}md} on the first line of \texttt{CLAUDE.\allowbreak{}md} still works and never loads the file twice. That import is also the fallback where direct reading is unavailable: a session with the built-in \texttt{AGENTS.\allowbreak{}md} plugin disabled, sometimes the first session after an upgrade, and — before 2.1.281 — sessions on Amazon Bedrock or with telemetry disabled.

\FloatBarrier
\setcounter{section}{1}
\section{A useful project file}

\begin{codeblock}{9.0}{10.6}
\cl{\#~Research~Brief~Workspace}
\cl{}
\cl{\#\#~Purpose}
\cl{Produce~concise,~evidence-led~briefs~from~approved~sources.}
\cl{}
\cl{\#\#~Directory~map}
\cl{-~\textasciigrave{}sources/\textasciigrave{}:~immutable~source~material}
\cl{-~\textasciigrave{}notes/\textasciigrave{}:~working~notes~and~extraction}
\cl{-~\textasciigrave{}drafts/\textasciigrave{}:~unapproved~manuscripts}
\cl{-~\textasciigrave{}deliverables/\textasciigrave{}:~human-approved~outputs~only}
\cl{-~\textasciigrave{}references/\textasciigrave{}:~retrieval-on-demand~policies}
\cl{-~\textasciigrave{}templates/\textasciigrave{}:~reusable~structures}
\cl{}
\cl{\#\#~Operating~rules}
\cl{-~Never~modify~files~in~\textasciigrave{}sources/\textasciigrave{}.}
\cl{-~Separate~source~claims~from~analyst~inference.}
\cl{-~Record~one~atomic~claim~per~evidence-ledger~row.}
\cl{-~Do~not~publish,~send,~push,~or~overwrite~a~deliverable~without~approval.}
\cl{-~Never~read~or~print~credentials.}
\cl{}
\cl{\#\#~Verification~commands}
\cl{-~Structure~check:~\textasciigrave{}python3~tools/check\_structure.py~drafts/<file>.md\textasciigrave{}}
\cl{-~Link~check:~~~~~~\textasciigrave{}python3~tools/check\_links.py~drafts/<file>.md\textasciigrave{}}
\cl{}
\cl{\#\#~Definition~of~done}
\cl{-~All~material~claims~carry~direct~source~references.}
\cl{-~Contradictions~and~gaps~are~visible~rather~than~resolved~silently.}
\cl{-~Links~and~rendering~have~been~checked~and~the~output~inspected.}
\cl{-~A~human~has~approved~promotion~from~\textasciigrave{}drafts/\textasciigrave{}~to~\textasciigrave{}deliverables/\textasciigrave{}.}
\cl{}
\cl{\#\#~References}
\cl{-~Source-quality~rules:~read~\textasciigrave{}references/source-policy.md\textasciigrave{}.}
\cl{-~House~style:~read~\textasciigrave{}references/editorial-style.md\textasciigrave{}~only~when~drafting.}
\end{codeblock}

\FloatBarrier
\setcounter{section}{2}
\section{Initialise, then edit}

\texttt{/\allowbreak{}init} generates a starter \texttt{CLAUDE.\allowbreak{}md} by analysing the project; if one exists it suggests improvements rather than overwriting [25]. It also reads Cursor and Copilot rule files and incorporates the relevant parts. Setting \texttt{CLAUDE\_\allowbreak{}CODE\_\allowbreak{}NEW\_\allowbreak{}INIT=\allowbreak{}1} enables an interactive multi-phase flow that explores with a subagent, asks follow-up questions and presents a reviewable proposal before writing anything [25].

The course warns against \texttt{/\allowbreak{}init} because generated files could become verbose [44]. Both positions reconcile: generate a draft, then edit it hard.

\begin{enumerate}
\item Run \texttt{/\allowbreak{}init}.
\item Read every line.
\item Delete generic description, changelog, transient state and duplicated documentation.
\item Keep operational facts, conventions that differ from tool defaults, pitfalls, and verification commands.
\item Run \texttt{/\allowbreak{}context all} and read the list under \textbf{Memory files}: that is what loaded at launch. \texttt{/\allowbreak{}memory} is the editor, not the evidence — it lists instruction-file locations, including files that do not exist yet. A subdirectory's file is absent from both until Claude works there, when a \texttt{Loaded} line names it [25].
\end{enumerate}

\textbf{Target under 200 lines.} Longer files consume context and reduce adherence. A file over 4 MiB is skipped entirely [25]. \texttt{/\allowbreak{}doctor} proposes trims for a checked-in file, cutting content Claude can derive from the codebase and keeping rationale and conventions (2.1.206 or later), and \texttt{/\allowbreak{}doctor prom\allowbreak{}pt-\allowbreak{}audit} reports instruction files that are outdated, refer to files or commands that do not exist, or contradict each other, proposing edits without applying them \emph{(requires 2.1.283 or later)}; it runs through the bundled \texttt{/\allowbreak{}claude-\allowbreak{}api} skill and is unavailable where that skill is turned off [25].

\FloatBarrier
\setcounter{section}{3}
\section{An index, not a warehouse}

Long reference material belongs in separate files with a retrieval rule, as in the \texttt{References} section above. Imports with \texttt{@path/\allowbreak{}to/\allowbreak{}file} are supported to a maximum depth of four hops, but \textbf{imported files load together with the file that imports them and do not reduce context occupancy} — they help organisation, not cost [25]. Import parsing skips code spans and fences, so `\texttt{ }@README\texttt{ }` in backticks stays literal.

An import in a \emph{project} memory file whose path resolves outside the working directory is external, and the first time Claude Code encounters one it shows an approval dialog listing the files; declining disables them permanently [25]. That dialog is a supply-chain control: it exists because someone else can commit an import into a shared repository. A symlink in \texttt{.\allowbreak{}claude/\allowbreak{}rules/\allowbreak{}} whose target is outside the working directory needs the same approval, and even then only linked rules with no \texttt{paths} field load. A link to a network path, such as a UNC share, is not followed at all; \texttt{\textbackslash{}\textbackslash{}wsl\$} paths do not count as network paths.

Block-level HTML comments are stripped before injection, so maintainer notes cost no context [25].

\FloatBarrier
\setcounter{section}{4}
\section{Modular rules}

Split large instruction sets into \texttt{.\allowbreak{}claude/\allowbreak{}rules/\allowbreak{}*.\allowbreak{}md}, discovered recursively [25]. A rule with no \texttt{paths} frontmatter loads at launch with the same priority as \texttt{.\allowbreak{}claude/\allowbreak{}CLAUDE.\allowbreak{}md}. A rule \emph{with} \texttt{paths} loads only when Claude reads, writes or edits a matching file, or views one with a single-file shell command such as \texttt{cat} or \texttt{head}:

\begin{codeblock}{9.0}{10.6}
\cl{-{}-{}-}
\cl{paths:}
\cl{~~-~"sources/**/*.md"}
\cl{~~-~"notes/**/*.\{md,txt\}"}
\cl{-{}-{}-}
\cl{}
\cl{\#~Source~handling}
\cl{}
\cl{-~Never~edit~a~file~under~\textasciigrave{}sources/\textasciigrave{}.}
\cl{-~Record~provenance~as~\textasciigrave{}source-id~·~retrieved-date~·~URL~or~physical~location\textasciigrave{}.}
\end{codeblock}

User rules in \texttt{\textasciitilde{}/\allowbreak{}.\allowbreak{}claude/\allowbreak{}rules/\allowbreak{}} load before project rules, so a project rule appears later in context. Neither overrides the other: where the two conflict Claude may follow either, so keep them consistent [25]. Symlinks are supported and circular links are handled. A rule's whole \texttt{paths} list shares a budget of 1,000 expanded patterns and 4 MiB; a pattern that would exceed it is used unexpanded and matches nothing [25]. \texttt{paths} is the only frontmatter field a rule has; any other is ignored without an error, and frontmatter that does not parse makes the rule load unconditionally.

A file that loads on demand — a subdirectory's \texttt{CLAUDE.\allowbreak{}md}, or a path-scoped rule — is also subject to a per-file limit: 40,000 characters in a 200K context window. Past it, Claude receives only the start of the file and a note to read the rest, and until it chooses to, the later instructions are not in context \emph{(requires 2.1.309 or later)} [25].

\FloatBarrier
\setcounter{section}{5}
\section{Maintenance trigger}

Review \texttt{CLAUDE.\allowbreak{}md} when Claude repeatedly searches for a documented command, when you correct the same rule twice, when structure changes, when two instructions conflict, or when \texttt{/\allowbreak{}context} shows disproportionate instruction overhead. Ask for a proposed diff with a reason for every addition and removal — never a blind "optimise this".

\textbf{Note on compaction.} Project-root \texttt{CLAUDE.\allowbreak{}md} survives \texttt{/\allowbreak{}compact}: Claude Code re-reads it from disk and re-injects it [25]. Nested files and path-scoped rules reload as Claude reads the files they apply to. An instruction that disappears after compaction was given only in conversation, lives in a nested file that has not reloaded yet, or is a path-scoped rule that has not matched a file since. The first is the argument for writing it down.

\subsection*{Exercise}

Create the project \texttt{CLAUDE.\allowbreak{}md}, start a new session, and ask Claude to state the directory rules without searching every file. Then ask which instructions are stable and which belong in a path-scoped rule or a skill.

\emph{A worked solution is given in Appendix M.}

\begingroup
\def\kprows{%
\item \texttt{CLAUDE.\allowbreak{}md} is delivered as context, not as enforced configuration, and so is everything under \texttt{.\allowbreak{}claude/\allowbreak{}rules/\allowbreak{}}; something that must hold belongs in a permission rule or a hook.
\item Files are concatenated from the filesystem root down, with \texttt{CLAUDE.\allowbreak{}local.\allowbreak{}md} appended last at each level. Under the default setting, \texttt{AGENTS.\allowbreak{}md} is read only where no \texttt{CLAUDE.\allowbreak{}md} is.
\item Target under 200 lines. Imports organise content but do not reduce context, because they load at launch.
\item Project-root \texttt{CLAUDE.\allowbreak{}md} survives \texttt{/\allowbreak{}compact}; nested files and path-scoped rules reload on demand, and an instruction given only in conversation is gone.
}%
\def\kpbody{\begin{minipage}{\textwidth}\subsection*{Key points}\begin{itemize}\kprows\end{itemize}\end{minipage}}%
\begingroup
\def\sloppy{\tolerance 9999\emergencystretch 3em\hfuzz 200pt\vfuzz 200pt}%
\hbadness=10000\vbadness=10000\hfuzz=200pt\vfuzz=200pt
\global\setbox\kpbox=\hbox{\kpbody}%
\endgroup
\par\addvspace{4.2mm}
\ifdim\dimexpr\ht\kpbox+\dp\kpbox\relax>0.30\textheight
  \typeout{HANDBOOK-KEYPOINTS broken \the\dimexpr\ht\kpbox+\dp\kpbox\relax}%
  \subsection*{Key points}
  \begin{itemize}\kprows\end{itemize}
\else
  \typeout{HANDBOOK-KEYPOINTS atomic \the\dimexpr\ht\kpbox+\dp\kpbox\relax}%
  \noindent\kpbody
\fi
\par\addvspace{1.4mm}
\endgroup

\FloatBarrier
\renewcommand{\chaptertitlelabel}{Part III \textperiodcentered\ Chapter 10}
\setcounter{chapter}{9}
\chapter{Global instructions and auto memory}

Persistent knowledge serves three different purposes, and conflating them produces bloated context and stale behaviour:

\begin{enumerate}
\item \textbf{Instructions} tell Claude how to behave — \texttt{CLAUDE.\allowbreak{}md}, rules.
\item \textbf{Reference files} store facts and procedures Claude retrieves on demand.
\item \textbf{Auto memory} stores concise notes Claude writes for itself.
\end{enumerate}

\FloatBarrier
\setcounter{section}{0}
\section{Global instructions}

Use \texttt{\textasciitilde{}/\allowbreak{}.\allowbreak{}claude/\allowbreak{}CLAUDE.\allowbreak{}md} sparingly. Suitable: your preferred working language, a universal safety reminder, a stable output convention, a pointer to a user rules directory. Unsuitable: project-specific commands, one client's house style, temporary priorities, or a catalogue of every tool you might one day use.

\FloatBarrier
\setcounter{section}{1}
\section{Auto memory}

Auto memory is on by default in local sessions. As it works, Claude saves four kinds of note, recorded as a \texttt{type} field in the memory file's frontmatter [25]:

\begin{itemize}
\item \texttt{user} — your role, expertise and working preferences;
\item \texttt{feedback} — corrections you gave and approaches you confirmed;
\item \texttt{project} — ongoing work, deadlines and decisions not derivable from the code or history;
\item \texttt{reference} — where to find information outside the project.
\end{itemize}

Claude skips anything derivable from the codebase and anything your \texttt{CLAUDE.\allowbreak{}md} already says.

\textbf{Storage.} Each project gets \texttt{\textasciitilde{}/\allowbreak{}.\allowbreak{}claude/\allowbreak{}projects/\allowbreak{}<project>/\allowbreak{}memory/\allowbreak{}}, where \texttt{<project>} is derived \textbf{from the git repository} by default, so every worktree and subdirectory of the same repository shares one memory directory; outside a repository the project root is used [25]. That keying is a default and not an identity: \texttt{auto\allowbreak{}Memory\allowbreak{}Directory} moves it, and \texttt{CLAUDE\_\allowbreak{}CODE\_\allowbreak{}PROJECT\_\allowbreak{}DIR\_\allowbreak{}NAME} set beside \texttt{CLAUDE\_\allowbreak{}CONFIG\_\allowbreak{}DIR} makes every project launched with that configuration directory share one \emph{(requires 2.1.234 or later)}. \texttt{MEMORY.\allowbreak{}md} is the index; topic files hold detail and are read on demand. Relocate storage with \texttt{auto\allowbreak{}Memory\allowbreak{}Directory}, an absolute or \texttt{\textasciitilde{}/\allowbreak{}}-prefixed path, honoured from any settings scope — though a value set in a project's \texttt{.\allowbreak{}claude/\allowbreak{}settings.\allowbreak{}json} or \texttt{.\allowbreak{}claude/\allowbreak{}settings.\allowbreak{}local.\allowbreak{}json} is honoured only under the same workspace-trust rule that governs hooks in settings files. While \texttt{permissions.\allowbreak{}block\allowbreak{}Reads\allowbreak{}Outside\allowbreak{}Working\allowbreak{}Directories} is on, no auto memory is loaded from or saved to a directory that a repository-supplied settings file chooses [25]. Auto memory is machine-local and is not shared across machines or cloud environments.

\textbf{Loading limit.} The first 200 lines of \texttt{MEMORY.\allowbreak{}md}, or the first 25 KB, whichever comes first, are loaded at the start of every conversation; content beyond that is not loaded [25]. When the index approaches a limit Claude Code prompts Claude to shorten it; when it exceeds one, the write succeeds but Claude Code returns an error instructing Claude to rewrite the index, because everything past the limit is dropped on the next load.

\textbf{Disabling.} Toggle auto memory in \texttt{/\allowbreak{}memory}, which writes \texttt{auto\allowbreak{}Memory\allowbreak{}Enabled} to \texttt{\textasciitilde{}/\allowbreak{}.\allowbreak{}claude/\allowbreak{}settings.\allowbreak{}json}; set \texttt{\{"auto\allowbreak{}Memory\allowbreak{}Enabled":\allowbreak{} false\}} in a project's settings to disable it for that project only; or set \texttt{CLAUDE\_\allowbreak{}CODE\_\allowbreak{}DISABLE\_\allowbreak{}AUTO\_\allowbreak{}MEMORY=\allowbreak{}1} [25]. This corrects M06 in Table 1.

\begin{calloutbox}{palebrass}{brassdark}{2.0mm}
\textbf{CAUTION — retention.} Claude Code deletes old session transcripts after the \texttt{cleanup\allowbreak{}Period\allowbreak{}Days} retention period but \textbf{excludes the memory directory from that sweep}. \texttt{MEMORY.\allowbreak{}md} and its topic files persist until you or Claude edits or deletes them [25]. If your retention obligations assume a 30-day local sweep, auto memory is outside it. This corrects E5.
\end{calloutbox}

Treat auto memory as a convenience, not a specification. It is a cache the model maintains for itself: Claude may remember something situational, or generalise a preference too far, and nothing reviews an entry before it is used. It is not the authoritative record of a decision, a deadline or a policy. Stable project truth belongs in reviewed files under version control.

\FloatBarrier
\setcounter{section}{2}
\section{A decision table}

\begingroup
\def\tblrows{%
"Never publish without approval" & A permission rule or hook on \textbf{each} route that can publish — shell, MCP server, browser, deployment — plus a \texttt{CLAUDE.\allowbreak{}md} reminder. No single rule means "publish" \\
Project directory map & Project \texttt{CLAUDE.\allowbreak{}md} \\
Editorial style guide & \texttt{references/\allowbreak{}editorial-\allowbreak{}style.\allowbreak{}md}, retrieved on demand \\
Rules that apply only to one file type & \texttt{.\allowbreak{}claude/\allowbreak{}rules/\allowbreak{}*.\allowbreak{}md} with \texttt{paths} frontmatter \\
A multi-step procedure & A skill (Chapter 15) \\
Current open tasks & \texttt{notes/\allowbreak{}task-\allowbreak{}state.\allowbreak{}md} or an issue tracker \\
A correction useful only this session & The conversation \\
A recurring preference Claude inferred & Auto memory, after your review \\
Full change history & Git \\
Something that must hold regardless of what Claude decides & A \texttt{Pre\allowbreak{}Tool\allowbreak{}Use} hook (Chapter 17) \\
}%
\def\tblbody{\begin{minipage}{\textwidth}\boxcaption{Table 7 — Where a given piece of knowledge belongs}
{\footnotesize\begin{tabular}{L{62.4mm}L{80.6mm}}
\toprule
\textbf{Information} & \textbf{Best home} \\
\midrule
\tblrows
\bottomrule\end{tabular}}\end{minipage}}%
\begingroup
\def\sloppy{\tolerance 9999\emergencystretch 3em\hfuzz 200pt\vfuzz 200pt}%
\hbadness=10000\vbadness=10000\hfuzz=200pt\vfuzz=200pt
\global\setbox\tblbox=\hbox{\tblbody}%
\endgroup
\par\addvspace{2.6mm}
\ifdim\dimexpr\ht\tblbox+\dp\tblbox\relax>0.55\textheight
  \typeout{HANDBOOK-TABLE broken \the\dimexpr\ht\tblbox+\dp\tblbox\relax}%
  \tabcaption{Table 7 — Where a given piece of knowledge belongs}
  {\footnotesize\begin{longtable}{L{62.4mm}L{80.6mm}}
  \toprule
\textbf{Information} & \textbf{Best home} \\
\midrule\endfirsthead
  \multicolumn{2}{@{}l@{}}{%
  \sffamily\footnotesize\itshape\color{inkgrey}Table 7 — Where a given piece of knowledge belongs \textemdash\ continued}\\[1.2mm]
  \toprule
\textbf{Information} & \textbf{Best home} \\
\midrule\endhead
  \bottomrule\endfoot
  \bottomrule\endlastfoot
  \tblrows
  \end{longtable}}%
\else
  \typeout{HANDBOOK-TABLE atomic \the\dimexpr\ht\tblbox+\dp\tblbox\relax}%
  \noindent\tblbody
\fi
\par\addvspace{2.6mm}
\endgroup

\FloatBarrier
\setcounter{section}{3}
\section{Memory audit}

Monthly for active workspaces, or after any surprising behaviour:

\begin{codeblock}{8.0}{9.4}
\cl{List~every~instruction~and~memory~file~affecting~this~project.~For~each,~state~its~scope,}
\cl{approximate~context~cost,~last~relevant~use,~and~any~conflict~or~stale~entry.~Propose~changes}
\cl{as~a~diff.~Do~not~apply~them.}
\end{codeblock}

Then verify the file list yourself with \texttt{/\allowbreak{}context all}, which shows what loaded at launch, and \texttt{/\allowbreak{}memory}, which lists the locations and opens them; files that load on demand announce themselves with a \texttt{Loaded} line instead. Use \texttt{/\allowbreak{}doctor} or \texttt{/\allowbreak{}hooks} when something configured is not taking effect [56]. The \texttt{Instruction\allowbreak{}s\allowbreak{}Loaded} hook logs which \texttt{CLAUDE.\allowbreak{}md} and rules files loaded, when and why — the best answer when path-scoped rules misbehave, with the one gap that it does not fire for an \texttt{AGENTS.\allowbreak{}md} read in place of a \texttt{CLAUDE.\allowbreak{}md} [25], [83].

An audit of instructions answers four different questions, and a pass on one is not a pass on the next: whether a file \textbf{loaded}; whether it \textbf{influenced} what Claude did; whether anything \textbf{enforced} it; and whether the \textbf{outcome} was right. For a consequential run, record the resolved instruction files and their versions — a commit hash, not a file name.

\begingroup
\def\kprows{%
\item Instructions, reference files and auto memory serve three different purposes; conflating them bloats context and staleness.
\item Auto memory is keyed to the git repository by default and shared across all its worktrees.
\item Only the first 200 lines or 25 KB of \texttt{MEMORY.\allowbreak{}md} load at session start.
\item Auto-memory files are excluded from the \texttt{cleanup\allowbreak{}Period\allowbreak{}Days} retention sweep and persist until deleted.
}%
\def\kpbody{\begin{minipage}{\textwidth}\subsection*{Key points}\begin{itemize}\kprows\end{itemize}\end{minipage}}%
\begingroup
\def\sloppy{\tolerance 9999\emergencystretch 3em\hfuzz 200pt\vfuzz 200pt}%
\hbadness=10000\vbadness=10000\hfuzz=200pt\vfuzz=200pt
\global\setbox\kpbox=\hbox{\kpbody}%
\endgroup
\par\addvspace{4.2mm}
\ifdim\dimexpr\ht\kpbox+\dp\kpbox\relax>0.30\textheight
  \typeout{HANDBOOK-KEYPOINTS broken \the\dimexpr\ht\kpbox+\dp\kpbox\relax}%
  \subsection*{Key points}
  \begin{itemize}\kprows\end{itemize}
\else
  \typeout{HANDBOOK-KEYPOINTS atomic \the\dimexpr\ht\kpbox+\dp\kpbox\relax}%
  \noindent\kpbody
\fi
\par\addvspace{1.4mm}
\endgroup

\FloatBarrier
\renewcommand{\chaptertitlelabel}{Part III \textperiodcentered\ Chapter 11}
\setcounter{chapter}{10}
\chapter{Engineer context deliberately}

Context is what the model has for the current turn: system prompt, tool descriptions, project instructions, attached files, tool results and conversation history. A file on disk is not "known" merely because Claude could read it.

\FloatBarrier
\setcounter{section}{0}
\section{Inspect the window}

\begin{codeblock}{9.0}{10.6}
\cl{/context}
\end{codeblock}

Use it to diagnose composition, not to chase a number: it breaks usage down by category and names the instruction and memory files that loaded [35]. The course cites 200,000–250,000 tokens as a threshold some users avoid; that is practitioner experience, not a product limit [44]. Windows, models and auto-compaction behaviour change, and \texttt{/\allowbreak{}autocompact} sets the auto-compact window explicitly [37], [63].

Compaction also happens without you. Claude Code compacts automatically as the window fills, and \textbf{idle compaction is on by default}: a conversation that has reached a threshold, 200,000 tokens unless you change it, and is then left for about an hour, is summarised while you are away, and your next message sends the summary instead of the conversation [61]. All of the conditions must hold: an interactive session, auto-compaction enabled, the one-hour prompt cache, and a computer that stays awake. The terminal marks it with a line beginning \texttt{Compacted w\allowbreak{}hile idle}. Set \texttt{idle\allowbreak{}Compaction} to \texttt{false} if a long conversation must survive a lunch break intact \emph{(requires 2.1.290 or later; an earlier release ignores the key, and turning off \texttt{auto\allowbreak{}Compact\allowbreak{}Enabled} is the way to stop it there)}.

\FloatBarrier
\setcounter{section}{1}
\section{Four ways to recover focus}

\textbf{1 — Persist durable state.} Findings, plans, decisions and open questions belong in project files. A conversation is a working surface, not a database.

\textbf{2 — Hand off to a clean session.}

\begin{codeblock}{8.0}{9.4}
\cl{Prepare~a~handoff~for~a~clean~session.~Include~the~objective,~completed~work,~approved}
\cl{decisions,~open~issues,~exact~file~references,~verification~already~performed,~and~the~next}
\cl{safe~action.~Exclude~conversational~history~that~does~not~affect~the~task.}
\end{codeblock}

A handoff is a manifest for continuation. It is not evidence, and it is not independence: a clean session that reads it inherits the author's account of what was verified, so anything that matters is re-checked against the artefact the handoff names, not taken from the handoff.

\textbf{3 — \texttt{/\allowbreak{}clear}.} Starts fresh while retaining project instructions and memory. It also clears any active goal, and it closes the Chrome tab group unless work that survives the clear is still running [21], [84]. Whether it clears session-scoped scheduled tasks is \textbf{UNVERIFIED} (Appendix J): cancel them explicitly rather than relying on \texttt{/\allowbreak{}clear}. Save your handoff first. The previous conversation is not lost: resume it with \texttt{/\allowbreak{}resume}, or reach it from the rewind menu's previous-session entry in the same process [31].

\textbf{4 — \texttt{/\allowbreak{}compact [in\allowbreak{}structions]}.} Summarises to free context. Compaction is lossy; say what must survive:

\begin{codeblock}{8.0}{9.4}
\cl{/compact~Preserve~all~approved~decisions,~unresolved~contradictions,~file~paths,~verification}
\cl{results,~and~the~human~approval~gate.~Remove~discarded~options.}
\end{codeblock}

To compact only part of a conversation, open \texttt{/\allowbreak{}rewind}, select a message and choose \textbf{Summarize from here} or \textbf{Summarize up to here} [35].

What comes back after a compaction is specific, and less than a reader tends to assume [35]. The system prompt, the project-root \texttt{CLAUDE.\allowbreak{}md}, unscoped rules and auto memory are re-injected from disk. Claude Code re-reads \textbf{up to five} of the files Claude had read or edited, most recently modified first, and a file over 5,000 tokens comes back as a path reference without its content. Invoked skills return truncated (Section 15.3). Path-scoped rules and nested \texttt{CLAUDE.\allowbreak{}md} files are \textbf{summarised away with the conversation} and reload only when Claude next touches a matching file, so a rule that must persist across a compaction should lose its \texttt{paths} or move to the project-root file. Context a hook added earlier is summarised with everything else. Background commands and subagents keep running.

Read the resulting summary when the task is consequential. Note the cost: \texttt{/\allowbreak{}compact} forces one summarisation request over the full history and then invalidates the conversation layer of the prompt cache, though the system prompt, and project context that has not changed, can stay cached [68]. Figure 3 sets out what each of these operations preserves and what it discards.

\begin{figure}[tbp]
\begin{codefig}{9.0}{10.6}
\cl{~~~~~~~~~~~~~~~~~~~~~~~~/compact~~~/clear~~~~resume†~~~/branch~~~/rewind§}
\cl{~~~\pmboxdrawuni{2500}\pmboxdrawuni{2500}\pmboxdrawuni{2500}\pmboxdrawuni{2500}\pmboxdrawuni{2500}\pmboxdrawuni{2500}\pmboxdrawuni{2500}\pmboxdrawuni{2500}\pmboxdrawuni{2500}\pmboxdrawuni{2500}\pmboxdrawuni{2500}\pmboxdrawuni{2500}\pmboxdrawuni{2500}\pmboxdrawuni{2500}\pmboxdrawuni{2500}\pmboxdrawuni{2500}\pmboxdrawuni{2500}\pmboxdrawuni{2500}\pmboxdrawuni{2500}\pmboxdrawuni{2500}\pmboxdrawuni{2500}\pmboxdrawuni{2500}\pmboxdrawuni{2500}\pmboxdrawuni{2500}\pmboxdrawuni{2500}\pmboxdrawuni{2500}\pmboxdrawuni{2500}\pmboxdrawuni{2500}\pmboxdrawuni{2500}\pmboxdrawuni{2500}\pmboxdrawuni{2500}\pmboxdrawuni{2500}\pmboxdrawuni{2500}\pmboxdrawuni{2500}\pmboxdrawuni{2500}\pmboxdrawuni{2500}\pmboxdrawuni{2500}\pmboxdrawuni{2500}\pmboxdrawuni{2500}\pmboxdrawuni{2500}\pmboxdrawuni{2500}\pmboxdrawuni{2500}\pmboxdrawuni{2500}\pmboxdrawuni{2500}\pmboxdrawuni{2500}\pmboxdrawuni{2500}\pmboxdrawuni{2500}\pmboxdrawuni{2500}\pmboxdrawuni{2500}\pmboxdrawuni{2500}\pmboxdrawuni{2500}\pmboxdrawuni{2500}\pmboxdrawuni{2500}\pmboxdrawuni{2500}\pmboxdrawuni{2500}\pmboxdrawuni{2500}\pmboxdrawuni{2500}\pmboxdrawuni{2500}\pmboxdrawuni{2500}\pmboxdrawuni{2500}\pmboxdrawuni{2500}\pmboxdrawuni{2500}\pmboxdrawuni{2500}\pmboxdrawuni{2500}\pmboxdrawuni{2500}\pmboxdrawuni{2500}\pmboxdrawuni{2500}\pmboxdrawuni{2500}\pmboxdrawuni{2500}}
\cl{~~~Conversation~history~~~~summary~~~~gone~~~reloaded~~~~copied~~cut~back}
\cl{~~~Project~CLAUDE.md~~~~~~re-read~~re-read~~~~re-read~~~in~place~in~place}
\cl{~~~Path-scoped~rules~~~~~~on~match~on~match~~~on~match~~in~place~~as~kept}
\cl{~~~Auto~memory~~~~~~~~~~~~re-read~~~~kept~~~~~~~kept~~~~~~kept~~~~~kept}
\cl{~~~Active~goal~~~~~~~~~~~~~~kept~~~~~gone~~~restored‡~~~~~kept~~~~~kept}
\cl{~~~Scheduled~tasks~~~~~~~~~~kept~~~~~~~~?~~~restored*~~~~~kept~~~~~kept}
\cl{~~~Session~permission}
\cl{~~~~~grants~~~~~~~~~~~~~~~~~kept~~~~~kept~~~~~~~gone~~~~kept**~~~~~kept}
\cl{~~~Files~on~disk~~~~~~~~~~~~kept~~~~~kept~~~~~~~kept~~~~~~kept~~~reverted}
\cl{~~~Git~history~~~~~~~~~~~~~~kept~~~~~kept~~~~~~~kept~~~~~~kept~~~~~kept}
\cl{~~~\pmboxdrawuni{2500}\pmboxdrawuni{2500}\pmboxdrawuni{2500}\pmboxdrawuni{2500}\pmboxdrawuni{2500}\pmboxdrawuni{2500}\pmboxdrawuni{2500}\pmboxdrawuni{2500}\pmboxdrawuni{2500}\pmboxdrawuni{2500}\pmboxdrawuni{2500}\pmboxdrawuni{2500}\pmboxdrawuni{2500}\pmboxdrawuni{2500}\pmboxdrawuni{2500}\pmboxdrawuni{2500}\pmboxdrawuni{2500}\pmboxdrawuni{2500}\pmboxdrawuni{2500}\pmboxdrawuni{2500}\pmboxdrawuni{2500}\pmboxdrawuni{2500}\pmboxdrawuni{2500}\pmboxdrawuni{2500}\pmboxdrawuni{2500}\pmboxdrawuni{2500}\pmboxdrawuni{2500}\pmboxdrawuni{2500}\pmboxdrawuni{2500}\pmboxdrawuni{2500}\pmboxdrawuni{2500}\pmboxdrawuni{2500}\pmboxdrawuni{2500}\pmboxdrawuni{2500}\pmboxdrawuni{2500}\pmboxdrawuni{2500}\pmboxdrawuni{2500}\pmboxdrawuni{2500}\pmboxdrawuni{2500}\pmboxdrawuni{2500}\pmboxdrawuni{2500}\pmboxdrawuni{2500}\pmboxdrawuni{2500}\pmboxdrawuni{2500}\pmboxdrawuni{2500}\pmboxdrawuni{2500}\pmboxdrawuni{2500}\pmboxdrawuni{2500}\pmboxdrawuni{2500}\pmboxdrawuni{2500}\pmboxdrawuni{2500}\pmboxdrawuni{2500}\pmboxdrawuni{2500}\pmboxdrawuni{2500}\pmboxdrawuni{2500}\pmboxdrawuni{2500}\pmboxdrawuni{2500}\pmboxdrawuni{2500}\pmboxdrawuni{2500}\pmboxdrawuni{2500}\pmboxdrawuni{2500}\pmboxdrawuni{2500}\pmboxdrawuni{2500}\pmboxdrawuni{2500}\pmboxdrawuni{2500}\pmboxdrawuni{2500}\pmboxdrawuni{2500}\pmboxdrawuni{2500}\pmboxdrawuni{2500}}
\cl{~~~~*~~tasks~scheduled~with~CronCreate:~recurring~ones~not~yet~expired,~one-shots}
\cl{~~~~~~~not~yet~due;~a~self-paced~/loop~is~not~restored}
\cl{~~~~**~kept~by~/branch,~lost~by~-{}-fork-session}
\cl{~~~~†~~a~transcript~loaded~in~a~new~process,~rebuilt~from~the~saved~conversation}
\cl{~~~~~~~and~from~live~settings.~Reattaching~to~a~session~that~is~still~running,}
\cl{~~~~~~~or~switching~with~/resume~inside~one~process,~is~a~different~operation}
\cl{~~~~‡~~the~goal~carries~over;~its~turn~count,~timer~and~token~baseline~reset}
\cl{~~~~?~~UNVERIFIED:~the~current~pages~do~not~say~whether~/clear~removes~them}
\cl{~~~~§~~restoring~conversation~and~code~together.~Either~can~be~restored~alone,}
\cl{~~~~~~~and~file~restoration~covers~files~Claude\textquotesingle{}s~file~tools~edited,~not}
\cl{~~~~~~~files~only~a~shell~command~changed}
\end{codefig}
\figcaption{Figure 3 — What each context operation keeps of conversation and session state. Only the last row survives all of them.}
\end{figure}

Not every cell of that figure is a quotation. The documentation states the \texttt{/\allowbreak{}compact} cells for instructions, rules and memory, the \texttt{/\allowbreak{}clear} and resume cells for goals, the resume cell for scheduled tasks, and the \texttt{/\allowbreak{}branch} cell for permission grants [26], [31], [35], [84]. The cells for goals, tasks and grants under \texttt{/\allowbreak{}compact}, \texttt{/\allowbreak{}branch} and \texttt{/\allowbreak{}rewind}, and the loss of grants on resume, are inferences from whether the operation stays inside one process. Treat them as the expected behaviour and check the ones your work depends on.

\FloatBarrier
\setcounter{section}{2}
\section{Rewind, branch, fork, subtask}

These solve different problems and are routinely confused.

\begingroup
\def\tblrows{%
\texttt{/\allowbreak{}rewind} & Returns the conversation, the code, or both to a checkpoint, or summarises part of the conversation & Later state is removed from whichever you restore \\
\texttt{/\allowbreak{}branch [nam\allowbreak{}e]} & Copies the conversation and switches you into the copy; the original is untouched on disk & Same process, so session permission grants carry over \\
\texttt{-\allowbreak{}-\allowbreak{}fork-\allowbreak{}session} & Same copy, in a \textbf{new process} & New process starts without "allow for this session" grants \\
\texttt{/\allowbreak{}fork} & Copies the conversation into a background session. With agent view turned off it starts a fork subagent instead & Runs separately \\
\texttt{/\allowbreak{}subtask} & Spawns a \textbf{fork subagent} inheriting the full conversation. Unavailable with agent view turned off, where \texttt{/\allowbreak{}fork} does this & Returns a result; tool calls stay out of your context \\
}%
\def\tblbody{\begin{minipage}{\textwidth}\boxcaption{Table 8 — Mechanisms that change the shape of a conversation}
{\footnotesize\begin{tabular}{L{28.7mm}L{71.0mm}L{40.1mm}}
\toprule
\textbf{Mechanism} & \textbf{What it does} & \textbf{Where the work goes} \\
\midrule
\tblrows
\bottomrule\end{tabular}}\end{minipage}}%
\begingroup
\def\sloppy{\tolerance 9999\emergencystretch 3em\hfuzz 200pt\vfuzz 200pt}%
\hbadness=10000\vbadness=10000\hfuzz=200pt\vfuzz=200pt
\global\setbox\tblbox=\hbox{\tblbody}%
\endgroup
\par\addvspace{2.6mm}
\ifdim\dimexpr\ht\tblbox+\dp\tblbox\relax>0.55\textheight
  \typeout{HANDBOOK-TABLE broken \the\dimexpr\ht\tblbox+\dp\tblbox\relax}%
  \tabcaption{Table 8 — Mechanisms that change the shape of a conversation}
  {\footnotesize\begin{longtable}{L{28.7mm}L{71.0mm}L{40.1mm}}
  \toprule
\textbf{Mechanism} & \textbf{What it does} & \textbf{Where the work goes} \\
\midrule\endfirsthead
  \multicolumn{3}{@{}l@{}}{%
  \sffamily\footnotesize\itshape\color{inkgrey}Table 8 — Mechanisms that change the shape of a conversation \textemdash\ continued}\\[1.2mm]
  \toprule
\textbf{Mechanism} & \textbf{What it does} & \textbf{Where the work goes} \\
\midrule\endhead
  \bottomrule\endfoot
  \bottomrule\endlastfoot
  \tblrows
  \end{longtable}}%
\else
  \typeout{HANDBOOK-TABLE atomic \the\dimexpr\ht\tblbox+\dp\tblbox\relax}%
  \noindent\tblbody
\fi
\par\addvspace{2.6mm}
\endgroup

\FloatBarrier
\setcounter{section}{3}
\section{Reduce standing overhead}

\begin{itemize}
\item Keep \texttt{CLAUDE.\allowbreak{}md} concise and rules conditional.
\item Disable unused MCP servers, plugins and agents — after measuring, not before (Chapter 13).
\item Keep large source material in files and reference precise sections.
\item Delegate bounded independent analysis, not every small action.
\item Remember that enabling Chrome by default loads browser tools into every session and increases context usage [21].
\end{itemize}

\subsection*{Exercise}

After drafting your brief, run \texttt{/\allowbreak{}context}. Save a handoff, start a clean session, and ask the new session to continue. Verify that it can name completed work, open issues and the next check without receiving the old transcript.

\emph{A worked solution is given in Appendix M.}

\begingroup
\def\kprows{%
\item Context is what the model has this turn; a file on disk is not known merely because it could be read.
\item Durable state belongs in files. A conversation is a working surface, not a database.
\item \texttt{/\allowbreak{}compact} is lossy and invalidates the cached conversation; say explicitly what must survive. After it, at most five files are re-read and path-scoped rules are gone until a matching file is touched.
\item Compaction also runs on its own: as the window fills, and by default on a long conversation left idle for about an hour.
\item Rewind, branch, fork and subtask solve four different problems and are routinely confused.
}%
\def\kpbody{\begin{minipage}{\textwidth}\subsection*{Key points}\begin{itemize}\kprows\end{itemize}\end{minipage}}%
\begingroup
\def\sloppy{\tolerance 9999\emergencystretch 3em\hfuzz 200pt\vfuzz 200pt}%
\hbadness=10000\vbadness=10000\hfuzz=200pt\vfuzz=200pt
\global\setbox\kpbox=\hbox{\kpbody}%
\endgroup
\par\addvspace{4.2mm}
\ifdim\dimexpr\ht\kpbox+\dp\kpbox\relax>0.30\textheight
  \typeout{HANDBOOK-KEYPOINTS broken \the\dimexpr\ht\kpbox+\dp\kpbox\relax}%
  \subsection*{Key points}
  \begin{itemize}\kprows\end{itemize}
\else
  \typeout{HANDBOOK-KEYPOINTS atomic \the\dimexpr\ht\kpbox+\dp\kpbox\relax}%
  \noindent\kpbody
\fi
\par\addvspace{1.4mm}
\endgroup

\FloatBarrier
\renewcommand{\chaptertitlelabel}{Part III \textperiodcentered\ Chapter 12}
\setcounter{chapter}{11}
\chapter{Sessions: resume, name, retain}

A session is a saved conversation tied to a project directory. The CLI, Desktop, web and VS Code each maintain their own history [31].

Resuming one is \textbf{rehydration, not the restoration of a snapshot}. What comes back is assembled from two sources at the moment you resume. From the saved transcript: the conversation, and a small amount of state recorded in it. From the present: the Claude Code build now installed, the settings files and managed policy now in force, the models now available, the repository as it now stands, and the agent, plugin and MCP definitions now on disk. A session resumed a week later is the old conversation in a new environment, and the second half of that sentence is the part that causes surprises.

\FloatBarrier
\setcounter{section}{0}
\section{Resume}

\begin{codeblock}{8.0}{9.4}
\cl{claude~-{}-continue~~~~~~~~~~~~\#~most~recent~interactive~session~here}
\cl{claude~-{}-resume~~~~~~~~~~~~~~\#~session~picker}
\cl{claude~-{}-resume~<name|id>~~~~\#~directly;~an~absolute~path~to~a~transcript~file~also~works}
\cl{claude~-{}-from-pr~<number>~~~~\#~picker~filtered~to~sessions~linked~to~that~PR}
\cl{claude~-n~research-brief-source-audit~~~\#~name~at~startup~(-{}-name)}
\end{codeblock}

Inside a session, \texttt{/\allowbreak{}resume} switches conversation and \texttt{/\allowbreak{}rename} renames [31], [37]. In the picker, \texttt{Ctrl+R} renames, \texttt{Ctrl+A} widens to all projects, \texttt{Ctrl+W} to all worktrees of the repository, \texttt{Ctrl+B} filters to the current branch, and pasting a pull-request URL finds the session that created it [31].

\textbf{From the transcript, a resumed session restores} conversation history, model, agent, permission mode, an active goal, and unexpired scheduled tasks [31]. Every one of those carries a condition, and the conditions are where the surprises live.

\textbf{The history} is not the whole of it once a session has been compacted: the conversation reopens at the last compaction's summary, with the messages that compaction left unsummarised and everything since, and the summarised messages are not loaded again. A tool call that was still running when the previous process ended is not rerun, and background subagents, commands and workflows that died with that process appear as notes that they did not finish [31].

\textbf{The mode} depends on the route you resume by, and the rule is not the same on each [31]. On a terminal resume — \texttt{-\allowbreak{}-\allowbreak{}continue}, or \texttt{-\allowbreak{}-\allowbreak{}resume} with an ID or a name that matches one session — the mode the session ended in is restored, with three exceptions. \texttt{bypass\allowbreak{}Permissions} is never restored: the session starts in the mode a new one would, and bypass has to be enabled again at launch. \texttt{auto} is restored only while the account still meets the auto-mode requirements. Manual is restored as Manual where a new session would start in \texttt{auto} from the built-in default, but where a \texttt{default\allowbreak{}Mode} in a settings file takes effect the session resumes in \emph{that} mode instead. \texttt{plan} \textbf{is} restored, unless \texttt{-\allowbreak{}-\allowbreak{}fork-\allowbreak{}session} is passed. On this route \texttt{-\allowbreak{}-\allowbreak{}permission-\allowbreak{}mode} or \texttt{-\allowbreak{}-\allowbreak{}dangerously-\allowbreak{}skip-\allowbreak{}permissions} overrides the restored mode.

The other routes restore less. A session chosen from the picker, and a non-interactive \texttt{claude -\allowbreak{}p -\allowbreak{}-\allowbreak{}resume}, start in the mode a new session or run would start in from the same command line; \texttt{/\allowbreak{}resume} inside a session continues in the mode the current session is in. The one stored mode all three still honour is plan, each on its own terms. From the picker, plan is restored unless \texttt{-\allowbreak{}-\allowbreak{}permission-\allowbreak{}mode}, \texttt{-\allowbreak{}-\allowbreak{}dangerously-\allowbreak{}skip-\allowbreak{}permissions} or \texttt{-\allowbreak{}-\allowbreak{}fork-\allowbreak{}session} was passed. On the non-interactive route it is restored under further conditions the page lists \emph{(non-interactive and VS Code: requires 2.1.246 or later)}. With \texttt{/\allowbreak{}resume} it is restored even if Claude Code was launched with \texttt{-\allowbreak{}-\allowbreak{}permission-\allowbreak{}mode} or \texttt{-\allowbreak{}-\allowbreak{}dangerously-\allowbreak{}skip-\allowbreak{}permissions} — unless that conversation had already been open earlier in the same run, in which case it continues in the current mode. On every route, a deny rule that removes the \texttt{Exit\allowbreak{}Plan\allowbreak{}Mode} tool prevents plan mode being restored, because Claude could not then present a plan for approval. A launch flag is therefore an override on a terminal resume and not a general one [31].

\textbf{The agent} of a session started with \texttt{-\allowbreak{}-\allowbreak{}agent} is looked for in the session's original directory, if you have trusted that workspace, and then in the directory you resume from. If it is in neither, the session resumes with the default tools and a warning naming the agent — so a resumed session can have \emph{more} tools than the one you left [31]. \textbf{The goal} carries over with its turn count, timer and token baseline reset.

\textbf{The model} is \textbf{not} restored when it has been retired or is disallowed by \texttt{available\allowbreak{}Models}, when a \texttt{-\allowbreak{}-\allowbreak{}model} flag or an \texttt{ANTHROPIC\_\allowbreak{}MODEL}-family environment variable picks one at launch, or on providers that use provider-specific deployment IDs [31]. \textbf{Scheduled tasks} are restored only if they were scheduled with \texttt{Cron\allowbreak{}Create} — recurring tasks that have not expired, and one-shot tasks whose time has not passed; a self-paced \texttt{/\allowbreak{}loop} is not restored and has to be started again, and background Bash and monitor tasks are not restored at all (Section 23.3) [26].

\textbf{It does not restore} \texttt{-\allowbreak{}-\allowbreak{}mcp-\allowbreak{}config}, \texttt{-\allowbreak{}-\allowbreak{}settings}, \texttt{-\allowbreak{}-\allowbreak{}plugin-\allowbreak{}dir}, \texttt{-\allowbreak{}-\allowbreak{}fallback-\allowbreak{}model}, or directories added with \texttt{-\allowbreak{}-\allowbreak{}add-\allowbreak{}dir}; pass those again. Directories added mid-session with \texttt{/\allowbreak{}add-\allowbreak{}dir} are not restored either, though the session picker still uses them to find the session. The standard settings files are re-read at launch, so what lives in them needs no re-passing — and is whatever those files say \emph{now} [31].

Before consequential work continues in a resumed session, record what it actually is, since none of it can be assumed from the session you remember: the session ID; the working directory, worktree and branch; the Claude Code version, model and provider; the instruction files and agent that loaded; the permission mode and the rules in force; whether the sandbox is on; and which connectors are attached. \texttt{/\allowbreak{}status}, \texttt{/\allowbreak{}context all}, \texttt{/\allowbreak{}permissions} and \texttt{/\allowbreak{}mcp} answer all of it in under a minute.

\textbf{A gotcha worth knowing:} sessions whose \emph{first} prompt was a \texttt{/\allowbreak{}loop} command do not appear in the picker and are skipped by \texttt{-\allowbreak{}-\allowbreak{}continue} [31]. If a scheduled task seems to have vanished, that is usually why. Sessions created by \texttt{claude -\allowbreak{}p} or the Agent SDK are left out of both as well, and are resumed by passing their ID. And if the conversation you resume belongs to a background session that is still running, Claude Code opens the running session itself and does not load a second copy from the transcript \emph{(requires 2.1.285 or later)}. That holds for a plain interactive resume. With piped input or output, or with flags that configure the session, read its output or limit the run, Claude Code sends nothing and exits with status 1, naming the \texttt{claude atta\allowbreak{}ch} command to use; \texttt{-\allowbreak{}-\allowbreak{}fork-\allowbreak{}session} resumes a copy instead.

\FloatBarrier
\setcounter{section}{1}
\section{Name by outcome}

Good names survive a week:

\begin{itemize}
\item \texttt{client-\allowbreak{}x-\allowbreak{}brief-\allowbreak{}source-\allowbreak{}audit}
\item \texttt{accessibili\allowbreak{}ty-\allowbreak{}remediation-\allowbreak{}plan}
\item \texttt{invoice-\allowbreak{}workflow-\allowbreak{}test-\allowbreak{}2026-\allowbreak{}08}
\end{itemize}

Avoid \texttt{test}, \texttt{new chat}, or whatever the first prompt happened to be. Unnamed interactive sessions receive a default display name such as \texttt{my-\allowbreak{}app-\allowbreak{}3f} — which is \textbf{not} a resume handle — plus a generated title from the first prompt, which is [31]. If two live sessions share a name, Claude Code renames the newer one with a two-word suffix and tells you (2.1.232 or later).

\FloatBarrier
\setcounter{section}{2}
\section{Retention, storage and privacy}

Transcripts are stored as \textbf{JSONL} at \texttt{\textasciitilde{}/\allowbreak{}.\allowbreak{}claude/\allowbreak{}projects/\allowbreak{}<project>/\allowbreak{}<session-\allowbreak{}id>.\allowbreak{}jsonl}, where \texttt{<project>} is the working directory path with non-alphanumeric characters replaced by hyphens [31]. The entry format is internal and changes between releases; do not build on it. Use \texttt{/\allowbreak{}export} for a rendered transcript, or the structured interfaces (\texttt{claude -\allowbreak{}p -\allowbreak{}-\allowbreak{}output-\allowbreak{}format json}, the \texttt{transcript\_\allowbreak{}path} supplied to hooks and status lines, or the Agent SDK) for scripts [31].

\begingroup
\def\tblrows{%
Move storage off \texttt{\textasciitilde{}/\allowbreak{}.\allowbreak{}claude} & \texttt{CLAUDE\_\allowbreak{}CONFIG\_\allowbreak{}DIR} & Environment variable \\
Name the project directory yourself (2.1.234+) & \texttt{CLAUDE\_\allowbreak{}CODE\_\allowbreak{}PROJECT\_\allowbreak{}DIR\_\allowbreak{}NAME} & Environment variable, alongside \texttt{CLAUDE\_\allowbreak{}CONFIG\_\allowbreak{}DIR} \\
Change the 30-day retention & \texttt{cleanup\allowbreak{}Period\allowbreak{}Days} & \texttt{settings.\allowbreak{}json} \\
Set an age limit for Claude Desktop and Cowork transcripts, which are otherwise kept at any age (2.1.248+) & \texttt{desktop\allowbreak{}Session\allowbreak{}Cleanup\allowbreak{}Period\allowbreak{}Days} & User settings, managed settings, or \texttt{-\allowbreak{}-\allowbreak{}settings} \\
Suppress transcript writes in all modes & \texttt{CLAUDE\_\allowbreak{}CODE\_\allowbreak{}SKIP\_\allowbreak{}PROMPT\_\allowbreak{}HISTORY} & Environment variable \\
Suppress writes for one non-interactive run & \texttt{-\allowbreak{}-\allowbreak{}no-\allowbreak{}session-\allowbreak{}persistence} & CLI flag with \texttt{claude -\allowbreak{}p} \\
}%
\def\tblbody{\begin{minipage}{\textwidth}\boxcaption{Table 9 — Controls over transcript location, retention and writing}
{\footnotesize\begin{tabular}{L{38.8mm}L{63.5mm}L{37.5mm}}
\toprule
\textbf{To} & \textbf{Set} & \textbf{Where} \\
\midrule
\tblrows
\bottomrule\end{tabular}}\end{minipage}}%
\begingroup
\def\sloppy{\tolerance 9999\emergencystretch 3em\hfuzz 200pt\vfuzz 200pt}%
\hbadness=10000\vbadness=10000\hfuzz=200pt\vfuzz=200pt
\global\setbox\tblbox=\hbox{\tblbody}%
\endgroup
\par\addvspace{2.6mm}
\ifdim\dimexpr\ht\tblbox+\dp\tblbox\relax>0.55\textheight
  \typeout{HANDBOOK-TABLE broken \the\dimexpr\ht\tblbox+\dp\tblbox\relax}%
  \tabcaption{Table 9 — Controls over transcript location, retention and writing}
  {\footnotesize\begin{longtable}{L{38.8mm}L{63.5mm}L{37.5mm}}
  \toprule
\textbf{To} & \textbf{Set} & \textbf{Where} \\
\midrule\endfirsthead
  \multicolumn{3}{@{}l@{}}{%
  \sffamily\footnotesize\itshape\color{inkgrey}Table 9 — Controls over transcript location, retention and writing \textemdash\ continued}\\[1.2mm]
  \toprule
\textbf{To} & \textbf{Set} & \textbf{Where} \\
\midrule\endhead
  \bottomrule\endfoot
  \bottomrule\endlastfoot
  \tblrows
  \end{longtable}}%
\else
  \typeout{HANDBOOK-TABLE atomic \the\dimexpr\ht\tblbox+\dp\tblbox\relax}%
  \noindent\tblbody
\fi
\par\addvspace{2.6mm}
\endgroup

The thirty days is not a deadline and does not cover everything [32]. A transcript of a session started or last continued in Claude Desktop or Cowork is kept \textbf{at any age} unless \texttt{desktop\allowbreak{}Session\allowbreak{}Cleanup\allowbreak{}Period\allowbreak{}Days} is set, and is then deleted only once it is older than \emph{both} limits, so a value of 7 beside the default of 30 still keeps it thirty days. Managed \texttt{cleanup\allowbreak{}Period\allowbreak{}Days} and the HIPAA configuration are among the cases that apply the ordinary limit to those transcripts instead. The sweep itself is skipped in a \texttt{-\allowbreak{}-\allowbreak{}bare} run, and paused where Claude Code cannot safely determine the retention period, which \texttt{/\allowbreak{}status} reports.

Retention here means one thing: how long local application data can still be recovered from this machine. It says nothing about whether what a session did was correct or authorised, and nothing about what a model provider or a cloud session keeps, which is the subject of Chapter 32. To delete a project's transcripts sooner than the sweep would, run \texttt{claude purg\allowbreak{}e}; deleting a background session with \texttt{claude rm} leaves its transcript on disk [31].

\begin{calloutbox}{palebrass}{brassdark}{2.0mm}
\textbf{CAUTION:} transcripts are plaintext on disk. Restrict filesystem access, shorten retention where your obligations require it, and never allow a secret into output on the assumption that history is encrypted. Remember that the auto-memory directory is excluded from the retention sweep (Section 10.2). Data governance is treated properly in Chapter 32.
\end{calloutbox}

\FloatBarrier
\setcounter{section}{3}
\section{When to start another session}

Start clean when the objective changes, when a decision needs independent review, when context is dominated by discarded work, when a bounded subtask can run in isolation, or when you need an alternative path without confusing the approved one.

Do not run parallel sessions that edit the same files. Use worktrees or strict file ownership — Chapter 29.

\begingroup
\def\kprows{%
\item Resume is rehydration: the conversation comes from the transcript, and the build, settings, policy, models and repository are whatever they are now.
\item A terminal resume restores history, model, agent, goal, eligible scheduled tasks and the permission mode — \texttt{plan} included, \texttt{bypass\allowbreak{}Permissions} never; the other resume routes restore less.
\item Configuration passed as flags is not restored; pass \texttt{-\allowbreak{}-\allowbreak{}mcp-\allowbreak{}config}, \texttt{-\allowbreak{}-\allowbreak{}settings}, \texttt{-\allowbreak{}-\allowbreak{}plugin-\allowbreak{}dir} and \texttt{-\allowbreak{}-\allowbreak{}add-\allowbreak{}dir} again.
\item A session whose first prompt was \texttt{/\allowbreak{}loop} is hidden from the picker and skipped by \texttt{-\allowbreak{}-\allowbreak{}continue}.
\item Transcripts are plaintext JSONL on disk; retention, location and whether they are written at all are configurable.
}%
\def\kpbody{\begin{minipage}{\textwidth}\subsection*{Key points}\begin{itemize}\kprows\end{itemize}\end{minipage}}%
\begingroup
\def\sloppy{\tolerance 9999\emergencystretch 3em\hfuzz 200pt\vfuzz 200pt}%
\hbadness=10000\vbadness=10000\hfuzz=200pt\vfuzz=200pt
\global\setbox\kpbox=\hbox{\kpbody}%
\endgroup
\par\addvspace{4.2mm}
\ifdim\dimexpr\ht\kpbox+\dp\kpbox\relax>0.30\textheight
  \typeout{HANDBOOK-KEYPOINTS broken \the\dimexpr\ht\kpbox+\dp\kpbox\relax}%
  \subsection*{Key points}
  \begin{itemize}\kprows\end{itemize}
\else
  \typeout{HANDBOOK-KEYPOINTS atomic \the\dimexpr\ht\kpbox+\dp\kpbox\relax}%
  \noindent\kpbody
\fi
\par\addvspace{1.4mm}
\endgroup

\breakrule

\FloatBarrier
\parttitle{Part IV}{Extending Claude Code}
\addcontentsline{toc}{part}{Part IV \textemdash\ Extending Claude Code}

Every chapter in this part enlarges what Claude Code can reach. Each is therefore also a supply-chain chapter. The organising principle is stated once here and applied throughout:

\begin{calloutbox}{palegrey}{quoterule}{1.2mm}
Anthropic does not security-audit MCP servers, and plugins and marketplaces can execute arbitrary code with your user privileges. Connectors are reviewed against listing criteria before entering the Anthropic Directory, but that is a listing review, not a security audit [40], [59].
\end{calloutbox}

The consequence is that trust in agentic work is not a perimeter property. Figure 4 enumerates the channels through which content the agent did not author reaches the session, and none of them is the network boundary.

\begin{figure}[tbp]
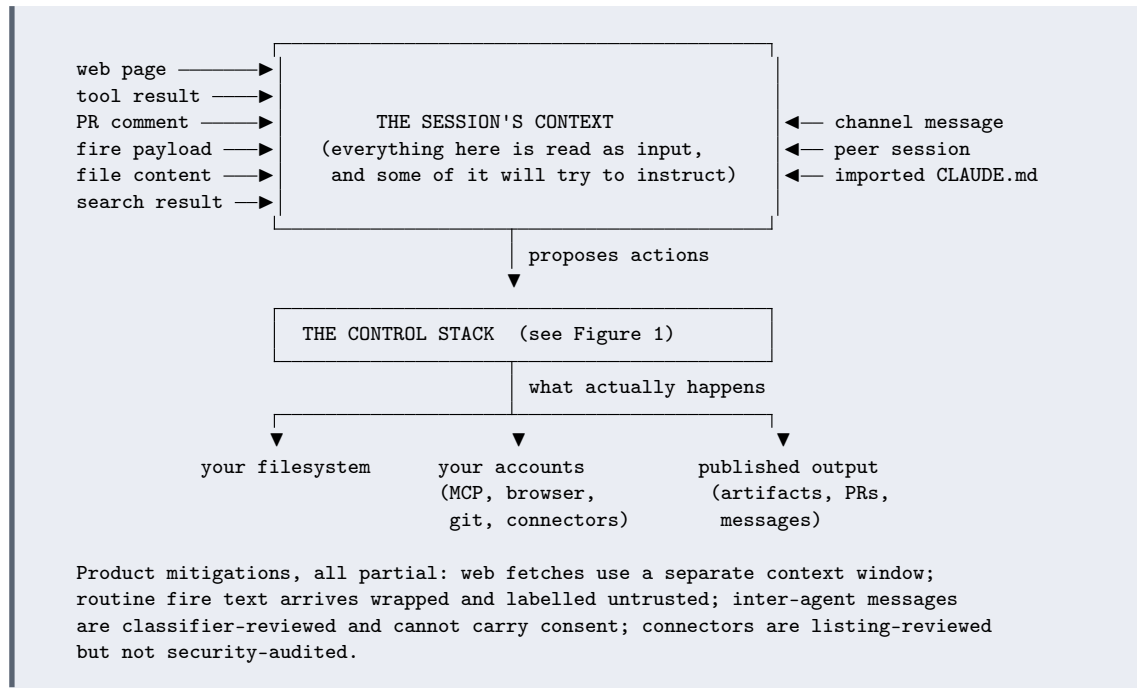

\begin{codefig}{8.0}{9.4}
\cl{~~~~~~~~~~~~~~~~~~~~\pmboxdrawuni{250C}\pmboxdrawuni{2500}\pmboxdrawuni{2500}\pmboxdrawuni{2500}\pmboxdrawuni{2500}\pmboxdrawuni{2500}\pmboxdrawuni{2500}\pmboxdrawuni{2500}\pmboxdrawuni{2500}\pmboxdrawuni{2500}\pmboxdrawuni{2500}\pmboxdrawuni{2500}\pmboxdrawuni{2500}\pmboxdrawuni{2500}\pmboxdrawuni{2500}\pmboxdrawuni{2500}\pmboxdrawuni{2500}\pmboxdrawuni{2500}\pmboxdrawuni{2500}\pmboxdrawuni{2500}\pmboxdrawuni{2500}\pmboxdrawuni{2500}\pmboxdrawuni{2500}\pmboxdrawuni{2500}\pmboxdrawuni{2500}\pmboxdrawuni{2500}\pmboxdrawuni{2500}\pmboxdrawuni{2500}\pmboxdrawuni{2500}\pmboxdrawuni{2500}\pmboxdrawuni{2500}\pmboxdrawuni{2500}\pmboxdrawuni{2500}\pmboxdrawuni{2500}\pmboxdrawuni{2500}\pmboxdrawuni{2500}\pmboxdrawuni{2500}\pmboxdrawuni{2500}\pmboxdrawuni{2500}\pmboxdrawuni{2500}\pmboxdrawuni{2500}\pmboxdrawuni{2500}\pmboxdrawuni{2500}\pmboxdrawuni{2500}\pmboxdrawuni{2510}}
\cl{~~~web~page~\pmboxdrawuni{2500}\pmboxdrawuni{2500}\pmboxdrawuni{2500}\pmboxdrawuni{2500}\pmboxdrawuni{2500}\pmboxdrawuni{2500}\pmboxdrawuni{2500}▶\pmboxdrawuni{2502}~~~~~~~~~~~~~~~~~~~~~~~~~~~~~~~~~~~~~~~~~~~\pmboxdrawuni{2502}}
\cl{~~~tool~result~\pmboxdrawuni{2500}\pmboxdrawuni{2500}\pmboxdrawuni{2500}\pmboxdrawuni{2500}▶\pmboxdrawuni{2502}~~~~~~~~~~~~~~~~~~~~~~~~~~~~~~~~~~~~~~~~~~~\pmboxdrawuni{2502}}
\cl{~~~PR~comment~\pmboxdrawuni{2500}\pmboxdrawuni{2500}\pmboxdrawuni{2500}\pmboxdrawuni{2500}\pmboxdrawuni{2500}▶\pmboxdrawuni{2502}~~~~~~~~THE~SESSION\textquotesingle{}S~CONTEXT~~~~~~~~~~~~~~\pmboxdrawuni{2502}◀\pmboxdrawuni{2500}\pmboxdrawuni{2500}~channel~message}
\cl{~~~fire~payload~\pmboxdrawuni{2500}\pmboxdrawuni{2500}\pmboxdrawuni{2500}▶\pmboxdrawuni{2502}~~~(everything~here~is~read~as~input,~~~~~~\pmboxdrawuni{2502}◀\pmboxdrawuni{2500}\pmboxdrawuni{2500}~peer~session}
\cl{~~~file~content~\pmboxdrawuni{2500}\pmboxdrawuni{2500}\pmboxdrawuni{2500}▶\pmboxdrawuni{2502}~~~~and~some~of~it~will~try~to~instruct)~~~\pmboxdrawuni{2502}◀\pmboxdrawuni{2500}\pmboxdrawuni{2500}~imported~CLAUDE.md}
\cl{~~~search~result~\pmboxdrawuni{2500}\pmboxdrawuni{2500}▶\pmboxdrawuni{2502}~~~~~~~~~~~~~~~~~~~~~~~~~~~~~~~~~~~~~~~~~~~\pmboxdrawuni{2502}}
\cl{~~~~~~~~~~~~~~~~~~~~\pmboxdrawuni{2514}\pmboxdrawuni{2500}\pmboxdrawuni{2500}\pmboxdrawuni{2500}\pmboxdrawuni{2500}\pmboxdrawuni{2500}\pmboxdrawuni{2500}\pmboxdrawuni{2500}\pmboxdrawuni{2500}\pmboxdrawuni{2500}\pmboxdrawuni{2500}\pmboxdrawuni{2500}\pmboxdrawuni{2500}\pmboxdrawuni{2500}\pmboxdrawuni{2500}\pmboxdrawuni{2500}\pmboxdrawuni{2500}\pmboxdrawuni{2500}\pmboxdrawuni{2500}\pmboxdrawuni{2500}\pmboxdrawuni{2500}\pmboxdrawuni{252C}\pmboxdrawuni{2500}\pmboxdrawuni{2500}\pmboxdrawuni{2500}\pmboxdrawuni{2500}\pmboxdrawuni{2500}\pmboxdrawuni{2500}\pmboxdrawuni{2500}\pmboxdrawuni{2500}\pmboxdrawuni{2500}\pmboxdrawuni{2500}\pmboxdrawuni{2500}\pmboxdrawuni{2500}\pmboxdrawuni{2500}\pmboxdrawuni{2500}\pmboxdrawuni{2500}\pmboxdrawuni{2500}\pmboxdrawuni{2500}\pmboxdrawuni{2500}\pmboxdrawuni{2500}\pmboxdrawuni{2500}\pmboxdrawuni{2500}\pmboxdrawuni{2500}\pmboxdrawuni{2518}}
\cl{~~~~~~~~~~~~~~~~~~~~~~~~~~~~~~~~~~~~~~~~~\pmboxdrawuni{2502}~proposes~actions}
\cl{~~~~~~~~~~~~~~~~~~~~~~~~~~~~~~~~~~~~~~~~~▼}
\cl{~~~~~~~~~~~~~~~~~~~~\pmboxdrawuni{250C}\pmboxdrawuni{2500}\pmboxdrawuni{2500}\pmboxdrawuni{2500}\pmboxdrawuni{2500}\pmboxdrawuni{2500}\pmboxdrawuni{2500}\pmboxdrawuni{2500}\pmboxdrawuni{2500}\pmboxdrawuni{2500}\pmboxdrawuni{2500}\pmboxdrawuni{2500}\pmboxdrawuni{2500}\pmboxdrawuni{2500}\pmboxdrawuni{2500}\pmboxdrawuni{2500}\pmboxdrawuni{2500}\pmboxdrawuni{2500}\pmboxdrawuni{2500}\pmboxdrawuni{2500}\pmboxdrawuni{2500}\pmboxdrawuni{2500}\pmboxdrawuni{2500}\pmboxdrawuni{2500}\pmboxdrawuni{2500}\pmboxdrawuni{2500}\pmboxdrawuni{2500}\pmboxdrawuni{2500}\pmboxdrawuni{2500}\pmboxdrawuni{2500}\pmboxdrawuni{2500}\pmboxdrawuni{2500}\pmboxdrawuni{2500}\pmboxdrawuni{2500}\pmboxdrawuni{2500}\pmboxdrawuni{2500}\pmboxdrawuni{2500}\pmboxdrawuni{2500}\pmboxdrawuni{2500}\pmboxdrawuni{2500}\pmboxdrawuni{2500}\pmboxdrawuni{2500}\pmboxdrawuni{2500}\pmboxdrawuni{2500}\pmboxdrawuni{2510}}
\cl{~~~~~~~~~~~~~~~~~~~~\pmboxdrawuni{2502}~~THE~CONTROL~STACK~~(see~Figure~1)~~~~~~~~\pmboxdrawuni{2502}}
\cl{~~~~~~~~~~~~~~~~~~~~\pmboxdrawuni{2514}\pmboxdrawuni{2500}\pmboxdrawuni{2500}\pmboxdrawuni{2500}\pmboxdrawuni{2500}\pmboxdrawuni{2500}\pmboxdrawuni{2500}\pmboxdrawuni{2500}\pmboxdrawuni{2500}\pmboxdrawuni{2500}\pmboxdrawuni{2500}\pmboxdrawuni{2500}\pmboxdrawuni{2500}\pmboxdrawuni{2500}\pmboxdrawuni{2500}\pmboxdrawuni{2500}\pmboxdrawuni{2500}\pmboxdrawuni{2500}\pmboxdrawuni{2500}\pmboxdrawuni{2500}\pmboxdrawuni{2500}\pmboxdrawuni{252C}\pmboxdrawuni{2500}\pmboxdrawuni{2500}\pmboxdrawuni{2500}\pmboxdrawuni{2500}\pmboxdrawuni{2500}\pmboxdrawuni{2500}\pmboxdrawuni{2500}\pmboxdrawuni{2500}\pmboxdrawuni{2500}\pmboxdrawuni{2500}\pmboxdrawuni{2500}\pmboxdrawuni{2500}\pmboxdrawuni{2500}\pmboxdrawuni{2500}\pmboxdrawuni{2500}\pmboxdrawuni{2500}\pmboxdrawuni{2500}\pmboxdrawuni{2500}\pmboxdrawuni{2500}\pmboxdrawuni{2500}\pmboxdrawuni{2500}\pmboxdrawuni{2500}\pmboxdrawuni{2518}}
\cl{~~~~~~~~~~~~~~~~~~~~~~~~~~~~~~~~~~~~~~~~~\pmboxdrawuni{2502}~what~actually~happens}
\cl{~~~~~~~~~~~~~~~~~~~~\pmboxdrawuni{250C}\pmboxdrawuni{2500}\pmboxdrawuni{2500}\pmboxdrawuni{2500}\pmboxdrawuni{2500}\pmboxdrawuni{2500}\pmboxdrawuni{2500}\pmboxdrawuni{2500}\pmboxdrawuni{2500}\pmboxdrawuni{2500}\pmboxdrawuni{2500}\pmboxdrawuni{2500}\pmboxdrawuni{2500}\pmboxdrawuni{2500}\pmboxdrawuni{2500}\pmboxdrawuni{2500}\pmboxdrawuni{2500}\pmboxdrawuni{2500}\pmboxdrawuni{2500}\pmboxdrawuni{2500}\pmboxdrawuni{2500}\pmboxdrawuni{2534}\pmboxdrawuni{2500}\pmboxdrawuni{2500}\pmboxdrawuni{2500}\pmboxdrawuni{2500}\pmboxdrawuni{2500}\pmboxdrawuni{2500}\pmboxdrawuni{2500}\pmboxdrawuni{2500}\pmboxdrawuni{2500}\pmboxdrawuni{2500}\pmboxdrawuni{2500}\pmboxdrawuni{2500}\pmboxdrawuni{2500}\pmboxdrawuni{2500}\pmboxdrawuni{2500}\pmboxdrawuni{2500}\pmboxdrawuni{2500}\pmboxdrawuni{2500}\pmboxdrawuni{2500}\pmboxdrawuni{2500}\pmboxdrawuni{2500}\pmboxdrawuni{2500}\pmboxdrawuni{2510}}
\cl{~~~~~~~~~~~~~~~~~~~~▼~~~~~~~~~~~~~~~~~~~~▼~~~~~~~~~~~~~~~~~~~~~~▼}
\cl{~~~~~~~~~~~~~~your~filesystem~~~~~~your~accounts~~~~~~~~~~published~output}
\cl{~~~~~~~~~~~~~~~~~~~~~~~~~~~~~~~~~~~(MCP,~browser,~~~~~~~~~~(artifacts,~PRs,}
\cl{~~~~~~~~~~~~~~~~~~~~~~~~~~~~~~~~~~~~git,~connectors)~~~~~~~~messages)}
\cl{}
\cl{~~~Product~mitigations,~all~partial:~most~web~fetches~go~through~a~separate~model~call;}
\cl{~~~routine~fire~text~arrives~wrapped~and~labelled~untrusted;~inter-agent~messages}
\cl{~~~are~classifier-reviewed~and~cannot~carry~consent;~connectors~are~listing-reviewed}
\cl{~~~but~not~security-audited.}
\end{codefig}
\figcaption{Figure 4 — Where untrusted content enters an agentic session. Every arrow is a documented injection channel; none of them is the network perimeter.}
\end{figure}

Appendix F maps each mechanism in this part to its threat model and the controls available.

\FloatBarrier
\renewcommand{\chaptertitlelabel}{Part IV \textperiodcentered\ Chapter 13}
\setcounter{chapter}{12}
\chapter{Connect external tools with MCP}

The Model Context Protocol is an open standard for connecting AI clients to external tools and data [85]. A server might search a knowledge base, read an issue tracker, query a database, or act on a service. It expands capability and trust simultaneously: a connected tool acts with whatever authority you gave it.

Five separate questions hide inside "is this server safe to connect", and the product answers each with a different mechanism. Keep them apart, because a yes to one is routinely mistaken for a yes to all:

\begin{enumerate}
\item \textbf{Configuration} — where the definition is discovered, and which definition wins.
\item \textbf{Activation} — whether that definition is allowed to load in this session.
\item \textbf{Identity} — which account authenticates, and what that account may do at the provider.
\item \textbf{Tool authority} — which of the server's tools Claude may call, and with what approval.
\item \textbf{Verification} — whether the effect at the far end was the one intended.
\end{enumerate}

Scope and approval settle the first two and nothing else.

\FloatBarrier
\setcounter{section}{0}
\section{Scopes}

\begingroup
\def\tblrows{%
Local (default) & \texttt{\textasciitilde{}/\allowbreak{}.\allowbreak{}claude.\allowbreak{}json}, under the project's path & The current project only & Private configuration, experiments, servers whose credentials must stay out of version control \\
Project & \texttt{.\allowbreak{}mcp.\allowbreak{}json} at the repository root & The current project only & Shared team configuration, committed \\
User & \texttt{\textasciitilde{}/\allowbreak{}.\allowbreak{}claude.\allowbreak{}json} & All your projects & Personal utilities you use everywhere \\
}%
\def\tblbody{\begin{minipage}{\textwidth}\boxcaption{Table 10 — MCP configuration scopes}
{\footnotesize\begin{tabular}{L{14.3mm}L{34.6mm}L{26.6mm}L{61.0mm}}
\toprule
\textbf{Scope} & \textbf{Storage} & \textbf{Loads in} & \textbf{Use} \\
\midrule
\tblrows
\bottomrule\end{tabular}}\end{minipage}}%
\begingroup
\def\sloppy{\tolerance 9999\emergencystretch 3em\hfuzz 200pt\vfuzz 200pt}%
\hbadness=10000\vbadness=10000\hfuzz=200pt\vfuzz=200pt
\global\setbox\tblbox=\hbox{\tblbody}%
\endgroup
\par\addvspace{2.6mm}
\ifdim\dimexpr\ht\tblbox+\dp\tblbox\relax>0.55\textheight
  \typeout{HANDBOOK-TABLE broken \the\dimexpr\ht\tblbox+\dp\tblbox\relax}%
  \tabcaption{Table 10 — MCP configuration scopes}
  {\footnotesize\begin{longtable}{L{14.3mm}L{34.6mm}L{26.6mm}L{61.0mm}}
  \toprule
\textbf{Scope} & \textbf{Storage} & \textbf{Loads in} & \textbf{Use} \\
\midrule\endfirsthead
  \multicolumn{4}{@{}l@{}}{%
  \sffamily\footnotesize\itshape\color{inkgrey}Table 10 — MCP configuration scopes \textemdash\ continued}\\[1.2mm]
  \toprule
\textbf{Scope} & \textbf{Storage} & \textbf{Loads in} & \textbf{Use} \\
\midrule\endhead
  \bottomrule\endfoot
  \bottomrule\endlastfoot
  \tblrows
  \end{longtable}}%
\else
  \typeout{HANDBOOK-TABLE atomic \the\dimexpr\ht\tblbox+\dp\tblbox\relax}%
  \noindent\tblbody
\fi
\par\addvspace{2.6mm}
\endgroup

When one server is defined in more than one place, Claude Code connects to it once and takes the whole entry from the highest-precedence source, without merging fields: local, then project, then user, then plugin-provided servers, then claude.ai connectors. Duplicates are matched by name across the three scopes and by endpoint for plugins and connectors. A server your organisation provides through \texttt{managed\allowbreak{}Mcp\allowbreak{}Servers} ranks above all of them \emph{(requires 2.1.259 or later)}, and the desktop app's Code tab is an exception to the order: for a stdio server named both in user scope and in \texttt{.\allowbreak{}mcp.\allowbreak{}json}, it uses the user-scope definition [34]. "Local" here does not mean what it means for settings: a local-scoped server lives in your home directory, not in the project's \texttt{.\allowbreak{}claude/\allowbreak{}settings.\allowbreak{}local.\allowbreak{}json}.

Project scope carries an approval mechanism that the course omitted and that the first edition under-described. A project-scoped server from \texttt{.\allowbreak{}mcp.\allowbreak{}json} appears as \texttt{⏸ Pending a\allowbreak{}pproval} until you run \texttt{claude} interactively and approve it; rejected servers are recorded in \texttt{disabled\allowbreak{}Mcpjson\allowbreak{}Servers} [34]. From 2.1.196, approvals are read only from settings files that are \textbf{not} checked into the repository until you trust the workspace, so a cloned repository cannot approve its own servers — \texttt{enable\allowbreak{}All\allowbreak{}Project\allowbreak{}Mcp\allowbreak{}Servers} or \texttt{enabled\allowbreak{}Mcpjson\allowbreak{}Servers} committed to a project's \texttt{.\allowbreak{}claude/\allowbreak{}settings.\allowbreak{}json} is ignored in an untrusted folder [34]. That is the control that stops a malicious pull request from silently connecting a server.

\begin{calloutbox}{palebrass}{brassdark}{2.0mm}
\textbf{CAUTION — the approval prompt exists only where a prompt can be shown.} In \texttt{claude -\allowbreak{}p} runs, Agent SDK sessions and cloud sessions outside a project's threads, Claude Code cannot ask, and it \textbf{loads project-scoped servers without asking}. It also skips the prompt in a session started in \texttt{bypass\allowbreak{}Permissions} mode with \texttt{skip\allowbreak{}Dangerous\allowbreak{}Mode\allowbreak{}Permission\allowbreak{}Prompt} set in user or managed settings [34]. A pipeline that checks out a branch and runs \texttt{claude -\allowbreak{}p} in it therefore connects whatever that branch's \texttt{.\allowbreak{}mcp.\allowbreak{}json} names. To keep a server out, list it in \texttt{disabled\allowbreak{}Mcpjson\allowbreak{}Servers}, which holds in every permission mode; exclude project settings with \texttt{-\allowbreak{}-\allowbreak{}setting-\allowbreak{}sources}; or start the run with \texttt{-\allowbreak{}-\allowbreak{}strict-\allowbreak{}mcp-\allowbreak{}config}, which uses only the servers passed with \texttt{-\allowbreak{}-\allowbreak{}mcp-\allowbreak{}config}.
\end{calloutbox}

\begin{calloutbox}{palebrass}{brassdark}{2.0mm}
\textbf{CAUTION:} never place a literal credential in \texttt{.\allowbreak{}mcp.\allowbreak{}json}. Use OAuth where the provider supports it, or \texttt{\$\{VAR\}} expansion referencing an environment variable, and use a least-privilege account. See M11 in Table 1. One class of variable will not expand at all: in a remote server's \texttt{url} and \texttt{headers}, Claude Code reads its own credentials, your cloud provider's, and others such as \texttt{HTTPS\_\allowbreak{}PROXY} and \texttt{NPM\_\allowbreak{}TOKEN} as empty, so that a repository's \texttt{.\allowbreak{}mcp.\allowbreak{}json} cannot send them to a server it names [34]. The symptom is a \texttt{401} with no warning. Copy the value into a variable with a name of your own if the server is meant to receive it.
\end{calloutbox}

\FloatBarrier
\setcounter{section}{1}
\section{Transports}

\begin{codeblock}{9.0}{10.6}
\cl{claude~mcp~add~-{}-transport~http~notion~https://mcp.notion.com/mcp~-{}-scope~project}
\cl{claude~mcp~list}
\cl{claude~mcp~get~notion}
\end{codeblock}

Four transports exist [34]:

\begin{itemize}
\item \textbf{HTTP} — preferred. In JSON, \texttt{type:\allowbreak{} "http"}; \texttt{streamable-\allowbreak{}http} is accepted as an alias so configurations copied from a server's own documentation work unmodified.
\item \textbf{SSE} — \textbf{deprecated}. Use HTTP where available. For a service that still exposes only SSE, add it with \texttt{-\allowbreak{}-\allowbreak{}transport h\allowbreak{}ttp} anyway: Claude Code tries HTTP first and switches to SSE when the server does not accept it \emph{(requires 2.1.265 or later)}.
\item \textbf{stdio} — a local process. Everything after \texttt{-\allowbreak{}-\allowbreak{}} is passed to the server untouched, so \texttt{claude mcp \allowbreak{}add -\allowbreak{}-\allowbreak{}env KEY=\allowbreak{}value -\allowbreak{}-\allowbreak{}transport s\allowbreak{}tdio myserv\allowbreak{}er -\allowbreak{}-\allowbreak{} python ser\allowbreak{}ver.\allowbreak{}py -\allowbreak{}-\allowbreak{}port 8080} runs the server with that argument and environment.
\item \textbf{WebSocket} (\texttt{type:\allowbreak{} "ws"}) — persistent bidirectional, for servers that push events. It supports neither OAuth nor the \texttt{-\allowbreak{}-\allowbreak{}transport} flag; authentication is header-only.
\end{itemize}

A JSON entry with a \texttt{url} but no \texttt{type} is a configuration error: Claude Code reads a type-less entry as stdio and skips it with an explicit message [34].

The transport is also the execution locus, and that matters more than the wire format. A remote server runs on someone else's machine and sees what you send it. \textbf{A stdio server is a program running on yours, as you}, with your files and your environment; no approval of its definition constrains what that process does once started.

Inside a session, \texttt{/\allowbreak{}mcp} inspects status, completes OAuth, lists a server's tools, and turns a server off for the current project without deleting its configuration. From the shell, \texttt{claude mcp} supports \texttt{add}, \texttt{add-\allowbreak{}json}, \texttt{list}, \texttt{get}, \texttt{remove}, \texttt{login}, \texttt{logout} and \texttt{reset-\allowbreak{}project-\allowbreak{}choices}, which clears the project approvals above [34], [55].

By default, an MCP tool call in the main conversation of an interactive session that is still running after two minutes moves to a background task rather than blocking the session, and its result arrives as a notification. Calls from subagents, calls to IDE servers and calls waiting on an open elicitation dialog do not move, and \texttt{CLAUDE\_\allowbreak{}CODE\_\allowbreak{}MCP\_\allowbreak{}AUTO\_\allowbreak{}BACKGROUND\_\allowbreak{}MS} changes the threshold or, at \texttt{0}, turns the behaviour off. Stopping that task cancels the call on your side only: the server may already have done some of the work, or may still finish it [34]. Do not read a stopped task as an undone action.

\FloatBarrier
\setcounter{section}{2}
\section{A safe installation procedure}

\begin{enumerate}
\item Identify the provider's official repository or documentation. Verify the URL character by character.
\item Define the minimum data and actions the task needs.
\item Use a test account or a least-privilege workspace.
\item Inspect source, publisher, package and network destinations where you can. For a stdio server, pin the version or digest you inspected; an \texttt{npx} command without one runs whatever is published on the day.
\item Choose the narrowest scope that works.
\item Keep secrets out of \texttt{.\allowbreak{}mcp.\allowbreak{}json} and out of version control.
\item Start a new session so configuration reloads.
\item Authenticate and inspect the exposed tools with \texttt{/\allowbreak{}mcp}. Record which account authenticated.
\item Exercise a read-only operation first.
\item Add explicit approval — an \texttt{ask} rule or a human gate — for writes.
\item Run one negative test: an operation the account should not be able to perform, confirmed refused at the provider.
\item Write down how the connection is removed, at both ends.
\end{enumerate}

Step 9 is good practice and proves less than it appears to. What a connection can do is the intersection of the credential's scopes, the provider's roles and resource permissions, what the server implements, which tools it exposes, and your Claude Code permission rules. "Read-only" is a claim about that whole intersection, and a successful read tests none of its limits. Step 11 does.

\begin{calloutbox}{palebrass}{brassdark}{2.0mm}
\textbf{CAUTION — tool output is untrusted input.} Content returned by an MCP server can contain instructions aimed at the agent, private data, or material designed to redirect the task. This is indirect prompt injection [4], [22], [23]. Treat every tool result as data, constrain what the agent may do next, and require confirmation before any external write [59].
\end{calloutbox}

\FloatBarrier
\setcounter{section}{3}
\section{Context cost, measured rather than assumed}

The course warns that tool definitions consume context [44]. That was true and is now largely mitigated: tool search is on by default and defers tool definitions until Claude needs them, so that only tool names and server instructions load at session start [34], [35]. The deferral is not universal. Older models, some third-party deployments and the HIPAA configuration load every tool upfront, as does a session that denies \texttt{Tool\allowbreak{}Search} or leaves it out of \texttt{-\allowbreak{}-\allowbreak{}tools}, and a server's author can mark a tool to load upfront. Measure with \texttt{/\allowbreak{}context} and \texttt{/\allowbreak{}usage} rather than assuming. Remove servers you do not need anyway — the argument is security and clarity, not tokens.

A server's \texttt{always\allowbreak{}Load:\allowbreak{} true} exempts its tools from deferral — except a tool whose author set \texttt{\_\allowbreak{}meta["anthr\allowbreak{}opic/\allowbreak{}always\allowbreak{}Load"]:\allowbreak{} false}, which stays deferred when the server comes from \texttt{-\allowbreak{}-\allowbreak{}mcp-\allowbreak{}config}, an Agent SDK application or a plugin \emph{(requires 2.1.285 or later)} — and \texttt{always\allowbreak{}Load:\allowbreak{} false} keeps them deferred even where the server's author marked a tool to load upfront \emph{(the second requires 2.1.287 or later)}; per-tool result-size overrides are available where a specific server needs different handling [34].

\FloatBarrier
\setcounter{section}{4}
\section{Organisational control}

By default anyone running Claude Code can connect any server, and Anthropic reviews directory connectors against listing criteria but does not security-audit or manage any MCP server [86]. An administrator has three mechanisms, which do different things and are easily confused:

\begingroup
\def\tblrows{%
\texttt{managed-\allowbreak{}mcp.\allowbreak{}json} & \textbf{Exclusive control.} Only its servers load, plus the few exceptions below; users cannot add others. An empty server map disables MCP & A file in the system directory only; it cannot be delivered through server-managed settings \\
\texttt{managed\allowbreak{}Mcp\allowbreak{}Servers} & \textbf{Provides} remote servers to every user, who keep their own \emph{(requires 2.1.259 or later)} & Any managed settings source \\
\texttt{allowed\allowbreak{}Mcp\allowbreak{}Servers} · \texttt{denied\allowbreak{}Mcp\allowbreak{}Servers} & \textbf{Filters} which configured servers may load. Neither list is a registry: a server must still be added by someone & Any settings scope; for enforcement, a managed source \\
}%
\def\tblbody{\begin{minipage}{\textwidth}\boxcaption{Table 11 — Three ways an organisation governs MCP servers}
{\footnotesize\begin{tabular}{L{34.8mm}L{65.0mm}L{40.0mm}}
\toprule
\textbf{Mechanism} & \textbf{What it does} & \textbf{Delivered by} \\
\midrule
\tblrows
\bottomrule\end{tabular}}\end{minipage}}%
\begingroup
\def\sloppy{\tolerance 9999\emergencystretch 3em\hfuzz 200pt\vfuzz 200pt}%
\hbadness=10000\vbadness=10000\hfuzz=200pt\vfuzz=200pt
\global\setbox\tblbox=\hbox{\tblbody}%
\endgroup
\par\addvspace{2.6mm}
\ifdim\dimexpr\ht\tblbox+\dp\tblbox\relax>0.55\textheight
  \typeout{HANDBOOK-TABLE broken \the\dimexpr\ht\tblbox+\dp\tblbox\relax}%
  \tabcaption{Table 11 — Three ways an organisation governs MCP servers}
  {\footnotesize\begin{longtable}{L{34.8mm}L{65.0mm}L{40.0mm}}
  \toprule
\textbf{Mechanism} & \textbf{What it does} & \textbf{Delivered by} \\
\midrule\endfirsthead
  \multicolumn{3}{@{}l@{}}{%
  \sffamily\footnotesize\itshape\color{inkgrey}Table 11 — Three ways an organisation governs MCP servers \textemdash\ continued}\\[1.2mm]
  \toprule
\textbf{Mechanism} & \textbf{What it does} & \textbf{Delivered by} \\
\midrule\endhead
  \bottomrule\endfoot
  \bottomrule\endlastfoot
  \tblrows
  \end{longtable}}%
\else
  \typeout{HANDBOOK-TABLE atomic \the\dimexpr\ht\tblbox+\dp\tblbox\relax}%
  \noindent\tblbody
\fi
\par\addvspace{2.6mm}
\endgroup

Source: [86].

Four properties of the filters decide whether a policy holds [86]:

\begin{itemize}
\item \textbf{A denylist match is final.} Nothing overrides it, it applies to servers from every source including the organisation's own, and denylists merge from every scope, so a user can always block a server for themselves. \texttt{denied\allowbreak{}Mcp\allowbreak{}Servers} can block \texttt{claude-\allowbreak{}in-\allowbreak{}chrome}, which is how an organisation turns off browser integration centrally [21].
\item \textbf{An allowlist is advisory until it is locked.} Without \texttt{allow\allowbreak{}Managed\allowbreak{}Mcp\allowbreak{}Servers\allowbreak{}Only}, allowlists from every scope merge, including the user's own, so a user can broaden yours. The lock makes the managed list the only one — and, as Section 31.2 warns, the lock with no list behind it admits everything, unless an embedding host supplies one.
\item \textbf{A \texttt{server\allowbreak{}Name} entry is not a security control.} The name is the label a user assigns, so a user can call any server \texttt{github}. Enforce with \texttt{server\allowbreak{}Url} or \texttt{server\allowbreak{}Command} entries; once an allowlist contains one of those, a name match no longer admits a server of that type.
\item \textbf{\texttt{server\allowbreak{}Command} matches the command and arguments exactly and ignores \texttt{env}}, and a \texttt{server\allowbreak{}Url} pattern with the hostname written in full matches the scheme's default port only. A deny entry written for a staging host without a port does not block that host on port 8443.
\end{itemize}

\begin{calloutbox}{palebrass}{brassdark}{2.0mm}
\textbf{CAUTION — an allowlist no longer filters your own managed file.} Servers in \texttt{managed-\allowbreak{}mcp.\allowbreak{}json} need no \texttt{allowed\allowbreak{}Mcp\allowbreak{}Servers} entry unless they use \texttt{\$\{VAR\}} expansion outside \texttt{env}. Before 2.1.305 a server using \texttt{\$\{VAR\}} only in \texttt{env} was checked as well, and before 2.1.259 every managed server was. An organisation that used the allowlist to keep some of its own managed servers from loading finds them \textbf{loading on each user's first launch of a newer client, with no prompt or notice} [86]. Use \texttt{denied\allowbreak{}Mcp\allowbreak{}Servers} entries, or a separate file per group, before the upgrade rather than after it.
\end{calloutbox}

Some servers are not reached by one or both filters, and an audit should name them [34], [86]. \textbf{Outside both lists:} in-process servers that the launching application registers — the VS Code extension's own, and the claude.ai connectors the desktop app delivers to its local and SSH sessions. No MCP setting or \texttt{managed-\allowbreak{}mcp.\allowbreak{}json} reaches them; those connectors are governed from the claude.ai organisation settings instead. \textbf{Outside the allowlist only, and still subject to the denylist:} built-in servers such as Claude in Chrome; every \texttt{managed\allowbreak{}Mcp\allowbreak{}Servers} entry; and a \texttt{managed-\allowbreak{}mcp.\allowbreak{}json} server that uses no \texttt{\$\{VAR\}} expansion outside \texttt{env}.

For claude.ai connectors an organisation can also set a control on an individual tool. A tool set to \texttt{ask} prompts on every call, even in \texttt{accept\allowbreak{}Edits}, \texttt{auto} and \texttt{bypass\allowbreak{}Permissions} modes and regardless of allow rules, and in \texttt{dont\allowbreak{}Ask} mode is denied; a tool set to \texttt{blocked} is removed before Claude sees it [34]. The exception is the desktop app's local and SSH sessions: there \texttt{blocked} tools are withheld, but \texttt{ask} does not reach Claude Code, which applies the session's ordinary permission rules instead [34]. That is the only MCP control in this chapter that acts on tool authority rather than on whether a server loads.

\subsection*{Exercise}

Do not install a live server yet. Ask Claude to produce a threat model for a hypothetical Notion connection: data reachable, write operations exposed, credential storage, scope, approval gates, logging, and revocation path. Approve installation only after replacing each assumption with a statement from the provider's documentation.

\emph{A worked solution is given in Appendix M.}

\begingroup
\def\kprows{%
\item A connected server acts with whatever authority you granted it, and Anthropic does not security-audit MCP servers.
\item Scope and approval decide where a definition is found and whether it loads. They say nothing about identity, tool authority or effect.
\item Project-scoped servers need per-user approval and workspace trust in an interactive session; \texttt{claude -\allowbreak{}p}, the Agent SDK and cloud sessions outside a project's threads load them without asking.
\item Prefer HTTP transport; SSE is deprecated, a \texttt{url} entry without a \texttt{type} is a configuration error, and a stdio server is a local program running as you.
\item Treat every tool result as untrusted data, and require confirmation before any external write.
}%
\def\kpbody{\begin{minipage}{\textwidth}\subsection*{Key points}\begin{itemize}\kprows\end{itemize}\end{minipage}}%
\begingroup
\def\sloppy{\tolerance 9999\emergencystretch 3em\hfuzz 200pt\vfuzz 200pt}%
\hbadness=10000\vbadness=10000\hfuzz=200pt\vfuzz=200pt
\global\setbox\kpbox=\hbox{\kpbody}%
\endgroup
\par\addvspace{4.2mm}
\ifdim\dimexpr\ht\kpbox+\dp\kpbox\relax>0.30\textheight
  \typeout{HANDBOOK-KEYPOINTS broken \the\dimexpr\ht\kpbox+\dp\kpbox\relax}%
  \subsection*{Key points}
  \begin{itemize}\kprows\end{itemize}
\else
  \typeout{HANDBOOK-KEYPOINTS atomic \the\dimexpr\ht\kpbox+\dp\kpbox\relax}%
  \noindent\kpbody
\fi
\par\addvspace{1.4mm}
\endgroup

\FloatBarrier
\renewcommand{\chaptertitlelabel}{Part IV \textperiodcentered\ Chapter 14}
\setcounter{chapter}{13}
\chapter{Environment variables and secrets}

A \texttt{.\allowbreak{}env} file is a developer convenience, not a security boundary. Commit only a template:

\begin{codeblock}{9.0}{10.6}
\cl{\#~.env.example~—~names~only,~never~values}
\cl{NOTION\_TOKEN=}
\cl{DATABASE\_URL=}
\end{codeblock}

\FloatBarrier
\setcounter{section}{0}
\section{Why "Claude cannot see it" is unsafe}

If Claude may read \texttt{.\allowbreak{}env}, its contents enter tool output. If it runs a command that prints an environment variable, the value enters the plaintext transcript (Section 12.3). A subprocess may transmit it. A sentence in \texttt{CLAUDE.\allowbreak{}md} saying "do not read secrets" is a preference, not enforcement [25]. This is M04 in Table 1, and it is the most common false belief among new users.

An environment variable is an injection mechanism, not a place of safety. A secret in Claude Code's environment is inherited by the shell commands it runs, sandboxed or not, and by the hooks, local MCP servers and helper commands it starts [28].

\FloatBarrier
\setcounter{section}{1}
\section{Controls, and the edge each one covers}

The controls below are usually listed as one stack. They are not one stack: each covers a different edge between a secret and a process, and a secret is exposed along whichever edge is left uncovered.

\begingroup
\def\tblrows{%
\texttt{Read} deny rule (Section 7.2) & Claude's built-in file tools, \texttt{@file} mentions, file commands Claude Code recognises in Bash such as \texttt{cat}, \texttt{head}, \texttt{sed} and \texttt{tee}, and redirections. With the sandbox's filesystem isolation on, the path is also added to the sandbox's \texttt{deny\allowbreak{}Read} & With the sandbox off: a command that reads without naming the file, such as a recursive \texttt{grep}, and a script or program that opens the file itself. Any process outside the sandbox \\
\texttt{sandbox.\allowbreak{}credentials}, \texttt{deny} (Section 7.4) & Sandboxed shell commands and their child processes: the variable is unset, and the file is unreadable while filesystem isolation is on & The built-in file tools; hooks, local MCP servers and helper commands, which run outside the sandbox with your full access; excluded commands and unsandboxed retries \\
\texttt{sandbox.\allowbreak{}credentials}, \texttt{mask} & The same commands, which see a placeholder; on macOS a masked \emph{file} is simply unreadable. Honoured only from user settings, managed settings and \texttt{-\allowbreak{}-\allowbreak{}settings}, and only with TLS termination configured & \textbf{Not redaction.} The sandbox proxy substitutes the real value on requests to the hosts you allow, so it authorises the credential's use there. Without TLS termination the placeholder is sent and authentication fails \\
\texttt{CLAUDE\_\allowbreak{}CODE\_\allowbreak{}SUBPROCESS\_\allowbreak{}ENV\_\allowbreak{}SCRUB=\allowbreak{}1} & The environment of every subprocess Claude Code starts, including hooks and stdio MCP servers. Heuristic: it recognises a credential by its name or by the shape of its value & Files; GitHub tokens, proxy settings and Git configuration variables, which it leaves in place; a secret whose name and value do not look like one \\
Sandbox \texttt{network.\allowbreak{}allowed\allowbreak{}Domains} & Where sandboxed shell commands may connect & The WebFetch tool; processes outside the sandbox \\
\texttt{Web\allowbreak{}Fetch(domai\allowbreak{}n:\allowbreak{}…)} rules & The WebFetch tool, and the sandbox's allowed or denied domain list, to which Claude Code adds the rule's domain & Shell commands that run outside the sandbox. A bare \texttt{Web\allowbreak{}Fetch} rule, with no \texttt{domain:\allowbreak{}} part, does not touch the sandbox lists \\
}%
\def\tblbody{\begin{minipage}{\textwidth}\boxcaption{Table 12 — What each secret control reaches, and what it does not}
{\footnotesize\begin{tabular}{L{69.7mm}L{34.3mm}L{35.8mm}}
\toprule
\textbf{Control} & \textbf{Reaches} & \textbf{Does not reach} \\
\midrule
\tblrows
\bottomrule\end{tabular}}\end{minipage}}%
\begingroup
\def\sloppy{\tolerance 9999\emergencystretch 3em\hfuzz 200pt\vfuzz 200pt}%
\hbadness=10000\vbadness=10000\hfuzz=200pt\vfuzz=200pt
\global\setbox\tblbox=\hbox{\tblbody}%
\endgroup
\par\addvspace{2.6mm}
\ifdim\dimexpr\ht\tblbox+\dp\tblbox\relax>0.55\textheight
  \typeout{HANDBOOK-TABLE broken \the\dimexpr\ht\tblbox+\dp\tblbox\relax}%
  \tabcaption{Table 12 — What each secret control reaches, and what it does not}
  {\footnotesize\begin{longtable}{L{69.7mm}L{34.3mm}L{35.8mm}}
  \toprule
\textbf{Control} & \textbf{Reaches} & \textbf{Does not reach} \\
\midrule\endfirsthead
  \multicolumn{3}{@{}l@{}}{%
  \sffamily\footnotesize\itshape\color{inkgrey}Table 12 — What each secret control reaches, and what it does not \textemdash\ continued}\\[1.2mm]
  \toprule
\textbf{Control} & \textbf{Reaches} & \textbf{Does not reach} \\
\midrule\endhead
  \bottomrule\endfoot
  \bottomrule\endlastfoot
  \tblrows
  \end{longtable}}%
\else
  \typeout{HANDBOOK-TABLE atomic \the\dimexpr\ht\tblbox+\dp\tblbox\relax}%
  \noindent\tblbody
\fi
\par\addvspace{2.6mm}
\endgroup

Source: [28], [29], [75].

With that table in hand, the layers are:

\begin{enumerate}
\item \textbf{Least privilege.} Issue a credential limited to the service, action and environment the task needs. It is the main control on damage after exposure, alongside the others that act on the credential rather than on Claude Code: short expiry, binding to one audience, rate and spend limits, and approval at the target system for anything consequential.
\item \textbf{Permission deny rules.} Block reads of secret files and directories (Section 7.2).
\item \textbf{Sandbox credential controls.} \texttt{sandbox.\allowbreak{}credentials.\allowbreak{}files} and \texttt{.\allowbreak{}env\allowbreak{}Vars} (Section 7.4). There is no built-in credential deny list: only what you name is protected [28]. Prefer \texttt{deny} unless a command must authenticate, and scope a \texttt{mask} entry with \texttt{inject\allowbreak{}Hosts}; without it the real value goes to every allowed domain.
\item \textbf{Network restriction.} Limit shell egress with the sandbox's \texttt{network.\allowbreak{}allowed\allowbreak{}Domains}, and the WebFetch tool with \texttt{Web\allowbreak{}Fetch(domai\allowbreak{}n:\allowbreak{}…)} rules. The relation runs one way: a domain-scoped WebFetch rule is also added to the sandbox's domain lists, while \texttt{allowed\allowbreak{}Domains} does nothing to WebFetch. A deny rule on \texttt{curl} and \texttt{wget} is worth having and is not a boundary: it does not match the same program called by path or inside \texttt{sh -\allowbreak{}c} [29], [59].
\item \textbf{Output discipline.} Never echo, log or interpolate a secret into a command the model will see.
\item \textbf{Git hygiene.} Ignore secret files and scan the staged diff before every commit.
\item \textbf{Rotation.} Revoke immediately if a credential reaches a transcript, a commit, a log, an artifact, or an untrusted service.
\end{enumerate}

Do not paste a security configuration without testing that it blocks. Ask Claude to attempt the denied read and confirm the denial, then check \texttt{/\allowbreak{}permissions} for the rule's source file. Then test the edge the rule does not cover, which is the point of the exercise below.

\FloatBarrier
\setcounter{section}{2}
\section{Incident response}

\begin{enumerate}
\item Stop the task.
\item Revoke or rotate the credential. Do this before anything else.
\item \textbf{Preserve what you need to scope the exposure before you clean anything}: the session transcript, the commit, the log line. Keep it access-controlled; it contains the secret.
\item Review where the value went: the local transcript, the provider's and the target system's own logs, the repository history, and — for organisations — whatever Claude Code telemetry you export (Chapter 33). What that telemetry holds depends on your exporter and on which detail and redaction settings were on at the time, so check before relying on it.
\item Then remove the value from files, logs and any published artifact, under your retention process. If it was committed, follow the hosting provider's sensitive-data removal procedure; deleting the latest line is not enough [24].
\item Correct the control that permitted exposure, and add a regression test for it.
\end{enumerate}

\subsection*{Exercise}

Use a synthetic canary, never a real credential. Put a made-up value such as \texttt{CANARY\_\allowbreak{}7f3a\_\allowbreak{}not\_\allowbreak{}a\_\allowbreak{}secret} in a \texttt{.\allowbreak{}env} file in a practice workspace, add \texttt{Read(.\allowbreak{}/\allowbreak{}.\allowbreak{}env)} to \texttt{permissions.\allowbreak{}deny}, and start a new session in Manual mode with the sandbox off. Ask Claude to read the file with its file tool and confirm the denial. Then ask it to run a short script that opens the file and prints its contents. Record which attempt was blocked, by what, and what a search of the session transcript for the canary finds afterwards. Then turn the sandbox on and run the script again.

\emph{A worked solution is given in Appendix M.}

\begingroup
\def\kprows{%
\item A \texttt{.\allowbreak{}env} file is a convenience, not a boundary. An instruction not to read secrets is not enforcement.
\item Each control covers one edge: with the sandbox off a \texttt{Read} deny rule does not stop a script that opens the file, sandbox credential controls reach only sandboxed shell commands, and the sandbox's domain list does not limit the WebFetch tool.
\item Masking a credential is not redaction: it authorises the sandbox proxy to send the real value to the hosts you allow.
\item Test that a deny rule actually blocks, and then test the edge it does not cover.
\item On exposure, rotate first, preserve the evidence, investigate, and only then clean up.
}%
\def\kpbody{\begin{minipage}{\textwidth}\subsection*{Key points}\begin{itemize}\kprows\end{itemize}\end{minipage}}%
\begingroup
\def\sloppy{\tolerance 9999\emergencystretch 3em\hfuzz 200pt\vfuzz 200pt}%
\hbadness=10000\vbadness=10000\hfuzz=200pt\vfuzz=200pt
\global\setbox\kpbox=\hbox{\kpbody}%
\endgroup
\par\addvspace{4.2mm}
\ifdim\dimexpr\ht\kpbox+\dp\kpbox\relax>0.30\textheight
  \typeout{HANDBOOK-KEYPOINTS broken \the\dimexpr\ht\kpbox+\dp\kpbox\relax}%
  \subsection*{Key points}
  \begin{itemize}\kprows\end{itemize}
\else
  \typeout{HANDBOOK-KEYPOINTS atomic \the\dimexpr\ht\kpbox+\dp\kpbox\relax}%
  \noindent\kpbody
\fi
\par\addvspace{1.4mm}
\endgroup

\FloatBarrier
\renewcommand{\chaptertitlelabel}{Part IV \textperiodcentered\ Chapter 15}
\setcounter{chapter}{14}
\chapter{Turn repeatable work into skills}

A skill is a reusable package of instructions and optional resources. Claude invokes it when its description matches the task, or you call it as \texttt{/\allowbreak{}skill-\allowbreak{}name}. Unlike \texttt{CLAUDE.\allowbreak{}md}, a skill's body loads only when used, so long reference material costs almost nothing until needed [36].

\begin{codeblock}{9.0}{10.6}
\cl{\textasciitilde{}/.claude/skills/<name>/SKILL.md~~~~~\#~user~scope}
\cl{.claude/skills/<name>/SKILL.md~~~~~~~\#~project~scope}
\end{codeblock}

Skills also arrive from the managed settings directory, from plugins, and from a claude.ai account. Two facts about location are easy to get wrong [36]. Where the same name exists in more than one place, the enterprise skill wins over the personal one and \textbf{the personal one wins over the project's}, so a skill in your home directory silently replaces the one your team committed. And a personal skill is local to the machine: cloud sessions and Cowork sessions do not read \texttt{\textasciitilde{}/\allowbreak{}.\allowbreak{}claude/\allowbreak{}skills/\allowbreak{}}.

A skill is more than a prompt fragment. It can act on five planes at once, and a review that reads only the prose has looked at one of them:

\begin{enumerate}
\item \textbf{Selection} — its \texttt{description} decides when Claude loads it.
\item \textbf{Context} — its body and supporting files enter the conversation.
\item \textbf{Authorisation} — \texttt{allowed-\allowbreak{}tools} pre-approves tool calls; \texttt{disallowed-\allowbreak{}tools} removes them.
\item \textbf{Host execution} — inline shell commands run on your machine while the skill renders.
\item \textbf{Lifecycle} — \texttt{context:\allowbreak{} fork} delegates the work, and \texttt{hooks} outlive the invocation.
\end{enumerate}

Custom commands have merged into skills: \texttt{.\allowbreak{}claude/\allowbreak{}commands/\allowbreak{}deploy.\allowbreak{}md} and \texttt{.\allowbreak{}claude/\allowbreak{}skills/\allowbreak{}deploy/\allowbreak{}SKILL.\allowbreak{}md} both create \texttt{/\allowbreak{}deploy}. Existing command files keep working; use skills for anything new [36].

\FloatBarrier
\setcounter{section}{0}
\section{Build one from observed work}

This is the course's strongest contribution and it is worth stating plainly [44]:

\begin{enumerate}
\item Complete the procedure manually, with Claude, once.
\item Identify inputs, tools, decisions, outputs and failure modes.
\item Ask Claude to write a minimal skill from what actually happened.
\item Review every file and every tool grant.
\item Test it on a genuinely different example.
\item Correct the output by hand.
\item Update the skill with the \textbf{general} lesson, not the incidental detail.
\end{enumerate}

\FloatBarrier
\setcounter{section}{1}
\section{Example project skill}

\begin{codeblock}{8.0}{9.4}
\cl{-{}-{}-}
\cl{name:~research-brief}
\cl{description:~Create~an~evidence-led~decision~brief~from~approved~sources~in~this~workspace.}
\cl{~~Use~when~the~user~asks~for~a~research~brief,~evidence~synthesis,~or~source-backed}
\cl{~~recommendation.}
\cl{allowed-tools:~Read,~Grep,~Glob}
\cl{-{}-{}-}
\cl{}
\cl{\#~Research~brief}
\cl{}
\cl{\#\#~Required~input}
\cl{-~Decision~question~and~audience}
\cl{-~Approved~source~scope}
\cl{-~Output~length~and~deadline}
\cl{}
\cl{\#\#~Procedure}
\cl{1.~Read~\textasciigrave{}CLAUDE.md\textasciigrave{}~and~the~approved~source~list.}
\cl{2.~Extract~one~atomic~claim~per~ledger~row.}
\cl{3.~Separate~observed~facts,~source~claims,~and~analyst~inference.}
\cl{4.~Search~for~contradictions~and~missing~evidence.}
\cl{5.~Draft~in~\textasciigrave{}drafts/\textasciigrave{};~never~write~directly~to~\textasciigrave{}deliverables/\textasciigrave{}.}
\cl{6.~Run~the~evidence-reviewer~subagent~if~available.}
\cl{7.~Report~gaps~and~request~human~approval.}
\cl{}
\cl{\#\#~Output}
\cl{Executive~answer~·~scope~and~method~·~findings~with~inline~citations~·~evidence~table~·}
\cl{contradictions~and~gaps~·~source~register.}
\end{codeblock}

The grant in that example is narrower than it may look. \texttt{allowed-\allowbreak{}tools:\allowbreak{} Read,\allowbreak{} Grep,\allowbreak{} Glob} pre-approves those three tools for the invoking turn; it does \textbf{not} restrict the skill to them. Every other tool stays callable under your ordinary permission settings, so step 5, writing the draft, prompts in Manual mode — which is the intended behaviour, not an omission.

\FloatBarrier
\setcounter{section}{2}
\section{Frontmatter that matters for safety}

\begingroup
\def\tblrows{%
\texttt{description} & Decides when Claude selects the skill. Cut at 1,536 characters in the listing, together with \texttt{when\_\allowbreak{}to\_\allowbreak{}use}, so put the key use case first \\
\texttt{allowed-\allowbreak{}tools} & Tools Claude may use \textbf{without asking} during the turn that invokes the skill. The grant clears when you send your next message. It pre-approves and does not restrict. Workspace trust does not gate it, even in a \texttt{-\allowbreak{}p} run in a folder you have never trusted. Ignored for project and personal skills under managed \texttt{allow\allowbreak{}Managed\allowbreak{}Permission\allowbreak{}Rules\allowbreak{}Only} (2.1.282+) \\
\texttt{disallowed-\allowbreak{}tools} & Tools removed from the pool while the skill is active; also clears on your next message. It cannot remove \texttt{End\allowbreak{}Conversatio\allowbreak{}n} while any other tool remains \\
\texttt{disable-\allowbreak{}model-\allowbreak{}invocation:\allowbreak{} true} & Claude cannot invoke the skill on its own. A \texttt{/\allowbreak{}name} you write in the middle of a message still counts as your permission for that message. Also prevents preloading into subagents, and prevents it running when a scheduled task fires with the skill as its prompt (2.1.196+) \\
\texttt{context:\allowbreak{} fork} & Runs the skill in a subagent that does not see your conversation — not a fork of it. Runs in the background by default, with the narrower background tool set, and a background run's edits to files Claude had not already edited fall outside your checkpoints. \texttt{background:\allowbreak{} false} runs it in the foreground, with the full tool set and normal rewinding \\
\texttt{user-\allowbreak{}invocable:\allowbreak{} false} & Only Claude can invoke it; hidden from the \texttt{/\allowbreak{}} menu \\
\texttt{shell} & \texttt{bash} (default) or \texttt{powershell} for inline shell in the skill body \\
\texttt{hooks} & Lifecycle hooks registered when the skill is invoked, active for the rest of the session. \texttt{once:\allowbreak{} true} removes a hook after its first \emph{successful} run; one that fails, blocks or times out stays and runs again \\
}%
\def\tblbody{\begin{minipage}{\textwidth}\boxcaption{Table 13 — Skill frontmatter fields with a security consequence}
{\footnotesize\begin{tabular}{L{51.3mm}L{91.8mm}}
\toprule
\textbf{Field} & \textbf{Effect} \\
\midrule
\tblrows
\bottomrule\end{tabular}}\end{minipage}}%
\begingroup
\def\sloppy{\tolerance 9999\emergencystretch 3em\hfuzz 200pt\vfuzz 200pt}%
\hbadness=10000\vbadness=10000\hfuzz=200pt\vfuzz=200pt
\global\setbox\tblbox=\hbox{\tblbody}%
\endgroup
\par\addvspace{2.6mm}
\ifdim\dimexpr\ht\tblbox+\dp\tblbox\relax>0.55\textheight
  \typeout{HANDBOOK-TABLE broken \the\dimexpr\ht\tblbox+\dp\tblbox\relax}%
  \tabcaption{Table 13 — Skill frontmatter fields with a security consequence}
  {\footnotesize\begin{longtable}{L{51.3mm}L{91.8mm}}
  \toprule
\textbf{Field} & \textbf{Effect} \\
\midrule\endfirsthead
  \multicolumn{2}{@{}l@{}}{%
  \sffamily\footnotesize\itshape\color{inkgrey}Table 13 — Skill frontmatter fields with a security consequence \textemdash\ continued}\\[1.2mm]
  \toprule
\textbf{Field} & \textbf{Effect} \\
\midrule\endhead
  \bottomrule\endfoot
  \bottomrule\endlastfoot
  \tblrows
  \end{longtable}}%
\else
  \typeout{HANDBOOK-TABLE atomic \the\dimexpr\ht\tblbox+\dp\tblbox\relax}%
  \noindent\tblbody
\fi
\par\addvspace{2.6mm}
\endgroup

Two of those rows depend on policy as well as on the file. Where managed settings lock hooks to plugins with \texttt{strict\allowbreak{}Plugin\allowbreak{}Only\allowbreak{}Customizati\allowbreak{}on}, a personal or project skill runs without any of its frontmatter hooks and, if it declares \texttt{hooks}, loses its \texttt{allowed-\allowbreak{}tools} pre-approval too; and each source of skills may register at most 1,024 hooks in a session \emph{(requires 2.1.304 or later)} [61], [83]. A guard written as a skill hook is therefore not a guard until \texttt{/\allowbreak{}hooks} lists it.

A field name must match exactly: a misspelled field is ignored without an error, and frontmatter is read only when the opening \texttt{-\allowbreak{}-\allowbreak{}-\allowbreak{}} is the file's first line [36].

Three behaviours are easy to miss. \texttt{\$\{CLAUDE\_\allowbreak{}SKILL\_\allowbreak{}DIR\}} and \texttt{\$\{CLAUDE\_\allowbreak{}PROJECT\_\allowbreak{}DIR\}} are substituted both in the body and in Bash rules inside \texttt{allowed-\allowbreak{}tools}, so a skill can run a bundled script without a prompt — \texttt{allowed-\allowbreak{}tools:\allowbreak{} Bash(\$\{C\allowbreak{}LAUDE\_\allowbreak{}SKILL\_\allowbreak{}DIR\}/\allowbreak{}scripts/\allowbreak{}render.\allowbreak{}sh *)} [36]. The rendered \texttt{SKILL.\allowbreak{}md} enters the conversation once and \textbf{stays for the rest of the session}; Claude Code does not re-read the file on later turns, so write standing guidance rather than one-time steps — and after a compaction only the first 5,000 tokens of each invoked skill are re-attached, within a combined budget of 25,000, most recent first [36]. And inline shell commands \textbf{never prompt} while a skill renders: each is checked against your permission rules, and outside auto mode anything that is not already allowed aborts the invocation. A skill can pre-approve its own injected command with \texttt{allowed-\allowbreak{}tools}, and where nothing else stops it that command runs on your machine, without a prompt, before Claude has read a word of the skill. Three things do stop it: a deny or ask rule, which overrides \texttt{allowed-\allowbreak{}tools}; \texttt{disable\allowbreak{}Skill\allowbreak{}Shell\allowbreak{}Execution}; and the skill being one synced from a claude.ai account, whose commands an ordinary local session does not execute.

Use \texttt{disable-\allowbreak{}model-\allowbreak{}invocation:\allowbreak{} true} for anything with side effects — \texttt{/\allowbreak{}commit}, \texttt{/\allowbreak{}deploy}, \texttt{/\allowbreak{}send-\allowbreak{}message}. You do not want Claude deciding your code looks ready to ship.

\FloatBarrier
\setcounter{section}{3}
\section{Skills synced from claude.ai}

A session signed in with a claude.ai account also loads the skills enabled for that account: the ones you create or turn on in your claude.ai settings, the ones your organisation provides there, and Anthropic's built-in skills such as \texttt{pdf} and \texttt{xlsx}, some of which always sync [36]. In the terminal \emph{(requires 2.1.273 or later)} Claude Code downloads them into \texttt{\textasciitilde{}/\allowbreak{}.\allowbreak{}claude/\allowbreak{}skills/\allowbreak{}synced/\allowbreak{}} in the background when the session starts, then checks for changes — about every ten minutes while you work, about every forty while the session is idle — and adds, updates or removes a skill in the running session. It syncs only in a session that uses a sign-in stored by \texttt{/\allowbreak{}login} and fetches feature flags. It does not sync with an API key, \texttt{ANTHROPIC\_\allowbreak{}AUTH\_\allowbreak{}TOKEN}, \texttt{CLAUDE\_\allowbreak{}CODE\_\allowbreak{}OAUTH\_\allowbreak{}TOKEN} or an \texttt{api\allowbreak{}Key\allowbreak{}Helper}; on Amazon Bedrock or with \texttt{CLAUDE\_\allowbreak{}CODE\_\allowbreak{}DISABLE\_\allowbreak{}NONESSENTIA\allowbreak{}L\_\allowbreak{}TRAFFIC}; in bare mode or with \texttt{-\allowbreak{}-\allowbreak{}safe-\allowbreak{}mode}; or where managed settings lock skills to plugin sources or a \texttt{-\allowbreak{}-\allowbreak{}setting-\allowbreak{}sources} list leaves out \texttt{user} [36].

For a security review, a synced skill is content nobody wrote on this machine, and Claude Code treats it so. In a session on your machine other than Cowork, it does not run a synced skill's \texttt{!} command lines, attach the files its \texttt{@} references name, or substitute the \texttt{\$\{CLAUDE\_\allowbreak{}PROJECT\_\allowbreak{}DIR\}} and \texttt{\$\{CLAUDE\_\allowbreak{}SESSION\_\allowbreak{}ID\}} placeholders; they reach Claude as literal text, and a \texttt{!} line reaches it as the \texttt{disable\allowbreak{}Skill\allowbreak{}Shell\allowbreak{}Execution} placeholder when that setting is on. It also sanitises the skill's description \emph{(2.1.228 or later)}. The frontmatter still applies, so an \texttt{allowed-\allowbreak{}tools} grant goes through the normal permission flow — and does not apply at all under \texttt{allow\allowbreak{}Managed\allowbreak{}Permission\allowbreak{}Rules\allowbreak{}Only} [36]. The sync is one-way: an edit under \texttt{\textasciitilde{}/\allowbreak{}.\allowbreak{}claude/\allowbreak{}skills/\allowbreak{}synced/\allowbreak{}} is not uploaded, and a later sync can overwrite or remove it. A synced skill whose short name another command already uses runs only as \texttt{/\allowbreak{}anthropic-\allowbreak{}skills:\allowbreak{}<name>}, and that namespace is reserved for synced skills in every session [36].

\texttt{/\allowbreak{}skills} lists what synced under \texttt{claude.\allowbreak{}ai sync}. To stop syncing on one machine, set \texttt{sync\allowbreak{}Claude\allowbreak{}Ai\allowbreak{}Skills} to \texttt{false} in user settings; the synced skills move to \texttt{\textasciitilde{}/\allowbreak{}.\allowbreak{}claude/\allowbreak{}skills/\allowbreak{}.\allowbreak{}trash/\allowbreak{}} at the next start. An organisation can set the same key in managed settings, or turn Skills off on claude.ai for everyone [36].

\FloatBarrier
\setcounter{section}{4}
\section{Skill security review}

Before installing or running a third-party skill, inspect:

\begin{itemize}
\item \texttt{SKILL.\allowbreak{}md} and every referenced file;
\item \texttt{allowed-\allowbreak{}tools}, \texttt{disallowed-\allowbreak{}tools} and any \texttt{hooks} block;
\item inline shell execution (`\texttt{ !}command\texttt{ }\texttt{ lines and }\texttt{ }`\texttt{! }\texttt{ blocks),\allowbreak{} together w\allowbreak{}ith any }allowed-tools` entry that pre-approves it;
\item the \texttt{description}, which decides when the skill is selected: an over-broad one loads the skill, and its grants, for tasks it was never reviewed for;
\item package installation and network calls;
\item paths read or written;
\item credential access;
\item any instruction that discourages review or suppresses output.
\end{itemize}

Organisational and personal controls: \texttt{disable\allowbreak{}Bundled\allowbreak{}Skills} turns off every bundled skill except \texttt{/\allowbreak{}doctor}; \texttt{skill\allowbreak{}Overrides} sets an individual skill to \texttt{"off"} or \texttt{"user-\allowbreak{}invocable-\allowbreak{}only"} without editing its file; a \texttt{Skill} deny rule restricts Claude's access; and \texttt{disable\allowbreak{}Skill\allowbreak{}Shell\allowbreak{}Execution} replaces inline shell command lines with a placeholder in user, project, plugin and additional-directory skills, though not in bundled or managed ones [36].

\begin{calloutbox}{palebrass}{brassdark}{2.0mm}
\textbf{CAUTION:} stars, download counts, a video recommendation, or an official-sounding name are not a security review. Pin a reviewed version and repeat the review after every update. This is M12 in Table 1.
\end{calloutbox}

\subsection*{Exercise}

Install the example skill locally, invoke \texttt{/\allowbreak{}research-\allowbreak{}brief} on a second source set, and record every place it made an assumption. Improve the skill only with a rule that generalises, and test the revision in a fresh session.

\emph{A worked solution is given in Appendix M.}

\begingroup
\def\kprows{%
\item Derive a skill from a procedure you have already performed manually, and generalise the lesson rather than the instance.
\item \texttt{allowed-\allowbreak{}tools} pre-approves tools for the invoking turn only, and clears on your next message. It grants; it does not restrict, and workspace trust does not gate it.
\item Use \texttt{disable-\allowbreak{}model-\allowbreak{}invocation:\allowbreak{} true} for anything with side effects.
\item The rendered skill stays in the conversation for the rest of the session and is not re-read on later turns, so evaluate a revised skill in a fresh session.
}%
\def\kpbody{\begin{minipage}{\textwidth}\subsection*{Key points}\begin{itemize}\kprows\end{itemize}\end{minipage}}%
\begingroup
\def\sloppy{\tolerance 9999\emergencystretch 3em\hfuzz 200pt\vfuzz 200pt}%
\hbadness=10000\vbadness=10000\hfuzz=200pt\vfuzz=200pt
\global\setbox\kpbox=\hbox{\kpbody}%
\endgroup
\par\addvspace{4.2mm}
\ifdim\dimexpr\ht\kpbox+\dp\kpbox\relax>0.30\textheight
  \typeout{HANDBOOK-KEYPOINTS broken \the\dimexpr\ht\kpbox+\dp\kpbox\relax}%
  \subsection*{Key points}
  \begin{itemize}\kprows\end{itemize}
\else
  \typeout{HANDBOOK-KEYPOINTS atomic \the\dimexpr\ht\kpbox+\dp\kpbox\relax}%
  \noindent\kpbody
\fi
\par\addvspace{1.4mm}
\endgroup

\FloatBarrier
\renewcommand{\chaptertitlelabel}{Part IV \textperiodcentered\ Chapter 16}
\setcounter{chapter}{15}
\chapter{Delegate bounded work to subagents}

A subagent is a specialised instance with its own context window, system prompt, tool access and permissions [27]. Delegate when independence, focus or parallelism improves the result — not to make a small action feel important.

\FloatBarrier
\setcounter{section}{0}
\section{Fresh context, forks, and why it matters}

An ordinary subagent starts fresh: it receives its own system prompt, the delegation message, the \texttt{CLAUDE.\allowbreak{}md} hierarchy, a git-status snapshot and any preloaded skills. It does \textbf{not} receive the parent's conversation history, output style, or auto memory [27]. Fresh is not the same as independent. Unless its definition sets \texttt{omit\allowbreak{}Claude\allowbreak{}Md} \emph{(requires 2.1.271 or later)}, it reads the same project instructions as the session it is checking, and it runs under the same credentials and, by default, on the same model.

A \textbf{fork} is different: it inherits the entire conversation, the parent's system prompt, tools and model, and shares the parent's prompt cache. \texttt{/\allowbreak{}subtask} creates one, whether or not fork mode is on; with agent view turned off \texttt{/\allowbreak{}subtask} is unavailable and \texttt{/\allowbreak{}fork} creates it instead [37]. It is the right mechanism for continuing the current reasoning somewhere else, and the wrong one for checking it.

\begin{calloutbox}{palebrass}{brassdark}{2.0mm}
\textbf{CAUTION — this is the correction that matters most in this chapter.} Fork mode is \textbf{on by default} — in an interactive terminal from 2.1.232, and from 2.1.302 in \texttt{-\allowbreak{}p}, the Agent SDK, Desktop and VS Code too, though \texttt{claude mcp \allowbreak{}serve} never offers it — controlled by the \texttt{CLAUDE\_\allowbreak{}CODE\_\allowbreak{}FORK\_\allowbreak{}SUBAGENT} environment variable (\texttt{1} on, \texttt{0} off) [27]. Fork mode does two things. It lets Claude \textbf{start a fork on its own}, so work you asked to have checked may be handed to a copy of the conversation that produced it. And while it is on, Claude cannot wait for a subagent it spawns from the main conversation of an interactive terminal, VS Code or desktop session: that subagent runs in the background and reports in a later turn, unless \texttt{CLAUDE\_\allowbreak{}CODE\_\allowbreak{}DISABLE\_\allowbreak{}BACKGROUND\_\allowbreak{}TASKS=\allowbreak{}1} is set. The documentation distinguishes a fork from every other subagent, which starts fresh from its definition, so a defined agent is not a fork; but naming an agent in prose leaves Claude to decide whether to delegate to it at all. If you need genuine independence — and the capstone of this book does — \textbf{@-mention} the agent, which the documentation says ensures that specific subagent runs, and confirm the reviewer's context from its output. To stop Claude forking at all, add \texttt{Agent(fork)} to \texttt{permissions.\allowbreak{}deny}. This is M03 in Table 1.
\end{calloutbox}

A skill with \texttt{context:\allowbreak{} fork} is a third thing, despite the word: it runs the skill in a subagent under the rules of Chapter 15, and is not a fork of the conversation.

\FloatBarrier
\setcounter{section}{1}
\section{Where subagents live, and which one wins}

\begingroup
\def\tblrows{%
Managed settings & Organisation-wide & 1 (highest) \\
\texttt{-\allowbreak{}-\allowbreak{}agents} CLI flag & Current session & 2 \\
\texttt{.\allowbreak{}claude/\allowbreak{}agents/\allowbreak{}} & Current project & 3 \\
\texttt{\textasciitilde{}/\allowbreak{}.\allowbreak{}claude/\allowbreak{}agents/\allowbreak{}} & All your projects & 4 \\
Plugin \texttt{agents/\allowbreak{}} & Where the plugin is enabled & 5 (lowest) \\
}%
\def\tblbody{\begin{minipage}{\textwidth}\boxcaption{Table 14 — Subagent definition precedence}
{\footnotesize\begin{tabular}{L{56.6mm}L{63.9mm}L{19.4mm}}
\toprule
\textbf{Source} & \textbf{Scope} & \textbf{Priority} \\
\midrule
\tblrows
\bottomrule\end{tabular}}\end{minipage}}%
\begingroup
\def\sloppy{\tolerance 9999\emergencystretch 3em\hfuzz 200pt\vfuzz 200pt}%
\hbadness=10000\vbadness=10000\hfuzz=200pt\vfuzz=200pt
\global\setbox\tblbox=\hbox{\tblbody}%
\endgroup
\par\addvspace{2.6mm}
\ifdim\dimexpr\ht\tblbox+\dp\tblbox\relax>0.55\textheight
  \typeout{HANDBOOK-TABLE broken \the\dimexpr\ht\tblbox+\dp\tblbox\relax}%
  \tabcaption{Table 14 — Subagent definition precedence}
  {\footnotesize\begin{longtable}{L{56.6mm}L{63.9mm}L{19.4mm}}
  \toprule
\textbf{Source} & \textbf{Scope} & \textbf{Priority} \\
\midrule\endfirsthead
  \multicolumn{3}{@{}l@{}}{%
  \sffamily\footnotesize\itshape\color{inkgrey}Table 14 — Subagent definition precedence \textemdash\ continued}\\[1.2mm]
  \toprule
\textbf{Source} & \textbf{Scope} & \textbf{Priority} \\
\midrule\endhead
  \bottomrule\endfoot
  \bottomrule\endlastfoot
  \tblrows
  \end{longtable}}%
\else
  \typeout{HANDBOOK-TABLE atomic \the\dimexpr\ht\tblbox+\dp\tblbox\relax}%
  \noindent\tblbody
\fi
\par\addvspace{2.6mm}
\endgroup

Project definitions are found by walking up from the working directory, and where two nested \texttt{.\allowbreak{}claude/\allowbreak{}agents/\allowbreak{}} directories define the same name the one closest to the working directory wins [27].

\texttt{/\allowbreak{}agents} no longer opens an agent editor \emph{(changed at 2.1.198)}; it directs you to ask Claude to create or update an agent, which you then read as Markdown yourself [37]. This is M13 in Table 1. \texttt{/\allowbreak{}list-\allowbreak{}agents} (alias \texttt{/\allowbreak{}peers}) lists subagents, teammates and other sessions [37].

\FloatBarrier
\setcounter{section}{2}
\section{An independent read-only reviewer}

\texttt{.\allowbreak{}claude/\allowbreak{}agents/\allowbreak{}evidence-\allowbreak{}reviewer.\allowbreak{}md}:

\begin{codeblock}{7.0}{8.3}
\cl{-{}-{}-}
\cl{name:~evidence-reviewer}
\cl{description:~Independently~review~a~draft~for~unsupported,~overstated,~stale,~or~contradictory}
\cl{~~claims.~Use~after~a~research~brief~is~drafted.}
\cl{tools:~Read,~Grep,~Glob}
\cl{permissionMode:~plan}
\cl{model:~inherit}
\cl{maxTurns:~25}
\cl{-{}-{}-}
\cl{}
\cl{You~are~an~independent~evidence~reviewer.~You~did~not~author~the~draft~and~you~have~no~access}
\cl{to~the~reasoning~that~produced~it.}
\cl{}
\cl{For~each~material~claim:}
\cl{1.~Locate~its~cited~source~in~the~workspace.}
\cl{2.~Classify~support~as~supports~/~partially~supports~/~contradicts~/~missing.}
\cl{3.~Check~that~date,~scope~and~attribution~match.}
\cl{4.~Report~the~exact~file~and~passage~requiring~correction.}
\cl{}
\cl{Do~not~edit~files.~Return~a~severity-ordered~review~and~a~final~pass/fail~verdict.}
\end{codeblock}

Two different lines make this reviewer read-only, and they are not equally strong. The \texttt{tools} line is an allowlist: a subagent given \texttt{Read,\allowbreak{} Grep,\allowbreak{} Glob} has no tool with which to write, in any permission mode. The \texttt{permission\allowbreak{}Mode} line is a second layer, and the weaker one.

\begin{calloutbox}{palebrass}{brassdark}{2.0mm}
\textbf{CAUTION:} a subagent's own \texttt{permission\allowbreak{}Mode} does not always apply. Three parent modes defeat it. A parent in \texttt{bypass\allowbreak{}Permissions} or \texttt{accept\allowbreak{}Edits} takes precedence and cannot be overridden. A parent in \texttt{auto} hands the subagent auto mode, and any \texttt{permission\allowbreak{}Mode} in the subagent's frontmatter is ignored: the classifier evaluates its calls under the parent's own block and allow rules. And \texttt{auto} is the mode an interactive terminal or VS Code session starts in on every plan from 2.1.283, unless settings or an organisation change it, so this is the ordinary case rather than an exotic one [27]. A parent in \texttt{plan} mode overrides it in the other direction: the subagent stays in plan mode whatever it asks for. So do not rely on \texttt{permission\allowbreak{}Mode} to make a reviewer read-only. Rely on the \texttt{tools} allowlist, and omit \texttt{Agent} from it so that the reviewer cannot hand the work to a subagent with more.
\end{calloutbox}

\FloatBarrier
\setcounter{section}{3}
\section{The frontmatter you should know}

Only \texttt{name} and \texttt{description} are required. The rest change behaviour materially [27]: \texttt{tools} and \texttt{disallowed\allowbreak{}Tools} (both accept MCP patterns such as \texttt{mcp\_\allowbreak{}\_\allowbreak{}server\_\allowbreak{}\_\allowbreak{}*}); \texttt{model} (\texttt{sonnet}, \texttt{opus}, \texttt{haiku}, \texttt{fable}, a full ID, or \texttt{inherit}); \texttt{permission\allowbreak{}Mode}; \texttt{max\allowbreak{}Turns}; \texttt{skills} to preload; \texttt{mcp\allowbreak{}Servers}; \texttt{hooks}; \texttt{memory} (\texttt{user}, \texttt{project} or \texttt{local}, each with its own \texttt{agent-\allowbreak{}memory} directory and the same 200-line/25 KB index limit); \texttt{background}; \texttt{effort}; \texttt{isolation:\allowbreak{} worktree} to run in a temporary git worktree; \texttt{omit\allowbreak{}Claude\allowbreak{}Md}; \texttt{auto\allowbreak{}Compact\allowbreak{}Window} \emph{(requires 2.1.296 or later)}; \texttt{color}; and \texttt{initial\allowbreak{}Prompt}.

Four behaviours of that list are easy to get wrong [27]:

\begin{itemize}
\item \textbf{A misspelled field is ignored without an error.} Field names are camelCase and must match exactly; \texttt{permissionm\allowbreak{}ode:\allowbreak{} plan} is not a restriction. From 2.1.296 the debug log names unrecognised fields. Outside plugins, a file with no \texttt{name}, no \texttt{description} or unparseable YAML is skipped silently, and \texttt{claude plug\allowbreak{}in validate\allowbreak{} .\allowbreak{}claude/\allowbreak{}agents} finds the ones that do not parse. A plugin's agent with no \texttt{name} still loads, under its filename. One whose frontmatter does not parse also loads, under its filename and with every field, tool restrictions included, ignored [27], [39].
\item \textbf{\texttt{Agent(worke\allowbreak{}r,\allowbreak{} researcher\allowbreak{})} restricts spawning only for an agent run as the main thread with \texttt{claude -\allowbreak{}-\allowbreak{}agent}.} In a subagent's own definition, listing \texttt{Agent} lets it spawn while the depth limit allows, and any type list inside the parentheses is ignored. To stop a subagent spawning, leave \texttt{Agent} out of \texttt{tools} or add it to \texttt{disallowed\allowbreak{}Tools}.
\item \textbf{A plugin's subagents do not get \texttt{permission\allowbreak{}Mode}, \texttt{mcp\allowbreak{}Servers}, \texttt{hooks} or \texttt{initial\allowbreak{}Prompt}.} Those four fields are ignored for a definition that arrives in a plugin.
\item \textbf{\texttt{disallowed\allowbreak{}Tools} removes whole tools.} An entry with a specifier, such as \texttt{Bash(git pu\allowbreak{}sh *)}, still removes \texttt{Bash} entirely.
\end{itemize}

The model a subagent runs on is resolved in a fixed order: a model Claude passes for that invocation, then the definition's \texttt{model} field, then the \texttt{CLAUDE\_\allowbreak{}CODE\_\allowbreak{}SUBAGENT\_\allowbreak{}MODEL} environment variable, then the main conversation's model. Before 2.1.251 the variable came first and overrode the other two; it is now a default, and it takes \texttt{CLAUDE\_\allowbreak{}CODE\_\allowbreak{}SUBAGENT\_\allowbreak{}MODEL\_\allowbreak{}FORCE=\allowbreak{}1} to apply one model to subagents, teammates and workflow agents \emph{(requires 2.1.257 or later)} — except a fork and a skill run in a subagent with \texttt{model:\allowbreak{} inherit}, which stay on the main conversation's model. Every one of those values is checked against an organisation's \texttt{available\allowbreak{}Models}, and a blocked value is substituted with a warning rather than refused [27]. Effort follows the same shape: the \texttt{effort} field, overridden by a per-invocation value \emph{(requires 2.1.292 or later)}, with \texttt{CLAUDE\_\allowbreak{}CODE\_\allowbreak{}EFFORT\_\allowbreak{}LEVEL} above both. A subagent with no \texttt{effort} of its own runs at the session's level only when it is on the session's model. \texttt{/\allowbreak{}tasks} shows the model and effort each running subagent actually has \emph{(requires 2.1.242 or later)}, which is the check to make rather than the configuration to trust.

\FloatBarrier
\setcounter{section}{4}
\section{Limits and defaults}

\begin{itemize}
\item \textbf{Background is the default.} A subagent Claude spawns from the main conversation starts in the background. Background subagents run concurrently, surface permission prompts in your main session, carry a smaller built-in tool set, and return results in a later turn. A grant you give one of those prompts that lasts beyond the single call applies to the whole session, your main conversation included [27].
\item \textbf{Nesting is three layers deep} by default; at the limit the \texttt{Agent} tool is withheld. \texttt{CLAUDE\_\allowbreak{}CODE\_\allowbreak{}MAX\_\allowbreak{}SUBAGENT\_\allowbreak{}SPAWN\_\allowbreak{}DEPTH=\allowbreak{}1} disables nesting entirely [27].
\item \textbf{Twenty concurrent subagents} by default, then \texttt{Concurrent \allowbreak{}subagent li\allowbreak{}mit reached}. \texttt{CLAUDE\_\allowbreak{}CODE\_\allowbreak{}MAX\_\allowbreak{}CONCURRENT\_\allowbreak{}SUBAGENTS} changes it. A \texttt{/\allowbreak{}subtask} fork and a resumed subagent take a slot without being blocked by the limit, so the running count can exceed it, and sessions with \texttt{ultracode} active do not enforce it at all [27].
\item \textbf{Built-ins} — \texttt{Explore} and \texttt{Plan}, both read-only, and \texttt{General-\allowbreak{}purpose}. \texttt{Explore} and \texttt{Plan} skip \texttt{CLAUDE.\allowbreak{}md} and git status for speed; \texttt{Explore} runs on the main conversation's model, except that when that model is Fable on a Claude subscription, a Console account or an LLM gateway it runs on Opus. Deny one with \texttt{permissions.\allowbreak{}deny:\allowbreak{} ["Agent(Ex\allowbreak{}plore)"]}, naming each type you mean to block, or remove both with \texttt{CLAUDE\_\allowbreak{}CODE\_\allowbreak{}DISABLE\_\allowbreak{}EXPLORE\_\allowbreak{}PLAN\_\allowbreak{}AGENTS=\allowbreak{}1} \emph{(requires 2.1.198 or later)} [27].
\end{itemize}

These mechanisms differ mainly in what they inherit. Figure 5 compares the five topologies on that axis.

\begin{figure}[tbp]
\begin{codefig}{9.0}{10.6}
\cl{~~~SUBAGENT~(non-fork)~~~~~~~~~~FORK~~(~/subtask~)~~~~~~~~~TEAMMATE}
\cl{~~~\pmboxdrawuni{250C}\pmboxdrawuni{2500}\pmboxdrawuni{2500}\pmboxdrawuni{2500}\pmboxdrawuni{2500}\pmboxdrawuni{2500}\pmboxdrawuni{2500}\pmboxdrawuni{2500}\pmboxdrawuni{2500}\pmboxdrawuni{2500}\pmboxdrawuni{2500}\pmboxdrawuni{2500}\pmboxdrawuni{2500}\pmboxdrawuni{2500}\pmboxdrawuni{2500}\pmboxdrawuni{2510}~~~~~~~~~~~~~\pmboxdrawuni{250C}\pmboxdrawuni{2500}\pmboxdrawuni{2500}\pmboxdrawuni{2500}\pmboxdrawuni{2500}\pmboxdrawuni{2500}\pmboxdrawuni{2500}\pmboxdrawuni{2500}\pmboxdrawuni{2500}\pmboxdrawuni{2500}\pmboxdrawuni{2500}\pmboxdrawuni{2500}\pmboxdrawuni{2500}\pmboxdrawuni{2500}\pmboxdrawuni{2500}\pmboxdrawuni{2510}~~~~~~~~~~~\pmboxdrawuni{250C}\pmboxdrawuni{2500}\pmboxdrawuni{2500}\pmboxdrawuni{2500}\pmboxdrawuni{2500}\pmboxdrawuni{2500}\pmboxdrawuni{2500}\pmboxdrawuni{2500}\pmboxdrawuni{2500}\pmboxdrawuni{2500}\pmboxdrawuni{2500}\pmboxdrawuni{2500}\pmboxdrawuni{2500}\pmboxdrawuni{2500}\pmboxdrawuni{2500}\pmboxdrawuni{2510}}
\cl{~~~\pmboxdrawuni{2502}~own~context~~\pmboxdrawuni{2502}~~~~~~~~~~~~~\pmboxdrawuni{2502}~FULL~parent~~\pmboxdrawuni{2502}~~~~~~~~~~~\pmboxdrawuni{2502}~own~context~~\pmboxdrawuni{2502}}
\cl{~~~\pmboxdrawuni{2502}~own~prompt~~~\pmboxdrawuni{2502}~~~~~~~~~~~~~\pmboxdrawuni{2502}~conversation~\pmboxdrawuni{2502}~~~~~~~~~~~\pmboxdrawuni{2502}~own~session~~\pmboxdrawuni{2502}}
\cl{~~~\pmboxdrawuni{2502}~own~tools~~~~\pmboxdrawuni{2502}~~~~~~~~~~~~~\pmboxdrawuni{2502}~parent~tools~\pmboxdrawuni{2502}~~~~~~~~~~~\pmboxdrawuni{2502}~lead\textquotesingle{}s~perms~\pmboxdrawuni{2502}}
\cl{~~~\pmboxdrawuni{2502}~CLAUDE.md~✓~~\pmboxdrawuni{2502}~~~~~~~~~~~~~\pmboxdrawuni{2502}~same~model~~~\pmboxdrawuni{2502}~~~~~~~~~~~\pmboxdrawuni{2502}~CLAUDE.md~✓~~\pmboxdrawuni{2502}}
\cl{~~~\pmboxdrawuni{2502}~history~~~✗~~\pmboxdrawuni{2502}~~~~~~~~~~~~~\pmboxdrawuni{2502}~shared~cache~\pmboxdrawuni{2502}~~~~~~~~~~~\pmboxdrawuni{2502}~history~~~✗~~\pmboxdrawuni{2502}}
\cl{~~~\pmboxdrawuni{2514}\pmboxdrawuni{2500}\pmboxdrawuni{2500}\pmboxdrawuni{2500}\pmboxdrawuni{2500}\pmboxdrawuni{2500}\pmboxdrawuni{2500}\pmboxdrawuni{252C}\pmboxdrawuni{2500}\pmboxdrawuni{2500}\pmboxdrawuni{2500}\pmboxdrawuni{2500}\pmboxdrawuni{2500}\pmboxdrawuni{2500}\pmboxdrawuni{2500}\pmboxdrawuni{2518}~~~~~~~~~~~~~\pmboxdrawuni{2514}\pmboxdrawuni{2500}\pmboxdrawuni{2500}\pmboxdrawuni{2500}\pmboxdrawuni{2500}\pmboxdrawuni{2500}\pmboxdrawuni{2500}\pmboxdrawuni{252C}\pmboxdrawuni{2500}\pmboxdrawuni{2500}\pmboxdrawuni{2500}\pmboxdrawuni{2500}\pmboxdrawuni{2500}\pmboxdrawuni{2500}\pmboxdrawuni{2500}\pmboxdrawuni{2518}~~~~~~~~~~~\pmboxdrawuni{2514}\pmboxdrawuni{2500}\pmboxdrawuni{2500}\pmboxdrawuni{2500}\pmboxdrawuni{2500}\pmboxdrawuni{2500}\pmboxdrawuni{2500}\pmboxdrawuni{252C}\pmboxdrawuni{2500}\pmboxdrawuni{2500}\pmboxdrawuni{2500}\pmboxdrawuni{2500}\pmboxdrawuni{2500}\pmboxdrawuni{2500}\pmboxdrawuni{2500}\pmboxdrawuni{2518}}
\cl{~~~~~~~~~~\pmboxdrawuni{2502}~returns~a~result~~~~~~~~~~~\pmboxdrawuni{2502}~returns~a~result~~~~~~~~~\pmboxdrawuni{2502}~messages}
\cl{~~~~~~~~~~▼~~~~~~~~~~~~~~~~~~~~~~~~~~~~▼~~~~~~~~~~~~~~~~~~~~~~~~~~▼}
\cl{~~~~~~main~conversation~~~~~~~~~~~main~conversation~~~~~~~~peers~+~task~list}
\cl{}
\cl{~~~WORKFLOW~AGENT~~~~~~~~~~~~~~~~~~~~~~~~~~AGENT~TEAM~LEAD}
\cl{~~~\pmboxdrawuni{250C}\pmboxdrawuni{2500}\pmboxdrawuni{2500}\pmboxdrawuni{2500}\pmboxdrawuni{2500}\pmboxdrawuni{2500}\pmboxdrawuni{2500}\pmboxdrawuni{2500}\pmboxdrawuni{2500}\pmboxdrawuni{2500}\pmboxdrawuni{2500}\pmboxdrawuni{2500}\pmboxdrawuni{2500}\pmboxdrawuni{2500}\pmboxdrawuni{2500}\pmboxdrawuni{2500}\pmboxdrawuni{2500}\pmboxdrawuni{2500}\pmboxdrawuni{2500}\pmboxdrawuni{2500}\pmboxdrawuni{2500}\pmboxdrawuni{2500}\pmboxdrawuni{2500}\pmboxdrawuni{2500}\pmboxdrawuni{2500}\pmboxdrawuni{2500}\pmboxdrawuni{2500}\pmboxdrawuni{2510}~~~~~~~~~~~~\pmboxdrawuni{250C}\pmboxdrawuni{2500}\pmboxdrawuni{2500}\pmboxdrawuni{2500}\pmboxdrawuni{2500}\pmboxdrawuni{2500}\pmboxdrawuni{2500}\pmboxdrawuni{2500}\pmboxdrawuni{2500}\pmboxdrawuni{2500}\pmboxdrawuni{2500}\pmboxdrawuni{2500}\pmboxdrawuni{2500}\pmboxdrawuni{2500}\pmboxdrawuni{2500}\pmboxdrawuni{2500}\pmboxdrawuni{2500}\pmboxdrawuni{2500}\pmboxdrawuni{2500}\pmboxdrawuni{2500}\pmboxdrawuni{2500}\pmboxdrawuni{2500}\pmboxdrawuni{2500}\pmboxdrawuni{2500}\pmboxdrawuni{2500}\pmboxdrawuni{2500}\pmboxdrawuni{2500}\pmboxdrawuni{2510}}
\cl{~~~\pmboxdrawuni{2502}~spawned~by~a~script~~~~~~\pmboxdrawuni{2502}~~~~~~~~~~~~\pmboxdrawuni{2502}~your~session~~~~~~~~~~~~~~\pmboxdrawuni{2502}}
\cl{~~~\pmboxdrawuni{2502}~mode~by~subagent~rules~~~\pmboxdrawuni{2502}~~~~~~~~~~~~\pmboxdrawuni{2502}~teammate~plans~approved~~~\pmboxdrawuni{2502}}
\cl{~~~\pmboxdrawuni{2502}~result~→~script~variable~\pmboxdrawuni{2502}~~~~~~~~~~~~\pmboxdrawuni{2502}~unreviewed~by~default~~~~~\pmboxdrawuni{2502}}
\cl{~~~\pmboxdrawuni{2502}~up~to~16~at~once,~1,000~~\pmboxdrawuni{2502}~~~~~~~~~~~~\pmboxdrawuni{2502}~one~team~per~session~~~~~~\pmboxdrawuni{2502}}
\cl{~~~\pmboxdrawuni{2502}~per~run~~~~~~~~~~~~~~~~~~\pmboxdrawuni{2502}~~~~~~~~~~~~\pmboxdrawuni{2502}~no~nested~teams~~~~~~~~~~~\pmboxdrawuni{2502}}
\cl{~~~\pmboxdrawuni{2514}\pmboxdrawuni{2500}\pmboxdrawuni{2500}\pmboxdrawuni{2500}\pmboxdrawuni{2500}\pmboxdrawuni{2500}\pmboxdrawuni{2500}\pmboxdrawuni{2500}\pmboxdrawuni{2500}\pmboxdrawuni{2500}\pmboxdrawuni{2500}\pmboxdrawuni{2500}\pmboxdrawuni{2500}\pmboxdrawuni{2500}\pmboxdrawuni{2500}\pmboxdrawuni{2500}\pmboxdrawuni{2500}\pmboxdrawuni{2500}\pmboxdrawuni{2500}\pmboxdrawuni{2500}\pmboxdrawuni{2500}\pmboxdrawuni{2500}\pmboxdrawuni{2500}\pmboxdrawuni{2500}\pmboxdrawuni{2500}\pmboxdrawuni{2500}\pmboxdrawuni{2500}\pmboxdrawuni{2518}~~~~~~~~~~~~\pmboxdrawuni{2514}\pmboxdrawuni{2500}\pmboxdrawuni{2500}\pmboxdrawuni{2500}\pmboxdrawuni{2500}\pmboxdrawuni{2500}\pmboxdrawuni{2500}\pmboxdrawuni{2500}\pmboxdrawuni{2500}\pmboxdrawuni{2500}\pmboxdrawuni{2500}\pmboxdrawuni{2500}\pmboxdrawuni{2500}\pmboxdrawuni{2500}\pmboxdrawuni{2500}\pmboxdrawuni{2500}\pmboxdrawuni{2500}\pmboxdrawuni{2500}\pmboxdrawuni{2500}\pmboxdrawuni{2500}\pmboxdrawuni{2500}\pmboxdrawuni{2500}\pmboxdrawuni{2500}\pmboxdrawuni{2500}\pmboxdrawuni{2500}\pmboxdrawuni{2500}\pmboxdrawuni{2500}\pmboxdrawuni{2518}}
\cl{}
\cl{~~~Independence~for~review~purposes:~~subagent~>~teammate~>~workflow~agent~>~fork}
\cl{~~~A~fork~carries~the~reasoning~it~would~be~reviewing.~Fork~mode~is~on~by~DEFAULT,}
\cl{~~~so~Claude~may~start~one~on~its~own.~@-mention~the~reviewer.}
\end{codefig}
\figcaption{Figure 5 — Five delegation topologies and what each one inherits.}
\end{figure}

\FloatBarrier
\setcounter{section}{5}
\section{Skill or subagent?}

\begingroup
\def\tblrows{%
Apply a procedure in the current context & Skill \\
Independent judgement, fresh context & Subagent (non-fork) \\
Continue the current reasoning elsewhere without polluting context & Fork / \texttt{/\allowbreak{}subtask} \\
Parallel bounded analysis & Several subagents \\
Deterministic file or command action & A script or hook — not a model \\
Reusable procedure that coordinates reviewers & Skill that invokes subagents \\
Dozens to hundreds of agents with codified orchestration & Dynamic workflow (Chapter 22) \\
Peers that talk to each other and share a task list & Agent teams (Chapter 30) \\
}%
\def\tblbody{\begin{minipage}{\textwidth}\boxcaption{Table 15 — Choosing a delegation mechanism}
{\footnotesize\begin{tabular}{L{97.6mm}L{45.4mm}}
\toprule
\textbf{Need} & \textbf{Prefer} \\
\midrule
\tblrows
\bottomrule\end{tabular}}\end{minipage}}%
\begingroup
\def\sloppy{\tolerance 9999\emergencystretch 3em\hfuzz 200pt\vfuzz 200pt}%
\hbadness=10000\vbadness=10000\hfuzz=200pt\vfuzz=200pt
\global\setbox\tblbox=\hbox{\tblbody}%
\endgroup
\par\addvspace{2.6mm}
\ifdim\dimexpr\ht\tblbox+\dp\tblbox\relax>0.55\textheight
  \typeout{HANDBOOK-TABLE broken \the\dimexpr\ht\tblbox+\dp\tblbox\relax}%
  \tabcaption{Table 15 — Choosing a delegation mechanism}
  {\footnotesize\begin{longtable}{L{97.6mm}L{45.4mm}}
  \toprule
\textbf{Need} & \textbf{Prefer} \\
\midrule\endfirsthead
  \multicolumn{2}{@{}l@{}}{%
  \sffamily\footnotesize\itshape\color{inkgrey}Table 15 — Choosing a delegation mechanism \textemdash\ continued}\\[1.2mm]
  \toprule
\textbf{Need} & \textbf{Prefer} \\
\midrule\endhead
  \bottomrule\endfoot
  \bottomrule\endlastfoot
  \tblrows
  \end{longtable}}%
\else
  \typeout{HANDBOOK-TABLE atomic \the\dimexpr\ht\tblbox+\dp\tblbox\relax}%
  \noindent\tblbody
\fi
\par\addvspace{2.6mm}
\endgroup

\FloatBarrier
\setcounter{section}{6}
\section{Delegation brief}

Every subagent prompt should state the exact question; the files or sources it may use; the tools and actions allowed; the required return format; the success criteria; and when to stop and report uncertainty. Never delegate "fix everything".

\begingroup
\def\kprows{%
\item An ordinary subagent starts fresh; a fork inherits everything, and fork mode — on by default — lets Claude start a fork on its own. @-mention a reviewer to be sure it runs.
\item A parent in \texttt{accept\allowbreak{}Edits}, \texttt{bypass\allowbreak{}Permissions} or \texttt{auto} defeats a subagent's own \texttt{permission\allowbreak{}Mode}; \texttt{auto} is where an interactive session starts unless something changes it. Make a reviewer read-only with its \texttt{tools} allowlist.
\item Background is the default, nesting goes three layers deep, and twenty concurrent subagents is the default limit, not enforced under \texttt{ultracode}.
\item A misspelled frontmatter field is ignored silently; check the running subagent's model and effort in \texttt{/\allowbreak{}tasks}.
\item Delegate a bounded question with named sources, allowed tools, a return format and a stop condition — never 'fix everything'.
}%
\def\kpbody{\begin{minipage}{\textwidth}\subsection*{Key points}\begin{itemize}\kprows\end{itemize}\end{minipage}}%
\begingroup
\def\sloppy{\tolerance 9999\emergencystretch 3em\hfuzz 200pt\vfuzz 200pt}%
\hbadness=10000\vbadness=10000\hfuzz=200pt\vfuzz=200pt
\global\setbox\kpbox=\hbox{\kpbody}%
\endgroup
\par\addvspace{4.2mm}
\ifdim\dimexpr\ht\kpbox+\dp\kpbox\relax>0.30\textheight
  \typeout{HANDBOOK-KEYPOINTS broken \the\dimexpr\ht\kpbox+\dp\kpbox\relax}%
  \subsection*{Key points}
  \begin{itemize}\kprows\end{itemize}
\else
  \typeout{HANDBOOK-KEYPOINTS atomic \the\dimexpr\ht\kpbox+\dp\kpbox\relax}%
  \noindent\kpbody
\fi
\par\addvspace{1.4mm}
\endgroup

\FloatBarrier
\renewcommand{\chaptertitlelabel}{Part IV \textperiodcentered\ Chapter 17}
\setcounter{chapter}{16}
\chapter{Event-driven behaviour with hooks}

A hook runs a configured action at a lifecycle event. Hooks are how you make something happen \emph{because of an event} rather than because the model remembered a sentence. What is deterministic about a hook is narrower than the word suggests: the event is dispatched to a matching handler every time. Whether the handler then does its job is a separate question. A \texttt{prompt} or \texttt{agent} handler is a model judgement; a \texttt{command}, \texttt{http} or \texttt{mcp\_\allowbreak{}tool} handler can be missing, time out or print something unparseable; and on most events, unless you say otherwise, a hook that fails lets the action go ahead (Section 17.3).

\FloatBarrier
\setcounter{section}{0}
\section{Events}

Claude Code exposes a large event surface. The ones that carry most of the practical weight [83]:

\begin{itemize}
\item \texttt{Session\allowbreak{}Start}, \texttt{Session\allowbreak{}End} — matchable by start type (\texttt{startup}, \texttt{resume}, \texttt{clear}, \texttt{compact}, \texttt{fork}) and by end reason.
\item \texttt{User\allowbreak{}Prompt\allowbreak{}Submit} — before Claude processes a prompt; exit 2 rejects it.
\item \texttt{Pre\allowbreak{}Tool\allowbreak{}Use} — before a tool call; exit 2 blocks it.
\item \texttt{Permission\allowbreak{}Request} — when a call needs a permission decision. Return a JSON decision rather than exit 2.
\item \texttt{Permission\allowbreak{}Denied} — after auto mode denies a call; may set \texttt{retry}.
\item \texttt{Post\allowbreak{}Tool\allowbreak{}Use}, \texttt{Post\allowbreak{}Tool\allowbreak{}Use\allowbreak{}Failure}, \texttt{Post\allowbreak{}Tool\allowbreak{}Batch} — after the fact; cannot undo an external side effect.
\item \texttt{Stop}, \texttt{Stop\allowbreak{}Failure}, \texttt{Subagent\allowbreak{}Start}, \texttt{Subagent\allowbreak{}Stop}, \texttt{Teammate\allowbreak{}Idle} — turn and agent lifecycle; exit 2 on \texttt{Stop} prevents stopping.
\item \texttt{Task\allowbreak{}Created}, \texttt{Task\allowbreak{}Completed} — shared task list; exit 2 blocks the operation.
\item \texttt{Pre\allowbreak{}Model\allowbreak{}Switch}, \texttt{Post\allowbreak{}Model\allowbreak{}Switch} — around a change of model; the first can block the switch.
\item \texttt{Pre\allowbreak{}Compact}, \texttt{Post\allowbreak{}Compact}, \texttt{Instruction\allowbreak{}s\allowbreak{}Loaded}, \texttt{Config\allowbreak{}Change}, \texttt{Cwd\allowbreak{}Changed}, \texttt{Directory\allowbreak{}Added}, \texttt{File\allowbreak{}Changed}, \texttt{Worktree\allowbreak{}Create}, \texttt{Worktree\allowbreak{}Remove}, \texttt{Notificatio\allowbreak{}n}, \texttt{Message\allowbreak{}Display}, \texttt{Elicitation}, \texttt{Elicitation\allowbreak{}Result}.
\end{itemize}

\texttt{Config\allowbreak{}Change} deserves a specific mention: it lets an organisation audit or block settings changes made during a session [59].

\FloatBarrier
\setcounter{section}{1}
\section{Handler types}

Five handler types are available [83]: \texttt{command} (shell or exec form, with \texttt{async} and \texttt{async\allowbreak{}Rewake}), \texttt{http} (POST, with \texttt{allowed\allowbreak{}Env\allowbreak{}Vars} header interpolation), \texttt{mcp\_\allowbreak{}tool} (call a tool on an already-connected server), \texttt{prompt} (single-turn model evaluation, \texttt{\$ARGUMENTS} placeholder, 30-second default), and \texttt{agent} (spawns a subagent; experimental).

Default timeouts are 600 seconds for \texttt{command}, \texttt{http} and \texttt{mcp\_\allowbreak{}tool} — 30 seconds under \texttt{User\allowbreak{}Prompt\allowbreak{}Submit}, \texttt{Pre\allowbreak{}Model\allowbreak{}Switch} and \texttt{Post\allowbreak{}Model\allowbreak{}Switch}, 10 under \texttt{Message\allowbreak{}Display} — 30 for \texttt{prompt} and 60 for \texttt{agent}. \texttt{Session\allowbreak{}End} hooks share a 1.5-second budget by default [83].

\texttt{\$\{CLAUDE\_\allowbreak{}PROJECT\_\allowbreak{}DIR\}}, \texttt{\$\{CLAUDE\_\allowbreak{}PLUGIN\_\allowbreak{}ROOT\}} and \texttt{\$\{CLAUDE\_\allowbreak{}PLUGIN\_\allowbreak{}DATA\}} are substituted in hook commands; \texttt{\$\{CLAUDE\_\allowbreak{}PROJECT\_\allowbreak{}DIR\}} stays at the project root even inside a worktree [83].

\FloatBarrier
\setcounter{section}{2}
\section{A blocking example}

\begin{codeblock}{9.0}{10.6}
\cl{\{}
\cl{~~"hooks":~\{}
\cl{~~~~"PreToolUse":~[}
\cl{~~~~~~\{}
\cl{~~~~~~~~"matcher":~"Bash",}
\cl{~~~~~~~~"hooks":~[}
\cl{~~~~~~~~~~\{}
\cl{~~~~~~~~~~~~"type":~"command",}
\cl{~~~~~~~~~~~~"if":~"Bash(rm~*)",}
\cl{~~~~~~~~~~~~"command":~"bash",}
\cl{~~~~~~~~~~~~"args":~["\$\{CLAUDE\_PROJECT\_DIR\}/.claude/hooks/guard-rm.sh"],}
\cl{~~~~~~~~~~~~"timeout":~20,}
\cl{~~~~~~~~~~~~"onFailure":~"block",}
\cl{~~~~~~~~~~~~"statusMessage":~"Checking~destructive~command…"}
\cl{~~~~~~~~~~\}}
\cl{~~~~~~~~]}
\cl{~~~~~~\}}
\cl{~~~~]}
\cl{~~\}}
\cl{\}}
\end{codeblock}

\begin{codeblock}{8.0}{9.4}
\cl{\#!/usr/bin/env~bash}
\cl{\#~.claude/hooks/guard-rm.sh~—~deny~rm~-rf;~stay~silent~otherwise}
\cl{set~-euo~pipefail}
\cl{COMMAND=\$(jq~-r~\textquotesingle{}.tool\_input.command\textquotesingle{})}
\cl{if~[[~"\$COMMAND"~==~*\textquotesingle{}rm~-rf\textquotesingle{}*~]];~then}
\cl{~~jq~-n~\textquotesingle{}\{hookSpecificOutput:\{hookEventName:"PreToolUse",}
\cl{~~~~~~~~~~permissionDecision:"deny",}
\cl{~~~~~~~~~~permissionDecisionReason:"Destructive~rm~-rf~blocked~by~project~policy"\}\}\textquotesingle{}}
\cl{else}
\cl{~~exit~0}
\cl{fi}
\end{codeblock}

Exec form (\texttt{args} set) spawns \texttt{command} directly with no shell, so the path needs no quoting even when it contains a space. On native Windows exec form needs a real executable on \texttt{PATH}, not a \texttt{.\allowbreak{}sh} script, so the hook names \texttt{bash} and passes the script as its argument [83].

\textbf{What happens when the guard itself fails} is the part of this example that matters most. A hook has three outcomes, not two [83]. Exit 0 is success, and any JSON decision it printed is applied. Exit 2 is a blocking error. Anything else — any other exit code, a script that is not there, a timeout, output that will not parse — is a \emph{non-blocking} error: Claude Code shows a notice and carries out the action. One exception: with another exit code but schema-valid JSON on stdout, Claude Code ignores the code and the JSON fields decide. A policy hook with a mistyped path therefore lets everything through. \texttt{"on\allowbreak{}Failure":\allowbreak{} "block"} reverses that for \texttt{command} and \texttt{http} hooks \emph{(requires 2.1.295 or later)}: a failure then does what exit 2 does on that event. The \texttt{set -\allowbreak{}euo pipefai\allowbreak{}l} line is what makes the setting bite here — without it a missing \texttt{jq} leaves \texttt{COMMAND} empty, the test finds nothing, and the script reports success. The setting has no effect on \texttt{Stop}, \texttt{Subagent\allowbreak{}Stop}, \texttt{Task\allowbreak{}Completed} and \texttt{Teammate\allowbreak{}Idle} hooks, nor on background hooks that set \texttt{async} or \texttt{async\allowbreak{}Rewake}. On Windows without Git Bash the default differs again, and a failing shell-form \texttt{Pre\allowbreak{}Tool\allowbreak{}Use} or \texttt{Permission\allowbreak{}Request} hook can block with the field unset; set it explicitly rather than rely on either default.

\textbf{Hooks do not bypass permission rules.} Deny and ask rules are evaluated regardless of what a \texttt{Pre\allowbreak{}Tool\allowbreak{}Use} hook returns: a matching deny still blocks, and a matching ask still prompts, even when the hook returned \texttt{allow} [29]. Conversely, a hook that exits 2 stops the call \emph{before} permission rules are evaluated, so it overrides an allow rule. The documentation builds a pattern on that asymmetry — allow \texttt{Bash} broadly, then reject specific commands in a hook [29] — and this book does not recommend it as a control. The guard above matches the string \texttt{rm -\allowbreak{}rf}; it does not match \texttt{rm -\allowbreak{}fr}, \texttt{rm -\allowbreak{}r -\allowbreak{}f}, a command assembled from a variable, or the same deletion written in another language. A filter of that kind is defence in depth. The control is still a narrow allow list, a deny rule and a sandbox (Chapter 7), with the hook in front of them rather than instead of them.

\FloatBarrier
\setcounter{section}{3}
\section{Safer creation workflow}

\begin{enumerate}
\item Define the exact event and the narrowest matcher.
\item Write a standalone script with deterministic input and output.
\item Test it manually against benign fixtures.
\item Ask Claude to propose the configuration as a diff.
\item Confirm paths, quoting, timeout and exit behaviour.
\item Test in the practice repository, including the failure path.
\item Add failure logging and write down how to remove it.
\end{enumerate}

Inspect what is actually configured with \texttt{/\allowbreak{}hooks}, which shows event, matcher, type, source file and full command [37], [56].

\FloatBarrier
\setcounter{section}{4}
\section{Threat model}

\begin{itemize}
\item Can its inputs contain untrusted filenames, prompt text or tool output?
\item Does it build a shell command by string concatenation?
\item Can it approve a permission request? (\texttt{Permission\allowbreak{}Request} hooks can. Treat them as privileged code.)
\item What happens if it runs twice, or in parallel?
\item Does it send data over the network? If so, is the URL on \texttt{allowed\allowbreak{}Http\allowbreak{}Hook\allowbreak{}Urls}?
\item Does a failure block work, or silently allow it? By default it allows; \texttt{on\allowbreak{}Failure} decides, for the handler types and events it applies to.
\item Is a post-action check being mistaken for prevention?
\end{itemize}

Organisational controls: \texttt{disable\allowbreak{}All\allowbreak{}Hooks} turns hooks off subject to settings precedence, and set anywhere but managed settings it cannot disable managed policy hooks — set \emph{in} managed settings it disables those too; \texttt{allow\allowbreak{}Managed\allowbreak{}Hooks\allowbreak{}Only} in managed settings blocks user, project and local hooks and the hooks of plugins that managed settings do not force-enable; \texttt{allowed\allowbreak{}Http\allowbreak{}Hook\allowbreak{}Urls} restricts HTTP hook destinations from every source [83].

\begin{calloutbox}{palebrass}{brassdark}{2.0mm}
\textbf{CAUTION:} a natural-language rule is probabilistic; a hook is executable automation. That makes its dispatch more reliable and the hook correspondingly more dangerous. Note also that \texttt{/\allowbreak{}goal} (Chapter 21) is implemented as a session-scoped prompt Stop hook, so \texttt{disable\allowbreak{}All\allowbreak{}Hooks} or \texttt{allow\allowbreak{}Managed\allowbreak{}Hooks\allowbreak{}Only} disables it too [84].
\end{calloutbox}

\subsection*{Exercise}

Create a project-only \texttt{Stop} hook that appends a timestamped, non-sensitive completion record to a local file. It must not make a network call, approve a permission, or include conversation content. Test success, failure and removal.

\emph{A worked solution is given in Appendix M.}

\begingroup
\def\kprows{%
\item A hook is executable automation: its dispatch is more reliable than a written rule, and it is correspondingly more dangerous.
\item On most events a hook that fails lets the action go ahead unless \texttt{on\allowbreak{}Failure} is \texttt{"block"}, a setting some events and background hooks ignore; test the failure path, not only the block.
\item A hook exiting 2 blocks a call before permission rules are evaluated; no hook decision bypasses a matching deny rule. Only a mod, where the built-in guard is not loaded, can approve past either (Section 18.3).
\item \texttt{Permission\allowbreak{}Request} hooks can influence approval and must be treated as privileged code.
\item A post-action hook cannot undo an external side effect. Prevention happens at \texttt{Pre\allowbreak{}Tool\allowbreak{}Use} or not at all.
}%
\def\kpbody{\begin{minipage}{\textwidth}\subsection*{Key points}\begin{itemize}\kprows\end{itemize}\end{minipage}}%
\begingroup
\def\sloppy{\tolerance 9999\emergencystretch 3em\hfuzz 200pt\vfuzz 200pt}%
\hbadness=10000\vbadness=10000\hfuzz=200pt\vfuzz=200pt
\global\setbox\kpbox=\hbox{\kpbody}%
\endgroup
\par\addvspace{4.2mm}
\ifdim\dimexpr\ht\kpbox+\dp\kpbox\relax>0.30\textheight
  \typeout{HANDBOOK-KEYPOINTS broken \the\dimexpr\ht\kpbox+\dp\kpbox\relax}%
  \subsection*{Key points}
  \begin{itemize}\kprows\end{itemize}
\else
  \typeout{HANDBOOK-KEYPOINTS atomic \the\dimexpr\ht\kpbox+\dp\kpbox\relax}%
  \noindent\kpbody
\fi
\par\addvspace{1.4mm}
\endgroup

\FloatBarrier
\renewcommand{\chaptertitlelabel}{Part IV \textperiodcentered\ Chapter 18}
\setcounter{chapter}{17}
\chapter{Evaluate and install plugins}

A plugin is a distributable bundle. A marketplace is a catalogue pointing Claude Code at plugins; it is not an app-store security review.

The word \emph{install} hides four separate decisions, and a reader who makes one has not made the others:

\begin{enumerate}
\item \textbf{Trust a catalogue and its update path} — adding a marketplace, and whether it auto-updates.
\item \textbf{Acquire an artefact} — which files, at which version, arrive on disk.
\item \textbf{Enable it in a scope} — user, project or local; a plugin can be installed and off.
\item \textbf{Authorise what its components do} — the processes, prompts, credentials and remote systems they reach, which no install step reviews.
\end{enumerate}

\FloatBarrier
\setcounter{section}{0}
\section{What a plugin can contain}

This inventory is the reason plugins deserve a gate [39], [40]. It is the documented set at 2.1.322 and it is not exhaustive: the list has grown in every release this book has been checked against.

\begingroup
\def\tblrows{%
\texttt{.\allowbreak{}claude-\allowbreak{}plugin/\allowbreak{}plugin.\allowbreak{}json} & Manifest: name (the skill namespace), description, version, author. Optional; only \texttt{name} is required inside it \\
\texttt{skills/\allowbreak{}} & Skills, invoked as \texttt{/\allowbreak{}plugin-\allowbreak{}name:\allowbreak{}skill-\allowbreak{}name} \\
\texttt{commands/\allowbreak{}} & Legacy flat-file skills \\
\texttt{agents/\allowbreak{}} & Subagent definitions \\
\texttt{hooks/\allowbreak{}hooks.\allowbreak{}json} & Lifecycle event handlers of every kind Chapter 17 describes — shell command, HTTP request, prompt and the rest; it can also point at a \textbf{mod}, JavaScript that runs inside Claude Code (Section 18.3) \\
\texttt{.\allowbreak{}mcp.\allowbreak{}json} & MCP server configurations; a stdio server is a process started on your machine \\
\texttt{.\allowbreak{}lsp.\allowbreak{}json} & Language-server configurations \\
\texttt{monitors/\allowbreak{}monitors.\allowbreak{}json} & \textbf{Background monitors}: shell commands Claude Code starts on its own at session start, on reload, or the first time a named skill runs \\
\texttt{bin/\allowbreak{}} & \textbf{Executables added to the Bash tool's \texttt{PATH}} while the plugin is enabled. They come after your own \texttt{PATH} entries, so a plugin cannot shadow \texttt{git} or another system command \\
\texttt{settings.\allowbreak{}json} & Default settings applied when enabled. Two keys take effect: \texttt{subagent\allowbreak{}Status\allowbreak{}Line}, and \texttt{agent}, which runs one of the plugin's agents as the \textbf{main thread} with that agent's system prompt and model. Plugin defaults are the lowest settings layer, so your own \texttt{agent} setting overrides it \\
\texttt{output-\allowbreak{}styles/\allowbreak{}} · \texttt{themes/\allowbreak{}} & Output styles, which a plugin may force with \texttt{force-\allowbreak{}for-\allowbreak{}plugin} (Section 19.3), and colour themes \\
\texttt{workflows/\allowbreak{}} & Dynamic workflow definitions (Chapter 22) \\
}%
\def\tblbody{\begin{minipage}{\textwidth}\boxcaption{Table 16 — Plugin components, all at the plugin root}
{\footnotesize\begin{tabular}{L{53.3mm}L{89.7mm}}
\toprule
\textbf{Directory or file} & \textbf{Contents} \\
\midrule
\tblrows
\bottomrule\end{tabular}}\end{minipage}}%
\begingroup
\def\sloppy{\tolerance 9999\emergencystretch 3em\hfuzz 200pt\vfuzz 200pt}%
\hbadness=10000\vbadness=10000\hfuzz=200pt\vfuzz=200pt
\global\setbox\tblbox=\hbox{\tblbody}%
\endgroup
\par\addvspace{2.6mm}
\ifdim\dimexpr\ht\tblbox+\dp\tblbox\relax>0.55\textheight
  \typeout{HANDBOOK-TABLE broken \the\dimexpr\ht\tblbox+\dp\tblbox\relax}%
  \tabcaption{Table 16 — Plugin components, all at the plugin root}
  {\footnotesize\begin{longtable}{L{53.3mm}L{89.7mm}}
  \toprule
\textbf{Directory or file} & \textbf{Contents} \\
\midrule\endfirsthead
  \multicolumn{2}{@{}l@{}}{%
  \sffamily\footnotesize\itshape\color{inkgrey}Table 16 — Plugin components, all at the plugin root \textemdash\ continued}\\[1.2mm]
  \toprule
\textbf{Directory or file} & \textbf{Contents} \\
\midrule\endhead
  \bottomrule\endfoot
  \bottomrule\endlastfoot
  \tblrows
  \end{longtable}}%
\else
  \typeout{HANDBOOK-TABLE atomic \the\dimexpr\ht\tblbox+\dp\tblbox\relax}%
  \noindent\tblbody
\fi
\par\addvspace{2.6mm}
\endgroup

Read that table twice. A plugin can place executables on your \texttt{PATH}, start background processes, change the system prompt of your main session, run code inside Claude Code, and register hooks — all on enable. And installing a plugin also enables it, unless its author set \texttt{default\allowbreak{}Enabled:\allowbreak{} false} [40].

Two properties decide how much of that your existing controls see [40]:

\begin{itemize}
\item \textbf{Permission rules and the sandbox cover Claude's tool calls, not the code a plugin runs by itself.} A call to one of the plugin's MCP tools, and a Bash command that runs something from its \texttt{bin/\allowbreak{}}, are tool calls and your rules apply. Hooks, monitors, MCP servers, LSP servers and the processes a mod starts run outside the sandbox, with your full user permissions.
\item \textbf{Skills, commands and agents are instructions}, and enter Claude's context. They change what Claude does with the tools it already has.
\end{itemize}

\FloatBarrier
\setcounter{section}{1}
\section{Installation gate}

\begin{codeblock}{9.0}{10.6}
\cl{/plugin~marketplace~add~owner/repository}
\cl{/plugin~install~plugin-name@marketplace-name}
\cl{/reload-plugins}
\end{codeblock}

\begin{codeblock}{7.0}{8.3}
\cl{claude~plugin~validate~./your-plugin}
\cl{claude~-{}-plugin-dir~./my-plugin~~~~~~~~\#~load~a~directory~for~one~session;~also~accepts~a~.zip}
\cl{claude~-{}-plugin-dir~./my-plugin~plugin~details~my-plugin~~~\#~inventory,~without~starting~a~session}
\end{codeblock}

Before adding a marketplace or plugin:

\begin{enumerate}
\item Confirm the exact publisher and repository. \texttt{claude plug\allowbreak{}in marketpl\allowbreak{}ace list} prints the source each marketplace was added from.
\item Read the manifest and inventory every component against Table 16. The details pane in \texttt{/\allowbreak{}plugin} lists what will install; it shows that a hook exists, not what it runs.
\item Read \texttt{hooks/\allowbreak{}hooks.\allowbreak{}json}, \texttt{.\allowbreak{}mcp.\allowbreak{}json} and every file in \texttt{bin/\allowbreak{}}, and review install scripts, monitors and tool grants.
\item Search for network calls, credential reads, persistence and automatic updates.
\item Pin what you reviewed, by identity and not by label. A \texttt{version} string in the manifest is a label its author controls; the artefact you read is a commit SHA or an archive digest. When neither the manifest nor the marketplace entry sets \texttt{version}, Claude Code computes one from the source — the commit SHA for a Git source, the SHA-256 digest for an archive, each shortened to twelve characters — and a \texttt{command} source derives its version from whatever the command produced, and re-runs once per session regardless of the marketplace's auto-update setting [87].
\item Decide the update path. Auto-update is on by default for Anthropic's official marketplaces, two of them excepted, and for a marketplace added from claude.ai; it is off for every other, the community marketplace included. Where it is on, \textbf{the files you reviewed can change on disk in the background} [40], [87].
\item Test in a disposable environment. \texttt{-\allowbreak{}-\allowbreak{}plugin-\allowbreak{}dir} is a way to load a directory for one session; it is not isolation, and the plugin runs with your permissions exactly as an installed one does.
\item Measure context and usage impact.
\item Record owner, purpose, version, artefact identity, approval date and removal procedure.
\end{enumerate}

\texttt{claude plug\allowbreak{}in validate} checks that manifests and frontmatter are well formed. It says nothing about what the plugin does when it runs, with one exception: for a mod it lists the events handled and the calls made (Section 18.3). A pre-install inventory is likewise necessary and not sufficient. It lists declared components; it does not prove the absence of undeclared helper code or of dependencies fetched at install.

Anthropic publishes two general-purpose marketplaces: \texttt{claude-\allowbreak{}plugins-\allowbreak{}official}, holding plugins Anthropic maintains together with plugins from partners and other authors, and added automatically the first time you start an interactive terminal session; and \texttt{claude-\allowbreak{}community}, holding third-party plugins their authors submitted, which you add yourself [88]. Claude Code accepts the official and community names only for marketplaces sourced from Anthropic's own GitHub organisation, so a third party cannot present its catalogue as one of them, and where the community catalogue pins a plugin to a commit SHA — which it does for nearly every entry — Claude Code refuses to install a different commit [40]. A marketplace's name tells you who publishes the catalogue, not what each plugin in it does. Neither is a warranty.

\FloatBarrier
\setcounter{section}{2}
\section{Mods: code that runs inside Claude Code}

A mod is a plugin made of JavaScript or TypeScript event handlers that Claude Code calls \emph{in its own process} when something happens: a tool call, a submitted prompt, a part of the interface being drawn [89]. Mods are on by default \emph{(requires 2.1.287 or later in the terminal, 2.1.286 in the desktop app)}. Everything else in this part works from outside Claude Code — a hook runs a script, a skill gives Claude text, an MCP server gives it tools. A mod is inside, and that changes what it can do.

Once it loads, a mod can [89]:

\begin{itemize}
\item \textbf{act on your machine as you} — read and write files anywhere your account can, start programs, make network requests;
\item \textbf{read your secrets} — environment variables and settings files, including an API key kept in either;
\item \textbf{see and change your session} — every prompt and every tool call; it can rewrite either, submit a prompt as if you had typed it, or message another of your sessions;
\item \textbf{approve a tool call before you are asked};
\item \textbf{spend your usage}, by calling a model on your plan or key.
\end{itemize}

Mods are not sandboxed. With sandboxing on, the sandbox isolates the Bash commands Claude runs; a process a mod starts runs outside it.

\begin{calloutbox}{palebrass}{brassdark}{2.0mm}
\textbf{CAUTION — on an unmanaged machine, a mod can approve what a deny rule refuses.} This book describes deny rules as enforced by the product and not widened from below (Section 7.2). That holds \textbf{where a built-in guard is loaded}, and the guard loads only when the machine has managed settings or a \texttt{managed-\allowbreak{}mcp.\allowbreak{}json} file, or the session runs on a Team or Enterprise \texttt{/\allowbreak{}login} [29], [90]. Without one of those — a personal machine, a session on an API key, or one on a long-lived \texttt{CLAUDE\_\allowbreak{}CODE\_\allowbreak{}OAUTH\_\allowbreak{}TOKEN} even where a Team login is stored — a mod that handles tool approval can approve a call that a deny rule refuses or that a \texttt{Pre\allowbreak{}Tool\allowbreak{}Use} hook blocked, and can skip the prompt an ask rule would have shown. In a local terminal session, \texttt{cc-\allowbreak{}plugin-\allowbreak{}sec-\allowbreak{}default} on the \textbf{Installed} tab of \texttt{/\allowbreak{}plugin} means the guard is loaded. If it is not there, your deny rules bind Claude and do not bind a mod.
\end{calloutbox}

With the guard loaded, the defaults are these [90]:

\begin{itemize}
\item A user's mod cannot approve a call that a deny rule refuses, that an ask rule prompts for, or that a \texttt{Pre\allowbreak{}Tool\allowbreak{}Use} hook blocked, unless an administrator sets \texttt{allow\allowbreak{}Mods\allowbreak{}To\allowbreak{}Override\allowbreak{}Deny\allowbreak{}Rules}.
\item A user's mod cannot change what managed hooks receive or decide, the system prompt, managed instructions, or the tools of managed MCP servers.
\item A \texttt{Pre\allowbreak{}Tool\allowbreak{}Use} hook in managed settings runs before any mod, and runs again on a call a mod has rewritten.
\item The guard fails closed: if it cannot read managed settings it refuses every user's mod.
\end{itemize}

Two limits remain even then, and both matter to a security review. \textbf{Permission rules and \texttt{Pre\allowbreak{}Tool\allowbreak{}Use} hooks apply to Claude's tool calls, not to a mod's own file and process calls.} With \texttt{Read(.\allowbreak{}env)} denied, a mod can still read that file itself, or start a program that does. And in auto mode a call a mod approves runs without the classifier's check, except in a session with the guard loaded that uses \texttt{-\allowbreak{}-\allowbreak{}permission-\allowbreak{}prompt-\allowbreak{}tool} or an Agent SDK permission callback, where the classifier checks the call first. A mod cannot restyle or alter the permission prompt, in any setup.

For an organisation the controls are, from narrowest to widest [90]:

\begingroup
\def\tblrows{%
\texttt{allow\allowbreak{}Managed\allowbreak{}Mods\allowbreak{}Only}, an option on the guard & No mod a user brings runs its hooks — installed, loaded with \texttt{-\allowbreak{}-\allowbreak{}plugin-\allowbreak{}dir}, or written by Claude during a session. Users' settings hooks, status lines and \texttt{/\allowbreak{}goal} keep working \\
\texttt{disable\allowbreak{}Sideload\allowbreak{}Flags} & Rejects \texttt{-\allowbreak{}-\allowbreak{}plugin-\allowbreak{}dir} and \texttt{-\allowbreak{}-\allowbreak{}plugin-\allowbreak{}url}, and keeps mods Claude writes in a session from loading (Section 31.2) \\
\texttt{allow\allowbreak{}Managed\allowbreak{}Hooks\allowbreak{}Only} & Wider: only the organisation's mods and built-in ones load, and users' settings hooks are blocked too \\
\texttt{disable\allowbreak{}All\allowbreak{}Hooks} in managed settings & Widest: stops every installed mod and every settings hook, \textbf{the organisation's own managed hooks included} \\
}%
\def\tblbody{\begin{minipage}{\textwidth}\boxcaption{Table 17 — Managed controls over mods}
{\footnotesize\begin{tabular}{L{43.0mm}L{100.0mm}}
\toprule
\textbf{Setting} & \textbf{Effect} \\
\midrule
\tblrows
\bottomrule\end{tabular}}\end{minipage}}%
\begingroup
\def\sloppy{\tolerance 9999\emergencystretch 3em\hfuzz 200pt\vfuzz 200pt}%
\hbadness=10000\vbadness=10000\hfuzz=200pt\vfuzz=200pt
\global\setbox\tblbox=\hbox{\tblbody}%
\endgroup
\par\addvspace{2.6mm}
\ifdim\dimexpr\ht\tblbox+\dp\tblbox\relax>0.55\textheight
  \typeout{HANDBOOK-TABLE broken \the\dimexpr\ht\tblbox+\dp\tblbox\relax}%
  \tabcaption{Table 17 — Managed controls over mods}
  {\footnotesize\begin{longtable}{L{43.0mm}L{100.0mm}}
  \toprule
\textbf{Setting} & \textbf{Effect} \\
\midrule\endfirsthead
  \multicolumn{2}{@{}l@{}}{%
  \sffamily\footnotesize\itshape\color{inkgrey}Table 17 — Managed controls over mods \textemdash\ continued}\\[1.2mm]
  \toprule
\textbf{Setting} & \textbf{Effect} \\
\midrule\endhead
  \bottomrule\endfoot
  \bottomrule\endlastfoot
  \tblrows
  \end{longtable}}%
\else
  \typeout{HANDBOOK-TABLE atomic \the\dimexpr\ht\tblbox+\dp\tblbox\relax}%
  \noindent\tblbody
\fi
\par\addvspace{2.6mm}
\endgroup

A mod Claude Code copies from a GitHub, git, URL or npm source counts as a user's, not as the organisation's; to count as the organisation's it must sit in a directory the administrator places on each machine. Some of Claude Code's own features are built-in mods — \texttt{AGENTS.\allowbreak{}md} support and the \texttt{/\allowbreak{}diff} pane among them — and \texttt{disable\allowbreak{}All\allowbreak{}Hooks}, \texttt{-\allowbreak{}-\allowbreak{}bare} and \texttt{-\allowbreak{}-\allowbreak{}safe-\allowbreak{}mode} do not stop those [89].

Before installing a mod, list what it does without running it [89]:

\begin{codeblock}{7.0}{8.3}
\cl{claude~plugin~validate~./some-mod~~~~~~\#~the~hooks:~and~calls:~lines~list~events~handled~and~calls~made}
\end{codeblock}

A mod has no way to reach a file, a process or the network except through the mods API, which is why that listing is possible. Read it for file, process and network calls, and for a handler on tool approval. To stop every installed mod for one session, start Claude Code with \texttt{-\allowbreak{}-\allowbreak{}safe-\allowbreak{}mode}.

\FloatBarrier
\setcounter{section}{3}
\section{Removal is a procedure}

Uninstalling removes less than the word suggests [40]:

\begin{enumerate}
\item Run \texttt{claude plug\allowbreak{}in uninstal\allowbreak{}l <plugin>} with the \texttt{-\allowbreak{}-\allowbreak{}scope} you installed it at. The shell command acts on one scope, so check the others. From the \textbf{Installed} tab or with \texttt{/\allowbreak{}plugin unin\allowbreak{}stall} in a session, a plugin that the repository's \texttt{.\allowbreak{}claude/\allowbreak{}settings.\allowbreak{}json} enables brings a question instead: disable it for you alone, or uninstall it for everyone.
\item Reload or restart, and confirm in \texttt{/\allowbreak{}plugin} that it is gone and that no forced output style or main-thread agent remains in effect.
\item Restart the session, so that monitors and server processes it started are gone.
\item Delete what stays on disk. The plugin's cached files remain under \texttt{\textasciitilde{}/\allowbreak{}.\allowbreak{}claude/\allowbreak{}plugins/\allowbreak{}cache/\allowbreak{}} for 14 days before a background sweep removes them — and after you uninstall your last plugin, until you install another. By default its persistent data directory is deleted when that was the last scope it was installed at, and kept with \texttt{-\allowbreak{}-\allowbreak{}keep-\allowbreak{}data}.
\item If you do not trust the marketplace's owner either, remove the marketplace, which uninstalls every plugin you installed from it.
\item Revoke what it was given: credentials, OAuth grants, and any integration created at the far end.
\item Run a negative test — invoke something the plugin provided and confirm that it is not found.
\end{enumerate}

\FloatBarrier
\setcounter{section}{4}
\section{Build instead of bundle}

If you need one procedure from a large plugin, writing a small local skill using the idea is usually safer and simpler than importing opaque code and unused connectors. Preserve licences and attribution when adapting third-party material.

\begin{calloutbox}{palebrass}{brassdark}{2.0mm}
\textbf{CAUTION:} official documentation states that a plugin you install can execute arbitrary code on your machine with your user privileges [40]. "Installed" means "trusted to run", not "visible in a menu". See M18 in Table 1.
\end{calloutbox}

\begingroup
\def\kprows{%
\item A plugin may ship executables placed on the shell path, background monitors, hooks, MCP and LSP servers, code that runs inside Claude Code, and a settings key that replaces the main session's agent.
\item 'Installed' means 'trusted to run', not 'visible in a menu'. Permission rules and the sandbox cover Claude's tool calls, not the code a plugin runs by itself.
\item A mod runs inside Claude Code as you. By default, without managed settings, a \texttt{managed-\allowbreak{}mcp.\allowbreak{}json} file or a Team or Enterprise login as the session's credential, a mod can approve a call that a deny rule refuses.
\item Pin the artefact you reviewed by commit or digest, not by its version label, and know whether its marketplace auto-updates.
\item Removal is a procedure: check every scope, stop what was started, delete the cache, revoke what was granted, and test that it is gone.
}%
\def\kpbody{\begin{minipage}{\textwidth}\subsection*{Key points}\begin{itemize}\kprows\end{itemize}\end{minipage}}%
\begingroup
\def\sloppy{\tolerance 9999\emergencystretch 3em\hfuzz 200pt\vfuzz 200pt}%
\hbadness=10000\vbadness=10000\hfuzz=200pt\vfuzz=200pt
\global\setbox\kpbox=\hbox{\kpbody}%
\endgroup
\par\addvspace{4.2mm}
\ifdim\dimexpr\ht\kpbox+\dp\kpbox\relax>0.30\textheight
  \typeout{HANDBOOK-KEYPOINTS broken \the\dimexpr\ht\kpbox+\dp\kpbox\relax}%
  \subsection*{Key points}
  \begin{itemize}\kprows\end{itemize}
\else
  \typeout{HANDBOOK-KEYPOINTS atomic \the\dimexpr\ht\kpbox+\dp\kpbox\relax}%
  \noindent\kpbody
\fi
\par\addvspace{1.4mm}
\endgroup

\FloatBarrier
\renewcommand{\chaptertitlelabel}{Part IV \textperiodcentered\ Chapter 19}
\setcounter{chapter}{18}
\chapter{Output styles: Claude Code beyond software engineering}

An output style is a set of instructions that sets Claude's role, tone and response format for every response in a session [91]. It is the named, persistent, switchable mechanism for register and persona, and a custom style leaves out the section of built-in software-engineering instructions that Claude Code's full system prompt carries. For a reader using Claude Code for research, operations, analysis or writing, this is the most direct fit between the product and the task, and it was absent from the first edition entirely.

It is prompt composition and nothing more. A style does not create an execution identity, a tool boundary, a permission mode, a credential scope, a sandbox or an independent reviewer. The documentation is explicit that a style gives Claude instructions to follow and does not guarantee that something always or never happens [91]; what must hold belongs in a hook (Chapter 17).

\FloatBarrier
\setcounter{section}{0}
\section{The built-in styles}

\begingroup
\def\tblrows{%
Default & No style selected: Claude Code's standard instructions, written for software engineering \\
Proactive & Starts work at once, makes reasonable assumptions about routine decisions rather than asking, and does not switch to plan mode unless asked \\
Concise & Leads with the result, skips preamble and narration, keeps responses short — while always preserving error reports, security warnings and destructive-action confirmations in full (2.1.237+) \\
Explanatory & Adds short \texttt{Insight} blocks in the conversation explaining the choices behind the work \\
Learning & The same \texttt{Insight} blocks, and leaves pieces with a real design decision for you to write, marked \texttt{TODO(human)} \\
}%
\def\tblbody{\begin{minipage}{\textwidth}\boxcaption{Table 18 — Built-in output styles}
{\footnotesize\begin{tabular}{H{22.5mm}L{120.5mm}}
\toprule
\textbf{Style} & \textbf{Behaviour} \\
\midrule
\tblrows
\bottomrule\end{tabular}}\end{minipage}}%
\begingroup
\def\sloppy{\tolerance 9999\emergencystretch 3em\hfuzz 200pt\vfuzz 200pt}%
\hbadness=10000\vbadness=10000\hfuzz=200pt\vfuzz=200pt
\global\setbox\tblbox=\hbox{\tblbody}%
\endgroup
\par\addvspace{2.6mm}
\ifdim\dimexpr\ht\tblbox+\dp\tblbox\relax>0.55\textheight
  \typeout{HANDBOOK-TABLE broken \the\dimexpr\ht\tblbox+\dp\tblbox\relax}%
  \tabcaption{Table 18 — Built-in output styles}
  {\footnotesize\begin{longtable}{H{22.5mm}L{120.5mm}}
  \toprule
\textbf{Style} & \textbf{Behaviour} \\
\midrule\endfirsthead
  \multicolumn{2}{@{}l@{}}{%
  \sffamily\footnotesize\itshape\color{inkgrey}Table 18 — Built-in output styles \textemdash\ continued}\\[1.2mm]
  \toprule
\textbf{Style} & \textbf{Behaviour} \\
\midrule\endhead
  \bottomrule\endfoot
  \bottomrule\endlastfoot
  \tblrows
  \end{longtable}}%
\else
  \typeout{HANDBOOK-TABLE atomic \the\dimexpr\ht\tblbox+\dp\tblbox\relax}%
  \noindent\tblbody
\fi
\par\addvspace{2.6mm}
\endgroup

Each of the four keeps the standard software-engineering instructions and adds its own; only a custom style leaves that section out [91].

\begin{calloutbox}{palebrass}{brassdark}{2.0mm}
\textbf{CAUTION — two separate axes.} Proactive changes what Claude \emph{tends to attempt}; a permission mode changes how a tool call is \emph{authorised}. Switching to Proactive does not change your permission mode, and permission prompts appear exactly as before [91]. The documentation presents Proactive as the way to get stronger autonomous behaviour in a mode that still prompts you [33], so Proactive with Manual mode is a reasonable combination and Proactive with \texttt{bypass\allowbreak{}Permissions} is not. The style also tells Claude to check with you before an action that deletes data or changes a shared or production system — and that check is an instruction Claude follows, not a permission prompt.
\end{calloutbox}

\FloatBarrier
\setcounter{section}{1}
\section{Setting and creating a style}

Run \texttt{/\allowbreak{}output-\allowbreak{}style <styl\allowbreak{}e>}, or \texttt{/\allowbreak{}config} and select \textbf{Output style}; either saves the choice as \texttt{output\allowbreak{}Style} in \texttt{.\allowbreak{}claude/\allowbreak{}settings.\allowbreak{}local.\allowbreak{}json} [91]. The command, withdrawn in an earlier release, is back and also works in non-interactive and Agent SDK sessions \emph{(requires 2.1.269 or later)}. Three behaviours of that setting differ from what Version 1.0 of this book described:

\begin{itemize}
\item \textbf{A switch applies from your next message} \emph{(requires 2.1.251 or later)}. Before that release it applied only after \texttt{/\allowbreak{}clear} or in a new session.
\item \textbf{A switch does not discard the prompt cache.} Claude Code delivers the new style's instructions as a message in the conversation, so the request still reads the system prompt and the earlier conversation from the cache [68].
\item \textbf{The value is case-sensitive in a settings file.} \texttt{Explanatory} selects the style; \texttt{explanatory} silently gives you Default. The command ignores case.
\end{itemize}

A custom style is a Markdown file in \texttt{\textasciitilde{}/\allowbreak{}.\allowbreak{}claude/\allowbreak{}output-\allowbreak{}styles}, the project's \texttt{.\allowbreak{}claude/\allowbreak{}output-\allowbreak{}styles}, or \texttt{.\allowbreak{}claude/\allowbreak{}output-\allowbreak{}styles} inside the managed settings directory:

\begin{codeblock}{7.0}{8.3}
\cl{-{}-{}-}
\cl{name:~Evidence~analyst}
\cl{description:~Research~and~analysis~register~with~explicit~provenance}
\cl{keep-coding-instructions:~false}
\cl{-{}-{}-}
\cl{}
\cl{You~are~a~research~analyst~producing~decision-grade~written~work.~You~are~not~writing~software.}
\cl{}
\cl{-~Lead~with~the~answer~to~the~decision~question,~then~the~evidence.}
\cl{-~Distinguish~observed~fact,~source~claim,~and~your~own~inference,~explicitly,~every~time.}
\cl{-~Never~assert~a~figure,~date~or~attribution~without~naming~where~it~came~from.}
\cl{-~State~what~you~could~not~verify~as~a~numbered~list~of~open~items.~Do~not~smooth~over~a~gap.}
\cl{-~Use~British~English~and~the~register~of~a~professional~briefing~note.}
\end{codeblock}

Set \texttt{keep-\allowbreak{}coding-\allowbreak{}instruction\allowbreak{}s:\allowbreak{} true} when you are changing how Claude communicates but still want it engineering the same way; leave it out when the work is not software at all [91]. The field acts only on the full system prompt. The shorter system prompt does not include that section in the first place, so there the field has no effect; set \texttt{CLAUDE\_\allowbreak{}CODE\_\allowbreak{}SIMPLE\_\allowbreak{}SYSTEM\_\allowbreak{}PROMPT=\allowbreak{}0} to select the full prompt if you rely on it. A misspelled field is ignored without an error, and in the terminal Claude Code reads style files when it starts, so a style created or edited during a session needs a restart to be picked up.

A style like this one improves how claims are \emph{presented}. It does not make them true. The provenance it asks for still has to be inspectable, and still has to be checked by something other than the session that wrote it (Chapter 34).

\FloatBarrier
\setcounter{section}{2}
\section{Where a style does and does not apply}

An output style applies to the main conversation and to a fork, which inherits the parent's whole conversation and system prompt. Every other subagent runs its own system prompt and is unaffected [27], [91]. If you rely on a style for tone or discipline in delegated work, you must put the same instruction in the agent definition.

Your selection is not the last word either. A plugin can ship a style marked \texttt{force-\allowbreak{}for-\allowbreak{}plugin:\allowbreak{} true}, which applies whenever the plugin is enabled and \textbf{overrides your \texttt{output\allowbreak{}Style} setting}; where several enabled plugins do so, the first one loaded wins [91]. That is one more reason to read what a plugin contains before enabling it (Chapter 18).

Compare the mechanisms: an output style sets role and register for the session; \texttt{CLAUDE.\allowbreak{}md} adds what Claude should know about the project and stays loaded whichever style is active; \texttt{-\allowbreak{}-\allowbreak{}append-\allowbreak{}system-\allowbreak{}prompt} appends to the system prompt for one invocation and \texttt{-\allowbreak{}-\allowbreak{}system-\allowbreak{}prompt} replaces it; an agent runs a separate prompt entirely; a skill loads task-specific instructions on demand [55], [91].

\begingroup
\def\kprows{%
\item An output style is prompt composition: role, tone and format for the session. It creates no boundary, permission or independent check.
\item Proactive changes what Claude attempts; the permission mode changes what is authorised. Neither changes the other.
\item A style switch applies from the next message and keeps the prompt cache; in a settings file the name is case-sensitive.
\item A style applies to the main conversation and to a fork; every other subagent runs its own system prompt.
\item A plugin style marked \texttt{force-\allowbreak{}for-\allowbreak{}plugin} overrides your own selection.
}%
\def\kpbody{\begin{minipage}{\textwidth}\subsection*{Key points}\begin{itemize}\kprows\end{itemize}\end{minipage}}%
\begingroup
\def\sloppy{\tolerance 9999\emergencystretch 3em\hfuzz 200pt\vfuzz 200pt}%
\hbadness=10000\vbadness=10000\hfuzz=200pt\vfuzz=200pt
\global\setbox\kpbox=\hbox{\kpbody}%
\endgroup
\par\addvspace{4.2mm}
\ifdim\dimexpr\ht\kpbox+\dp\kpbox\relax>0.30\textheight
  \typeout{HANDBOOK-KEYPOINTS broken \the\dimexpr\ht\kpbox+\dp\kpbox\relax}%
  \subsection*{Key points}
  \begin{itemize}\kprows\end{itemize}
\else
  \typeout{HANDBOOK-KEYPOINTS atomic \the\dimexpr\ht\kpbox+\dp\kpbox\relax}%
  \noindent\kpbody
\fi
\par\addvspace{1.4mm}
\endgroup

\breakrule

\FloatBarrier
\parttitle{Part V}{Browser use, autonomy and orchestration}
\addcontentsline{toc}{part}{Part V \textemdash\ Browser use, autonomy and orchestration}

\FloatBarrier
\renewcommand{\chaptertitlelabel}{Part V \textperiodcentered\ Chapter 20}
\setcounter{chapter}{19}
\chapter{Use Claude in Chrome safely}

Claude in Chrome lets Claude inspect and interact with browser tabs. Claude opens new tabs for browser tasks, collected into a tab group tied to the session, and \textbf{shares your browser's login state}, so it can reach any site you are already signed into [21]. That is the feature and the risk in one sentence. The actions run in a visible window, and at a login page or a CAPTCHA Claude pauses and asks you to handle it [21].

\FloatBarrier
\setcounter{section}{0}
\section{Prerequisites and connection}

You need Chrome, Edge, or another Chromium browser (Brave, Arc, Vivaldi, Opera are detected); the Claude in Chrome extension version 1.0.36 or later; and a direct Anthropic plan — Pro, Max, Ultra, Team or Enterprise [21]. Three constraints catch people out: \textbf{WSL is not supported}; \textbf{API-key or long-lived-token sessions cannot use Chrome integration at all}, even with \texttt{-\allowbreak{}-\allowbreak{}chrome}, because the extension cannot authenticate with those credentials; and it is not available through Amazon Bedrock, Google Cloud's Agent Platform or Microsoft Foundry [21]. Sign in with \texttt{/\allowbreak{}login}. If you belong to more than one organisation, the extension must be signed in to the same one as Claude Code, or the browser tools report that the extension is not connected [21].

In an Enterprise organisation with HIPAA enabled the integration is off until an Owner turns it on, and the Business Associate Agreement does not cover data sent to third-party sites through it [21].

\begin{codeblock}{9.0}{10.6}
\cl{claude~-{}-chrome}
\end{codeblock}

\begin{codeblock}{9.0}{10.6}
\cl{/chrome}
\end{codeblock}

\texttt{/\allowbreak{}chrome} checks connection status, manages permissions, reconnects the extension, and selects among connected browsers. The integration is working when the panel shows \emph{Status: Enabled} and \emph{Extension: Installed}.

\textbf{PRACTITIONER NOTE:} enabling Chrome by default loads browser tools into every session and increases context usage [21]. Prefer \texttt{-\allowbreak{}-\allowbreak{}chrome} when you need it.

A repository cannot switch the integration on for you. A project's \texttt{.\allowbreak{}claude/\allowbreak{}settings.\allowbreak{}json} or \texttt{.\allowbreak{}claude/\allowbreak{}settings.\allowbreak{}local.\allowbreak{}json} that sets \texttt{CLAUDE\_\allowbreak{}CODE\_\allowbreak{}ENABLE\_\allowbreak{}CFC} is ignored with a warning, because a checked-out repository must not be able to connect Claude to your browser; only \texttt{-\allowbreak{}-\allowbreak{}chrome}, \texttt{/\allowbreak{}chrome}, your user settings or \texttt{CLAUDE\_\allowbreak{}CODE\_\allowbreak{}ENABLE\_\allowbreak{}CFC=\allowbreak{}1} in your own environment turn it on \emph{(requires 2.1.290 or later)} [21], [57], [75].

\FloatBarrier
\setcounter{section}{1}
\section{Use a dedicated browser profile}

Create a profile for agent work with only the accounts the task requires; no personal password manager unlocked; no unrelated tabs; restricted site permissions; test rather than production accounts where possible; and downloads directed to a controlled folder. An incognito window is not account isolation if the extension and login state are still available.

Site-level permissions are inherited from the Chrome extension and managed in its settings [21]. They are not the gate in every mode. In auto mode, when the classifier itself approves a browser call to a site, the extension skips its own per-site check for that call, unless your permission rules deny some site to Claude in Chrome \emph{(requires 2.1.275 or later)} [21], [57]. Auto mode is what a session starts in when nothing configures a mode (Section 7.1), so on an unconfigured machine the classifier, not the extension's site list, decides most browser calls. By default it blocks browser actions that could send page content, cookies or credentials off-origin, and allows navigation to a trusted internal domain, to localhost, or to a URL you named [33]. One surface differs: a VS Code session that connected to the browser at start because \textbf{Enabled by default} is on asks before actions on a site you have not allowed in Manual, Edit automatically, Auto and Bypass permissions modes, and in auto and bypass until you type \texttt{@browser} in that session [21]. A site you want refused in every mode belongs in a deny rule, not only in the extension's settings.

\FloatBarrier
\setcounter{section}{2}
\section{Escalate capability in stages}

\begin{enumerate}
\item \textbf{Observe} — "Describe this page; take no action."
\item \textbf{Navigate} — "Open the settings page; change nothing."
\item \textbf{Draft} — "Fill the form but do not submit."
\item \textbf{Verify} — "Show the entered values and the validation state."
\item \textbf{Act} — a human approves the final submission.
\end{enumerate}

The staging is yours to enforce; no mode enforces it for you. Version 1.0 of this book said Plan mode did, with read-only calls running and clicks, typing and navigation prompting. The page no longer says that. It now documents a prompt in Plan mode before Claude records a GIF or opens a new tab, and for a \texttt{tabs\_\allowbreak{}context\_\allowbreak{}mcp} call that sets \texttt{create\allowbreak{}If\allowbreak{}Empty} or a \texttt{browser\_\allowbreak{}batch} that includes either; and in an interactive CLI session with bypass permissions available and feature-flag fetching off, even those run without a prompt [21]. Whether a click, typing or a navigation in Plan mode prompts is \textbf{UNVERIFIED}: the current page is silent and the earlier statement has been withdrawn, so Appendix J gives the check. Outside Plan mode, on a site you have allowed and with auto mode unavailable, screenshots and page reads run without a prompt while typing, navigation and JavaScript still ask \emph{(requires 2.1.288 or later)} [57].

Each stage is therefore a separate prompt that you send after reading the result of the last, and the final submission is an action you take yourself.

\FloatBarrier
\setcounter{section}{3}
\section{What the browser can carry out}

File uploads are subject to four restrictions worth stating: Claude can upload a file only when the session is permitted to read it, so a \texttt{Read} deny rule also blocks uploading it; a single upload is capped at 10 MB in total; files with multiple hard links are refused, which is common inside \texttt{node\_\allowbreak{}modules}; and a file whose name or folder is one credentials are kept under, such as \texttt{.\allowbreak{}env}, a \texttt{.\allowbreak{}pem} or \texttt{.\allowbreak{}key} file, or anything under \texttt{.\allowbreak{}ssh}, is refused \emph{(requires 2.1.293 or later)} [21]. Claude Code reads the file and sends its contents to the browser, so the upload works in remote sessions as well as local ones [21]. The credential-name check is a filename check: a secret saved under an innocuous name is not caught by it, which is why the \texttt{Read} deny rule is the control that matters (Chapter 14).

\begin{calloutbox}{palebrass}{brassdark}{2.0mm}
\textbf{CAUTION:} screenshots, GIF recordings, downloads and page text may contain sensitive information, and a recording captures everything visible on a logged-in page. These artefacts enter the session transcript or the filesystem [21]. Review before sharing.
\end{calloutbox}

\FloatBarrier
\setcounter{section}{4}
\section{Prompt injection and authenticated action}

A web page is untrusted input. Text on a page can instruct an agent to abandon your task, disclose data, or navigate elsewhere; this is the canonical indirect prompt injection channel [4], [22]. Keep the objective and the allowed domains explicit, restrict network access, and require confirmation before any external write.

\begin{calloutbox}{palebrass}{brassdark}{2.0mm}
\textbf{CAUTION:} do not use browser automation for financial transfers, irreversible account operations, legal acceptance, or safety-critical control without a purpose-built system and an independent human confirmation path.
\end{calloutbox}

\FloatBarrier
\setcounter{section}{5}
\section{Computer use}

Where a task cannot be done in a browser, computer use lets Claude open applications, click, type and see the screen. From the CLI it is a research preview on macOS only, on Pro, Max or Ultra plans, not Team or Enterprise, and only in an interactive session, never with \texttt{-\allowbreak{}p}; the Desktop app offers it on macOS and Windows [92]. It is the broadest and slowest route, so Claude uses an MCP server, Bash or Claude in Chrome first when one of them can do the task [92].

It is a built-in MCP server, \texttt{computer-\allowbreak{}use}, off until you enable it in \texttt{/\allowbreak{}mcp}; the setting persists per project [92]. Enabling it grants no app. The first time Claude needs a given app in a session you are asked to allow or deny it, and the approval lasts for that session only. Terminals and IDEs carry a warning that approving them is equivalent to shell access, Finder that it can read or write any file, System Settings that it can change system settings; they are warned about, not blocked. Control also varies by category: browsers and trading platforms are view-only, terminals and IDEs click-only, everything else full control [92].

Only one session at a time can hold the computer, and it keeps the lock until it exits, not until the task ends. \texttt{Esc} anywhere aborts the current action, and the key press is consumed so that on-screen content cannot use it to dismiss a dialog. Your terminal is hidden from Claude's screenshots, so prompts in your own session cannot feed back into the model [92].

The same staging discipline applies, with a wider blast radius. Unlike the sandboxed Bash tool, computer use runs on your actual desktop with the apps you approve, and the page says plainly that the trust boundary is different [92]. A browser profile can be isolated; a desktop session generally cannot.

\subsection*{Exercise}

Use a harmless public page or a local HTML fixture. Ask Claude to extract a small table, cite the page location, and take no action. Then ask it to navigate to one linked page. Inspect the browser history and the transcript to confirm the exact scope of what happened.

\emph{A worked solution is given in Appendix M.}

\begingroup
\def\kprows{%
\item Chrome integration shares your browser's authenticated state, so it can reach anything you are signed into.
\item Escalate in stages: observe, navigate, draft, verify, act — and let a human perform the final submission.
\item No mode enforces the staging. Plan mode documents prompts only for GIFs and new tabs, and in auto mode the classifier, not the extension's site list, decides most calls.
\item A web page is untrusted input. Keep the objective and the allowed domains explicit.
}%
\def\kpbody{\begin{minipage}{\textwidth}\subsection*{Key points}\begin{itemize}\kprows\end{itemize}\end{minipage}}%
\begingroup
\def\sloppy{\tolerance 9999\emergencystretch 3em\hfuzz 200pt\vfuzz 200pt}%
\hbadness=10000\vbadness=10000\hfuzz=200pt\vfuzz=200pt
\global\setbox\kpbox=\hbox{\kpbody}%
\endgroup
\par\addvspace{4.2mm}
\ifdim\dimexpr\ht\kpbox+\dp\kpbox\relax>0.30\textheight
  \typeout{HANDBOOK-KEYPOINTS broken \the\dimexpr\ht\kpbox+\dp\kpbox\relax}%
  \subsection*{Key points}
  \begin{itemize}\kprows\end{itemize}
\else
  \typeout{HANDBOOK-KEYPOINTS atomic \the\dimexpr\ht\kpbox+\dp\kpbox\relax}%
  \noindent\kpbody
\fi
\par\addvspace{1.4mm}
\endgroup

\FloatBarrier
\renewcommand{\chaptertitlelabel}{Part V \textperiodcentered\ Chapter 21}
\setcounter{chapter}{20}
\chapter{Keep work moving with goals}

\texttt{/\allowbreak{}goal} sets a completion condition. After each turn your configured small fast model checks whether the condition holds; if not, Claude starts another turn instead of returning control to you [84]. One goal is active per session — setting a second replaces the first — and the condition may be up to 4,000 characters. The point of the design is that completion is judged by a fresh model and not by the one doing the work; the limit of it is the subject of the next section.

\begin{codeblock}{7.0}{8.3}
\cl{/goal~<condition>~~~~~~\#~set;~starts~a~turn~immediately}
\cl{/goal~~~~~~~~~~~~~~~~~~\#~status:~condition,~duration,~turns~evaluated,~token~spend,~last~reason}
\cl{/goal~clear~~~~~~~~~~~~\#~remove~(aliases:~stop,~off,~reset,~none,~cancel)}
\end{codeblock}

\begin{codeblock}{9.0}{10.6}
\cl{claude~-p~"/goal~CHANGELOG.md~has~an~entry~for~every~PR~merged~this~week"}
\end{codeblock}

In a \texttt{-\allowbreak{}p} run the loop runs to completion in one invocation, and with the default text output nothing prints until it ends, so a long goal can look stuck; add \texttt{-\allowbreak{}-\allowbreak{}output-\allowbreak{}format stre\allowbreak{}am-\allowbreak{}json -\allowbreak{}-\allowbreak{}verbose} to see each message [84].

\FloatBarrier
\setcounter{section}{0}
\section{What the evaluator can and cannot see}

\textbf{The evaluator does not run commands or read files.} It judges only what Claude has surfaced in the conversation [84]. This single fact determines what a good condition looks like: "all tests in \texttt{test/\allowbreak{}auth} pass" works because Claude runs the tests and the output lands in the transcript where the evaluator can read it. "Finish the project" does not, because it is satisfiable by an assertion.

The evaluator returns one of three verdicts: \emph{not yet met} (Claude continues, taking the reason as guidance), \emph{met} (goal cleared, achieved entry recorded), or \emph{impossible} (goal cleared, failed entry recorded with the reason). Press \texttt{Ctrl+O} to see the reason behind a verdict [84]. To evaluate on a different model, set \texttt{ANTHROPIC\_\allowbreak{}DEFAULT\_\allowbreak{}HAIKU\_\allowbreak{}MODEL} — knowing that Claude Code reads that variable everywhere it uses the small fast model, so it also moves the \texttt{haiku} alias and session-title generation.

\texttt{/\allowbreak{}goal} is implemented as a session-scoped \textbf{prompt-based Stop hook}, which has two consequences: it is unavailable when \texttt{disable\allowbreak{}All\allowbreak{}Hooks} is true after settings precedence, or when \texttt{allow\allowbreak{}Managed\allowbreak{}Hooks\allowbreak{}Only} is set in managed settings; and it is subject to the same workspace-trust rule as hooks in settings files [83], [84].

\FloatBarrier
\setcounter{section}{1}
\section{Anatomy of a safe goal}

Include a concrete deliverable; the checks required; the evidence to report; explicit exclusions; human gates; a turn, time or resource bound; and a condition for stopping when blocked.

\begin{codeblock}{7.0}{8.3}
\cl{/goal~Create~drafts/brief-v2.md~from~the~approved~sources.~The~goal~is~met~only~when~every}
\cl{material~claim~has~a~source-ledger~row,~the~evidence-reviewer~subagent~reports~no~high-severity}
\cl{unsupported~claim,~the~local~link~check~exits~0~with~its~output~shown,~and~the~final~message}
\cl{lists~the~commands~run~and~the~files~changed~as~evidence.~Do~not~publish,~push,~install}
\cl{software,~or~access~accounts.~Stop~and~ask~if~blocked,~or~after~20~turns.}
\end{codeblock}

\FloatBarrier
\setcounter{section}{2}
\section{Behaviour you should expect}

\begin{itemize}
\item \textbf{A goal does not change your permission mode} [84]. In Manual mode Claude still asks before tool calls your settings do not allow. To let goal turns run unattended, combine it with auto mode — deliberately, and with sandboxing.
\item \textbf{Background work defers evaluation.} If a subagent or background shell command is still running when a turn ends, evaluation is skipped until a turn finishes with nothing running [84].
\item \textbf{Check-ins back off.} Once background work has kept the goal waiting 30 minutes, Claude Code delivers a check-in asking Claude to read the running tasks' output and fix or stop anything stuck; later check-ins double the interval up to four times the first. Set \texttt{CLAUDE\_\allowbreak{}CODE\_\allowbreak{}GOAL\_\allowbreak{}CHECKIN\_\allowbreak{}MINUTES} to change it, or \texttt{0} to turn check-ins off (2.1.234+; idle check-ins 2.1.236+; before 2.1.239 only idle check-ins backed off, and one delivered at a turn end recurred at the first interval) [84].
\item \textbf{A no-progress guard exists.} If Claude answers the evaluator repeatedly without tool use, Claude Code stops the loop, warns, and returns control with the goal still set [84].
\item \textbf{Idle check-ins are capped.} In an interactive session Claude Code starts a turn on its own to deliver a due check-in, at most three times per goal between your prompts \emph{(the cap requires 2.1.246 or later)} [84].
\item \textbf{An error you have to fix clears the goal}: an authentication failure or a disk too full to renew the login, where Claude Code manages its own credentials; an exhausted credit balance; a context overflow auto-compaction could not clear; and an unavailable model. The warning ends \texttt{Run /\allowbreak{}goal again \allowbreak{}to continue} [84].
\item \textbf{Any other error leaves the goal set}, and in an interactive session Claude Code then either retries or pauses \emph{(requires 2.1.269 or later)}. It retries after a failure that tends to clear by itself, such as an overloaded server, and pauses after three such retries. It pauses at once after a failure a retry would only repeat — a rate limit, a usage limit, a hook that ended the turn. A paused goal waits for your next message, except that a session set to continue when a usage limit resets resumes work toward the goal then. Setting \texttt{CLAUDE\_\allowbreak{}CODE\_\allowbreak{}GOAL\_\allowbreak{}CHECKIN\_\allowbreak{}MINUTES} to \texttt{0} turns the automatic retries off along with the check-ins [84].
\item \textbf{Goals survive resume}, on every resume route \emph{(the picker from 2.1.239)}, with turn count, timer and token baseline reset. A goal already achieved or cleared is not restored [84].
\end{itemize}

\FloatBarrier
\setcounter{section}{3}
\section{Context is the real multiplier}

The course's long-running agent example works not because \texttt{/\allowbreak{}goal} is magical but because it combines a detailed proposal, tool documentation, learned procedures, a plan, tests and a measurable finished state [44]. A goal cannot repair missing inputs, and it cannot make an unsafe access safe. Clear a stale goal when the objective changes; \texttt{/\allowbreak{}clear} also removes it.

\begingroup
\def\kprows{%
\item The evaluator does not run commands or read files; it judges only what Claude surfaced in the conversation.
\item A good condition names a deliverable, the checks, the evidence to report, the exclusions and a bound.
\item A goal does not change your permission mode, and background work defers evaluation.
\item \texttt{/\allowbreak{}goal} is a prompt Stop hook, so \texttt{disable\allowbreak{}All\allowbreak{}Hooks} or \texttt{allow\allowbreak{}Managed\allowbreak{}Hooks\allowbreak{}Only} disables it.
}%
\def\kpbody{\begin{minipage}{\textwidth}\subsection*{Key points}\begin{itemize}\kprows\end{itemize}\end{minipage}}%
\begingroup
\def\sloppy{\tolerance 9999\emergencystretch 3em\hfuzz 200pt\vfuzz 200pt}%
\hbadness=10000\vbadness=10000\hfuzz=200pt\vfuzz=200pt
\global\setbox\kpbox=\hbox{\kpbody}%
\endgroup
\par\addvspace{4.2mm}
\ifdim\dimexpr\ht\kpbox+\dp\kpbox\relax>0.30\textheight
  \typeout{HANDBOOK-KEYPOINTS broken \the\dimexpr\ht\kpbox+\dp\kpbox\relax}%
  \subsection*{Key points}
  \begin{itemize}\kprows\end{itemize}
\else
  \typeout{HANDBOOK-KEYPOINTS atomic \the\dimexpr\ht\kpbox+\dp\kpbox\relax}%
  \noindent\kpbody
\fi
\par\addvspace{1.4mm}
\endgroup

\FloatBarrier
\renewcommand{\chaptertitlelabel}{Part V \textperiodcentered\ Chapter 22}
\setcounter{chapter}{21}
\chapter{Dynamic workflows and ultracode}

A dynamic workflow is a JavaScript script that orchestrates subagents at scale. Claude writes the script for the task you describe, and a runtime executes it in the background while your session stays responsive [19]. Requires 2.1.154 or later. It is available on every paid plan, with Anthropic API access, and on Amazon Bedrock, Google Cloud's Agent Platform and Microsoft Foundry; on Pro, enable it from the Dynamic workflows row in \texttt{/\allowbreak{}config} [19].

The distinction that matters is \textbf{who holds the plan}. With subagents, skills and agent teams, Claude decides turn by turn what to spawn next and every result lands in a context window. A workflow script holds the loop, the branching and the intermediate results, so Claude's context holds only the final answer [19]. The script is plain JavaScript: \texttt{agent()} spawns one subagent, \texttt{pipeline()} runs one per item in a list, \texttt{parallel()} runs a set at once, and the runtime makes \texttt{Date.\allowbreak{}now()}, \texttt{Math.\allowbreak{}random()} and a no-argument \texttt{new Date()} throw, so that a relaunched run repeats the same calls [19].

\FloatBarrier
\setcounter{section}{0}
\section{Entry points}

\begin{codeblock}{7.0}{8.3}
\cl{/deep-research~<question>~~~~~\#~bundled~workflow:~fan-out~search,~cross-check,~cited~report}
\cl{/workflows~~~~~~~~~~~~~~~~~~~~\#~progress~view}
\cl{/effort~ultracode~~~~~~~~~~~~~\#~automatic~workflow~orchestration,~at~the~session\textquotesingle{}s~effort~level}
\cl{/effort~ultracode~off~~~~~~~~~\#~turn~it~off~again~(2.1.284+)}
\end{codeblock}

\begin{codeblock}{7.0}{8.3}
\cl{claude~-{}-effort~ultracode~~~~~\#~ultracode~on~and~effort~xhigh,~where~both~are~available~(2.1.203+)}
\end{codeblock}

Ultracode is a Claude Code setting, not an effort level \emph{(requires 2.1.284 or later; before that release turning it on set the session to \texttt{xhigh})}. With it on, Claude plans a workflow for each substantive task at whichever effort level the session runs at, and turning it on or off with \texttt{/\allowbreak{}effort} leaves that level unchanged. The launch flag and the Agent SDK's \texttt{effort\allowbreak{}Level:\allowbreak{} "ultracode\allowbreak{}"} also set \texttt{xhigh}; where ultracode is unavailable, as with dynamic workflows off, the flag starts the session with it off at the highest level the model allows. The \texttt{ultracode} settings key makes it persistent; the \texttt{effort\allowbreak{}Level} setting and \texttt{CLAUDE\_\allowbreak{}CODE\_\allowbreak{}EFFORT\_\allowbreak{}LEVEL} do not accept it [63]. Ultracode is unavailable when workflows are turned off or the model does not support \texttt{xhigh} [63]. On an installation with dynamic workflows not enabled, \texttt{/\allowbreak{}effort ultr\allowbreak{}acode} was refused with a message saying so. Neither \texttt{claude -\allowbreak{}-\allowbreak{}help} nor the \texttt{/\allowbreak{}effort} usage string lists the word, though both surfaces take it (Section 4.2).

Turning ultracode on is an opt-in to large runs, and three checks stand down while it is on: the \texttt{Large workf\allowbreak{}low} warning, the session's concurrent-subagent limit for agents Claude spawns with the Agent tool, and, in auto mode, the approval of the first workflow launch [19]. No managed setting rules out ultracode alone. An organisation that wants it gone turns workflows off; an effort cap lowers the level a session with ultracode on runs at, but does not turn it off [19].

Including the keyword \texttt{ultracode} in a prompt you type yourself opts a single task in without changing session effort; asking in your own words works too [19]. The keyword is an opt-in \textbf{only in a prompt you type} — at the interactive prompt, in an IDE extension panel, in a Remote Control client, or in an SDK message stamped as human input. It does not start a workflow from \texttt{-\allowbreak{}p}, from an SDK message not so stamped, from a scheduled task prompt, or from a webhook or pull-request comment relayed into the conversation (from 2.1.210). Before 2.1.160 the literal trigger word was \texttt{workflow}. \texttt{Alt+W} (\texttt{Option+W} on macOS) dismisses the highlight for one prompt, and the Ultracode keyword trigger row in \texttt{/\allowbreak{}config} turns the keyword off [19]. The keyword chooses how the work is structured and nothing else: the agents' tool calls get the same permission checks and sandboxing as any other call in the session [19].

\FloatBarrier
\setcounter{section}{1}
\section{Limits and permissions}

\begingroup
\def\tblrows{%
Concurrent agents & Up to 16 by default, fewer with fewer CPUs; \texttt{CLAUDE\_\allowbreak{}CODE\_\allowbreak{}WORKFLOW\_\allowbreak{}MAX\_\allowbreak{}CONCURRENT\_\allowbreak{}AGENTS} takes 1 to 256 (2.1.269+) \\
Total agents per run & 1,000 \\
Items in one \texttt{parallel()} or \texttt{pipeline()} call & 4,096; a longer list is rejected with an error \\
Mid-run user input & Not possible; a run pauses only for agent permission prompts and a usage-limit wait \\
Filesystem or shell from the script & None; agents do the work, the script coordinates \\
Module loading & A script containing \texttt{import()} fails before the run starts \\
Prompt-cache stagger & Matching agents start up to \texttt{CLAUDE\_\allowbreak{}CODE\_\allowbreak{}WORKFLOW\_\allowbreak{}PREFIX\_\allowbreak{}STAGGER\_\allowbreak{}MS} ms after the first, default 5000 \\
}%
\def\tblbody{\begin{minipage}{\textwidth}\boxcaption{Table 19 — Dynamic workflow runtime constraints}
{\footnotesize\begin{tabular}{L{39.5mm}L{103.5mm}}
\toprule
\textbf{Constraint} & \textbf{Value} \\
\midrule
\tblrows
\bottomrule\end{tabular}}\end{minipage}}%
\begingroup
\def\sloppy{\tolerance 9999\emergencystretch 3em\hfuzz 200pt\vfuzz 200pt}%
\hbadness=10000\vbadness=10000\hfuzz=200pt\vfuzz=200pt
\global\setbox\tblbox=\hbox{\tblbody}%
\endgroup
\par\addvspace{2.6mm}
\ifdim\dimexpr\ht\tblbox+\dp\tblbox\relax>0.55\textheight
  \typeout{HANDBOOK-TABLE broken \the\dimexpr\ht\tblbox+\dp\tblbox\relax}%
  \tabcaption{Table 19 — Dynamic workflow runtime constraints}
  {\footnotesize\begin{longtable}{L{39.5mm}L{103.5mm}}
  \toprule
\textbf{Constraint} & \textbf{Value} \\
\midrule\endfirsthead
  \multicolumn{2}{@{}l@{}}{%
  \sffamily\footnotesize\itshape\color{inkgrey}Table 19 — Dynamic workflow runtime constraints \textemdash\ continued}\\[1.2mm]
  \toprule
\textbf{Constraint} & \textbf{Value} \\
\midrule\endhead
  \bottomrule\endfoot
  \bottomrule\endlastfoot
  \tblrows
  \end{longtable}}%
\else
  \typeout{HANDBOOK-TABLE atomic \the\dimexpr\ht\tblbox+\dp\tblbox\relax}%
  \noindent\tblbody
\fi
\par\addvspace{2.6mm}
\endgroup

\begin{calloutbox}{palebrass}{brassdark}{2.0mm}
\textbf{CAUTION:} The subagents a workflow spawns use your permission rules and take their mode by the rules for any subagent [19], [27]. Under a session in \texttt{bypass\allowbreak{}Permissions}, \texttt{accept\allowbreak{}Edits} or \texttt{auto} they run in that mode; under \texttt{plan} they stay in plan mode, and a \texttt{dont\allowbreak{}Ask} agent there is denied what plan mode would ask; under Manual or \texttt{dont\allowbreak{}Ask} an agent type's own \texttt{permission\allowbreak{}Mode} applies, unless it asks for \texttt{bypass\allowbreak{}Permissions}, or for \texttt{auto} where auto mode is unavailable, and an unset field inherits the session's mode. Version 1.0 said they always run in \texttt{accept\allowbreak{}Edits}, which the page no longer states. In auto mode — what a session starts in when nothing configures a mode — the classifier decides the agents' actions, and in \texttt{accept\allowbreak{}Edits} their file edits inside the working directory are approved without asking. Each mode's own prompts, denials and the actions no mode auto-approves still apply [33], so a long run can stop on a prompt; add the tools the agents need to your allow rules first [19]. Either way, commit a clean baseline before starting a run.
\end{calloutbox}

How the launch itself is approved depends on the mode [19]. In auto mode you are asked at the first launch only, and any \textbf{Yes} records consent in your user settings; with ultracode on, not even then. In Manual and \texttt{accept\allowbreak{}Edits} you are asked on every run, unless you chose \textbf{Yes, and don't ask again} for a bundled, saved or plugin workflow in this project. In \texttt{bypass\allowbreak{}Permissions} the run starts at once. The prompt offers \textbf{View raw script}, and \texttt{Ctrl+G} opens the script in your editor; read it before approving a script Claude wrote for this task.

In \texttt{claude -\allowbreak{}p} and the Agent SDK no prompt is shown: the Workflow tool call goes through the same evaluation as any other call, so deny rules, ask rules and \texttt{dont\allowbreak{}Ask} apply. A launch there needs an allow rule (\texttt{Workflow}, or \texttt{Workflow(<n\allowbreak{}ame>)} for one saved workflow), auto mode, bypass, a \texttt{Pre\allowbreak{}Tool\allowbreak{}Use} hook that returns \texttt{allow}, or the host's own approval; in plan mode a hook's approval alone does not start it [19]. Two further limits hold everywhere. Claude can start a workflow only from a script file the session may already read [19]. And in auto mode the prompt a script passes to \texttt{agent()} does not count as a request from you when the classifier reviews that agent's actions, because Claude Code marks it as text the script computed [19] — so a script cannot launder an instruction into your authority.

Size guidance is separate from the hard caps. \texttt{workflow\allowbreak{}Size\allowbreak{}Guideline} takes \texttt{unrestricte\allowbreak{}d}, \texttt{small} (fewer than 5 agents), \texttt{medium} (fewer than 10), or \texttt{large} (fewer than 50). The default is \texttt{medium}, or \texttt{small} when signed in on a Pro plan \emph{(requires 2.1.271 or later)}. Set it in \texttt{/\allowbreak{}config} or in any settings file from 2.1.219; a settings file wins and hides the \texttt{/\allowbreak{}config} row [19]. Version 1.0 gave \texttt{medium} as fewer than 15, which was the page's figure in August. The guideline is advice to Claude, not a cap, and a prompt that asks for a different scale overrides it.

Claude Code shows a \texttt{Large workf\allowbreak{}low} warning when a run schedules more than 25 agents or projects more than 1.5 million tokens — advisory only, and suppressed when ultracode is on. \textbf{The 25 is the built-in default's threshold, not a constant, and not \texttt{medium}'s count:} choose a guideline yourself and its own agent count replaces the threshold, so \texttt{small} warns above five, \texttt{medium} chosen explicitly above ten, and \texttt{large} above fifty [19].

\FloatBarrier
\setcounter{section}{2}
\section{Resume semantics, and why they matter for cost}

Pause and resume a run from \texttt{/\allowbreak{}workflows} with \texttt{p}; a run you stopped is relaunched by asking Claude to run the same script again. Replay follows the order agents \emph{started}, and each agent either returns its saved result or runs again [19]:

\begin{itemize}
\item \textbf{Completed} returns its saved result — until the first agent whose prompt differs from the previous run, because you edited the script or an earlier agent returned something different. That agent and every one after it run again, completed or not.
\item \textbf{Still running when you stopped} starts over. Stopping the whole run counts no agent as failed, so the completed agents after it keep their results.
\item \textbf{Failed} runs again, and so does every agent started after it, even ones that completed. Stopping one agent on its own with \texttt{x} counts as failing it.
\item \textbf{Failed at a usage limit} runs again, and completed agents keep their results \emph{(requires 2.1.318 or later)} — unless the script called another agent after the failure and that agent returned a result, in which case every agent started after the failed one runs again too.
\end{itemize}

Version 1.0 said that cached results stop at the first agent that did not finish. That is now true only of a \emph{failed} agent or a changed prompt, which is where the cost lies: one failure in the middle of a fan-out reruns everything that started after it. Completed agents could also rerun on an unedited relaunch whose \texttt{pipeline()} items each ran more than one agent; that was fixed \emph{(requires 2.1.305 or later)} [19].

A run does not continue on its own after you leave. If you background the session, ask Claude there to relaunch it; if you exit and stop tasks, the saved results stay under the session's directory in \texttt{\textasciitilde{}/\allowbreak{}.\allowbreak{}claude/\allowbreak{}projects/\allowbreak{}}, so \texttt{claude -\allowbreak{}-\allowbreak{}resume} followed by a relaunch replays them, while a fresh session has no earlier run and starts over [19]. When the saved results cannot be found, the relaunch fails with \texttt{nothing to \allowbreak{}resume} rather than starting over silently.

Two automatic recoveries change the cost picture. When an agent hits a claude.ai usage limit, the run pauses instead of failing, and continues shortly after the reset \emph{(requires 2.1.271 or later; 2.1.319 or later in the Desktop Code tab)}. The conditions are all of these: a claude.ai subscription without usage-based billing; an interactive terminal session or a local session started in the Code tab; \texttt{auto\allowbreak{}Continue\allowbreak{}At\allowbreak{}Usage\allowbreak{}Limit} on and, in the Code tab, the \textbf{Auto-continue when limits reset} box left checked; a reset within 24 hours; and a run that has not already waited twice. A terminal run in \texttt{-\allowbreak{}p}, the SDK, a background, Remote Control or teammate session does not pause; nor, in the Code tab, does a run in an SSH, WSL, Dispatch or scheduled-task session, one started from Remote Control or through a corporate launcher, or one with \texttt{CLAUDE\_\allowbreak{}CODE\_\allowbreak{}DISABLE\_\allowbreak{}NONESSENTIA\allowbreak{}L\_\allowbreak{}TRAFFIC} set. There the agent fails. Managed settings that set \texttt{auto\allowbreak{}Continue\allowbreak{}At\allowbreak{}Usage\allowbreak{}Limit} to \texttt{true} keep Code-tab runs pausing even with the box unchecked [19]. And an agent whose output stops arriving restarts from the same prompt, at most five times; files the stalled attempt changed stay changed, and its tokens stay in the total [19].

\FloatBarrier
\setcounter{section}{3}
\section{When a workflow earns its cost}

Use one when the task has genuinely independent lanes: several source families researched in parallel; multiple reviewers testing the same proposal independently; a large change divisible by non-overlapping ownership; research, challenge and synthesis benefiting from separate contexts.

Do not use one to format a file, answer a simple question, or manufacture the appearance of corroboration by asking identical agents the same question from the same evidence.

Before a large task, run the workflow on a small slice first; \texttt{/\allowbreak{}workflows} shows each agent's token usage as it goes, and runs count toward your plan's usage and rate limits [19]. Each agent's model is chosen in the order used for subagents, with a model the script names for a stage counting as the per-invocation model, and a model blocked by an \texttt{available\allowbreak{}Models} allowlist is substituted by the subagent rules [19]. A workflow agent's prompt cache holds for five minutes by default, even on a subscription, unless \texttt{subagent\allowbreak{}Prompt\allowbreak{}Cache\allowbreak{}Ttl} is set to \texttt{1h} [19].

\FloatBarrier
\setcounter{section}{4}
\section{A research workflow pattern}

The pattern separates gathering from checking and never lets one context do both. Figure 6 shows it end to end, including the release gate that a person, not an agent, has to pass.

\begin{figure}[tbp]
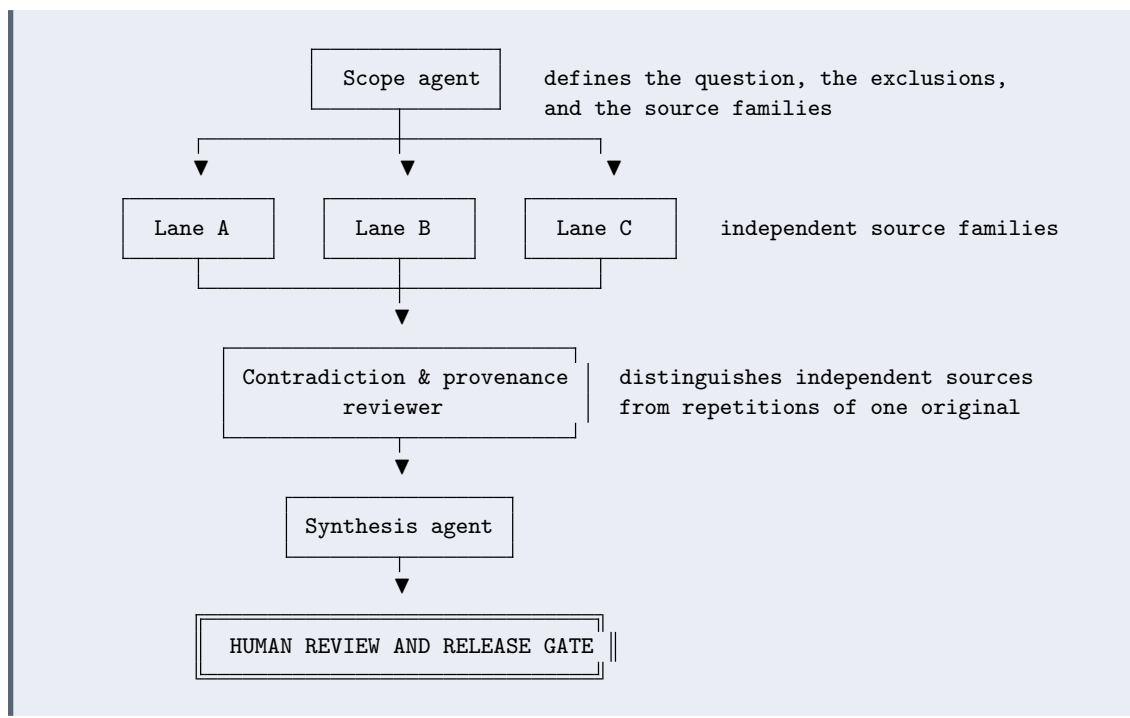

\begin{codefig}{9.0}{10.6}
\cl{~~~~~~~~~~~~~~~~~~~~~\pmboxdrawuni{250C}\pmboxdrawuni{2500}\pmboxdrawuni{2500}\pmboxdrawuni{2500}\pmboxdrawuni{2500}\pmboxdrawuni{2500}\pmboxdrawuni{2500}\pmboxdrawuni{2500}\pmboxdrawuni{2500}\pmboxdrawuni{2500}\pmboxdrawuni{2500}\pmboxdrawuni{2500}\pmboxdrawuni{2500}\pmboxdrawuni{2500}\pmboxdrawuni{2500}\pmboxdrawuni{2510}}
\cl{~~~~~~~~~~~~~~~~~~~~~\pmboxdrawuni{2502}~~Scope~agent~\pmboxdrawuni{2502}~~~defines~the~question,~the~exclusions,}
\cl{~~~~~~~~~~~~~~~~~~~~~\pmboxdrawuni{2514}\pmboxdrawuni{2500}\pmboxdrawuni{2500}\pmboxdrawuni{2500}\pmboxdrawuni{2500}\pmboxdrawuni{2500}\pmboxdrawuni{2500}\pmboxdrawuni{252C}\pmboxdrawuni{2500}\pmboxdrawuni{2500}\pmboxdrawuni{2500}\pmboxdrawuni{2500}\pmboxdrawuni{2500}\pmboxdrawuni{2500}\pmboxdrawuni{2500}\pmboxdrawuni{2518}~~~and~the~source~families}
\cl{~~~~~~~~~~~~\pmboxdrawuni{250C}\pmboxdrawuni{2500}\pmboxdrawuni{2500}\pmboxdrawuni{2500}\pmboxdrawuni{2500}\pmboxdrawuni{2500}\pmboxdrawuni{2500}\pmboxdrawuni{2500}\pmboxdrawuni{2500}\pmboxdrawuni{2500}\pmboxdrawuni{2500}\pmboxdrawuni{2500}\pmboxdrawuni{2500}\pmboxdrawuni{2500}\pmboxdrawuni{2500}\pmboxdrawuni{2500}\pmboxdrawuni{253C}\pmboxdrawuni{2500}\pmboxdrawuni{2500}\pmboxdrawuni{2500}\pmboxdrawuni{2500}\pmboxdrawuni{2500}\pmboxdrawuni{2500}\pmboxdrawuni{2500}\pmboxdrawuni{2500}\pmboxdrawuni{2500}\pmboxdrawuni{2500}\pmboxdrawuni{2500}\pmboxdrawuni{2500}\pmboxdrawuni{2500}\pmboxdrawuni{2500}\pmboxdrawuni{2500}\pmboxdrawuni{2510}}
\cl{~~~~~~~~~~~~▼~~~~~~~~~~~~~~~▼~~~~~~~~~~~~~~~▼}
\cl{~~~~~~\pmboxdrawuni{250C}\pmboxdrawuni{2500}\pmboxdrawuni{2500}\pmboxdrawuni{2500}\pmboxdrawuni{2500}\pmboxdrawuni{2500}\pmboxdrawuni{2500}\pmboxdrawuni{2500}\pmboxdrawuni{2500}\pmboxdrawuni{2500}\pmboxdrawuni{2500}\pmboxdrawuni{2500}\pmboxdrawuni{2510}~~~\pmboxdrawuni{250C}\pmboxdrawuni{2500}\pmboxdrawuni{2500}\pmboxdrawuni{2500}\pmboxdrawuni{2500}\pmboxdrawuni{2500}\pmboxdrawuni{2500}\pmboxdrawuni{2500}\pmboxdrawuni{2500}\pmboxdrawuni{2500}\pmboxdrawuni{2500}\pmboxdrawuni{2500}\pmboxdrawuni{2510}~~~\pmboxdrawuni{250C}\pmboxdrawuni{2500}\pmboxdrawuni{2500}\pmboxdrawuni{2500}\pmboxdrawuni{2500}\pmboxdrawuni{2500}\pmboxdrawuni{2500}\pmboxdrawuni{2500}\pmboxdrawuni{2500}\pmboxdrawuni{2500}\pmboxdrawuni{2500}\pmboxdrawuni{2500}\pmboxdrawuni{2510}}
\cl{~~~~~~\pmboxdrawuni{2502}~~Lane~A~~~\pmboxdrawuni{2502}~~~\pmboxdrawuni{2502}~~Lane~B~~~\pmboxdrawuni{2502}~~~\pmboxdrawuni{2502}~~Lane~C~~~\pmboxdrawuni{2502}~~~independent~source~families}
\cl{~~~~~~\pmboxdrawuni{2514}\pmboxdrawuni{2500}\pmboxdrawuni{2500}\pmboxdrawuni{2500}\pmboxdrawuni{2500}\pmboxdrawuni{2500}\pmboxdrawuni{252C}\pmboxdrawuni{2500}\pmboxdrawuni{2500}\pmboxdrawuni{2500}\pmboxdrawuni{2500}\pmboxdrawuni{2500}\pmboxdrawuni{2518}~~~\pmboxdrawuni{2514}\pmboxdrawuni{2500}\pmboxdrawuni{2500}\pmboxdrawuni{2500}\pmboxdrawuni{2500}\pmboxdrawuni{2500}\pmboxdrawuni{252C}\pmboxdrawuni{2500}\pmboxdrawuni{2500}\pmboxdrawuni{2500}\pmboxdrawuni{2500}\pmboxdrawuni{2500}\pmboxdrawuni{2518}~~~\pmboxdrawuni{2514}\pmboxdrawuni{2500}\pmboxdrawuni{2500}\pmboxdrawuni{2500}\pmboxdrawuni{2500}\pmboxdrawuni{2500}\pmboxdrawuni{252C}\pmboxdrawuni{2500}\pmboxdrawuni{2500}\pmboxdrawuni{2500}\pmboxdrawuni{2500}\pmboxdrawuni{2500}\pmboxdrawuni{2518}}
\cl{~~~~~~~~~~~~\pmboxdrawuni{2514}\pmboxdrawuni{2500}\pmboxdrawuni{2500}\pmboxdrawuni{2500}\pmboxdrawuni{2500}\pmboxdrawuni{2500}\pmboxdrawuni{2500}\pmboxdrawuni{2500}\pmboxdrawuni{2500}\pmboxdrawuni{2500}\pmboxdrawuni{2500}\pmboxdrawuni{2500}\pmboxdrawuni{2500}\pmboxdrawuni{2500}\pmboxdrawuni{2500}\pmboxdrawuni{2500}\pmboxdrawuni{253C}\pmboxdrawuni{2500}\pmboxdrawuni{2500}\pmboxdrawuni{2500}\pmboxdrawuni{2500}\pmboxdrawuni{2500}\pmboxdrawuni{2500}\pmboxdrawuni{2500}\pmboxdrawuni{2500}\pmboxdrawuni{2500}\pmboxdrawuni{2500}\pmboxdrawuni{2500}\pmboxdrawuni{2500}\pmboxdrawuni{2500}\pmboxdrawuni{2500}\pmboxdrawuni{2500}\pmboxdrawuni{2518}}
\cl{~~~~~~~~~~~~~~~~~~~~~~~~~~~~▼}
\cl{~~~~~~~~~~~~~~\pmboxdrawuni{250C}\pmboxdrawuni{2500}\pmboxdrawuni{2500}\pmboxdrawuni{2500}\pmboxdrawuni{2500}\pmboxdrawuni{2500}\pmboxdrawuni{2500}\pmboxdrawuni{2500}\pmboxdrawuni{2500}\pmboxdrawuni{2500}\pmboxdrawuni{2500}\pmboxdrawuni{2500}\pmboxdrawuni{2500}\pmboxdrawuni{2500}\pmboxdrawuni{2500}\pmboxdrawuni{2500}\pmboxdrawuni{2500}\pmboxdrawuni{2500}\pmboxdrawuni{2500}\pmboxdrawuni{2500}\pmboxdrawuni{2500}\pmboxdrawuni{2500}\pmboxdrawuni{2500}\pmboxdrawuni{2500}\pmboxdrawuni{2500}\pmboxdrawuni{2500}\pmboxdrawuni{2500}\pmboxdrawuni{2500}\pmboxdrawuni{2510}}
\cl{~~~~~~~~~~~~~~\pmboxdrawuni{2502}~Contradiction~\&~provenance~\pmboxdrawuni{2502}~~distinguishes~independent~sources}
\cl{~~~~~~~~~~~~~~\pmboxdrawuni{2502}~~~~~~~~~reviewer~~~~~~~~~~~\pmboxdrawuni{2502}~~from~repetitions~of~one~original}
\cl{~~~~~~~~~~~~~~\pmboxdrawuni{2514}\pmboxdrawuni{2500}\pmboxdrawuni{2500}\pmboxdrawuni{2500}\pmboxdrawuni{2500}\pmboxdrawuni{2500}\pmboxdrawuni{2500}\pmboxdrawuni{2500}\pmboxdrawuni{2500}\pmboxdrawuni{2500}\pmboxdrawuni{2500}\pmboxdrawuni{2500}\pmboxdrawuni{2500}\pmboxdrawuni{2500}\pmboxdrawuni{252C}\pmboxdrawuni{2500}\pmboxdrawuni{2500}\pmboxdrawuni{2500}\pmboxdrawuni{2500}\pmboxdrawuni{2500}\pmboxdrawuni{2500}\pmboxdrawuni{2500}\pmboxdrawuni{2500}\pmboxdrawuni{2500}\pmboxdrawuni{2500}\pmboxdrawuni{2500}\pmboxdrawuni{2500}\pmboxdrawuni{2500}\pmboxdrawuni{2518}}
\cl{~~~~~~~~~~~~~~~~~~~~~~~~~~~~▼}
\cl{~~~~~~~~~~~~~~~~~~~\pmboxdrawuni{250C}\pmboxdrawuni{2500}\pmboxdrawuni{2500}\pmboxdrawuni{2500}\pmboxdrawuni{2500}\pmboxdrawuni{2500}\pmboxdrawuni{2500}\pmboxdrawuni{2500}\pmboxdrawuni{2500}\pmboxdrawuni{2500}\pmboxdrawuni{2500}\pmboxdrawuni{2500}\pmboxdrawuni{2500}\pmboxdrawuni{2500}\pmboxdrawuni{2500}\pmboxdrawuni{2500}\pmboxdrawuni{2500}\pmboxdrawuni{2500}\pmboxdrawuni{2510}}
\cl{~~~~~~~~~~~~~~~~~~~\pmboxdrawuni{2502}~Synthesis~agent~\pmboxdrawuni{2502}}
\cl{~~~~~~~~~~~~~~~~~~~\pmboxdrawuni{2514}\pmboxdrawuni{2500}\pmboxdrawuni{2500}\pmboxdrawuni{2500}\pmboxdrawuni{2500}\pmboxdrawuni{2500}\pmboxdrawuni{2500}\pmboxdrawuni{2500}\pmboxdrawuni{2500}\pmboxdrawuni{252C}\pmboxdrawuni{2500}\pmboxdrawuni{2500}\pmboxdrawuni{2500}\pmboxdrawuni{2500}\pmboxdrawuni{2500}\pmboxdrawuni{2500}\pmboxdrawuni{2500}\pmboxdrawuni{2500}\pmboxdrawuni{2518}}
\cl{~~~~~~~~~~~~~~~~~~~~~~~~~~~~▼}
\cl{~~~~~~~~~~~~\pmboxdrawuni{2554}\pmboxdrawuni{2550}\pmboxdrawuni{2550}\pmboxdrawuni{2550}\pmboxdrawuni{2550}\pmboxdrawuni{2550}\pmboxdrawuni{2550}\pmboxdrawuni{2550}\pmboxdrawuni{2550}\pmboxdrawuni{2550}\pmboxdrawuni{2550}\pmboxdrawuni{2550}\pmboxdrawuni{2550}\pmboxdrawuni{2550}\pmboxdrawuni{2550}\pmboxdrawuni{2550}\pmboxdrawuni{2550}\pmboxdrawuni{2550}\pmboxdrawuni{2550}\pmboxdrawuni{2550}\pmboxdrawuni{2550}\pmboxdrawuni{2550}\pmboxdrawuni{2550}\pmboxdrawuni{2550}\pmboxdrawuni{2550}\pmboxdrawuni{2550}\pmboxdrawuni{2550}\pmboxdrawuni{2550}\pmboxdrawuni{2550}\pmboxdrawuni{2550}\pmboxdrawuni{2550}\pmboxdrawuni{2550}\pmboxdrawuni{2557}}
\cl{~~~~~~~~~~~~\pmboxdrawuni{2551}~~HUMAN~REVIEW~AND~RELEASE~GATE~\pmboxdrawuni{2551}}
\cl{~~~~~~~~~~~~\pmboxdrawuni{255A}\pmboxdrawuni{2550}\pmboxdrawuni{2550}\pmboxdrawuni{2550}\pmboxdrawuni{2550}\pmboxdrawuni{2550}\pmboxdrawuni{2550}\pmboxdrawuni{2550}\pmboxdrawuni{2550}\pmboxdrawuni{2550}\pmboxdrawuni{2550}\pmboxdrawuni{2550}\pmboxdrawuni{2550}\pmboxdrawuni{2550}\pmboxdrawuni{2550}\pmboxdrawuni{2550}\pmboxdrawuni{2550}\pmboxdrawuni{2550}\pmboxdrawuni{2550}\pmboxdrawuni{2550}\pmboxdrawuni{2550}\pmboxdrawuni{2550}\pmboxdrawuni{2550}\pmboxdrawuni{2550}\pmboxdrawuni{2550}\pmboxdrawuni{2550}\pmboxdrawuni{2550}\pmboxdrawuni{2550}\pmboxdrawuni{2550}\pmboxdrawuni{2550}\pmboxdrawuni{2550}\pmboxdrawuni{2550}\pmboxdrawuni{255D}}
\end{codefig}
\figcaption{Figure 6 — A cross-checked research workflow with a human release gate}
\end{figure}

The contradiction reviewer is the load-bearing element. Three pages repeating one press release are one source, and a workflow that counts them as three has manufactured confidence rather than evidence. The bundled \texttt{/\allowbreak{}deep-\allowbreak{}research} workflow already votes on each claim and filters out those that do not survive cross-checking, reporting a claim as \emph{unverified} rather than refuted when a verifier could not check it [19].

\FloatBarrier
\setcounter{section}{5}
\section{Saving and distributing a workflow}

Run \texttt{/\allowbreak{}workflows}, select the run, press \texttt{s}, and choose \texttt{.\allowbreak{}claude/\allowbreak{}workflows/\allowbreak{}} (shared with the repository) or \texttt{\textasciitilde{}/\allowbreak{}.\allowbreak{}claude/\allowbreak{}workflows/\allowbreak{}} (yours alone). It then runs as \texttt{/\allowbreak{}<name>}, and a saved workflow takes input through the \texttt{args} global [19]. Claude Code refuses to write through a symlink — before 2.1.216 it followed the link — but the two locations are checked differently: at the project location it refuses if \texttt{.\allowbreak{}claude}, \texttt{.\allowbreak{}claude/\allowbreak{}workflows} or the target file is a symlink, at the personal location only if the target file is [19]. In a monorepo, saving writes to the closest existing \texttt{.\allowbreak{}claude/\allowbreak{}workflows/\allowbreak{}}, workflows load from every one along the path with the closest winning, and a project workflow beats a personal one of the same name [19]. \textbf{A saved project workflow is therefore code a clone brings with it}, and a name collision resolves in the repository's favour; review \texttt{.\allowbreak{}claude/\allowbreak{}workflows/\allowbreak{}} as you would any other script in the tree. To edit one, load the \texttt{/\allowbreak{}workflow-\allowbreak{}authoring} skill first \emph{(requires 2.1.248 or later)} and run \texttt{/\allowbreak{}reload-\allowbreak{}skills} after the change [19]. Distribute across teams by placing the script in a plugin's \texttt{workflows/\allowbreak{}} directory, where it is namespaced as \texttt{/\allowbreak{}plugin-\allowbreak{}name:\allowbreak{}workflow-\allowbreak{}name}.

Turn workflows off with the \texttt{/\allowbreak{}config} toggle, \texttt{"disable\allowbreak{}Workflows":\allowbreak{} true}, or \texttt{CLAUDE\_\allowbreak{}CODE\_\allowbreak{}DISABLE\_\allowbreak{}WORKFLOWS=\allowbreak{}1}; organisations use managed settings or the toggle on the Claude Code admin settings page [19]. Turning them off removes \texttt{/\allowbreak{}workflows}, the workflow commands and the \texttt{/\allowbreak{}workflow-\allowbreak{}authoring} skill, stops the keyword, and makes ultracode unavailable; a run already in progress keeps going [19].

\subsection*{Exercise}

Run \texttt{/\allowbreak{}deep-\allowbreak{}research} on a non-sensitive topic with a \texttt{small} size guideline. Require direct sources, a contradiction pass and a source register. Compare the result with a single-agent attempt, and use \texttt{/\allowbreak{}usage} to decide whether the workflow added decision value or only cost.

\emph{A worked solution is given in Appendix M.}

\begingroup
\def\kprows{%
\item A workflow moves the orchestration into a script, so the conversation holds only the final answer.
\item Workflow agents take their mode by the subagent rules, so under auto, \texttt{accept\allowbreak{}Edits} or bypass they run in the session's mode and their edits go through without you. Commit a clean baseline first.
\item Up to 16 concurrent agents by default (1 to 256 by setting) and 1,000 in total per run; the size guideline is advice, the limits are not.
\item Use one only where the lanes are genuinely independent — identical agents on identical evidence manufacture confidence, not corroboration.
}%
\def\kpbody{\begin{minipage}{\textwidth}\subsection*{Key points}\begin{itemize}\kprows\end{itemize}\end{minipage}}%
\begingroup
\def\sloppy{\tolerance 9999\emergencystretch 3em\hfuzz 200pt\vfuzz 200pt}%
\hbadness=10000\vbadness=10000\hfuzz=200pt\vfuzz=200pt
\global\setbox\kpbox=\hbox{\kpbody}%
\endgroup
\par\addvspace{4.2mm}
\ifdim\dimexpr\ht\kpbox+\dp\kpbox\relax>0.30\textheight
  \typeout{HANDBOOK-KEYPOINTS broken \the\dimexpr\ht\kpbox+\dp\kpbox\relax}%
  \subsection*{Key points}
  \begin{itemize}\kprows\end{itemize}
\else
  \typeout{HANDBOOK-KEYPOINTS atomic \the\dimexpr\ht\kpbox+\dp\kpbox\relax}%
  \noindent\kpbody
\fi
\par\addvspace{1.4mm}
\endgroup

\FloatBarrier
\renewcommand{\chaptertitlelabel}{Part V \textperiodcentered\ Chapter 23}
\setcounter{chapter}{22}
\chapter{Session-scoped scheduling with \texttt{/\allowbreak{}loop}}

\texttt{/\allowbreak{}loop} is a bundled skill (alias \texttt{/\allowbreak{}proactive}) that re-runs a prompt while the session stays open [26], [37]. It is not cron. This chapter carries three of the six first-edition corrections, so read it even if you know the feature.

\FloatBarrier
\setcounter{section}{0}
\section{Three forms, three behaviours}

\begingroup
\def\tblrows{%
Interval and prompt & \texttt{/\allowbreak{}loop 5m che\allowbreak{}ck the depl\allowbreak{}oy} & Runs on a fixed cron schedule \\
Prompt only & \texttt{/\allowbreak{}loop check \allowbreak{}the deploy} & Runs at an interval Claude chooses each iteration, between one minute and one hour, printing the delay and the reason \\
Interval only, or nothing & \texttt{/\allowbreak{}loop} & Runs the \textbf{built-in maintenance prompt}, or your \texttt{loop.\allowbreak{}md} if one exists \\
}%
\def\tblbody{\begin{minipage}{\textwidth}\boxcaption{Table 20 — What \texttt{/\allowbreak{}loop} does with each combination of arguments}
{\footnotesize\begin{tabular}{L{26.4mm}L{23.6mm}L{89.9mm}}
\toprule
\textbf{You provide} & \textbf{Example} & \textbf{What happens} \\
\midrule
\tblrows
\bottomrule\end{tabular}}\end{minipage}}%
\begingroup
\def\sloppy{\tolerance 9999\emergencystretch 3em\hfuzz 200pt\vfuzz 200pt}%
\hbadness=10000\vbadness=10000\hfuzz=200pt\vfuzz=200pt
\global\setbox\tblbox=\hbox{\tblbody}%
\endgroup
\par\addvspace{2.6mm}
\ifdim\dimexpr\ht\tblbox+\dp\tblbox\relax>0.55\textheight
  \typeout{HANDBOOK-TABLE broken \the\dimexpr\ht\tblbox+\dp\tblbox\relax}%
  \tabcaption{Table 20 — What \texttt{/\allowbreak{}loop} does with each combination of arguments}
  {\footnotesize\begin{longtable}{L{26.4mm}L{23.6mm}L{89.9mm}}
  \toprule
\textbf{You provide} & \textbf{Example} & \textbf{What happens} \\
\midrule\endfirsthead
  \multicolumn{3}{@{}l@{}}{%
  \sffamily\footnotesize\itshape\color{inkgrey}Table 20 — What \texttt{/\allowbreak{}loop} does with each combination of arguments \textemdash\ continued}\\[1.2mm]
  \toprule
\textbf{You provide} & \textbf{Example} & \textbf{What happens} \\
\midrule\endhead
  \bottomrule\endfoot
  \bottomrule\endlastfoot
  \tblrows
  \end{longtable}}%
\else
  \typeout{HANDBOOK-TABLE atomic \the\dimexpr\ht\tblbox+\dp\tblbox\relax}%
  \noindent\tblbody
\fi
\par\addvspace{2.6mm}
\endgroup

\begin{calloutbox}{palebrass}{brassdark}{2.0mm}
\textbf{CAUTION — M01.} A bare \texttt{/\allowbreak{}loop} does \textbf{not} list your scheduled tasks. It starts an autonomous maintenance loop that continues unfinished work, tends the current branch's pull request — review comments, failed CI, merge conflicts — and runs cleanup passes such as bug hunts when nothing else is pending [26]. Each run goes through your session's permission checks before it pushes or deletes anything, so what it may do unattended is whatever your permission mode allows — and this is still autonomous execution you did not ask for. To list tasks, ask Claude in natural language, or use \texttt{Cron\allowbreak{}List}.
\end{calloutbox}

Replace the built-in prompt with your own by creating \texttt{.\allowbreak{}claude/\allowbreak{}loop.\allowbreak{}md} (project, takes precedence) or \texttt{\textasciitilde{}/\allowbreak{}.\allowbreak{}claude/\allowbreak{}loop.\allowbreak{}md} (user). It defines one default prompt for a bare \texttt{/\allowbreak{}loop}, and is ignored whenever you supply a prompt yourself. Edits take effect on the next iteration; content beyond 25,000 bytes is truncated [26].

\FloatBarrier
\setcounter{section}{1}
\section{Operational constraints}

\begin{itemize}
\item Claude Code must be \textbf{running and idle}. A scheduled prompt fires between turns, never mid-response.
\item Minimum interval one minute. Units are \texttt{s}, \texttt{m}, \texttt{h}, \texttt{d}; seconds round up to the nearest minute; intervals that do not map to a clean cron step, such as \texttt{7m} or \texttt{90m}, are rounded and Claude tells you what it chose.
\item \textbf{Jitter is specified, not random.} A recurring task starts each run late by up to half the gap to its next run, and by at most 30 minutes; a task with evenly spaced runs is late by the same amount every time. The exception is a task that runs every five minutes, written \texttt{*/\allowbreak{}5 * * * *}, which is what \texttt{/\allowbreak{}loop 5m} creates: it is not delayed, and each run starts 4 minutes 45 seconds after the previous one so that a five-minute prompt cache has not expired. A one-shot scheduled for \texttt{:\allowbreak{}00} or \texttt{:\allowbreak{}30} runs up to 90 seconds early, and no other one-shot is shifted. \emph{Mitigation, for one-shots only:} schedule off the hour and half hour, for example \texttt{3 9 * * *}. For a recurring task, moving the minute moves the run and the delay is still added on top. The rows under a \texttt{Cron\allowbreak{}Create} or \texttt{Cron\allowbreak{}List} call show the next run time \emph{(requires 2.1.299 or later)}.
\item \textbf{No catch-up.} A task whose time passes while Claude is busy fires once when Claude becomes idle, not once per missed interval.
\item \textbf{Seven-day expiry.} Recurring tasks expire seven days after creation, firing one final time. It applies to a self-paced loop too.
\item \textbf{Fifty tasks} per session, each with an eight-character ID.
\item Times are interpreted in \textbf{your local timezone}.
\item With feature-flag fetching off, a task you ask to keep across sessions is stored in the project's \texttt{.\allowbreak{}claude/\allowbreak{}scheduled\_\allowbreak{}tasks.\allowbreak{}json}; scheduling it fails if that directory or file is a symlink, and a copy of the file in another folder is listed there but does not run.
\end{itemize}

All of the above from [26].

\FloatBarrier
\setcounter{section}{2}
\section{Lifecycle, and what actually stops a loop}

This is E2. Session-scoped does not mean session-bound:

\begin{itemize}
\item \texttt{Esc} while a \textbf{self-paced} \texttt{/\allowbreak{}loop} is waiting stops it, and a \texttt{Pending wak\allowbreak{}eup cancell\allowbreak{}ed} notice confirms it. \texttt{Ctrl+C} while Claude is idle stops it too, except in a background session you have attached to: there, pressing \texttt{Ctrl+C} twice detaches and leaves the loop running, and \texttt{Esc} is what stops it \emph{(requires 2.1.295 or later)}. A \textbf{fixed-interval} loop such as \texttt{/\allowbreak{}loop 5m} is not stopped by \texttt{Esc} or \texttt{Ctrl+C}, and neither is a task you scheduled by asking Claude directly: ask Claude to cancel it, or it runs until its seven days elapse.
\item \textbf{Backgrounding the session carries \texttt{/\allowbreak{}loop} tasks into a background session that keeps running without a terminal.}
\item Starting a fresh conversation with \texttt{/\allowbreak{}clear} cleared all session-scoped tasks according to the page Version 1.0 was built from; the current page no longer says so, so treat it as \textbf{UNVERIFIED} (Appendix J) [26].
\item \texttt{claude -\allowbreak{}-\allowbreak{}resume} or \texttt{-\allowbreak{}-\allowbreak{}continue} \textbf{restores} tasks scheduled with \texttt{Cron\allowbreak{}Create}: recurring tasks that have not expired and one-shot tasks whose time has not passed. A self-paced \texttt{/\allowbreak{}loop} is not restored; run \texttt{/\allowbreak{}loop} again. Background Bash and monitor tasks are never restored, and a restored recurring task does not fire for the runs it missed.
\item In a background session, a self-paced loop stops if its next wakeup falls due while the session's process is not running, as after a crash; the transcript says so when it restarts.
\item In self-paced mode Claude can end the loop itself by calling \texttt{Schedule\allowbreak{}Wakeup} with \texttt{stop:\allowbreak{} true}. What happens when an iteration ends without rescheduling or stopping is \textbf{UNVERIFIED}, because two primary sources disagree. The scheduled-tasks page describes one fallback wakeup about 20 minutes later, the loop ending if that iteration does not reschedule either [26]. The release notes for 2.1.319 say the loop now ends when a turn finishes without rescheduling, with \texttt{CLAUDE\_\allowbreak{}CODE\_\allowbreak{}LOOP\_\allowbreak{}KEEPALIVE=\allowbreak{}1} keeping the fallback [57]. Do not rely on either; Appendix J gives the check.
\end{itemize}

Underneath, Claude uses \texttt{Cron\allowbreak{}Create}, \texttt{Cron\allowbreak{}List} and \texttt{Cron\allowbreak{}Delete}, which accept standard five-field cron expressions; extended syntax such as \texttt{L}, \texttt{W}, \texttt{?} and name aliases is not supported [26]. \texttt{CLAUDE\_\allowbreak{}CODE\_\allowbreak{}DISABLE\_\allowbreak{}CRON=\allowbreak{}1} disables the scheduler entirely, and \texttt{/\allowbreak{}loop} with it: tasks already scheduled stop firing.

\textbf{Provider differences.} Self-paced intervals and the built-in maintenance prompt now work on every provider and with feature-flag fetching turned off. On Amazon Bedrock, Claude Platform on AWS, Google Cloud's Agent Platform and Microsoft Foundry, and with fetching off, both need 2.1.248 or later; an earlier client there ran a prompt with no interval on a fixed schedule and answered a bare \texttt{/\allowbreak{}loop} with its usage message [26].

\textbf{Skills on a scheduled fire.} Only skills Claude may invoke on its own actually execute. Built-in commands, skills marked \texttt{disable-\allowbreak{}model-\allowbreak{}invocation:\allowbreak{} true} — including the bundled \texttt{/\allowbreak{}verify} — skills withheld by \texttt{skill\allowbreak{}Overrides} or a \texttt{Skill} deny rule, and MCP prompts all arrive as plain text instead [26].

\FloatBarrier
\setcounter{section}{3}
\section{The Monitor alternative}

When you ask for a dynamic \texttt{/\allowbreak{}loop} schedule, Claude may use the \textbf{Monitor tool} instead, in a session where that tool is available. Monitor runs a background script and streams each output line back, avoiding polling altogether; it is usually more token-efficient and more responsive than re-running a prompt on an interval [26], [72]. If you find yourself polling a log, a build, or a queue, ask for a monitor rather than a loop.

\FloatBarrier
\setcounter{section}{4}
\section{Suitable and unsuitable uses}

Suitable: checking a long build; re-running a read-only local verification; reminding the active session to inspect a queue; reporting a non-critical condition during an attended work period.

Unsuitable: anything with a service-level commitment; security, medical or emergency monitoring; payments or trading; anything needing exact timing or catch-up; unattended destructive repair.

Always state what to do when the condition is missing, ambiguous or unsafe. "Report only" is a good default.

\begingroup
\def\kprows{%
\item A bare \texttt{/\allowbreak{}loop} starts an autonomous maintenance loop; it does not list your tasks.
\item Backgrounding carries loops into a background session, and resume restores unexpired scheduled ones, but not a self-paced loop.
\item Jitter is specified: a recurring task runs late by up to half its interval, at most 30 minutes — a five-minute task excepted — and only a one-shot can avoid its shift by being scheduled off \texttt{:\allowbreak{}00} and \texttt{:\allowbreak{}30}.
\item \texttt{Esc} stops a self-paced loop and not a fixed-interval one; cancel that by asking.
\item For a stream to watch rather than a state to poll, ask for the Monitor tool instead.
}%
\def\kpbody{\begin{minipage}{\textwidth}\subsection*{Key points}\begin{itemize}\kprows\end{itemize}\end{minipage}}%
\begingroup
\def\sloppy{\tolerance 9999\emergencystretch 3em\hfuzz 200pt\vfuzz 200pt}%
\hbadness=10000\vbadness=10000\hfuzz=200pt\vfuzz=200pt
\global\setbox\kpbox=\hbox{\kpbody}%
\endgroup
\par\addvspace{4.2mm}
\ifdim\dimexpr\ht\kpbox+\dp\kpbox\relax>0.30\textheight
  \typeout{HANDBOOK-KEYPOINTS broken \the\dimexpr\ht\kpbox+\dp\kpbox\relax}%
  \subsection*{Key points}
  \begin{itemize}\kprows\end{itemize}
\else
  \typeout{HANDBOOK-KEYPOINTS atomic \the\dimexpr\ht\kpbox+\dp\kpbox\relax}%
  \noindent\kpbody
\fi
\par\addvspace{1.4mm}
\endgroup

\FloatBarrier
\renewcommand{\chaptertitlelabel}{Part V \textperiodcentered\ Chapter 24}
\setcounter{chapter}{23}
\chapter{Cloud routines and desktop scheduled tasks}

Three scheduling mechanisms exist and they are not interchangeable. Choose by what must be true when the work runs.

A naming note first. The documentation now calls a cloud routine a \textbf{cloud scheduled task}, "formerly a routine", and the earlier name survives in the web address \texttt{claude.\allowbreak{}ai/\allowbreak{}code/\allowbreak{}routines}, the \texttt{/\allowbreak{}routines} alias, the \texttt{/\allowbreak{}fire} endpoint path and its beta header, the \texttt{<routine-\allowbreak{}fire-\allowbreak{}payload>} tag and the organisation toggle [20]. This chapter keeps \emph{routine} for the cloud kind, because the bare phrase "scheduled task" now names both it and the Desktop kind.

\begingroup
\def\tblrows{%
Runs on & Anthropic cloud, or your organisation's self-hosted environment & Your machine, while the app is open and the machine awake & Your machine \\
Requires machine on & No & Yes & Yes \\
Requires open session & No & No & Yes \\
Persists across restarts & Yes & Yes & Restored on \texttt{-\allowbreak{}-\allowbreak{}resume} if unexpired; a self-paced loop is not \\
Access to local files & No — a fresh clone & Yes & Yes \\
MCP & Connectors configured per task & Config files and connectors & Inherits from the session \\
Permission prompts & \textbf{Auto mode by default} where available — the classifier decides; ask rules and repeated blocks are among what stops a run & Configurable per task & Inherits from the session \\
Minimum interval & \textbf{1 hour} & 1 hour in the forms; 1 minute when Claude sets it & 1 minute \\
}%
\def\tblbody{\begin{minipage}{\textwidth}\boxcaption{Table 21 — Comparing the three scheduling mechanisms}
{\footnotesize\begin{tabular}{L{24.6mm}L{48.9mm}L{35.7mm}L{27.3mm}}
\toprule
\textbf{} & \textbf{Cloud routine} & \textbf{Desktop scheduled task} & \textbf{\texttt{/\allowbreak{}loop}} \\
\midrule
\tblrows
\bottomrule\end{tabular}}\end{minipage}}%
\begingroup
\def\sloppy{\tolerance 9999\emergencystretch 3em\hfuzz 200pt\vfuzz 200pt}%
\hbadness=10000\vbadness=10000\hfuzz=200pt\vfuzz=200pt
\global\setbox\tblbox=\hbox{\tblbody}%
\endgroup
\par\addvspace{2.6mm}
\ifdim\dimexpr\ht\tblbox+\dp\tblbox\relax>0.55\textheight
  \typeout{HANDBOOK-TABLE broken \the\dimexpr\ht\tblbox+\dp\tblbox\relax}%
  \tabcaption{Table 21 — Comparing the three scheduling mechanisms}
  {\footnotesize\begin{longtable}{L{24.6mm}L{48.9mm}L{35.7mm}L{27.3mm}}
  \toprule
\textbf{} & \textbf{Cloud routine} & \textbf{Desktop scheduled task} & \textbf{\texttt{/\allowbreak{}loop}} \\
\midrule\endfirsthead
  \multicolumn{4}{@{}l@{}}{%
  \sffamily\footnotesize\itshape\color{inkgrey}Table 21 — Comparing the three scheduling mechanisms \textemdash\ continued}\\[1.2mm]
  \toprule
\textbf{} & \textbf{Cloud routine} & \textbf{Desktop scheduled task} & \textbf{\texttt{/\allowbreak{}loop}} \\
\midrule\endhead
  \bottomrule\endfoot
  \bottomrule\endlastfoot
  \tblrows
  \end{longtable}}%
\else
  \typeout{HANDBOOK-TABLE atomic \the\dimexpr\ht\tblbox+\dp\tblbox\relax}%
  \noindent\tblbody
\fi
\par\addvspace{2.6mm}
\endgroup

Source: [20], [26], [93].

Figure 7 reduces the choice to the question that actually settles it: what has to be true already, at the moment the work runs, with nobody watching.

\begin{figure}[tbp]
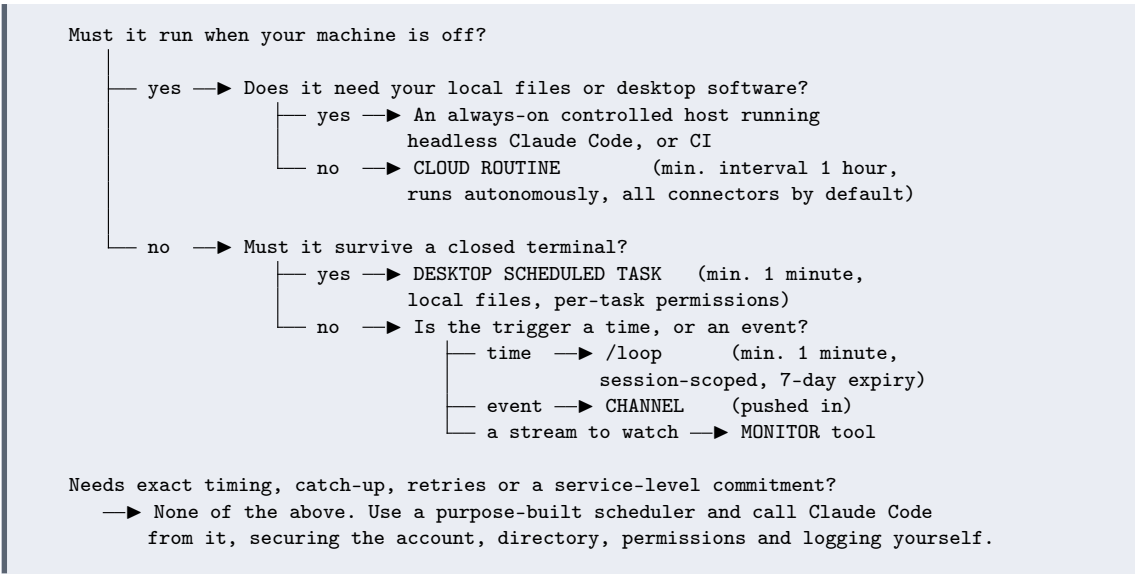

\begin{codefig}{8.0}{9.4}
\cl{~~~Must~it~run~when~your~machine~is~off?}
\cl{~~~~~~\pmboxdrawuni{2502}}
\cl{~~~~~~\pmboxdrawuni{251C}\pmboxdrawuni{2500}\pmboxdrawuni{2500}~yes~\pmboxdrawuni{2500}\pmboxdrawuni{2500}▶~Does~it~need~your~local~files~or~desktop~software?}
\cl{~~~~~~\pmboxdrawuni{2502}~~~~~~~~~~~~~~\pmboxdrawuni{251C}\pmboxdrawuni{2500}\pmboxdrawuni{2500}~yes~\pmboxdrawuni{2500}\pmboxdrawuni{2500}▶~An~always-on~controlled~host~running}
\cl{~~~~~~\pmboxdrawuni{2502}~~~~~~~~~~~~~~\pmboxdrawuni{2502}~~~~~~~~~~~headless~Claude~Code,~or~CI}
\cl{~~~~~~\pmboxdrawuni{2502}~~~~~~~~~~~~~~\pmboxdrawuni{2514}\pmboxdrawuni{2500}\pmboxdrawuni{2500}~no~~\pmboxdrawuni{2500}\pmboxdrawuni{2500}▶~CLOUD~ROUTINE~~~~~~~~(min.~interval~1~hour,}
\cl{~~~~~~\pmboxdrawuni{2502}~~~~~~~~~~~~~~~~~~~~~~~~~~auto~mode~where~available,~all~connectors)}
\cl{~~~~~~\pmboxdrawuni{2502}}
\cl{~~~~~~\pmboxdrawuni{2514}\pmboxdrawuni{2500}\pmboxdrawuni{2500}~no~~\pmboxdrawuni{2500}\pmboxdrawuni{2500}▶~Must~it~survive~a~closed~terminal?}
\cl{~~~~~~~~~~~~~~~~~~~~~\pmboxdrawuni{251C}\pmboxdrawuni{2500}\pmboxdrawuni{2500}~yes~\pmboxdrawuni{2500}\pmboxdrawuni{2500}▶~DESKTOP~SCHEDULED~TASK~~~(app~open,~machine~awake,}
\cl{~~~~~~~~~~~~~~~~~~~~~\pmboxdrawuni{2502}~~~~~~~~~~~local~files,~per-task~permissions)}
\cl{~~~~~~~~~~~~~~~~~~~~~\pmboxdrawuni{2514}\pmboxdrawuni{2500}\pmboxdrawuni{2500}~no~~\pmboxdrawuni{2500}\pmboxdrawuni{2500}▶~Is~the~trigger~a~time,~or~an~event?}
\cl{~~~~~~~~~~~~~~~~~~~~~~~~~~~~~~~~~~~~\pmboxdrawuni{251C}\pmboxdrawuni{2500}\pmboxdrawuni{2500}~time~~\pmboxdrawuni{2500}\pmboxdrawuni{2500}▶~/loop~~~~~~(min.~1~minute,}
\cl{~~~~~~~~~~~~~~~~~~~~~~~~~~~~~~~~~~~~\pmboxdrawuni{2502}~~~~~~~~~~~~~session-scoped,~7-day~expiry)}
\cl{~~~~~~~~~~~~~~~~~~~~~~~~~~~~~~~~~~~~\pmboxdrawuni{251C}\pmboxdrawuni{2500}\pmboxdrawuni{2500}~event~\pmboxdrawuni{2500}\pmboxdrawuni{2500}▶~CHANNEL~~~~(pushed~in)}
\cl{~~~~~~~~~~~~~~~~~~~~~~~~~~~~~~~~~~~~\pmboxdrawuni{2514}\pmboxdrawuni{2500}\pmboxdrawuni{2500}~a~stream~to~watch~\pmboxdrawuni{2500}\pmboxdrawuni{2500}▶~MONITOR~tool}
\cl{}
\cl{~~~Needs~exact~timing,~catch-up,~retries~or~a~service-level~commitment?}
\cl{~~~~~~\pmboxdrawuni{2500}\pmboxdrawuni{2500}▶~None~of~the~above.~Use~a~purpose-built~scheduler~and~call~Claude~Code}
\cl{~~~~~~~~~~from~it,~securing~the~account,~directory,~permissions~and~logging~yourself.}
\end{codefig}
\figcaption{Figure 7 — Choosing a scheduling mechanism. Start from what must be true when the work runs, not from how often it runs.}
\end{figure}

\FloatBarrier
\setcounter{section}{0}
\section{What a routine actually is}

\begin{calloutbox}{palegrey}{quoterule}{1.2mm}
\textbf{Routines are in research preview.} Behaviour, limits and the API surface may change [20].
\end{calloutbox}

A routine is a saved configuration — a prompt, one or more repositories, an environment, and a set of connectors — with one or more triggers attached [20]:

\begin{itemize}
\item \textbf{Scheduled} — hourly, daily, weekly or a custom cron expression, or a single future time.
\item \textbf{API} — an HTTP POST to the routine's own \texttt{/\allowbreak{}fire} endpoint with a bearer token.
\item \textbf{GitHub} — pull-request, release and issue events, with pull-request filters on author, title, body, base and head branch, labels, draft state and merged state.
\end{itemize}

A single routine can combine all three, and a routine needs no schedule at all. This is M19 in Table 1: routines are widely described as scheduled sessions, which is a third of the feature. They are available on Pro, Max, Ultra, Team and Enterprise plans, run on Anthropic-managed infrastructure or, when routed there, on the organisation's self-hosted environment, and belong to your individual account: they are not shared with teammates [20].

Create them at \texttt{claude.\allowbreak{}ai/\allowbreak{}code/\allowbreak{}routines}, in the Desktop app's \textbf{Code} tab under \textbf{Scheduled} (choosing \textbf{Cloud}, not \textbf{Local}, where the app asks), or from the CLI with \texttt{/\allowbreak{}schedule} (alias \texttt{/\allowbreak{}routines}), which also supports \texttt{/\allowbreak{}schedule li\allowbreak{}st}, \texttt{/\allowbreak{}schedule up\allowbreak{}date} and \texttt{/\allowbreak{}schedule ru\allowbreak{}n} [20]. Custom cron intervals are set in the edit dialog or through \texttt{/\allowbreak{}schedule up\allowbreak{}date}; expressions more frequent than hourly are rejected. A five-field expression is read in UTC unless the field carries a \texttt{CRON\_\allowbreak{}TZ=\allowbreak{}} prefix, and a run set exactly on the hour can start several minutes late [20]. An API trigger and its token can be managed only on the web, not from the CLI [20].

The GitHub trigger needs the Claude GitHub App installed on the repository; \texttt{/\allowbreak{}web-\allowbreak{}setup} grants clone access but does not install the app or enable webhook delivery. Every event starts its own session, and webhook events beyond the per-routine and per-account hourly caps are dropped [20].

\FloatBarrier
\setcounter{section}{1}
\section{Why a routine deserves a design review}

\textbf{A routine run is unattended, and you do not choose its permission mode.} A run that starts a new session on Anthropic's cloud begins in auto mode unless auto mode is unavailable, and the classifier allows or blocks what would otherwise wait for you [20]. A run still stops and waits in two cases: an action that prompts even in auto mode, such as one matching an ask rule, and blocks that reach auto mode's repeated-block thresholds, after which a prompt waits in the run's session until someone opens it and answers. Where auto mode is unavailable, because the organisation turned it off or the model does not support it, the run keeps the mode it launched in [20]. In a project, a run can continue a thread instead of starting a session, and then keeps whatever mode the thread is in; a task in a self-hosted environment follows that environment's own approval configuration [20].

Version 1.0 of this book said there are no approval prompts during a run, which was what the page said in August. It is now wrong in both directions: the classifier stands between the run and most actions, and an ask rule can stop a run that nobody is watching. \textbf{An ask rule is therefore the checkpoint you can configure inside a routine}, and a run that hits it waits silently until someone looks. It is not the only thing that can stop a run: the page's list of stops is given as examples, and the actions no mode auto-approves — connector tools an organisation set to \texttt{ask}, MCP tools that require user interaction, and the rest — prompt in auto mode too [20], [33]. The saved prompt is the run's assignment, but it is not live input from you and cannot act as approval or consent for anything during the run [20].

Four consequences follow, and all four are easy to miss:

\begin{enumerate}
\item \textbf{All of your connected connectors are included by default.} Claude can use an included connector during a run, writes included, subject to your permission rules and the classifier. Remove everything the routine does not need [20].
\item \textbf{It acts as you.} Commits and pull requests carry your GitHub user; Slack messages, tickets and other connector actions use your linked accounts [20].
\item \textbf{Branch protection is GitHub's, applied to the access you connected.} Claude pushes to a \texttt{claude/\allowbreak{}}-prefixed branch unless the prompt directs it elsewhere, and the control over which branches a run can push to is a branch protection rule or ruleset on GitHub. For runs on Anthropic's infrastructure, GitHub applies those rules to your connected access, \textbf{so a rule that access can bypass does not block a run's push} [20]. Version 1.0 described a check of Claude Code's own that rejected a push to a branch someone else had an open pull request from; the current page describes no such check.
\item \textbf{Environment variables are visible to anyone who uses the environment}, and so is the setup script. In an Anthropic-hosted environment, store an API key as a \emph{network secret}, which the agent proxy attaches to requests for the hosts you list without the key entering the session [94].
\end{enumerate}

The \textbf{Default} environment uses \emph{Limited} network access with \textbf{Allow package managers} checked, reaching a default allowlist of package registries, cloud provider APIs, container registries and common development domains; requests outside it fail with \texttt{403} and \texttt{x-\allowbreak{}deny-\allowbreak{}reason:\allowbreak{} host\_\allowbreak{}not\_\allowbreak{}allowed}. Connector traffic routes through Anthropic's servers and does not need allowlist changes [20], [94]. An environment saved earlier as \emph{Trusted} opens as Limited with package managers allowed [94].

During a run in auto mode Claude also publishes an artifact without asking once the classifier allows the publish; a new artifact is visible only to you until you share it [20].

\FloatBarrier
\setcounter{section}{2}
\section{Fire-payload text is untrusted by design}

Text supplied with an API trigger or with \textbf{Run now} arrives wrapped in a \texttt{<routine-\allowbreak{}fire-\allowbreak{}payload>} block that labels it as untrusted data and tells Claude not to follow instructions inside it unless the routine's own prompt says to [20]. A routine must therefore \emph{opt in} — "investigate the alert described in the routine-fire-payload block" — or the text is inert context. This is a deliberate injection control: anyone holding the bearer token can send \texttt{text}, and the wrapper makes a leaked token's text arrive labelled as untrusted data. It is an instruction to the model, not an enforced separation, so a leaked token still matters.

The \texttt{/\allowbreak{}fire} endpoint ships under the \texttt{experimenta\allowbreak{}l-\allowbreak{}cc-\allowbreak{}routine-\allowbreak{}2026-\allowbreak{}04-\allowbreak{}01} beta header; request and response shapes may change, breaking changes ship behind new dated header versions, and the two most recent previous versions keep working [20]. Each routine has its own token, scoped to firing that routine and shown once; rotating it with \textbf{Regenerate} stops the old one at once, and deleting it removes the API trigger [20].

\FloatBarrier
\setcounter{section}{3}
\section{Before you activate one}

\begin{enumerate}
\item Remove every connector the task does not need.
\item Use read-only or least-privilege service accounts.
\item Specify the repository and the branch explicitly.
\item Prohibit direct production push and external publication in the prompt.
\item Define the safe failure state.
\item Configure notifications.
\item Run the prompt interactively, against test data, first.
\item Inspect the first several scheduled runs by opening them.
\end{enumerate}

\begin{calloutbox}{palebrass}{brassdark}{2.0mm}
\textbf{CAUTION:} a run marked \textbf{Succeeded} means the session ended without Claude Code reporting a failure — no infrastructure error, nothing more. It does not mean the task succeeded [20]. Blocked network requests, missing connector tools and work Claude could not complete appear inside the transcript, not on the row. Never treat the outcome as assurance. A run whose session you archive is relabelled \textbf{Cancelled}, even if it succeeded.
\end{calloutbox}

Routines draw down subscription usage like interactive sessions, and each way of starting a run also has an hourly limit with no overage: 100 scheduled runs an hour per account, one-off runs included; 30 an hour per routine shared by \textbf{Run now}, API fires and re-arming a one-off; 100 an hour per account for \textbf{Run now} and re-arming together; and a separate 100 an hour per account for API fires [20]. Version 1.0 described a daily per-account cap that one-off runs did not count against; the page no longer does. A scheduled run over its limit waits; a \textbf{Run now} or an API fire over its limit fails [20].

If the GitHub connection is missing or expired when a run is due, the routine skips runs for up to 72 hours and then turns itself off; it has to be turned back on after you reconnect [20]. A paused subscription puts routines on hold. Team and Enterprise Owners can turn routines off organisation-wide from the \textbf{Remote sessions} capabilities, which stops existing ones and hides \texttt{/\allowbreak{}schedule} in the CLI from 2.1.227; in an Enterprise organisation with custom roles, a member also needs a role that includes the permission [20].

\FloatBarrier
\setcounter{section}{4}
\section{Desktop scheduled tasks and headless automation}

A Desktop scheduled task runs on your machine with access to local files and tools, with its own permission mode set per task [93]. It is the right choice when the work needs local resources and must survive a closed terminal — but not a closed app: tasks run only while the Desktop app is open and the computer is awake, and a run the machine sleeps through is skipped [93]. On waking, Desktop starts exactly one catch-up run for the most recent time missed in the last seven days and discards the rest, so a task set for 9am can run at 11pm; put the timing guard in the prompt [93].

The forms reject a custom cron expression more frequent than hourly, but Claude, asked in a session, is not held to that limit and can set an interval of minutes [93]. By default a run works against the folder as it stands, uncommitted changes included; select \textbf{Worktree} in the task's edit form to give each run its own Git worktree. Allow rules from \texttt{\textasciitilde{}/\allowbreak{}.\allowbreak{}claude/\allowbreak{}settings.\allowbreak{}json} apply to these sessions as well, and the task's prompt lives on disk as \texttt{\textasciitilde{}/\allowbreak{}.\allowbreak{}claude/\allowbreak{}scheduled-\allowbreak{}tasks/\allowbreak{}<task-\allowbreak{}name>/\allowbreak{}SKILL.\allowbreak{}md}, which survives deleting the task [93]. A run that reaches a permission prompt stalls until you open its session and answer, and an MCP tool marked \texttt{requires\allowbreak{}User\allowbreak{}Interaction} stalls every run that calls it [93].

A separate architecture is a local operating-system scheduler invoking headless Claude Code (\texttt{claude -\allowbreak{}p}) [95]. That is a legitimate design, and everything about it — the account it uses, its working directory, its permission rules, its logging, its retries, its cost, and the fact that trust verification is disabled under \texttt{-\allowbreak{}p} [59] — must be secured explicitly. It is not a scheduling feature; it is a system you are building.

\begin{calloutbox}{palebrass}{brassdark}{2.0mm}
\textbf{CAUTION:} never make a research-preview routine the only control for business-critical or safety-critical work.
\end{calloutbox}

\begingroup
\def\kprows{%
\item Routines, now called cloud scheduled tasks, are a research preview that trigger on schedules, API calls and GitHub events.
\item A run that starts a new cloud session begins in auto mode where it is available, with every connected connector by default, under your identity; an ask rule is the checkpoint you can configure, and a run that hits one waits unseen.
\item A run marked Succeeded means no infrastructure error, not that the task succeeded.
\item Minimum interval is one hour, and fire-payload text arrives explicitly labelled as untrusted.
}%
\def\kpbody{\begin{minipage}{\textwidth}\subsection*{Key points}\begin{itemize}\kprows\end{itemize}\end{minipage}}%
\begingroup
\def\sloppy{\tolerance 9999\emergencystretch 3em\hfuzz 200pt\vfuzz 200pt}%
\hbadness=10000\vbadness=10000\hfuzz=200pt\vfuzz=200pt
\global\setbox\kpbox=\hbox{\kpbody}%
\endgroup
\par\addvspace{4.2mm}
\ifdim\dimexpr\ht\kpbox+\dp\kpbox\relax>0.30\textheight
  \typeout{HANDBOOK-KEYPOINTS broken \the\dimexpr\ht\kpbox+\dp\kpbox\relax}%
  \subsection*{Key points}
  \begin{itemize}\kprows\end{itemize}
\else
  \typeout{HANDBOOK-KEYPOINTS atomic \the\dimexpr\ht\kpbox+\dp\kpbox\relax}%
  \noindent\kpbody
\fi
\par\addvspace{1.4mm}
\endgroup

\breakrule

\FloatBarrier
\parttitle{Part VI}{Distributed and cloud operation}
\addcontentsline{toc}{part}{Part VI \textemdash\ Distributed and cloud operation}

\FloatBarrier
\renewcommand{\chaptertitlelabel}{Part VI \textperiodcentered\ Chapter 25}
\setcounter{chapter}{24}
\chapter{Continue local work with Remote Control}

Remote Control lets another signed-in device supervise a Claude Code process running on your own machine. \textbf{Execution stays local}: commands run and files are read and written on your computer. What travels is the conversation — your messages, Claude's responses and the tool activity, which can carry code and file contents — relayed through the Anthropic API over TLS and stored on Anthropic servers while connected, so that it syncs across devices, under the data-usage retention policy [41], [59].

\begin{codeblock}{9.0}{10.6}
\cl{/remote-control~~~~~~~~~~\#~alias~/rc}
\end{codeblock}

\begin{codeblock}{9.0}{10.6}
\cl{claude~-{}-remote-control~research-brief~~~~\#~alias~-{}-rc}
\cl{claude~remote-control~~~~~~~~~~~~~~~~~~~~~\#~server~mode}
\end{codeblock}

It needs a Pro, Max, Ultra, Team or Enterprise plan and a claude.ai login; API keys are not supported, and on Team and Enterprise an Owner must first turn on the Remote Control toggle [41]. Where Enterprise custom roles govern access, your roles need both the cloud sessions permission and the \textbf{Claude Code remote control} permission [41]. It is unavailable through Amazon Bedrock, Google Cloud's Agent Platform or Microsoft Foundry, when \texttt{ANTHROPIC\_\allowbreak{}BASE\_\allowbreak{}URL} points anywhere but \texttt{api.\allowbreak{}anthropic.\allowbreak{}com} (an LLM gateway or a proxy), behind a Claude apps gateway, and when \texttt{CLAUDE\_\allowbreak{}CODE\_\allowbreak{}DISABLE\_\allowbreak{}NONESSENTIA\allowbreak{}L\_\allowbreak{}TRAFFIC}, \texttt{CLAUDE\_\allowbreak{}CODE\_\allowbreak{}DISABLE\_\allowbreak{}FEATURE\_\allowbreak{}FLAGS} or its older name \texttt{DISABLE\_\allowbreak{}GROWTHBOOK} is set. \texttt{DISABLE\_\allowbreak{}TELEMETRY} or \texttt{DO\_\allowbreak{}NOT\_\allowbreak{}TRACK} alone no longer blocks it \emph{(requires 2.1.283 or later; 2.1.304 or later in the Desktop app)}, unless the organisation requires Trusted Devices — but with either set, sessions do not auto-connect through an admin or built-in default [41]. In a directory you have not trusted, server mode asks before it starts, and in the home directory the answer is never saved [41].

Server mode serves several sessions from one process, up to \texttt{-\allowbreak{}-\allowbreak{}capacity} (32 by default). Its \texttt{-\allowbreak{}-\allowbreak{}spawn} mode decides where they run: \texttt{same-\allowbreak{}dir}, the default, puts every session in the current directory, where they can conflict over the same files; \texttt{worktree} gives each on-demand session its own git worktree; \texttt{session} serves exactly one [41]. \texttt{-\allowbreak{}-\allowbreak{}permission-\allowbreak{}mode} sets the sessions' starting mode. The \texttt{-\allowbreak{}-\allowbreak{}sandbox} and \texttt{-\allowbreak{}-\allowbreak{}no-\allowbreak{}sandbox} flags were removed in 2.1.300, so sandboxing for server sessions is turned on in a settings file, and a global flag that would change what the sessions can do, such as \texttt{-\allowbreak{}-\allowbreak{}settings}, makes the server refuse to start rather than be silently dropped [41]. Name sessions clearly so a phone cannot confuse two projects; a session renamed from claude.ai or the Claude app is renamed in the CLI too, from 2.1.221 [31].

The connection is outbound HTTPS only, opens no inbound port, and uses multiple short-lived, narrowly scoped credentials, each limited to a purpose and expiring independently, to limit the blast radius of any single compromise [41], [59]. \texttt{disable\allowbreak{}Remote\allowbreak{}Control} turns the feature off entirely, and organisations with Zero Data Retention or the HIPAA configuration cannot enable it [41].

\textbf{Auto-connect cannot be planted by a repository.} \textbf{Enable Remote Control for all sessions} in \texttt{/\allowbreak{}config}, or \texttt{remote\allowbreak{}Control\allowbreak{}At\allowbreak{}Startup:\allowbreak{} true} in user or managed settings, connects every interactive session at start. In project or local settings Claude Code honours a \texttt{false} and ignores a \texttt{true}, so a checked-in file can turn auto-connect off for a repository but never on. A managed \texttt{true} outranks a \texttt{false} in your user settings, but a project or local \texttt{false} turns auto-connect off even over a managed \texttt{true} [41]. Auto-connect signs in as you, and a session it starts appears only in your own account's apps.

\textbf{Trusted Devices} (beta, off by default) ties viewing or steering a session to an enrolled device and a sign-in no more than 18 hours old, refreshed with Face ID, Touch ID, Windows Hello or a passkey. An Owner turns it on for a Team or Enterprise organisation; on Pro, Max and Ultra you turn it on yourself. It protects only sessions started after it is enabled, and a lost device is revoked from account settings [41]. For a team that lets members steer local sessions from phones, it is the control that makes a stolen, still-signed-in phone less than a remote shell.

\FloatBarrier
\setcounter{section}{0}
\section{Operating rules}

\begin{itemize}
\item Start from a narrow project directory, never a home directory.
\item Keep your normal permission mode and sandboxing. Remoteness is not a reason to widen access. The Claude app does not offer Bypass permissions for a Remote Control session [51].
\item Do not expose a broad shell merely because you are away from the desk.
\item Confirm which machine, repository, branch and account the session controls before you approve anything.
\item Close remote control when supervision is over.
\item Treat the relayed transcript as cloud-stored data for as long as you are connected (Chapter 32).
\end{itemize}

Remote Control is good for answering a question, unblocking a decision, or approving a safe local edit. It is poor for inspecting large diffs, reviewing sensitive commands, or making high-impact approvals on a small screen. Defer those until you can review properly — the small screen is not a reason to lower the bar, it is a reason to wait.

The local Claude Code process must remain running. Close the terminal, quit the app that runs the session, or stop \texttt{claude}, and the session goes offline even though the mobile or web client stays open [41]. Sessions of a server you stopped can be brought back in the same directory for about four hours: \texttt{claude remo\allowbreak{}te-\allowbreak{}control} restores all of them, \texttt{claude remo\allowbreak{}te-\allowbreak{}control -\allowbreak{}-\allowbreak{}continue} the one it started with, and \texttt{claude remo\allowbreak{}te-\allowbreak{}control -\allowbreak{}-\allowbreak{}session-\allowbreak{}id <id>} one by ID. That holds only if no other server ran in the directory and this one was not started with \texttt{-\allowbreak{}-\allowbreak{}no-\allowbreak{}create-\allowbreak{}session-\allowbreak{}in-\allowbreak{}dir}. A network outage is different: an awake server that cannot reach the network for about ten minutes exits, and you start a new session. A session started with \texttt{-\allowbreak{}-\allowbreak{}remote-\allowbreak{}control} or \texttt{/\allowbreak{}remote-\allowbreak{}control} is resumed like any conversation [41].

A connected device can do more than talk. It can stop the session's background subagents and workflows, which stops them on your machine; change the model and the effort level, unless \texttt{CLAUDE\_\allowbreak{}CODE\_\allowbreak{}EFFORT\_\allowbreak{}LEVEL} pins it; read a diff of your changes; and attach files, which Claude Code downloads to your machine [41]. Permission prompts and \texttt{Ask\allowbreak{}User\allowbreak{}Question} questions stay open until answered, but other dialogs forwarded to the device close after five minutes with their no-action default, adjustable with \texttt{dialog\allowbreak{}Expiry} \emph{(requires 2.1.224 or later)} [41]. A line typed on the device that starts with \texttt{!} reaches Claude as a message; it does not run in the shell [41].

\FloatBarrier
\setcounter{section}{1}
\section{Mobile}

The Claude app for iOS and Android is a client for Claude Code sessions, not a place where code runs. From it you reach cloud sessions and projects, a session on your own machine through Remote Control, or the Desktop app through Dispatch, which starts work on your machine from a message and needs a Pro, Max or Ultra plan [51]. \texttt{/\allowbreak{}mobile} (aliases \texttt{/\allowbreak{}ios}, \texttt{/\allowbreak{}android}) shows a download QR code [37]. With Remote Control active, Claude can push a notification to your phone when a task finishes or needs a decision, configured by two \texttt{/\allowbreak{}config} toggles [41]. The same operating rules apply, with the review caveat above sharpened: a phone is an excellent notification surface and a poor audit surface.

\subsection*{Exercise}

Start Remote Control on the practice workspace. From a second device, request a read-only status report. Confirm the output appears locally, then disconnect. Grant no new permissions during the exercise.

\emph{A worked solution is given in Appendix M.}

\begingroup
\def\kprows{%
\item Execution stays local; the transcript is relayed and stored server-side while connected.
\item Keep your normal permission mode — remoteness is not a reason to widen access.
\item A phone is an excellent notification surface and a poor audit surface. Defer high-impact approvals.
\item Closing the local process takes the session offline even though the client stays open; a project file can turn auto-connect off but never on.
}%
\def\kpbody{\begin{minipage}{\textwidth}\subsection*{Key points}\begin{itemize}\kprows\end{itemize}\end{minipage}}%
\begingroup
\def\sloppy{\tolerance 9999\emergencystretch 3em\hfuzz 200pt\vfuzz 200pt}%
\hbadness=10000\vbadness=10000\hfuzz=200pt\vfuzz=200pt
\global\setbox\kpbox=\hbox{\kpbody}%
\endgroup
\par\addvspace{4.2mm}
\ifdim\dimexpr\ht\kpbox+\dp\kpbox\relax>0.30\textheight
  \typeout{HANDBOOK-KEYPOINTS broken \the\dimexpr\ht\kpbox+\dp\kpbox\relax}%
  \subsection*{Key points}
  \begin{itemize}\kprows\end{itemize}
\else
  \typeout{HANDBOOK-KEYPOINTS atomic \the\dimexpr\ht\kpbox+\dp\kpbox\relax}%
  \noindent\kpbody
\fi
\par\addvspace{1.4mm}
\endgroup

\FloatBarrier
\renewcommand{\chaptertitlelabel}{Part VI \textperiodcentered\ Chapter 26}
\setcounter{chapter}{25}
\chapter{Cloud sessions, \texttt{-\allowbreak{}-\allowbreak{}cloud} and \texttt{-\allowbreak{}-\allowbreak{}teleport}}

\begin{calloutbox}{palegrey}{quoterule}{1.2mm}
Cloud sessions are available on Pro, Max, Ultra and Team plans, and to Enterprise users with premium or Chat + Claude Code seats. Where custom roles govern access, starting one also needs a role with the cloud sessions permission [50].
\end{calloutbox}

A cloud session runs on Anthropic-managed infrastructure, or on your organisation's self-hosted environment when routed there. You start one from the browser at \texttt{claude.\allowbreak{}ai/\allowbreak{}code}, the Code tab of the Claude app, the Desktop app's \textbf{Cloud} environments, \texttt{claude -\allowbreak{}-\allowbreak{}cloud}, or a cloud scheduled task. It keeps running after you close your laptop, can be checked or steered from any device, and counts toward your plan's usage limits with no separate charge for the VM [50]. This is a different mechanism from Remote Control, and confusing the two is the most common error in this area: \texttt{-\allowbreak{}-\allowbreak{}cloud} creates a cloud session that runs elsewhere; \texttt{-\allowbreak{}-\allowbreak{}remote-\allowbreak{}control} exposes a \textbf{local} session for monitoring [50].

\FloatBarrier
\setcounter{section}{0}
\section{From terminal to cloud}

\begin{codeblock}{9.0}{10.6}
\cl{claude~-{}-cloud~"Fix~the~authentication~bug~in~src/auth/login.ts"}
\end{codeblock}

The cloud VM clones your current directory's \textbf{GitHub remote at your current branch}, not your local checkout — push first if you have local commits [50]. One repository at a time; the older \texttt{-\allowbreak{}-\allowbreak{}remote} spelling is a deprecated alias. From a repository with no git remote, or a github.com repository the Claude GitHub App is not installed on — even if you connected GitHub with \texttt{/\allowbreak{}web-\allowbreak{}setup} — Claude Code bundles the local repository and uploads it: for a full clone, history across all branches plus uncommitted changes to tracked files, untracked files excluded; on macOS, Linux and WSL a \texttt{-\allowbreak{}-\allowbreak{}filter} partial clone uploads as a working-tree snapshot without history. Over 100 MB it falls back to the current branch, then to a squashed snapshot of the working tree, and fails if that is still too large. The session can push back only where your GitHub connection has push access. Force bundling with \texttt{CCR\_\allowbreak{}FORCE\_\allowbreak{}BUNDLE=\allowbreak{}1} [50].

\begin{calloutbox}{palebrass}{brassdark}{2.0mm}
\textbf{CAUTION:} on macOS, Linux and WSL the upload leaves out uncommitted changes to files named like credentials or keys — \texttt{.\allowbreak{}env}, \texttt{*.\allowbreak{}tfvars}, \texttt{id\_\allowbreak{}rsa}, \texttt{*.\allowbreak{}pem} — and of files a git filter such as Git LFS manages, and names them in a \texttt{Left on thi\allowbreak{}s machine:\allowbreak{}} notice. \textbf{On native Windows it does not}: uncommitted changes to tracked files upload as they are, whatever the file is called. Stash or revert such an edit first [50].
\end{calloutbox}

On macOS, Linux and WSL the upload also needs git 2.31 or later and a plain checkout: a submodule, a \texttt{-\allowbreak{}-\allowbreak{}separate-\allowbreak{}git-\allowbreak{}dir}, \texttt{-\allowbreak{}-\allowbreak{}shared} or \texttt{-\allowbreak{}-\allowbreak{}reference} clone, \texttt{core.\allowbreak{}worktree}, reftable refs, a sparse linked worktree or git configuration kept inside the working tree all stop it with a \texttt{Not uploadi\allowbreak{}ng this wor\allowbreak{}king tree:\allowbreak{}} error naming the cause [50].

Send a follow-up to a running cloud session from any machine where you are logged in. The command queues one message and exits without waiting for a reply:

\begin{codeblock}{9.0}{10.6}
\cl{claude~-p~"your~message"~-{}-cloud~<session-id>}
\end{codeblock}

Without \texttt{-\allowbreak{}p}, \texttt{-\allowbreak{}-\allowbreak{}cloud <sess\allowbreak{}ion-\allowbreak{}id>} is refused with \texttt{Attaching t\allowbreak{}o an existi\allowbreak{}ng cloud se\allowbreak{}ssion is no\allowbreak{}t enabled f\allowbreak{}or your acc\allowbreak{}ount.\allowbreak{}} \texttt{-\allowbreak{}-\allowbreak{}output-\allowbreak{}format json} returns \texttt{\{ok,\allowbreak{} session\_\allowbreak{}id,\allowbreak{} url\}}. \texttt{-\allowbreak{}-\allowbreak{}cloud} needs an Anthropic account and is unavailable through Amazon Bedrock, Google Cloud's Agent Platform or another third-party provider; a gateway set only through \texttt{ANTHROPIC\_\allowbreak{}BASE\_\allowbreak{}URL} does not count as a provider, but you must still sign in with \texttt{claude auth\allowbreak{} login}. The organisation's \texttt{allow\_\allowbreak{}remote\_\allowbreak{}sessions} policy must be on [50].

\FloatBarrier
\setcounter{section}{1}
\section{From cloud to terminal}

\begin{codeblock}{9.0}{10.6}
\cl{claude~-{}-teleport~~~~~~~~~~~~~~~\#~interactive~picker}
\cl{claude~-{}-teleport~<session-id>~~\#~directly}
\end{codeblock}

\begin{codeblock}{9.0}{10.6}
\cl{/teleport~~~~~~\#~alias~/tp,~from~inside~a~session}
\end{codeblock}

\texttt{/\allowbreak{}tasks} then \texttt{t}, or \textbf{Open in > Terminal} on claude.ai/code, reach the same place, and \texttt{/\allowbreak{}teleport} typed inside the cloud session replies with the exact command to run [50].

\texttt{-\allowbreak{}-\allowbreak{}teleport} verifies the repository, fetches and checks out the cloud session's branch, and loads the conversation locally — except what the cloud session had already compacted, which arrives as the summary only [50]. It is \textbf{distinct from \texttt{-\allowbreak{}-\allowbreak{}resume}}: \texttt{-\allowbreak{}-\allowbreak{}resume} reopens a conversation from this machine's local history and does not list cloud sessions [50]. The terminal gets its own copy: new work stays local and does not appear back in the cloud session. To keep steering from a phone afterwards, start \texttt{/\allowbreak{}remote-\allowbreak{}control} in the local session [50].

Teleport requires a clean git state, a checkout of the same repository (not a fork), the branch pushed to the remote, and the same claude.ai account. The branch fetch never waits for input: if git or ssh would ask for a password, a passphrase or a new host key, the fetch fails, so load the key into \texttt{ssh-\allowbreak{}agent} and fetch once by hand first [50].

\FloatBarrier
\setcounter{section}{2}
\section{GitHub access, and what it actually grants}

Two methods, and they reach different repositories [50]:

\begin{itemize}
\item \textbf{The Claude GitHub App}, authorised during web onboarding, reaches any public repository and the private repositories the App is installed on.
\item \textbf{\texttt{/\allowbreak{}web-\allowbreak{}setup}} sends your local \texttt{gh} CLI token to your Claude account and reaches any repository that token can access, whether or not the App is installed.
\end{itemize}

\begin{calloutbox}{palebrass}{brassdark}{2.0mm}
\textbf{CAUTION:} \textbf{\texttt{/\allowbreak{}web-\allowbreak{}setup} makes the token's reach the session's reach.} Connected that way, cloud sessions can be started on any repository your \texttt{gh} token can access, so a broad token exposes every private repository it covers. For installation-based control over private repositories, use the App and install it only where cloud sessions should work. Version 1.0 said that either method reached everything the connecting account could see; the November page scopes the App to public repositories and its installations [50]. Auto-fix and project threads need the App on the repository whichever method you used.
\end{calloutbox}

Inside a running session the reach is narrower again: the GitHub proxy serves API requests only for the repositories attached to that session, and answers others with a 403 reading \texttt{Git\allowbreak{}Hub access \allowbreak{}to … is not\allowbreak{} enabled fo\allowbreak{}r this sess\allowbreak{}ion} — source archives included, even of a public repository. Committed files of public repositories still arrive through \texttt{raw.\allowbreak{}githubuserc\allowbreak{}ontent.\allowbreak{}com} where the environment's network access allows it [94].

In Anthropic-hosted environments the GitHub credential is stored encrypted on Anthropic's servers and never enters the session's VM: GitHub traffic goes through a proxy that attaches it on the server side [50], [59].

On Team and Enterprise, an organisation setting called \textbf{Quick setup}, off by default, governs the shortcuts: while it is off \texttt{/\allowbreak{}web-\allowbreak{}setup} is hidden and onboarding prompts for the App and shows the environment form; an Owner turns it on in the organisation's Claude Code settings [50]. Organisations with Zero Data Retention or the HIPAA configuration cannot use \texttt{/\allowbreak{}web-\allowbreak{}setup} or any other cloud session feature, and those using customer-managed encryption keys cannot start cloud sessions [42], [50].

\FloatBarrier
\setcounter{section}{3}
\section{Auto-fix, and a real hazard}

Claude can watch a pull request and respond automatically to CI failures and review comments. It needs the Claude GitHub App on the repository. Turn it on from the pull request's three-dot menu in the session's \textbf{Changes} card, with \texttt{/\allowbreak{}autofix-\allowbreak{}pr} from the branch, by asking in the mobile app, or by pasting a PR URL into a session and asking Claude to auto-fix it [50]. Claude makes a change when it is confident and the change does not conflict with earlier instructions, asks when a comment is ambiguous or architecturally significant, and notes duplicates.

\begin{calloutbox}{palebrass}{brassdark}{2.0mm}
\textbf{CAUTION:} Claude may reply to review-comment threads on GitHub using \textbf{your} account, labelled as coming from Claude Code. If your repository uses comment-triggered automation — Atlantis, Terraform Cloud, or custom Actions running on \texttt{issue\_\allowbreak{}comment} — Claude replying on your behalf can trigger those workflows. Review your repository automation before enabling auto-fix, and disable it where a comment can deploy infrastructure [50].
\end{calloutbox}

GitHub emits no webhook when the base branch advances into a merge conflict, so auto-fix cannot react to conflicts; open the session and ask for a rebase [50].

\FloatBarrier
\setcounter{section}{4}
\section{Sharing a cloud session}

On Enterprise and Team accounts the options are \textbf{Private} and \textbf{Team}, with repository-access verification enabled by default. On Max, Ultra and Pro they are \textbf{Private} and \textbf{Public}, where public means visible to any signed-in claude.ai user and \textbf{repository access verification is not enabled by default} [50].

\begin{calloutbox}{palebrass}{brassdark}{2.0mm}
\textbf{CAUTION:} a session may contain code and credentials from private repositories. Check before sharing. The settings that require repository access and hide your name are under Settings → Claude Code → Sharing settings, or under Privacy where the Settings menu shows Code instead [50].
\end{calloutbox}

\FloatBarrier
\setcounter{section}{5}
\section{Isolation and limits}

Each Anthropic-hosted session runs in an isolated VM; network access is limited by default and can be disabled; git credentials, signing keys and network secrets stay outside the sandbox, a proxy authenticating on the session's behalf with scoped credentials; and all operations are logged for audit [50], [59]. \textbf{The proxy does not confine a push to the working branch.} It rejects branch deletions and pushes of anything but a branch, such as a tag; which branches a session may update is decided by GitHub, applying your branch protection rules and rulesets to the access you connected, and a rule that access can bypass does not block the push [59]. Version 1.0 said pushes were restricted to the current branch. Self-hosted environments move all of this to your deployment.

\textbf{When network access is disabled, Claude Code can still reach the Anthropic API, which may allow data to exit the VM} [50]. Anthropic-hosted sessions and Code Review reach your repositories from Anthropic-managed infrastructure, so a GitHub Enterprise Cloud organisation that restricts access by IP must enable allow-list inheritance for installed GitHub Apps \textbf{and} add Anthropic's outbound addresses, because inheritance covers only what the App does as an installation, not what it does on a user's behalf; otherwise starting a session stops with \texttt{Your Git\allowbreak{}Hub organiz\allowbreak{}ation has a\allowbreak{}n IP allowl\allowbreak{}ist that is\allowbreak{} blocking C\allowbreak{}laude} [60], [96]. Self-hosted sessions connect from inside your network. A session idle long enough has its VM reclaimed: reopening restores the conversation, not shell commands, subagents or session-scoped scheduled tasks that were still running, and a session waiting on an MCP approval or sign-in counts as idle [50]. In the session's mode dropdown the Claude app offers Accept edits, Plan and Auto, never Bypass permissions [51].

\FloatBarrier
\setcounter{section}{6}
\section{Self-hosted environments}

On Team and Enterprise plans an organisation can run cloud sessions on its own infrastructure. This is a public beta, off by default: an Owner turns on \textbf{Allow self-hosted environments} on the Cloud environments admin page, which needs cloud sessions enabled, and it is unavailable with Zero Data Retention or the HIPAA configuration [97]. An \emph{environment} is a named group of \emph{runners}, created in claude.ai admin settings; a runner is a long-lived process on your hosts that polls Anthropic's control plane over outbound HTTPS, claims a queued session, clones the repository and starts a Claude Code process for it. Anthropic never connects into your network [97].

What moves and what does not is the point to get right. Repository checkouts, build artifacts, secrets and the files a session creates stay on your machines, and sessions reach internal services from inside the network. \textbf{The conversation does not move.} Prompts, responses and tool results still go to \texttt{api.\allowbreak{}anthropic.\allowbreak{}com}, Anthropic stores the transcript so the session can be resumed from another surface, and orchestration, queueing and the claude.ai interface stay Anthropic-hosted. A runner configured to send model requests to Amazon Bedrock or Google Cloud's Agent Platform still streams the conversation to Anthropic, and on such a runner server-managed settings and claude.ai organisation policies do not reach the session [97].

Two operating rules carry security weight. A runner serves one owner at a time, locking to the owner of the first session it claims, so checked-out code never mixes between owners; at the default \texttt{-\allowbreak{}-\allowbreak{}drain-\allowbreak{}grace-\allowbreak{}sec} of \texttt{0} it exits when its sessions finish, so the orchestrator can restart it on a fresh disk [97]. That separation holds only if you deploy for it: a fresh container or VM for each runner process, destroyed when the process exits, with \texttt{-\allowbreak{}-\allowbreak{}capacity 1}, and no filesystem reused between restarts or ever across owners [98]. The owner of a Claude Tag channel session is the Tag agent, not the person who wrote the Slack message, so a runner locked to it serves sessions different people started once its capacity is above one or its drain grace is positive [97].

\textbf{Owner locking is not an admission boundary.} Dispatch has no per-environment access control: any member of the organisation can dispatch a session to any of its environments, and if an Owner routes Claude Tag channels to one, anyone Claude Tag admits can start sessions that run there — by default anyone in the connected Slack workspace, with or without a Claude account. \texttt{-\allowbreak{}-\allowbreak{}lock-\allowbreak{}to-\allowbreak{}account} bounds which account's sessions a host executes, not who can dispatch, so put on a runner host only data and credentials that everyone who can dispatch to it may read [98]. And not every surface routes there: sessions started from claude.ai/code, the apps, cloud scheduled tasks and \texttt{claude -\allowbreak{}-\allowbreak{}cloud} can; Claude Security and Code Review sessions do not yet [97]. In exchange you build and maintain the runner image, operate the fleet and control its network, which is why Anthropic-hosted environments remain the better default for most teams.

Slack is a further way in. Claude Tag runs \texttt{@Claude} as the organisation's shared identity with admin-configured access; the earlier Claude Code in Slack, which runs each cloud session under an individual's account, answers only Pro, Max and Ultra accounts, and only in workspaces no organisation has connected to Claude Tag [99]. Claude Tag's governance — what it retains, and that it receives no server-managed settings — is in Chapter 31 and Chapter 32.

\begingroup
\def\kprows{%
\item \texttt{-\allowbreak{}-\allowbreak{}cloud} creates a session that runs elsewhere; \texttt{-\allowbreak{}-\allowbreak{}remote-\allowbreak{}control} exposes a local one. They are unrelated.
\item The GitHub App reaches public repositories and its private installations; \texttt{/\allowbreak{}web-\allowbreak{}setup} reaches whatever your \texttt{gh} token can.
\item \texttt{-\allowbreak{}-\allowbreak{}teleport} pulls a cloud session and its branch into the terminal; \texttt{-\allowbreak{}-\allowbreak{}resume} does not list cloud sessions.
\item Auto-fix can reply to review threads using your GitHub account, which can trigger comment-driven CI automation.
}%
\def\kpbody{\begin{minipage}{\textwidth}\subsection*{Key points}\begin{itemize}\kprows\end{itemize}\end{minipage}}%
\begingroup
\def\sloppy{\tolerance 9999\emergencystretch 3em\hfuzz 200pt\vfuzz 200pt}%
\hbadness=10000\vbadness=10000\hfuzz=200pt\vfuzz=200pt
\global\setbox\kpbox=\hbox{\kpbody}%
\endgroup
\par\addvspace{4.2mm}
\ifdim\dimexpr\ht\kpbox+\dp\kpbox\relax>0.30\textheight
  \typeout{HANDBOOK-KEYPOINTS broken \the\dimexpr\ht\kpbox+\dp\kpbox\relax}%
  \subsection*{Key points}
  \begin{itemize}\kprows\end{itemize}
\else
  \typeout{HANDBOOK-KEYPOINTS atomic \the\dimexpr\ht\kpbox+\dp\kpbox\relax}%
  \noindent\kpbody
\fi
\par\addvspace{1.4mm}
\endgroup

\FloatBarrier
\renewcommand{\chaptertitlelabel}{Part VI \textperiodcentered\ Chapter 27}
\setcounter{chapter}{26}
\chapter{Push events into a session with channels}

\begin{calloutbox}{palegrey}{quoterule}{1.2mm}
Channels are a research preview. They require Anthropic authentication through claude.ai or a Console API key, and are unavailable on Amazon Bedrock, Google Cloud's Agent Platform and Microsoft Foundry. Team and Enterprise organisations must explicitly enable them [54].
\end{calloutbox}

A channel is an MCP server that \textbf{pushes} events into a running session, rather than waiting to be queried [54]. Where a standard connector gives Claude on-demand access to a system, a channel lets a system reach into the session you already have open — a CI failure, a monitoring alert, a chat message — while Claude still has your files open and remembers what you were debugging.

Events arrive only while the session is open, so an always-on setup means a background process or a persistent terminal [54]. A channel can be two-way: Claude replies through it, and the terminal then shows the inbound message and the reply tool call, while the reply text itself appears only on the other platform [54].

\FloatBarrier
\setcounter{section}{0}
\section{Enabling one}

Channels install as plugins and require Bun. Telegram, Discord and iMessage are included in the preview, with \texttt{fakechat} as a local demo [54].

\begin{codeblock}{9.0}{10.6}
\cl{/plugin~install~telegram@claude-plugins-official}
\cl{/telegram:configure~<token>}
\end{codeblock}

\begin{codeblock}{9.0}{10.6}
\cl{claude~-{}-channels~plugin:telegram@claude-plugins-official}
\end{codeblock}

\texttt{-\allowbreak{}-\allowbreak{}channels} takes several plugins, space-separated. \textbf{Being present in \texttt{.\allowbreak{}mcp.\allowbreak{}json} is not enough}: a server must also be named in \texttt{-\allowbreak{}-\allowbreak{}channels} for that session [54]. During the preview \texttt{-\allowbreak{}-\allowbreak{}channels} accepts only plugins on the Anthropic-maintained allowlist, or on your organisation's list where one is set; anything else lets Claude Code start normally, but the channel does not register and the startup notice says why. A channel you are building is loaded with \texttt{-\allowbreak{}-\allowbreak{}dangerously-\allowbreak{}load-\allowbreak{}development-\allowbreak{}channels}. Neither flag appears in \texttt{claude -\allowbreak{}-\allowbreak{}help} while the feature is in preview; both work regardless [54].

A channel does not make a session unattended. \textbf{A permission prompt pauses the session until someone answers it}, at the terminal or, where the channel declares permission relay, through the channel; \texttt{-\allowbreak{}-\allowbreak{}dangerously-\allowbreak{}skip-\allowbreak{}permissions} removes most prompts but not the actions no mode auto-approves, and belongs only in an environment you trust [33], [54].

\FloatBarrier
\setcounter{section}{1}
\section{The security model, and the part that matters}

Every approved channel plugin maintains a \textbf{sender allowlist}; messages from anyone else are silently dropped. Telegram and Discord bootstrap it by pairing — message the bot, receive a code, approve it in session with \texttt{/\allowbreak{}telegram:\allowbreak{}access pair\allowbreak{} <code>} (or the Discord equivalent), then lock down with \texttt{policy allo\allowbreak{}wlist}. iMessage, macOS only, lets your own messages through automatically and takes other contacts by handle with \texttt{/\allowbreak{}imessage:\allowbreak{}access allo\allowbreak{}w} [54].

\begin{calloutbox}{palebrass}{brassdark}{2.0mm}
\textbf{CAUTION — read this before enabling a channel.} The allowlist also gates \textbf{permission relay} where a channel declares it. \textbf{Anyone who can reply through the channel can approve or deny tool use in your session} [54]. Add to a channel allowlist only senders you would trust with that authority, and understand that a compromised messaging account becomes a compromised permission gate.
\end{calloutbox}

Organisational controls are two managed settings that users cannot override: \texttt{channels\allowbreak{}Enabled}, the master switch, and \texttt{allowed\allowbreak{}Channel\allowbreak{}Plugins}, which replaces the Anthropic-maintained plugin allowlist with your own [54]. claude.ai Team and Enterprise organisations are blocked until an Owner enables channels, in the organisation's Claude Code settings or with \texttt{channels\allowbreak{}Enabled:\allowbreak{} true}; Console organisations with API-key authentication are permitted by default unless managed settings are deployed; Pro, Max and Ultra users without an organisation skip the checks [54]. While the switch is off the MCP server still connects and its tools work, but channel messages do not arrive. An empty \texttt{allowed\allowbreak{}Channel\allowbreak{}Plugins} array blocks allowlisted plugins but the development flag can still bypass it — to block channels entirely, leave \texttt{channels\allowbreak{}Enabled} unset [54]. In every case no channel runs until a user names it in \texttt{-\allowbreak{}-\allowbreak{}channels}.

In non-interactive mode with \texttt{-\allowbreak{}p}, tools that need terminal input — multiple-choice questions and plan approval — are disabled so the session never stalls [54].

\FloatBarrier
\setcounter{section}{2}
\section{When a channel is the right shape}

\begingroup
\def\tblrows{%
Cloud sessions & Run a task in a fresh cloud sandbox cloned from GitHub \\
Claude in Slack & Spawns a cloud session from an \texttt{@Claude} mention \\
Standard MCP server & Claude queries it during a task; nothing is pushed \\
Remote Control & You drive your local session from claude.ai or mobile \\
\textbf{Channels} & An external system pushes an event into your already-running local session \\
}%
\def\tblbody{\begin{minipage}{\textwidth}\boxcaption{Table 22 — Where channels sit among the ways to reach a session}
{\footnotesize\begin{tabular}{L{30.7mm}L{112.3mm}}
\toprule
\textbf{Mechanism} & \textbf{What it does} \\
\midrule
\tblrows
\bottomrule\end{tabular}}\end{minipage}}%
\begingroup
\def\sloppy{\tolerance 9999\emergencystretch 3em\hfuzz 200pt\vfuzz 200pt}%
\hbadness=10000\vbadness=10000\hfuzz=200pt\vfuzz=200pt
\global\setbox\tblbox=\hbox{\tblbody}%
\endgroup
\par\addvspace{2.6mm}
\ifdim\dimexpr\ht\tblbox+\dp\tblbox\relax>0.55\textheight
  \typeout{HANDBOOK-TABLE broken \the\dimexpr\ht\tblbox+\dp\tblbox\relax}%
  \tabcaption{Table 22 — Where channels sit among the ways to reach a session}
  {\footnotesize\begin{longtable}{L{30.7mm}L{112.3mm}}
  \toprule
\textbf{Mechanism} & \textbf{What it does} \\
\midrule\endfirsthead
  \multicolumn{2}{@{}l@{}}{%
  \sffamily\footnotesize\itshape\color{inkgrey}Table 22 — Where channels sit among the ways to reach a session \textemdash\ continued}\\[1.2mm]
  \toprule
\textbf{Mechanism} & \textbf{What it does} \\
\midrule\endhead
  \bottomrule\endfoot
  \bottomrule\endlastfoot
  \tblrows
  \end{longtable}}%
\else
  \typeout{HANDBOOK-TABLE atomic \the\dimexpr\ht\tblbox+\dp\tblbox\relax}%
  \noindent\tblbody
\fi
\par\addvspace{2.6mm}
\endgroup

Source: [54].

Two shapes recur: a \textbf{chat bridge}, where you ask a question from your phone and the work runs on your machine against real files; and a \textbf{webhook receiver}, where CI, an error tracker or a deploy pipeline delivers an event to a session that already has context.

\begingroup
\def\kprows{%
\item A channel is an MCP server that pushes events into a session you already have open.
\item A server must be named in \texttt{-\allowbreak{}-\allowbreak{}channels} for that session; being in \texttt{.\allowbreak{}mcp.\allowbreak{}json} is not enough.
\item The sender allowlist also gates permission relay: anyone who can reply through the channel can approve tool use in your session.
\item Organisations control availability with \texttt{channels\allowbreak{}Enabled} and \texttt{allowed\allowbreak{}Channel\allowbreak{}Plugins} in managed settings.
}%
\def\kpbody{\begin{minipage}{\textwidth}\subsection*{Key points}\begin{itemize}\kprows\end{itemize}\end{minipage}}%
\begingroup
\def\sloppy{\tolerance 9999\emergencystretch 3em\hfuzz 200pt\vfuzz 200pt}%
\hbadness=10000\vbadness=10000\hfuzz=200pt\vfuzz=200pt
\global\setbox\kpbox=\hbox{\kpbody}%
\endgroup
\par\addvspace{4.2mm}
\ifdim\dimexpr\ht\kpbox+\dp\kpbox\relax>0.30\textheight
  \typeout{HANDBOOK-KEYPOINTS broken \the\dimexpr\ht\kpbox+\dp\kpbox\relax}%
  \subsection*{Key points}
  \begin{itemize}\kprows\end{itemize}
\else
  \typeout{HANDBOOK-KEYPOINTS atomic \the\dimexpr\ht\kpbox+\dp\kpbox\relax}%
  \noindent\kpbody
\fi
\par\addvspace{1.4mm}
\endgroup

\FloatBarrier
\renewcommand{\chaptertitlelabel}{Part VI \textperiodcentered\ Chapter 28}
\setcounter{chapter}{27}
\chapter{Publish work as artifacts}

\begin{calloutbox}{palegrey}{quoterule}{1.2mm}
Artifacts need a Pro, Max, Ultra, Team or Enterprise plan and a session signed in to claude.ai with \texttt{/\allowbreak{}login}; an API key, a gateway token or a cloud-provider credential cannot publish [53].
\end{calloutbox}

An artifact is a live, interactive web page published from a session to a private URL on claude.ai, updating in place as the session continues [53]. For the research workflow of this book it is the natural delivery surface for anything easier to look at than to read: an annotated diff, a dashboard, a comparison of options, an investigation timeline that fills in while a long task runs.

Claude may publish one on its own when the output suits a page, or you can ask. Unless you name a location, Claude writes an HTML or Markdown file to a temporary directory \textbf{outside} your project and publishes it; the page stays private to you until you share it [53].

\textbf{A new artifact published in answer to a prompt you typed goes through without a permission prompt or a classifier review.} Version 1.0 said the first publish of each artifact asked. Beyond any ask rule that matches the publish [20], Claude Code asks first — or, in auto mode, has the classifier review the publish — in these cases [53]:

\begin{itemize}
\item the session is in plan mode (before the first publish of each artifact);
\item the publish declares runtime capabilities, such as connector calls or file downloads;
\item the artifact is shared publicly (once per conversation before changing it, and before each reply to a comment on it);
\item the turn was started by a session-scoped scheduled task rather than by you (before the first publish of each artifact);
\item Claude cannot read the file without your approval, which an allow rule for \texttt{Artifact} does not skip; or
\item in Manual or \texttt{accept\allowbreak{}Edits} mode, the publish carries an edit to a file Claude could not otherwise change without asking.
\end{itemize}

Because an ordinary private publish may go through without a prompt, review what the session can read \textbf{and} what it publishes; do not rely on a publishing prompt as the gate. A one-step update — an edit to the page's file made inside the publish, announced as \texttt{Also makes \allowbreak{}1 edit to t\allowbreak{}hat file on\allowbreak{} disk.\allowbreak{}} — can be turned off with \texttt{CLAUDE\_\allowbreak{}CODE\_\allowbreak{}ARTIFACT\_\allowbreak{}PUBLISH\_\allowbreak{}EDITS=\allowbreak{}0}, after which you must quit and relaunch, because a running session does not reread it [53].

\texttt{Ctrl+]} reopens the session's most recent artifact; \texttt{/\allowbreak{}artifacts} lists every artifact you own and every one shared with you, this session's first: \texttt{Enter} opens one already attached or attaches it, \texttt{o} opens it, \texttt{c} copies the link, and \texttt{d} permanently deletes one you own, outside a cloud session, after you confirm — its link then stops working for everyone [53].

To update an artifact from a different session, give Claude its URL or attach it with \texttt{/\allowbreak{}artifacts}. Without either, a new session creates a \textbf{new} artifact instead of updating the existing one [53].

\FloatBarrier
\setcounter{section}{0}
\section{What an artifact is not}

It is a capture of work, not an application: one self-contained, static page with no backend and no routes [53].

\begingroup
\def\tblrows{%
External requests & A strict Content Security Policy allows typefaces from Google Fonts and scripts from five CDN hosts (cdnjs, unpkg, the Tailwind and jQuery CDNs, selected jsDelivr paths); it blocks every external image and all other external scripts, stylesheets and fonts, and \texttt{fetch}, XHR and WebSocket reach only the page's own origin and the Google Fonts hosts. Connector calls go through claude.ai \\
No backend & A static page; cannot authenticate viewers \\
Downloads & The page cannot start one itself; a download control needs the declared downloads capability \\
Single page & Relative links do not resolve; use in-page anchors \\
Source file & \texttt{.\allowbreak{}html}, \texttt{.\allowbreak{}htm} or \texttt{.\allowbreak{}md}, valid UTF-8 (or UTF-16LE with a byte-order mark); a path naming a network host is refused unread \\
Rendered size & 16 MiB or smaller \\
}%
\def\tblbody{\begin{minipage}{\textwidth}\boxcaption{Table 23 — Artifact page constraints}
{\footnotesize\begin{tabular}{L{24.2mm}L{118.8mm}}
\toprule
\textbf{Constraint} & \textbf{Effect} \\
\midrule
\tblrows
\bottomrule\end{tabular}}\end{minipage}}%
\begingroup
\def\sloppy{\tolerance 9999\emergencystretch 3em\hfuzz 200pt\vfuzz 200pt}%
\hbadness=10000\vbadness=10000\hfuzz=200pt\vfuzz=200pt
\global\setbox\tblbox=\hbox{\tblbody}%
\endgroup
\par\addvspace{2.6mm}
\ifdim\dimexpr\ht\tblbox+\dp\tblbox\relax>0.55\textheight
  \typeout{HANDBOOK-TABLE broken \the\dimexpr\ht\tblbox+\dp\tblbox\relax}%
  \tabcaption{Table 23 — Artifact page constraints}
  {\footnotesize\begin{longtable}{L{24.2mm}L{118.8mm}}
  \toprule
\textbf{Constraint} & \textbf{Effect} \\
\midrule\endfirsthead
  \multicolumn{2}{@{}l@{}}{%
  \sffamily\footnotesize\itshape\color{inkgrey}Table 23 — Artifact page constraints \textemdash\ continued}\\[1.2mm]
  \toprule
\textbf{Constraint} & \textbf{Effect} \\
\midrule\endhead
  \bottomrule\endfoot
  \bottomrule\endlastfoot
  \tblrows
  \end{longtable}}%
\else
  \typeout{HANDBOOK-TABLE atomic \the\dimexpr\ht\tblbox+\dp\tblbox\relax}%
  \noindent\tblbody
\fi
\par\addvspace{2.6mm}
\endgroup

\FloatBarrier
\setcounter{section}{1}
\section{Sharing, and who sees what}

A new artifact is visible only to you. On Team and Enterprise plans you can grant access to specific people or to everyone in the organisation, move it into a shared folder whose members can open it and whose owner and editors can change it, and make someone an \textbf{editor} who can publish new versions. A public link opens to anyone on the internet with no sign-in; on Team and Enterprise, public sharing is off until an Owner enables \textbf{External sharing} [53]. If an editor publishes while Claude holds an older copy, Claude Code refuses the stale publish and Claude merges onto the newer version [53].

Each publish becomes a version, and the Share control chooses which version viewers see. Organisation viewers see your name; a viewer opening a public link without signing in, or from outside your organisation, sees \texttt{Content is \allowbreak{}user-\allowbreak{}generated a\allowbreak{}nd unverifi\allowbreak{}ed.\allowbreak{}} instead [53].

\FloatBarrier
\setcounter{section}{2}
\section{Comments, and pages written by someone else}

People you share with inside the organisation can comment on the page, and Claude can read and answer the threads \emph{(requires 2.1.221 or later)}. Claude may reply to or resolve only a thread that someone who can edit the artifact has activated, with \textbf{Send to Claude} or an \texttt{@claude} mention; replies show as Claude, via you [53]. After your session publishes an artifact it watches for such comments and, in a permission mode that lets it post without asking, replies on its own \emph{(requires 2.1.228 or later)}; otherwise it waits with \texttt{Comments ar\allowbreak{}e waiting o\allowbreak{}n Artifact:\allowbreak{} <name>}, and in plan mode it does not reply at all. It stops on its own after 60 comments or activations on one artifact in an hour. \texttt{Ctrl+C} once at an idle prompt pauses replies, stopping the watch in \texttt{/\allowbreak{}tasks} stops them for that artifact, and \texttt{Ctrl+X Ctrl\allowbreak{}+K} twice stops them for the session [53]. Comment watching, like reading another organisation's artifact, needs feature-flag fetching on.

\begin{calloutbox}{palebrass}{brassdark}{2.0mm}
\textbf{CAUTION:} \textbf{an activated comment is input to your session from whoever can edit the page.} In a mode that posts without asking, an editor's comment reaches Claude and is answered while you are away. Share edit rights only with people you would let prompt your session, and leave auto-reply off — by staying in a mode that asks, or stopping the watch — where that is not acceptable.
\end{calloutbox}

A page someone else wrote is read like a web page: Claude gets a summary of what it asked about, which reports instructions written into the page rather than relaying them, and Claude Code saves the full source locally for when the exact content is needed [53]. Claude Code asks before reading another organisation's public artifact even in auto mode, and in \texttt{bypass\allowbreak{}Permissions}, where it cannot ask, Claude cannot read one; plan mode asks before reading one made by someone else in your organisation [53].

\FloatBarrier
\setcounter{section}{3}
\section{Live data through connectors}

A page can call MCP connectors each time someone views it, so it shows current data rather than a snapshot (2.1.209+). Claude declares which connectors the page may call, and the page cannot call others [53]. The access model is the important part:

\begin{itemize}
\item \textbf{Each viewer uses their own connectors.} Calls go through the viewing account's connections, so two people opening the same dashboard can see different data. The page never sees anyone's credentials.
\item \textbf{Viewers approve access first}, and a decline lasts for that page load.
\item \textbf{Actions use the viewer's account too.} A control that posts a message or updates an issue acts as whoever clicked it.
\end{itemize}

Only connectors from your claude.ai account qualify: a local server from \texttt{.\allowbreak{}mcp.\allowbreak{}json} can feed data while Claude builds the page, but the published page cannot call it [53]. A connector-backed page can be shared within the organisation or publicly, as your plan and settings allow, but \textbf{its connector calls do not run for a viewer who opens the public link without signing in, or from outside your organisation} — that viewer sees the page without its live sections [53]. Version 1.0 said such a page could not be shared publicly at all. Ask Claude to include a fallback message in each live section naming the connector it needs, so a viewer without the connection sees what to connect rather than an empty panel.

A page can also offer a file it generates, such as a CSV export, through a download control. The viewer blocks any download the page starts itself, \texttt{data:\allowbreak{}} and \texttt{blob:\allowbreak{}} links included, so the control needs the downloads capability, which claude.ai enables per account [53].

\FloatBarrier
\setcounter{section}{4}
\section{Templates, and a database the page can keep}

Instead of a page built from scratch, Claude can start an artifact from a template on your claude.ai account: Claude Slides for a presentation, Claude Design for a visual design, or Claude Docs for a document other people will read and edit. Each opens in its own editor on claude.ai, where you and your teammates change it directly or ask Claude to, and exports to formats such as PowerPoint, PDF or Word [53]. \texttt{/\allowbreak{}slides} and \texttt{/\allowbreak{}design} start a deck or a design canvas from a brief \emph{(2.1.265 or later)}; Claude Docs reaches Claude Code as a claude.ai connector, which \texttt{/\allowbreak{}mcp} lists as \texttt{claude.\allowbreak{}ai Claude D\allowbreak{}ocs}. Templates are in beta, on by default on Pro, Max, Ultra and Team plans; on Enterprise an Owner turns each one on under \textbf{Organization settings > Artifacts} [53]. The governance consequence follows from the editor: a template artifact can change in a claude.ai editor without any session republishing it, so review the version you are about to rely on, not the one Claude published.

The release notes also describe an artifact \emph{database}: collections a published page reads and writes, with access rules that can refuse a write [57]. The artifacts page fetched on 12 November 2026 does not document it, so this book says no more than that it exists. Rows in such a database are written by the page's viewers, so treat what a session reads from one as data, not as instructions (Section 28.3).

\FloatBarrier
\setcounter{section}{5}
\section{Governance}

For a regulated organisation, the controls are these [53]: artifacts are \textbf{on by default} on Team and Enterprise, and an Owner can turn them off, or on Enterprise with role-based access control scope them to roles; connector calls from artifacts have a \textbf{separate} toggle; public sharing is separately controlled and turning it off blocks existing public links without changing each artifact's audience; retention periods can be set independently for private and shared artifacts; publishing, sharing and deleting appear in the audit log under \texttt{claude\_\allowbreak{}artifact\_\allowbreak{}*} event types; and the Compliance API can list, retrieve and delete artifacts organisation-wide.

Artifacts are unavailable where the organisation has customer-managed encryption keys, HIPAA configuration, or Zero Data Retention enabled, and on third-party model providers [53]. They are off by default in the Agent SDK, the GitHub Action, MCP-server contexts and \texttt{-\allowbreak{}p} runs, and with \texttt{CLAUDE\_\allowbreak{}CODE\_\allowbreak{}DISABLE\_\allowbreak{}NONESSENTIA\allowbreak{}L\_\allowbreak{}TRAFFIC} set; and the organisation policy that allows them is loaded from \texttt{api.\allowbreak{}anthropic.\allowbreak{}com}, so where that cannot be reached they are unavailable [53]. The viewer loads pages from a sandboxed \texttt{*.\allowbreak{}claudeuserc\allowbreak{}ontent.\allowbreak{}com} origin, which must be allowlisted if you restrict outbound access, and a page that uses a library loads it from the five CDN hosts, with no fallback if they are blocked [53], [96].

Disable artifacts for your own sessions with the \textbf{Artifacts} row in \texttt{/\allowbreak{}config} or \texttt{"enable\allowbreak{}Artifact":\allowbreak{} false} (the older \texttt{"disable\allowbreak{}Artifact":\allowbreak{} true} still works), \texttt{CLAUDE\_\allowbreak{}CODE\_\allowbreak{}DISABLE\_\allowbreak{}ARTIFACT=\allowbreak{}1}, or by adding \texttt{Artifact} to \texttt{permissions.\allowbreak{}deny}. A \texttt{false} in a project's settings turns them off for that project, and a \texttt{true} there does not turn them back on \emph{(requires 2.1.242 or later)}; once a \texttt{-\allowbreak{}-\allowbreak{}settings} file, the variable or managed settings turn them off, no settings file re-enables them [53].

\begin{calloutbox}{palebrass}{brassdark}{2.0mm}
\textbf{CAUTION:} publishing is distribution. An artifact built from a session that touched private repositories, customer data or internal figures carries that content to a hosted page. Review the page, not the prompt, before sharing — and remember that an editor on Team or Enterprise can publish a new version you have not seen.
\end{calloutbox}

\begingroup
\def\kprows{%
\item An artifact is a hosted page, so publishing is distribution — review the page, not the prompt.
\item Connector-backed pages call connectors as the \emph{viewer}, so two people can see different data.
\item A new artifact publishes without a prompt in most modes; an activated comment from an editor is input to your session.
\item A connector-backed page can be public, but its live sections run only for signed-in viewers in your organisation.
\item Publishing, sharing and deletion are audit-logged, and retention is separately configurable for private and shared artifacts.
}%
\def\kpbody{\begin{minipage}{\textwidth}\subsection*{Key points}\begin{itemize}\kprows\end{itemize}\end{minipage}}%
\begingroup
\def\sloppy{\tolerance 9999\emergencystretch 3em\hfuzz 200pt\vfuzz 200pt}%
\hbadness=10000\vbadness=10000\hfuzz=200pt\vfuzz=200pt
\global\setbox\kpbox=\hbox{\kpbody}%
\endgroup
\par\addvspace{4.2mm}
\ifdim\dimexpr\ht\kpbox+\dp\kpbox\relax>0.30\textheight
  \typeout{HANDBOOK-KEYPOINTS broken \the\dimexpr\ht\kpbox+\dp\kpbox\relax}%
  \subsection*{Key points}
  \begin{itemize}\kprows\end{itemize}
\else
  \typeout{HANDBOOK-KEYPOINTS atomic \the\dimexpr\ht\kpbox+\dp\kpbox\relax}%
  \noindent\kpbody
\fi
\par\addvspace{1.4mm}
\endgroup

\breakrule

\FloatBarrier
\parttitle{Part VII}{Operating at scale}
\addcontentsline{toc}{part}{Part VII \textemdash\ Operating at scale}

\FloatBarrier
\renewcommand{\chaptertitlelabel}{Part VII \textperiodcentered\ Chapter 29}
\setcounter{chapter}{28}
\chapter{Parallel sessions without collisions}

Parallel agents reduce elapsed time and increase token use, coordination overhead and — the binding constraint — human decision load. The course's warning is the right one: three interactive sessions asking frequent questions feel like three colleagues interrupting at once [44].

\FloatBarrier
\setcounter{section}{0}
\section{The mechanisms}

\begingroup
\def\tblrows{%
Background the current session & \texttt{/\allowbreak{}background}, \texttt{/\allowbreak{}bg} & Detach work and free the terminal; carries \texttt{/\allowbreak{}loop} tasks with it \\
New background session & \texttt{claude -\allowbreak{}-\allowbreak{}bg "<prompt\allowbreak{}>"} & Start bounded work separately \\
Fork the conversation & \texttt{/\allowbreak{}fork} & Copy the conversation into another background session, by default told to make its own worktree before changing code \\
Fork subagent & \texttt{/\allowbreak{}subtask} & Bounded result returned to the parent, full context inherited; \texttt{/\allowbreak{}fork} does this when agent view is off \\
Subagent & Delegation & Fresh-context worker (Chapter 16) \\
Agent view & \texttt{claude agen\allowbreak{}ts} & Monitor and dispatch many background sessions from one screen; a research preview \\
Worktree & \texttt{claude -\allowbreak{}-\allowbreak{}worktree <n\allowbreak{}ame>} & Isolated git working tree and branch \\
Conversation branch & \texttt{/\allowbreak{}branch} & Explore an alternative in the current interface \\
Parallel change & \texttt{/\allowbreak{}batch <inst\allowbreak{}ruction>} & A bundled skill: 5 to 30 units, one background subagent per unit, each in its own worktree \\
Agent team & See Chapter 30 & Peer sessions with shared tasks and messaging \\
Dynamic workflow & See Chapter 22 & Codified orchestration of dozens to hundreds of agents \\
}%
\def\tblbody{\begin{minipage}{\textwidth}\boxcaption{Table 24 — Ways to run more than one thing at a time}
{\footnotesize\begin{tabular}{L{24.6mm}L{26.6mm}L{88.6mm}}
\toprule
\textbf{Mechanism} & \textbf{Entry point} & \textbf{Use} \\
\midrule
\tblrows
\bottomrule\end{tabular}}\end{minipage}}%
\begingroup
\def\sloppy{\tolerance 9999\emergencystretch 3em\hfuzz 200pt\vfuzz 200pt}%
\hbadness=10000\vbadness=10000\hfuzz=200pt\vfuzz=200pt
\global\setbox\tblbox=\hbox{\tblbody}%
\endgroup
\par\addvspace{2.6mm}
\ifdim\dimexpr\ht\tblbox+\dp\tblbox\relax>0.55\textheight
  \typeout{HANDBOOK-TABLE broken \the\dimexpr\ht\tblbox+\dp\tblbox\relax}%
  \tabcaption{Table 24 — Ways to run more than one thing at a time}
  {\footnotesize\begin{longtable}{L{24.6mm}L{26.6mm}L{88.6mm}}
  \toprule
\textbf{Mechanism} & \textbf{Entry point} & \textbf{Use} \\
\midrule\endfirsthead
  \multicolumn{3}{@{}l@{}}{%
  \sffamily\footnotesize\itshape\color{inkgrey}Table 24 — Ways to run more than one thing at a time \textemdash\ continued}\\[1.2mm]
  \toprule
\textbf{Mechanism} & \textbf{Entry point} & \textbf{Use} \\
\midrule\endhead
  \bottomrule\endfoot
  \bottomrule\endlastfoot
  \tblrows
  \end{longtable}}%
\else
  \typeout{HANDBOOK-TABLE atomic \the\dimexpr\ht\tblbox+\dp\tblbox\relax}%
  \noindent\tblbody
\fi
\par\addvspace{2.6mm}
\endgroup

Sources: [27], [37], [76], [100], [101].

By default a session dispatched from agent view or started with \texttt{claude -\allowbreak{}-\allowbreak{}bg} moves into a worktree of its own before it edits files. It does not when it was an open session you backgrounded, when it is already in a linked worktree, when the write is outside the working directory, outside a git repository with no \texttt{Worktree\allowbreak{}Create} hook, or with \texttt{worktree.\allowbreak{}bg\allowbreak{}Isolation} set to \texttt{"none"} [76]. \textbf{Agent teams do not isolate teammates in worktrees at all}, so a team needs the work partitioned by file instead [30], [100].

\FloatBarrier
\setcounter{section}{1}
\section{Prevent file collisions}

Two sessions editing the same checkout can overwrite each other, conflict, or validate against a moving target. Choose one pattern and state it in \texttt{CLAUDE.\allowbreak{}md}:

\begin{enumerate}
\item \textbf{Single writer} — one session edits; the others only review.
\item \textbf{File ownership} — each session owns explicitly non-overlapping paths.
\item \textbf{Git worktrees} — each implementation session gets its own branch and working tree [101]. \texttt{.\allowbreak{}worktreeinc\allowbreak{}lude} controls what is carried into a new worktree, and a subagent can be given \texttt{isolation:\allowbreak{} worktree} to work in a temporary one [27], [101].
\item \textbf{Sequential integration} — a separate session or a human reviews and merges completed units.
\end{enumerate}

Worktrees are the safest default for concurrent change. Each branch runs its own verification before integration, and integration tests run after the merge — not instead of it.

The isolation is enforced, within limits worth knowing. While a session — or any subagent it spawns — is in a worktree, Claude Code blocks an \texttt{Edit}, \texttt{Write} or \texttt{Notebook\allowbreak{}Edit} that targets the main checkout; a Bash, PowerShell or Monitor command whose working directory is the main checkout or cannot be verified to stay outside it; and, for Bash and Monitor only, a command that redirects git there (\texttt{git -\allowbreak{}C}, \texttt{-\allowbreak{}-\allowbreak{}git-\allowbreak{}dir}, \texttt{GIT\_\allowbreak{}DIR}, a \texttt{cd} first) or whose git use it cannot verify from the text. \textbf{PowerShell gets only the working-directory check} [101]. \textbf{None of these checks follows what a shell command writes}, so \texttt{cp} or a redirect into the main checkout is governed only by your permission rules and sandbox. And a worktree is not isolated from everything: it shares the repository's \texttt{.\allowbreak{}git} directory and project-scope plugins, and a "don't ask again" Bash approval given in a worktree is saved to the main checkout's local settings and applies in every worktree of the repository — except on Windows and wherever else Claude Code does not use the repository root, where it stays with that worktree [101].

Note the interaction with hooks: \texttt{\$\{CLAUDE\_\allowbreak{}PROJECT\_\allowbreak{}DIR\}} stays at the project root inside a worktree while the hook input's \texttt{cwd} shows the worktree path, so a hook script must decide which it means [83].

\FloatBarrier
\setcounter{section}{2}
\section{Cross-session messaging}

Claude can list and message your other sessions on this machine \emph{(requires 2.1.224 or later; 2.1.234 on native Windows; 2.1.248 on a third-party provider or with feature-flag fetching off)} [102]. It can also reach your cloud sessions and sessions on your other machines, but only from a session connected to Remote Control with a claude.ai sign-in as its active authentication, and starting a conversation with another machine \emph{requires 2.1.225 or later} [102]. \texttt{/\allowbreak{}list-\allowbreak{}agents} (alias \texttt{/\allowbreak{}peers}) enumerates them [37]. A message to a session on this machine travels over a local socket or named pipe and never through Anthropic's servers; one to another machine or a cloud session goes through them, the former over that machine's Remote Control connection [102].

An incoming message cannot approve anything, is not to change permission settings, \texttt{CLAUDE.\allowbreak{}md} or other configuration because another session asked, and cannot run a command: a \texttt{/\allowbreak{}compact} in its text arrives as text [102]. What a session does with arrivals is \texttt{cross\allowbreak{}Session\allowbreak{}Inbound}: \texttt{accept}, \texttt{hold} (a notice, no delivery) or \texttt{refuse}. With no value set, Claude Code sorts sessions into two classes: those that bypass permission prompts — which includes plan mode in an interactive terminal session where bypass is available — and those that prompt, which includes auto, \texttt{accept\allowbreak{}Edits} and \texttt{dont\allowbreak{}Ask}. A prompting session accepts messages unless the sender bypasses prompts; a bypassing session holds every message for your approval unless the sender bypasses too [102]. \texttt{isolate\allowbreak{}Peer\allowbreak{}Machines:\allowbreak{} true} makes every message that would leave the machine wait for your approval, even in \texttt{bypass\allowbreak{}Permissions}, and a checked-in project file can turn it on but not off; denying \texttt{Send\allowbreak{}Message} and \texttt{List\allowbreak{}Agents} stops sending and listing, and also removes messaging to subagents and teammates [102].

\begin{calloutbox}{palebrass}{brassdark}{2.0mm}
\textbf{CAUTION:} an incoming message is told it came from another Claude session, not from you. A peer session \textbf{cannot approve a permission prompt or supply consent on your behalf}, and a session denied an action cannot relay it to another session to bypass the check. In auto mode the classifier treats a relayed approval claim as untrusted input and reviews each message before delivery, and a message it blocks never arrives [30], [102]. Treat messages as coordination input, never as privileged control.
\end{calloutbox}

\FloatBarrier
\setcounter{section}{3}
\section{Protect your attention}

\begin{itemize}
\item Keep at most two or three interactive sessions until you know your own limit.
\item Use background work for anything that does not need frequent decisions.
\item Name sessions by outcome (Section 12.2).
\item Maintain a small dashboard: owner, branch, task, status, next human decision.
\item Batch reviews rather than checking every minute.
\item Stop starting work when integration becomes the bottleneck. This is the real ceiling, and it arrives sooner than the token limit.
\end{itemize}

\subsection*{A parallel task brief \{\#sec:parallel-brief\}}

\begin{codeblock}{7.0}{8.3}
\cl{Task:~~~~~~Review~only~sources/source-set-b/~for~contradictory~claims.}
\cl{Ownership:~Read~source-set-b.~Write~notes/review-b.md~and~nothing~else.}
\cl{Do~not:~~~~Edit~drafts,~change~configuration,~message~external~systems,~or~push.}
\cl{Return:~~~~Claim~IDs,~the~contradiction,~source~passages,~confidence,~open~questions.}
\cl{Done:~~~~~~Every~file~in~source-set-b~is~accounted~for~and~the~output~passes~its~schema~check.}
\end{codeblock}

\begingroup
\def\kprows{%
\item Parallelism is bounded by your decision load, not by tokens; integration becomes the bottleneck first.
\item Choose one collision pattern — single writer, file ownership, worktrees or sequential integration — and write it into \texttt{CLAUDE.\allowbreak{}md}.
\item Worktrees are the safest default for concurrent change.
\item A peer session cannot approve a permission or supply consent on your behalf; a session that bypasses prompts holds incoming messages for you by default.
}%
\def\kpbody{\begin{minipage}{\textwidth}\subsection*{Key points}\begin{itemize}\kprows\end{itemize}\end{minipage}}%
\begingroup
\def\sloppy{\tolerance 9999\emergencystretch 3em\hfuzz 200pt\vfuzz 200pt}%
\hbadness=10000\vbadness=10000\hfuzz=200pt\vfuzz=200pt
\global\setbox\kpbox=\hbox{\kpbody}%
\endgroup
\par\addvspace{4.2mm}
\ifdim\dimexpr\ht\kpbox+\dp\kpbox\relax>0.30\textheight
  \typeout{HANDBOOK-KEYPOINTS broken \the\dimexpr\ht\kpbox+\dp\kpbox\relax}%
  \subsection*{Key points}
  \begin{itemize}\kprows\end{itemize}
\else
  \typeout{HANDBOOK-KEYPOINTS atomic \the\dimexpr\ht\kpbox+\dp\kpbox\relax}%
  \noindent\kpbody
\fi
\par\addvspace{1.4mm}
\endgroup

\FloatBarrier
\renewcommand{\chaptertitlelabel}{Part VII \textperiodcentered\ Chapter 30}
\setcounter{chapter}{29}
\chapter{Agent teams}

\begin{calloutbox}{palegrey}{quoterule}{1.2mm}
\textbf{Agent teams are experimental and disabled by default.} Enable them by setting \texttt{CLAUDE\_\allowbreak{}CODE\_\allowbreak{}EXPERIMENTA\allowbreak{}L\_\allowbreak{}AGENT\_\allowbreak{}TEAMS=\allowbreak{}1} in \texttt{settings.\allowbreak{}json} or the environment. Without it no team is set up, no team directories are written, and Claude does not spawn or propose teammates [30].
\end{calloutbox}

An agent team is a lead session coordinating peer sessions. Teammates work independently, each in its own context window, share a task list where they have the Task tools — newer models lack them by default, and then coordinate by message — and message each other directly [72]. Unlike subagents, you can interact with an individual teammate without going through the lead [30]. A teammate loads the project's \texttt{CLAUDE.\allowbreak{}md}, MCP servers and skills, and the lead's spawn prompt, but not the lead's conversation, so the spawn prompt has to carry the task's context [30]. Teams need an interactive session: in \texttt{-\allowbreak{}p} and the Agent SDK, Claude spawns no teammates and a named subagent runs as an ordinary one [30].

\FloatBarrier
\setcounter{section}{0}
\section{The behaviour change you must understand before enabling}

\begin{calloutbox}{palebrass}{brassdark}{2.0mm}
\textbf{CAUTION:} enabling agent teams changes ordinary delegation. Claude names subagents on its own so it can message them later, and \textbf{while agent teams are enabled a named subagent launches as a teammate}, without asking you — unless the call is a fork or sets \texttt{isolation} — so teams can form during delegation you never framed as team work [30]. What changes is the shape of the work: a teammate is a separate session with the lead's permission mode, no worktree, and the limitations listed below. Version 1.0 also said a teammate's idle notification carried no output; the November page says it includes the teammate's final answer [30].
\end{calloutbox}

Set the variable to \texttt{0} to restore subagent behaviour; no restart is needed, because settings-file \texttt{env} values are reapplied to the running session on save [30]. Note the precedence trap: \texttt{0} in your user settings is overridden by \texttt{1} in project settings, local settings, a \texttt{-\allowbreak{}-\allowbreak{}settings} payload, or managed settings.

\FloatBarrier
\setcounter{section}{1}
\section{Architecture}

\begingroup
\def\tblrows{%
Team lead & The main session; spawns teammates and coordinates & The session you are in \\
Teammates & Separate Claude Code instances & Own context windows \\
Team config & Runtime state — session IDs, pane IDs & \texttt{\textasciitilde{}/\allowbreak{}.\allowbreak{}claude/\allowbreak{}teams/\allowbreak{}\{team\}/\allowbreak{}config.\allowbreak{}json} \\
Mailbox & Per-agent message queue & \texttt{\textasciitilde{}/\allowbreak{}.\allowbreak{}claude/\allowbreak{}teams/\allowbreak{}\{team\}/\allowbreak{}inboxes/\allowbreak{}\{agent\}.\allowbreak{}json} \\
Task list & Shared work items & \texttt{\textasciitilde{}/\allowbreak{}.\allowbreak{}claude/\allowbreak{}tasks/\allowbreak{}\{team\}/\allowbreak{}} \\
}%
\def\tblbody{\begin{minipage}{\textwidth}
{\footnotesize\begin{tabular}{L{21.8mm}L{41.1mm}H{76.9mm}}
\toprule
\textbf{Component} & \textbf{Role} & \textbf{Location} \\
\midrule
\tblrows
\bottomrule\end{tabular}}\end{minipage}}%
\begingroup
\def\sloppy{\tolerance 9999\emergencystretch 3em\hfuzz 200pt\vfuzz 200pt}%
\hbadness=10000\vbadness=10000\hfuzz=200pt\vfuzz=200pt
\global\setbox\tblbox=\hbox{\tblbody}%
\endgroup
\par\addvspace{2.6mm}
\ifdim\dimexpr\ht\tblbox+\dp\tblbox\relax>0.55\textheight
  \typeout{HANDBOOK-TABLE broken \the\dimexpr\ht\tblbox+\dp\tblbox\relax}%
  
  {\footnotesize\begin{longtable}{L{21.8mm}L{41.1mm}H{76.9mm}}
  \toprule
\textbf{Component} & \textbf{Role} & \textbf{Location} \\
\midrule\endfirsthead
  \multicolumn{3}{@{}l@{}}{%
  \sffamily\footnotesize\itshape\color{inkgrey}Continued}\\[1.2mm]
  \toprule
\textbf{Component} & \textbf{Role} & \textbf{Location} \\
\midrule\endhead
  \bottomrule\endfoot
  \bottomrule\endlastfoot
  \tblrows
  \end{longtable}}%
\else
  \typeout{HANDBOOK-TABLE atomic \the\dimexpr\ht\tblbox+\dp\tblbox\relax}%
  \noindent\tblbody
\fi
\par\addvspace{2.6mm}
\endgroup

The team name is \texttt{session-\allowbreak{}} plus the first eight characters of the session ID. The config directory is removed when the session ends; the task list stays on disk, kept for the period \texttt{cleanup\allowbreak{}Period\allowbreak{}Days} sets [30]. Claude Code validates every mailbox entry and removes malformed ones, and reports a message as sent only once the write to the recipient's mailbox succeeds [30]. Do not hand-edit the config: it is overwritten on the next state update, and there is no project-level equivalent.

\FloatBarrier
\setcounter{section}{2}
\section{Control}

Describe the task and the teammates in natural language. Display mode defaults to \texttt{in-\allowbreak{}process}, where all teammates run in your terminal and the agent panel below the prompt selects among them: arrow keys to select, \texttt{Enter} to open a teammate's transcript and message it, \texttt{Esc} to clear the selection or, while viewing a teammate, interrupt its turn, \texttt{x} to stop it, \texttt{Ctrl+T} to toggle the task list [30]. While you view a teammate, plain text and skills go to it but built-in commands go to the lead, and \texttt{/\allowbreak{}compact}, \texttt{/\allowbreak{}clear} and \texttt{/\allowbreak{}rewind} ask first because they act on the lead's conversation. Split panes require tmux or iTerm2 with the \texttt{it2} CLI and are unsupported in the VS Code integrated terminal, Windows Terminal and Ghostty; set \texttt{teammate\allowbreak{}Mode} or pass \texttt{-\allowbreak{}-\allowbreak{}teammate-\allowbreak{}mode} [30].

A teammate's model is the first of: the model your spawn prompt names; for a teammate spawned from a subagent definition, the definition's \texttt{model}; \texttt{CLAUDE\_\allowbreak{}CODE\_\allowbreak{}SUBAGENT\_\allowbreak{}MODEL}; the lead's model \emph{(this order requires 2.1.251 or later)}. An installed mod that sets a model in its \texttt{agent.\allowbreak{}spawn} hook replaces the first source [30]. \texttt{CLAUDE\_\allowbreak{}CODE\_\allowbreak{}SUBAGENT\_\allowbreak{}MODEL\_\allowbreak{}FORCE=\allowbreak{}1} skips the first two \emph{(requires 2.1.257 or later)}, \texttt{teammate\allowbreak{}Default\allowbreak{}Model} was removed in 2.1.234, and the organisation's \texttt{available\allowbreak{}Models} allowlist is applied to whatever is chosen [30]. A teammate on the lead's model inherits the lead's effort level [30].

A teammate spawned from a subagent definition takes its \texttt{tools} list — to which an in-process teammate gets \texttt{Send\allowbreak{}Message} added, and the Task tools too where the session has them — its \texttt{model}, and its body, appended to the system prompt in-process and replacing it in a split pane. In-process teammates also take \texttt{disallowed\allowbreak{}Tools}, though the tools Claude Code adds survive it, \texttt{effort}, and \texttt{auto\allowbreak{}Compact\allowbreak{}Window} \emph{(requires 2.1.296 or later)}. \texttt{skills} is never applied, and \texttt{mcp\allowbreak{}Servers} is applied only to a split-pane teammate [30]. A teammate Claude brings back gets a definition from a project's \texttt{.\allowbreak{}claude/\allowbreak{}agents/\allowbreak{}} only if you trusted the folder the file is in [30].

\textbf{Plan mode in a team is not a review gate by default.} A teammate spawned while the lead is in plan mode plans first, but when it sends its plan — if you have not changed that teammate's permission mode since it was spawned — Claude Code approves it at once, without the lead reviewing it, and passes the teammate the lead's mode, Manual if that is plan mode. The documented way to get a review is to put a running in-process teammate into plan mode yourself, with \texttt{Shift+Tab} while viewing it: then the lead reviews that plan, and the lead decides, not you [30]. Version 1.0 said the lead always approved autonomously and could be steered with criteria in the prompt. \textbf{For a plan you must approve yourself, keep the planning in the lead session}, in plan mode, where the plan comes to you. A subagent is not a substitute while teams are on: a named subagent launches as a teammate, and even an ordinary subagent's plan returns to Claude, not to you, so the lead would have to present it and wait for your approval before anything is implemented.

\FloatBarrier
\setcounter{section}{3}
\section{Permissions}

\begin{calloutbox}{palebrass}{brassdark}{2.0mm}
\textbf{CAUTION:} a new teammate starts in the lead's mode, and per-teammate modes cannot be set at spawn time. If the lead runs with \texttt{-\allowbreak{}-\allowbreak{}dangerously-\allowbreak{}skip-\allowbreak{}permissions}, \textbf{every teammate does too}; a lead in \texttt{dont\allowbreak{}Ask} starts teammates in Manual \emph{(for a split-pane teammate, 2.1.317 or later)} [30]. Worse, \textbf{an in-process teammate that is not in plan mode runs in the lead's mode whenever the lead is in \texttt{accept\allowbreak{}Edits}, auto or \texttt{bypass\allowbreak{}Permissions} — switching the teammate to Manual does not change that}, even though its status bar shows Manual. To make it ask before editing, switch the lead to Manual [30]. Teammate permission prompts surface in the lead session, which becomes a bottleneck at team scale; pre-approve genuinely safe operations before spawning rather than approving under pressure afterwards.
\end{calloutbox}

An in-process teammate runs under the lead's live allow and deny rules, so \texttt{-\allowbreak{}-\allowbreak{}disallowed\allowbreak{}Tools}, \texttt{-\allowbreak{}-\allowbreak{}tools} and \texttt{-\allowbreak{}-\allowbreak{}allowed\allowbreak{}Tools} all reach it. A split-pane teammate gets the first two \emph{(requires 2.1.309 or later)} but not \texttt{-\allowbreak{}-\allowbreak{}allowed\allowbreak{}Tools}; put such rules under \texttt{permissions.\allowbreak{}allow} in a \texttt{-\allowbreak{}-\allowbreak{}settings} payload instead [30].

The inter-agent messaging rules of Section 29.3 apply in full: a teammate cannot approve a prompt or supply consent for you, and the auto-mode classifier reviews every message before delivery, including structured protocol messages such as shutdown requests and plan-approval responses.

\FloatBarrier
\setcounter{section}{4}
\section{Quality gates}

Three hooks are specific to teams and are the mechanism for enforcing standards across peers [30], [83]:

\begin{itemize}
\item \texttt{Teammate\allowbreak{}Idle} — runs when a teammate is about to go idle; exit 2 sends feedback and keeps it working.
\item \texttt{Task\allowbreak{}Created} — exit 2 prevents creation and returns feedback.
\item \texttt{Task\allowbreak{}Completed} — exit 2 prevents completion and returns feedback.
\end{itemize}

A \texttt{Task\allowbreak{}Completed} hook that requires evidence before a task may be closed is the team-scale equivalent of the release gate in Chapter 34.

\FloatBarrier
\setcounter{section}{5}
\section{Limitations to plan around}

From [30]: \texttt{/\allowbreak{}resume} and \texttt{/\allowbreak{}rewind} do not restore in-process teammates, and the lead may try to message teammates that no longer exist; task status can lag, blocking dependents; shutdown waits for the current tool call; there is exactly one team per session and no nested teams; the lead is fixed for the session's lifetime; and an in-process teammate's own subagents run in the foreground, so a definition with \texttt{background:\allowbreak{} true} returns an error.

Start with three to five teammates. Token cost scales linearly with teammates, coordination overhead scales worse, and three focused teammates routinely outperform five scattered ones [30], [67]. An in-process teammate's prompt cache lasts five minutes by default, even on a subscription; \texttt{subagent\allowbreak{}Prompt\allowbreak{}Cache\allowbreak{}Ttl:\allowbreak{} "1h"} extends it at the higher one-hour write rate [30].

\FloatBarrier
\setcounter{section}{6}
\section{When a team beats the alternatives}

Use a team when teammates need to \textbf{share findings and challenge each other}: parallel review with distinct lenses, competing hypotheses in a debugging investigation, cross-layer work with clear ownership. Use subagents when only the result matters. Use a workflow when the orchestration itself should be repeatable (Chapter 22). Use worktrees when the issue is file isolation rather than reasoning (Section 29.2).

The adversarial pattern is the one that earns its cost:

\begin{codeblock}{7.0}{8.3}
\cl{Users~report~the~app~exits~after~one~message~instead~of~staying~connected.~Spawn~five~teammates}
\cl{to~investigate~different~hypotheses.~Have~them~message~each~other~to~try~to~disprove~each}
\cl{other\textquotesingle{}s~theories,~like~a~scientific~debate.~Update~the~findings~document~with~the~consensus}
\cl{that~emerges,~and~record~which~theories~were~eliminated~and~by~what~evidence.}
\end{codeblock}

Sequential investigation anchors on the first plausible explanation. Independent investigators actively trying to falsify each other produce a surviving theory that is much more likely to be the actual cause — the same epistemics that make the contradiction reviewer load-bearing in Section 22.5.

\begingroup
\def\kprows{%
\item Agent teams are experimental and off by default; enabling them changes ordinary delegation.
\item While enabled, a subagent Claude names launches as a teammate, without asking, so teams form during ordinary delegation.
\item Teammates start in the lead's mode; an in-process teammate not in plan mode stays in it while the lead is in \texttt{accept\allowbreak{}Edits}, auto or bypass, whatever its own status bar shows.
\item A plan-mode teammate's plan is approved automatically; putting a running teammate into plan mode yourself gets the lead's review, not yours.
}%
\def\kpbody{\begin{minipage}{\textwidth}\subsection*{Key points}\begin{itemize}\kprows\end{itemize}\end{minipage}}%
\begingroup
\def\sloppy{\tolerance 9999\emergencystretch 3em\hfuzz 200pt\vfuzz 200pt}%
\hbadness=10000\vbadness=10000\hfuzz=200pt\vfuzz=200pt
\global\setbox\kpbox=\hbox{\kpbody}%
\endgroup
\par\addvspace{4.2mm}
\ifdim\dimexpr\ht\kpbox+\dp\kpbox\relax>0.30\textheight
  \typeout{HANDBOOK-KEYPOINTS broken \the\dimexpr\ht\kpbox+\dp\kpbox\relax}%
  \subsection*{Key points}
  \begin{itemize}\kprows\end{itemize}
\else
  \typeout{HANDBOOK-KEYPOINTS atomic \the\dimexpr\ht\kpbox+\dp\kpbox\relax}%
  \noindent\kpbody
\fi
\par\addvspace{1.4mm}
\endgroup

\breakrule

\FloatBarrier
\parttitle{Part VIII}{Governance, assurance and compliance}
\addcontentsline{toc}{part}{Part VIII \textemdash\ Governance, assurance and compliance}

This part is written for readers who must answer to somebody else: an audit committee, a regulator, a customer's security questionnaire, or a board. It assumes the operating discipline of Parts I–III and asks a different question: what can be \emph{enforced}, what can be \emph{evidenced}, and what remains a matter of trust.

\FloatBarrier
\renewcommand{\chaptertitlelabel}{Part VIII \textperiodcentered\ Chapter 31}
\setcounter{chapter}{30}
\chapter{Enterprise configuration and managed policy}

Everything in Chapters 7, 13, 15, 17 and 18 is a control a \emph{user} can set and therefore a control a user can unset. Managed settings are the tier above: Claude Code applies them above every other level, and no user, project, local or \texttt{-\allowbreak{}-\allowbreak{}settings} value overrides them, apart from a small set of exceptions where a \textbf{stricter} value from a lower level still counts [70], [71].

\FloatBarrier
\setcounter{section}{0}
\section{Delivery mechanisms}

\begingroup
\def\tblrows{%
Server-managed & claude.ai admin console, or a self-hosted Claude apps gateway & Fetched at startup, polled hourly & One place to change policy for a claude.ai organisation \\
MDM or OS policy & macOS configuration profile (\texttt{com.\allowbreak{}anthropic.\allowbreak{}claudecode}) or Windows \texttt{HKLM\textbackslash{}SOFTWARE\textbackslash{}Policies\textbackslash{}Claude\allowbreak{}Code} & Startup, re-checked every 30 minutes & You already manage devices with Jamf, Intune or Group Policy \\
File-based & \texttt{managed-\allowbreak{}settings.\allowbreak{}json} in a system directory & Startup, reloaded on change & Machines without MDM, Linux hosts, images you build \\
\texttt{HKCU} registry & \texttt{HKCU\textbackslash{}SOFTWARE\textbackslash{}Policies\textbackslash{}Claude\allowbreak{}Code} & Startup, re-checked every 30 minutes; used only when no admin document is present above it & You cannot write the machine-level key \\
}%
\def\tblbody{\begin{minipage}{\textwidth}\boxcaption{Table 25 — How a managed policy reaches a machine}
{\scriptsize\begin{tabular}{L{26.6mm}L{62.7mm}L{20.8mm}L{26.5mm}}
\toprule
\textbf{Mechanism} & \textbf{Delivered as} & \textbf{Read when} & \textbf{Use when} \\
\midrule
\tblrows
\bottomrule\end{tabular}}\end{minipage}}%
\begingroup
\def\sloppy{\tolerance 9999\emergencystretch 3em\hfuzz 200pt\vfuzz 200pt}%
\hbadness=10000\vbadness=10000\hfuzz=200pt\vfuzz=200pt
\global\setbox\tblbox=\hbox{\tblbody}%
\endgroup
\par\addvspace{2.6mm}
\ifdim\dimexpr\ht\tblbox+\dp\tblbox\relax>0.55\textheight
  \typeout{HANDBOOK-TABLE broken \the\dimexpr\ht\tblbox+\dp\tblbox\relax}%
  \tabcaption{Table 25 — How a managed policy reaches a machine}
  {\scriptsize\begin{longtable}{L{26.6mm}L{62.7mm}L{20.8mm}L{26.5mm}}
  \toprule
\textbf{Mechanism} & \textbf{Delivered as} & \textbf{Read when} & \textbf{Use when} \\
\midrule\endfirsthead
  \multicolumn{4}{@{}l@{}}{%
  \sffamily\footnotesize\itshape\color{inkgrey}Table 25 — How a managed policy reaches a machine \textemdash\ continued}\\[1.2mm]
  \toprule
\textbf{Mechanism} & \textbf{Delivered as} & \textbf{Read when} & \textbf{Use when} \\
\midrule\endhead
  \bottomrule\endfoot
  \bottomrule\endlastfoot
  \tblrows
  \end{longtable}}%
\else
  \typeout{HANDBOOK-TABLE atomic \the\dimexpr\ht\tblbox+\dp\tblbox\relax}%
  \noindent\tblbody
\fi
\par\addvspace{2.6mm}
\endgroup

Source: [71].

File paths: \texttt{/\allowbreak{}Library/\allowbreak{}Application\allowbreak{} Support/\allowbreak{}Claude\allowbreak{}Code/\allowbreak{}} on macOS, \texttt{/\allowbreak{}etc/\allowbreak{}claude-\allowbreak{}code/\allowbreak{}} on Linux and WSL, \texttt{C:\allowbreak{}\textbackslash{}Program Fil\allowbreak{}es\textbackslash{}Claude\allowbreak{}Code\textbackslash{}} on Windows. The legacy Windows path \texttt{C:\allowbreak{}\textbackslash{}Program\allowbreak{}Data\textbackslash{}Claude\allowbreak{}Code\textbackslash{}managed-\allowbreak{}settings.\allowbreak{}json} is \textbf{not} read [71]. A \texttt{managed-\allowbreak{}settings.\allowbreak{}d/\allowbreak{}} directory alongside the main file lets several teams own parts of one policy; files merge alphabetically, single values replace (a \texttt{fallback\allowbreak{}Model} chain and a \texttt{model\allowbreak{}Picker} lineup whole, as does a same-name \texttt{extra\allowbreak{}Known\allowbreak{}Marketplace\allowbreak{}s} or \texttt{managed\allowbreak{}Mcp\allowbreak{}Servers} entry), lists combine with duplicates removed, and nested blocks merge key by key. The first three mechanisms are \textbf{admin sources}. The \texttt{HKCU} key is user-writable and is not one, which is why Claude Code never applies it beneath an admin document that is present — including one it cannot read.

\begin{calloutbox}{palebrass}{brassdark}{2.0mm}
\textbf{CAUTION — the precedence rule that surprises administrators.} When more than one managed source delivers a policy, Claude Code by default uses \textbf{the highest-ranked source that delivers at least one policy key and ignores the rest} — it does not merge them, and it shows no warning for the sources it skipped [71]. The order is remote settings, then MDM or OS policy, then managed settings files, then \texttt{HKCU}. Setting \texttt{managed\allowbreak{}Sources\allowbreak{}Behavior} to \texttt{"merge"} in the highest-ranked source you deploy makes every admin source contribute instead: lists combine, locks take the strictest value, and restriction allowlists are taken whole from the highest source that sets one \emph{(requires 2.1.242 or later)}. Merging adds a lower source's \texttt{permissions.\allowbreak{}allow} rules and hooks to the policy, so turn it on only where every lower source is under an administrator's control. Run \texttt{/\allowbreak{}status} on the machine: its \texttt{Setting sou\allowbreak{}rces} line names the source selected, and from the same release a \texttt{Skipped sou\allowbreak{}rces} line names each source found and not used.
\end{calloutbox}

Three groups of keys do not follow the first-wins rule, and each has caught an administrator who tested on one machine and deployed to another [71]:

\begin{itemize}
\item \textbf{Keys read from every admin source.} The sandbox locks \texttt{allow\allowbreak{}Managed\allowbreak{}Domains\allowbreak{}Only} and \texttt{allow\allowbreak{}Managed\allowbreak{}Read\allowbreak{}Paths\allowbreak{}Only}, \texttt{allow\allowbreak{}All\allowbreak{}Claude\allowbreak{}Ai\allowbreak{}Mcps}, \texttt{force\allowbreak{}Remote\allowbreak{}Settings\allowbreak{}Refresh}, and \texttt{env}, merged per variable \emph{(requires 2.1.223 or later)}. From 2.1.273 the MCP lock \texttt{allow\allowbreak{}Managed\allowbreak{}Mcp\allowbreak{}Servers\allowbreak{}Only}, \texttt{denied\allowbreak{}Mcp\allowbreak{}Servers} and \texttt{disable\allowbreak{}Claude\allowbreak{}Ai\allowbreak{}Connectors} joined them, and from 2.1.267 \texttt{max\allowbreak{}Effort\allowbreak{}Level}, where the lowest cap in any admin source applies.
\item \textbf{Login keys.} On a machine with an MDM policy or a managed settings file, \texttt{force\allowbreak{}Login\allowbreak{}Org\allowbreak{}UUID} and \texttt{force\allowbreak{}Login\allowbreak{}Method} are taken from that policy or file, ahead of remote settings. Set them there, and in the claude.ai console as well until every client runs 2.1.299 or later.
\item \textbf{The gateway login keys.} \texttt{force\allowbreak{}Login\allowbreak{}Gateway\allowbreak{}Url}, \texttt{gateway\allowbreak{}Internal\allowbreak{}Networks} and the \texttt{"gateway"} value of \texttt{force\allowbreak{}Login\allowbreak{}Method} are never read from remote settings at all.
\end{itemize}

\textbf{Which sessions a policy reaches} depends on where the session runs, not on who is signed in [71]. On a developer's machine the terminal, the VS Code and JetBrains extensions, the desktop app's Code tab and Agent SDK sessions read every source, and the SDK loads managed settings even when \texttt{setting\allowbreak{}Sources} excludes the user, project and local files. A cloud session in an Anthropic-hosted environment reads no device profile or file, so its policy must come from server-managed settings; a session in a self-hosted environment also reads the managed settings file in its runner image. Claude Tag sessions do \textbf{not} receive server-managed settings. Nor do Cowork sessions in the desktop app: on the user's machine they read the device's MDM policy and managed settings file, and inside a full VM sandbox or on an Anthropic-managed VM they find no device policy to read. If your control over one of those surfaces is the claude.ai console alone, you do not have one.

\FloatBarrier
\setcounter{section}{1}
\section{A minimum viable enterprise policy}

\begin{codeblock}{9.0}{10.6}
\cl{\{}
\cl{~~"permissions":~\{}
\cl{~~~~"deny":~[}
\cl{~~~~~~"Read(./.env)",}
\cl{~~~~~~"Read(./.env.*)",}
\cl{~~~~~~"Read(./secrets/**)",}
\cl{~~~~~~"Read(\textasciitilde{}/.ssh/**)",}
\cl{~~~~~~"Read(\textasciitilde{}/.aws/**)"}
\cl{~~~~],}
\cl{~~~~"disableBypassPermissionsMode":~"disable"}
\cl{~~\},}
\cl{~~"allowManagedPermissionRulesOnly":~true,}
\cl{~~"allowManagedHooksOnly":~true,}
\cl{~~"allowManagedMcpServersOnly":~true,}
\cl{~~"allowedMcpServers":~[}
\cl{~~~~\{~"serverCommand":~["/usr/local/bin/company-mcp-server"]~\}}
\cl{~~],}
\cl{~~"sandbox":~\{}
\cl{~~~~"enabled":~true,}
\cl{~~~~"network":~\{~"allowManagedDomainsOnly":~true~\},}
\cl{~~~~"credentials":~\{}
\cl{~~~~~~"files":~~~[\{~"path":~"\textasciitilde{}/.aws/credentials",~"mode":~"deny"~\}],}
\cl{~~~~~~"envVars":~[\{~"name":~"GITHUB\_TOKEN",~"mode":~"deny"~\}]}
\cl{~~~~\}}
\cl{~~\},}
\cl{~~"strictKnownMarketplaces":~[}
\cl{~~~~\{~"source":~"github",~"repo":~"<your-org>/approved-plugins"~\}}
\cl{~~],}
\cl{~~"disableSideloadFlags":~true,}
\cl{~~"forceLoginMethod":~"claudeai",}
\cl{~~"forceLoginOrgUUID":~"<your-organization-uuid>",}
\cl{~~"cleanupPeriodDays":~7,}
\cl{~~"requiredMinimumVersion":~"2.1.241"}
\cl{\}}
\end{codeblock}

Each key answers a threat from an earlier chapter. \texttt{allow\allowbreak{}Managed\allowbreak{}Permission\allowbreak{}Rules\allowbreak{}Only} closes the gap that a developer can widen their own permission rules (Chapter 7): Claude Code then ignores \texttt{allow}, \texttt{ask} and \texttt{deny} rules in every other file, ignores \texttt{-\allowbreak{}-\allowbreak{}allowed\allowbreak{}Tools}, hides the always-allow choices in prompts, and — from 2.1.282 — ignores the \texttt{allowed-\allowbreak{}tools} frontmatter of skills and commands from the repository, the user's own directories and \texttt{-\allowbreak{}-\allowbreak{}add-\allowbreak{}dir} [61]. \texttt{allow\allowbreak{}Managed\allowbreak{}Hooks\allowbreak{}Only} blocks user, project and local hooks and the hooks of every plugin your managed settings do not force-enable; managed hooks, hooks the Agent SDK registers, and hooks from plugins force-enabled in managed \texttt{enabled\allowbreak{}Plugins} still run. As noted in Chapter 21, it disables \texttt{/\allowbreak{}goal} as a side effect, which is a trade-off to make consciously. \texttt{disable\allowbreak{}Sideload\allowbreak{}Flags} rejects \texttt{-\allowbreak{}-\allowbreak{}plugin-\allowbreak{}dir}, \texttt{-\allowbreak{}-\allowbreak{}plugin-\allowbreak{}url}, \texttt{-\allowbreak{}-\allowbreak{}agents} and \texttt{-\allowbreak{}-\allowbreak{}mcp-\allowbreak{}config} at startup on a workstation, closing the local-testing path of Chapter 18 (2.1.193+); a \texttt{-\allowbreak{}-\allowbreak{}mcp-\allowbreak{}config} whose servers are all in-process SDK entries is still accepted, and a cloud session starts and drops the server-delivered entries instead of refusing [61].

\texttt{force\allowbreak{}Login\allowbreak{}Method} and \texttt{force\allowbreak{}Login\allowbreak{}Org\allowbreak{}UUID} restrict which claude.ai organisation a developer may sign in to, which is one part of keeping traffic inside a ZDR organisation (Chapter 32) and not the whole of it [58], [71]. The organisation check applies to claude.ai logins. It does not verify which organisation a Console credential belongs to; the terminal's interactive login screen pre-selects the forced method without enforcing it; \texttt{claude setu\allowbreak{}p-\allowbreak{}token} and \texttt{/\allowbreak{}install-\allowbreak{}github-\allowbreak{}app} enforce the method only; and a session on a cloud provider authenticates against that provider and is not checked at all. Pair the two keys with \texttt{allowed\allowbreak{}Providers}, below, if the requirement is that traffic goes nowhere else.

As printed, the policy locks the sandbox's network allowlist to managed sources and supplies no domains, so a sandboxed command reaches no network at all; add your registries and hosts under \texttt{sandbox.\allowbreak{}network.\allowbreak{}allowed\allowbreak{}Domains} in the same file.

Two keys in that policy are \textbf{allowlists, and their type is the control} [61]:

\begin{itemize}
\item \texttt{strict\allowbreak{}Known\allowbreak{}Marketplace\allowbreak{}s} is an array of marketplace source objects. Unset, users can add any marketplace; an empty array blocks every source, the official Anthropic marketplace included. \texttt{blocked\allowbreak{}Marketplace\allowbreak{}s} is the matching blocklist, checked before download so a blocked source never touches the filesystem.
\item \texttt{allow\allowbreak{}Managed\allowbreak{}Mcp\allowbreak{}Servers\allowbreak{}Only} is a lock on a list, not a list. It makes the managed \texttt{allowed\allowbreak{}Mcp\allowbreak{}Servers} the only allowlist that counts, and \textbf{if no admin source sets that list, every server that passes the denylist loads}, unless an embedding host supplies a list of its own as parent settings [71]. The lock alone admits everything. Write the list with \texttt{server\allowbreak{}Url} or \texttt{server\allowbreak{}Command} entries: a \texttt{server\allowbreak{}Name} entry matches a label the user chose (Section 13.5).
\end{itemize}

\begin{calloutbox}{palebrass}{brassdark}{2.0mm}
\textbf{CAUTION — an erratum to Version 1.0.} The policy printed in Version 1.0 of this book set \texttt{"strict\allowbreak{}Known\allowbreak{}Marketplace\allowbreak{}s":\allowbreak{} true} and \texttt{allow\allowbreak{}Managed\allowbreak{}Mcp\allowbreak{}Servers\allowbreak{}Only} with no \texttt{allowed\allowbreak{}Mcp\allowbreak{}Servers} beside it. The first is not a value the key accepts: a client at 2.1.277 or later enforces an invalid value as an empty allowlist, which locks out every marketplace, and an earlier client drops the key and enforces nothing [71]. The second deploys a lock with nothing behind it. If you copied that policy, replace both lines with the forms above and confirm with \texttt{claude doct\allowbreak{}or}.
\end{calloutbox}

Other managed-only keys worth knowing: \texttt{strict\allowbreak{}Plugin\allowbreak{}Only\allowbreak{}Customizati\allowbreak{}on} blocks skills, agents, hooks and MCP servers from user and project sources; \texttt{plugin\allowbreak{}Trust\allowbreak{}Message} appends your own wording to the plugin trust warning; \texttt{allowed\allowbreak{}Channel\allowbreak{}Plugins} replaces the Anthropic channel allowlist, and takes effect only with \texttt{channels\allowbreak{}Enabled:\allowbreak{} true}, without which channels stay blocked on Team and Enterprise plans; \texttt{allowed\allowbreak{}Providers} lists the services a machine may reach Claude through and refuses a session on any other \emph{(requires 2.1.285 or later)} [61], [71].

Model choice needs three keys, because \textbf{a managed \texttt{model} value is a default, not a lock}, and \texttt{-\allowbreak{}-\allowbreak{}model} or \texttt{ANTHROPIC\_\allowbreak{}MODEL} still picks the model for a session [71]. \texttt{available\allowbreak{}Models} constrains \texttt{/\allowbreak{}model}, \texttt{-\allowbreak{}-\allowbreak{}model} and the \texttt{model} key in a developer's own files, but with its default prefix matching an entry such as \texttt{claude-\allowbreak{}opus-\allowbreak{}5} also admits later versions that extend it, and the picker's Default option can still resolve to a model outside the list. An app that embeds Claude Code and sets \texttt{CLAUDE\_\allowbreak{}CODE\_\allowbreak{}PROVIDER\_\allowbreak{}MANAGED\_\allowbreak{}BY\_\allowbreak{}HOST} overrides managed \texttt{model}, \texttt{fallback\allowbreak{}Model}, \texttt{model\allowbreak{}Picker} and \texttt{model\allowbreak{}Overrides}; a managed \texttt{available\allowbreak{}Models} stays in force unless the app supplies its own [63], [70]. \texttt{enforce\allowbreak{}Available\allowbreak{}Models} closes the Default gap; \texttt{available\allowbreak{}Models\allowbreak{}Match:\allowbreak{} "exact"} makes each entry admit only the version it names; \texttt{denied\allowbreak{}Models} blocks a model even where the allowlist admits it \emph{(the last two require 2.1.283 or later)} [61].

\FloatBarrier
\setcounter{section}{2}
\section{The Claude apps gateway}

For an organisation that must route inference through its own cloud provider, Anthropic ships a gateway inside the \texttt{claude} binary: \texttt{claude gate\allowbreak{}way -\allowbreak{}-\allowbreak{}config gate\allowbreak{}way.\allowbreak{}yaml} runs a self-hosted service between developers' clients and Amazon Bedrock, Claude Platform on AWS, Google Cloud's Agent Platform, Microsoft Foundry or the Anthropic API, with failover between them [103]. Developers sign in with the corporate OIDC identity provider and receive short-lived bearer tokens; the upstream credential stays in your infrastructure, and a user disabled in the identity provider loses access when the next refresh fails, within an hour by default. The gateway enforces model access server-side by identity-provider group, delivers each group's managed settings itself, and relays OTLP telemetry to your collector. Gateway sessions do not fetch the claude.ai console's server-managed settings, so the gateway's own policy block takes that role [103].

Four properties matter for policy:

\begin{itemize}
\item \textbf{It is a source of managed settings, so it is guarded like one.} A trusted gateway can push settings that run commands on developer machines. Claude Code therefore connects by default only to a gateway at a private address. A managed \texttt{gateway\allowbreak{}Internal\allowbreak{}Networks} list admits a public IPv4 block the organisation owns \emph{(2.1.268 or later)}, and then only when every address the gateway's name resolves to is inside that block, the developer's machine connects from inside the same block, and the connection is direct. Claude Code pins the gateway's TLS certificate on first connect, and checks the pin again at sign-in, on refresh and on each settings fetch. A request routed through an HTTPS proxy skips the pin check, so put the gateway host in \texttt{NO\_\allowbreak{}PROXY} [103].
\item \textbf{It replaces the claude.ai login.} Claude Code uses one sign-in at a time, and a gateway sign-in takes the place of a Team or Enterprise login. The login keys — \texttt{force\allowbreak{}Login\allowbreak{}Method} set to \texttt{"gateway"} and \texttt{force\allowbreak{}Login\allowbreak{}Gateway\allowbreak{}Url} — belong in the managed settings file you deploy by MDM; leave \texttt{force\allowbreak{}Login\allowbreak{}Method} and \texttt{force\allowbreak{}Login\allowbreak{}Org\allowbreak{}UUID} out of the gateway's own policy block, or some signed-in sessions exit at startup [103].
\item \textbf{Startup waits for the gateway's policy.} At startup a signed-in session exits after about ten seconds if the gateway is unreachable, rather than starting without its settings. That is a property of startup, not of the gateway as a whole: by default spend enforcement fails open when the gateway's database is unreachable, so inference stays up, unless \texttt{fail\_\allowbreak{}closed\_\allowbreak{}on\_\allowbreak{}error} is set [104]. There is no service-token flow, so a CI job authenticates against the provider directly; on a machine whose managed settings force the gateway, a session authenticating with an API key or \texttt{api\allowbreak{}Key\allowbreak{}Helper} exits at startup [103].
\item \textbf{It removes features as well as adding control.} Server-side web search, Remote Control, \texttt{/\allowbreak{}design-\allowbreak{}sync}, and everything that needs feature-flag fetching, such as \texttt{/\allowbreak{}import}, are unavailable on gateway sessions; the one-hour cache TTL is an opt-in [103].
\end{itemize}

The gateway does not log or store prompt or completion content. Usage telemetry carries identity, token counts, model and latency; logs and traces, which can include commands and file paths, are a per-destination choice. Its data plane sends nothing to Anthropic unless the Anthropic API is a configured upstream [103]. One deployment detail widens the attack surface: once \texttt{parent\allowbreak{}Settings\allowbreak{}Behavior:\allowbreak{} "merge"} is deployed, any host process that launches Claude Code — Claude Desktop, an Agent SDK application, an IDE extension — can supply parent settings, and without the \texttt{allow\allowbreak{}Managed*Onl\allowbreak{}y} locks a host's allow rules and sandbox allowlists still apply. A parent-supplied \texttt{available\allowbreak{}Models}, \texttt{allowed\allowbreak{}Mcp\allowbreak{}Servers} or \texttt{strict\allowbreak{}Known\allowbreak{}Marketplace\allowbreak{}s} also applies wherever your own managed sources leave that key unset, so set those keys yourself [103].

\FloatBarrier
\setcounter{section}{3}
\section{Validation behaviour, and which way each key fails}

A managed policy can be wrong in two ways that look the same from the console: the \emph{source} can be unusable, or a \emph{value} inside a usable source can be invalid. Claude Code treats them differently, and — the part Version 1.0 understated — it does not treat every key the same way [71]. For a value that fails schema validation it first skips each individual entry it can repair, such as one malformed permission rule, and warns. What it does with a value that still fails depends on the key.

\begingroup
\def\tblrows{%
A managed settings file, drop-in file, MDM plist or \texttt{HKLM} value that is present but is not a JSON object & \textbf{closed} & Claude Code refuses to start and names the source, even when another admin source delivers a valid policy \emph{(requires 2.1.259 or later)} \\
A managed source the operating system refuses to let Claude Code read, with no other admin source supplying a policy & \textbf{open} & Every session starts without that source's policies; \texttt{/\allowbreak{}status}, \texttt{claude doct\allowbreak{}or} and stderr under \texttt{-\allowbreak{}p} record it \emph{(requires 2.1.285 or later)} \\
A managed source that cannot be read for any other reason, such as an I/O error, with no other admin source supplying a policy & \textbf{closed} & Every session exits at startup \emph{(requires 2.1.285 or later)} \\
\texttt{policy\allowbreak{}Helper} output with a schema violation left after entry repairs & \textbf{closed} & The whole helper run fails and Claude Code refuses to start \\
A top-level key with a single restrictive value, such as \texttt{allow\allowbreak{}Managed\allowbreak{}Permission\allowbreak{}Rules\allowbreak{}Only} or \texttt{disable\allowbreak{}Auto\allowbreak{}Mode}, and a lock inside \texttt{permissions} such as \texttt{disable\allowbreak{}Bypass\allowbreak{}Permissions\allowbreak{}Mode} & \textbf{closed} & Reads as its restrictive value until fixed \emph{(requires 2.1.282 or later)} \\
\texttt{disable\allowbreak{}All\allowbreak{}Hooks} & \textbf{open} & Dropped with a warning, because enforcing \texttt{true} would unload the managed hooks too \emph{(requires 2.1.282 or later)} \\
An unreadable \texttt{deny} or \texttt{ask} list in \texttt{permissions} & \textbf{closed} & \texttt{allow} and \texttt{additional\allowbreak{}Directories} are withheld, so no grant applies without the restrictions written beside it \emph{(requires 2.1.282 or later)} \\
\texttt{allowed\allowbreak{}Mcp\allowbreak{}Servers} & \textbf{closed} & Empty allowlist: no server a user adds is admitted. \texttt{managed\allowbreak{}Mcp\allowbreak{}Servers} entries still load, and \texttt{managed-\allowbreak{}mcp.\allowbreak{}json} servers that skip the allowlist check \\
\texttt{allowed\allowbreak{}Providers} & \textbf{closed} & Empty allowlist: every provider is refused and Claude Code does not start \emph{(requires 2.1.285 or later)} \\
\texttt{allowed\allowbreak{}Http\allowbreak{}Hook\allowbreak{}Urls} · \texttt{http\allowbreak{}Hook\allowbreak{}Allowed\allowbreak{}Env\allowbreak{}Vars} & \textbf{closed} & Empty managed allowlist, but entries in user, project and local settings still apply, because these two lists merge across files \emph{(requires 2.1.267 or later)} \\
\texttt{allowed\allowbreak{}Channel\allowbreak{}Plugins} & \textbf{closed} & Empty allowlist: no channel plugin is admitted \emph{(requires 2.1.267 or later)} \\
\texttt{strict\allowbreak{}Known\allowbreak{}Marketplace\allowbreak{}s} & \textbf{closed} & Empty allowlist: no marketplace source is admitted, the official one included \emph{(requires 2.1.277 or later)} \\
\texttt{available\allowbreak{}Models} & \textbf{closed} & Empty allowlist, leaving only the Default model \\
\texttt{available\allowbreak{}Models\allowbreak{}Match} & \textbf{closed} & Reads as \texttt{exact}; a valid model allowlist beside it is unaffected \\
\texttt{force\allowbreak{}Login\allowbreak{}Org\allowbreak{}UUID} & \textbf{closed} & No organisation may log in \\
\texttt{strict\allowbreak{}Plugin\allowbreak{}Only\allowbreak{}Customizati\allowbreak{}on}, when neither a Boolean nor an array & \textbf{closed} & Reads as \texttt{true}, locking all four surfaces \emph{(requires 2.1.282 or later)} \\
A wholly invalid \texttt{denied\allowbreak{}Mcp\allowbreak{}Servers} or \texttt{denied\allowbreak{}Models} & \textbf{open} & Dropped with a warning and blocks nothing until fixed; an individual invalid entry is stripped and the rest enforced \\
A wholly invalid \texttt{blocked\allowbreak{}Marketplace\allowbreak{}s} & \textbf{open} & Dropped with a warning. An invalid entry is stripped; one that parses but can never match, such as a pattern that does not compile, is kept with a warning and blocks nothing \emph{(requires 2.1.277 or later)} \\
An invalid Boolean inside \texttt{sandbox}, other than \texttt{fail\allowbreak{}If\allowbreak{}Unavailable} & \textbf{closed} & Reads as whichever value keeps the sandbox strictest \emph{(requires 2.1.283 or later)} \\
An invalid \texttt{sandbox.\allowbreak{}network.\allowbreak{}denied\allowbreak{}Domains}, \texttt{filesystem.\allowbreak{}deny\allowbreak{}Read} or \texttt{filesystem.\allowbreak{}deny\allowbreak{}Write}, or any entry in one & \textbf{closed} & The matching allowlists are withheld until the deny list is fixed \emph{(requires 2.1.283 or later)} \\
A recoverable invalid entry in \texttt{sandbox.\allowbreak{}credentials} & \textbf{closed} & Degraded to \texttt{mode:\allowbreak{} "deny"}: the credential stays blocked, not masked \emph{(requires 2.1.221 or later)} \\
\texttt{sandbox.\allowbreak{}fail\allowbreak{}If\allowbreak{}Unavailable} & \textbf{open} & Dropped, so an unreadable value never stops sessions starting across a fleet \emph{(requires 2.1.283 or later)} \\
\texttt{required\allowbreak{}Minimum\allowbreak{}Version} · \texttt{required\allowbreak{}Maximum\allowbreak{}Version} & \textbf{open} & Dropped rather than enforced, by design \\
}%
\def\tblbody{\begin{minipage}{\textwidth}\boxcaption{Table 26 — Which way an invalid managed policy fails}
{\footnotesize\begin{tabular}{L{63.5mm}L{12.3mm}L{64.0mm}}
\toprule
\textbf{What is invalid} & \textbf{Fails} & \textbf{What Claude Code does} \\
\midrule
\tblrows
\bottomrule\end{tabular}}\end{minipage}}%
\begingroup
\def\sloppy{\tolerance 9999\emergencystretch 3em\hfuzz 200pt\vfuzz 200pt}%
\hbadness=10000\vbadness=10000\hfuzz=200pt\vfuzz=200pt
\global\setbox\tblbox=\hbox{\tblbody}%
\endgroup
\par\addvspace{2.6mm}
\ifdim\dimexpr\ht\tblbox+\dp\tblbox\relax>0.55\textheight
  \typeout{HANDBOOK-TABLE broken \the\dimexpr\ht\tblbox+\dp\tblbox\relax}%
  \tabcaption{Table 26 — Which way an invalid managed policy fails}
  {\footnotesize\begin{longtable}{L{63.5mm}L{12.3mm}L{64.0mm}}
  \toprule
\textbf{What is invalid} & \textbf{Fails} & \textbf{What Claude Code does} \\
\midrule\endfirsthead
  \multicolumn{3}{@{}l@{}}{%
  \sffamily\footnotesize\itshape\color{inkgrey}Table 26 — Which way an invalid managed policy fails \textemdash\ continued}\\[1.2mm]
  \toprule
\textbf{What is invalid} & \textbf{Fails} & \textbf{What Claude Code does} \\
\midrule\endhead
  \bottomrule\endfoot
  \bottomrule\endlastfoot
  \tblrows
  \end{longtable}}%
\else
  \typeout{HANDBOOK-TABLE atomic \the\dimexpr\ht\tblbox+\dp\tblbox\relax}%
  \noindent\tblbody
\fi
\par\addvspace{2.6mm}
\endgroup

Source: [61], [71]. \textbf{closed} means the restrictive reading is enforced or Claude Code does not start; \textbf{open} means the value is dropped and nothing is enforced in its place.

Read the table for its exceptions, not its rule. Three of them matter in an audit:

\begin{itemize}
\item \textbf{A denied read fails open; an unparseable file fails closed.} A \texttt{managed-\allowbreak{}settings.\allowbreak{}json} that is malformed JSON stops every session on the machine. The same file made unreadable by its permissions lets every session start without it. Restrictive file permissions on the policy are therefore a way to deploy no policy at all: a file pushed with permissions the developer's account cannot read is skipped, and the only trace is a line in \texttt{/\allowbreak{}status} and \texttt{claude doct\allowbreak{}or}.
\item \textbf{Deny lists fail open and allowlists fail closed.} A wholly invalid \texttt{denied\allowbreak{}Mcp\allowbreak{}Servers}, \texttt{denied\allowbreak{}Models} or \texttt{blocked\allowbreak{}Marketplace\allowbreak{}s} blocks nothing. If a control must hold when its own value is mistyped, express it as an allowlist.
\item \textbf{Two allowlists are closed only on the managed side.} \texttt{allowed\allowbreak{}Http\allowbreak{}Hook\allowbreak{}Urls} and \texttt{http\allowbreak{}Hook\allowbreak{}Allowed\allowbreak{}Env\allowbreak{}Vars} merge across settings files, so while the managed list is empty an entry in a user or project file still applies.
\end{itemize}

A quoted Boolean on a lock, \texttt{"true"} for \texttt{true}, is read as the Boolean it spells, with a notice in \texttt{/\allowbreak{}status}; a \texttt{null} removes the key. This tolerance belongs to managed settings alone: a user, project or local file whose JSON or top-level shape fails validation is rejected whole [71]. Find dropped and reinterpreted entries in the startup dialog, on stderr under \texttt{-\allowbreak{}p}, or with \texttt{claude doct\allowbreak{}or}, which lists each with its source and field.

\FloatBarrier
\setcounter{section}{4}
\section{What a managed policy does not reach}

Be explicit with stakeholders about the boundary. The documentation's own description of server-managed settings is that they are a client-side control and not a security boundary: on an unmanaged device a user needs no administrator rights to get round them [71], [105]:

\begin{itemize}
\item \textbf{A local administrator can edit the managed source itself.} MDM redeployment on a schedule is the mitigation, not an assumption of integrity.
\item \textbf{The server-managed cache can be edited locally.} The edit applies at the next startup and lasts until the next successful fetch, and longer for keys that apply only at launch, such as \texttt{model}, which stay in effect until the session is restarted. Claude Code withholds cached proxy, endpoint and credential variables until the server confirms the payload, so a tampered cache cannot redirect the fetch that would correct it [105].
\item \textbf{A session can start before its policy does.} Server-managed settings are fetched at startup, and unless Anthropic requires them for the login, Claude Code does not block startup on the fetch: it waits up to five seconds when the developer signs in at startup and otherwise not at all. A session signed in to a gateway does wait, and exits if the gateway cannot be reached. \texttt{force\allowbreak{}Remote\allowbreak{}Settings\allowbreak{}Refresh} blocks startup until a fresh fetch succeeds and exits if it does not; it is honoured from any admin source, so set it in an MDM profile or the managed settings file as well [61].
\item \textbf{Choosing another provider skips the fetch.} Claude Code fetches remote settings only when the session authenticates to Anthropic's API directly with an eligible credential or signs in to a gateway. On another provider, or when \texttt{ANTHROPIC\_\allowbreak{}BASE\_\allowbreak{}URL} points anywhere else, it starts at the next source down. A key returned by an \texttt{api\allowbreak{}Key\allowbreak{}Helper} does not trigger the fetch either, and a CI job on a Console API key receives the Console organisation's settings, not those of your claude.ai organisation [105]. Policy that must hold on those sessions belongs in a file or an MDM profile.
\item \textbf{Some settings wait for the developer.} A server-managed change to a setting that needs approval, such as a hook or an \texttt{env} variable, waits for the developer to accept a dialog in an interactive session, and applies for the current run only in a session that an IDE extension or the Agent SDK hosts.
\item \textbf{Managed settings bind Claude Code only.} A developer calling the API from another tool is not under them.
\item \textbf{Session-level choices} escape a managed default: the model, unless \texttt{available\allowbreak{}Models} is deployed, and the auto-compact window, which \texttt{-\allowbreak{}-\allowbreak{}autocompact} still sets.
\end{itemize}

Two organisation-level toggles live outside settings files entirely: Remote Control and cloud sessions are enabled or disabled organisation-wide in the claude.ai Claude Code admin settings, with \texttt{disable\allowbreak{}Remote\allowbreak{}Control} available per device and no per-device key for cloud sessions. Turning Remote Control off also disconnects sessions already connected, at their next policy refresh, about an hour at most \emph{(requires 2.1.286 or later)}. \texttt{claude doct\allowbreak{}or} reports on its \texttt{Organizatio\allowbreak{}n policy} line whether those settings reached a machine \emph{(requires 2.1.261 or later)} [71].

\begingroup
\def\kprows{%
\item Managed settings are the only tier a user cannot unset, and no other level overrides them.
\item When several managed sources exist, the highest-ranked with a policy key wins and the rest are ignored with no warning, unless \texttt{managed\allowbreak{}Sources\allowbreak{}Behavior} is \texttt{"merge"}. Check \texttt{/\allowbreak{}status}.
\item A lock is not a list: \texttt{allow\allowbreak{}Managed\allowbreak{}Mcp\allowbreak{}Servers\allowbreak{}Only} without \texttt{allowed\allowbreak{}Mcp\allowbreak{}Servers} admits every server, and \texttt{strict\allowbreak{}Known\allowbreak{}Marketplace\allowbreak{}s} takes an array, not \texttt{true}.
\item Invalid values do not fail one way. Allowlists and locks fail closed, deny lists and version bounds fail open, and a policy file the operating system will not let Claude Code read fails open where an unparseable one stops the machine.
\item Server-managed settings do not reach every session: one on another provider never fetches them, and Claude Tag and Cowork sessions do not receive them.
}%
\def\kpbody{\begin{minipage}{\textwidth}\subsection*{Key points}\begin{itemize}\kprows\end{itemize}\end{minipage}}%
\begingroup
\def\sloppy{\tolerance 9999\emergencystretch 3em\hfuzz 200pt\vfuzz 200pt}%
\hbadness=10000\vbadness=10000\hfuzz=200pt\vfuzz=200pt
\global\setbox\kpbox=\hbox{\kpbody}%
\endgroup
\par\addvspace{4.2mm}
\ifdim\dimexpr\ht\kpbox+\dp\kpbox\relax>0.30\textheight
  \typeout{HANDBOOK-KEYPOINTS broken \the\dimexpr\ht\kpbox+\dp\kpbox\relax}%
  \subsection*{Key points}
  \begin{itemize}\kprows\end{itemize}
\else
  \typeout{HANDBOOK-KEYPOINTS atomic \the\dimexpr\ht\kpbox+\dp\kpbox\relax}%
  \noindent\kpbody
\fi
\par\addvspace{1.4mm}
\endgroup

\FloatBarrier
\renewcommand{\chaptertitlelabel}{Part VIII \textperiodcentered\ Chapter 32}
\setcounter{chapter}{31}
\chapter{Data governance, residency and retention}

This chapter states what leaves the machine, what is retained and for how long, what an organisation can turn off, and where the boundaries of Zero Data Retention actually fall. For a reader operating across Australia, the European Union and Asia-Pacific, the mapping to obligations under the Privacy Act, the GDPR and the EU AI Act is a matter of applying these mechanics to your own processing purposes [17], [106], [107]; this chapter supplies the mechanics, not the legal conclusion.

\FloatBarrier
\setcounter{section}{0}
\section{What leaves the machine}

Claude Code runs locally and sends prompts and model outputs over TLS 1.2 or later to the configured provider [43]. Beyond inference, five other flows exist and each has its own control:

\begingroup
\def\tblrows{%
Metrics & Latency, reliability, usage patterns. \textbf{Never} code, prompts or file paths & \texttt{DISABLE\_\allowbreak{}TELEMETRY=\allowbreak{}1} \\
Error reports & Error messages and stack traces from Claude Code internals, with known secret patterns, file paths, email addresses and personal information redacted before leaving the machine & \texttt{DISABLE\_\allowbreak{}ERROR\_\allowbreak{}REPORTING=\allowbreak{}1} \\
\texttt{/\allowbreak{}feedback}, \texttt{/\allowbreak{}bug}, \texttt{/\allowbreak{}share} & A copy of the conversation \textbf{including code}; you choose current session, or the project's last 24 hours or 7 days & \texttt{DISABLE\_\allowbreak{}FEEDBACK\_\allowbreak{}COMMAND=\allowbreak{}1} \\
Session quality survey & The rating only. A separate opt-in follow-up uploads the transcript, subagent transcripts and raw session log, with known API-key and token patterns redacted & \texttt{CLAUDE\_\allowbreak{}CODE\_\allowbreak{}DISABLE\_\allowbreak{}FEEDBACK\_\allowbreak{}SURVEY=\allowbreak{}1}, or \texttt{feedback\allowbreak{}Survey\allowbreak{}Rate} \\
WebFetch domain safety check & The \textbf{hostname only}, to \texttt{api.\allowbreak{}anthropic.\allowbreak{}com}, checked against a blocklist; cached five minutes & \texttt{skip\allowbreak{}Web\allowbreak{}Fetch\allowbreak{}Preflight:\allowbreak{} true} \\
}%
\def\tblbody{\begin{minipage}{\textwidth}\boxcaption{Table 27 — Non-inference data flows and their opt-outs}
{\footnotesize\begin{tabular}{L{20.5mm}L{42.4mm}H{76.9mm}}
\toprule
\textbf{Flow} & \textbf{Contents} & \textbf{Opt-out} \\
\midrule
\tblrows
\bottomrule\end{tabular}}\end{minipage}}%
\begingroup
\def\sloppy{\tolerance 9999\emergencystretch 3em\hfuzz 200pt\vfuzz 200pt}%
\hbadness=10000\vbadness=10000\hfuzz=200pt\vfuzz=200pt
\global\setbox\tblbox=\hbox{\tblbody}%
\endgroup
\par\addvspace{2.6mm}
\ifdim\dimexpr\ht\tblbox+\dp\tblbox\relax>0.55\textheight
  \typeout{HANDBOOK-TABLE broken \the\dimexpr\ht\tblbox+\dp\tblbox\relax}%
  \tabcaption{Table 27 — Non-inference data flows and their opt-outs}
  {\footnotesize\begin{longtable}{L{20.5mm}L{42.4mm}H{76.9mm}}
  \toprule
\textbf{Flow} & \textbf{Contents} & \textbf{Opt-out} \\
\midrule\endfirsthead
  \multicolumn{3}{@{}l@{}}{%
  \sffamily\footnotesize\itshape\color{inkgrey}Table 27 — Non-inference data flows and their opt-outs \textemdash\ continued}\\[1.2mm]
  \toprule
\textbf{Flow} & \textbf{Contents} & \textbf{Opt-out} \\
\midrule\endhead
  \bottomrule\endfoot
  \bottomrule\endlastfoot
  \tblrows
  \end{longtable}}%
\else
  \typeout{HANDBOOK-TABLE atomic \the\dimexpr\ht\tblbox+\dp\tblbox\relax}%
  \noindent\tblbody
\fi
\par\addvspace{2.6mm}
\endgroup

Source: [43].

Error reports are sent only for a Pro, Max or Ultra sign-in on 2.1.198 or later, connecting directly to the Claude API, in an organisation with no ZDR or HIPAA agreement. A feedback report Claude drafts waits on your machine until you choose to send it, then takes the \texttt{/\allowbreak{}feedback} path. The survey is also off whenever \texttt{DISABLE\_\allowbreak{}TELEMETRY} or \texttt{DO\_\allowbreak{}NOT\_\allowbreak{}TRACK} is set; with \texttt{CLAUDE\_\allowbreak{}CODE\_\allowbreak{}ENABLE\_\allowbreak{}FEEDBACK\_\allowbreak{}SURVEY\_\allowbreak{}FOR\_\allowbreak{}OTEL=\allowbreak{}1} it comes back under any of the three blocking variables, logging ratings to your own collector only, with the transcript follow-up still off [43], [75].

\texttt{CLAUDE\_\allowbreak{}CODE\_\allowbreak{}DISABLE\_\allowbreak{}NONESSENTIA\allowbreak{}L\_\allowbreak{}TRAFFIC} disables all of the above at once \textbf{except} the WebFetch check and official marketplace auto-install, each of which has its own opt-out [43]. \textbf{The provider changes the defaults.} On Amazon Bedrock, Google Cloud's Agent Platform, Microsoft Foundry and Claude Platform on AWS, metrics, error reports and \texttt{/\allowbreak{}feedback} uploads are off by default — except that metrics default to on when a host platform sets \texttt{CLAUDE\_\allowbreak{}CODE\_\allowbreak{}PROVIDER\_\allowbreak{}MANAGED\_\allowbreak{}BY\_\allowbreak{}HOST} — and the survey and the WebFetch check still run. There, and with no Anthropic credentials configured, \texttt{/\allowbreak{}feedback} and a \textbf{Yes} to the survey's transcript follow-up write to a local archive under \texttt{\textasciitilde{}/\allowbreak{}.\allowbreak{}claude/\allowbreak{}feedback-\allowbreak{}bundles/\allowbreak{}} instead, and nothing leaves the machine until you forward the file. A signed-in Claude apps gateway session has usage analytics, error reporting and survey ratings to Anthropic disabled by its credential, with no setting to re-enable them [43]. Note the coupling: setting \texttt{DISABLE\_\allowbreak{}TELEMETRY} or \texttt{CLAUDE\_\allowbreak{}CODE\_\allowbreak{}DISABLE\_\allowbreak{}NONESSENTIA\allowbreak{}L\_\allowbreak{}TRAFFIC} also disables feature-flag evaluation, which can make Remote Control unavailable [43]. With \texttt{CLAUDE\_\allowbreak{}CODE\_\allowbreak{}DISABLE\_\allowbreak{}NONESSENTIA\allowbreak{}L\_\allowbreak{}TRAFFIC} it is unavailable; with only \texttt{DISABLE\_\allowbreak{}TELEMETRY} or \texttt{DO\_\allowbreak{}NOT\_\allowbreak{}TRACK} it stays available from 2.1.283 (2.1.304 in the Desktop app) unless the organisation requires Trusted Devices [41]. Other features that need fetched flags, such as account-synced skills and plugins and \texttt{claude impo\allowbreak{}rt}, stop working with any of these set [75].

\begin{calloutbox}{palebrass}{brassdark}{2.0mm}
\textbf{CAUTION:} if you disable the WebFetch preflight, combine it with \texttt{Web\allowbreak{}Fetch(domai\allowbreak{}n:\allowbreak{}…)} permission rules. Disabling the check means WebFetch will attempt any URL without consulting the blocklist [43].
\end{calloutbox}

\FloatBarrier
\setcounter{section}{1}
\section{Retention}

\begingroup
\def\tblrows{%
Free, Pro, Max, Ultra — data use for model improvement \textbf{on} & 5 years \\
Free, Pro, Max, Ultra — data use \textbf{off} & 30 days \\
Team, Enterprise, API — standard & 30 days \\
Claude for Enterprise with ZDR & Not stored after the response is returned, except as required by law or to address misuse \\
Transcripts shared via \texttt{/\allowbreak{}feedback}, \texttt{/\allowbreak{}bug}, \texttt{/\allowbreak{}share} & 5 years \\
Transcripts shared through the session-quality follow-up & Up to 6 months \\
Sessions flagged for a policy violation, even under ZDR & Up to 2 years \\
}%
\def\tblbody{\begin{minipage}{\textwidth}\boxcaption{Table 28 — Retention by account type}
{\footnotesize\begin{tabular}{L{94.2mm}L{48.7mm}}
\toprule
\textbf{Account} & \textbf{Server-side retention} \\
\midrule
\tblrows
\bottomrule\end{tabular}}\end{minipage}}%
\begingroup
\def\sloppy{\tolerance 9999\emergencystretch 3em\hfuzz 200pt\vfuzz 200pt}%
\hbadness=10000\vbadness=10000\hfuzz=200pt\vfuzz=200pt
\global\setbox\tblbox=\hbox{\tblbody}%
\endgroup
\par\addvspace{2.6mm}
\ifdim\dimexpr\ht\tblbox+\dp\tblbox\relax>0.55\textheight
  \typeout{HANDBOOK-TABLE broken \the\dimexpr\ht\tblbox+\dp\tblbox\relax}%
  \tabcaption{Table 28 — Retention by account type}
  {\footnotesize\begin{longtable}{L{94.2mm}L{48.7mm}}
  \toprule
\textbf{Account} & \textbf{Server-side retention} \\
\midrule\endfirsthead
  \multicolumn{2}{@{}l@{}}{%
  \sffamily\footnotesize\itshape\color{inkgrey}Table 28 — Retention by account type \textemdash\ continued}\\[1.2mm]
  \toprule
\textbf{Account} & \textbf{Server-side retention} \\
\midrule\endhead
  \bottomrule\endfoot
  \bottomrule\endlastfoot
  \tblrows
  \end{longtable}}%
\else
  \typeout{HANDBOOK-TABLE atomic \the\dimexpr\ht\tblbox+\dp\tblbox\relax}%
  \noindent\tblbody
\fi
\par\addvspace{2.6mm}
\endgroup

Source: [42], [43].

\textbf{Training.} Anthropic does not train generative models on code or prompts sent to Claude Code under commercial terms unless the customer opts in, for example through the Development Partner Program. Consumer accounts have a choice, and the setting governs Claude Code usage from those accounts too [43].

\textbf{Cloud sessions can be deleted one by one}, and deleting one permanently removes its event data [43].

\textbf{Local caching} is separate from all of the above: transcripts sit in plaintext under \texttt{\textasciitilde{}/\allowbreak{}.\allowbreak{}claude/\allowbreak{}projects/\allowbreak{}} for 30 days by default, adjustable with \texttt{cleanup\allowbreak{}Period\allowbreak{}Days} [32], [43]. Transcripts of sessions started or last continued in Claude Desktop or Cowork are the exception \emph{(requires 2.1.248 or later)}: they are kept at any age unless \texttt{desktop\allowbreak{}Session\allowbreak{}Cleanup\allowbreak{}Period\allowbreak{}Days} sets a limit, managed settings set \texttt{cleanup\allowbreak{}Period\allowbreak{}Days}, or the HIPAA configuration applies, among other cases (Section 12.3) [32]. Remember that auto memory is excluded from that sweep (Section 10.2) and that agent-team task lists are included in it (Section 30.2).

\textbf{Encryption at rest depends on the provider} [43]: infrastructure-level AES-256 on the Anthropic API; AES-256 with AWS-managed or customer-managed KMS keys on Amazon Bedrock; Google-managed keys with CMEK available on Google Cloud's Agent Platform; and on Microsoft Foundry it depends on the hosting option — for \emph{Hosted on Azure}, prompts and completions remain within Azure and only usage metadata and safety-flagged content egress to Anthropic.

\FloatBarrier
\setcounter{section}{2}
\section{Zero Data Retention, and its exact boundary}

ZDR for Claude Code is available to qualified accounts on Claude for Enterprise. It is \textbf{not} part of the standard Enterprise plan, cannot be enabled from admin settings, and is enabled per organisation by Anthropic after eligibility review — including for each new organisation created under the same account. It applies only to Anthropic's direct platform; on Amazon Bedrock, Google Cloud's Agent Platform or Microsoft Foundry, that platform's retention policy governs [42]. A Claude for Enterprise organisation with HIPAA enabled can bring the CLI and the Desktop Code tab under its Business Associate Agreement without ZDR, once the HIPAA configuration is applied to Claude Code and Cowork in local mode; without that configuration, BAA coverage of Claude Code still needs ZDR [42].

\textbf{Covered:} model inference calls made through Claude Code on Claude for Enterprise.

\textbf{Not covered, even for a ZDR organisation} [42]: chat on claude.ai; Cowork sessions; artifacts in chat and Cowork, which are kept until deleted unless the organisation sets a retention period; Claude Code Analytics, which stores no prompts or responses but does collect productivity metadata such as account emails and usage statistics, and shows no contribution metrics; user and seat management data; and anything processed by third-party tools, MCP servers or external integrations.

\textbf{Disabled under ZDR}, at the backend and regardless of what the client displays [42]:

\begin{itemize}
\item cloud sessions, including those started from the Desktop app;
\item the unified Chat and Cowork experience, which requires cloud sessions;
\item Claude Tag, which retains channel memory and session transcripts;
\item artifacts in Claude Code, and Claude Docs;
\item feedback submission (\texttt{/\allowbreak{}feedback}, \texttt{/\allowbreak{}bug}, \texttt{/\allowbreak{}share});
\item \textbf{Remote Control}, because it stores the transcript on Anthropic servers to sync across devices.
\end{itemize}

A disabled feature may still appear in the terminal; using it returns an error saying the organisation's policies do not allow it, and future features that store prompts or completions may join the list [42].

Claude Fable 5.2, 5.1 and 5 require data retention by default. \textbf{Whether a ZDR organisation can use them is decided by Anthropic's Covered Models policies, not by Claude Code}; where it cannot, they are absent from \texttt{/\allowbreak{}model} or shown disabled, and the server rejects requests for them. Fable models are not the default, and the \texttt{best} alias resolves to Opus [42]. Version 1.0 said Fable 5 was unavailable under ZDR outright; the page now makes it a policy question.

\begin{calloutbox}{palebrass}{brassdark}{2.0mm}
\textbf{CAUTION — the routing gap.} ZDR applies to requests that authenticate \textbf{into} the ZDR-enabled organisation. A developer signing in with a personal account, or with an API key from a different organisation, is not covered. Deploying \texttt{force\allowbreak{}Login\allowbreak{}Method} and \texttt{force\allowbreak{}Login\allowbreak{}Org\allowbreak{}UUID} through managed settings is the documented step [42], but it narrows the gap rather than closing it: the organisation check applies to claude.ai logins, not to which organisation a Console credential belongs to, and the interactive login screen pre-selects the forced method without enforcing it [58]. Chapter 31 lists the remaining paths and pairs the keys with \texttt{allowed\allowbreak{}Providers}. An organisation that believes it has ZDR but has deployed none of this has a policy, not a control.
\end{calloutbox}

\FloatBarrier
\setcounter{section}{3}
\section{Cloud execution}

In an Anthropic-hosted cloud session the repository is cloned to an isolated VM and Anthropic stores the session transcript so you can return to it; GitHub credentials are stored encrypted on Anthropic's servers and never enter the VM, because an Anthropic proxy attaches them to GitHub traffic server-side; \textbf{all outbound traffic passes through a security proxy for audit logging and abuse prevention}; and session data follows the retention policy for your account type [43], [59]. Claude reaches only the repository the session was started in, not every repository you have connected [43]. A self-hosted environment moves isolation, egress control and git credentials to your own deployment.

Remote Control follows the \emph{local} data flow, because execution is local — with the qualification that while connected the transcript is stored on Anthropic servers to sync across devices [41], [43]. That single sentence is why Remote Control is disabled under ZDR, and it is the fact most often missed when a team assumes "execution is local" means "nothing leaves".

\FloatBarrier
\setcounter{section}{4}
\section{A data-governance questionnaire}

Answer these before an organisational rollout, and record the answers:

\begin{enumerate}
\item Which plan and provider is each cohort of developers on, and therefore which retention row of Table 28 applies?
\item Is training opt-out contractual (commercial terms) or a user setting (consumer)?
\item Is \texttt{force\allowbreak{}Login\allowbreak{}Org\allowbreak{}UUID} deployed, so that sessions actually authenticate into the intended organisation?
\item Which of the flows in Table 27 are disabled, and by which delivery mechanism?
\item What is \texttt{cleanup\allowbreak{}Period\allowbreak{}Days}, and does anyone rely on the sweep covering auto memory? (It does not.)
\item Are artifacts, channels, Remote Control and web sessions enabled, and who approved each?
\item Where do published artifacts live, what is their retention, and who can share them publicly?
\item Which MCP servers and connectors are permitted, and what data can each reach?
\item Is \texttt{claude\_\allowbreak{}artifact\_\allowbreak{}*} and Claude Code audit-log data being collected and reviewed?
\item What is the incident procedure when a credential reaches a transcript (Section 14.3)?
\end{enumerate}

\begingroup
\def\kprows{%
\item Five non-inference flows leave the machine, each with its own opt-out; one variable disables all but the WebFetch check, and marketplace auto-install needs its own opt-out too.
\item Retention differs by plan, by feature and by consent, and local caching is separate from all of it.
\item ZDR covers inference on Claude for Enterprise and disables cloud sessions, Claude Tag, artifacts in Claude Code, feedback and Remote Control; whether Fable models are usable is a policy question.
\item ZDR applies to requests authenticating into the ZDR organisation; \texttt{force\allowbreak{}Login\allowbreak{}Org\allowbreak{}UUID} narrows the gap for claude.ai logins but does not check Console credentials.
}%
\def\kpbody{\begin{minipage}{\textwidth}\subsection*{Key points}\begin{itemize}\kprows\end{itemize}\end{minipage}}%
\begingroup
\def\sloppy{\tolerance 9999\emergencystretch 3em\hfuzz 200pt\vfuzz 200pt}%
\hbadness=10000\vbadness=10000\hfuzz=200pt\vfuzz=200pt
\global\setbox\kpbox=\hbox{\kpbody}%
\endgroup
\par\addvspace{4.2mm}
\ifdim\dimexpr\ht\kpbox+\dp\kpbox\relax>0.30\textheight
  \typeout{HANDBOOK-KEYPOINTS broken \the\dimexpr\ht\kpbox+\dp\kpbox\relax}%
  \subsection*{Key points}
  \begin{itemize}\kprows\end{itemize}
\else
  \typeout{HANDBOOK-KEYPOINTS atomic \the\dimexpr\ht\kpbox+\dp\kpbox\relax}%
  \noindent\kpbody
\fi
\par\addvspace{1.4mm}
\endgroup

\FloatBarrier
\renewcommand{\chaptertitlelabel}{Part VIII \textperiodcentered\ Chapter 33}
\setcounter{chapter}{32}
\chapter{Observability, cost control and accessibility}

\FloatBarrier
\setcounter{section}{0}
\section{OpenTelemetry}

Claude Code exports metrics, logs and — in beta — traces through OpenTelemetry [108]. This is the mechanism that turns "we use an AI agent" into something an assurance function can examine.

\begin{codeblock}{9.0}{10.6}
\cl{export~CLAUDE\_CODE\_ENABLE\_TELEMETRY=1}
\cl{export~OTEL\_METRICS\_EXPORTER=otlp~~~~~~~~~~\#~otlp~|~prometheus~|~console~|~none}
\cl{export~OTEL\_LOGS\_EXPORTER=otlp~~~~~~~~~~~~~\#~otlp~|~console~|~none}
\cl{export~OTEL\_EXPORTER\_OTLP\_PROTOCOL=grpc~~~~\#~grpc~|~http/json~|~http/protobuf}
\cl{export~OTEL\_EXPORTER\_OTLP\_ENDPOINT=http://collector.example.com:4317}
\cl{export~OTEL\_EXPORTER\_OTLP\_HEADERS="Authorization=Bearer~<token>"}
\end{codeblock}

Deploy it fleet-wide through the \texttt{env} block of managed settings (Chapter 31). \textbf{A repository cannot do it for you}: Claude Code ignores the exporter variables in a project's \texttt{.\allowbreak{}claude/\allowbreak{}settings.\allowbreak{}json} and \texttt{.\allowbreak{}claude/\allowbreak{}settings.\allowbreak{}local.\allowbreak{}json}, so a repository can neither turn telemetry on, nor choose where it goes, nor capture content; it can only turn a signal off by setting its exporter to \texttt{none}, and not even that when managed settings, a \texttt{-\allowbreak{}-\allowbreak{}settings} file or the launching environment sets the variable [108]. When managed settings set an \texttt{OTEL\_\allowbreak{}EXPORTER\_\allowbreak{}OTLP\_\allowbreak{}*} endpoint, protocol or credential, Claude Code removes the conflicting developer-set variables at startup \emph{(requires 2.1.217 or later)}: a managed generic endpoint removes every developer-set per-signal endpoint, and a managed credential removes developer-set endpoints too, so a developer cannot point one signal at another collector. Until 2.1.251 a developer-set \texttt{BETA\_\allowbreak{}TRACING\_\allowbreak{}ENDPOINT} could still redirect the logs and traces that detailed beta tracing exports. The exporter selectors still follow ordinary precedence, so lock them in managed settings too if they matter [108]. Claude Code does not pass \texttt{OTEL\_\allowbreak{}*} variables to the commands, hooks and servers it spawns [108].

Three surfaces are configured elsewhere [108]. Desktop Code tab sessions read managed settings, but the Cowork form under \textbf{Monitoring} in the admin console applies to Cowork only. Cloud sessions take the variables from server-managed settings or from the cloud environment's own variables, and Claude Tag sessions only from the latter, because they do not receive server-managed settings; anyone using an environment can read its variables, so a collector credential belongs in server-managed settings, and the collector's domain must be on the environment's network allowlist.

\begingroup
\def\tblrows{%
\texttt{claude\_\allowbreak{}code.\allowbreak{}session.\allowbreak{}count} & CLI sessions started & — \\
\texttt{claude\_\allowbreak{}code.\allowbreak{}lines\_\allowbreak{}of\_\allowbreak{}code.\allowbreak{}count} & Lines of code modified & — \\
\texttt{claude\_\allowbreak{}code.\allowbreak{}pull\_\allowbreak{}request.\allowbreak{}count} & Pull requests created & — \\
\texttt{claude\_\allowbreak{}code.\allowbreak{}commit.\allowbreak{}count} & Commits created & — \\
\texttt{claude\_\allowbreak{}code.\allowbreak{}cost.\allowbreak{}usage} & Session cost & USD \\
\texttt{claude\_\allowbreak{}code.\allowbreak{}token.\allowbreak{}usage} & Tokens used & tokens \\
\texttt{claude\_\allowbreak{}code.\allowbreak{}code\_\allowbreak{}edit\_\allowbreak{}tool.\allowbreak{}decision} & Code-edit permission decisions & — \\
\texttt{claude\_\allowbreak{}code.\allowbreak{}active\_\allowbreak{}time.\allowbreak{}total} & Active time & s \\
}%
\def\tblbody{\begin{minipage}{\textwidth}\boxcaption{Table 29 — Exported metrics}
{\footnotesize\begin{tabular}{L{76.7mm}L{50.8mm}L{12.3mm}}
\toprule
\textbf{Metric} & \textbf{Meaning} & \textbf{Unit} \\
\midrule
\tblrows
\bottomrule\end{tabular}}\end{minipage}}%
\begingroup
\def\sloppy{\tolerance 9999\emergencystretch 3em\hfuzz 200pt\vfuzz 200pt}%
\hbadness=10000\vbadness=10000\hfuzz=200pt\vfuzz=200pt
\global\setbox\tblbox=\hbox{\tblbody}%
\endgroup
\par\addvspace{2.6mm}
\ifdim\dimexpr\ht\tblbox+\dp\tblbox\relax>0.55\textheight
  \typeout{HANDBOOK-TABLE broken \the\dimexpr\ht\tblbox+\dp\tblbox\relax}%
  \tabcaption{Table 29 — Exported metrics}
  {\footnotesize\begin{longtable}{L{76.7mm}L{50.8mm}L{12.3mm}}
  \toprule
\textbf{Metric} & \textbf{Meaning} & \textbf{Unit} \\
\midrule\endfirsthead
  \multicolumn{3}{@{}l@{}}{%
  \sffamily\footnotesize\itshape\color{inkgrey}Table 29 — Exported metrics \textemdash\ continued}\\[1.2mm]
  \toprule
\textbf{Metric} & \textbf{Meaning} & \textbf{Unit} \\
\midrule\endhead
  \bottomrule\endfoot
  \bottomrule\endlastfoot
  \tblrows
  \end{longtable}}%
\else
  \typeout{HANDBOOK-TABLE atomic \the\dimexpr\ht\tblbox+\dp\tblbox\relax}%
  \noindent\tblbody
\fi
\par\addvspace{2.6mm}
\endgroup

Events carry the audit signal. \texttt{claude\_\allowbreak{}code.\allowbreak{}tool\_\allowbreak{}decision} records a permission decision with its \texttt{source}; \texttt{claude\_\allowbreak{}code.\allowbreak{}permission\_\allowbreak{}mode\_\allowbreak{}changed} records \texttt{from\_\allowbreak{}mode}, \texttt{to\_\allowbreak{}mode} and \texttt{trigger}; \texttt{claude\_\allowbreak{}code.\allowbreak{}mcp\_\allowbreak{}server\_\allowbreak{}connection} records connection status and transport; \texttt{claude\_\allowbreak{}code.\allowbreak{}plugin\_\allowbreak{}installed} and \texttt{claude\_\allowbreak{}code.\allowbreak{}plugin\_\allowbreak{}loaded} record plugin name and marketplace, for a third-party marketplace only with \texttt{OTEL\_\allowbreak{}LOG\_\allowbreak{}TOOL\_\allowbreak{}DETAILS=\allowbreak{}1}; \texttt{claude\_\allowbreak{}code.\allowbreak{}auth} records login and logout; \texttt{claude\_\allowbreak{}code.\allowbreak{}api\_\allowbreak{}refusal} records model refusals [108]. Events produced while processing a prompt share its \texttt{prompt.\allowbreak{}id}, so filtering by it traces every action that prompt triggered — precisely the query an incident review needs. Startup events such as \texttt{managed\_\allowbreak{}settings\_\allowbreak{}resolved} and \texttt{plugin\_\allowbreak{}loaded} belong to no prompt; correlate those by session, timestamp and \texttt{event.\allowbreak{}sequence} [108].

\begin{calloutbox}{palebrass}{brassdark}{2.0mm}
\textbf{CAUTION — content logging is off by default and should usually stay off.} \texttt{OTEL\_\allowbreak{}LOG\_\allowbreak{}USER\_\allowbreak{}PROMPTS}, \texttt{OTEL\_\allowbreak{}LOG\_\allowbreak{}ASSISTANT\_\allowbreak{}RESPONSES}, \texttt{OTEL\_\allowbreak{}LOG\_\allowbreak{}TOOL\_\allowbreak{}DETAILS}, \texttt{OTEL\_\allowbreak{}LOG\_\allowbreak{}TOOL\_\allowbreak{}CONTENT} and \texttt{OTEL\_\allowbreak{}LOG\_\allowbreak{}RAW\_\allowbreak{}API\_\allowbreak{}BODIES} are all disabled by default [108]. \texttt{OTEL\_\allowbreak{}LOG\_\allowbreak{}ASSISTANT\_\allowbreak{}RESPONSES} falls back to \texttt{OTEL\_\allowbreak{}LOG\_\allowbreak{}USER\_\allowbreak{}PROMPTS} when unset, and raw API bodies carry the whole conversation, so enabling them implies everything the other flags reveal; \texttt{file:\allowbreak{}<dir>} writes them untruncated to local files, which need their own collection and retention controls. Under detailed beta tracing the gates cross: model output and tool results on request spans follow \texttt{OTEL\_\allowbreak{}LOG\_\allowbreak{}USER\_\allowbreak{}PROMPTS}, so turning off \texttt{OTEL\_\allowbreak{}LOG\_\allowbreak{}ASSISTANT\_\allowbreak{}RESPONSES} or \texttt{OTEL\_\allowbreak{}LOG\_\allowbreak{}TOOL\_\allowbreak{}CONTENT} does not suppress them. A collector rule that drops prompt text must name both \texttt{prompt} and \texttt{prompt\_\allowbreak{}text} [108]. Enabling any of them moves prompt and code content into your telemetry pipeline, which changes that pipeline's classification, retention obligations and access controls. Decide that deliberately, with your privacy function, not as a debugging convenience.
\end{calloutbox}

Cardinality and identity controls matter for both cost and privacy: \texttt{OTEL\_\allowbreak{}METRICS\_\allowbreak{}INCLUDE\_\allowbreak{}SESSION\_\allowbreak{}ID} (default \texttt{true}), \texttt{OTEL\_\allowbreak{}METRICS\_\allowbreak{}INCLUDE\_\allowbreak{}ACCOUNT\_\allowbreak{}UUID} (default \texttt{true}), \texttt{OTEL\_\allowbreak{}METRICS\_\allowbreak{}INCLUDE\_\allowbreak{}VERSION} (default \texttt{false}) and \texttt{OTEL\_\allowbreak{}METRICS\_\allowbreak{}INCLUDE\_\allowbreak{}ENTRYPOINT} (default \texttt{false}) [108]. Standard attributes include \texttt{user.\allowbreak{}email} and \texttt{organizatio\allowbreak{}n.\allowbreak{}id} when authenticated — personal data by most definitions — and they go only to the endpoint you configure, never to Anthropic. With an API key, or on Amazon Bedrock, Google Cloud's Agent Platform or Microsoft Foundry, there is no Claude account in the session: only an installation identifier and the session identifier are populated, and you attach user identity yourself through \texttt{OTEL\_\allowbreak{}RESOURCE\_\allowbreak{}ATTRIBUTES} [108]. In a Claude Tag channel session Claude acts as the organisation's shared identity, so the \texttt{user.\allowbreak{}*} attributes do not say who tagged it [108].

Two limits for an assurance reading. Claude Code emits the raw event stream only; anomaly detection, baselining and alerting are your SIEM's job. And the \texttt{retention\_\allowbreak{}sweep} event is how you check that machines apply a managed \texttt{cleanup\allowbreak{}Period\allowbreak{}Days}; its absence on a machine is not, on its own, a failure [108].

The analytics dashboard serves Claude for Teams and Enterprise, with usage metrics, contribution metrics through the GitHub app, a leaderboard and CSV export; API customers have a Console dashboard of usage and spend. Contribution metrics are unavailable under ZDR and for repositories on GitHub Enterprise Server [42], [109]. Usage on Amazon Bedrock, Google Cloud's Agent Platform or Microsoft Foundry appears in neither dashboard; OpenTelemetry or a gateway is the way to attribute it per user [67].

\FloatBarrier
\setcounter{section}{1}
\section{Cost control}

Three levers, in order of effect [67]:

\begin{enumerate}
\item \textbf{Context discipline.} Every request in a long session carries the whole conversation, so usage climbs faster than activity suggests. Handoffs and clean sessions are a cost control, not only a quality one (Chapter 11).
\item \textbf{Architecture.} An agent team costs roughly linearly in teammates; a dynamic workflow can spawn up to 1,000 agents; a fork shares the parent's prompt cache and is cheaper than a fresh subagent (Chapter 30;Chapter 22;Section 16.1). Choose the cheapest structure that answers the question.
\item \textbf{Model and effort.} \texttt{CLAUDE\_\allowbreak{}CODE\_\allowbreak{}SUBAGENT\_\allowbreak{}MODEL} sets the default model for delegated work, beneath a subagent's own \texttt{model} field and a model Claude passes (Section 16.4); \texttt{workflow\allowbreak{}Size\allowbreak{}Guideline} guides the fan-out Claude asks for, and the runtime caps bound it; \texttt{/\allowbreak{}model} before a large run is worth the two seconds.
\end{enumerate}

Usage also accrues while you are idle: session recap, idle compaction, session-scoped scheduled tasks, cross-session messages delivered as new turns, and goal check-ins each send context without a prompt from you [67]. The session-cost estimates in \texttt{/\allowbreak{}usage}, the status line and OpenTelemetry are at list price unless a \texttt{model\allowbreak{}Pricing} table is delivered through managed settings \emph{(requires 2.1.242 or later)}, which changes what is reported, not what is charged; the \texttt{/\allowbreak{}model} picker's prices stay at list price [67].

Watch the \texttt{Large workf\allowbreak{}low} warning, which appears when a run schedules more than 25 agents or projects more than 1.5 million tokens, and remember it is advisory, suppressed under ultracode, and pegged to the size guideline in force rather than to 25 — set one and its agent count becomes the threshold [19]. \texttt{/\allowbreak{}usage} and \texttt{/\allowbreak{}insights} are the in-session views — on a subscription plan \texttt{/\allowbreak{}usage} attributes recent usage to skills, subagents, plugins and MCP servers and recommends changes, and \texttt{f} applies those it can \emph{(requires 2.1.305 or later)}, though not while Remote Control is connected. That breakdown, like \texttt{/\allowbreak{}insights}, reads only this machine's history; the plan-limit bars are fetched from the server. The analytics dashboard and OTel metrics are the organisational views [37], [67], [108], [109].

\FloatBarrier
\setcounter{section}{2}
\section{Accessibility}

Claude Code has a screen reader mode that replaces the visual interface with labelled, linear text for screen readers such as VoiceOver and NVDA: menus become numbered lists, diffs read line by line, and the terminal bell rings when Claude needs you [110]. \textbf{It is opt-in and does not start on its own when a screen reader is running.} Turn it on with \texttt{claude -\allowbreak{}-\allowbreak{}ax-\allowbreak{}screen-\allowbreak{}reader}, \texttt{CLAUDE\_\allowbreak{}AX\_\allowbreak{}SCREEN\_\allowbreak{}READER=\allowbreak{}1} or \texttt{"ax\allowbreak{}Screen\allowbreak{}Reader":\allowbreak{} true} in user settings \emph{(requires 2.1.181 or later)}. A project's settings cannot turn it on \emph{(from 2.1.305)}, and Claude Code says so at startup [57], [110]. Screen magnifiers, reduced motion and colourblind-friendly themes have their own settings — \texttt{CLAUDE\_\allowbreak{}CODE\_\allowbreak{}ACCESSIBILI\allowbreak{}TY}, \texttt{prefers\allowbreak{}Reduced\allowbreak{}Motion} and the daltonized themes [110]. Related controls: \texttt{/\allowbreak{}keybindings} opens the keyboard-shortcuts file [111]; \texttt{/\allowbreak{}statusline} configures a custom status bar showing context usage, cost and git state [112]; terminal configuration covers \texttt{Shift+Enter} for newlines, a terminal bell when Claude finishes, tmux behaviour and theme matching [113]; and the Concise output style materially reduces the volume of text a screen-reader user must traverse while preserving error reports and destructive-action confirmations in full (Chapter 19) [91].

An accessibility statement belongs in any organisational rollout plan. A tool that a colleague cannot use is not deployed; it is deployed to some people.

\begingroup
\def\kprows{%
\item OpenTelemetry turns 'we use an agent' into something an assurance function can examine.
\item Events raised while processing a prompt carry its identifier, so one instruction's full activity can be reconstructed; startup events need session and sequence instead.
\item Content logging is off by default; enabling it moves prompts and code into your telemetry pipeline and changes its classification.
\item Context discipline, then architecture, then model choice — in that order of effect on cost.
}%
\def\kpbody{\begin{minipage}{\textwidth}\subsection*{Key points}\begin{itemize}\kprows\end{itemize}\end{minipage}}%
\begingroup
\def\sloppy{\tolerance 9999\emergencystretch 3em\hfuzz 200pt\vfuzz 200pt}%
\hbadness=10000\vbadness=10000\hfuzz=200pt\vfuzz=200pt
\global\setbox\kpbox=\hbox{\kpbody}%
\endgroup
\par\addvspace{4.2mm}
\ifdim\dimexpr\ht\kpbox+\dp\kpbox\relax>0.30\textheight
  \typeout{HANDBOOK-KEYPOINTS broken \the\dimexpr\ht\kpbox+\dp\kpbox\relax}%
  \subsection*{Key points}
  \begin{itemize}\kprows\end{itemize}
\else
  \typeout{HANDBOOK-KEYPOINTS atomic \the\dimexpr\ht\kpbox+\dp\kpbox\relax}%
  \noindent\kpbody
\fi
\par\addvspace{1.4mm}
\endgroup

\breakrule

\FloatBarrier
\parttitle{Part IX}{Capstone}
\addcontentsline{toc}{part}{Part IX \textemdash\ Capstone}

\FloatBarrier
\renewcommand{\chaptertitlelabel}{Part IX \textperiodcentered\ Chapter 34}
\setcounter{chapter}{33}
\chapter{Build a verified research-brief agent}

This capstone combines every essential mechanism without requiring production access. Complete it in the practice repository of “The running project”. The measure of success is not that Claude says it is finished; it is that the evidence exists.

\FloatBarrier
\setcounter{section}{0}
\section{Target outcome}

Given an approved question and source set, the workspace produces a draft decision brief, an \textbf{independent} evidence review, a rendered deliverable, and a verification report. It is not meant to publish on its own, and Stage 1's deny rules make that a boundary rather than a wish.

\FloatBarrier
\setcounter{section}{1}
\section{Stage 1 — Establish control}

\begin{enumerate}
\item Run \texttt{claude -\allowbreak{}-\allowbreak{}version} and \texttt{claude doct\allowbreak{}or}; record the version and any diagnostics.
\item Confirm the version, model and account with \texttt{/\allowbreak{}status} [37], and the project directory and git branch with \texttt{pwd} and \texttt{git status}.
\item Review \texttt{.\allowbreak{}gitignore} and permission rules; run \texttt{/\allowbreak{}permissions}, which shows the rules by scope, and note where each one comes from [37].
\item Enable sandboxing with \texttt{/\allowbreak{}sandbox} where supported, including credential deny entries. Deny \texttt{Edit(/\allowbreak{}deliverable\allowbreak{}s/\allowbreak{}**)} and \texttt{Artifact} in \texttt{.\allowbreak{}claude/\allowbreak{}settings.\allowbreak{}json}, so that promotion and publishing are yours [29]. The leading \texttt{/\allowbreak{}} anchors the path at the project root; a \texttt{.\allowbreak{}/\allowbreak{}} path is relative to the current directory and stops protecting the folder once Claude changes directory [29]. A path rule binds Claude's file tools and the file commands it recognises in the shell, not a script that writes files itself. The sandbox carries the same deny into its write limits, but it closes that route only where it is supported, with Strict sandbox mode on (\texttt{allow\allowbreak{}Unsandboxed\allowbreak{}Commands:\allowbreak{} false}, with \texttt{fail\allowbreak{}If\allowbreak{}Unavailable}) and no \texttt{excluded\allowbreak{}Commands} entry covering the renderer: otherwise a matching allow rule also approves an unsandboxed retry, which runs with no prompt [28].
\item Commit a clean baseline.
\end{enumerate}

\textbf{Evidence:} version string, \texttt{git status} output, permission summary, sandbox status, baseline commit hash.

\FloatBarrier
\setcounter{section}{2}
\section{Stage 2 — Create project knowledge}

\begin{enumerate}
\item Finalise \texttt{CLAUDE.\allowbreak{}md} under 200 lines (Chapter 9).
\item Add \texttt{references/\allowbreak{}source-\allowbreak{}policy.\allowbreak{}md} with a retrieval rule rather than inlining it.
\item Add a path-scoped rule for \texttt{sources/\allowbreak{}**} that forbids modification.
\item Add \texttt{templates/\allowbreak{}claim-\allowbreak{}ledger.\allowbreak{}md} and \texttt{templates/\allowbreak{}brief.\allowbreak{}md}.
\item Inspect auto memory with \texttt{/\allowbreak{}memory} and remove anything inaccurate or over-generalised.
\end{enumerate}

\textbf{Evidence:} \texttt{/\allowbreak{}context all} lists, under Memory files, the instruction files you expect to load at launch and no others; the instruction files do not contradict each other.

\FloatBarrier
\setcounter{section}{3}
\section{Stage 3 — Choose the register}

Create the \emph{Evidence analyst} output style of Section 19.2, restart Claude Code so the new file is read, and select it with \texttt{/\allowbreak{}output-\allowbreak{}style} \emph{(2.1.269 or later)} or \texttt{/\allowbreak{}config} [37].

\textbf{Evidence:} a new session's first response distinguishes source claim from inference without being asked.

\FloatBarrier
\setcounter{section}{4}
\section{Stage 4 — Make the reusable procedure}

\begin{enumerate}
\item Complete one brief manually, with Claude.
\item Create the \texttt{research-\allowbreak{}brief} skill from the observed procedure (Section 15.1).
\item Review every file and every tool grant. Set \texttt{allowed-\allowbreak{}tools} to the minimum.
\item Test it on a genuinely different source set.
\end{enumerate}

\textbf{Evidence:} the second run follows the same output contract without importing topic-specific assumptions from the first.

\FloatBarrier
\setcounter{section}{5}
\section{Stage 5 — Add an independent reviewer}

\begin{enumerate}
\item Create the read-only \texttt{evidence-\allowbreak{}reviewer} subagent of Section 16.3.
\item \textbf{Confirm it is not a fork, and not a teammate.} Fork mode is on by default (Section 16.1); @-mention the agent so that it, and not a fork, runs, and verify from its opening output that it has no knowledge of the drafting conversation. Confirm \texttt{CLAUDE\_\allowbreak{}CODE\_\allowbreak{}EXPERIMENTA\allowbreak{}L\_\allowbreak{}AGENT\_\allowbreak{}TEAMS} is unset or \texttt{0}, because while teams are on a subagent Claude names launches as a teammate (Section 30.1).
\item Confirm the parent session is not in \texttt{accept\allowbreak{}Edits}, \texttt{bypass\allowbreak{}Permissions} or \texttt{auto}, each of which overrides the reviewer's \texttt{permission\allowbreak{}Mode}; a parent in plan mode keeps the reviewer in plan mode too [27]. \texttt{auto} is where an interactive session starts unless something changes it, so this step is not a formality.
\item Pass only the draft, the ledger and the source files.
\item Require a severity-ordered report and a pass/fail verdict.
\item Correct the draft and re-run the reviewer.
\end{enumerate}

\textbf{Evidence:} no unresolved high-severity unsupported claim; the reviewer's disagreements remain visible in the record rather than being silently reconciled.

\FloatBarrier
\setcounter{section}{6}
\section{Stage 6 — Add exactly one piece of bounded automation}

Choose \textbf{one}:

\begin{itemize}
\item a \texttt{Stop} hook that appends a timestamped, non-sensitive completion record (Chapter 17);
\item a \texttt{/\allowbreak{}loop} running a read-only local check during an attended session (Chapter 23);
\item a dynamic workflow with three genuinely independent research lanes (Chapter 22).
\end{itemize}

Do not add a plugin, a live MCP server, a cloud routine and browser access in order to "use every feature". Capability must answer a need. Every mechanism you add is one you must also review, monitor and be able to remove.

\textbf{Evidence:} a written threat model, a test result including the failure path, a removal procedure, and the measured \texttt{/\allowbreak{}usage} impact.

\FloatBarrier
\setcounter{section}{7}
\section{Stage 7 — Render and verify}

Generate the final draft in an editable source format, render it, then run:

\begin{itemize}
\item \textbf{structural checks} — required sections present, heading order correct, tables complete;
\item \textbf{source checks} — every material claim maps to a ledger row and a named source;
\item \textbf{link checks} — each URL resolves or is explicitly marked inaccessible;
\item \textbf{content checks} — contradictions, dates, scope and attribution;
\item \textbf{file checks} — opens successfully, expected page count, no clipping or overflow;
\item \textbf{security checks} — no secret in files, diff, logs, artifacts or quoted transcript excerpts;
\item \textbf{git checks} — reviewed diff containing only intended files, and a clean commit.
\end{itemize}

Run the checks as commands whose output lands in the transcript, so that a goal evaluator can see them (Section 21.1) and so that you can archive them.

\FloatBarrier
\setcounter{section}{8}
\section{Stage 8 — Human release gate}

Promotion from \texttt{drafts/\allowbreak{}} to \texttt{deliverable\allowbreak{}s/\allowbreak{}} happens only after a human confirms:

\begin{codeblock}{8.0}{9.4}
\cl{[~]~The~decision~question~is~answered.}
\cl{[~]~All~material~claims~are~supported,~or~visibly~qualified.}
\cl{[~]~The~independent~review~passed,~and~it~was~genuinely~independent.}
\cl{[~]~Commands~and~links~were~actually~executed,~and~their~output~was~inspected.}
\cl{[~]~External~side~effects~are~enumerated.}
\cl{[~]~No~credential~or~private~data~appears~in~any~output,~artifact~or~transcript~excerpt.}
\cl{[~]~The~rendered~document~was~opened~and~read.}
\cl{[~]~Publication,~sending~or~sharing~is~separately~approved.}
\end{codeblock}

\FloatBarrier
\setcounter{section}{9}
\section{A final goal}

\begin{codeblock}{7.0}{8.3}
\cl{/goal~Produce~drafts/final-brief.md~and~drafts/final-brief.pdf~from~the~approved~source~set.}
\cl{The~goal~is~met~only~when:~the~evidence-reviewer~subagent~reports~no~unresolved~high-severity}
\cl{finding;~the~structure,~link~and~security~checks~each~exit~0~with~their~output~shown~in~this}
\cl{conversation;~a~render~inspection~report~exists~at~notes/render-check.md;~git~diff~contains~only}
\cl{the~intended~files;~and~the~completion~message~lists~the~exact~commands~run~and~the~files}
\cl{changed.~Do~not~publish,~push,~use~external~accounts,~install~software,~or~promote~anything~to}
\cl{deliverables/.~Stop~and~ask~if~blocked,~on~any~missing~permission,~source~or~credential,~or}
\cl{after~25~turns.}
\end{codeblock}

Note what that condition does: every clause is something Claude must \textbf{surface in the conversation}, because the evaluator cannot run commands or read files (Section 21.1). A condition the evaluator cannot see is a condition it will accept on assertion.

The capstone is complete when the evidence exists — not when Claude says it is complete.

\begingroup
\def\kprows{%
\item Evidence-Gated Delivery: control, knowledge, register, procedure, independent review, bounded automation, verification, release.
\item Confirm the reviewer is genuinely independent and that the parent mode does not override it.
\item Add exactly one automation mechanism, with a threat model, a tested failure path and a removal procedure.
\item Run checks as commands whose output lands in the transcript, so the evidence is both evaluable and archivable.
}%
\def\kpbody{\begin{minipage}{\textwidth}\subsection*{Key points}\begin{itemize}\kprows\end{itemize}\end{minipage}}%
\begingroup
\def\sloppy{\tolerance 9999\emergencystretch 3em\hfuzz 200pt\vfuzz 200pt}%
\hbadness=10000\vbadness=10000\hfuzz=200pt\vfuzz=200pt
\global\setbox\kpbox=\hbox{\kpbody}%
\endgroup
\par\addvspace{4.2mm}
\ifdim\dimexpr\ht\kpbox+\dp\kpbox\relax>0.30\textheight
  \typeout{HANDBOOK-KEYPOINTS broken \the\dimexpr\ht\kpbox+\dp\kpbox\relax}%
  \subsection*{Key points}
  \begin{itemize}\kprows\end{itemize}
\else
  \typeout{HANDBOOK-KEYPOINTS atomic \the\dimexpr\ht\kpbox+\dp\kpbox\relax}%
  \noindent\kpbody
\fi
\par\addvspace{1.4mm}
\endgroup

\breakrule

\FloatBarrier
\frontchapter{Closing perspective}

The most capable Claude Code configuration is not the one with the most plugins, connectors, agents and permissions. It is the one that repeatedly produces an inspectable result from a clear objective, trusted context, narrow access and observed evidence.

Three propositions are worth carrying out of this book.

\textbf{Capability is not the constraint; verification is.} Every part of this system has grown faster than the human capacity to supervise it. Sixteen concurrent agents by default, a thousand per run, routines that fire on webhooks, channels that push events into a live session, teams that spawn without being asked. The scarce resource is not agent time. It is the number of things a person can meaningfully check, and every design decision should be read as a claim on that budget.

\textbf{Trust boundaries have moved inward.} The interesting attack surface is no longer the network perimeter but the content the agent reads: a web page, a tool result, a pull-request comment, a message from a peer session, a fire payload from a leaked token. Anthropic's own controls reflect this — untrusted-input wrappers, a separate model call for most web fetches, auto mode's classifier review of inter-agent messages, the rule that a teammate cannot consent on your behalf [30], [33]. Design your own workflows on the same assumption, and treat zero trust as the operating model rather than a slogan [4], [8].

\textbf{Governance is now a build-time concern.} Managed policy, retention, residency, audit and accessibility are not paperwork applied after a tool is adopted. They are settings, and settings have precedence orders, failure modes and gaps — a ZDR organisation without \texttt{force\allowbreak{}Login\allowbreak{}Org\allowbreak{}UUID}, a managed file silently skipped because a higher-priority source exists, an auto-memory directory outside the retention sweep. The frameworks are converging on the same demand: document the system, evidence the controls, and name the human who is accountable [6], [11].

Start small. Preserve what works in reviewed files. Add autonomy only where you can define both completion and failure. Make external actions deliberate. And verify the artefact, not the confidence of the final message.

\breakrule

\FloatBarrier
\appendix
\setcounter{chapter}{0}
\setcounter{section}{0}
\appendixchapter{A}{In-session command reference}
Commands are recognised at the start of a message. Type \texttt{/\allowbreak{}} to see what is available for your plan, platform and installed release; the table is a working selection, not the full list, and a few available commands, such as \texttt{/\allowbreak{}heapdump}, stay out of the \texttt{/\allowbreak{}} menu until you type the whole name [37]. Entries marked \textbf{[S]} are bundled skills rather than built-in commands; a scheduled task can only fire a skill Claude is permitted to invoke on its own (Section 23.3). Verified against [37] on 12 November 2026.

\begingroup
\def\tblrows{%
\texttt{/\allowbreak{}help} & Show help and available commands & First stop when the interface differs from this book \\
\texttt{/\allowbreak{}status} & Session, account and setting-source status & Shows which managed source was selected (Chapter 31) \\
\texttt{/\allowbreak{}doctor} \textbf{[S]} & Setup checkup; diagnose and fix & Alias \texttt{/\allowbreak{}checkup}; \texttt{claude doct\allowbreak{}or} is its read-only terminal counterpart: it can list fixes but applies none, and runs no \texttt{CLAUDE.\allowbreak{}md} checkup \\
\texttt{/\allowbreak{}config} & Open settings or set an individual key & Alias \texttt{/\allowbreak{}settings}. Home of output style, workflows, ultracode keyword \\
\texttt{/\allowbreak{}output-\allowbreak{}style} & List output styles, or switch to one & Applies from the next message; saved as \texttt{output\allowbreak{}Style} (Chapter 19) \\
\texttt{/\allowbreak{}login} · \texttt{/\allowbreak{}logout} & Authenticate, switch account, sign out & Several features need a claude.ai login, not an API key \\
\texttt{/\allowbreak{}privacy-\allowbreak{}settings} & View and update privacy settings & Pro, Max and Ultra only. See Chapter 32 \\
\texttt{/\allowbreak{}model} & Switch model and save as default & Availability changes \\
\texttt{/\allowbreak{}effort} & Set effort: \texttt{low}, \texttt{medium}, \texttt{high}, \texttt{xhigh}, \texttt{max} and \texttt{auto} & Availability differs by surface — see Table 3. \texttt{/\allowbreak{}effort ultr\allowbreak{}acode} and \texttt{/\allowbreak{}effort ultr\allowbreak{}acode off} switch workflow orchestration, which is a setting and not a level \\
\texttt{/\allowbreak{}fast} & Toggle fast mode & Model-dependent \\
\texttt{/\allowbreak{}advisor} & Enable or disable a second model Claude consults for guidance & Experimental; Anthropic API or Claude Platform on AWS only [65] \\
\texttt{/\allowbreak{}usage} & Cost, limits and activity & \texttt{/\allowbreak{}cost} and \texttt{/\allowbreak{}stats} are both documented aliases; \texttt{/\allowbreak{}stats} opens on the Stats tab \\
\texttt{/\allowbreak{}insights} & HTML report analysing recent sessions & Not in cloud sessions \\
\texttt{/\allowbreak{}context} & Context composition & Diagnose; do not chase a threshold \\
\texttt{/\allowbreak{}autocompact} & Set the auto-compact window &  \\
\texttt{/\allowbreak{}compact [in\allowbreak{}structions]} & Summarise the conversation & Lossy; say what must survive \\
\texttt{/\allowbreak{}clear} & Empty context, new conversation & Aliases \texttt{/\allowbreak{}reset}, \texttt{/\allowbreak{}new}. Removes an active goal [84]; whether it also clears session-scoped tasks is \textbf{UNVERIFIED} (Appendix J) \\
\texttt{/\allowbreak{}rewind} & Roll code and conversation to a checkpoint, or summarise from a message & Aliases \texttt{/\allowbreak{}checkpoint}, \texttt{/\allowbreak{}undo}. Cannot undo external side effects \\
\texttt{/\allowbreak{}plan} & Enter plan mode & Optional task description \\
\texttt{/\allowbreak{}permissions} & Manage allow/ask/deny rules & Alias \texttt{/\allowbreak{}allowed-\allowbreak{}tools}. Deny → ask → allow \\
\texttt{/\allowbreak{}sandbox} & Toggle sandbox mode & Not on native Windows; use WSL2 [28] \\
\texttt{/\allowbreak{}hooks} & View configured hooks & Read-only; shows source file \\
\texttt{/\allowbreak{}init} & Draft a starter \texttt{CLAUDE.\allowbreak{}md} & Review and shorten \\
\texttt{/\allowbreak{}memory} & Edit instruction files; toggle auto memory & Writes \texttt{auto\allowbreak{}Memory\allowbreak{}Enabled} \\
\texttt{/\allowbreak{}add-\allowbreak{}dir} & Add a working directory & Confirm the target first \\
\texttt{/\allowbreak{}cd} & Move the session to a new directory & Relocates session storage \\
\texttt{/\allowbreak{}diff} & Interactive diff viewer & Review before commit \\
\texttt{/\allowbreak{}export} & Export the conversation as plain text & The supported audit artefact \\
\texttt{/\allowbreak{}recap} & One-line session summary &  \\
\texttt{/\allowbreak{}btw} & Side question, not added to history &  \\
\texttt{/\allowbreak{}focus} & Show only the last prompt and response & Fullscreen rendering only; VS Code has its own Focus view \\
\texttt{/\allowbreak{}resume} · \texttt{/\allowbreak{}rename} · \texttt{/\allowbreak{}branch} & Session control & See Chapter 12 \\
\texttt{/\allowbreak{}fork} · \texttt{/\allowbreak{}subtask} & Copy to a background session; fork a subagent & Forks inherit the conversation. With agent view off, \texttt{/\allowbreak{}subtask} is unavailable and \texttt{/\allowbreak{}fork} spawns the forked subagent \\
\texttt{/\allowbreak{}background} & Detach the session & Alias \texttt{/\allowbreak{}bg}; carries \texttt{/\allowbreak{}loop} tasks with it \\
\texttt{/\allowbreak{}tasks} & Background work, finished subagents and watched artifacts & Alias \texttt{/\allowbreak{}bashes} \\
\texttt{/\allowbreak{}list-\allowbreak{}agents} & Subagents, teammates, other sessions & Alias \texttt{/\allowbreak{}peers}; only where cross-session messaging is on \\
\texttt{/\allowbreak{}stop} & Stop the attached background session & Transcript and worktree are kept \\
\texttt{/\allowbreak{}agents} & Prints a reminder to ask Claude, or to edit the agent files & The interactive editor ended at 2.1.197 \\
\texttt{/\allowbreak{}mcp} & Inspect and authenticate MCP servers & Review tools and scopes \\
\texttt{/\allowbreak{}plugin} · \texttt{/\allowbreak{}reload-\allowbreak{}plugins} & Plugin management & Review arbitrary-code risk first \\
\texttt{/\allowbreak{}import} & Bring Codex, Gemini CLI or Cursor configuration in & Requires 2.1.213+, Cursor 2.1.265+. Not on cloud providers, through an apps gateway, or with feature-flag fetching off \\
\texttt{/\allowbreak{}chrome} & Configure the browser connection & Shares authenticated browser state \\
\texttt{/\allowbreak{}artifacts} & List, attach, open, copy a link or delete & To update one from another session, attach it or give Claude its URL; \texttt{d} deletes one you own, outside cloud sessions \\
\texttt{/\allowbreak{}goal} · \texttt{/\allowbreak{}goal clear} & Persistent completion condition & Implemented as a prompt Stop hook \\
\texttt{/\allowbreak{}loop [inter\allowbreak{}val] [promp\allowbreak{}t]} \textbf{[S]} & Session-scoped recurring prompt & Alias \texttt{/\allowbreak{}proactive}. \textbf{Bare \texttt{/\allowbreak{}loop} runs a maintenance prompt} \\
\texttt{/\allowbreak{}schedule} & Create and manage cloud scheduled tasks, formerly routines & Alias \texttt{/\allowbreak{}routines}; research preview \\
\texttt{/\allowbreak{}remote-\allowbreak{}control} & Attach remote access & Alias \texttt{/\allowbreak{}rc} \\
\texttt{/\allowbreak{}desktop} & Continue the session in Desktop & Alias \texttt{/\allowbreak{}app}; macOS or x64 Windows, with a Claude subscription \\
\texttt{/\allowbreak{}teleport} & Pull a cloud session into the terminal & Alias \texttt{/\allowbreak{}tp}; needs a claude.ai subscription \\
\texttt{/\allowbreak{}mobile} & QR code for the mobile app & Aliases \texttt{/\allowbreak{}ios}, \texttt{/\allowbreak{}android} \\
\texttt{/\allowbreak{}workflows} & Workflow progress view & \texttt{s} saves a run as a command \\
\texttt{/\allowbreak{}deep-\allowbreak{}research <q\allowbreak{}uestion>} & Bundled research workflow & Requires WebSearch \\
\texttt{/\allowbreak{}batch} \textbf{[S]} & Large-scale parallel changes &  \\
\texttt{/\allowbreak{}code-\allowbreak{}review} \textbf{[S]} & Review a diff or PR & Alias \texttt{/\allowbreak{}review}; \texttt{ultra} for the cloud pass \\
\texttt{/\allowbreak{}security-\allowbreak{}review} & Security pass over the branch's diff against origin's default branch & A built-in command Claude can also invoke as a skill; needs an \texttt{origin} remote. Not a substitute for specialist review [36] \\
\texttt{/\allowbreak{}verify} \textbf{[S]} & Run and observe verification & \texttt{disable-\allowbreak{}model-\allowbreak{}invocation:\allowbreak{} true} \\
\texttt{/\allowbreak{}fewer-\allowbreak{}permission-\allowbreak{}prompts} \textbf{[S]} & Propose an allowlist from transcripts & Review before applying \\
\texttt{/\allowbreak{}update-\allowbreak{}config} \textbf{[S]} & Describe a settings change; Claude edits the settings file & Review the diff: it writes permission rules and hooks \\
\texttt{/\allowbreak{}skills} · \texttt{/\allowbreak{}reload-\allowbreak{}skills} & List skills and set their visibility; rescan from disk & Plugin skills, \texttt{disable-\allowbreak{}model-\allowbreak{}invocation} skills and managed overrides cannot be cycled \\
\texttt{/\allowbreak{}auto-\allowbreak{}mode-\allowbreak{}setup} & Draft \texttt{auto\allowbreak{}Mode.\allowbreak{}environment} entries & See [114] \\
\texttt{/\allowbreak{}debug} \textbf{[S]} · \texttt{/\allowbreak{}heapdump} & Diagnostics &  \\
\texttt{/\allowbreak{}voice} & Voice dictation & Proofread negations and numbers \\
\texttt{/\allowbreak{}keybindings} · \texttt{/\allowbreak{}color} · \texttt{/\allowbreak{}theme} & Interface configuration & See [111] \\
\texttt{/\allowbreak{}feedback} · \texttt{/\allowbreak{}bug} · \texttt{/\allowbreak{}share} & Send feedback or a transcript & Sends conversation content after a consent screen; on a third-party provider writes a local archive instead — see Table 27 \\
\texttt{/\allowbreak{}install-\allowbreak{}github-\allowbreak{}app} · \texttt{/\allowbreak{}install-\allowbreak{}slack-\allowbreak{}app} & Integration installers & The GitHub app works only with github.com repositories \\
\texttt{/\allowbreak{}autofix-\allowbreak{}pr} & Watch a PR and push fixes & Can trigger comment-driven automation \\
\texttt{/\allowbreak{}powerup} & Interactive feature lessons &  \\
\texttt{/\allowbreak{}restart} & Restart and resume, applying an installed update & Alias \texttt{/\allowbreak{}update}; 2.1.288+; Pro, Max or Ultra; native install on macOS or Linux, not WSL \\
\texttt{/\allowbreak{}exit} & Exit & Alias \texttt{/\allowbreak{}quit} \\
}%
\def\tblbody{\begin{minipage}{\textwidth}
{\scriptsize\begin{tabular}{L{48.6mm}L{40.8mm}L{50.5mm}}
\toprule
\textbf{Command} & \textbf{Purpose} & \textbf{Note} \\
\midrule
\tblrows
\bottomrule\end{tabular}}\end{minipage}}%
\begingroup
\def\sloppy{\tolerance 9999\emergencystretch 3em\hfuzz 200pt\vfuzz 200pt}%
\hbadness=10000\vbadness=10000\hfuzz=200pt\vfuzz=200pt
\global\setbox\tblbox=\hbox{\tblbody}%
\endgroup
\par\addvspace{2.6mm}
\ifdim\dimexpr\ht\tblbox+\dp\tblbox\relax>0.55\textheight
  \typeout{HANDBOOK-TABLE broken \the\dimexpr\ht\tblbox+\dp\tblbox\relax}%
  
  {\scriptsize\begin{longtable}{L{48.6mm}L{40.8mm}L{50.5mm}}
  \toprule
\textbf{Command} & \textbf{Purpose} & \textbf{Note} \\
\midrule\endfirsthead
  \multicolumn{3}{@{}l@{}}{%
  \sffamily\footnotesize\itshape\color{inkgrey}Continued}\\[1.2mm]
  \toprule
\textbf{Command} & \textbf{Purpose} & \textbf{Note} \\
\midrule\endhead
  \bottomrule\endfoot
  \bottomrule\endlastfoot
  \tblrows
  \end{longtable}}%
\else
  \typeout{HANDBOOK-TABLE atomic \the\dimexpr\ht\tblbox+\dp\tblbox\relax}%
  \noindent\tblbody
\fi
\par\addvspace{2.6mm}
\endgroup

\FloatBarrier
\setcounter{chapter}{1}
\setcounter{section}{0}
\appendixchapter{B}{Shell commands and flags}
Run \texttt{claude -\allowbreak{}-\allowbreak{}help} and \texttt{claude mcp -\allowbreak{}-\allowbreak{}help} before scripting: interactive commands and shell flags are different interfaces, and a few flags are deliberately absent from \texttt{-\allowbreak{}-\allowbreak{}help} while their feature is in preview. Verified against [55] on 12 November 2026.

\texttt{claude -\allowbreak{}-\allowbreak{}help} reports 67 long options in option position and 18 subcommands on 2.1.322, against 62 and 13 on 2.1.246. Three further flags — \texttt{-\allowbreak{}-\allowbreak{}allowed-\allowbreak{}tools}, \texttt{-\allowbreak{}-\allowbreak{}background} and \texttt{-\allowbreak{}-\allowbreak{}disallowed-\allowbreak{}tools} — are named only inside other flags' descriptions rather than in option position, and are not counted among the 67. The listing below is a practitioner's selection from that set, not a reproduction of it, and \texttt{-\allowbreak{}-\allowbreak{}effort} is one of the flags whose accepted values exceed what \texttt{-\allowbreak{}-\allowbreak{}help} enumerates (Table 3).

\begin{codeblock}{7.0}{8.3}
\cl{claude~~~~~~~~~~~~~~~~~~~~~~~~~~~~~~~~~~\#~start~interactive}
\cl{claude~-{}-version~~~~~~~~~~~~~~~~~~~~~~~~\#~version}
\cl{claude~doctor~~~~~~~~~~~~~~~~~~~~~~~~~~~\#~diagnostics}
\cl{claude~-p~"<prompt>"~~~~~~~~~~~~~~~~~~~~\#~non-interactive~(alias~-{}-print)}
\cl{claude~-p~-{}-output-format~json~~~~~~~~~~\#~structured~result:~JSON~or~stream-json}
\cl{claude~-{}-continue~~~~~~~~~~~~~~~~~~~~~~~\#~resume~most~recent~here~(alias~-c)}
\cl{claude~-{}-resume~[name|id]~~~~~~~~~~~~~~~\#~picker,~or~resume~directly~(alias~-r)}
\cl{claude~-{}-from-pr~<number>~~~~~~~~~~~~~~~\#~picker~filtered~to~a~pull~request}
\cl{claude~-n~<name>~~~~~~~~~~~~~~~~~~~~~~~~\#~name~the~session~(alias~-{}-name)}
\cl{claude~-{}-resume~<id>~-{}-fork-session~~~~~\#~resume~a~copy:~new~ID,~new~process}
\cl{claude~-p~-{}-no-session-persistence~~~~~~\#~print~mode~only:~no~transcript~saved}
\cl{claude~-{}-model~<model>~~~~~~~~~~~~~~~~~~\#~model~for~this~session}
\cl{claude~-{}-effort~ultracode~~~~~~~~~~~~~~~\#~ultracode~on~and~effort~xhigh,~where~available~(2.1.203+)}
\cl{claude~-{}-permission-mode~plan~~~~~~~~~~~\#~start~in~a~specific~permission~mode}
\cl{claude~-{}-dangerously-skip-permissions~~~\#~isolated~environments~only}
\cl{claude~-{}-add-dir~<path>~~~~~~~~~~~~~~~~~\#~additional~working~directory}
\cl{claude~-{}-settings~<file>~~~~~~~~~~~~~~~~\#~explicit~settings~payload}
\cl{claude~-{}-setting-sources~<list>~~~~~~~~~\#~restrict~which~settings~sources~load}
\cl{claude~-{}-append-system-prompt~"<text>"~~\#~one-invocation~system~prompt~addition}
\cl{claude~-{}-agent~<name>~~~~~~~~~~~~~~~~~~~\#~run~the~session~as~a~subagent~definition}
\cl{claude~-{}-agents~\textquotesingle{}<json>\textquotesingle{}~~~~~~~~~~~~~~~~\#~define~session-scoped~subagents}
\cl{claude~-{}-bg~"<prompt>"~~~~~~~~~~~~~~~~~~\#~background~session~(alias~-{}-background)}
\cl{claude~agents~~~~~~~~~~~~~~~~~~~~~~~~~~~\#~agent~view:~monitor~background~sessions}
\cl{claude~-{}-worktree~<name>~~~~~~~~~~~~~~~~\#~isolated~git~worktree~and~branch}
\cl{claude~-{}-teammate-mode~auto~~~~~~~~~~~~~\#~agent-team~display~mode~(experimental)}
\cl{claude~-{}-chrome~/~-{}-no-chrome~~~~~~~~~~~\#~browser~integration~for~this~session}
\cl{claude~-{}-remote-control~[name]~~~~~~~~~~\#~expose~this~local~session~(alias~-{}-rc)}
\cl{claude~remote-control~~~~~~~~~~~~~~~~~~~\#~server~mode}
\cl{claude~-{}-cloud~"<task>"~~~~~~~~~~~~~~~~~\#~create~a~cloud~session}
\cl{claude~-p~"<msg>"~-{}-cloud~<session-id>~\#~queue~a~message~into~a~cloud~session}
\cl{claude~-{}-teleport~[session-id]~~~~~~~~~~\#~pull~a~cloud~session~into~the~terminal}
\cl{claude~-{}-channels~plugin:<name>@<mkt>~~~\#~enable~channel~servers~for~this~session}
\cl{claude~-{}-plugin-dir~<path|zip>~~~~~~~~~~\#~load~a~local~plugin}
\cl{claude~-{}-plugin-url~<url>~~~~~~~~~~~~~~~\#~load~a~plugin~archive~for~this~session}
\cl{claude~-{}-mcp-config~<file>~~~~~~~~~~~~~~\#~session~MCP~configuration}
\cl{claude~mcp~add~-{}-transport~http~<n>~<u>~\#~add~an~MCP~server}
\cl{claude~mcp~list~|~get~|~remove~|~login~|~logout}
\cl{claude~plugin~init~|~validate~|~details~<name>}
\cl{claude~plugin~marketplace~add~<owner/repo>}
\cl{claude~auth~login~~~~~~~~~~~~~~~~~~~~~~~\#~account~authentication}
\cl{claude~setup-token~~~~~~~~~~~~~~~~~~~~~~\#~long-lived~token~(blocks~Chrome,~artifacts)}
\end{codeblock}

\texttt{claude -\allowbreak{}-\allowbreak{}effort auto} is refused: the flag warns and the session runs at the default effort. \texttt{claude -\allowbreak{}-\allowbreak{}permission-\allowbreak{}mode} is the one flag whose value set this book states in full, in Table 5.

Flags not restored on \texttt{-\allowbreak{}-\allowbreak{}resume}: \texttt{-\allowbreak{}-\allowbreak{}mcp-\allowbreak{}config}, \texttt{-\allowbreak{}-\allowbreak{}settings}, \texttt{-\allowbreak{}-\allowbreak{}plugin-\allowbreak{}dir}, \texttt{-\allowbreak{}-\allowbreak{}fallback-\allowbreak{}model}, \texttt{-\allowbreak{}-\allowbreak{}add-\allowbreak{}dir} [31].

\FloatBarrier
\setcounter{chapter}{2}
\setcounter{section}{0}
\appendixchapter{C}{Files, directories and scope}
\texttt{\textasciitilde{}} is the current user's home directory. Windows paths resolve under the user profile; use the live settings documentation for exact platform expansion [32], [70].

\begingroup
\def\tblrows{%
\texttt{CLAUDE.\allowbreak{}md} & Project instructions & Yes, after review \\
\texttt{.\allowbreak{}claude/\allowbreak{}CLAUDE.\allowbreak{}md} & Alternative project instruction location & Yes, after review \\
\texttt{CLAUDE.\allowbreak{}local.\allowbreak{}md} & Private project instructions & No; keep ignored \\
\texttt{\textasciitilde{}/\allowbreak{}.\allowbreak{}claude/\allowbreak{}CLAUDE.\allowbreak{}md} & User-wide instructions & No \\
\texttt{/\allowbreak{}Library/\allowbreak{}Application\allowbreak{} Support/\allowbreak{}Claude\allowbreak{}Code/\allowbreak{}CLAUDE.\allowbreak{}md} (macOS) · \texttt{/\allowbreak{}etc/\allowbreak{}claude-\allowbreak{}code/\allowbreak{}CLAUDE.\allowbreak{}md} (Linux, WSL) · \texttt{C:\allowbreak{}\textbackslash{}Program Fil\allowbreak{}es\textbackslash{}Claude\allowbreak{}Code\textbackslash{}CLAUDE.\allowbreak{}md} (Windows) & Managed policy instructions; cannot be excluded & Deployed by MDM \\
\texttt{.\allowbreak{}claude/\allowbreak{}rules/\allowbreak{}*.\allowbreak{}md} & Modular project rules, optionally path-scoped & Usually yes \\
\texttt{\textasciitilde{}/\allowbreak{}.\allowbreak{}claude/\allowbreak{}rules/\allowbreak{}*.\allowbreak{}md} & User rules; load before project rules & No \\
\texttt{.\allowbreak{}claude/\allowbreak{}settings.\allowbreak{}json} & Shared project settings & Yes, if secret-free \\
\texttt{.\allowbreak{}claude/\allowbreak{}settings.\allowbreak{}local.\allowbreak{}json} & Personal project settings and saved approvals & No \\
\texttt{\textasciitilde{}/\allowbreak{}.\allowbreak{}claude/\allowbreak{}settings.\allowbreak{}json} & User settings & No \\
\texttt{managed-\allowbreak{}settings.\allowbreak{}json} and \texttt{managed-\allowbreak{}settings.\allowbreak{}d/\allowbreak{}*.\allowbreak{}json} in the system directory & Organisation policy & Deployed, not committed \\
\texttt{managed-\allowbreak{}mcp.\allowbreak{}json} & Managed MCP configuration & Deployed \\
\texttt{.\allowbreak{}mcp.\allowbreak{}json} & Shared project MCP configuration & Yes only if secret-free and reviewed \\
\texttt{\textasciitilde{}/\allowbreak{}.\allowbreak{}claude.\allowbreak{}json} & User state; local and user MCP configuration & No \\
\texttt{.\allowbreak{}claude/\allowbreak{}skills/\allowbreak{}<name>/\allowbreak{}SKILL.\allowbreak{}md} & Project skill & Yes, after security review \\
\texttt{\textasciitilde{}/\allowbreak{}.\allowbreak{}claude/\allowbreak{}skills/\allowbreak{}<name>/\allowbreak{}SKILL.\allowbreak{}md} & User skill & No \\
\texttt{.\allowbreak{}claude/\allowbreak{}commands/\allowbreak{}<name>.\allowbreak{}md} & Project command, the older single-file form of a skill & Yes, after the same review as a skill \\
\texttt{.\allowbreak{}claude/\allowbreak{}agents/\allowbreak{}<name>.\allowbreak{}md} & Project subagent & Yes, after tool review \\
\texttt{\textasciitilde{}/\allowbreak{}.\allowbreak{}claude/\allowbreak{}agents/\allowbreak{}<name>.\allowbreak{}md} & User subagent & No \\
\texttt{.\allowbreak{}claude/\allowbreak{}workflows/\allowbreak{}*.\allowbreak{}js} · \texttt{\textasciitilde{}/\allowbreak{}.\allowbreak{}claude/\allowbreak{}workflows/\allowbreak{}*.\allowbreak{}js} & Saved dynamic workflows & Project ones, after review \\
\texttt{.\allowbreak{}claude/\allowbreak{}output-\allowbreak{}styles/\allowbreak{}} · \texttt{\textasciitilde{}/\allowbreak{}.\allowbreak{}claude/\allowbreak{}output-\allowbreak{}styles/\allowbreak{}} & Output styles & Project ones, yes \\
\texttt{.\allowbreak{}claude/\allowbreak{}loop.\allowbreak{}md} · \texttt{\textasciitilde{}/\allowbreak{}.\allowbreak{}claude/\allowbreak{}loop.\allowbreak{}md} & Default prompt for a bare \texttt{/\allowbreak{}loop} & Project one, after review \\
\texttt{.\allowbreak{}claude/\allowbreak{}hooks/\allowbreak{}} & Hook scripts & Yes, after review \\
\texttt{\textasciitilde{}/\allowbreak{}.\allowbreak{}claude/\allowbreak{}projects/\allowbreak{}<project>/\allowbreak{}<session-\allowbreak{}id>.\allowbreak{}jsonl} & Session transcripts, plaintext JSONL & No \\
\texttt{\textasciitilde{}/\allowbreak{}.\allowbreak{}claude/\allowbreak{}projects/\allowbreak{}<project>/\allowbreak{}memory/\allowbreak{}} & Auto memory; \textbf{excluded from the retention sweep} & No \\
\texttt{\textasciitilde{}/\allowbreak{}.\allowbreak{}claude/\allowbreak{}agent-\allowbreak{}memory/\allowbreak{}<name>/\allowbreak{}} · \texttt{.\allowbreak{}claude/\allowbreak{}agent-\allowbreak{}memory[-\allowbreak{}local]/\allowbreak{}<name>/\allowbreak{}} & Subagent persistent memory & Project one, deliberately \\
\texttt{\textasciitilde{}/\allowbreak{}.\allowbreak{}claude/\allowbreak{}teams/\allowbreak{}<team>/\allowbreak{}} · \texttt{\textasciitilde{}/\allowbreak{}.\allowbreak{}claude/\allowbreak{}tasks/\allowbreak{}<team>/\allowbreak{}} & Agent-team config, mailboxes and task list & No \\
\texttt{\textasciitilde{}/\allowbreak{}.\allowbreak{}claude/\allowbreak{}channels/\allowbreak{}<name>/\allowbreak{}.\allowbreak{}env} & Channel plugin credentials & \textbf{Never} \\
\texttt{\textasciitilde{}/\allowbreak{}.\allowbreak{}claude/\allowbreak{}feedback-\allowbreak{}bundles/\allowbreak{}} & Feedback archives written on third-party providers or without Anthropic credentials & No \\
\texttt{.\allowbreak{}env} & Local environment values & \textbf{Never} \\
\texttt{.\allowbreak{}env.\allowbreak{}example} & Variable names only & Yes \\
\texttt{.\allowbreak{}gitignore} · \texttt{.\allowbreak{}worktreeinc\allowbreak{}lude} & Exclusion and worktree rules & Yes \\
}%
\def\tblbody{\begin{minipage}{\textwidth}
{\scriptsize\begin{tabular}{L{76.9mm}L{38.3mm}L{24.6mm}}
\toprule
\textbf{Path} & \textbf{Scope and purpose} & \textbf{Commit?} \\
\midrule
\tblrows
\bottomrule\end{tabular}}\end{minipage}}%
\begingroup
\def\sloppy{\tolerance 9999\emergencystretch 3em\hfuzz 200pt\vfuzz 200pt}%
\hbadness=10000\vbadness=10000\hfuzz=200pt\vfuzz=200pt
\global\setbox\tblbox=\hbox{\tblbody}%
\endgroup
\par\addvspace{2.6mm}
\ifdim\dimexpr\ht\tblbox+\dp\tblbox\relax>0.55\textheight
  \typeout{HANDBOOK-TABLE broken \the\dimexpr\ht\tblbox+\dp\tblbox\relax}%
  
  {\scriptsize\begin{longtable}{L{76.9mm}L{38.3mm}L{24.6mm}}
  \toprule
\textbf{Path} & \textbf{Scope and purpose} & \textbf{Commit?} \\
\midrule\endfirsthead
  \multicolumn{3}{@{}l@{}}{%
  \sffamily\footnotesize\itshape\color{inkgrey}Continued}\\[1.2mm]
  \toprule
\textbf{Path} & \textbf{Scope and purpose} & \textbf{Commit?} \\
\midrule\endhead
  \bottomrule\endfoot
  \bottomrule\endlastfoot
  \tblrows
  \end{longtable}}%
\else
  \typeout{HANDBOOK-TABLE atomic \the\dimexpr\ht\tblbox+\dp\tblbox\relax}%
  \noindent\tblbody
\fi
\par\addvspace{2.6mm}
\endgroup

\FloatBarrier
\section*{Settings precedence}
\markboth{Settings precedence}{Settings precedence}

Highest to lowest [70]:

\begin{codeblock}{9.0}{10.6}
\cl{Managed~settings~~~(managed-settings.json,~MDM,~or~the~claude.ai~console)}
\cl{~~~~~~~~↓}
\cl{Command~line~~~~~~~(claude~-{}-settings)}
\cl{~~~~~~~~↓}
\cl{Project~local~~~~~~(.claude/settings.local.json)}
\cl{~~~~~~~~↓}
\cl{Shared~project~~~~~(.claude/settings.json)}
\cl{~~~~~~~~↓}
\cl{User~~~~~~~~~~~~~~~(\textasciitilde{}/.claude/settings.json)}
\end{codeblock}

Environment variables are not a level in this stack: where a behaviour has both a variable and a key, which one applies is documented per pair [70].

Two cautions. \textbf{No level, including command-line arguments, overrides a managed permission rule} [29]. And permission-rule evaluation has its own precedence — deny, then ask, then allow — which is independent of settings precedence: do not assume a narrower allow defeats a broader deny (Section 7.2).

\FloatBarrier
\setcounter{chapter}{3}
\setcounter{section}{0}
\appendixchapter{D}{Settings and variables of security consequence}
A working set for a security review. Full definitions are in [61] and [75].

\begingroup
\def\tblrows{%
\texttt{permissions.\allowbreak{}deny} / \texttt{ask} / \texttt{allow} & Tool access. Deny → ask → allow \\
\texttt{permissions.\allowbreak{}disable\allowbreak{}Bypass\allowbreak{}Permissions\allowbreak{}Mode} & Prevents \texttt{bypass\allowbreak{}Permissions} \\
\texttt{allow\allowbreak{}Managed\allowbreak{}Permission\allowbreak{}Rules\allowbreak{}Only} & Only managed settings supply permission rules; \texttt{-\allowbreak{}-\allowbreak{}disallowed\allowbreak{}Tools} and the session's own deny and ask rules still restrict \\
\texttt{allow\allowbreak{}Managed\allowbreak{}Hooks\allowbreak{}Only} & Blocks user, project and local hooks and hooks of plugins not force-enabled by managed settings — and \texttt{/\allowbreak{}goal} \\
\texttt{allow\allowbreak{}Managed\allowbreak{}Mcp\allowbreak{}Servers\allowbreak{}Only} · \texttt{allowed\allowbreak{}Mcp\allowbreak{}Servers} · \texttt{denied\allowbreak{}Mcp\allowbreak{}Servers} & MCP allow/deny at organisation level \\
\texttt{disable\allowbreak{}All\allowbreak{}Hooks} & Turns hooks off. Outside managed settings, managed hooks keep running; in managed settings, every configured hook stops (SDK in-process hooks excepted) \\
\texttt{allowed\allowbreak{}Http\allowbreak{}Hook\allowbreak{}Urls} · \texttt{http\allowbreak{}Hook\allowbreak{}Allowed\allowbreak{}Env\allowbreak{}Vars} & Restrict HTTP hook destinations and header interpolation \\
\texttt{sandbox.\allowbreak{}enabled} · \texttt{sandbox.\allowbreak{}filesystem.\allowbreak{}deny\allowbreak{}Read} / \texttt{deny\allowbreak{}Write} / \texttt{allow\allowbreak{}Read} & Filesystem isolation \\
\texttt{sandbox.\allowbreak{}credentials.\allowbreak{}files} / \texttt{.\allowbreak{}env\allowbreak{}Vars} & Credential deny or mask \\
\texttt{sandbox.\allowbreak{}network.\allowbreak{}allow\allowbreak{}Managed\allowbreak{}Domains\allowbreak{}Only} · \texttt{strict\allowbreak{}Allowlist} & Network allowlist locks \\
\texttt{sandbox.\allowbreak{}auto\allowbreak{}Allow\allowbreak{}Bash\allowbreak{}If\allowbreak{}Sandboxed} & Defaults to \texttt{true}; \texttt{false} sends sandboxed commands through the regular permission flow, where your rules and mode decide \\
\texttt{strict\allowbreak{}Known\allowbreak{}Marketplace\allowbreak{}s} · \texttt{blocked\allowbreak{}Marketplace\allowbreak{}s} · \texttt{plugin\allowbreak{}Trust\allowbreak{}Message} & Plugin supply chain \\
\texttt{strict\allowbreak{}Plugin\allowbreak{}Only\allowbreak{}Customizati\allowbreak{}on} & Blocks skills, agents, hooks and MCP from user and project sources: \texttt{true} for all four, or an array naming the ones to lock \\
\texttt{disable\allowbreak{}Sideload\allowbreak{}Flags} & Rejects \texttt{-\allowbreak{}-\allowbreak{}plugin-\allowbreak{}dir}, \texttt{-\allowbreak{}-\allowbreak{}plugin-\allowbreak{}url}, \texttt{-\allowbreak{}-\allowbreak{}agents}, \texttt{-\allowbreak{}-\allowbreak{}mcp-\allowbreak{}config} at startup, except an \texttt{-\allowbreak{}-\allowbreak{}mcp-\allowbreak{}config} of in-process SDK servers; in cloud sessions it filters the delivered servers instead \\
\texttt{disable\allowbreak{}Bundled\allowbreak{}Skills} · \texttt{skill\allowbreak{}Overrides} · \texttt{disable\allowbreak{}Skill\allowbreak{}Shell\allowbreak{}Execution} & Skill exposure and inline shell \\
\texttt{channels\allowbreak{}Enabled} · \texttt{allowed\allowbreak{}Channel\allowbreak{}Plugins} & Channels, which can relay permission decisions \\
\texttt{enable\allowbreak{}Artifact} · \texttt{CLAUDE\_\allowbreak{}CODE\_\allowbreak{}DISABLE\_\allowbreak{}ARTIFACT} & Artifact publishing (\texttt{disable\allowbreak{}Artifact} deprecated) \\
\texttt{disable\allowbreak{}Workflows} · \texttt{CLAUDE\_\allowbreak{}CODE\_\allowbreak{}DISABLE\_\allowbreak{}WORKFLOWS} · \texttt{workflow\allowbreak{}Size\allowbreak{}Guideline} & Dynamic workflows \\
\texttt{disable\allowbreak{}Remote\allowbreak{}Control} & Turns Remote Control off everywhere it can start; in managed settings, per-device enforcement \\
\texttt{force\allowbreak{}Login\allowbreak{}Method} · \texttt{force\allowbreak{}Login\allowbreak{}Org\allowbreak{}UUID} & The first selects the login method; the second, \textbf{from a managed source}, restricts claude.ai logins to the intended organisation — elsewhere it only pre-selects one. Part of ZDR routing; Console credentials are not checked \\
\texttt{available\allowbreak{}Models} · \texttt{enforce\allowbreak{}Available\allowbreak{}Models} & Model allowlist. A managed \texttt{model} is only a default \\
\texttt{cleanup\allowbreak{}Period\allowbreak{}Days} & Local transcript retention. Does \textbf{not} cover auto memory \\
\texttt{auto\allowbreak{}Memory\allowbreak{}Enabled} · \texttt{auto\allowbreak{}Memory\allowbreak{}Directory} · \texttt{CLAUDE\_\allowbreak{}CODE\_\allowbreak{}DISABLE\_\allowbreak{}AUTO\_\allowbreak{}MEMORY} & Auto memory \\
\texttt{claude\allowbreak{}Md} · \texttt{claude\allowbreak{}Md\allowbreak{}Excludes} & Managed instruction content and monorepo exclusions \\
\texttt{required\allowbreak{}Minimum\allowbreak{}Version} · \texttt{required\allowbreak{}Maximum\allowbreak{}Version} & Version floor and ceiling: outside the range Claude Code refuses to start, but an invalid value is ignored and the key is ignored outside managed settings \\
\texttt{CLAUDE\_\allowbreak{}CODE\_\allowbreak{}EXPERIMENTA\allowbreak{}L\_\allowbreak{}AGENT\_\allowbreak{}TEAMS} & Agent teams; changes ordinary delegation when on \\
\texttt{CLAUDE\_\allowbreak{}CODE\_\allowbreak{}FORK\_\allowbreak{}SUBAGENT} & Fork mode; \textbf{on by default} (interactive terminal from 2.1.232; \texttt{-\allowbreak{}p}, Agent SDK, Desktop and VS Code from 2.1.302; never under \texttt{claude mcp \allowbreak{}serve}) \\
\texttt{CLAUDE\_\allowbreak{}CODE\_\allowbreak{}MAX\_\allowbreak{}SUBAGENT\_\allowbreak{}SPAWN\_\allowbreak{}DEPTH} · \texttt{.\allowbreak{}.\allowbreak{}.\allowbreak{}\_\allowbreak{}MAX\_\allowbreak{}CONCURRENT\_\allowbreak{}SUBAGENTS} & Subagent nesting and concurrency \\
\texttt{CLAUDE\_\allowbreak{}CODE\_\allowbreak{}DISABLE\_\allowbreak{}CRON} & Disables the scheduler and \texttt{/\allowbreak{}loop} \\
\texttt{CLAUDE\_\allowbreak{}CODE\_\allowbreak{}SKIP\_\allowbreak{}PROMPT\_\allowbreak{}HISTORY} · \texttt{CLAUDE\_\allowbreak{}CONFIG\_\allowbreak{}DIR} · \texttt{CLAUDE\_\allowbreak{}CODE\_\allowbreak{}PROJECT\_\allowbreak{}DIR\_\allowbreak{}NAME} & Transcript writing and location \\
\texttt{DISABLE\_\allowbreak{}TELEMETRY} · \texttt{DISABLE\_\allowbreak{}ERROR\_\allowbreak{}REPORTING} · \texttt{DISABLE\_\allowbreak{}FEEDBACK\_\allowbreak{}COMMAND} · \texttt{CLAUDE\_\allowbreak{}CODE\_\allowbreak{}DISABLE\_\allowbreak{}FEEDBACK\_\allowbreak{}SURVEY} & Individual egress flows \\
\texttt{CLAUDE\_\allowbreak{}CODE\_\allowbreak{}DISABLE\_\allowbreak{}NONESSENTIA\allowbreak{}L\_\allowbreak{}TRAFFIC} & All of the above at once; also disables Remote Control feature flags and \texttt{/\allowbreak{}schedule} \\
\texttt{skip\allowbreak{}Web\allowbreak{}Fetch\allowbreak{}Preflight} & Skips the WebFetch hostname safety check, so the blocklist is not consulted; pair it with \texttt{Web\allowbreak{}Fetch} rules \\
\texttt{OTEL\_\allowbreak{}LOG\_\allowbreak{}USER\_\allowbreak{}PROMPTS} · \texttt{OTEL\_\allowbreak{}LOG\_\allowbreak{}ASSISTANT\_\allowbreak{}RESPONSES} · \texttt{OTEL\_\allowbreak{}LOG\_\allowbreak{}TOOL\_\allowbreak{}CONTENT} · \texttt{OTEL\_\allowbreak{}LOG\_\allowbreak{}RAW\_\allowbreak{}API\_\allowbreak{}BODIES} & Content logging into telemetry. Off by default; leave them off unless you have decided otherwise with your privacy function \\
}%
\def\tblbody{\begin{minipage}{\textwidth}
{\footnotesize\begin{tabular}{H{78.7mm}L{64.3mm}}
\toprule
\textbf{Key or variable} & \textbf{Effect} \\
\midrule
\tblrows
\bottomrule\end{tabular}}\end{minipage}}%
\begingroup
\def\sloppy{\tolerance 9999\emergencystretch 3em\hfuzz 200pt\vfuzz 200pt}%
\hbadness=10000\vbadness=10000\hfuzz=200pt\vfuzz=200pt
\global\setbox\tblbox=\hbox{\tblbody}%
\endgroup
\par\addvspace{2.6mm}
\ifdim\dimexpr\ht\tblbox+\dp\tblbox\relax>0.55\textheight
  \typeout{HANDBOOK-TABLE broken \the\dimexpr\ht\tblbox+\dp\tblbox\relax}%
  
  {\footnotesize\begin{longtable}{H{78.7mm}L{64.3mm}}
  \toprule
\textbf{Key or variable} & \textbf{Effect} \\
\midrule\endfirsthead
  \multicolumn{2}{@{}l@{}}{%
  \sffamily\footnotesize\itshape\color{inkgrey}Continued}\\[1.2mm]
  \toprule
\textbf{Key or variable} & \textbf{Effect} \\
\midrule\endhead
  \bottomrule\endfoot
  \bottomrule\endlastfoot
  \tblrows
  \end{longtable}}%
\else
  \typeout{HANDBOOK-TABLE atomic \the\dimexpr\ht\tblbox+\dp\tblbox\relax}%
  \noindent\tblbody
\fi
\par\addvspace{2.6mm}
\endgroup

\FloatBarrier
\setcounter{chapter}{4}
\setcounter{section}{0}
\appendixchapter{E}{Security release checklist}
\FloatBarrier
\section*{Before the task}
\markboth{Before the task}{Before the task}

\begin{codeblock}{7.0}{8.3}
\cl{[~]~Correct~machine,~account,~organisation,~project,~repository~and~branch~confirmed}
\cl{[~]~Working~directory~is~narrow;~not~a~home~directory}
\cl{[~]~Clean~baseline~or~recoverable~backup~exists}
\cl{[~]~No~secret~in~prompts,~in~files~in~scope,~or~in~version~control}
\cl{[~]~Permission~rules~follow~least~privilege,~and~you~have~tested~that~a~deny~actually~blocks}
\cl{[~]~Sandboxing~and~credential~protection~enabled~where~supported}
\cl{[~]~External~services~use~test~or~least-privilege~accounts}
\cl{[~]~Third-party~skills,~plugins,~marketplaces,~MCP~servers,~channels,~hooks~and~scripts~reviewed}
\cl{[~]~Human~approval~gates~written~into~the~plan}
\cl{[~]~Token,~time,~turn~and~agent~bounds~explicit}
\cl{[~]~Fork~mode~and~permission-mode~inheritance~understood~for~any~delegated~reviewer}
\end{codeblock}

\FloatBarrier
\section*{During the task}
\markboth{During the task}{During the task}

\begin{codeblock}{9.0}{10.6}
\cl{[~]~Initial~tool~calls~match~the~intended~direction}
\cl{[~]~Every~permission~request~understood~before~approval}
\cl{[~]~Unexpected~network,~credential,~package~or~account~access~stops~the~task}
\cl{[~]~Browser~actions~stay~within~the~approved~profile~and~domain~set}
\cl{[~]~Parallel~sessions~have~isolated~ownership~or~worktrees}
\cl{[~]~Material~decisions~and~findings~are~being~written~to~files}
\cl{[~]~No~stale~goal~or~loop~is~competing~with~the~current~task}
\cl{[~]~Inter-agent~messages~are~treated~as~coordination,~never~as~approval}
\end{codeblock}

\FloatBarrier
\section*{Before release}
\markboth{Before release}{Before release}

\begin{codeblock}{9.0}{10.6}
\cl{[~]~Diff~reviewed~file~by~file}
\cl{[~]~Automated~checks~actually~ran,~and~their~output~was~inspected}
\cl{[~]~Behaviour~or~rendering~observed~directly}
\cl{[~]~Independent~review~completed,~and~confirmed~to~be~independent}
\cl{[~]~External~side~effects~enumerated,~including~anything~published~or~messaged}
\cl{[~]~Links~and~citations~checked}
\cl{[~]~No~secret~in~output,~logs,~artifacts,~or~transcript~excerpts~quoted~into~files}
\cl{[~]~Publication,~merge,~push,~payment~or~account~change~separately~approved}
\cl{[~]~Rollback~route~and~monitoring~owner~known}
\end{codeblock}

\FloatBarrier
\setcounter{chapter}{5}
\setcounter{section}{0}
\appendixchapter{F}{Threat model and control map}
\begingroup
\def\tblrows{%
Indirect prompt injection & Web page, tool output, PR comment, fire payload, channel message & A separate model call for most web fetches; untrusted-input wrappers; auto mode's classifier review of inter-agent messages & Explicit objective and domain allowlist; human gate on external writes & Chapter 20, Chapter 13, Chapter 24 \\
Credential exposure & \texttt{.\allowbreak{}env} read, command output, transcript, artifact & Deny rules; sandbox credential deny/mask & Least-privilege keys; rotation; output discipline & Chapter 14 \\
Malicious or careless third-party code & Skill, plugin, marketplace, MCP server, channel plugin & Review before install; pinned versions; \texttt{strict\allowbreak{}Known\allowbreak{}Marketplace\allowbreak{}s} & Isolated test profile; \texttt{disable\allowbreak{}Sideload\allowbreak{}Flags} & Chapter 15, Chapter 18, Chapter 13 \\
Unreviewed autonomous action & Routine, \texttt{/\allowbreak{}loop}, goal, auto-fix, agent team & Human gates in the prompt; least-privilege connectors; bounded turns & Notifications; inspection of the first several runs & Chapter 23, Chapter 24, Chapter 21 \\
Loss of oversight through delegation & Fork inheriting authoring context; teammate inheriting permissive parent & Explicit agent definitions; verify independence & \texttt{permission\allowbreak{}Mode}; parent mode discipline & Chapter 16, Chapter 30 \\
Data leaving the boundary & Telemetry, feedback, artifacts, cloud sessions, Remote Control transcript & Egress opt-outs; ZDR where eligible & \texttt{force\allowbreak{}Login\allowbreak{}Org\allowbreak{}UUID}; artifact and sharing controls & Chapter 32 \\
Irreversible external action & Push, publish, payment, deploy, account change & Human gate; branch protection; \texttt{ask} rules & Separate approval step; scoped credentials & Chapter 8, Appendix E \\
Concurrent-edit corruption & Two sessions, one checkout & Worktrees or file ownership & Sequential integration & Chapter 29 \\
Configuration drift and shadow policy & User settings, plugin settings, unselected managed source & \texttt{allow\allowbreak{}Managed*Onl\allowbreak{}y} locks; \texttt{/\allowbreak{}status} verification & \texttt{Config\allowbreak{}Change} hooks; \texttt{claude doct\allowbreak{}or} & Chapter 31 \\
Over-trust in agent self-report & "Done" without evidence & Evidence-based completion conditions & Independent review; release checklist & Chapter 21, Chapter 34 \\
}%
\def\tblbody{\begin{minipage}{\textwidth}
{\scriptsize\begin{tabular}{L{25.3mm}L{21.8mm}L{38.7mm}L{33.7mm}L{13.9mm}}
\toprule
\textbf{Threat} & \textbf{Vector} & \textbf{Primary control} & \textbf{Secondary control} & \textbf{Chapter} \\
\midrule
\tblrows
\bottomrule\end{tabular}}\end{minipage}}%
\begingroup
\def\sloppy{\tolerance 9999\emergencystretch 3em\hfuzz 200pt\vfuzz 200pt}%
\hbadness=10000\vbadness=10000\hfuzz=200pt\vfuzz=200pt
\global\setbox\tblbox=\hbox{\tblbody}%
\endgroup
\par\addvspace{2.6mm}
\ifdim\dimexpr\ht\tblbox+\dp\tblbox\relax>0.55\textheight
  \typeout{HANDBOOK-TABLE broken \the\dimexpr\ht\tblbox+\dp\tblbox\relax}%
  
  {\scriptsize\begin{longtable}{L{25.3mm}L{21.8mm}L{38.7mm}L{33.7mm}L{13.9mm}}
  \toprule
\textbf{Threat} & \textbf{Vector} & \textbf{Primary control} & \textbf{Secondary control} & \textbf{Chapter} \\
\midrule\endfirsthead
  \multicolumn{5}{@{}l@{}}{%
  \sffamily\footnotesize\itshape\color{inkgrey}Continued}\\[1.2mm]
  \toprule
\textbf{Threat} & \textbf{Vector} & \textbf{Primary control} & \textbf{Secondary control} & \textbf{Chapter} \\
\midrule\endhead
  \bottomrule\endfoot
  \bottomrule\endlastfoot
  \tblrows
  \end{longtable}}%
\else
  \typeout{HANDBOOK-TABLE atomic \the\dimexpr\ht\tblbox+\dp\tblbox\relax}%
  \noindent\tblbody
\fi
\par\addvspace{2.6mm}
\endgroup

\FloatBarrier
\setcounter{chapter}{6}
\setcounter{section}{0}
\appendixchapter{G}{Troubleshooting}
\begingroup
\def\tblrows{%
\texttt{claude:\allowbreak{} command no\allowbreak{}t found} & Installer path not on \texttt{PATH} & Restart the terminal; run \texttt{claude doct\allowbreak{}or}; see [115] \\
Login repeats, or the wrong account appears & Stored credentials, or an API key taking precedence over a claude.ai login & \texttt{/\allowbreak{}status}, then \texttt{/\allowbreak{}login}. Check \texttt{ANTHROPIC\_\allowbreak{}API\_\allowbreak{}KEY}, \texttt{ANTHROPIC\_\allowbreak{}AUTH\_\allowbreak{}TOKEN}, \texttt{api\allowbreak{}Key\allowbreak{}Helper} \\
A documented command is missing & Authentication path, provider, plan, or an organisation policy & \texttt{/\allowbreak{}status}; check whether the feature needs a claude.ai login \\
Claude cannot find files & Wrong project folder or working directory & \texttt{pwd}; use \texttt{/\allowbreak{}add-\allowbreak{}dir} only after confirming the target \\
Repeated permission prompts & Rule too narrow, not persistent, or saved in another scope & \texttt{/\allowbreak{}permissions}; inspect the source settings file \\
A \texttt{CLAUDE.\allowbreak{}md} edit had no effect & Instructions are read at session start and cached & Start a new session; see [68] \\
An instruction disappeared after \texttt{/\allowbreak{}compact} & It was given only in conversation, or sits in a nested file or path-scoped rule that has not reloaded & Add it to \texttt{CLAUDE.\allowbreak{}md}; project-root files survive compaction, the others reload when Claude works with a matching file \\
A rule is not loading & Path-scoped rule has not matched a file yet & \texttt{/\allowbreak{}context}; add an \texttt{Instruction\allowbreak{}s\allowbreak{}Loaded} hook \\
Managed policy is not applying & A higher-priority managed source was selected, or the file is invalid & \texttt{/\allowbreak{}status} \texttt{Setting sou\allowbreak{}rces} line; \texttt{claude doct\allowbreak{}or} \\
An MCP server is disconnected & Session did not reload, OAuth expired, or transport changed & New session; \texttt{/\allowbreak{}mcp}; \texttt{claude mcp \allowbreak{}get <name>} \\
A project MCP server never connects & Awaiting approval, or the workspace is untrusted & \texttt{claude mcp \allowbreak{}list} for \texttt{⏸ Pending a\allowbreak{}pproval}; run \texttt{claude} interactively \\
An MCP tool is absent & Tool Search deferred it, or the scope differs & Ask for it explicitly; inspect \texttt{/\allowbreak{}mcp} \\
\texttt{/\allowbreak{}agents} does not open an editor & Behaviour changed in 2.1.198 & Edit \texttt{.\allowbreak{}claude/\allowbreak{}agents/\allowbreak{}} directly \\
A "reviewer" already knows the argument & It was a fork & Check \texttt{CLAUDE\_\allowbreak{}CODE\_\allowbreak{}FORK\_\allowbreak{}SUBAGENT}; use an explicit agent definition \\
A subagent ignores its \texttt{permission\allowbreak{}Mode} & Parent is in \texttt{accept\allowbreak{}Edits}, \texttt{bypass\allowbreak{}Permissions} or \texttt{auto} & Change the parent's mode \\
Claude spawned teammates unasked & Agent teams enabled; a named subagent launches as a teammate & Set \texttt{CLAUDE\_\allowbreak{}CODE\_\allowbreak{}EXPERIMENTA\allowbreak{}L\_\allowbreak{}AGENT\_\allowbreak{}TEAMS=\allowbreak{}0}; check settings precedence \\
A workflow looks expensive & Too many lanes, or duplicated work & \texttt{/\allowbreak{}workflows}, \texttt{/\allowbreak{}usage}; pause; set \texttt{workflow\allowbreak{}Size\allowbreak{}Guideline} \\
A loop did not run on time & Session busy or closed, jitter, expiry, or a missed interval & A recurring task runs late by design; only a one-shot is helped by avoiding \texttt{:\allowbreak{}00} and \texttt{:\allowbreak{}30}. Use a durable scheduler if timing matters \\
A loop is running something you did not ask for & A bare \texttt{/\allowbreak{}loop} started the maintenance prompt & \texttt{Esc} while waiting; check for \texttt{loop.\allowbreak{}md} \\
A session vanished from the picker & Its first prompt was \texttt{/\allowbreak{}loop} & \texttt{claude -\allowbreak{}-\allowbreak{}resume <id>}; widening the picker with \texttt{Ctrl+A} does not show it \\
A routine "succeeded" but did nothing & \textbf{Succeeded} means no infrastructure error & Open the run and read the transcript \\
A routine run is stuck and nobody was told & An ask rule, or auto mode's repeated-block threshold, left a prompt waiting & Open the run from the routine's detail page and answer it \\
A routine cannot reach local software & It runs on Anthropic-managed cloud infrastructure & Run it in a self-hosted environment that can reach the software, or use a Desktop scheduled task or a secured local scheduler \\
Chrome will not connect & Extension, native messaging host, or an API-key or \texttt{setup-\allowbreak{}token} session & \texttt{/\allowbreak{}chrome} → Reconnect; restart Chrome; confirm \texttt{/\allowbreak{}login} \\
An artifact created a second page & The session had no URL for the original & Attach with \texttt{/\allowbreak{}artifacts}, or pass the URL \\
A secret was printed & A tool read it, or a command echoed it & Stop, rotate, preserve the evidence, investigate, clean up, then fix the control (Section 14.3) \\
Two sessions changed the same file & Shared checkout collision & Stop writers; inspect git; move to worktrees \\
}%
\def\tblbody{\begin{minipage}{\textwidth}
{\scriptsize\begin{tabular}{L{29.7mm}L{35.7mm}L{74.4mm}}
\toprule
\textbf{Symptom} & \textbf{Likely cause} & \textbf{Safe next check} \\
\midrule
\tblrows
\bottomrule\end{tabular}}\end{minipage}}%
\begingroup
\def\sloppy{\tolerance 9999\emergencystretch 3em\hfuzz 200pt\vfuzz 200pt}%
\hbadness=10000\vbadness=10000\hfuzz=200pt\vfuzz=200pt
\global\setbox\tblbox=\hbox{\tblbody}%
\endgroup
\par\addvspace{2.6mm}
\ifdim\dimexpr\ht\tblbox+\dp\tblbox\relax>0.55\textheight
  \typeout{HANDBOOK-TABLE broken \the\dimexpr\ht\tblbox+\dp\tblbox\relax}%
  
  {\scriptsize\begin{longtable}{L{29.7mm}L{35.7mm}L{74.4mm}}
  \toprule
\textbf{Symptom} & \textbf{Likely cause} & \textbf{Safe next check} \\
\midrule\endfirsthead
  \multicolumn{3}{@{}l@{}}{%
  \sffamily\footnotesize\itshape\color{inkgrey}Continued}\\[1.2mm]
  \toprule
\textbf{Symptom} & \textbf{Likely cause} & \textbf{Safe next check} \\
\midrule\endhead
  \bottomrule\endfoot
  \bottomrule\endlastfoot
  \tblrows
  \end{longtable}}%
\else
  \typeout{HANDBOOK-TABLE atomic \the\dimexpr\ht\tblbox+\dp\tblbox\relax}%
  \noindent\tblbody
\fi
\par\addvspace{2.6mm}
\endgroup

Use \texttt{claude doct\allowbreak{}or}, \texttt{/\allowbreak{}doctor}, \texttt{/\allowbreak{}hooks}, \texttt{/\allowbreak{}context} and [56] before applying a fix found in a forum, and [60] to look up an exact error string.

\FloatBarrier
\setcounter{chapter}{7}
\setcounter{section}{0}
\appendixchapter{H}{Standards and regulatory crosswalk}
This crosswalk maps the controls described in this book to seventeen external frameworks. It is offered as a starting point for an assurance conversation, not as a certification claim: a control being present is not evidence that it is effective, and mapping is a matter of interpretation. Use it to answer the question an auditor, a regulator or a customer security questionnaire will actually ask — \emph{which of your controls addresses this requirement, and what evidence do you hold?}

\begin{calloutbox}{palegrey}{quoterule}{1.2mm}
\textbf{UNVERIFIED — ISO/IEC 27001:2022 Annex A identifiers.} The NIST, IEEE, SSDF and ISO/IEC 42001 clause identifiers in this crosswalk were confirmed against primary or publisher sources. The Annex A control numbers were not: ISO standards are paywalled, and only secondary summaries were available at the verification date. Confirm them against your own copy of the standard before relying on this mapping in an audit response. Note also that ISO/IEC 27001:2022 carries Amendment 1:2024 [12].
\end{calloutbox}

\FloatBarrier
\section*{Control-to-framework map}
\markboth{Control-to-framework map}{Control-to-framework map}

\begingroup
\def\tblrows{%
Permission rules and modes & Chapter 7 & GOVERN 1.2, MANAGE 2.2 & 8.1 operational planning and control & A.8.2 privileged access; A.8.3 information access restriction & NIST SP 800-207 §3 policy enforcement point [8] \\
Sandboxing and isolation & Chapter 7 & MANAGE 2.3 & 8.1 & A.8.22 segregation; A.8.31 separation of environments & Complements, does not implement, ASD Essential Eight application control [18] \\
Managed policy tier & Chapter 31 & GOVERN 1.1, 2.1 & 5.2 policy; 9.2 internal audit & A.5.1 policies; A.8.9 configuration management & NIST CSF 2.0 GV.PO, PR.PS [10] \\
Credential handling and rotation & Chapter 14 & MANAGE 2.2 & 8.1 & A.5.17 authentication information; A.8.24 cryptography & Essential Eight: restrict admin privileges [18] \\
Supply-chain review of skills, plugins, MCP, channels & Chapter 15;Chapter 18;Chapter 13;Chapter 27 & MAP 4.1, MANAGE 3.1, 3.2 & 8.3 AI risk treatment & A.5.19–A.5.23 supplier and cloud services; A.8.30 outsourced development & NIST SP 800-218 PO.3, PS.3, PW.4 [9]; SLSA provenance [116] \\
Untrusted-input handling and injection defence & Chapter 20;Chapter 13;Chapter 24 & MAP 5.1, MEASURE 2.7, MANAGE 4.1 & 6.1.2 risk assessment & A.8.26 application security requirements & OWASP LLM01 [4]; MITRE ATLAS AML.T0051 [5]; [22] \\
Human approval gates & Chapter 34 & GOVERN 3.2, MANAGE 4.1 & 8.4 AI system impact assessment & A.5.4 management responsibilities & EU AI Act Art. 14 human oversight [17]; IEEE 7000-2021 [16] \\
Evidence-Gated Delivery and verification & Chapter 34;“Scope and method” & MEASURE 2.5, 2.9, 3.3 & 8.1; 9.1 monitoring & A.8.29 security testing; A.8.32 change management & IEEE Std 1012-2024 V\&V [15]; ISO/IEC 25010 [14] \\
Version control and rollback & Chapter 8 & MANAGE 4.3 & 8.1 & A.8.32 change management; A.8.13 backup & NIST SP 800-218 PS.1, PS.2 [9] \\
Telemetry, audit and traceability & Chapter 33 & MEASURE 1.3, 2.8, MANAGE 4.1 & 9.1; 10.2 nonconformity & A.8.15 logging; A.8.16 monitoring & NIST CSF 2.0 DE.CM, DE.AE [10]; EU AI Act Art. 12 record-keeping [17] \\
Data retention, residency and ZDR & Chapter 32 & GOVERN 6.1, MAP 4.2 & 7.5 documented information & A.5.33 protection of records; A.5.34 privacy and PII & GDPR Arts. 5, 25, 32 [107]; Australian Privacy Principles 8, 11 [106] \\
Threat modelling of the agentic environment & Appendix F & MAP 1.1, 2.3, 5.1 & 6.1.2 & A.8.25 secure development lifecycle & ISO/IEC 23894 §6 [13]; MITRE ATLAS [5] \\
Accessibility of the toolchain & Chapter 33 & GOVERN 5.1, MAP 1.6 & 4.2 interested parties & — & IEEE 7000-2021 value elicitation [16] \\
Re-verification against a moving baseline & Appendix K & MEASURE 2.4, MANAGE 4.2 & 9.3 management review; 10.1 improvement & A.8.8 technical vulnerability management & Secure AI system development, "secure operation" [117] \\
}%
\def\tblbody{\begin{minipage}{\textwidth}\boxcaption{Table 30 — Book controls mapped to external frameworks}
{\scriptsize\begin{tabular}{L{25.7mm}L{17.1mm}L{12.1mm}L{22.3mm}L{27.4mm}L{25.6mm}}
\toprule
\textbf{Control in this book} & \textbf{Chapter} & \textbf{NIST AI RMF 1.0 [6]} & \textbf{ISO/IEC 42001 [11]} & \textbf{ISO/IEC 27001:2022 Annex A [12]} & \textbf{Other} \\
\midrule
\tblrows
\bottomrule\end{tabular}}\end{minipage}}%
\begingroup
\def\sloppy{\tolerance 9999\emergencystretch 3em\hfuzz 200pt\vfuzz 200pt}%
\hbadness=10000\vbadness=10000\hfuzz=200pt\vfuzz=200pt
\global\setbox\tblbox=\hbox{\tblbody}%
\endgroup
\par\addvspace{2.6mm}
\ifdim\dimexpr\ht\tblbox+\dp\tblbox\relax>0.55\textheight
  \typeout{HANDBOOK-TABLE broken \the\dimexpr\ht\tblbox+\dp\tblbox\relax}%
  \tabcaption{Table 30 — Book controls mapped to external frameworks}
  {\scriptsize\begin{longtable}{L{25.7mm}L{17.1mm}L{12.1mm}L{22.3mm}L{27.4mm}L{25.6mm}}
  \toprule
\textbf{Control in this book} & \textbf{Chapter} & \textbf{NIST AI RMF 1.0 [6]} & \textbf{ISO/IEC 42001 [11]} & \textbf{ISO/IEC 27001:2022 Annex A [12]} & \textbf{Other} \\
\midrule\endfirsthead
  \multicolumn{6}{@{}l@{}}{%
  \sffamily\footnotesize\itshape\color{inkgrey}Table 30 — Book controls mapped to external frameworks \textemdash\ continued}\\[1.2mm]
  \toprule
\textbf{Control in this book} & \textbf{Chapter} & \textbf{NIST AI RMF 1.0 [6]} & \textbf{ISO/IEC 42001 [11]} & \textbf{ISO/IEC 27001:2022 Annex A [12]} & \textbf{Other} \\
\midrule\endhead
  \bottomrule\endfoot
  \bottomrule\endlastfoot
  \tblrows
  \end{longtable}}%
\else
  \typeout{HANDBOOK-TABLE atomic \the\dimexpr\ht\tblbox+\dp\tblbox\relax}%
  \noindent\tblbody
\fi
\par\addvspace{2.6mm}
\endgroup

\FloatBarrier
\section*{Reading the map for three common obligations}
\markboth{Reading the map for three common obligations}{Reading the map for three common obligations}

\textbf{EU AI Act human oversight (Art. 14).} Article 14(4) enumerates five capabilities the natural person must be enabled to have: to understand the system's capacities and limitations; to remain aware of automation bias — "the possible tendency of automatically relying or over-relying on the output"; to correctly interpret the output; \textbf{to decide, in any particular situation, not to use the system or to otherwise disregard, override or reverse its output}; and to intervene or interrupt the system through a stop control or equivalent [17]. In the terms of this book: Part 0 addresses capacities and limitations; the evidence-label convention and the UNVERIFIED class address automation bias directly; the completion-condition discipline of Chapter 21 addresses correct interpretation of output; the human release gate of Chapter 34 is precisely the fourth capability, the decision to disregard or override an agent's output; and \texttt{Esc}, \texttt{/\allowbreak{}goal clear}, \texttt{/\allowbreak{}rewind} and the managed \texttt{disable\allowbreak{}Bypass\allowbreak{}Permissions\allowbreak{}Mode} key address intervention and stop. The residual gap is that an agent's external actions — a push, a published artifact, a sent message — are not reversible by any local control, which is why they are gated rather than monitored.

\textbf{ISO/IEC 42001 clause 8.1, operational planning and control.} Clause 8 is \emph{Operation}, and its sub-clauses are 8.1 operational planning and control, 8.2 AI risk assessment, 8.3 AI risk treatment and 8.4 AI system impact assessment [11]. The obligation under 8.1 is to plan, implement and control the processes needed to meet the management system's requirements. The managed-policy tier of Chapter 31 is the only mechanism in the product that survives a user's contrary intent, and only on a device the user does not administer: it is a client-side control, not a security boundary (Chapter 31). So any 42001 control that must hold organisation-wide starts there, with device, identity and provider-side controls beside it, or is not implemented. Appendix L sequences this into maturity levels.

\textbf{Secure development (NIST SP 800-218).} SSDF practice groups map cleanly: PO.3 supporting toolchains covers the plugin and MCP review of Chapter 18 and Chapter 13; PS.1–PS.3 protecting software covers version control and provenance in Chapter 8; PW.4 reusing existing software covers the supply-chain gate on third-party skills; and RV.1 identifying vulnerabilities covers \texttt{/\allowbreak{}security-\allowbreak{}review}, the security-guidance plugin and the re-verification procedure [9], [79].

\FloatBarrier
\setcounter{chapter}{8}
\setcounter{section}{0}
\appendixchapter{I}{Glossary}
\begin{deflist}
\item \textbf{Agent} \textemdash\ a model operating in a loop with tools, state and a task.  
\item \textbf{Agent team} \textemdash\ a lead session coordinating peer sessions with a shared task list and mailboxes.  
\item \textbf{Allow / ask / deny} \textemdash\ permission-rule outcomes, evaluated in that reverse order: deny first.  
\item \textbf{Artifact} \textemdash\ a live page published from a session to a URL on claude.ai.  
\item \textbf{Auto memory} \textemdash\ repository-scoped notes Claude maintains for future sessions.  
\item \textbf{Auto mode} \textemdash\ permission mode in which a classifier reviews actions instead of you.  
\item \textbf{Channel} \textemdash\ an MCP server that pushes events into a running session.  
\item \textbf{Checkpoint} \textemdash\ saved conversation and file state used by rewind.  
\item \textbf{Connector} \textemdash\ a claude.ai account integration, distinct from a locally configured MCP server.  
\item \textbf{Context engineering} \textemdash\ deliberate selection, organisation, persistence and removal of context.  
\item \textbf{Effort} \textemdash\ configured reasoning allocation for a model.  
\item \textbf{Fork} \textemdash\ a subagent that inherits the parent conversation. Fork mode, which lets Claude start one on its own, is on by default.  
\item \textbf{Goal} \textemdash\ a persistent completion condition evaluated after each turn by a separate model.  
\item \textbf{Handoff} \textemdash\ compact state prepared for another session or agent.  
\item \textbf{Headless mode} \textemdash\ non-interactive execution with \texttt{claude -\allowbreak{}p}.  
\item \textbf{Hook} \textemdash\ an executable action bound to a lifecycle event.  
\item \textbf{Human gate} \textemdash\ a point that cannot proceed without explicit human confirmation.  
\item \textbf{Managed settings} \textemdash\ organisation policy applied above every other settings level.  
\item \textbf{MCP} \textemdash\ Model Context Protocol, an open standard for exposing tools and resources to AI clients.  
\item \textbf{Output style} \textemdash\ instructions that set role, tone and response format for a session; prompt composition, not a control.  
\item \textbf{Prompt injection} \textemdash\ untrusted content attempting to redirect a model or its tool use.  
\item \textbf{Routine} \textemdash\ a saved cloud configuration fired by schedule, API call or GitHub event; now called a cloud scheduled task.  
\item \textbf{Sandbox} \textemdash\ operating-system-enforced filesystem and network constraints on tool execution.  
\item \textbf{Subagent} \textemdash\ a specialised instance delegated a bounded task in its own context window.  
\item \textbf{Surface} \textemdash\ an interface through which Claude Code runs: terminal, IDE, Desktop, web, mobile.  
\item \textbf{Teleport} \textemdash\ pulling a cloud session and its branch into the local terminal.  
\item \textbf{Tool Search} \textemdash\ deferred loading of tool definitions until they are needed.  
\item \textbf{Transcript} \textemdash\ locally stored conversation and tool activity, plaintext JSONL.  
\item \textbf{Ultracode} \textemdash\ a setting that turns on automatic dynamic-workflow orchestration at the session's effort level; the launch flag also sets \texttt{xhigh}.  
\item \textbf{Workflow} \textemdash\ a script that orchestrates subagents, executed by a runtime outside the conversation.  
\item \textbf{Worktree} \textemdash\ a separate git working tree and branch isolating concurrent change.  
\item \textbf{Zero Data Retention} \textemdash\ an enterprise configuration under which inference data is not stored.  
\end{deflist}

\FloatBarrier
\setcounter{chapter}{9}
\setcounter{section}{0}
\appendixchapter{J}{Verification and evidence ledger}
\FloatBarrier
\section*{Verification baseline}
\markboth{Verification baseline}{Verification baseline}

\begingroup
\def\tblrows{%
Installed CLI & \texttt{2.\allowbreak{}1.\allowbreak{}322 (Claude\allowbreak{} Code)}, 12 Nov. 2026. Version 1.0: \texttt{2.\allowbreak{}1.\allowbreak{}241}, 23 Aug. 2026, re-verified against \texttt{2.\allowbreak{}1.\allowbreak{}246}, 26 Aug. 2026 & High \\
Official documentation & Pages under \texttt{code.\allowbreak{}claude.\allowbreak{}com/\allowbreak{}docs/\allowbreak{}en/\allowbreak{}}, enumerated from the published index [1] & High for behaviour as of the date \\
Traceable-origin material & Introductory video course, publication date and duration confirmed; used only to attribute the origin of superseded claims [44] & High \\
Practitioner technique & Labelled PRACTITIONER NOTE throughout; judgement, not product behaviour & Medium unless independently verified \\
Third-party components & None installed or endorsed; risk depends on implementation and version & Insufficient until individually reviewed \\
}%
\def\tblbody{\begin{minipage}{\textwidth}
{\footnotesize\begin{tabular}{L{32.8mm}L{80.4mm}L{26.6mm}}
\toprule
\textbf{Item} & \textbf{Observed result} & \textbf{Confidence} \\
\midrule
\tblrows
\bottomrule\end{tabular}}\end{minipage}}%
\begingroup
\def\sloppy{\tolerance 9999\emergencystretch 3em\hfuzz 200pt\vfuzz 200pt}%
\hbadness=10000\vbadness=10000\hfuzz=200pt\vfuzz=200pt
\global\setbox\tblbox=\hbox{\tblbody}%
\endgroup
\par\addvspace{2.6mm}
\ifdim\dimexpr\ht\tblbox+\dp\tblbox\relax>0.55\textheight
  \typeout{HANDBOOK-TABLE broken \the\dimexpr\ht\tblbox+\dp\tblbox\relax}%
  
  {\footnotesize\begin{longtable}{L{32.8mm}L{80.4mm}L{26.6mm}}
  \toprule
\textbf{Item} & \textbf{Observed result} & \textbf{Confidence} \\
\midrule\endfirsthead
  \multicolumn{3}{@{}l@{}}{%
  \sffamily\footnotesize\itshape\color{inkgrey}Continued}\\[1.2mm]
  \toprule
\textbf{Item} & \textbf{Observed result} & \textbf{Confidence} \\
\midrule\endhead
  \bottomrule\endfoot
  \bottomrule\endlastfoot
  \tblrows
  \end{longtable}}%
\else
  \typeout{HANDBOOK-TABLE atomic \the\dimexpr\ht\tblbox+\dp\tblbox\relax}%
  \noindent\tblbody
\fi
\par\addvspace{2.6mm}
\endgroup

\FloatBarrier
\section*{Method}
\markboth{Method}{Method}

Each atomic product claim was adjudicated against the raw text of the primary documentation page that governs it, not against a summary of that page. Where an automated summarisation pass and the raw text disagreed — which happened, notably over the existence of \texttt{/\allowbreak{}sandbox}, \texttt{/\allowbreak{}workflows} and \texttt{-\allowbreak{}-\allowbreak{}worktree} — the raw text was taken as authoritative and the summary discarded. Claims that could be settled only against a running CLI are not asserted \textbf{from documentation}. Where one was later settled by exercising the product, it is recorded below as configuration-specific; where it was not, it appears as UNVERIFIED in the open list at the end of this appendix, with the local check that closes it.

\FloatBarrier
\section*{Misconception ledger}
\markboth{Misconception ledger}{Misconception ledger}

Each row corresponds to an entry in Part 0. The verdict records \emph{how} the claim failed, which is the point of the table: a claim that is simply false and a claim that is true but imprecise call for different responses from a reader who holds it.

\begingroup
\def\tblrows{%
\textbf{REJECTED} & The claim is false. The mechanism does something else entirely, and acting on the claim produces an outcome the reader did not intend. \\
\textbf{CORRECTED} & The claim names a real mechanism but states it wrongly — a detail, a quantifier or a release. The correction is a matter of precision. \\
\textbf{QUALIFIED} & The claim is true only under a condition the reader is unlikely to have met, and false as a general statement. \\
\textbf{REFRAMED} & The claim is factually accurate but its framing understates a risk that governs how the feature should be used. \\
}%
\def\tblbody{\begin{minipage}{\textwidth}
{\footnotesize\begin{tabular}{L{18.4mm}L{124.5mm}}
\toprule
\textbf{Verdict} & \textbf{What it means} \\
\midrule
\tblrows
\bottomrule\end{tabular}}\end{minipage}}%
\begingroup
\def\sloppy{\tolerance 9999\emergencystretch 3em\hfuzz 200pt\vfuzz 200pt}%
\hbadness=10000\vbadness=10000\hfuzz=200pt\vfuzz=200pt
\global\setbox\tblbox=\hbox{\tblbody}%
\endgroup
\par\addvspace{2.6mm}
\ifdim\dimexpr\ht\tblbox+\dp\tblbox\relax>0.55\textheight
  \typeout{HANDBOOK-TABLE broken \the\dimexpr\ht\tblbox+\dp\tblbox\relax}%
  
  {\footnotesize\begin{longtable}{L{18.4mm}L{124.5mm}}
  \toprule
\textbf{Verdict} & \textbf{What it means} \\
\midrule\endfirsthead
  \multicolumn{2}{@{}l@{}}{%
  \sffamily\footnotesize\itshape\color{inkgrey}Continued}\\[1.2mm]
  \toprule
\textbf{Verdict} & \textbf{What it means} \\
\midrule\endhead
  \bottomrule\endfoot
  \bottomrule\endlastfoot
  \tblrows
  \end{longtable}}%
\else
  \typeout{HANDBOOK-TABLE atomic \the\dimexpr\ht\tblbox+\dp\tblbox\relax}%
  \noindent\tblbody
\fi
\par\addvspace{2.6mm}
\endgroup

\begingroup
\def\tblrows{%
M01 & Run \texttt{/\allowbreak{}loop} with no arguments to see your scheduled tasks & REJECTED & [26] & Chapter 23 \\
M02 & Closing the session stops a loop & QUALIFIED & [26] & Chapter 23 \\
M03 & A subagent gives you an independent second opinion & QUALIFIED & [27] & Chapter 16 \\
M04 & \texttt{.\allowbreak{}env} keeps credentials out of Claude's reach & REJECTED & [25], [28], [29] & Chapter 14 \\
M05 & Scheduling jitter is unpredictable & CORRECTED & [26] & Chapter 23 \\
M06 & \texttt{CLAUDE\_\allowbreak{}CODE\_\allowbreak{}DISABLE\_\allowbreak{}AUTO\_\allowbreak{}MEMORY} is how you turn auto memory off & CORRECTED & [25] & Chapter 10 \\
M07 & The 30-day cleanup period governs local Claude Code state & CORRECTED & [25], [30] & Chapter 10 \\
M08 & Session history is capped at one month & CORRECTED & [31], [32] & Chapter 12 \\
M09 & \texttt{bypass\allowbreak{}Permissions} is a reasonable productivity setting & REFRAMED & [19], [30], [33] & Chapter 7 \\
M10 & Every MCP tool definition always consumes context & CORRECTED & [34], [35] & Chapter 13 \\
M11 & Copying \texttt{.\allowbreak{}mcp.\allowbreak{}json} transfers an authenticated server & REJECTED & [34] & Chapter 13 \\
M12 & A popular third-party skill is effectively safe & REJECTED & [36] & Chapter 15 \\
M13 & \texttt{/\allowbreak{}agents} opens an agent editor & CORRECTED & [27], [37] & Chapter 16 \\
M14 & Workflows are beta and cannot be monitored in an IDE & CORRECTED & [19] & Chapter 22 \\
M15 & Workflow agents inherit the session's permission mode & CORRECTED & [19], [27] & Chapter 22 \\
M16 & Claude in Chrome is a beta feature & CORRECTED & [21], [38] & Chapter 20 \\
M17 & \texttt{/\allowbreak{}loop} is equivalent to operating-system cron & REJECTED & [26] & Chapter 23 \\
M18 & Plugins are convenient feature packages & REFRAMED & [39], [40] & Chapter 18 \\
M19 & Routines are simply sessions on a schedule & CORRECTED & [20] & Chapter 24 \\
M20 & A green status on a routine run means the task succeeded & CORRECTED & [20] & Chapter 24 \\
M21 & Remote Control keeps everything local & QUALIFIED & [41]–[43] & Chapter 25 \\
}%
\def\tblbody{\begin{minipage}{\textwidth}\boxcaption{Table 31 — Misconception adjudication, with verdict and primary source}
{\footnotesize\begin{tabular}{L{7.5mm}L{71.8mm}L{18.4mm}L{16.9mm}L{18.7mm}}
\toprule
\textbf{} & \textbf{Claim} & \textbf{Verdict} & \textbf{Primary source} & \textbf{Treated in} \\
\midrule
\tblrows
\bottomrule\end{tabular}}\end{minipage}}%
\begingroup
\def\sloppy{\tolerance 9999\emergencystretch 3em\hfuzz 200pt\vfuzz 200pt}%
\hbadness=10000\vbadness=10000\hfuzz=200pt\vfuzz=200pt
\global\setbox\tblbox=\hbox{\tblbody}%
\endgroup
\par\addvspace{2.6mm}
\ifdim\dimexpr\ht\tblbox+\dp\tblbox\relax>0.55\textheight
  \typeout{HANDBOOK-TABLE broken \the\dimexpr\ht\tblbox+\dp\tblbox\relax}%
  \tabcaption{Table 31 — Misconception adjudication, with verdict and primary source}
  {\footnotesize\begin{longtable}{L{7.5mm}L{71.8mm}L{18.4mm}L{16.9mm}L{18.7mm}}
  \toprule
\textbf{} & \textbf{Claim} & \textbf{Verdict} & \textbf{Primary source} & \textbf{Treated in} \\
\midrule\endfirsthead
  \multicolumn{5}{@{}l@{}}{%
  \sffamily\footnotesize\itshape\color{inkgrey}Table 31 — Misconception adjudication, with verdict and primary source \textemdash\ continued}\\[1.2mm]
  \toprule
\textbf{} & \textbf{Claim} & \textbf{Verdict} & \textbf{Primary source} & \textbf{Treated in} \\
\midrule\endhead
  \bottomrule\endfoot
  \bottomrule\endlastfoot
  \tblrows
  \end{longtable}}%
\else
  \typeout{HANDBOOK-TABLE atomic \the\dimexpr\ht\tblbox+\dp\tblbox\relax}%
  \noindent\tblbody
\fi
\par\addvspace{2.6mm}
\endgroup

This table and the register of Table 1 are generated from a single record, so a claim is stated once and cannot drift between them.

\FloatBarrier
\section*{Reference and standards verification}
\markboth{Reference and standards verification}{Reference and standards verification}

A dedicated verification pass on 24 August 2026 checked the two reference classes a link checker cannot see: entries cited by document number or DOI, and assertions about what a standard or regulation actually requires. Twenty-three primary-source checks were made against publisher and issuing-body sources.

\textbf{Confirmed without change.} NIST AI 100-1, AI 600-1, SP 800-207 (including author list), SP 800-218 (authors and date), and CSWP 29; ISO/IEC 25010:2023, 23894:2023 and 42001:2023 as cited; the title, edition and publication date of ISO/IEC 27001; IEEE Std 1012-2016 and 7000-2021, including both DOIs against their IEEE Xplore document numbers (the DOI of 1012-2016 was right, but the pass missed that IEEE Std 1012-2024 had already superseded that edition; see the check of 14 November 2026 below); Greshake \emph{et al.}, Perez and Ribeiro, and Amodei \emph{et al.}; \textbf{all twenty-six NIST AI RMF category identifiers} used in the crosswalk; and \textbf{all six SSDF practice identifiers} — PO.3, PS.1, PS.2, PS.3, PW.4 and RV.1.

\textbf{Corrected.}

\begingroup
\def\tblrows{%
V1 & ISO/IEC 42001 clause numbers were wrong in seven crosswalk rows and the worked example. Clause 8 is \emph{Operation}: 8.1 operational planning and control, 8.2 AI risk assessment, 8.3 AI risk treatment, 8.4 AI system impact assessment & Renumbered throughout Appendix H; the sub-clause structure is now stated explicitly \\
V2 & The Article 14 mapping gave four of the five capabilities in Art. 14(4), omitting the decision not to use the system or to disregard, override or reverse its output & Restored, and mapped to the human release gate of Chapter 34 \\
V3 & ISO/IEC 27001:2022 carries Amendment 1:2024 & Noted in the reference entry \\
V4 & The OWASP LLM Top 10 has moved to the OWASP GenAI Security Project, with a 2026 edition published before this book's verification date & Reference updated to the current project, edition and canonical URL \\
V5 & SLSA cited at v1.0; current is v1.2, maintained by a vendor-neutral steering committee within the OpenSSF ecosystem & Reference corrected \\
V6 & Two self-describing counts were stale: the automated check count and the number of frameworks in the crosswalk & Corrected to 96 and seventeen \\
V7 & One internal reference pointed at the pre-renumbering appendix letter & Corrected \\
}%
\def\tblbody{\begin{minipage}{\textwidth}
{\footnotesize\begin{tabular}{L{4.8mm}L{99.7mm}L{35.3mm}}
\toprule
\textbf{ID} & \textbf{Finding} & \textbf{Resolution} \\
\midrule
\tblrows
\bottomrule\end{tabular}}\end{minipage}}%
\begingroup
\def\sloppy{\tolerance 9999\emergencystretch 3em\hfuzz 200pt\vfuzz 200pt}%
\hbadness=10000\vbadness=10000\hfuzz=200pt\vfuzz=200pt
\global\setbox\tblbox=\hbox{\tblbody}%
\endgroup
\par\addvspace{2.6mm}
\ifdim\dimexpr\ht\tblbox+\dp\tblbox\relax>0.55\textheight
  \typeout{HANDBOOK-TABLE broken \the\dimexpr\ht\tblbox+\dp\tblbox\relax}%
  
  {\footnotesize\begin{longtable}{L{4.8mm}L{99.7mm}L{35.3mm}}
  \toprule
\textbf{ID} & \textbf{Finding} & \textbf{Resolution} \\
\midrule\endfirsthead
  \multicolumn{3}{@{}l@{}}{%
  \sffamily\footnotesize\itshape\color{inkgrey}Continued}\\[1.2mm]
  \toprule
\textbf{ID} & \textbf{Finding} & \textbf{Resolution} \\
\midrule\endhead
  \bottomrule\endfoot
  \bottomrule\endlastfoot
  \tblrows
  \end{longtable}}%
\else
  \typeout{HANDBOOK-TABLE atomic \the\dimexpr\ht\tblbox+\dp\tblbox\relax}%
  \noindent\tblbody
\fi
\par\addvspace{2.6mm}
\endgroup

\textbf{Recorded as unverified.} The ISO/IEC 27001:2022 Annex A control identifiers used in the crosswalk could not be checked: ISO standards are paywalled and only secondary summaries were reachable. A caveat now appears in Appendix H rather than an implication that they were verified. The page range for Greshake \emph{et al.} was unconfirmed — the ACM Digital Library refuses automated clients — though title, venue and DOI were confirmed; the check of 14 November 2026 closed it. The official expansion of the MITRE ATLAS acronym varied between the project's materials; the check of 15 November 2026 closed it.

\textbf{A methodological result.} During this pass an automated summarisation of a national cyber security agency's page asserted that an international secure-AI guideline had been issued by one agency alone. The issuing bodies' own announcements record a joint publication by two national agencies with seventeen further international partners; the original citation was correct and the summary was wrong. This is the second confirmed false negative from automated summarisation in the preparation of this book, and it is why “Scope and method” treats raw-source adjudication as a rule rather than a preference.

\textbf{Re-checked on 14 November 2026.} Every standard, regulation and paper the book cites was re-checked against its issuer for identity, edition and currency: each DOI through the DOI resolver and the registrant's metadata, both arXiv identifiers through arXiv, the ISO and IEEE entries in their issuers' catalogues, and the two EU regulations in the Publications Office's repository. What each source requires, and the crosswalk's mapping to it, was not re-adjudicated. One finding was material. IEEE Std 1012-2016, cited since Version 1.0, had been superseded by IEEE Std 1012-2024 before Version 1.0 was printed, and the pass of 24 August confirmed its DOI without noticing; the reference now names the current edition, and “Changes since Version 1.0” lists the error. Two references printed a link to a parent page with the rest of the address following it as plain text; both now link the full address. The OWASP entry now cites the 2026 edition's own page and date, the Australian Privacy Principles entry the guidelines rather than the regulator's home page, and the Model Context Protocol entry the title its page now carries. The page range of Greshake \emph{et al.}, pp. 79–90, is confirmed from the publisher's metadata. For Regulation (EU) 2016/679 the regulation, its number and its Official Journal issue were confirmed that day; EUR-Lex did not answer, so the date and closing page were read on 15 November 2026 from the Official Journal text the Publications Office holds, and both stand as cited. On the same day the expansion of the MITRE ATLAS acronym was taken from the maintainers' own data release, which names it Adversarial Threat Landscape for AI Systems; the reference follows it. Four entries the reference database held but the text never cited were removed; none was printed. The ISO/IEC 27001:2022 Annex A identifiers remain as recorded above.

\FloatBarrier
\section*{Corrections since the first release}
\markboth{Corrections since the first release}{Corrections since the first release}

The book's account of Claude Code has been corrected twice since its first release, and the two passes were different in kind.

The re-verification of 26 August 2026 against Claude Code 2.1.246 re-adjudicated ten heavily cited pages of Version 1.0 and changed the statements below. Each was adjudicated against the raw text of the page named, and each is a case where the 23 August 2026 adjudication had reached a different answer or had stopped short of a condition the page attaches.

\begingroup
\def\tblrows{%
R1 & Chapter 7 stated that \texttt{bypass\allowbreak{}Permissions} propagates into a workflow's subagents. They always run in \texttt{accept\allowbreak{}Edits} and inherit your tool allowlist, whatever the session's mode & [19] \\
R2 & \texttt{/\allowbreak{}stats} is a documented alias of \texttt{/\allowbreak{}usage} that opens on the Stats tab, closing the last of the three UNVERIFIED items this book shipped with & [37] \\
R3 & The \texttt{Shift+Tab} cycle was recorded from a session capture and was wrong in order and in kind: the base cycle is three modes, and two more slot in only when each is enabled & [33] \\
R4 & The \texttt{Large workf\allowbreak{}low} warning fires above 25 agents rather than at 25, and a size guideline chosen by the reader replaces that number with its own & [19] \\
R5 & A resumed session has three permission-mode exceptions rather than one, conditions on restoring the model, and tasks and mid-session directories that are not restored at all & [31] \\
R6 & \texttt{auto\allowbreak{}Memory\allowbreak{}Directory} set in a project settings file is honoured only under the workspace-trust rule that governs hooks in settings files & [25] \\
R7 & A parent in \texttt{auto} also defeats a subagent's own \texttt{permission\allowbreak{}Mode}, and \texttt{auto} was then the starting mode on the Pro, Max and Team plans & [27] \\
}%
\def\tblbody{\begin{minipage}{\textwidth}\boxcaption{Table 32 — Corrections applied at the 26 August 2026 re-verification}
{\footnotesize\begin{tabular}{L{4.1mm}L{118.8mm}L{16.9mm}}
\toprule
\textbf{} & \textbf{What the re-verification changed} & \textbf{Primary source} \\
\midrule
\tblrows
\bottomrule\end{tabular}}\end{minipage}}%
\begingroup
\def\sloppy{\tolerance 9999\emergencystretch 3em\hfuzz 200pt\vfuzz 200pt}%
\hbadness=10000\vbadness=10000\hfuzz=200pt\vfuzz=200pt
\global\setbox\tblbox=\hbox{\tblbody}%
\endgroup
\par\addvspace{2.6mm}
\ifdim\dimexpr\ht\tblbox+\dp\tblbox\relax>0.55\textheight
  \typeout{HANDBOOK-TABLE broken \the\dimexpr\ht\tblbox+\dp\tblbox\relax}%
  \tabcaption{Table 32 — Corrections applied at the 26 August 2026 re-verification}
  {\footnotesize\begin{longtable}{L{4.1mm}L{118.8mm}L{16.9mm}}
  \toprule
\textbf{} & \textbf{What the re-verification changed} & \textbf{Primary source} \\
\midrule\endfirsthead
  \multicolumn{3}{@{}l@{}}{%
  \sffamily\footnotesize\itshape\color{inkgrey}Table 32 — Corrections applied at the 26 August 2026 re-verification \textemdash\ continued}\\[1.2mm]
  \toprule
\textbf{} & \textbf{What the re-verification changed} & \textbf{Primary source} \\
\midrule\endhead
  \bottomrule\endfoot
  \bottomrule\endlastfoot
  \tblrows
  \end{longtable}}%
\else
  \typeout{HANDBOOK-TABLE atomic \the\dimexpr\ht\tblbox+\dp\tblbox\relax}%
  \noindent\tblbody
\fi
\par\addvspace{2.6mm}
\endgroup

The adjudication of 12 November 2026 against Claude Code 2.1.322 is this edition's baseline. It worked chapter by chapter through the whole book, eleven weeks and more than sixty releases after the first, and found more. The table lists the principal correction in each chapter; the chapters carry the rest, each stamped with the release that changed the behaviour. Most rows record a product that moved after Version 1.0 was printed. Others — the example policy in N4 is one — record a statement that was wrong when printed and that the first adjudication missed; “Changes since Version 1.0” lists those separately for a reader of Version 1.0.

\begingroup
\def\tblrows{%
N1 & \texttt{auto} is the built-in starting mode of an interactive session on every plan, not on three; \texttt{claude -\allowbreak{}p}, the Agent SDK and the HIPAA configuration start in Manual (Chapter 7) & [33] \\
N2 & \texttt{plan} \textbf{is} restored when a session is resumed in the terminal. Version 1.0 said it never was (Chapter 12) & [31] \\
N3 & On most events a hook that fails lets the action proceed unless \texttt{on\allowbreak{}Failure} is \texttt{"block"}, which \texttt{Stop}-class and background hooks ignore, and on Windows without Git Bash a failing shell-form \texttt{Pre\allowbreak{}Tool\allowbreak{}Use} hook can block with it unset; what a hook makes deterministic is narrower than the word suggests (Chapter 17) & [83] \\
N4 & The example managed policy set \texttt{strict\allowbreak{}Known\allowbreak{}Marketplace\allowbreak{}s} to \texttt{true}, which the key does not accept, and deployed \texttt{allow\allowbreak{}Managed\allowbreak{}Mcp\allowbreak{}Servers\allowbreak{}Only} with no allowlist, which admits every server. Invalid managed values fail closed or open key by key, not by one rule (Chapter 31) & [61], [71] \\
N5 & Project-scoped MCP servers load without an approval prompt in \texttt{claude -\allowbreak{}p}, the Agent SDK and cloud sessions outside a project's threads, and an allowlist no longer filters an organisation's own \texttt{managed-\allowbreak{}mcp.\allowbreak{}json} servers unless they use variable expansion outside \texttt{env} (Chapter 13) & [34], [86] \\
N6 & Fork mode lets Claude start a fork and stops it waiting for a subagent; it does not turn a named subagent into a fork. An \texttt{Agent(type)} list is ignored in a subagent's own definition (Chapter 16) & [27] \\
N7 & \texttt{AGENTS.\allowbreak{}md} is read where no \texttt{CLAUDE.\allowbreak{}md} exists. Version 1.0 said it was never read. Project rules do not outrank user rules (Chapters 9 and 10) & [25] \\
N8 & Each secret control reaches a different edge: sandbox credential controls cover sandboxed shell commands only, and masking a credential authorises its use rather than redacting it (Chapter 14) & [28], [29] \\
N9 & A personal skill wins over a project skill of the same name, and \texttt{allowed-\allowbreak{}tools} pre-approves without restricting (Chapter 15) & [36] \\
N10 & The \texttt{/\allowbreak{}output-\allowbreak{}style} command exists, a style switch applies from the next message, and it keeps the prompt cache. Version 1.0 said the opposite of all three (Chapters 4 and 19) & [68], [91] \\
N11 & A plugin can carry a mod, code that runs inside Claude Code. By default, without managed settings, a \texttt{managed-\allowbreak{}mcp.\allowbreak{}json} file or a Team or Enterprise login as the session's credential, a mod can approve a call that a deny rule refuses, which qualifies this book's account of enforced deny rules (Chapters 7 and 18) & [29], [89], [90] \\
N12 & \texttt{ultracode} is a setting, not an effort level, and \texttt{/\allowbreak{}effort} accepts \texttt{auto} (Chapters 4 and 22) & [63] \\
N13 & \texttt{/\allowbreak{}security-\allowbreak{}review} is a built-in command, not a bundled skill; managed version bounds stop Claude Code starting rather than failing open; and \texttt{disable\allowbreak{}All\allowbreak{}Hooks} set in managed settings stops the organisation's own hooks too (Appendices A and D) & [37], [61] \\
N14 & Fork mode is on by default beyond the interactive terminal from 2.1.302, and \texttt{claude doct\allowbreak{}or} is a read-only checkup: the \texttt{CLAUDE.\allowbreak{}md} trims Version 1.0 gave it belong to \texttt{/\allowbreak{}doctor}, as they already did when Version 1.0 was printed, so the misattribution is also an erratum (Chapters 2 and 16) & [27], [52] \\
}%
\def\tblbody{\begin{minipage}{\textwidth}\boxcaption{Table 33 — Principal corrections applied at the 12 November 2026 adjudication}
{\footnotesize\begin{tabular}{L{6.1mm}L{116.7mm}L{16.9mm}}
\toprule
\textbf{} & \textbf{What the re-verification changed} & \textbf{Primary source} \\
\midrule
\tblrows
\bottomrule\end{tabular}}\end{minipage}}%
\begingroup
\def\sloppy{\tolerance 9999\emergencystretch 3em\hfuzz 200pt\vfuzz 200pt}%
\hbadness=10000\vbadness=10000\hfuzz=200pt\vfuzz=200pt
\global\setbox\tblbox=\hbox{\tblbody}%
\endgroup
\par\addvspace{2.6mm}
\ifdim\dimexpr\ht\tblbox+\dp\tblbox\relax>0.55\textheight
  \typeout{HANDBOOK-TABLE broken \the\dimexpr\ht\tblbox+\dp\tblbox\relax}%
  \tabcaption{Table 33 — Principal corrections applied at the 12 November 2026 adjudication}
  {\footnotesize\begin{longtable}{L{6.1mm}L{116.7mm}L{16.9mm}}
  \toprule
\textbf{} & \textbf{What the re-verification changed} & \textbf{Primary source} \\
\midrule\endfirsthead
  \multicolumn{3}{@{}l@{}}{%
  \sffamily\footnotesize\itshape\color{inkgrey}Table 33 — Principal corrections applied at the 12 November 2026 adjudication \textemdash\ continued}\\[1.2mm]
  \toprule
\textbf{} & \textbf{What the re-verification changed} & \textbf{Primary source} \\
\midrule\endhead
  \bottomrule\endfoot
  \bottomrule\endlastfoot
  \tblrows
  \end{longtable}}%
\else
  \typeout{HANDBOOK-TABLE atomic \the\dimexpr\ht\tblbox+\dp\tblbox\relax}%
  \noindent\tblbody
\fi
\par\addvspace{2.6mm}
\endgroup

\FloatBarrier
\section*{Closed by direct observation}
\markboth{Closed by direct observation}{Closed by direct observation}

Three items carried as UNVERIFIED in earlier drafts were reopened by exercising the product directly, most recently on Claude Code 2.1.322 on 12 November 2026, and settled against the documentation that governs each. Each verdict is \textbf{configuration-specific}: it records what one installation did on one date, on the surface named, and asserts nothing more general. What each observation did not reach stays in the open list below.

\begingroup
\def\tblrows{%
The \texttt{/\allowbreak{}effort} vocabulary & Five levels accepted everywhere, and \texttt{auto}, which the command accepts and the flag refuses (Table 3). \texttt{ultracode} is taken by both and enumerated by neither & The CLI command through \texttt{claude -\allowbreak{}p} and the \texttt{claude -\allowbreak{}-\allowbreak{}effort} flag, on 2.1.322; the VS Code extension on 2.1.246 & CONFIRMED (configuration-specific) \\
The \texttt{Shift+Tab} permission-mode cycle & The cycle never reaches \texttt{dont\allowbreak{}Ask}; order and conditions are given in Table 5 & Terminal CLI on 2.1.246, then adjudicated against [33], and re-adjudicated against it on 12 November 2026 & CONFIRMED \\
The \texttt{claude -\allowbreak{}-\allowbreak{}help} flag set & 67 long options in option position and 18 subcommands on 2.1.322, against 62 and 13 on 2.1.246 & The installed binary of each release & CONFIRMED \\
}%
\def\tblbody{\begin{minipage}{\textwidth}\boxcaption{Table 34 — Items closed by direct observation, with the release each was observed on}
{\footnotesize\begin{tabular}{L{30.7mm}L{29.5mm}L{31.2mm}L{45.1mm}}
\toprule
\textbf{Item} & \textbf{Observation} & \textbf{Surface observed} & \textbf{Verdict} \\
\midrule
\tblrows
\bottomrule\end{tabular}}\end{minipage}}%
\begingroup
\def\sloppy{\tolerance 9999\emergencystretch 3em\hfuzz 200pt\vfuzz 200pt}%
\hbadness=10000\vbadness=10000\hfuzz=200pt\vfuzz=200pt
\global\setbox\tblbox=\hbox{\tblbody}%
\endgroup
\par\addvspace{2.6mm}
\ifdim\dimexpr\ht\tblbox+\dp\tblbox\relax>0.55\textheight
  \typeout{HANDBOOK-TABLE broken \the\dimexpr\ht\tblbox+\dp\tblbox\relax}%
  \tabcaption{Table 34 — Items closed by direct observation, with the release each was observed on}
  {\footnotesize\begin{longtable}{L{30.7mm}L{29.5mm}L{31.2mm}L{45.1mm}}
  \toprule
\textbf{Item} & \textbf{Observation} & \textbf{Surface observed} & \textbf{Verdict} \\
\midrule\endfirsthead
  \multicolumn{4}{@{}l@{}}{%
  \sffamily\footnotesize\itshape\color{inkgrey}Table 34 — Items closed by direct observation, with the release each was observed on \textemdash\ continued}\\[1.2mm]
  \toprule
\textbf{Item} & \textbf{Observation} & \textbf{Surface observed} & \textbf{Verdict} \\
\midrule\endhead
  \bottomrule\endfoot
  \bottomrule\endlastfoot
  \tblrows
  \end{longtable}}%
\else
  \typeout{HANDBOOK-TABLE atomic \the\dimexpr\ht\tblbox+\dp\tblbox\relax}%
  \noindent\tblbody
\fi
\par\addvspace{2.6mm}
\endgroup

The \texttt{claude -\allowbreak{}-\allowbreak{}help} row closes an item that named a single release. Comparing two releases directly is cheaper than re-capturing one, and the delta is the useful result. Between 2.1.241 and 2.1.246 it was zero. Over the eleven weeks to 2.1.322 the same count gained five options and five subcommands, which is what a reader re-verifying against a later release needs to know: the flag surface is stable over days and not over months.

\FloatBarrier
\section*{A help string is not an enumeration}
\markboth{A help string is not an enumeration}{A help string is not an enumeration}

Exercising those two vocabularies produced a result worth recording as method rather than as content. \textbf{Twice, the product accepted a value its own help text does not list.} \texttt{claude -\allowbreak{}-\allowbreak{}effort} accepts \texttt{ultracode} while \texttt{-\allowbreak{}-\allowbreak{}help} enumerates five levels. \texttt{claude -\allowbreak{}-\allowbreak{}permission-\allowbreak{}mode} accepts \texttt{default} while the error message it prints when refusing a value enumerates six and omits it — so the under-reporting appeared in the most authoritative-looking enumeration the command line offers, the one shown at the moment of refusal.

The second case is the more instructive, because adjudication later explained it. \texttt{manual} is a documented alias for \texttt{default} (2.1.200 or later), and the validator lists the alias rather than the canonical value that settings files and the SDK use [33]. The enumeration was never wrong; it was incomplete in a way no amount of staring at it could reveal. Had this book treated the omission of \texttt{default} as evidence that \texttt{default} was not accepted, it would have printed a false statement drawn from an authoritative-looking source — which is precisely the failure the rule below prevents.

The consequence is a rule rather than a caveat. \textbf{A help string is evidence that a value is supported; it is never evidence that an unlisted value is refused.} Every statement in this book of the form \emph{the accepted values are X, Y and Z} is therefore stated for the surface it was observed on, and an absence from an enumeration is recorded as unestablished rather than as a refusal — unless the refusal was itself observed, as \texttt{claude -\allowbreak{}-\allowbreak{}effort auto} was. This strengthens the UNVERIFIED discipline of “Scope and method” rather than weakening it: the two vocabularies were larger than the documentation of them, and only exercising the product showed it.

\FloatBarrier
\section*{Residual uncertainty}
\markboth{Residual uncertainty}{Residual uncertainty}

Claude Code changes weekly. Model availability, interface labels, plan entitlements and preview features may all have moved since the verification date. Third-party MCP servers, plugins, skills and channels require separate per-version assessment; this book does not assert that the absence of a documented incident makes any component safe. Two classes of statement in this edition are explicitly weaker than the rest: anything labelled PRACTITIONER NOTE, which is judgement rather than product behaviour, and anything labelled UNVERIFIED, listed below.

\textbf{Open UNVERIFIED items at publication:}

\begingroup
\def\tblrows{%
The exact \texttt{/\allowbreak{}effort} menu on a given plan & Menu contents are plan- and model-dependent. Table 3 closes this for one configuration only & Observe \texttt{/\allowbreak{}effort} locally on your own plan and model \\
What the interactive terminal \texttt{/\allowbreak{}effort} selector offers & \texttt{/\allowbreak{}effort auto} was accepted in a \texttt{claude -\allowbreak{}p} run on 2.1.322, which closes the earlier question of whether the command takes it; the interactive selector's contents were not captured & Run a bare \texttt{/\allowbreak{}effort} in an interactive terminal session and read the selector \\
Whether the VS Code \texttt{/\allowbreak{}effort} accepts \texttt{ultracode} & Absent from that surface's usage string on 2.1.246, which — see above — is not evidence of refusal; not re-observed on 2.1.322 & Type \texttt{/\allowbreak{}effort ultr\allowbreak{}acode} in the extension and observe the result \\
Whether a Claude in Chrome click, typing or navigation prompts in Plan mode & The page fetched for Version 1.0 said state-changing browser calls prompt in Plan mode; the page fetched on 12 November 2026 documents prompts only for GIF recording and opening a new tab, and is silent on the rest & In Plan mode on a local fixture, ask Claude to type into a field and click a button, and record whether each prompts; repeat with auto mode unavailable \\
Whether \texttt{/\allowbreak{}clear} still clears session-scoped scheduled tasks & The page fetched for Version 1.0 said a fresh conversation clears them all; the page fetched on 12 November 2026 no longer says so, and does not say the opposite & Schedule a one-minute \texttt{/\allowbreak{}loop}, run \texttt{/\allowbreak{}clear}, then ask Claude to list scheduled tasks and wait two minutes \\
What a self-paced \texttt{/\allowbreak{}loop} does when an iteration ends without rescheduling & The scheduled-tasks page describes one fallback wakeup about 20 minutes later; the 2.1.319 release notes say the loop ends at once unless \texttt{CLAUDE\_\allowbreak{}CODE\_\allowbreak{}LOOP\_\allowbreak{}KEEPALIVE=\allowbreak{}1} is set. The two were fetched on the same day and disagree & Start \texttt{/\allowbreak{}loop} with a prompt whose first iteration neither reschedules nor stops, and watch whether a wakeup is scheduled; repeat with the variable set \\
}%
\def\tblbody{\begin{minipage}{\textwidth}
{\footnotesize\begin{tabular}{L{28.7mm}L{70.4mm}L{40.7mm}}
\toprule
\textbf{Item} & \textbf{Why it is open} & \textbf{How to close it} \\
\midrule
\tblrows
\bottomrule\end{tabular}}\end{minipage}}%
\begingroup
\def\sloppy{\tolerance 9999\emergencystretch 3em\hfuzz 200pt\vfuzz 200pt}%
\hbadness=10000\vbadness=10000\hfuzz=200pt\vfuzz=200pt
\global\setbox\tblbox=\hbox{\tblbody}%
\endgroup
\par\addvspace{2.6mm}
\ifdim\dimexpr\ht\tblbox+\dp\tblbox\relax>0.55\textheight
  \typeout{HANDBOOK-TABLE broken \the\dimexpr\ht\tblbox+\dp\tblbox\relax}%
  
  {\footnotesize\begin{longtable}{L{28.7mm}L{70.4mm}L{40.7mm}}
  \toprule
\textbf{Item} & \textbf{Why it is open} & \textbf{How to close it} \\
\midrule\endfirsthead
  \multicolumn{3}{@{}l@{}}{%
  \sffamily\footnotesize\itshape\color{inkgrey}Continued}\\[1.2mm]
  \toprule
\textbf{Item} & \textbf{Why it is open} & \textbf{How to close it} \\
\midrule\endhead
  \bottomrule\endfoot
  \bottomrule\endlastfoot
  \tblrows
  \end{longtable}}%
\else
  \typeout{HANDBOOK-TABLE atomic \the\dimexpr\ht\tblbox+\dp\tblbox\relax}%
  \noindent\tblbody
\fi
\par\addvspace{2.6mm}
\endgroup

\FloatBarrier
\setcounter{chapter}{10}
\setcounter{section}{0}
\appendixchapter{K}{Re-verification and version decay}
\FloatBarrier
\section*{Why this appendix exists}
\markboth{Why this appendix exists}{Why this appendix exists}

Between the release the course was recorded against and the release this book was verified against, documented behaviour changed at 2.1.154, 2.1.160, 2.1.181, 2.1.193, 2.1.196, 2.1.197, 2.1.198, 2.1.200, 2.1.206, 2.1.209, 2.1.210, 2.1.213, 2.1.216, 2.1.217, 2.1.219, 2.1.221, 2.1.223, 2.1.224, 2.1.225, 2.1.227, 2.1.228, 2.1.232, 2.1.234, 2.1.236, 2.1.237, 2.1.239, 2.1.242, 2.1.246, 2.1.248, 2.1.251, 2.1.257, 2.1.259, 2.1.261, 2.1.265, 2.1.267, 2.1.268, 2.1.269, 2.1.271, 2.1.273, 2.1.275, 2.1.277, 2.1.281, 2.1.282, 2.1.283, 2.1.284, 2.1.285, 2.1.286, 2.1.287, 2.1.288, 2.1.290, 2.1.292, 2.1.293, 2.1.295, 2.1.296, 2.1.299, 2.1.300, 2.1.302, 2.1.304, 2.1.305, 2.1.309, 2.1.317, 2.1.318 and 2.1.319 — sixty-three release-specific behaviour notes appear in the pages cited by this book, which is every release this book stamps inline and is checked against them. A reference work in this domain has a half-life measured in weeks, and the honest response is a procedure rather than a disclaimer.

\FloatBarrier
\section*{The procedure}
\markboth{The procedure}{The procedure}

Perform this quarterly, or before relying on the book for a consequential decision.

\begin{enumerate}
\item \textbf{Establish the delta.} Read the changelog and the weekly digest from the verification date forward [38], [57].
\item \textbf{Re-fetch the index.} Compare the current documentation index against the reference list of this edition [1]. New pages indicate new capability the book does not cover; removed pages indicate deprecation.
\item \textbf{Re-verify the version-stamped claims.} Every statement in this book carrying a release number is a candidate for change.
\item \textbf{Re-run the local evidence script} below, and reconcile it with Appendix A and Appendix B.
\item \textbf{Re-test the security controls that matter to you.} A deny rule that blocked last quarter should be tested again, not assumed.
\item \textbf{Update the ledger.} Record what changed, in Appendix J, with the source and the date.
\end{enumerate}

\FloatBarrier
\section*{Local evidence script}
\markboth{Local evidence script}{Local evidence script}

Run this on the machine whose behaviour you want to record. It writes a plain-text bundle you can attach to a review. It changes no configuration, but it is not passive: \texttt{claude mcp \allowbreak{}list} health-checks the servers it lists by connecting to them, which starts each approved stdio server as a local process with your permissions [34], [40]. Remove that line where starting the servers is itself the risk you are recording.

\begin{codeblock}{6.0}{7.1}
\cl{\#!/usr/bin/env~bash}
\cl{\#~collect-claude-evidence.sh~—~record~the~locally~observable~state~of~a~Claude~Code~install.}
\cl{set~-uo~pipefail}
\cl{OUT="claude-evidence-\$(date~-u~+\%Y\%m\%dT\%H\%M\%SZ).txt"}
\cl{\{}
\cl{~~echo~"===~collected~(UTC)~===";~date~-u}
\cl{~~echo;~echo~"===~claude~-{}-version~===";~claude~-{}-version~2>\&1}
\cl{~~echo;~echo~"===~claude~doctor~===";~claude~doctor~2>\&1}
\cl{~~echo;~echo~"===~claude~-{}-help~===";~claude~-{}-help~2>\&1}
\cl{~~echo;~echo~"===~claude~mcp~-{}-help~===";~claude~mcp~-{}-help~2>\&1}
\cl{~~echo;~echo~"===~claude~mcp~list~===";~claude~mcp~list~2>\&1}
\cl{~~echo;~echo~"===~claude~plugin~-{}-help~===";~claude~plugin~-{}-help~2>\&1}
\cl{~~echo;~echo~"===~settings~files~present~==="}
\cl{~~for~f~in~\textasciitilde{}/.claude/settings.json~.claude/settings.json~.claude/settings.local.json~\textbackslash{}}
\cl{~~~~~~~~~~~/etc/claude-code/managed-settings.json~\textbackslash{}}
\cl{~~~~~~~~~~~"/Library/Application~Support/ClaudeCode/managed-settings.json";~do}
\cl{~~~~[~-f~"\$f"~]~\&\&~\{~echo~"-{}-{}-~\$f";~sed~-e~\textquotesingle{}s/\textbackslash{}("[A-Za-z\_]*[Tt]oken"[[:space:]]*:[[:space:]]*\textbackslash{}).*/\textbackslash{}1"<redacted>"/\textquotesingle{}~\textbackslash{}}
\cl{~~~~~~~~~~~~~~~~~~~~~~~~~~~~~~~~~~~~~~-e~\textquotesingle{}s/\textbackslash{}("[A-Za-z\_]*[Kk]ey"[[:space:]]*:[[:space:]]*\textbackslash{}).*/\textbackslash{}1"<redacted>"/\textquotesingle{}~"\$f";~\}}
\cl{~~done}
\cl{~~echo;~echo~"===~instruction~and~memory~files~==="}
\cl{~~ls~-la~\textasciitilde{}/.claude/rules/~.claude/rules/~2>/dev/null}
\cl{~~ls~-la~\textasciitilde{}/.claude/projects/*/memory/~2>/dev/null~|~head~-40}
\cl{~~echo;~echo~"===~skills,~agents,~workflows~==="}
\cl{~~ls~-la~\textasciitilde{}/.claude/skills/~.claude/skills/~\textasciitilde{}/.claude/agents/~.claude/agents/~\textbackslash{}}
\cl{~~~~~~~~~\textasciitilde{}/.claude/workflows/~.claude/workflows/~2>/dev/null}
\cl{\}~>~"\$OUT"~2>\&1}
\cl{echo~"Wrote~\$OUT~—~review~it~for~secrets~before~sharing."}
\end{codeblock}

Then, inside a session, capture what only the running product can tell you: \texttt{/\allowbreak{}status}, \texttt{/\allowbreak{}permissions}, \texttt{/\allowbreak{}context}, \texttt{/\allowbreak{}memory}, \texttt{/\allowbreak{}hooks}, \texttt{/\allowbreak{}mcp}, \texttt{/\allowbreak{}usage}, and the \texttt{/\allowbreak{}effort} menu. Use \texttt{/\allowbreak{}export} to save the result [31], [37].

\begin{calloutbox}{palebrass}{brassdark}{2.0mm}
\textbf{CAUTION:} the redaction in the script above is a convenience, not a guarantee. Read the bundle before you send it anywhere.
\end{calloutbox}

\FloatBarrier
\setcounter{chapter}{11}
\setcounter{section}{0}
\appendixchapter{L}{Adoption maturity model}
A five-level model for organisational adoption of agentic development tooling. Each level states the controls that must be \emph{enforced} rather than requested, the evidence that demonstrates the level is real, and the role accountable for it. The model is the author's; it is offered as a sequencing aid, not as a standard.

The organising principle is stated in the executive summary and repeated here because it is the only rule that matters: \textbf{adopt the capability at the pace at which you can evidence its output.} A level is reached when the evidence exists, not when the intention is announced.

\begingroup
\def\tblrows{%
0 & \textbf{Unmanaged} & None. Individuals install and configure independently; defaults apply. & None available. The organisation cannot state which permission modes, connectors or plugins are in use. & Nobody \\
1 & \textbf{Visible} & Inventory only. No policy is deployed, but usage is known. & A list of users, surfaces and plans. \texttt{/\allowbreak{}status} output sampled across the fleet. Telemetry enabled with content logging off. & Engineering manager \\
2 & \textbf{Bounded} & Managed settings deployed and verified: deny rules for credential paths, \texttt{disable\allowbreak{}Bypass\allowbreak{}Permissions\allowbreak{}Mode}, \texttt{allow\allowbreak{}Managed\allowbreak{}Permission\allowbreak{}Rules\allowbreak{}Only}, an MCP allowlist, a marketplace policy. Sandboxing enabled where the platform supports it. & \texttt{/\allowbreak{}status} shows the intended managed source on sampled machines. \texttt{claude doct\allowbreak{}or} shows no dropped entries. A tested deny rule demonstrably blocks. & Security engineering \\
3 & \textbf{Evidenced} & Level 2, plus: telemetry exported to a collector with retention defined; audit review of \texttt{tool\_\allowbreak{}decision}, \texttt{permission\_\allowbreak{}mode\_\allowbreak{}changed}, \texttt{plugin\_\allowbreak{}installed} and \texttt{mcp\_\allowbreak{}server\_\allowbreak{}connection} events; a release gate for anything an agent publishes, merges or sends. & Traceability from a single prompt identifier to every action it caused. A completed release checklist per consequential deliverable. A named reviewer independent of the author. & Head of engineering, with security assurance \\
4 & \textbf{Governed} & Level 3, plus: data residency and retention decided per cohort with \texttt{force\allowbreak{}Login\allowbreak{}Org\allowbreak{}UUID} deployed; autonomous execution — routines, unattended goals, agent teams — permitted only under named approval; supply-chain review with pinned versions and a re-review trigger; accessibility position stated. & A control-to-framework map maintained (Appendix H). A quarterly re-verification record (Appendix K). An incident procedure exercised at least once. A register of approved autonomous workloads with owners. & Executive risk owner, reported to the board \\
}%
\def\tblbody{\begin{minipage}{\textwidth}\boxcaption{Table 35 — Adoption maturity levels for agentic development tooling}
{\scriptsize\begin{tabular}{L{9.7mm}L{14.8mm}L{52.8mm}L{34.6mm}L{21.4mm}}
\toprule
\textbf{Level} & \textbf{Name} & \textbf{Enforced controls} & \textbf{Evidence that the level is real} & \textbf{Accountable} \\
\midrule
\tblrows
\bottomrule\end{tabular}}\end{minipage}}%
\begingroup
\def\sloppy{\tolerance 9999\emergencystretch 3em\hfuzz 200pt\vfuzz 200pt}%
\hbadness=10000\vbadness=10000\hfuzz=200pt\vfuzz=200pt
\global\setbox\tblbox=\hbox{\tblbody}%
\endgroup
\par\addvspace{2.6mm}
\ifdim\dimexpr\ht\tblbox+\dp\tblbox\relax>0.55\textheight
  \typeout{HANDBOOK-TABLE broken \the\dimexpr\ht\tblbox+\dp\tblbox\relax}%
  \tabcaption{Table 35 — Adoption maturity levels for agentic development tooling}
  {\scriptsize\begin{longtable}{L{9.7mm}L{14.8mm}L{52.8mm}L{34.6mm}L{21.4mm}}
  \toprule
\textbf{Level} & \textbf{Name} & \textbf{Enforced controls} & \textbf{Evidence that the level is real} & \textbf{Accountable} \\
\midrule\endfirsthead
  \multicolumn{5}{@{}l@{}}{%
  \sffamily\footnotesize\itshape\color{inkgrey}Table 35 — Adoption maturity levels for agentic development tooling \textemdash\ continued}\\[1.2mm]
  \toprule
\textbf{Level} & \textbf{Name} & \textbf{Enforced controls} & \textbf{Evidence that the level is real} & \textbf{Accountable} \\
\midrule\endhead
  \bottomrule\endfoot
  \bottomrule\endlastfoot
  \tblrows
  \end{longtable}}%
\else
  \typeout{HANDBOOK-TABLE atomic \the\dimexpr\ht\tblbox+\dp\tblbox\relax}%
  \noindent\tblbody
\fi
\par\addvspace{2.6mm}
\endgroup

\FloatBarrier
\section*{Level transitions, and the failure at each one}
\markboth{Level transitions, and the failure at each one}{Level transitions, and the failure at each one}

\textbf{0 → 1.} The failure is believing an inventory exists because a procurement record exists. Ask for \texttt{/\allowbreak{}status} output, not a licence count.

\textbf{1 → 2.} The failure is writing standards into \texttt{CLAUDE.\allowbreak{}md}. That documents an intention; a managed settings file deploys a control (Figure 1). The second failure is deploying a policy that is silently ignored because a higher-priority managed source is present (Section 31.1).

\textbf{2 → 3.} The failure is enabling telemetry without deciding who reads it. An unread audit trail is an expense, not a control. The second failure is an "independent" reviewer that is a fork (Section 16.1).

\textbf{3 → 4.} The failure is permitting autonomous execution because it is available rather than because a named person accepted the residual risk. A routine runs with your connectors, under your identity, and where auto mode is available a classifier rather than a person approves most of its actions (Chapter 24) [20]; somebody must own that.

\textbf{A note on speed.} Level 4 is not the goal for every organisation. A two-person research group operating at Level 2 with disciplined release gates is in better shape than a large enterprise at Level 3 that has never tested a deny rule. The model sequences capability against evidence; it does not argue that more governance is always better.

\FloatBarrier
\setcounter{chapter}{12}
\setcounter{section}{0}
\appendixchapter{M}{Worked solutions to the exercises}

Each of the fourteen exercises has one worked solution here. They are defensible answers, not the only answers: every solution shows a route that survives inspection and states the evidence that would demonstrate the exercise was completed correctly rather than merely attempted. The exercises were written so that the failure modes are the lesson, so read the solution after your own attempt, not before it.

Three conventions run through all of them. Where an instruction can be \emph{enforced} rather than \emph{requested}, the solution enforces it — a permission rule, a hook, a sandbox — because a sentence in a prompt is a preference and a deny rule is a control (Chapter 7). Where a solution produces an artefact, the artefact has a stated schema, so that "done" is decidable by someone who was not watching. And every solution ends with a check that can fail; a check that cannot fail is a description, not a test.

\FloatBarrier
\section*{Chapter 1 — a four-element brief you can run}
\markboth{Chapter 1 — a four-element brief you can run}{Chapter 1 — a four-element brief you can run}

\textbf{Asked:} write a four-line brief in Outcome / Context / Boundaries / Evidence form and keep it for Chapter 5.

The four elements each close a specific failure mode: an unbounded outcome, an assumed context, an unstated prohibition, and an unverifiable result. A worked version, for a practice workspace holding \texttt{notes/\allowbreak{}}, \texttt{sources/\allowbreak{}} and an empty \texttt{drafts/\allowbreak{}}:

\begin{codeblock}{7.0}{8.3}
\cl{Outcome:~~~~A~brief~of~at~most~600~words,~built~only~from~the~files~in~notes/,~answering~two}
\cl{~~~~~~~~~~~~questions:~what~does~this~material~claim,~and~what~does~it~not~establish?}
\cl{Context:~~~~notes/~holds~fourteen~markdown~files~written~between~March~and~August~2026.}
\cl{~~~~~~~~~~~~sources/~holds~the~PDFs~they~cite.~Treat~notes/~as~claims~and~sources/~as~evidence.}
\cl{~~~~~~~~~~~~Identify~evidence~gaps;~do~not~fill~them.}
\cl{Boundaries:~Read-only~outside~drafts/.~Do~not~edit,~move,~rename~or~delete~anything~in~notes/}
\cl{~~~~~~~~~~~~or~sources/.~No~network~access.~No~shell~command~that~writes~outside~drafts/.}
\cl{Evidence:~~~Write~drafts/brief-v1.md~containing~a~claim-to-source~table~with~one~row~per~claim}
\cl{~~~~~~~~~~~~and~the~columns:~Claim~|~notes/~file~and~line~|~sources/~file~and~page~|~Supported.}
\cl{~~~~~~~~~~~~Record~every~unsupported~claim~as~UNSUPPORTED~rather~than~omitting~it.}
\end{codeblock}

Three properties make this runnable rather than aspirational. The outcome carries a length bound \emph{and} a question, so completion is decidable. The boundary prohibits writes by name rather than asking for care, and it names the operations — edit, move, rename, delete — that a request to "not change anything" leaves ambiguous. The evidence clause fixes the artefact and its schema, so the session cannot end with a plausible summary and no audit trail. Note the deliberate instruction to record gaps rather than close them: an agent asked for a complete brief will produce one, and the missing evidence is the finding you actually want.

Enforce what settings can enforce, and know what they leave open. In \texttt{.\allowbreak{}claude/\allowbreak{}settings.\allowbreak{}json}:

\begin{codeblock}{9.0}{10.6}
\cl{\{}
\cl{~~"permissions":~\{}
\cl{~~~~"deny":~[}
\cl{~~~~~~"Edit(./notes/**)",}
\cl{~~~~~~"Edit(./sources/**)",}
\cl{~~~~~~"Bash",}
\cl{~~~~~~"PowerShell",}
\cl{~~~~~~"WebFetch",}
\cl{~~~~~~"WebSearch"}
\cl{~~~~]}
\cl{~~\}}
\cl{\}}
\end{codeblock}

The two path rules refuse edits to \texttt{notes/\allowbreak{}} and \texttt{sources/\allowbreak{}} through Claude's file tools and the file commands it recognises in the shell; they do not reach a script that opens files itself [29]. Denying \texttt{Bash} and \texttt{Power\allowbreak{}Shell} closes both shell routes, and their network access with them, which the brief does not need. A Bash deny alone switches PowerShell off only on Windows with Git Bash, and not even there when \texttt{CLAUDE\_\allowbreak{}CODE\_\allowbreak{}USE\_\allowbreak{}POWERSHELL\_\allowbreak{}TOOL=\allowbreak{}1} or a scoped \texttt{Power\allowbreak{}Shell} rule keeps it on [72]. Hooks, MCP servers and plugins need their own controls. What remains unenforced is the rest of "read-only outside \texttt{drafts/\allowbreak{}}": the file tools can still write elsewhere in the workspace, so that part of the boundary is held by the instruction alone, and the check below tests it.

\textbf{Check.} Ask one question of the brief: could two competent people disagree about whether it was completed? If they could, the element that permits the disagreement is the one to tighten. Then confirm the controls fire — ask for an edit inside \texttt{notes/\allowbreak{}} and for a shell command, and watch both refused — and look for any file written outside \texttt{drafts/\allowbreak{}}. A deny rule you have never seen block anything is a rule you have not tested (Section 7.2).

\FloatBarrier
\section*{Chapter 3 — a read-only inspection in Plan mode}
\markboth{Chapter 3 — a read-only inspection in Plan mode}{Chapter 3 — a read-only inspection in Plan mode}

\textbf{Asked:} in Plan mode, have the workspace inspected and a smallest safe next step proposed, without modification and without leaving the folder.

Run the inspection from the practice folder and nowhere else. The prompt in the exercise is already well formed; what the exercise tests is whether you can tell the difference between a plan that respected the boundary and one that merely reported respecting it.

A good returned plan has four properties. It describes structure it could only have learned by reading — file counts, naming patterns, date ranges — rather than restating your prompt back. It separates what is missing into things you must decide and things it can determine. Its proposed next step is genuinely small: one file, one directory, one reversible operation. And it states what it did \emph{not} look at, which is the part most plans omit.

The mistake the exercise is built around is accepting a plan because it is well written. Plan mode constrains what Claude may do, not what it may claim (Section 6.1). Verify the boundary from outside the transcript:

\begin{codeblock}{9.0}{10.6}
\cl{\#~Nothing~under~the~workspace~changed~during~the~inspection.}
\cl{find~.~-newermt~\textquotesingle{}-15~minutes\textquotesingle{}~-type~f~-not~-path~\textquotesingle{}./.git/*\textquotesingle{}}
\cl{}
\cl{\#~And~with~Git~initialised,~the~stronger~statement:}
\cl{git~status~-{}-porcelain}
\end{codeblock}

Both should be empty. If \texttt{find} reports files that Claude only read, check whether an editor or indexer touched them — atime is not mtime, and a non-empty result here is more often your own tooling than the agent.

\textbf{Check.} Three questions, answerable from the plan alone: which specific files does it cite as evidence for its description of the structure; what did it decline to inspect and why; and is the proposed next step reversible without a backup? A plan that cannot answer all three has not inspected the workspace, it has described it plausibly.

\FloatBarrier
\section*{Chapter 5 — running the brief, and steering without interrupting}
\markboth{Chapter 5 — running the brief, and steering without interrupting}{Chapter 5 — running the brief, and steering without interrupting}

\textbf{Asked:} run the Chapter 1 brief, queue one non-urgent steering message mid-task requiring findings to be ordered by decision relevance, and interrupt only if a boundary is crossed.

The exercise separates two things that novices conflate: \emph{steering}, which joins the queue and is read as soon as the running tool calls finish, often within the same turn, and \emph{interruption}, which discards the current turn's remaining work. Steering is cheap; interruption throws away tokens already spent and forces re-derivation (Section 5.3).

Start the session with the brief exactly as written, then send this as a queued message while work is in progress — not after it stops:

\begin{codeblock}{7.0}{8.3}
\cl{Additional~requirement,~no~need~to~restart:~order~the~findings~by~decision~relevance~rather}
\cl{than~by~source~order.~A~finding~is~decision-relevant~if~acting~on~it~would~change~what~someone}
\cl{does~next.~Put~the~ones~that~change~a~decision~first,~and~say~for~each~what~decision~it~changes.}
\end{codeblock}

Two details make it a steering message rather than an interruption. It states explicitly that no restart is needed, which removes the ambiguity that causes an agent to discard completed work. And it defines the ordering criterion instead of naming it — "decision relevance" is otherwise interpreted as "importance", and you will get the same list in a different order.

Reserve interruption for boundary crossings: a write outside \texttt{drafts/\allowbreak{}}, a network call, an attempt to resolve an evidence gap by inference rather than recording it. When one occurs, interrupt and say what was crossed, rather than restating the boundary — the boundary was already stated, and repeating it teaches you nothing about why it failed.

\textbf{Check.} Two observations. First, did the steering message change the output without causing re-derivation of work already done? Compare the claim-to-source table before and after: the rows should be reordered and annotated, not regenerated. Second, count your interruptions. More than zero on a well-bounded brief means either the boundary was too loose or you interrupted for a disagreement rather than a violation, and the two have different remedies.

\FloatBarrier
\section*{Chapter 6 — a plan that satisfies the plan contract}
\markboth{Chapter 6 — a plan that satisfies the plan contract}{Chapter 6 — a plan that satisfies the plan contract}

\textbf{Asked:} plan a second brief that adds web research, requiring a query strategy, source-quality criteria, a claim ledger, a source register, a contradiction pass, and explicit approval before any external publication — and do not execute until the plan satisfies all eight points of Section 6.2.

The point of the exercise is that you review the plan against the contract, item by item, rather than against your impression of it. Ask for the plan in the contract's own shape so the review is mechanical:

\begin{codeblock}{8.0}{9.4}
\cl{Plan~only~—~do~not~execute.~This~brief~extends~drafts/brief-v1.md~with~web~research.}
\cl{}
\cl{Return~the~plan~under~these~headings,~and~under~each~state~what~you~will~do~and~how~I~would}
\cl{know~it~was~done:~Outcome;~Inputs~and~assumptions;~Steps~in~order;~Tools~and~permissions}
\cl{required;~What~could~go~wrong~and~the~detection~for~each;~What~you~will~not~do;~Evidence}
\cl{produced;~Approval~gates.}
\cl{}
\cl{Requirements~the~plan~must~satisfy:}
\cl{-~A~query~strategy:~the~search~terms,~why~those~terms,~and~what~would~make~you~change~them.}
\cl{-~Source-quality~criteria~stated~before~searching,~not~chosen~after~seeing~results.}
\cl{-~A~claim~ledger:~one~row~per~claim,~with~the~source~that~supports~it~and~its~status.}
\cl{-~A~source~register:~one~row~per~source,~with~its~class,~date~and~why~it~qualifies.}
\cl{-~A~contradiction~pass:~an~explicit~step~that~looks~for~sources~that~disagree,~and~records}
\cl{~~the~disagreement~rather~than~resolving~it~silently.}
\cl{-~No~external~publication,~message,~commit~or~push~without~separate~approval.}
\end{codeblock}

Reviewing it, the two contract items most often failed are \emph{what could go wrong and the detection for each} — plans list risks without detections, which is a list of worries, not a control — and \emph{what you will not do}, which is usually absent because it is the only item that requires the plan to constrain itself. Send the plan back for those two rather than accepting a plan that is strong on the first six.

The source-quality criteria must be fixed before searching. Criteria chosen after seeing results are a rationalisation of the results, and the failure is invisible in the finished brief. State them concretely: primary or issuing-body sources preferred over commentary; publication or revision date required; a named author or institution; and for any standard or regulation, the issuing body's own text rather than a summary of it (“Scope and method”).

\textbf{Check.} Go through the eight contract points and mark each satisfied or not, in writing. If you cannot mark a point without re-reading the plan twice, that point is not satisfied — the contract is a checklist precisely so that satisfaction is legible at a glance. Then confirm the approval gate is real by checking that publication appears in the plan as a step \emph{you} perform, not as a step Claude performs after asking.

\FloatBarrier
\section*{Chapter 8 — a conservative \texttt{.\allowbreak{}gitignore}, explained pattern by pattern}
\markboth{Chapter 8 — a conservative \texttt{.\allowbreak{}gitignore}, explained pattern by pattern}{Chapter 8 — a conservative \texttt{.\allowbreak{}gitignore}, explained pattern by pattern}

\textbf{Asked:} initialise Git in the practice workspace, have a conservative \texttt{.\allowbreak{}gitignore} drafted with every pattern explained, show the staged diff, and commit only after confirming that nothing sensitive or unrelated is included.

Order matters: \texttt{.\allowbreak{}gitignore} must exist before the first commit, because \texttt{.\allowbreak{}gitignore} does not untrack what is already tracked (Section 8.1). A file committed once remains in history after you ignore it, and removing it from history is a rewrite, not a fix.

\begin{codeblock}{9.0}{10.6}
\cl{git~init}
\cl{\#~draft~.gitignore,~review~it,~and~only~then:}
\cl{git~add~-A}
\cl{git~status~-{}-porcelain~~~~~~~~~~\#~what~will~be~committed,~as~a~list}
\cl{git~diff~-{}-staged~~~~~~~~~~~~~~~\#~what~will~be~committed,~as~content}
\cl{git~commit~-m~"Initial~commit:~practice~workspace"}
\end{codeblock}

A conservative starting point for this workspace:

\begin{codeblock}{9.0}{10.6}
\cl{\#~local~secrets,~never~committed}
\cl{.env}
\cl{\#~per-environment~secrets:~.env.local,~.env.production}
\cl{.env.*}
\cl{\#~the~template~is~safe~and~belongs~in~the~repository}
\cl{!.env.example}
\cl{\#~machine-local~permission~overrides,~not~shared~policy}
\cl{.claude/settings.local.json}
\cl{\#~personal~project~notes,~not~team~instructions}
\cl{CLAUDE.local.md}
\cl{\#~anything~filed~here~is~excluded~by~location,~not~by~name}
\cl{secrets/}
\cl{\#~intermediate~output;~keep~the~finished~drafts}
\cl{drafts/*.tmp}
\cl{\#~macOS~directory~metadata}
\cl{.DS\_Store}
\cl{\#~build~products,~regenerable}
\cl{\_\_pycache\_\_/}
\cl{\#~logs~may~quote~transcripts,~tokens~or~paths}
\cl{*.log}
\end{codeblock}

Every comment sits on its own line, and that is not a matter of style. Git treats \texttt{\#\allowbreak{}} as a comment only at the start of a line; \texttt{.\allowbreak{}env  \#\allowbreak{} local secr\allowbreak{}ets} is a single pattern that includes the explanation, matches no file, and leaves \texttt{.\allowbreak{}env} unignored without any warning.

Two patterns deserve the explanation the exercise asks for. \texttt{!.\allowbreak{}env.\allowbreak{}example} is a negation, and its position matters: a negation cannot resurrect a file whose \emph{parent directory} is excluded, so \texttt{secrets/\allowbreak{}} plus \texttt{!secrets/\allowbreak{}example.\allowbreak{}json} does not work — the directory is never descended into. And \texttt{secrets/\allowbreak{}} excludes by location rather than by name, which is the more robust of the two strategies: a rule that depends on someone naming a file correctly fails the first time someone does not.

The instruction to have every pattern explained is not politeness. An unexplained \texttt{.\allowbreak{}gitignore} is usually copied from a template for a different stack, and it will either exclude something you need or fail to exclude something you have. Ask for the explanation in the file itself, as above, so it survives into the repository.

\textbf{Check.} Before committing, run \texttt{git diff -\allowbreak{}-\allowbreak{}staged -\allowbreak{}-\allowbreak{}stat} and read the file list end to end. Ask of each entry: do I know why this is here? Then run a negative test — create \texttt{.\allowbreak{}env} with a dummy value and confirm \texttt{git status -\allowbreak{}-\allowbreak{}porcelain} does not list it. Finally, confirm push is not part of the flow: push is a separate, visible step that a human owns and the agent is not authorised to perform (Section 8.2).

\FloatBarrier
\section*{Chapter 9 — a \texttt{CLAUDE.\allowbreak{}md} that is an index, not a warehouse}
\markboth{Chapter 9 — a \texttt{CLAUDE.\allowbreak{}md} that is an index, not a warehouse}{Chapter 9 — a \texttt{CLAUDE.\allowbreak{}md} that is an index, not a warehouse}

\textbf{Asked:} create the project \texttt{CLAUDE.\allowbreak{}md}, start a new session, and have Claude state the directory rules without searching every file. Then determine which instructions are stable and which belong in a path-scoped rule or a skill.

The exercise tests one property: whether the file is an index into the workspace or a container for everything you know about it (Section 9.4). A warehouse file is loaded in full at every session start and consumes context that the task needs.

\begin{codeblock}{9.0}{10.6}
\cl{\#~Practice~research~workspace}
\cl{}
\cl{\#\#~Layout}
\cl{-~\textasciigrave{}notes/\textasciigrave{}~—~source~claims.~Read-only.~Never~edited~by~an~agent.}
\cl{-~\textasciigrave{}sources/\textasciigrave{}~—~primary~evidence,~PDFs.~Read-only.}
\cl{-~\textasciigrave{}drafts/\textasciigrave{}~—~the~only~writable~directory.~Output~goes~here.}
\cl{}
\cl{\#\#~Rules}
\cl{-~Cite~by~file~and~line~for~notes,~by~file~and~page~for~sources.}
\cl{-~Record~an~unsupported~claim~as~UNSUPPORTED;~do~not~resolve~it~by~inference.}
\cl{-~No~network~access~unless~the~task~brief~grants~it~explicitly.}
\cl{-~Publication,~commit~and~push~are~human~actions,~not~steps~in~a~plan.}
\cl{}
\cl{\#\#~Where~to~look}
\cl{-~Claim-to-source~schema:~\textasciigrave{}drafts/SCHEMA.md\textasciigrave{}}
\cl{-~Source-quality~criteria:~\textasciigrave{}drafts/CRITERIA.md\textasciigrave{}}
\cl{-~Prior~briefs~and~their~review~notes:~\textasciigrave{}drafts/archive/\textasciigrave{}}
\end{codeblock}

The three "where to look" pointers are what make it an index. The schema and criteria are stable documents that change rarely and are needed only by tasks that produce briefs; inlining them costs context in every session, including the ones that never write a brief.

Verify in a clean session, without leading the answer:

\begin{codeblock}{8.0}{9.4}
\cl{Without~reading~the~contents~of~notes/~or~sources/,~state~the~rules~that~apply~to~this}
\cl{workspace~and~where~each~one~comes~from.}
\end{codeblock}

If the answer arrives immediately and cites \texttt{CLAUDE.\allowbreak{}md}, the file loaded and reads as instructions. That is all the recital shows. It does not show that the rules will be followed under pressure from a task, and it is no evidence that anything enforces them: confirm loading with \texttt{/\allowbreak{}context all}, and test a rule that matters by asking for the thing it forbids. If Claude searches first, the rules are either not in the file or are buried below the point where they read as instructions.

The second half of the exercise is the classification. A rule is stable if it would still be true after the current task changes; it belongs in \texttt{CLAUDE.\allowbreak{}md}. A rule that applies only inside a subtree belongs in a path-scoped rule so it loads only when that subtree is in play. A rule that is really a \emph{procedure} — a sequence you repeat with variation — is a skill, not an instruction, and leaving it in \texttt{CLAUDE.\allowbreak{}md} both wastes context and makes it probabilistic (Chapter 15).

\begingroup
\def\tblrows{%
Directory layout and write boundary & Yes & \texttt{CLAUDE.\allowbreak{}md} \\
Citation format for notes and sources & Yes & \texttt{CLAUDE.\allowbreak{}md} \\
"Never edit a PDF in place" & Yes, but scoped & Path-scoped rule on \texttt{sources/\allowbreak{}} \\
The eight-step brief procedure & It is a procedure & A skill (Section 15.1) \\
"Use British spelling in drafts" & Yes, but scoped & Path-scoped rule on \texttt{drafts/\allowbreak{}} \\
}%
\def\tblbody{\begin{minipage}{\textwidth}
{\footnotesize\begin{tabular}{L{72.0mm}L{24.7mm}L{43.1mm}}
\toprule
\textbf{Instruction} & \textbf{Stable} & \textbf{Belongs in} \\
\midrule
\tblrows
\bottomrule\end{tabular}}\end{minipage}}%
\begingroup
\def\sloppy{\tolerance 9999\emergencystretch 3em\hfuzz 200pt\vfuzz 200pt}%
\hbadness=10000\vbadness=10000\hfuzz=200pt\vfuzz=200pt
\global\setbox\tblbox=\hbox{\tblbody}%
\endgroup
\par\addvspace{2.6mm}
\ifdim\dimexpr\ht\tblbox+\dp\tblbox\relax>0.55\textheight
  \typeout{HANDBOOK-TABLE broken \the\dimexpr\ht\tblbox+\dp\tblbox\relax}%
  
  {\footnotesize\begin{longtable}{L{72.0mm}L{24.7mm}L{43.1mm}}
  \toprule
\textbf{Instruction} & \textbf{Stable} & \textbf{Belongs in} \\
\midrule\endfirsthead
  \multicolumn{3}{@{}l@{}}{%
  \sffamily\footnotesize\itshape\color{inkgrey}Continued}\\[1.2mm]
  \toprule
\textbf{Instruction} & \textbf{Stable} & \textbf{Belongs in} \\
\midrule\endhead
  \bottomrule\endfoot
  \bottomrule\endlastfoot
  \tblrows
  \end{longtable}}%
\else
  \typeout{HANDBOOK-TABLE atomic \the\dimexpr\ht\tblbox+\dp\tblbox\relax}%
  \noindent\tblbody
\fi
\par\addvspace{2.6mm}
\endgroup

\textbf{Check.} Two measurements. Count the lines: a project file that has grown past roughly two hundred lines has almost certainly absorbed something that belongs elsewhere (Section 9.2). Then run \texttt{/\allowbreak{}context} in a fresh session and look at what the instruction load costs before any work has been done. If the project file is a visible fraction of the window at turn zero, move the stable-but-narrow material into path-scoped rules and the procedures into skills.

\FloatBarrier
\section*{Chapter 11 — a handoff that survives a clean session}
\markboth{Chapter 11 — a handoff that survives a clean session}{Chapter 11 — a handoff that survives a clean session}

\textbf{Asked:} after drafting the brief, run \texttt{/\allowbreak{}context}, save a handoff, start a clean session, and have the new session continue — verifying that it can name completed work, open issues and the next check without receiving the old transcript.

This is the exercise that distinguishes context \emph{engineering} from context \emph{hoarding}. The test is not whether the new session can be brought up to speed; it is whether the handoff you wrote carries the state, or whether you unconsciously relied on the transcript.

Run \texttt{/\allowbreak{}context} first and record the numbers, because the comparison afterwards is the finding (Section 11.1). Then ask for the handoff with a schema, not as a summary:

\begin{codeblock}{8.0}{9.4}
\cl{Write~drafts/HANDOFF.md~for~a~session~that~will~have~none~of~this~conversation.~Include:}
\cl{}
\cl{1.~Goal~—~the~outcome~in~one~sentence,~in~the~terms~of~the~original~brief.}
\cl{2.~Done~—~completed~work,~each~item~with~the~artefact~that~evidences~it.}
\cl{3.~Not~done~—~remaining~work,~in~the~order~it~should~be~attempted.}
\cl{4.~Open~issues~—~decisions~I~must~make,~each~with~the~options~and~what~each~would~cost.}
\cl{5.~Rejected~—~approaches~already~tried~and~abandoned,~with~the~reason.}
\cl{6.~Next~check~—~the~single~next~verification,~and~what~its~passing~would~prove.}
\cl{7.~Boundaries~—~the~constraints~in~force,~restated~in~full.}
\cl{}
\cl{Write~no~history.~A~new~session~must~not~need~to~know~what~happened,~only~what~is~true.}
\end{codeblock}

Item 5 is the one that is routinely omitted and the one that pays for the file. Without it the new session re-proposes a rejected approach, and you spend the saved context re-rejecting it. Item 7 is restated in full rather than referenced because a boundary that lives only in the old transcript does not survive the handoff — which is precisely the failure the exercise induces.

Start the clean session and ask it to continue from \texttt{drafts/\allowbreak{}HANDOFF.\allowbreak{}md} alone. Do not paste the brief, do not summarise, do not correct it until it has committed to a next step.

\textbf{Check.} Three answers from the new session, before you supply anything: what is done and what evidences it; what decision is currently blocking; what is the next check and what would its passing prove. Any of the three that requires you to fill in a gap identifies a missing section in the handoff. Run \texttt{/\allowbreak{}context} again in the new session and compare with the number you recorded: a handoff that reproduces the working state at a fraction of the context cost is the demonstration the exercise is after.

\FloatBarrier
\section*{Chapter 13 — a threat model before an MCP connection}
\markboth{Chapter 13 — a threat model before an MCP connection}{Chapter 13 — a threat model before an MCP connection}

\textbf{Asked:} without installing anything, produce a threat model for a hypothetical Notion connection covering data reachable, write operations exposed, credential storage, scope, approval gates, logging and revocation — then replace each assumption with a statement from the provider's documentation before approving installation.

The exercise is deliberately ordered so that the model is written while you still have no stake in the answer. A threat model written after installation is a justification.

Ask for it as a table with an explicit assumption column, because the assumptions are the deliverable:

\begin{codeblock}{8.0}{9.4}
\cl{Do~not~install~anything.~Produce~a~threat~model~for~connecting~a~Notion~MCP~server~to~this}
\cl{workspace,~as~a~table~with~the~columns:~Question~|~Current~answer~|~Is~this~an~assumption~or}
\cl{a~documented~fact~|~Source.~Cover:~what~data~becomes~reachable;~which~write~operations~are}
\cl{exposed;~where~the~credential~is~stored~and~what~can~read~it;~what~scope~the~credential}
\cl{carries;~which~operations~require~approval~and~which~do~not;~what~is~logged~and~where;~and}
\cl{how~access~is~revoked~and~how~I~would~confirm~revocation~took~effect.}
\cl{}
\cl{Mark~every~row~you~cannot~support~from~the~provider\textquotesingle{}s~own~documentation~as~ASSUMPTION.}
\cl{Do~not~fill~a~gap~by~inference.}
\end{codeblock}

The rows that most often come back as assumptions, and which matter most:

\begin{itemize}
\item \textbf{Scope.} Integration tokens frequently carry workspace-wide reach rather than page-level reach, and "connected to a page" describes what you selected in a picker, not what the token permits.
\item \textbf{Write operations.} Read-only is not a property of the connector's name, nor of the token alone: it is what remains after the token's scopes, the provider's roles, the tools the server exposes and your own permission rules have each been applied (Section 13.3).
\item \textbf{Revocation.} Revoking in the provider's console and removing the server locally are two different operations, and either alone leaves the other in place.
\item \textbf{Logging.} Whether the provider records the agent's reads, and whether you can retrieve that record, are separate questions.
\end{itemize}

Then do the second half, which is the actual work: replace each ASSUMPTION with a quotation from the provider's documentation, with its URL and the date you read it. Where the documentation does not answer the question, the row stays ASSUMPTION — an unanswered question recorded as unanswered is a finding, and treating silence as permission is the failure mode this book returns to repeatedly (Section 13.1).

An MCP server is untrusted input as well as a capability: its tool descriptions and its returned content both reach the model, so a compromised or careless server is an injection surface, not only a data-access surface (Section 13.4).

\textbf{Check.} Count the rows still marked ASSUMPTION. If any of scope, write operations or revocation remains an assumption, the connection is not ready to approve regardless of how useful it would be. Then state, in one sentence and without looking, what the worst outcome of a compromised token would be for this workspace. If you cannot, the model has been produced but not read.

\FloatBarrier
\section*{Chapter 14 — a canary, a deny rule, and the edge it leaves open}
\markboth{Chapter 14 — a canary, a deny rule, and the edge it leaves open}{Chapter 14 — a canary, a deny rule, and the edge it leaves open}

\textbf{Asked:} put a synthetic canary in \texttt{.\allowbreak{}env}, deny \texttt{Read(.\allowbreak{}/\allowbreak{}.\allowbreak{}env)}, and in Manual mode with the sandbox off confirm that Claude's file tool is refused; then have Claude run a script that opens the file itself; record what was blocked, by what, and what the transcript holds afterwards; then turn the sandbox on and repeat.

The exercise has two halves because the control has two sides. The first half shows the rule working. The second shows where it stops, and the second is the one that changes how you deploy it. The conditions matter: start with the sandbox \textbf{off}, in Manual mode, with no allow rule that would approve the script, and with session transcripts being saved as they are by default.

\begin{codeblock}{9.0}{10.6}
\cl{\{}
\cl{~~"permissions":~\{}
\cl{~~~~"deny":~["Read(./.env)",~"Read(./.env.*)"]}
\cl{~~\}}
\cl{\}}
\end{codeblock}

The first request — read \texttt{.\allowbreak{}env} — is refused, and \texttt{/\allowbreak{}permissions} names the settings file the rule came from. Asking for \texttt{cat .\allowbreak{}env} is refused as well, because \texttt{cat} is one of the file commands Claude Code recognises in Bash (Section 7.2).

The second request is a three-line script that opens \texttt{.\allowbreak{}env} and prints it. You are prompted for the command, and the prompt shows a script name, not a file read. If you approve it, the canary is printed: with no sandbox, a \texttt{Read} deny rule does not apply to a program that opens the file itself. Nothing malfunctioned. The rule matches Claude's file tools and the commands it can recognise, and a script is neither.

Now search the session transcript for the canary. It is there, in plaintext, in the tool result — and it will stay there for the retention period (Section 12.3). That is the reason the chapter's incident procedure rotates first: by the time you are reading the transcript the value has already been copied.

Close the edge. Turn the sandbox on with \texttt{/\allowbreak{}sandbox}, leave its filesystem isolation enabled, and run the same script again. It now fails to open the file, and you have changed nothing else: Claude Code adds the paths of your \texttt{Read} deny rules to the sandbox's own read-deny list, and the operating system refuses that read to every sandboxed process whatever its name (Section 7.4). Notice also that you were probably not prompted this time, because a command that runs inside the sandbox is approved automatically. An entry under \texttt{sandbox.\allowbreak{}credentials.\allowbreak{}files} with \texttt{"mode":\allowbreak{} "deny"} produces the same block and keeps credential rules in one place; to see it act on its own, repeat the run with the permission rule removed.

\textbf{Check.} You should be able to state three things without looking: which control stopped the file tool, which control stopped the script, and which process on your machine would still be able to read the file with both in place. If the answer to the third is "none", read Table 12 again: a hook, a local MCP server, a command you type yourself and an unsandboxed retry all run outside the sandbox.

\FloatBarrier
\section*{Chapter 15 — a skill improved only by rules that generalise}
\markboth{Chapter 15 — a skill improved only by rules that generalise}{Chapter 15 — a skill improved only by rules that generalise}

\textbf{Asked:} install the example skill locally, invoke \texttt{/\allowbreak{}research-\allowbreak{}brief} on a second source set, record every place it made an assumption, and improve the skill only with a rule that generalises.

The constraint in the last clause is the whole exercise. The natural response to a bad output is to add a sentence forbidding exactly that output, and a skill maintained that way becomes a list of past failures that grows without ever getting better.

Run it on a second source set — one with a different shape from the first, ideally with a missing date, a source that contradicts another, and one document that is not what its filename says. Record assumptions as they happen, in a file, with the evidence:

\begin{codeblock}{7.0}{8.3}
\cl{|~\#~|~Where~it~assumed~|~What~it~assumed~|~What~it~should~have~done~|}
\cl{|-{}-{}-|-{}-{}-|-{}-{}-|-{}-{}-|}
\cl{|~1~|~sources/report.pdf~|~that~the~undated~file~was~current~|~recorded~the~date~as~unknown~|}
\cl{|~2~|~two~disagreeing~notes~|~picked~the~more~recent~one~silently~|~recorded~the~disagreement~|}
\cl{|~3~|~a~filename~ending~.md~|~that~the~content~matched~the~name~|~checked~before~classifying~|}
\end{codeblock}

Now apply the test. For each row, ask: is the fix a rule about \emph{this input}, or a rule about a \emph{class of input}? Row 1's specific fix — "treat report.pdf as undated" — generalises to a rule about missing metadata. Row 2's specific fix — "prefer the note from August" — is a rule about two files and generalises to a rule about contradiction. Row 3 generalises to a rule about trusting declared type over observed content.

The generalised rules, which is what actually goes into the skill:

\begin{codeblock}{7.0}{8.3}
\cl{\#\#~Evidence~rules}
\cl{-~A~source~with~no~determinable~date~is~recorded~as~date-unknown~and~may~not~be~used~to}
\cl{~~establish~precedence~over~a~dated~source.}
\cl{-~When~two~sources~disagree,~record~the~disagreement~in~the~claim~ledger~with~both~positions.}
\cl{~~Do~not~resolve~it~by~recency,~by~source~order,~or~by~which~is~easier~to~summarise.}
\cl{-~Classify~a~source~by~inspecting~its~content,~not~by~its~filename~or~extension.~Where~the~two}
\cl{~~conflict,~record~the~conflict.}
\end{codeblock}

Each of those would have prevented the observed failure \emph{and} prevents failures you have not seen yet. That is the test for whether a rule belongs in a skill at all: a rule that can only fire on the input that produced it belongs in the task brief, not in the reusable procedure.

Keep the skill's frontmatter description precise while you are at it. A skill is selected by its description, so a vague one is either never invoked or invoked for the wrong task (Section 15.3).

\textbf{Check.} Do not re-run in the session where you edited the skill. The rendered skill stays in a conversation for the rest of the session (Section 15.3), so that session now holds the old instructions, the new ones and your discussion of the difference, and a run there tests all three at once. Start a fresh session for each run, keep the model, settings and brief the same, and run the revised skill on three inputs: the case that failed, the \emph{first} source set, and one it has never seen. Compare named properties rather than wording — was the undated source recorded as date-unknown, was the disagreement recorded with both positions. On the first source set those properties should be unchanged except where a new rule legitimately applies; a rule that alters the first result in some unrelated way has changed the procedure rather than closed a gap.

Then count: if the skill grew by more lines than the number of generalised rules you identified, something specific has probably been smuggled in. That is a prompt to look, not a proof. And check what the edit did not touch — the tool grants, any hooks, and where the skill writes — because a skill can pass every test of its prose and still have been given more authority than the procedure needs.

\FloatBarrier
\section*{Chapter 17 — a project-only \texttt{Stop} hook that records completion}
\markboth{Chapter 17 — a project-only \texttt{Stop} hook that records completion}{Chapter 17 — a project-only \texttt{Stop} hook that records completion}

\textbf{Asked:} create a project-only \texttt{Stop} hook that appends a timestamped, non-sensitive completion record to a local file, making no network call, approving no permission, and including no conversation content. Test success, failure and removal.

Follow the creation sequence rather than writing the configuration first: define the event and matcher, write a standalone script with deterministic input and output, test it against benign fixtures by hand, and only then wire it in (Section 17.4).

The script, tested on its own before Claude Code ever calls it:

\begin{codeblock}{8.0}{9.4}
\cl{\#!/usr/bin/env~bash}
\cl{\#~.claude/hooks/record-stop.sh~—~append~a~completion~record.~No~network.~No~content.}
\cl{set~-euo~pipefail}
\cl{LOG="\$\{CLAUDE\_PROJECT\_DIR\}/drafts/session-log.tsv"}
\cl{mkdir~-p~"\$(dirname~"\$LOG")"}
\cl{printf~\textquotesingle{}\%s\textbackslash{}t\%s\textbackslash{}t\%s\textbackslash{}n\textquotesingle{}~\textbackslash{}}
\cl{~~"\$(date~-u~+\%Y-\%m-\%dT\%H:\%M:\%SZ)"~\textbackslash{}}
\cl{~~"stop"~\textbackslash{}}
\cl{~~"\$(basename~"\$\{CLAUDE\_PROJECT\_DIR\}")"~>{}>~"\$LOG"}
\cl{exit~0}
\end{codeblock}

Three deliberate omissions. It reads nothing from standard input, so no conversation content can reach the log by accident — the requirement is met structurally rather than by care. It writes a fixed three-field record, so the log cannot grow a field that later carries something sensitive. And its last line is \texttt{exit 0}, so a successful run can never be read as a block. That is not the same as exiting 0 unconditionally: under \texttt{set -\allowbreak{}euo pipefai\allowbreak{}l} a failing \texttt{mkdir} or an unwritable log ends the script with another status before that line is reached. For this hook that is the right failure, because any status other than 0 or 2 is a non-blocking error which Claude Code reports and then ignores [83]. What the script must never do is exit 2: on \texttt{Stop} that sends Claude back to keep working instead of letting the turn end, which turns a logging failure into a turn that will not finish.

What a row in the log proves is narrow. \texttt{Stop} fires when the main agent finishes a turn, so a row shows that a turn ended and that the hook ran. It is not evidence that the task was completed, or that the session was.

The configuration, in the project's \texttt{.\allowbreak{}claude/\allowbreak{}settings.\allowbreak{}json} so that it is project-only and visible in review:

\begin{codeblock}{9.0}{10.6}
\cl{\{}
\cl{~~"hooks":~\{}
\cl{~~~~"Stop":~[}
\cl{~~~~~~\{}
\cl{~~~~~~~~"hooks":~[}
\cl{~~~~~~~~~~\{}
\cl{~~~~~~~~~~~~"type":~"command",}
\cl{~~~~~~~~~~~~"command":~"bash",}
\cl{~~~~~~~~~~~~"args":~["\$\{CLAUDE\_PROJECT\_DIR\}/.claude/hooks/record-stop.sh"],}
\cl{~~~~~~~~~~~~"timeout":~10,}
\cl{~~~~~~~~~~~~"statusMessage":~"Recording~completion…"}
\cl{~~~~~~~~~~\}}
\cl{~~~~~~~~]}
\cl{~~~~~~\}}
\cl{~~~~]}
\cl{~~\}}
\cl{\}}
\end{codeblock}

\texttt{\$\{CLAUDE\_\allowbreak{}PROJECT\_\allowbreak{}DIR\}} rather than a relative path, because it stays at the project root even inside a worktree; and \texttt{bash} with the script in \texttt{args} for exec form, in which the path needs no quoting even when it contains a space, and which native Windows can spawn only through a real executable such as \texttt{bash} on \texttt{PATH} [83]. A relative path fails another way: it resolves against a working directory you do not control (Section 17.2).

Now the three tests the exercise names.

\emph{Success.} Run any short task, let it finish, and inspect \texttt{drafts/\allowbreak{}session-\allowbreak{}log.\allowbreak{}tsv}. One new row, three fields, no prose.

\emph{Failure.} Make the script fail deterministically — point \texttt{LOG} at an unwritable path — and run another task. Confirm the turn still ends and the failure is visible rather than silent. This is the test people skip, and it is the one that matters: a logging hook that fails closed will eventually stop your work, and one that fails invisibly will leave you believing you have a record you do not have.

\emph{Removal.} Undo the deliberate failure first, so that a hook which has merely stopped working is not mistaken for one that has gone. Then delete the \texttt{Stop} block, start a new session, and confirm with \texttt{/\allowbreak{}hooks} that no Stop hook is configured — then run a task and confirm no new row appears. Verifying removal from the configuration file alone is insufficient, because hooks arrive from several sources and settings precedence decides which one is live (Section 17.5).

\textbf{Check.} Read the log file as an adversary would. Does any field carry a path, a prompt, a filename or a token? Then confirm the negative properties directly rather than by inspection of intent: the script contains no \texttt{curl}, \texttt{wget} or redirect to a URL, and returns no \texttt{permission\allowbreak{}Decision}. A hook that \emph{can} approve a permission is privileged code, and this one must not be able to.

\FloatBarrier
\section*{Chapter 20 — a bounded browser task, verified from outside the transcript}
\markboth{Chapter 20 — a bounded browser task, verified from outside the transcript}{Chapter 20 — a bounded browser task, verified from outside the transcript}

\textbf{Asked:} on a harmless public page or a local fixture, extract a small table with its location cited and take no action; then navigate to one linked page; then inspect browser history and the transcript to confirm the exact scope of what happened.

Use a local fixture if you have one — it removes the network from the first half of the exercise entirely, and it lets you know the ground truth of what the page contains. Whatever you use, the browser profile is the control surface that matters: a browsing agent operating in your everyday profile is operating with your logged-in sessions (Section 20.2).

The two-stage prompt, kept deliberately separate so that the second stage is a decision:

\begin{codeblock}{8.0}{9.4}
\cl{Stage~1.~Read~the~table~on~this~page.~Return~it~as~markdown,~and~for~each~row~state~where~on}
\cl{the~page~it~came~from.~Take~no~action:~no~click,~no~form~entry,~no~navigation,~no~download.}
\cl{}
\cl{Stage~2.~(Send~only~after~reviewing~stage~1.)~Navigate~to~the~single~link~labelled~"…"~and}
\cl{report~the~page~title~and~heading~structure.~Do~not~follow~any~further~link.}
\end{codeblock}

The verification is the point of the exercise, and it must come from outside the transcript, because the transcript is a record of what was reported, not of what occurred:

\begin{codeblock}{9.0}{10.6}
\cl{\#~Chrome~history,~read~from~a~copy~—~the~live~database~is~locked~while~Chrome~runs.}
\cl{cp~"\$HOME/.config/google-chrome/Default/History"~/tmp/h.db}
\cl{sqlite3~/tmp/h.db~\textbackslash{}}
\cl{~~"SELECT~datetime(v.visit\_time/1000000-11644473600,\textquotesingle{}unixepoch\textquotesingle{}),~u.url}
\cl{~~~FROM~visits~v~JOIN~urls~u~ON~u.id~=~v.url}
\cl{~~~ORDER~BY~v.visit\_time~DESC~LIMIT~20;"}
\end{codeblock}

Adjust the profile path for your platform. The query reads \texttt{visits}, one row per visit, not \texttt{urls}, which folds repeat visits to one address into a single row. Compare the visits since the exercise began against the transcript: exactly one navigation should appear for stage 2, and none for stage 1. Extra entries are the finding — commonly a redirect, a prefetch, or a link that resolved somewhere other than its label.

Page content is untrusted input. Text on a page can address the agent directly, and an instruction that arrives inside fetched content is indistinguishable in form from one you wrote (Section 20.5). Stage 1 asks for extraction and nothing else precisely so that a page attempting to induce an action produces a visible refusal rather than a quiet compliance.

\textbf{Check.} Three comparisons. The extracted table against the page — is every row present and correctly located, or has a merged cell been silently flattened? The history against the transcript — same count, same URLs. And the transcript against the stage boundary — did anything in stage 1 attempt an action, and if so, what on the page induced it?

\FloatBarrier
\section*{Chapter 22 — a small workflow, and whether it earned its cost}
\markboth{Chapter 22 — a small workflow, and whether it earned its cost}{Chapter 22 — a small workflow, and whether it earned its cost}

\textbf{Asked:} run \texttt{/\allowbreak{}deep-\allowbreak{}research} on a non-sensitive topic with a \texttt{small} size guideline, requiring direct sources, a contradiction pass and a source register; compare the result with a single-agent attempt; and use \texttt{/\allowbreak{}usage} to decide whether the workflow added decision value or only cost.

Set the guideline before the run, not during it — a change takes effect on the next prompt, and \texttt{small} asks Claude for fewer than five agents, which is advice rather than a cap (Section 22.2). Set \texttt{workflow\allowbreak{}Size\allowbreak{}Guideline} to \texttt{small} in \texttt{/\allowbreak{}config} or in a settings file. \texttt{/\allowbreak{}deep-\allowbreak{}research} also needs the WebSearch tool to be available.

Then commit a clean baseline. This is not general hygiene; it is specific to workflows. The subagents a workflow spawns use your permission rules and take their mode by the rules for any subagent [19], [27]. Under a session in \texttt{bypass\allowbreak{}Permissions}, \texttt{accept\allowbreak{}Edits} or \texttt{auto} they run in that mode; under \texttt{plan} they stay in plan mode, and a \texttt{dont\allowbreak{}Ask} agent there is denied what plan mode would ask; under Manual or \texttt{dont\allowbreak{}Ask} an agent type's own \texttt{permission\allowbreak{}Mode} applies, unless it asks for \texttt{bypass\allowbreak{}Permissions}, or for \texttt{auto} where auto mode is unavailable, and an unset field inherits the session's mode. Version 1.0 said they always run in \texttt{accept\allowbreak{}Edits}, which the page no longer states. In auto mode, the default when nothing configures a mode, the agents' edits are approved without asking you.

\begin{codeblock}{9.0}{10.6}
\cl{git~add~-A~\&\&~git~commit~-m~"Baseline~before~workflow~run"}
\end{codeblock}

Pick a topic with genuine disagreement in the literature, because a contradiction pass over a settled topic proves nothing. Then:

\begin{codeblock}{8.0}{9.4}
\cl{/deep-research~What~is~the~current~evidence~on~the~effectiveness~of~<topic>?~Requirements:}
\cl{cite~issuing~bodies~and~primary~studies~rather~than~commentary~about~them;~produce~a~source}
\cl{register~with~one~row~per~source~giving~its~class,~date~and~why~it~qualifies;~and~run~an}
\cl{explicit~contradiction~pass~that~reports~sources~which~disagree~and~states~the~disagreement}
\cl{rather~than~resolving~it.~Where~the~evidence~does~not~settle~a~question,~say~so.}
\end{codeblock}

Watch \texttt{/\allowbreak{}workflows} while it runs. Then run the same question as a single agent in a fresh session, with the same requirements, and put the two outputs side by side.

The comparison has a specific shape. Ask three questions of the pair, in this order. Did the workflow surface a source the single agent did not, or the same sources organised differently? Did the contradiction pass find a real disagreement, or manufacture one by quoting two sources that are not in conflict? And would a decision-maker choose differently given one output rather than the other? Only the third is decisive. The first two can both be yes while the answer to the third is no, and that is the outcome the exercise is designed to make visible.

Then read \texttt{/\allowbreak{}usage} for the session (Section 4.4). Divide the cost by the answer to the third question. A workflow that produced a better-organised version of the same conclusion at several times the cost did not add decision value; it added confidence, which is not the same thing and is sometimes worse than nothing.

If you stop the run mid fan-out and relaunch it, note what reruns: an agent that was still running starts over and completed agents keep their results, but stopping a \emph{single} agent with \texttt{x} counts as failing it, and every agent started after a failed one runs again even if it had completed (Section 22.3). It is worth doing both once, deliberately, on a small run, so that the difference is a measurement rather than a surprise.

\textbf{Check.} Write two sentences before looking at the cost: what the workflow told you that the single agent did not, and what decision that changes. Then look at \texttt{/\allowbreak{}usage}. If the two sentences were hard to write, the run was an exercise in throughput rather than in research — which is a legitimate finding, and the one this exercise most often produces.

\FloatBarrier
\section*{Chapter 25 — a read-only status report from a second device}
\markboth{Chapter 25 — a read-only status report from a second device}{Chapter 25 — a read-only status report from a second device}

\textbf{Asked:} start Remote Control on the practice workspace, request a read-only status report from a second device, confirm the output appears locally, then disconnect — granting no new permissions during the exercise.

The clause that carries the lesson is the last one. Remoteness creates pressure to widen access — the review surface is worse, so the temptation is to approve rather than inspect. The exercise exists to establish, once, that you can complete useful remote work without loosening anything (Section 25.1).

Start from the practice directory and name the session so that a phone cannot confuse it with another project:

\begin{codeblock}{9.0}{10.6}
\cl{cd~\textasciitilde{}/practice-workspace}
\cl{claude~-{}-remote-control~research-brief}
\end{codeblock}

Confirm before connecting the second device: correct machine, correct directory, correct branch, correct account. On a small screen these are the four facts you will not otherwise re-establish, and the phone shows a conversation, not a working directory.

From the second device, ask for something that cannot require a permission decision:

\begin{codeblock}{7.0}{8.3}
\cl{Read-only~status~report.~Do~not~edit,~create~or~delete~any~file,~and~do~not~run~any~command}
\cl{that~writes.~Report:~current~branch~and~whether~the~tree~is~clean;~which~files~in~drafts/}
\cl{changed~most~recently~and~when;~any~claim~in~drafts/brief-v1.md~still~marked~UNSUPPORTED;~and}
\cl{the~single~next~check~from~drafts/HANDOFF.md.~If~anything~you~would~need~requires~a~write~or~a}
\cl{new~permission,~stop~and~say~what~it~was.}
\end{codeblock}

The final sentence is the instrument. If the report comes back complete with no such statement, the task genuinely stayed read-only. If it comes back with a request, you have learned where the read-only boundary actually sits for this workspace — and the correct response during this exercise is to decline it and note it, not to approve it from a phone.

Confirm the output appears in the local terminal as well as on the device. Execution stays local: commands run and files change on the machine, and what travels is the conversation, tool activity included, relayed and stored while connected so it can sync across devices. Two consequences follow that are easy to miss — the local process must keep running, because closing it takes the session offline even though the mobile client stays open; and the relayed transcript is cloud-stored data for as long as you are connected, so anything you would not put in a cloud-stored conversation does not belong in a remote session (Chapter 32).

Then disconnect deliberately rather than by closing a lid: run \texttt{/\allowbreak{}remote-\allowbreak{}control} again and disconnect from its status panel, which leaves the local session running (Section 25.1). Confirm the local session's state is what you expect afterwards.

\textbf{Check.} Four confirmations. The report's facts match what you observe locally — branch, cleanliness, file times. \texttt{git status -\allowbreak{}-\allowbreak{}porcelain} is unchanged from before the exercise. No permission was granted: no approval prompt was accepted during the exercise. A settings-file diff finds only persisted rules; a one-time or session-only approval, including any file-modification approval, is never written to a file [29]. Where OpenTelemetry is on, \texttt{tool\_\allowbreak{}decision} events with a \texttt{user\_\allowbreak{}temporary} or \texttt{user\_\allowbreak{}permanent} source show any that were [108]. And the session is actually closed, verified on the machine rather than by the absence of the session on the phone.

\FloatBarrier
\section*{A note on the solutions}
\markboth{A note on the solutions}{A note on the solutions}

Most of the fourteen solutions end in a verification performed \emph{outside} the transcript — a filesystem check, a browser history, a \texttt{git status}, a settings file or a produced artefact read directly. That is not a stylistic preference. The transcript records what was reported, and every failure mode this book treats as serious is one in which the report and the reality differ. Where an exercise can be marked complete from the conversation alone, it is testing comprehension; where it requires an external check, it is testing the practice. The second kind is the one worth repeating on real work.

\FloatBarrier
\setcounter{chapter}{13}
\setcounter{section}{0}
\appendixchapter{N}{Cross-reference matrix and index}
\FloatBarrier
\section*{Chapter dependency matrix}
\markboth{Chapter dependency matrix}{Chapter dependency matrix}

\begingroup
\def\tblrows{%
1 & What Claude Code is, and what it is not & 5, 18, 21 \\
2 & Install, verify and authenticate & — \\
3 & Surfaces, projects and boundaries & — \\
4 & Modes, models, effort and usage & 7, 16, 22 \\
5 & Start and steer a session & 1, 7, 16 \\
6 & Plan, comment, approve, execute & 7, 20 \\
7 & Permissions, sandboxing and checkpoints & 8, 12, 14, 17, 18, 22, 31 \\
8 & Protect work with Git and GitHub & 17 \\
9 & Build an effective \texttt{CLAUDE.\allowbreak{}md} & 7, 17 \\
10 & Global instructions and auto memory & 15, 17 \\
11 & Engineer context deliberately & 13, 15 \\
12 & Sessions: resume, name, retain & 10, 23, 29, 32 \\
13 & Connect external tools with MCP & 31 \\
14 & Environment variables and secrets & 7, 12, 33 \\
15 & Turn repeatable work into skills & — \\
16 & Delegate bounded work to subagents & 15, 22, 30 \\
17 & Event-driven behaviour with hooks & 7, 18, 21 \\
18 & Evaluate and install plugins & 7, 17, 19, 22, 31 \\
19 & Output styles: Claude Code beyond software engineering & 17, 18, 34 \\
20 & Use Claude in Chrome safely & 7, 14 \\
21 & Keep work moving with goals & — \\
22 & Dynamic workflows and ultracode & 4 \\
23 & Session-scoped scheduling with \texttt{/\allowbreak{}loop} & — \\
24 & Cloud routines and desktop scheduled tasks & — \\
25 & Continue local work with Remote Control & 32 \\
26 & Cloud sessions, \texttt{-\allowbreak{}-\allowbreak{}cloud} and \texttt{-\allowbreak{}-\allowbreak{}teleport} & 31, 32 \\
27 & Push events into a session with channels & — \\
28 & Publish work as artifacts & — \\
29 & Parallel sessions without collisions & 12, 16, 22, 30 \\
30 & Agent teams & 22, 29, 34 \\
31 & Enterprise configuration and managed policy & 7, 13, 18, 21, 32 \\
32 & Data governance, residency and retention & 10, 12, 14, 30, 31 \\
33 & Observability, cost control and accessibility & 11, 16, 19, 22, 30, 31 \\
34 & Build a verified research-brief agent & 1, 3, 4, 5, 6, 7, 8, 9, 10, 11, 12, 13, 14, 15, 16, 17, 18, 19, 20, 21, 22, 23, 24, 25, 27, 29, 30, 31, 32, 33 \\
}%
\def\tblbody{\begin{minipage}{\textwidth}
{\footnotesize\begin{tabular}{L{16.9mm}L{91.9mm}L{31.0mm}}
\toprule
\textbf{Chapter} & \textbf{Title} & \textbf{Depends on / refers to} \\
\midrule
\tblrows
\bottomrule\end{tabular}}\end{minipage}}%
\begingroup
\def\sloppy{\tolerance 9999\emergencystretch 3em\hfuzz 200pt\vfuzz 200pt}%
\hbadness=10000\vbadness=10000\hfuzz=200pt\vfuzz=200pt
\global\setbox\tblbox=\hbox{\tblbody}%
\endgroup
\par\addvspace{2.6mm}
\ifdim\dimexpr\ht\tblbox+\dp\tblbox\relax>0.55\textheight
  \typeout{HANDBOOK-TABLE broken \the\dimexpr\ht\tblbox+\dp\tblbox\relax}%
  
  {\footnotesize\begin{longtable}{L{16.9mm}L{91.9mm}L{31.0mm}}
  \toprule
\textbf{Chapter} & \textbf{Title} & \textbf{Depends on / refers to} \\
\midrule\endfirsthead
  \multicolumn{3}{@{}l@{}}{%
  \sffamily\footnotesize\itshape\color{inkgrey}Continued}\\[1.2mm]
  \toprule
\textbf{Chapter} & \textbf{Title} & \textbf{Depends on / refers to} \\
\midrule\endhead
  \bottomrule\endfoot
  \bottomrule\endlastfoot
  \tblrows
  \end{longtable}}%
\else
  \typeout{HANDBOOK-TABLE atomic \the\dimexpr\ht\tblbox+\dp\tblbox\relax}%
  \noindent\tblbody
\fi
\par\addvspace{2.6mm}
\endgroup

\FloatBarrier
\section*{Index}
\markboth{Index}{Index}

\begin{multicols}{2}\footnotesize
\begin{deflist}
\item \textbf{accessibility} \textemdash\ 12.2, 33.3, 34.10, H, L, N  
\item \textbf{agent teams} \textemdash\ 16.6, 22, 29.1, 30, 30.1, 30.7, D, G, L, N  
\item \textbf{artifacts} \textemdash\ 2.1, 2.3, 2.4, 7.6, 8.6, 26.7, 28, 28.5, 28.6, 32.3, 32.5, 34.8  
\item \textbf{auto memory} \textemdash\ 10, 10.2, 10.3, 10.4, 16.1, 32.2, 32.5, 34.3, A, C, D, I  
\item \textbf{auto mode} \textemdash\ 3.2, 6.4, 7.1, 7.5, 15.3, 16.3, 17.1, 18.3, 20.2, 20.3, 20.6, 21.3  
\item \textbf{bypassPermissions} \textemdash\ 6.4, 7.1, 7.5, 7.6, 12.1, 12.4, 13.1, 13.5, 16.3, 16.7, 19.1, 22.2  
\item \textbf{channels} \textemdash\ 2.1, 12.4, 26.7, 27, 27.1, 27.2, 27.3, 31.2, 32.5, 34.10, C, D  
\item \textbf{checkpoint} \textemdash\ 5.3, 7.6, 11.3, 15.3, 24.2, 24.5, A, I, N  
\item \textbf{CLAUDE.md} \textemdash\ 1.3, 2.3, 3.3, 4.4, 4.5, 7, 9, 9.1, 9.3, 9.5, 9.6, 10  
\item \textbf{cleanupPeriodDays} \textemdash\ 10.2, 10.4, 12.3, 30.2, 32.2, 32.5, 33.1, D  
\item \textbf{cloud environment} \textemdash\ 10.2, 26.7, 33.1  
\item \textbf{compaction} \textemdash\ 9.6, 11.1, 11.2, 11.4, 12.1, 15.3, 21.3, 33.2, G  
\item \textbf{context window} \textemdash\ 3.3, 9.5, 16, 22, 30, 30.2, I  
\item \textbf{cron} \textemdash\ 12.1, 23, 23.1, 23.2, 23.3, 24.1, 24.5, D, J  
\item \textbf{deep-research} \textemdash\ 22.5, 22.6, A, M  
\item \textbf{deny rule} \textemdash\ 7.2, 7.6, 12.1, 14.2, 14.3, 15.5, 17.3, 17.5, 18.3, 18.5, 20.2, 20.4  
\item \textbf{dev container} \textemdash\ 7.4  
\item \textbf{effort} \textemdash\ 4.1, 4.2, 4.3, 4.5, 5.3, 8.4, 16.4, 16.7, 22.1, 25.1, 30.3, 31.1  
\item \textbf{fork} \textemdash\ 8.5, 11.3, 11.4, 12.1, 15, 15.3, 16.1, 16.4, 16.5, 16.6, 16.7, 17.1  
\item \textbf{goal} \textemdash\ 11.2, 12.1, 12.4, 17.5, 18.3, 21, 21.1, 21.3, 21.4, 31.2, 33.2, 34.8  
\item \textbf{hooks} \textemdash\ 3.2, 7.2, 7.4, 9.1, 10.2, 10.4, 12.3, 14.1, 14.2, 15, 15.3, 15.5  
\item \textbf{human gate} \textemdash\ 7.6, 13.3, 21.2, F, I  
\item \textbf{managed settings} \textemdash\ 7.2, 7.4, 7.5, 9.1, 12.3, 13.5, 14.2, 15, 15.3, 16.2, 17.5, 18.3  
\item \textbf{MCP} \textemdash\ 1.3, 3.2, 4.4, 7.2, 7.4, 10.3, 11.4, 12, 12.1, 12.4, 13.1, 13.2  
\item \textbf{output styles} \textemdash\ 18.1, 19.1, A, C, N  
\item \textbf{permission mode} \textemdash\ 4.1, 4.5, 6.4, 6.5, 7.1, 7.4, 12.1, 12.4, 13.1, 16.3, 19, 19.1  
\item \textbf{plan mode} \textemdash\ 3.3, 4.5, 6.1, 6.4, 6.5, 7.1, 7.4, 16.3, 19.1, 20.3, 20.6, 22.2  
\item \textbf{plugin} \textemdash\ 1.3, 4.4, 4.5, 8.4, 9.1, 11.4, 12, 12.1, 12.4, 13.1, 13.4, 15  
\item \textbf{prompt caching} \textemdash\ 4.4  
\item \textbf{prompt injection} \textemdash\ 13.3, 20.5, F, I  
\item \textbf{Remote Control} \textemdash\ 1.3, 22.1, 22.3, 25, 25.1, 25.2, 26, 27.3, 29.3, 31.3, 31.5, 32.1  
\item \textbf{routines} \textemdash\ 2.1, 2.3, 2.4, 24, 24.1, 24.4, 24.5, 34.10, A, J, L, N  
\item \textbf{rules} \textemdash\ 3.3, 7, 7.1, 7.2, 7.4, 7.6, 9.1, 9.4, 9.5, 9.6, 10, 10.1  
\item \textbf{sandbox} \textemdash\ 1.3, 2.1, 2.4, 3.2, 3.3, 6.1, 6.4, 7.1, 7.4, 7.6, 12.1, 14.1  
\item \textbf{skills} \textemdash\ 1.3, 3.3, 11.2, 15, 15.3, 15.4, 15.5, 16.1, 16.4, 18.1, 22, 22.6  
\item \textbf{subagent} \textemdash\ 4.4, 7.1, 7.5, 7.6, 9.3, 11.2, 11.3, 12.1, 13.2, 15.3, 16, 16.1  
\item \textbf{teleport} \textemdash\ 2.3, 2.4, 26.2, 26.7, A, I, N  
\item \textbf{Tool Search} \textemdash\ 13.4, G, I  
\item \textbf{transcript} \textemdash\ 1.3, 5.4, 5.5, 10.2, 11.4, 12, 12.1, 12.3, 12.4, 14.1, 14.2, 14.3  
\item \textbf{ultracode} \textemdash\ 4.1, 4.2, 16.5, 16.7, 22.1, 22.2, 22.6, 33.2, A, I, J, N  
\item \textbf{workflow} \textemdash\ 4.3, 7.1, 8.6, 12.1, 12.2, 16.4, 16.6, 18.1, 22, 22.1, 22.2, 22.3  
\item \textbf{worktree} \textemdash\ 3.3, 10.2, 10.4, 12.1, 12.4, 16.4, 17.1, 17.2, 24.5, 25, 26.1, 29.1  
\item \textbf{zero data retention} \textemdash\ 25, 26.3, 28.6, 32, I  
\end{deflist}
\end{multicols}

\breakrule

\FloatBarrier
\backmatter
\frontchapter{References}

The reference list is numbered in order of first citation in the text. Documentation pages were fetched, hashed and adjudicated on 12 November 2026. Other online sources were accessed on 23 August 2026 unless stated otherwise, and all 107 unique URLs in the text were re-checked for reachability on 12 November 2026. Each cited standard, regulation and paper was re-checked against its issuer for identity, edition and currency on 14 November 2026. Product documentation describes behaviour observed on Claude Code 2.1.322 and is subject to change; see “Why this appendix exists” for the re-verification procedure.

\begin{refslist}
\item[1.] Anthropic, “Claude Code documentation index,” \emph{llms.txt}. [Online]. Available: \url{https://code.claude.com/docs/llms.txt}. [accessed 12 Nov. 2026].
\end{refslist}

\begin{refslist}
\item[2.] Anthropic, “Overview,” \emph{Claude Code Documentation}. [Online]. Available: \url{https://code.claude.com/docs/en/overview}. [accessed 12 Nov. 2026].
\end{refslist}

\begin{refslist}
\item[3.] Anthropic, “How Claude Code works,” \emph{Claude Code Documentation}. [Online]. Available: \url{https://code.claude.com/docs/en/how-claude-code-works}. [accessed 12 Nov. 2026].
\end{refslist}

\begin{refslist}
\item[4.] OWASP GenAI Security Project, \emph{OWASP Top 10 for LLM Applications}, 2026 ed., published 3 Aug. 2026. [Online]. Available: \url{https://genai.owasp.org/resource/owasp-genai-llm-top-10-2026/}. [accessed 14 Nov. 2026]. The earlier project page at owasp.org is retained by OWASP as a legacy entry point.
\end{refslist}

\begin{refslist}
\item[5.] MITRE, \emph{ATLAS: Adversarial Threat Landscape for AI Systems}. [Online]. Available: \url{https://atlas.mitre.org/}. [accessed 15 Nov. 2026].
\end{refslist}

\begin{refslist}
\item[6.] National Institute of Standards and Technology, \emph{Artificial Intelligence Risk Management Framework (AI RMF 1.0)}, NIST AI 100-1. Gaithersburg, MD, USA: NIST, Jan. 2023. doi: 10.6028/NIST.AI.100-1.
\end{refslist}

\begin{refslist}
\item[7.] National Institute of Standards and Technology, \emph{Artificial Intelligence Risk Management Framework: Generative Artificial Intelligence Profile}, NIST AI 600-1. Gaithersburg, MD, USA: NIST, Jul. 2024. doi: 10.6028/NIST.AI.600-1.
\end{refslist}

\begin{refslist}
\item[8.] S. Rose, O. Borchert, S. Mitchell and S. Connelly, \emph{Zero Trust Architecture}, NIST Special Publication 800-207. Gaithersburg, MD, USA: NIST, Aug. 2020. doi: 10.6028/NIST.SP.800-207.
\end{refslist}

\begin{refslist}
\item[9.] M. Souppaya, K. Scarfone and D. Dodson, \emph{Secure Software Development Framework (SSDF) Version 1.1: Recommendations for Mitigating the Risk of Software Vulnerabilities}, NIST Special Publication 800-218. Gaithersburg, MD, USA: NIST, Feb. 2022. doi: 10.6028/NIST.SP.800-218.
\end{refslist}

\begin{refslist}
\item[10.] National Institute of Standards and Technology, \emph{The NIST Cybersecurity Framework (CSF) 2.0}, NIST CSWP 29. Gaithersburg, MD, USA: NIST, Feb. 2024. doi: 10.6028/NIST.CSWP.29.
\end{refslist}

\begin{refslist}
\item[11.] International Organization for Standardization, \emph{ISO/IEC 42001:2023 — Information technology — Artificial intelligence — Management system}. Geneva, Switzerland: ISO, 2023.
\end{refslist}

\begin{refslist}
\item[12.] International Organization for Standardization, \emph{ISO/IEC 27001:2022 — Information security, cybersecurity and privacy protection — Information security management systems — Requirements}, 3rd ed. Geneva, Switzerland: ISO, 25 Oct. 2022. Amended by ISO/IEC 27001:2022/Amd 1:2024.
\end{refslist}

\begin{refslist}
\item[13.] International Organization for Standardization, \emph{ISO/IEC 23894:2023 — Information technology — Artificial intelligence — Guidance on risk management}. Geneva, Switzerland: ISO, 2023.
\end{refslist}

\begin{refslist}
\item[14.] International Organization for Standardization, \emph{ISO/IEC 25010:2023 — Systems and software engineering — Systems and software Quality Requirements and Evaluation (SQuaRE) — Product quality model}. Geneva, Switzerland: ISO, 2023.
\end{refslist}

\begin{refslist}
\item[15.] IEEE, \emph{IEEE Standard for System, Software, and Hardware Verification and Validation}, IEEE Std 1012-2024. New York, NY, USA: IEEE, 2025. doi: 10.1109/IEEESTD.2025.11134780.
\end{refslist}

\begin{refslist}
\item[16.] IEEE, \emph{IEEE Standard Model Process for Addressing Ethical Concerns during System Design}, IEEE Std 7000-2021. New York, NY, USA: IEEE, 2021. doi: 10.1109/IEEESTD.2021.9536679.
\end{refslist}

\begin{refslist}
\item[17.] European Parliament and Council of the European Union, “Regulation (EU) 2024/1689 laying down harmonised rules on artificial intelligence (Artificial Intelligence Act),” \emph{Official Journal of the European Union}, L series, 12 Jul. 2024.
\end{refslist}

\begin{refslist}
\item[18.] Australian Signals Directorate, Australian Cyber Security Centre, \emph{Essential Eight Maturity Model}. [Online]. Available: \url{https://www.cyber.gov.au/resources-business-and-government/essential-cyber-security/essential-eight}. [accessed 14 Nov. 2026].
\end{refslist}

\begin{refslist}
\item[19.] Anthropic, “Orchestrate subagents at scale with dynamic workflows,” \emph{Claude Code Documentation}. [Online]. Available: \url{https://code.claude.com/docs/en/workflows}. [accessed 12 Nov. 2026].
\end{refslist}

\begin{refslist}
\item[20.] Anthropic, “Automate work with scheduled tasks,” \emph{Claude Code Documentation}. [Online]. Available: \url{https://code.claude.com/docs/en/routines}. [accessed 12 Nov. 2026].
\end{refslist}

\begin{refslist}
\item[21.] Anthropic, “Use Claude Code with Chrome,” \emph{Claude Code Documentation}. [Online]. Available: \url{https://code.claude.com/docs/en/chrome}. [accessed 12 Nov. 2026].
\end{refslist}

\begin{refslist}
\item[22.] K. Greshake, S. Abdelnabi, S. Mishra, C. Endres, T. Holz and M. Fritz, “Not what you’ve signed up for: Compromising real-world LLM-integrated applications with indirect prompt injection,” in \emph{Proc. 16th ACM Workshop on Artificial Intelligence and Security (AISec ’23)}, Copenhagen, Denmark, Nov. 2023, pp. 79–90. doi: 10.1145/3605764.3623985.
\end{refslist}

\begin{refslist}
\item[23.] F. Perez and I. Ribeiro, “Ignore previous prompt: Attack techniques for language models,” in \emph{NeurIPS ML Safety Workshop}, New Orleans, LA, USA, Dec. 2022. arXiv:2211.09527.
\end{refslist}

\begin{refslist}
\item[24.] GitHub, “Removing sensitive data from a repository,” \emph{GitHub Docs}. [Online]. Available: \url{https://docs.github.com/en/authentication/keeping-your-account-and-data-secure/removing-sensitive-data-from-a-repository}. [accessed 14 Nov. 2026].
\end{refslist}

\begin{refslist}
\item[25.] Anthropic, “How Claude remembers your project,” \emph{Claude Code Documentation}. [Online]. Available: \url{https://code.claude.com/docs/en/memory}. [accessed 12 Nov. 2026].
\end{refslist}

\begin{refslist}
\item[26.] Anthropic, “Run prompts on a schedule,” \emph{Claude Code Documentation}. [Online]. Available: \url{https://code.claude.com/docs/en/scheduled-tasks}. [accessed 12 Nov. 2026].
\end{refslist}

\begin{refslist}
\item[27.] Anthropic, “Create custom subagents,” \emph{Claude Code Documentation}. [Online]. Available: \url{https://code.claude.com/docs/en/sub-agents}. [accessed 12 Nov. 2026].
\end{refslist}

\begin{refslist}
\item[28.] Anthropic, “Configure the sandboxed Bash tool,” \emph{Claude Code Documentation}. [Online]. Available: \url{https://code.claude.com/docs/en/sandboxing}. [accessed 12 Nov. 2026].
\end{refslist}

\begin{refslist}
\item[29.] Anthropic, “Configure permissions,” \emph{Claude Code Documentation}. [Online]. Available: \url{https://code.claude.com/docs/en/permissions}. [accessed 12 Nov. 2026].
\end{refslist}

\begin{refslist}
\item[30.] Anthropic, “Orchestrate teams of Claude Code sessions,” \emph{Claude Code Documentation}. [Online]. Available: \url{https://code.claude.com/docs/en/agent-teams}. [accessed 12 Nov. 2026].
\end{refslist}

\begin{refslist}
\item[31.] Anthropic, “Manage sessions,” \emph{Claude Code Documentation}. [Online]. Available: \url{https://code.claude.com/docs/en/sessions}. [accessed 12 Nov. 2026].
\end{refslist}

\begin{refslist}
\item[32.] Anthropic, “Explore the .claude directory,” \emph{Claude Code Documentation}. [Online]. Available: \url{https://code.claude.com/docs/en/claude-directory}. [accessed 12 Nov. 2026].
\end{refslist}

\begin{refslist}
\item[33.] Anthropic, “Choose a permission mode,” \emph{Claude Code Documentation}. [Online]. Available: \url{https://code.claude.com/docs/en/permission-modes}. [accessed 12 Nov. 2026].
\end{refslist}

\begin{refslist}
\item[34.] Anthropic, “Connect Claude Code to tools via MCP,” \emph{Claude Code Documentation}. [Online]. Available: \url{https://code.claude.com/docs/en/mcp}. [accessed 12 Nov. 2026].
\end{refslist}

\begin{refslist}
\item[35.] Anthropic, “Explore the context window,” \emph{Claude Code Documentation}. [Online]. Available: \url{https://code.claude.com/docs/en/context-window}. [accessed 12 Nov. 2026].
\end{refslist}

\begin{refslist}
\item[36.] Anthropic, “Extend Claude with skills,” \emph{Claude Code Documentation}. [Online]. Available: \url{https://code.claude.com/docs/en/skills}. [accessed 12 Nov. 2026].
\end{refslist}

\begin{refslist}
\item[37.] Anthropic, “Commands,” \emph{Claude Code Documentation}. [Online]. Available: \url{https://code.claude.com/docs/en/commands}. [accessed 12 Nov. 2026].
\end{refslist}

\begin{refslist}
\item[38.] Anthropic, “What’s new,” \emph{Claude Code Documentation}. [Online]. Available: \url{https://code.claude.com/docs/en/whats-new/index}. [accessed 12 Nov. 2026].
\end{refslist}

\begin{refslist}
\item[39.] Anthropic, “Add components to a plugin,” \emph{Claude Code Documentation}. [Online]. Available: \url{https://code.claude.com/docs/en/plugins/components}. [accessed 12 Nov. 2026].
\end{refslist}

\begin{refslist}
\item[40.] Anthropic, “Plugin security and trust,” \emph{Claude Code Documentation}. [Online]. Available: \url{https://code.claude.com/docs/en/plugins/security}. [accessed 12 Nov. 2026].
\end{refslist}

\begin{refslist}
\item[41.] Anthropic, “Continue local sessions from any device with Remote Control,” \emph{Claude Code Documentation}. [Online]. Available: \url{https://code.claude.com/docs/en/remote-control}. [accessed 12 Nov. 2026].
\end{refslist}

\begin{refslist}
\item[42.] Anthropic, “Zero data retention,” \emph{Claude Code Documentation}. [Online]. Available: \url{https://code.claude.com/docs/en/zero-data-retention}. [accessed 12 Nov. 2026].
\end{refslist}

\begin{refslist}
\item[43.] Anthropic, “Data usage,” \emph{Claude Code Documentation}. [Online]. Available: \url{https://code.claude.com/docs/en/data-usage}. [accessed 12 Nov. 2026].
\end{refslist}

\begin{refslist}
\item[44.] SOS Automazioni, \emph{Claude Code: il corso completo per fare tutto, non solo programmare}. [Video]. Published 10 Jul. 2026. Available: https://www.youtube.com/watch?v=w6LDEPJglHw. [accessed 23 Aug. 2026].
\end{refslist}

\begin{refslist}
\item[45.] Anthropic, “Platforms and integrations,” \emph{Claude Code Documentation}. [Online]. Available: \url{https://code.claude.com/docs/en/platforms}. [accessed 12 Nov. 2026].
\end{refslist}

\begin{refslist}
\item[46.] Anthropic, “Let Claude coordinate ongoing work with Projects,” \emph{Claude Code Documentation}. [Online]. Available: \url{https://code.claude.com/docs/en/claude-projects}. [accessed 12 Nov. 2026].
\end{refslist}

\begin{refslist}
\item[47.] Anthropic, “Use Claude Code in VS Code,” \emph{Claude Code Documentation}. [Online]. Available: \url{https://code.claude.com/docs/en/vs-code}. [accessed 12 Nov. 2026].
\end{refslist}

\begin{refslist}
\item[48.] Anthropic, “JetBrains IDEs,” \emph{Claude Code Documentation}. [Online]. Available: \url{https://code.claude.com/docs/en/jetbrains}. [accessed 12 Nov. 2026].
\end{refslist}

\begin{refslist}
\item[49.] Anthropic, “Desktop application,” \emph{Claude Code Documentation}. [Online]. Available: \url{https://code.claude.com/docs/en/desktop}. [accessed 12 Nov. 2026].
\end{refslist}

\begin{refslist}
\item[50.] Anthropic, “Use Claude Code in the cloud,” \emph{Claude Code Documentation}. [Online]. Available: \url{https://code.claude.com/docs/en/claude-code-on-the-web}. [accessed 12 Nov. 2026].
\end{refslist}

\begin{refslist}
\item[51.] Anthropic, “Claude Code on mobile,” \emph{Claude Code Documentation}. [Online]. Available: \url{https://code.claude.com/docs/en/mobile}. [accessed 12 Nov. 2026].
\end{refslist}

\begin{refslist}
\item[52.] Anthropic, “Advanced setup,” \emph{Claude Code Documentation}. [Online]. Available: \url{https://code.claude.com/docs/en/setup}. [accessed 12 Nov. 2026].
\end{refslist}

\begin{refslist}
\item[53.] Anthropic, “Share session output as artifacts,” \emph{Claude Code Documentation}. [Online]. Available: \url{https://code.claude.com/docs/en/artifacts}. [accessed 12 Nov. 2026].
\end{refslist}

\begin{refslist}
\item[54.] Anthropic, “Push events into a running session with channels,” \emph{Claude Code Documentation}. [Online]. Available: \url{https://code.claude.com/docs/en/channels}. [accessed 12 Nov. 2026].
\end{refslist}

\begin{refslist}
\item[55.] Anthropic, “CLI reference,” \emph{Claude Code Documentation}. [Online]. Available: \url{https://code.claude.com/docs/en/cli-reference}. [accessed 12 Nov. 2026].
\end{refslist}

\begin{refslist}
\item[56.] Anthropic, “Debug your configuration,” \emph{Claude Code Documentation}. [Online]. Available: \url{https://code.claude.com/docs/en/debug-your-config}. [accessed 12 Nov. 2026].
\end{refslist}

\begin{refslist}
\item[57.] Anthropic, “Claude Code changelog,” \emph{Claude Code Documentation}. [Online]. Available: \url{https://code.claude.com/docs/en/changelog}. [accessed 12 Nov. 2026].
\end{refslist}

\begin{refslist}
\item[58.] Anthropic, “Authentication,” \emph{Claude Code Documentation}. [Online]. Available: \url{https://code.claude.com/docs/en/authentication}. [accessed 12 Nov. 2026].
\end{refslist}

\begin{refslist}
\item[59.] Anthropic, “Security,” \emph{Claude Code Documentation}. [Online]. Available: \url{https://code.claude.com/docs/en/security}. [accessed 12 Nov. 2026].
\end{refslist}

\begin{refslist}
\item[60.] Anthropic, “Error reference,” \emph{Claude Code Documentation}. [Online]. Available: \url{https://code.claude.com/docs/en/errors}. [accessed 12 Nov. 2026].
\end{refslist}

\begin{refslist}
\item[61.] Anthropic, “All settings,” \emph{Claude Code Documentation}. [Online]. Available: \url{https://code.claude.com/docs/en/settings-reference}. [accessed 12 Nov. 2026].
\end{refslist}

\begin{refslist}
\item[62.] Anthropic, “Set up Claude Code in a monorepo or large codebase,” \emph{Claude Code Documentation}. [Online]. Available: \url{https://code.claude.com/docs/en/large-codebases}. [accessed 12 Nov. 2026].
\end{refslist}

\begin{refslist}
\item[63.] Anthropic, “Model configuration,” \emph{Claude Code Documentation}. [Online]. Available: \url{https://code.claude.com/docs/en/model-config}. [accessed 12 Nov. 2026].
\end{refslist}

\begin{refslist}
\item[64.] Anthropic, “Speed up responses with fast mode,” \emph{Claude Code Documentation}. [Online]. Available: \url{https://code.claude.com/docs/en/fast-mode}. [accessed 12 Nov. 2026].
\end{refslist}

\begin{refslist}
\item[65.] Anthropic, “Escalate hard decisions with the advisor tool,” \emph{Claude Code Documentation}. [Online]. Available: \url{https://code.claude.com/docs/en/advisor}. [accessed 12 Nov. 2026].
\end{refslist}

\begin{refslist}
\item[66.] Anthropic, “Interactive mode,” \emph{Claude Code Documentation}. [Online]. Available: \url{https://code.claude.com/docs/en/interactive-mode}. [accessed 12 Nov. 2026].
\end{refslist}

\begin{refslist}
\item[67.] Anthropic, “Manage costs effectively,” \emph{Claude Code Documentation}. [Online]. Available: \url{https://code.claude.com/docs/en/costs}. [accessed 12 Nov. 2026].
\end{refslist}

\begin{refslist}
\item[68.] Anthropic, “How Claude Code uses prompt caching,” \emph{Claude Code Documentation}. [Online]. Available: \url{https://code.claude.com/docs/en/prompt-caching}. [accessed 12 Nov. 2026].
\end{refslist}

\begin{refslist}
\item[69.] Anthropic, “Voice dictation,” \emph{Claude Code Documentation}. [Online]. Available: \url{https://code.claude.com/docs/en/voice-dictation}. [accessed 12 Nov. 2026].
\end{refslist}

\begin{refslist}
\item[70.] Anthropic, “Settings files and precedence,” \emph{Claude Code Documentation}. [Online]. Available: \url{https://code.claude.com/docs/en/settings}. [accessed 12 Nov. 2026].
\end{refslist}

\begin{refslist}
\item[71.] Anthropic, “Deploy managed settings,” \emph{Claude Code Documentation}. [Online]. Available: \url{https://code.claude.com/docs/en/managed-settings}. [accessed 12 Nov. 2026].
\end{refslist}

\begin{refslist}
\item[72.] Anthropic, “Tools reference,” \emph{Claude Code Documentation}. [Online]. Available: \url{https://code.claude.com/docs/en/tools-reference}. [accessed 12 Nov. 2026].
\end{refslist}

\begin{refslist}
\item[73.] Anthropic, “Choose a sandbox environment,” \emph{Claude Code Documentation}. [Online]. Available: \url{https://code.claude.com/docs/en/sandbox-environments}. [accessed 12 Nov. 2026].
\end{refslist}

\begin{refslist}
\item[74.] Anthropic, “Development containers,” \emph{Claude Code Documentation}. [Online]. Available: \url{https://code.claude.com/docs/en/devcontainer}. [accessed 12 Nov. 2026].
\end{refslist}

\begin{refslist}
\item[75.] Anthropic, “Environment variables,” \emph{Claude Code Documentation}. [Online]. Available: \url{https://code.claude.com/docs/en/env-vars}. [accessed 12 Nov. 2026].
\end{refslist}

\begin{refslist}
\item[76.] Anthropic, “Manage multiple agents with agent view,” \emph{Claude Code Documentation}. [Online]. Available: \url{https://code.claude.com/docs/en/agent-view}. [accessed 12 Nov. 2026].
\end{refslist}

\begin{refslist}
\item[77.] Anthropic, “Checkpointing,” \emph{Claude Code Documentation}. [Online]. Available: \url{https://code.claude.com/docs/en/checkpointing}. [accessed 12 Nov. 2026].
\end{refslist}

\begin{refslist}
\item[78.] Anthropic, “Code Review,” \emph{Claude Code Documentation}. [Online]. Available: \url{https://code.claude.com/docs/en/code-review}. [accessed 12 Nov. 2026].
\end{refslist}

\begin{refslist}
\item[79.] Anthropic, “Catch security issues as Claude writes code,” \emph{Claude Code Documentation}. [Online]. Available: \url{https://code.claude.com/docs/en/security-guidance}. [accessed 12 Nov. 2026].
\end{refslist}

\begin{refslist}
\item[80.] Anthropic, “Scan your codebase for vulnerabilities,” \emph{Claude Code Documentation}. [Online]. Available: \url{https://code.claude.com/docs/en/claude-security}. [accessed 12 Nov. 2026].
\end{refslist}

\begin{refslist}
\item[81.] Anthropic, “Find bugs with ultrareview,” \emph{Claude Code Documentation}. [Online]. Available: \url{https://code.claude.com/docs/en/ultrareview}. [accessed 12 Nov. 2026].
\end{refslist}

\begin{refslist}
\item[82.] Anthropic, “Customize Code Review,” \emph{Claude Code Documentation}. [Online]. Available: \url{https://code.claude.com/docs/en/code-review-customize}. [accessed 12 Nov. 2026].
\end{refslist}

\begin{refslist}
\item[83.] Anthropic, “Hooks reference,” \emph{Claude Code Documentation}. [Online]. Available: \url{https://code.claude.com/docs/en/hooks}. [accessed 12 Nov. 2026].
\end{refslist}

\begin{refslist}
\item[84.] Anthropic, “Keep Claude working toward a goal,” \emph{Claude Code Documentation}. [Online]. Available: \url{https://code.claude.com/docs/en/goal}. [accessed 12 Nov. 2026].
\end{refslist}

\begin{refslist}
\item[85.] Model Context Protocol, “What is the Model Context Protocol (MCP)?” [Online]. Available: \url{https://modelcontextprotocol.io/introduction}. [accessed 23 Aug. 2026].
\end{refslist}

\begin{refslist}
\item[86.] Anthropic, “Control MCP server access for your organization,” \emph{Claude Code Documentation}. [Online]. Available: \url{https://code.claude.com/docs/en/managed-mcp}. [accessed 12 Nov. 2026].
\end{refslist}

\begin{refslist}
\item[87.] Anthropic, “Plugin loading reference,” \emph{Claude Code Documentation}. [Online]. Available: \url{https://code.claude.com/docs/en/plugins/loading}. [accessed 12 Nov. 2026].
\end{refslist}

\begin{refslist}
\item[88.] Anthropic, “Anthropic’s marketplaces,” \emph{Claude Code Documentation}. [Online]. Available: \url{https://code.claude.com/docs/en/plugins/anthropic-marketplaces}. [accessed 12 Nov. 2026].
\end{refslist}

\begin{refslist}
\item[89.] Anthropic, “Mods overview,” \emph{Claude Code Documentation}. [Online]. Available: \url{https://code.claude.com/docs/en/plugins/mods/overview}. [accessed 12 Nov. 2026].
\end{refslist}

\begin{refslist}
\item[90.] Anthropic, “Manage mods for your organization,” \emph{Claude Code Documentation}. [Online]. Available: \url{https://code.claude.com/docs/en/plugins/mods/admin}. [accessed 12 Nov. 2026].
\end{refslist}

\begin{refslist}
\item[91.] Anthropic, “Output styles,” \emph{Claude Code Documentation}. [Online]. Available: \url{https://code.claude.com/docs/en/output-styles}. [accessed 12 Nov. 2026].
\end{refslist}

\begin{refslist}
\item[92.] Anthropic, “Let Claude use your computer from the CLI,” \emph{Claude Code Documentation}. [Online]. Available: \url{https://code.claude.com/docs/en/computer-use}. [accessed 12 Nov. 2026].
\end{refslist}

\begin{refslist}
\item[93.] Anthropic, “Schedule recurring tasks in Claude Code Desktop,” \emph{Claude Code Documentation}. [Online]. Available: \url{https://code.claude.com/docs/en/desktop-scheduled-tasks}. [accessed 12 Nov. 2026].
\end{refslist}

\begin{refslist}
\item[94.] Anthropic, “Configure cloud environments,” \emph{Claude Code Documentation}. [Online]. Available: \url{https://code.claude.com/docs/en/cloud-environments}. [accessed 12 Nov. 2026].
\end{refslist}

\begin{refslist}
\item[95.] Anthropic, “Run Claude Code programmatically,” \emph{Claude Code Documentation}. [Online]. Available: \url{https://code.claude.com/docs/en/headless}. [accessed 12 Nov. 2026].
\end{refslist}

\begin{refslist}
\item[96.] Anthropic, “Enterprise network configuration,” \emph{Claude Code Documentation}. [Online]. Available: \url{https://code.claude.com/docs/en/network-config}. [accessed 12 Nov. 2026].
\end{refslist}

\begin{refslist}
\item[97.] Anthropic, “Self-hosted environments,” \emph{Claude Code Documentation}. [Online]. Available: \url{https://code.claude.com/docs/en/self-hosted-environments}. [accessed 12 Nov. 2026].
\end{refslist}

\begin{refslist}
\item[98.] Anthropic, “Deploy self-hosted environments to production,” \emph{Claude Code Documentation}. [Online]. Available: \url{https://code.claude.com/docs/en/self-hosted-environments-deploy}. [accessed 12 Nov. 2026].
\end{refslist}

\begin{refslist}
\item[99.] Anthropic, “Claude Code in Slack,” \emph{Claude Code Documentation}. [Online]. Available: \url{https://code.claude.com/docs/en/slack}. [accessed 12 Nov. 2026].
\end{refslist}

\begin{refslist}
\item[100.] Anthropic, “Run agents in parallel,” \emph{Claude Code Documentation}. [Online]. Available: \url{https://code.claude.com/docs/en/agents}. [accessed 12 Nov. 2026].
\end{refslist}

\begin{refslist}
\item[101.] Anthropic, “Run parallel sessions with worktrees,” \emph{Claude Code Documentation}. [Online]. Available: \url{https://code.claude.com/docs/en/worktrees}. [accessed 12 Nov. 2026].
\end{refslist}

\begin{refslist}
\item[102.] Anthropic, “Message your other Claude Code sessions,” \emph{Claude Code Documentation}. [Online]. Available: \url{https://code.claude.com/docs/en/cross-session-messaging}. [accessed 12 Nov. 2026].
\end{refslist}

\begin{refslist}
\item[103.] Anthropic, “Claude apps gateway for Amazon Bedrock, Claude Platform on AWS, Google Cloud, and Microsoft Foundry,” \emph{Claude Code Documentation}. [Online]. Available: \url{https://code.claude.com/docs/en/claude-apps-gateway}. [accessed 12 Nov. 2026].
\end{refslist}

\begin{refslist}
\item[104.] Anthropic, “Claude apps gateway configuration,” \emph{Claude Code Documentation}. [Online]. Available: \url{https://code.claude.com/docs/en/claude-apps-gateway-config}. [accessed 12 Nov. 2026].
\end{refslist}

\begin{refslist}
\item[105.] Anthropic, “Configure server-managed settings,” \emph{Claude Code Documentation}. [Online]. Available: \url{https://code.claude.com/docs/en/server-managed-settings}. [accessed 12 Nov. 2026].
\end{refslist}

\begin{refslist}
\item[106.] Office of the Australian Information Commissioner, \emph{Australian Privacy Principles guidelines}, Privacy Act 1988 (Cth). [Online]. Available: \url{https://www.oaic.gov.au/privacy/australian-privacy-principles/australian-privacy-principles-guidelines}. [accessed 14 Nov. 2026].
\end{refslist}

\begin{refslist}
\item[107.] European Parliament and Council of the European Union, “Regulation (EU) 2016/679 (General Data Protection Regulation),” \emph{Official Journal of the European Union}, L 119, pp. 1–88, 4 May 2016.
\end{refslist}

\begin{refslist}
\item[108.] Anthropic, “Monitoring,” \emph{Claude Code Documentation}. [Online]. Available: \url{https://code.claude.com/docs/en/monitoring-usage}. [accessed 12 Nov. 2026].
\end{refslist}

\begin{refslist}
\item[109.] Anthropic, “Track team usage with analytics,” \emph{Claude Code Documentation}. [Online]. Available: \url{https://code.claude.com/docs/en/analytics}. [accessed 12 Nov. 2026].
\end{refslist}

\begin{refslist}
\item[110.] Anthropic, “Use Claude Code with a screen reader,” \emph{Claude Code Documentation}. [Online]. Available: \url{https://code.claude.com/docs/en/accessibility}. [accessed 12 Nov. 2026].
\end{refslist}

\begin{refslist}
\item[111.] Anthropic, “Customize keyboard shortcuts,” \emph{Claude Code Documentation}. [Online]. Available: \url{https://code.claude.com/docs/en/keybindings}. [accessed 12 Nov. 2026].
\end{refslist}

\begin{refslist}
\item[112.] Anthropic, “Customize your status line,” \emph{Claude Code Documentation}. [Online]. Available: \url{https://code.claude.com/docs/en/statusline}. [accessed 12 Nov. 2026].
\end{refslist}

\begin{refslist}
\item[113.] Anthropic, “Configure your terminal for Claude Code,” \emph{Claude Code Documentation}. [Online]. Available: \url{https://code.claude.com/docs/en/terminal-config}. [accessed 12 Nov. 2026].
\end{refslist}

\begin{refslist}
\item[114.] Anthropic, “Configure auto mode,” \emph{Claude Code Documentation}. [Online]. Available: \url{https://code.claude.com/docs/en/auto-mode-config}. [accessed 12 Nov. 2026].
\end{refslist}

\begin{refslist}
\item[115.] Anthropic, “Troubleshoot installation and login,” \emph{Claude Code Documentation}. [Online]. Available: \url{https://code.claude.com/docs/en/troubleshoot-install}. [accessed 12 Nov. 2026].
\end{refslist}

\begin{refslist}
\item[116.] SLSA steering committee, \emph{Supply-chain Levels for Software Artifacts (SLSA)}, v1.2, a project within the Open Source Security Foundation. [Online]. Available: \url{https://slsa.dev/}. [accessed 23 Aug. 2026].
\end{refslist}

\begin{refslist}
\item[117.] National Cyber Security Centre (UK), Cybersecurity and Infrastructure Security Agency (US) and international partners, \emph{Guidelines for Secure AI System Development}, Nov. 2023. [Online]. Available: \url{https://www.ncsc.gov.uk/collection/guidelines-secure-ai-system-development}. [accessed 23 Aug. 2026].
\end{refslist}

\breakrule

\FloatBarrier
\frontchapter{About the author}

\textbf{David Soldani} (Senior Member, IEEE) received the M.Sc. degree (\emph{magna cum laude}) in engineering from the University of Florence, Italy, in 1994, and the D.Sc. degree in technology (Hons.) from the Helsinki University of Technology, Finland, in 2006.

Throughout his career he has held academic positions including Visiting Professor at the University of Surrey, U.K. (2014), Industry Professor at the University of Technology Sydney (UTS), Australia (2016), and Adjunct Professor at the University of New South Wales (UNSW) (2018).

In his professional roles he has served as Chief Information and Security Officer (CISO) and SVP Innovation and Advanced Research at Rakuten; Chief Technology and Cyber Security Officer with Huawei Asia Pacific; Head of 5G Technology at Nokia; and Head of the Central Research Institute and VP Strategic Research and Innovation in Europe at Huawei European Research Center. He is currently SVP Advanced Research and Innovation with Rakuten Mobile Inc., Tokyo, Japan.

Recognised internationally as one of the pioneers of 5G and 6G, he works across multi-disciplinary frontier research and innovation, with expertise spanning 6G, 5G-Advanced, cyber security, cloud computing, eBPF, artificial intelligence and big data. He has received numerous awards for leadership and professional contribution across the organisations he has worked in. He was granted a Distinguished Talent visa by the Australian Government in 2016.

He can be reached at \href{https://www.linkedin.com/in/dr-david-soldani/}{linkedin.com/in/dr-david-soldani}.

\emph{The views expressed in this book are the author's own and do not represent those of any current or former employer. No employer sponsored, reviewed or endorsed this work.}

\end{document}